%% file: ThesisPDF.tex
\documentclass[12pt]{report}

\makeatletter
\def\cleardoublepage{%
  \clearpage
  \if@openright
    \ifodd\c@page\else
      \hbox{}\thispagestyle{empty}\newpage
    \fi
  \fi
}
\makeatother
\input{settings}

\begin{document}

\includepdf[pages=-]{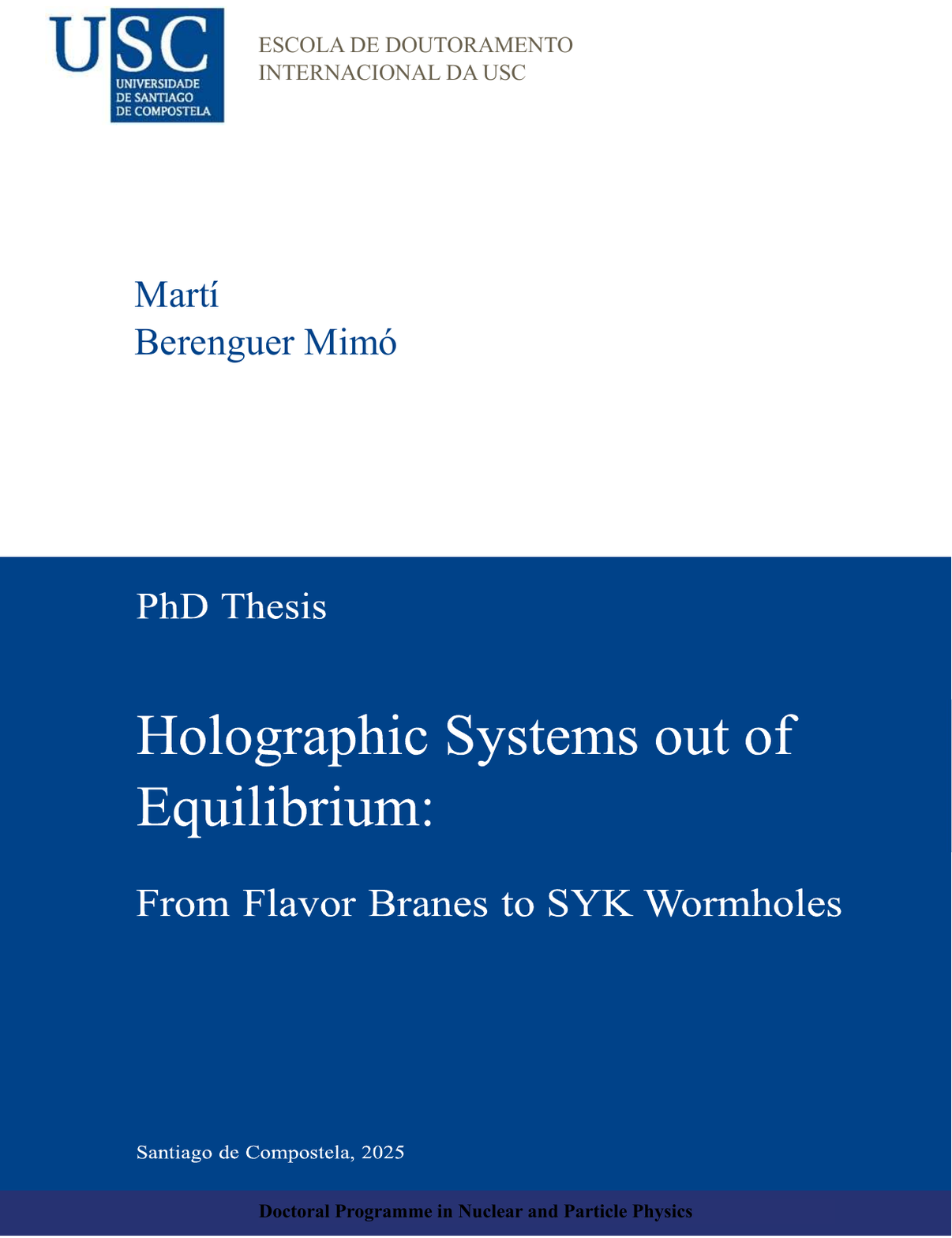}

\pagenumbering{roman}

\includepdf[pages=-]{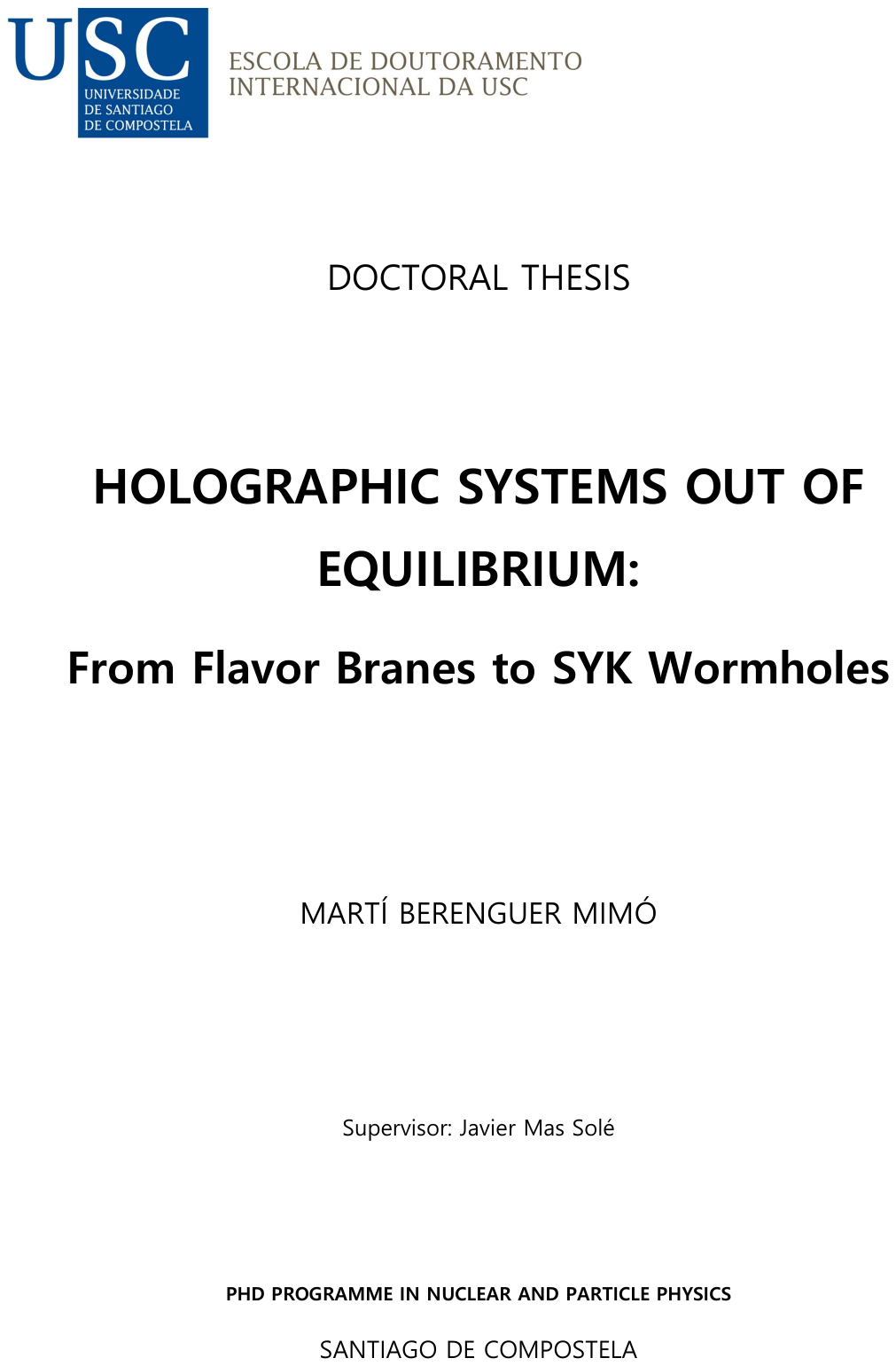}

\thispagestyle{empty} 
\vspace*{\fill}
\begin{flushright}
\textit{A la meva família.}
\end{flushright}
\vspace*{\fill}

\include{Part0/Acknowledgments}

\hypersetup{linkcolor=black}
\tableofcontents

\hypersetup{linkcolor=blue}

\newcounter{partnumber}
\renewcommand{\thepartnumber}{\Roman{partnumber}}

\include{Part0/About}

\phantomsection
\setcounter{partnumber}{\value{partnumber}+1} 
\cftaddtitleline{toc}{part}{\thepartnumber\hspace{1em}Background}{} 
\part*{\centering Part \thepartnumber \\ \vspace{0.5cm} Background}

\include{Part1/AdSCFT/AdSCFT}

\include{Part1/Braneintersections/Braneintersections}
\include{Part1/Braneintersections/BraneintersectionsAppendices}

\include{Part1/SYKwormholes/SYKwormholes}

\include{Part1/SYKwormholes/SYKwormholesAppendices}


\phantomsection
\setcounter{partnumber}{\value{partnumber}+1} 
\cftaddtitleline{toc}{part}{\thepartnumber\hspace{1em}Research and results}{}
\part*{\centering Part \thepartnumber \\ \vspace{0.5cm}Research and results}

\include{Part2/FloquetII/FloquetII}

\include{Part2/FloquetII/FloquetIIAppendices}

\include{Part2/HelicalB/HelicalB}

\include{Part2/HelicalB/HelicalBAppendices}

\include{Part2/FloquetSYK/FloquetSYK}

\include{Part2/FloquetSYK/FloquetSYKAppendices}


\phantomsection
\setcounter{partnumber}{\value{partnumber}+1} 
\cftaddtitleline{toc}{part}{\thepartnumber\hspace{1em}Discussion and conclusions}{}
\part*{\centering Part \thepartnumber \\ \vspace{0.5cm} Discussion and conclusions}

\include{Part3/Conclusions}

\include{Part3/Publications}

\include{Part3/Resumo}

\phantomsection
\addcontentsline{toc}{chapter}{Bibliography}
\bibliographystyle{JHEP}
\bibliography{references.bib}

\includepdf[pages=-]{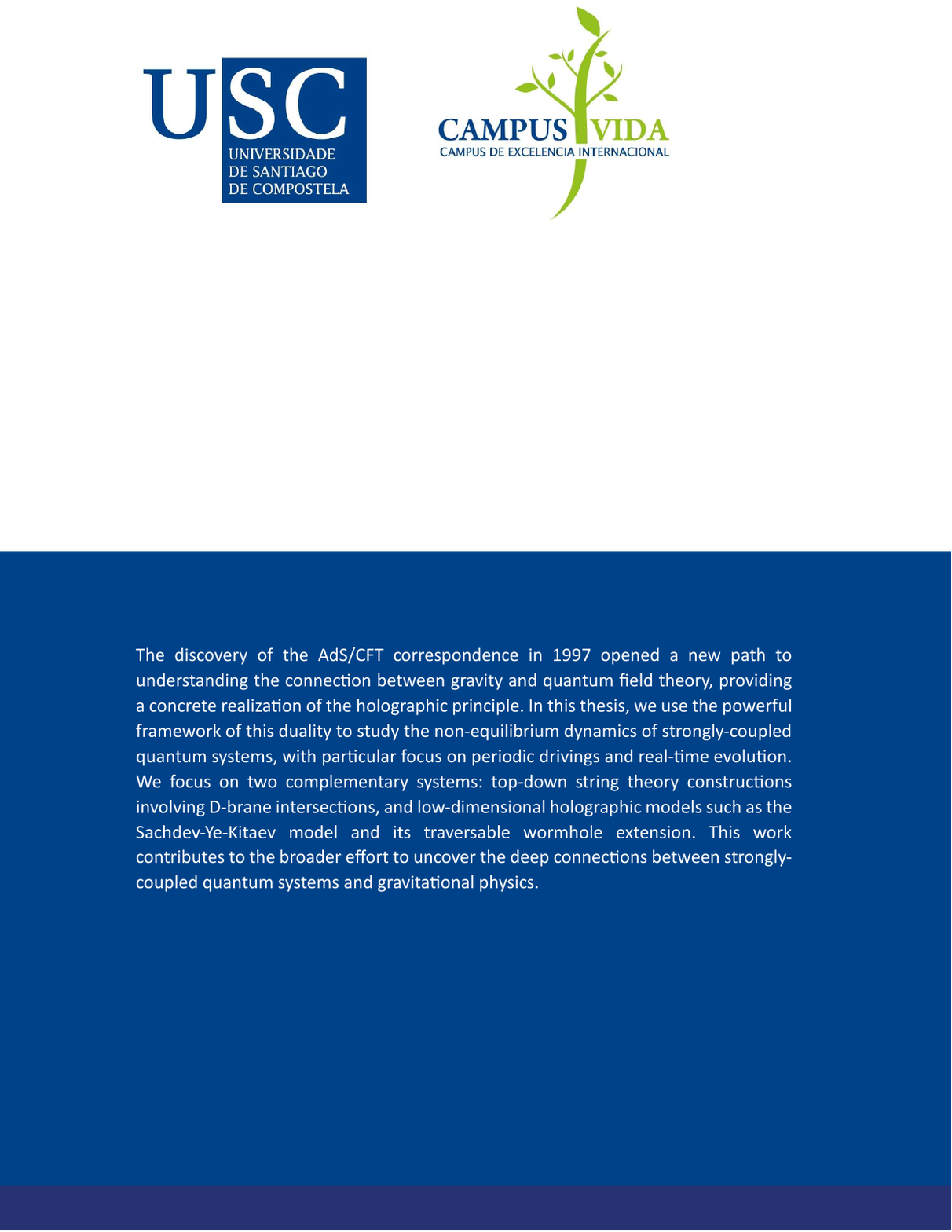}

\end{document}

%% file: settings.tex
\pdfoutput=1

\usepackage[utf8]{inputenc}

\usepackage[a4paper,width=160mm,top=25mm,bottom=35mm]{geometry}     
\usepackage{indentfirst}                        
\usepackage{csquotes}                           
\MakeOuterQuote{"}                              
\usepackage[multiple]{footmisc}                 
\usepackage{relsize}                            
\usepackage[table,svgnames,dvipsnames,xcdraw]{xcolor}  
\usepackage{soul}                       		
\usepackage{comment}                    		
\usepackage{pifont}
\usepackage{fancybox}                           

\usepackage{float}                              
\usepackage[pdftex]{graphicx}                   
\usepackage[small]{caption}                     
\usepackage{subcaption}                         
\usepackage{empheq}                             
\usepackage{wrapfig}                    		
\usepackage{sidecap}                    		
\usepackage{transparent}
\usepackage{pdfpages}

\let\origdoublepage\cleardoublepage
\renewcommand{\cleardoublepage}{\clearpage{\pagestyle{empty}\origdoublepage}}
\usepackage{lipsum}

\usepackage[numbers,sort&compress]{natbib}      

\usepackage{tikz}                               
\usetikzlibrary{decorations.pathreplacing}      
\usetikzlibrary{decorations.pathmorphing}       
\usepackage{pgfplots} \pgfplotsset{compat=1.18}
\usepackage{pgfmath}
\usetikzlibrary{arrows.meta}

\usepackage{tabularx}                           
\usepackage[thinlines]{easytable}               
\usepackage{longtable}                          
\usepackage{array}                      		
\usepackage{makecell}
\usepackage{multirow}
\usepackage{tabularray}

\usepackage{enumitem}                           

\usepackage{latexsym,amsmath,amsfonts,amssymb}  
\usepackage{mathtools}                  		
\usepackage{bbm}                                
\usepackage{braket}                     		
\usepackage{physics}                    		
\usepackage{tensor}                     		
\usepackage{cancel}                     		
\usepackage{slashed}                            

\numberwithin{equation}{chapter}                

\usepackage[toc,page]{appendix}                 

\usepackage{listings}                           

\usepackage{pdfscreen}                          
\usepackage[display]{texpower}                  

\usepackage{tocloft}

\usepackage{nonumonpart}                        

\newcommand{\lss}{\ell_s}                       
\newcommand{\gYM}{g_\text{YM}}
\newcommand{\ie}{\textit{i.e.}}
\newcommand{\h}{\hspace{0.5mm}}
\newcommand{\Sch}{\text{Sch}}                   
\DeclareMathOperator{\sgn}{sgn}

\usepackage[colorlinks=true
  ,linkcolor=blue
  ,urlcolor=blue
  ,anchorcolor=blue
  ,citecolor=blue
  ,filecolor=blue
  ,menucolor=blue
  ,linktocpage=true
  ,pdfproducer=medialab
  ,pdfa=true
  ,linktoc=all
]{hyperref}
\hypersetup{bookmarksnumbered=true}

\newcounter{appendixsection}
\newcommand{\chapterappendix}[1]{%
    \phantomsection
    \chapter*{Appendices for Chapter #1}%
    
    \hypertarget{appendix-ch#1}{}

    \addtocontents{toc}{\protect\vskip 0.5em}
    \addtocontents{toc}{\protect\noindent\protect\hyperlink{appendix-ch#1}{\textbf{\hspace{1.3em}Appendices for Chapter #1}}\protect\hfill\thepage\par}
    \addtocontents{toc}{\protect\vskip 0.3em}

    \setcounter{appendixsection}{0}
    \setcounter{equation}{0}

    \global\let\oldtheequation\theequation 

    \renewcommand{\thesection}{\thechapter.\Alph{appendixsection}}
    \renewcommand{\theequation}{\thechapter.\Alph{appendixsection}.\arabic{equation}}
}

\newcommand{\restoredefaultsectioning}{
    \setcounter{section}{0} 
    \renewcommand{\thesection}{\thechapter.\arabic{section}} 
    \renewcommand{\thesubsection}{\thesection.\arabic{subsection}} 
}

\newcommand{\restoredefaultnumbering}{%
\renewcommand{\thesection}{\thechapter.\arabic{section}} 
\global\let\theequation\oldtheequation 
}

\usepackage{fancyhdr}

\fancypagestyle{myfancy}{
  \fancyhf{}
  \fancyhead[C]{\vspace{5pt}\nouppercase{\rightmark}}
  \fancyfoot[C]{\thepage}
  
  \setlength{\headheight}{20pt}
}

\fancypagestyle{noheader}{
  \fancyhf{}
  \fancyfoot[C]{\thepage}
  
  \setlength{\headheight}{15pt}
}

\usepackage{lmodern} 

\usepackage{etoolbox}
\newcommand{\chapterquote}{} 

\makeatletter

\def\@makechapterhead#1{%
    \vspace*{50pt}%
    \phantomsection%
    \noindent
    \begin{tikzpicture}[baseline={(current bounding box.north)}]
        \node[anchor=west, text=gray!70, font=\bfseries\fontsize{90}{100}\selectfont] at (0,0) {\thechapter};
        \node[anchor=west, font=\Huge\bfseries, text width=\dimexpr\textwidth-3cm, align=left] at (3,0) {#1};
    \end{tikzpicture}%
    \par\nobreak\vskip 20pt
    \noindent\rule{\textwidth}{0.4pt}\par\nobreak\vskip 30pt
    \begingroup
        \edef\tempquote{\unexpanded\expandafter{\chapterquote}}%
        \ifx\tempquote\@empty
        \else
            \begin{flushright}
                {\small \tempquote}
            \end{flushright}
            \vspace{2em}
        \fi
    \endgroup
}

\def\@makeschapterhead#1{%
    \vspace*{50pt}%
    \phantomsection%
    \noindent
    \makebox[0pt][l]{\parbox[t]{\textwidth}{\Huge\bfseries #1}}%
    \par\nobreak\vskip 20pt
    \vskip 30pt
    \begingroup
        \edef\tempquote{\unexpanded\expandafter{\chapterquote}}%
        \ifx\tempquote\@empty
        \else
            \begin{flushright}
                {\small \tempquote}
            \end{flushright}
            \vspace{2em}
        \fi
    \endgroup
}

\makeatother

%% file: Part0/Acknowledgments.tex
\renewcommand{\chapterquote}{\textit{" [...] So do all who live to see such times, but that is not for them to decide.\\
All we have to decide is what to do with the time that is given to us."
}\\[0.5em]
\normalfont— J.R.R. Tolkien, \textit{The Fellowship of the Ring}.}

\chapter*{Acknowledgments}
\markright{Acknowledgments}

This thesis marks the end of a journey that began in September 2020, in unusual times because of the COVID-19 pandemic. I had to start working from home, in Cerdanyola, nearly a thousand kilometers away from Santiago de Compostela, adapting to a new and strange situation. Although it was a difficult beginning, these years have been full of learning, and I am deeply grateful to all those who have supported me along the way.

En primer lugar, quiero dar las gracias a mi supervisor, Javier Mas Solé, por confiar en mí desde el principio, acogiéndome como doctorando sin conocerme personalmente y animándome a empezar a pesar del complicado contexto. Gracias por tu paciencia durante el primer año a distancia, y por tu dedicación y guía constantes a lo largo de estos años.

También quiero agradecer a Alfonso Vázquez Ramallo, que para mí ha sido como un segundo director. Siempre has estado ahí para ayudar, discutir ideas y compartir todo tu conocimiento. Ha sido un privilegio aprender contigo.

Quiero agradecer a toda la gente del IGFAE con la que he tenido el placer de trabajar durante estos años. Primero, a Ana Garbayo, por su ayuda durante el primer año y por ser mi enlace con el resto de estudiantes cuando todo era aún remoto; y más tarde, a Juan Santos, con quien hemos pasado incontables horas intentando hacer funcionar los códigos de SYK. To all the string theory group, thank you. Special thanks to Tim, Eggon, and Kostas, always ready to help, and for carefully reading through this thesis.

También agradezco la labor del personal del instituto, en particular a la \textit{Management Unit} y en especial a Ricardo, Sonia y Manuel, que siempre hacen que todo sea un poco más fácil. Pero, sobre todo, gracias a Elena, con quien he compartido tantas charlas madrugadoras (incluso a las siete de la mañana), cuando el IGFAE todavía estaba en silencio y a oscuras. Empezar el día así era mucho más agradable, y todavía más si el Barça había ganado.

I also want to thank all the wonderful people I’ve met in Santiago during these years. Three of them quickly became close friends after my arrival: Jam, Sibylle, and Niels. Thank you for opening your doors and making me feel at home when everything was still new.

Many others have joined along the way. Some, like Luigi, Leander, Andrew, Aurélie, have already moved on from Santiago, while others, such as Praveen, Vero, Alex, and, especially, Sam, are still here. To all of you, and all the other people I have had the privilege to meet, thank you for the great moments and the friendship that has made this journey unforgettable.

I’m also very grateful to all the people I met during my three-month stay in Southampton. First of all, I want to thank Nick Evans for making the visit possible and for the time we shared, from scientific discussions to walks in the countryside. I’m also thankful to Felix Haehl for the engaging conversations and for inviting me to join one of his ongoing projects. And to all the PhD students, postdocs, and senior researchers I had the chance to meet there: thank you for making my time in Southampton so enjoyable, both inside and outside the department. Orestis, we will keep suffering but we will win the Champions League soon.

També vull donar les gràcies a totes les persones que, d’una manera o altra, han format part de les diferents etapes educatives de la meva vida. Gràcies, Antonio, per despertar el meu interès per la física a les classes de 4t d’ESO, i gràcies també a tots els professors i professores que m’han fet créixer, tant al Col·legi Montserrat com a l’Institut Pere Calders. També vull agrair als professors del Grau de Física de la UAB i del Màster en Astrofísica, Física de Partícules i Cosmologia de la UB.

D’aquest llarg recorregut també em quedo amb moltes amistats que han resistit el pas del temps. Gracias Miriam, Anaid, y Antonio. Gràcies, Carla, per una increïble amistat que va començar abans de començar la carrera, i que ha sabut perdurar al llarg del temps i la distància. I gràcies a les dues bèsties amb qui vaig tenir la sort de compartir el grau de física, el lab, i el “LAB”, en Dani i en Marc. Gràcies per l’amistat i per buscar qualsevol excusa per fer una visita a Santiago. Del máster quiero agradecer especialmente a otras dos bestias: Joseba y Óscar. Que año tan loco. Todavía no sé como sobrevivimos a esas "Fiebre de martes noche" (y de algunos miércoles, aunque muy diferentes...).

Hi ha gent que, tot i haver conegut més recentment, s’han convertit en persones molt importants per a mi. Gràcies, Sergi, per la gran amistat que vam començar fa només uns anys i que no ha deixat de créixer, i gràcies Patry, per aparèixer amb força fa poc més d'un any. No entendria les visites a Cerdanyola (i a Galícia per veure l'hoquei) sense vosaltres.

I, en aquest punt, també vull dedicar un record molt especial al Joan, que ens va deixar l’any passat. Sempre seràs amb nosaltres.

Finalment, vull agrair de tot cor el suport incondicional de la meva família al llarg de totes les etapes d’aquest camí, i especialment durant aquests últims anys viscuts en la distància. Gràcies eternes als meus pares, Lluís i Eulàlia, i als meus germans, Pau i Marc. També a l’àvia Pilar, el tiet Jaume, la Magda, i la resta de tiets i cosins. També als qui malauradament ja no hi són, però han estat sempre molt presents: l’avi Alfons, la iaia Lolita i l’avi Juanito. Aquesta tesi també és per vosaltres.

And finally, I want to thank my partner, Yesi. There are no words in English, Spanish, or Catalan to express how grateful I am. Thank you for always being there, a pesar de la distancia y las dificultades. T'estimo.

\vspace{1.2cm}
\begin{flushright}
    Santiago de Compostela, 2 de juliol de 2025.
\end{flushright}

\vspace{5em}

\begin{flushright}
\begin{minipage}{0.45\textwidth}
\color[gray]{0.6}
\textit{In my homeland, the Vallès,}\\
\textit{three hills are a mountain range,}\\
\textit{four pines, a deep forest,}\\
\textit{five acres, lots of land.}\\
\textit{“There's nowhere like the Vallès”.}
\end{minipage}
\begin{minipage}{0.4\textwidth}
\raggedleft
\textit{En ma terra del Vallès}\\
\textit{tres turons fan una serra,}\\
\textit{quatre pins un bosc espès,}\\
\textit{cinc quarteres massa terra.}\\
\textit{“Com el Vallès no hi ha res”.}
\end{minipage}

\vspace{2em}

\normalfont— Pere Quart (Joan Oliver), \textit{Corrandes d'exili}.
\end{flushright}

%% file: Part0/About.tex
\renewcommand{\chapterquote}{\textit{“Theoretical physics is fun. Most of us indulge in it for the same reason a painter paints or a dancer dances \h\textemdash\h the process itself is so enjoyable! Occasionally, there are additional benefits like fame and glory and even practical uses; but most good theoretical physicists will agree that these are not the primary reasons why they are doing it. The fun in figuring out the solutions to Nature’s brain teasers is a reward in itself.”} \\[0.5em] \normalfont— Thanu Padmanabhan, \textit{Sleeping Beauties in Theoretical Physics}.}
\chapter*{About this thesis}
\markright{About this thesis}
\addcontentsline{toc}{chapter}{About this thesis}
\pagestyle{myfancy}

\section*{Introduction}
\addcontentsline{toc}{section}{Introduction}

String theory was originally proposed in the late 1960s as a potential theory of the strong nuclear force. However, the emergence of quantum chromodynamics (QCD) as a more promising framework for describing hadronic interactions shifted attention away from string theory. Rather than being discarded, string theory evolved into a promising candidate for a consistent theory of quantum gravity, unifying the fundamental forces of Nature. A key milestone in this evolution was the formulation of the AdS/CFT correspondence by Juan Maldacena in 1997 \cite{Maldacena_1997}.

The AdS/CFT correspondence establishes a duality between a gravitational theory in a $(d+1)$-dimensional asymptotically Anti-de Sitter (AdS) spacetime and a conformal field theory (CFT) living on its $d$-dimensional boundary. One of its most powerful features is that it maps strongly coupled quantum field theories, which are typically very difficult to study using conventional methods, to classical or semi-classical gravitational systems, where computations are often more tractable. This duality provides a remarkable computational window into strongly interacting quantum systems.

The most well-understood example of this duality is the correspondence between $\mathcal{N}=4$ supersymmetric Yang-Mills (SYM) theory in four dimensions and type IIB string theory on 
AdS$_5\times S^5$. This duality has played a central role in many theoretical developments over the last three decades, and will be central to the first part of this thesis, along with several of its known deformations.

Beyond its computational power, the AdS/CFT correspondence provides a concrete realization of the holographic principle \cite{tHooft_1993,Susskind_1994}: the idea that a gravitational system can be fully described by degrees of freedom living on its lower-dimensional boundary. Even in situations where the exact dual gravitational description is unknown, holographic intuition has proven to be a useful guiding principle. One notable example is the Sachdev-Ye-Kitaev (SYK) model \cite{KitaevTalk1,KitaevTalk2}, a quantum mechanical system of randomly interacting Majorana fermions in $0+1$ dimensions. In a suitable large-$N$ and low-energy limit, the SYK model is dual to Jackiw-Teitelboim (JT) gravity \cite{Jackiw_1984,Teitelboim_1983}, a dilaton gravity theory in two dimensions. This duality makes the SYK model one of the few known examples of a solvable model of quantum gravity and a powerful theoretical laboratory to probe quantum chaos, entanglement dynamics, and information scrambling in strongly interacting systems. The second part of this thesis will be devoted to the study of the SYK model and a particularly interesting extension that admits a gravitational interpretation in terms of a traversable wormhole \cite{Maldacena_2018}.

A central topic throughout this thesis is the study of non-equilibrium dynamics in strongly coupled quantum systems, particularly under periodic drivings. The physics of periodically driven quantum systems has been the subject of intense study in recent years (see the reviews \cite{Bukov_2014,Eckardt_2016,Weinberg_2016,Holthaus_2015}, and references therein). Several reasons back up this interest, a very relevant one being technological in origin: the possibility of manipulating quantum systems in a controlled way by using time-periodic external fields. This approach goes under the name of Floquet engineering \cite{Oka_2018,Giovannini_2019,Rudner_2020}, following from the Floquet theorem, a temporal analogue of the Bloch theorem\footnote{\textit{Floquet de Neu} (Snowflake) (1964-2003), who remains the only known albino gorilla to date, lived in the Barcelona Zoo and became an iconic figure of Catalan culture. Remembered fondly by a generation, he remains a symbol of the city and a presence in the childhood of many.}. The artificial setup involves mainly irradiating the system with a circularly polarized laser, or shaking it around. With appropriate periodic drivings, new phases of quantum materials have been created and non-equilibrium phenomena have emerged. Examples include light-induced superconductivity \cite{Fausti_2011,Mitrano_2016}, Floquet topological insulators \cite{Oka_2010, Lindner_2010,Kitagawa_2011,Cayssol_2012,Rechtsman_2012,Wang_2013,Jotzu_2014,Dehghani_2015}, and artificial Weyl semimetals \cite{Hubener_2016,Zhang_2016,Bucciantini_2016}.

Of particular interest is the case of solutions where energy injection and dissipation balance, thereby reaching a Floquet type of {\em non-equilibrium steady state} (NESS). The AdS/CFT correspondence provides powerful computational techniques to study these systems from a complementary perspective to the usual one in condensed matter physics. By applying holographic methods to strongly coupled, time-dependent systems, one gains access to non-perturbative regimes that are otherwise difficult to explore. At the same time, the growing interest in complex, out-of-equilibrium phenomena in many-body physics motivates further development of the AdS/CFT correspondence beyond its traditional equilibrium applications. These challenges help test its limits and suggest new directions in which holography can be extended, making it more suitable for capturing real-time dynamics in strongly coupled systems.

\section*{Objectives and methodology}
\addcontentsline{toc}{section}{Objectives and methodology}

The goal of this thesis is to explore the role of periodic drivings in two types of holographic systems. The first type involves top-down constructions based on the D3/D5 and D3/D7 brane intersections, where electric and magnetic fields, usually time-dependent, are introduced. These models have the advantage that the gravitational dual of the gauge theory under consideration is explicitly known within string theory. This makes it possible to investigate the response of the system to external fields in detail (and in some cases analytically), even beyond the regime of linear response theory.

The second type of systems is based on a version of the SYK model involving two coupled copies, known as the Maldacena-Qi model \cite{Maldacena_2018}, which exhibits traversable wormhole behavior at low temperatures. The motivation to study this problem is twofold. On one hand, we are interested in studying how this gravitational phase responds to external perturbations that are periodic in time, with the hope that this may shed light on its quantum gravitational microstructure. On the other hand, we explore whether periodic drivings could contribute to a better understanding, and potentially an improvement, of the wormhole-inspired teleportation protocols that have recently captured the interest of the quantum computing community \cite{Brown_2019, Nezami_2021}.

In summary, this thesis aims to deepen our understanding of periodically driven strongly coupled quantum systems, making use of the AdS/CFT correspondence and holography.

Regarding the methodology, the technical development of the thesis relies on a combination of analytic and numerical approaches. In several cases, we are able to obtain exact results through analytical methods. However, many of the systems considered, particularly those involving time-dependent dynamics, require significant numerical effort. To this end, Wolfram Mathematica has been extensively employed, both for symbolic manipulations and for solving differential equations numerically.

In the part of the thesis devoted to the study of the SYK model, a second software has been used, in parallel to Mathematica. This is the NESSi (Non-Equilibrium Systems Simulation \cite{Schuler_2019}) library, tailored for solving the Schwinger-Dyson equations in non-equilibrium systems. These calculations were carried out using computational resources provided by CESGA (Centro de Supercomputación de Galicia), whose support has been instrumental for handling the numerical workload.

\section*{Outline}
\addcontentsline{toc}{section}{Outline}

The structure of the thesis is divided into three main Parts.

Part I provides the theoretical background on which the thesis is based. Chapter \ref{chap:AdSCFT} reviews the AdS/CFT correspondence as it is commonly formulated, starting from its string theoretic origin. We begin with the role of D-branes in string theory, explore how gauge theories emerge on their worldvolumes, and explain how, in the appropriate decoupling limit, the duality arises.

Chapter \ref{chap:Branes} continues in this line but focuses on how flavor degrees of freedom can be described within the AdS/CFT correspondence. We focus more specifically on the setups that will later be studied in detail: the D3/D5 and D3/D7 brane intersections. We discuss how these configurations behave in the presence of external, time-dependent electric and magnetic fields. Special attention is given to the physical mechanisms at play in these systems, and to the qualitative lessons that will become important in later chapters. 

Chapter \ref{chap:SYKwormholes} turns to a different class of models: the Sachdev-Ye-Kitaev (SYK) model and its connection to traversable wormholes, with a focus on the Maldacena-Qi model. While this change in topic may appear abrupt, its motivation will be developed throughout the bulk of the thesis.

Part II contains the original research. Chapter \ref{chap:FloquetII} presents a study of the D3/D5 intersection at finite temperature subjected to a rotating electric field. We map out the non-equilibrium phase diagram of the system, which includes conductive and insulating phases, and analyze the response currents induced by the driving. In particular, we identify characteristic resonances that go under the name of \textit{vector meson Floquet condensates} and \textit{Floquet suppression points}. We also study several notions of conductivity (both linear and non-linear) and obtain analytical results in specific limits. As we will discuss, the intrinsically non-equilibrium nature of the system introduces conceptual and technical subtleties that end up motivating the shift towards simpler systems, like the SYK model. The precise motivation for this shift is discussed in Section \ref{sec:problemsnoneq}.

However, before delving into the SYK model, we present Chapter \ref{chap:HelicalB}, which follows up on the ideas used in the previous Chapter by considering a similar setup in the D3/D7 intersection, now involving a spatially inhomogeneous magnetic field with helical structure. We analyze how these helical magnetic fields affect chiral symmetry breaking, identifying various types of D7-brane embeddings and showing that the helical component can in fact counteract the symmetry-breaking effect of a uniform magnetic field, leading to symmetry restoration. This Chapter is somewhat distinct from the rest of the thesis, in the sense that it is the only one in which the system is treated in equilibrium, but shares with the previous Chapter the methods and techniques used to solve it.

Finally, Chapter \ref{chap:FloquetSYK} turns to the SYK model and its gravitational interpretation, focusing on the real-time dynamics of the eternal traversable wormhole of Maldacena and Qi. We consider time-dependent deformations of the model, particularly periodic drivings, and study how the wormhole background responds to these perturbations. We find an interesting pattern of resonances that allow to suppress or enhance the traversability of the wormhole geometry. Our results suggest that certain unstable regions of the phase diagram can be dynamically accessed through non-equilibrium processes. Our numerical simulations provide preliminary evidence in this direction, which we use to study the chaotic properties of the unstable phase.

The thesis concludes with Part III, which summarizes the main results and outlines possible directions for future research.

%% file: Part1/AdSCFT/AdSCFT.tex
\pagenumbering{arabic}
\setcounter{page}{1} 

\renewcommand{\chapterquote}{\textit{“For obviously two truths
cannot contradict one another.”} \\[0.5em] \normalfont— Galileo Galilei, \textit{Dialogue}.}
\chapter{Introduction to the AdS/CFT correspondence}\label{chap:AdSCFT}

This Chapter provides a gentle introductory path towards the AdS/CFT correspondence, starting from where it all started: string theory. We begin by reviewing the key aspects of the bosonic string before moving on to superstring theory. In this context, D-branes emerge as fundamental objects, which will play a central role throughout this thesis. We then discuss the decoupling limit and the arguments that led Juan Maldacena, in 1997, to formulate the AdS/CFT correspondence. The Chapter concludes with an overview of how field theory observables can be computed from the dual gravitational description. The material covered here is well established and can be found in standard references on the subject, including \cite{Polchinski_1998v1,Polchinski_1998v2,Argurio_1998,MAGOO_1999,Johnson_2003,Zwiebach_2004,Mateos_2007review,Tong_2008,Polchinski_2010,CasalderreySolana_2011,Ramallo_2013,Hubeny_2014,Ammon_2015,Liu_2018}. References within the main text will generally refer to the original works.

\vspace{1cm}

\section{Bosonic string theory}

The fundamental building blocks of string theory are one-dimensional extended objects known as strings, which can oscillate in directions transverse to their length. Unlike point particles that trace one-dimensional worldlines through spacetime, strings sweep out a two-dimensional surface called the worldsheet.

Strings can be classified into two main types: open and closed. The worldsheet is described by two coordinates, $\tau$ and $\sigma$, corresponding to the proper time and the spatial extent of the string, respectively. For open strings, it is convenient to parametrize them as $\sigma \in [0,\pi]$, while for closed strings, $\sigma \in [0,2\pi]$.

The dynamics of a relativistic string is governed by the Nambu-Goto action, a natural extension of the action for a relativistic point particle. While the action for a point particle measures the proper length of its worldline, the Nambu-Goto action quantifies the area of the two-dimensional worldsheet swept out by the string as it evolves in spacetime. This area is determined by the embedding functions $X^M(\tau,\sigma)$, which describe how the worldsheet is mapped into the target spacetime. The Nambu-Goto action takes the form
\begin{equation}
    S_{\text{NG}}=-\frac{1}{2\pi \alpha'}\int d^2\xi \sqrt{-\det h_{ab}}~,
    \label{eq:SNG}
\end{equation}
where $\xi=(\tau,\sigma)$ are the worldsheet coordinates, and $h_{ab}$ is the induced metric on the worldhseet, given by
\begin{equation}
    h_{ab}=\frac{\partial X^M}{\partial \xi^a}\frac{\partial X^N}{\partial \xi^b}G_{MN}~.
    \label{eq:indmetricworldsheet}
\end{equation}
Here, $G_{MN}$ is the metric of the target spacetime, which we take to be the Minkowski metric, $\eta_{MN}$, for now. In general, capital indices take values $M,N=0,...,D-1$, with $D$ being the total spacetime dimension. The number of spatial dimensions will be denoted by $d=D-1$. The prefactor in the action represents the string tension, which has dimensions of energy per unit length:
\begin{equation}
    T=\frac{1}{2\pi \alpha'}=\frac{1}{2\pi \lss^2}~.
\end{equation}
The parameter $\lss$, known as string length, sets the scale at which stringy effects become significant. Its square, $\alpha'=\lss^2$, is often referred to as the Regge slope parameter. The low-energy limit of string theory corresponds to the regime where $\lss$ is much smaller than any other relevant length scale of the theory.

\subsection{String interactions and genus expansion}

In the 1970s, Gerard 't Hooft analyzed a particular limit of $SU(N_c)$ gauge theories, now known as the 't Hooft limit \cite{tHooft_1973}. This limit consists of taking $N_c \to \infty$ while keeping the \textit{'t Hooft coupling}, defined as $\lambda \equiv \gYM^2 N_c$, fixed, and treat the theory perturbatively in $1/N_c$. He observed that Feynman diagrams in such theories can be conveniently represented using the double-line notation, in which gluons are depicted as a pair of lines, one associated with a quark and the other with an antiquark. This notation makes the organization of diagrams in powers of $N_c$ and $\lambda$ manifest.

For fixed $\lambda$, the expansion of Feynman diagrams is naturally classified by the topology of the surfaces on which they can be drawn without crossings. The contribution of a given diagram scales as $N_c^\chi$, where $\chi$ is the Euler characteristic of the associated Riemann surface,
\begin{equation}
\chi = 2 - 2g - b~,
\end{equation}
with $g$ denoting the genus (the number of handles of the surface) and $b$ the number of boundaries, which arise from external fundamental quarks, represented as single lines. Thus, in the large-$N_c$ limit, the perturbative expansion of the gauge theory is a topological expansion.

The result is an amplitude that can be expressed as a sum of the form
\begin{equation}
    \mathcal{A}=\sum_{g,b=0}^\infty N_c^\chi\sum_{n=0}^\infty c_{g,b,n}\lambda^n~,
\end{equation}
where $c_{g,b,n}$ are numerical coeficients. A crucial feature of this expansion is that, in the large-$N_c$ limit, the dominant contributions come from diagrams with the highest value of $\chi$. Since $\chi$ is maximized for $g=0$ and $b=0$, the leading-order contribution arises from planar diagrams—those that can be drawn on a plane without lines crossing. This regime defines the \textit{planar limit} of the gauge theory.

Remarkably, a structurally identical expansion appears in perturbative string theory. Since the string worldsheet is a two-dimensional surface, string interactions are naturally classified by their genus, with higher-genus contributions corresponding to string loop corrections. Boundaries on the worldsheet correspond to open-string degrees of freedom, closely mirroring the role of fundamental quarks in the gauge theory expansion. This striking similarity suggests the identification
\begin{equation}
g_s \sim \frac{1}{N_c}~,
\end{equation}
where $g_s$ is the string coupling. Under this correspondence, the planar limit of the gauge theory maps to the classical limit of string theory, while non-planar diagrams correspond to quantum corrections in the string description.

From this discussion we learn two important lessons that will play a key role in the formulation of the AdS/CFT correspondence:
\begin{itemize}
    \item The large-$N_c$ expansion of gauge theories corresponds to the loop expansion in string theory.
    \item The planar limit, $N_c \to \infty$, corresponds to classical string theory, where quantum corrections—controlled by powers of $g_s$—are suppressed.
\end{itemize}

\subsection{String quantization and spectrum}

The square root in \eqref{eq:SNG} makes it complicated to quantize the Nambu-Goto action.  A more tractable approach involves introducing an auxiliary field, which eliminates the square root and leads to the Polyakov action. While the equations of motion derived from this action are equivalent to those of the Nambu-Goto action, the Polyakov formulation imposes an additional constraint: the vanishing of the worldsheet energy-momentum tensor. The action takes the form
\begin{align}
    S_\text{P}&=-\frac{1}{4\pi\alpha'}\int d^2\xi \sqrt{-g}g^{ab}\partial_a X^M\partial_b  X^N\eta_{MN}~, \label{eq:SP}\\
    T_{ab}&\equiv-\frac{4\pi\alpha'}{\sqrt{-g}}\frac{\delta S_{\text{P}}}{\delta g^{ab}}=0~,
\end{align}
where $g_{ab}$ is the worldsheet metric, playing the role of the auxiliary field. Its determinant is denoted by $g$. The constraints imposed by the second equation go under the name of \textit{Virasoro constraints}. 

In the conformal gauge, the worldsheet metric is diagonal, simplifying the equations of motion for $X^M(\tau,\sigma)$ derived from \eqref{eq:SP}. These equations reduce to a relativistic wave equation, whose solutions can be expanded in Fourier modes. For open strings, the form of the solutions depends on the boundary conditions imposed at the string endpoints. Denoting the endpoints by $\Bar{\sigma}=(0,\pi)$, two types of boundary conditions are possible:
\begin{itemize}
    \item Neumann boundary conditions: $\partial_\sigma X^M(\tau,\Bar{\sigma})=0$, allowing the endpoint to move freely in the corresponding direction.
    \item Dirichlet boundary conditions: $\delta X^M(\tau,\Bar{\sigma})=0$, which fixes the endpoint at a constant position, $X^M(\tau,\Bar{\sigma})=\Bar{x}_0^M$, constraining the string to end at a specific location in that direction.
\end{itemize}

Neumann and Dirichlet boundary conditions can be imposed independently for each spatial direction and each endpoint of the string. The time direction always satisfies Neumann boundary conditions. When Dirichlet boundary conditions are imposed on $p$ spatial directions, the string endpoints are restricted to a $(p+1)$-dimensional subspace of the full spacetime. These hypersurfaces, to which open-string endpoints are confined, are known as \textit{Dirichlet branes}, \textit{D-branes}, or \textit{D$p$-branes}. These objects are fundamental in string theory and play a crucial role in the AdS/CFT correspondence. The presence of Dirichlet boundary conditions leads to momentum non-conservation at the string endpoints, implying that D$p$-branes must absorb this momentum and thus behave as dynamical objects. We will return to the study of D-branes in more detail later, as they will be central to Chapters \ref{chap:FloquetII} and \ref{chap:HelicalB} of this thesis.

Expanding $X^M(\tau,\sigma)$ in Fourier modes allows to canonically quantize the string, wherein the classical embedding coordinates and their conjugate momenta are promoted to operators satisfying standard commutation relations. A striking consequence of this procedure is that Lorentz invariance at the quantum level requires the spacetime dimension to be $D=26$.

Canonical quantization yields a set of harmonic oscillators, with creation and annihilation operators corresponding to the vibrational modes of the string. These oscillators determine the quantum states of the string, and their energies contribute to the mass of the associated particle states. The resulting spectrum consists of a finite number of massless states along with an infinite tower of massive states, with a mass gap set by $1/\lss$. Consequently, in the low-energy limit $\lss\to 0$, the theory is described only by the massless states of the string.

It is worth pausing to examine the massless states of the bosonic string. In the open-string sector, the massless states correspond to $(D-2)$ states, each carrying a transverse Lorentz index, thereby being photon states. The bosonic closed string spectrum comes with a surprise: the massless states form a rank-two tensor, which can be decomposed into a symmetric traceless tensor, an antisymmetric tensor and a scalar. These correspond to the graviton, the Kalb-Ramond field and the dilaton, respectively. The appearance of the graviton within the string spectrum is a profound result: what began as a quantum theory of relativistic strings has ultimately revealed itself to be a quantum theory of gravity.

However, not everything is satisfactory: the spectrum also contains tachyons, \ie, particles with negative mass-squared, both in the open- and closed- string sector, indicating an inherent instability of the theory. Furthermore, a realistic theory of nature should accommodate fermions, which are absent in the purely bosonic formulation. Both problems are resolved by introducing fermionic degrees of freedom on the worldsheet, extending the Nambu-Goto action to a supersymmetric version. This leads to the development of \textit{superstring theory}.

\section{Superstring theory}

The introduction of fermions in the worldsheet theory is achieved by introducing two new dynamical variables, $\psi_\alpha^M(\tau,\sigma)$, with $\alpha=1,2$. These fermionic fields complement the existing bosonic ones, leading to the total action:
\begin{equation}
    S=-\frac{1}{4\pi \alpha'}\int d^2\xi~ \eta^{\alpha\beta}\left(\partial_\alpha X^M \partial_\beta X^N+i\Bar{\Psi}^M\gamma_\alpha\partial_\beta\Psi^N\right)G_{MN}(X)~,
    \label{eq:actionsuperstring}
\end{equation}
where the two fermionic components are combined into a Majorana spinor $\Psi^M=\left(\psi_1^M,\psi_2^M\right)^T$, and $\gamma^\alpha$ denotes worldsheet Dirac matrices. Unlike the bosonic case, the action contains only first derivatives, leading to different boundary conditions for $\psi_\alpha^M(\tau,\sigma)$. In particular, after fixing $\psi_1^M(\tau,0)=\psi_2^M(\tau,0)$ at one of the endpoints, two distinct choices are possible at the other endpoint:
\begin{equation}
    \psi_1^M(\tau,\pi)=\pm~\psi_2^M(\tau,\pi)~,
\end{equation}
which gives rise to two distinct sectors in the open-superstring theory: the Ramond (R) sector for the upper sign choice and the Neveu-Schwarz (NS) sector for the lower sign choice.

The addition of fermions modifies the dimension of spacetime needed for consistency. In superstring theory, $D=10$. As in the bosonic case, the states of the theory are constructed by acting with creation operators\h\textemdash\h now arising from both the bosonic and fermionic variables of \eqref{eq:actionsuperstring} \h\textemdash\h on the vacuum. The fermionic or bosonic nature of a state is determined by the operator $(-1)^F$, which effectively counts the number of fermionic oscillators acting on the vacuum. States with eigenvalue $(-1)^F=+1$ are bosonic, while those with $(-1)^F=-1$ are fermionic.

Among the infinite tower of excitations, the spectrum in the two sectors contains:
\begin{itemize}
    \item In the NS sector, the ground state is a tachyon with $(-1)^F=-1$. At the massless level, this sector contains 8 bosonic states.
    \item In the R sector, the ground states are massless and consist of 16 states: 8 bosonic and 8 fermionic, with opposite chirality. At each mass level, the R sector exhibits an equal number of bosonic and fermionic states, signaling the emergence of worldsheet supersymmetry.
\end{itemize}

However, the bosonic/fermionic classification of states described above is limited to the worldsheet and does not imply spacetime supersymmetry. Moreover, the presence of a tachyon in the NS sector remains problematic. The solution to this issue was proposed by Gliozzi, Scherk and Olive \cite{Gliozzi_1976}, and is known as the GSO projection. This procedure involves truncating both sectors into four distinct sub-sectors: R$+$, R$-$, NS$+$, NS$-$, where $+$ and $-$ denote the bosonic/fermionic character of the states. For example, the NS$+$ sector contains the massless states, while the tachyon resides in the NS$-$ sector. By disregarding the NS$-$ sector and combining the R and NS$+$ sector leads to a supersymmetric spectrum without tachyons, as desired.

Closed strings are roughly obtained by combining multiplicatively left-moving and right-moving copies of an open-string theory. Consequently, closed superstring theory can be divided into four sectors, based on the sectors chosen for the two copies of the open superstring. These four possible combinations are (NS,NS), (NS,R), (R,NS) and (R,R).

To ensure a consistent supersymmetric theory, the GSO projection constrains the possible combinations of the truncated sectors. Two particularly important truncations lead to Type IIA and Type IIB superstring theory:
\begin{align}
\begin{split}
    \text{Type IIA:} & \quad \text{(NS$+$,NS$+$), (NS$+$,R$+$), (R$-$,NS$+$), (R$-$,R$+$)} \\
    \text{Type IIB:} & \quad \text{(NS$+$,NS$+$), (NS$+$,R$-$), (R$-$,NS$+$), (R$-$,R$-$)}
\end{split}
\end{align}

Spacetime bosons arise from the NS-NS sector and R-R sector, while spacetime fermions originate from the NS-R and R-NS sectors. The (NS$+$,NS$+$) sector is identical in both Type IIA and Type IIB theories, corresponding to the graviton, the dilaton, and the Kalb-Ramond field $B_{MN}$. However, the bosons arising from the R-R sectors differ between the two theories. Specifically, each theory contains fields with an odd or even number of indices, as shown below:
\begin{align}
\begin{split}
    \text{Type IIA:} & \quad C_{(1)}~,~ C_{(3)}~. \\
    \text{Type IIB:} & \quad C_{(0)}~,~C_{(2)}~,~C_{(4)}~.
    \label{eq:RRfields}
\end{split}
\end{align}

The fermions also differ between Type IIA and Type IIB, as the R-sectors in each pairing have the same chirality in Type IIB, but opposite chirality in Type IIA. Other supersymmetric string theories emerge from different truncations, such as the \textit{$E_8\times E_8$ heterotic}, \textit{$SO(32)$ heterotic}, and \textit{Type I} string theories, though these will not be discussed in this thesis.

\section{D-branes and low-energy effective actions}

It was previously noted that imposing Dirichlet boundary conditions on $p$ spatial directions confines the endpoints of open strings to a $(p+1)$-dimensional hypersurface, known as \textit{D$p$-brane}. The quantization of open strings attached to a D$p$-brane follows the same procedure described earlier, with the modification that some directions obey Dirichlet rather than Neumann boundary conditions. This difference is reflected in the labeling of the oscillators in the Fourier expansion of the coordinates, which must be distinguished between the $p$ directions with Dirichlet boundary conditions, and the $D-p$ directions with Neumann boundary conditions.

The resulting states, created by the action of these oscillators, are naturally organized into those associated with directions parallel to the brane and those transverse to it. Aside from the ubiquitous tachyonic states, the massless spectrum in the parallel directions contains $(p+1)-2$ states that transform as a Lorentz vector on the brane, corresponding to photon states; the associated field is a Maxwell field residing on the brane worldvolume. The $d-p$ operators corresponding to the transverse directions give rise to $d-p$ massless states that transform as Lorentz scalars on the brane. These two results are important enough to be highlighted:

\begin{center}
    \Ovalbox{\parbox{0.9\textwidth}{\vspace{5pt} 
    \centering A D$p$-brane has a Maxwell field living on its worldvolume and  \par a massless scalar for each normal direction.
    \vspace{5pt}}}
\end{center}

The $d-p$ massless scalars are interpreted as excitations of the brane, parametrizing its deformations in the transverse directions. The presence of a Maxwell field indicates that, in the low-energy limit, the brane's worldvolume dynamics is governed by an abelian $U(1)$ gauge theory.

Consider, for example, the case of two parallel D$p$-branes separated by a distance $L$. In addition to open strings with both endpoints on the same brane\h\textemdash\h which yield two Maxwell fields, one for each brane\h\textemdash\h there exist open strings stretched between the two branes. These stretched strings give rise to two additional vector fields that are massive, with a mass determined by the string tension times the separation, $M=\frac{L}{2\pi \alpha'}$.

An interesting scenario emerges when the branes are brought very close together. In the limit $L\to 0$, the states originating from strings with endpoints on different branes become massless. In this regime, the effective worldvolume theory for the two branes is a $U(2)$ Yang-Mills theory. More generally, for a stack of $N_c$ D$p$-branes, the gauge group is promoted to $U(N_c)$. From the field theory perspective, the $U(1)$ part of the $U(N_c)$ group always decouples. From the perspective of the D-branes, this $U(1)$ factor can be understood as rigid motions of the brane's center of mass. The resulting interacting theory is always refered to as an $SU(N_c)$ gauge theory in $p+1$ dimensions.

But D$p$-branes are not merely extended objects on which open strings end; they also carry charges under the R-R antisymmetric fields present in the string theory under consideration. Analogous to the way a point particle acquires electric charge by coupling to a Maxwell gauge field $A_M$ through an interaction term of the form $q\int A_M \dot{X}^M d\tau$, a string naturally couples to the Kalb-Ramond field $B_{MN}$ with an interaction term of the form
\begin{equation}
    \int d\tau d\sigma~\partial_\tau X^M\partial_\sigma X^N B_{MN}(X(\tau,\sigma))~.
    \label{eq:couplingstring}
\end{equation}
Thus, in analogy with the point-particle case, one says that the string carries an electric Kalb-Ramond charge. However, due to the string's extended nature, its charge density is not a scalar but rather a vector tangent to the string. This observation prompts a natural question: what is the behavior of the string charge at the endpoints of the string? Moreover, if the string endpoints are attached to a D-brane, is the string charge conserved?

It turns out that the endpoints of the string are charged under the Maxwell field living on the D-brane, and the electric field lines emanating from these endpoints carry string charge. Nevertheless, since the string in the ambient space couples to the Kalb-Ramond field $B_{MN}$, the conservation of string charge and gauge invariance are ensured only if the physically relevant field strength on the brane is the combination
\begin{equation}
    \mathcal{F}_{MN}\equiv B_{MN}+2\pi\alpha'F_{MN}~.
    \label{eq:FplusB}
\end{equation}

Finally, the generalization from point-like particles to strings naturally generalizes to D$p$-branes. In analogy with the coupling given in \eqref{eq:couplingstring}, a D$p$-brane is said to be electrically charged if it couples to a $(p+1)$-dimensional massless antisymmetric tensor field $C_{(p+1)}$. Consequently, in type IIA and type IIB superstring theories, D$p$-branes can acquire charge by coupling to the tensor fields present in each theory, as listed in \eqref{eq:RRfields}. In addition, magnetically charged branes exist, which couple to the Hodge duals of the forms specified in \eqref{eq:RRfields}, although they are not called D-branes because those are not surfaces where strings can end.

Charge and energy conservation imply that a charged object is stable if no lighter particles exist that can carry the same charge. Unlike the D-branes in bosonic string theory, which are unstable by the presence of tachyons, the charged branes in type IIA and type IIB string theory are stable. The stable D$p$-branes in each string theory are summarized as:
\begin{align}
\begin{split}
    \text{Type IIA:} & \quad D0,~D2,~D4,~D6,~D8~\text{branes}. \\
    \text{Type IIB:} & \quad D(-1),~D1,~D3,~D5,~D7~\text{branes}.
    \label{eq:branesIIAIIB}
\end{split}
\end{align}

\subsection{Low-energy effective actions}

In the low-energy limit, $\lss\to 0$, the massive states become infinitely massive and only the massless states remain. It is useful to write a low-energy effective action for these modes. Several reasonings lead to the same answer: for instance, for type IIA string theory we should ask what action we can write down to describe a supersymmetric theory of gravity in 10 dimensions with two supercharges of opposite chirality. It turns out this theory is completely fixed and it corresponds to type IIA supergravity. Another option would be to use string theory amplitudes to compute the interactions between the massless fields and see what action reproduces them. A third option is to write the worldsheet action \eqref{eq:actionsuperstring} in a curved space background and demand that conformal invariance is not broken at the quantum level. It turns out the three procedures are consistent among them.

The procedure is analogous for type IIB string theory, now asking for a supersymmetric theory with two supercharges of the same chirality. A minor difference is that the self-duality condition of the 5-form has to be imposed by hand, leading to the action of type IIB supergravity. The bosonic part of the action is given by\footnote{This is the action Murph is shown writing on the board during the moment of climax in the movie \textit{Interstellar}.}
\begin{align}
\begin{split}
    S_{\text{IIB}}=\frac{1}{2\tilde{\kappa}_{10}^2}\Bigg[\int d^{10}X\sqrt{-G}\Big[e^{-2\Phi}&\left(\mathcal{R}+4\partial_M\Phi\partial^M\Phi-\frac{1}{2}|H_{(3)}|^2\right)-\frac{1}{2}\abs{F_{(1)}}^2 \Big. \Bigg. \\
    & \Bigg. \Big.-\frac{1}{2}|\Tilde{F}_{(3)}|^2 -\frac{1}{4}|\Tilde{F}_{(5)}|^2\Big]-\frac{1}{2}C_{(4)}\wedge H_{(3)}\wedge F_{(3)}\Bigg]~,
\end{split}
\end{align}
with the supplementary constraint $^*\Tilde{F}_{(5)}=\Tilde{F}_{(5)}$. $\Tilde{\kappa}_{10}$ is the ten-dimensional gravitational constant, $2\Tilde{\kappa}_{10}=(2\pi)^7\alpha'^4$. The field strenght tensors are given by
\begin{align}
\begin{split}
    F_{(p)}&=d C_{(p-1)}~,\quad H_{(3)}  = dB_{(2)}~,\quad \Tilde{F}_{(3)}  = F_{(3)}-C_{(0)}H_{(3)}~,\\
    \Tilde{F}_{(5)}&=F_{(5)}-\frac{1}{2}C_{(2)}\wedge H_{(3)}+\frac{1}{2}B_{(2)}\wedge F_{(3)}~.
\end{split}
\end{align}

The low-energy effective action governing the dynamics of open strings attached to D$p$-branes can be obtained from similar arguments. By considering the dynamics of the $U(1)$ gauge field that lives on the worldvolume of the brane, it is seen that in order to retain conformal invariance, the condition that the field strength $F_{ab}$ has to satisfy corresponds precisely to the equation of motion of a well-known theory that was built as a non-linear generalization of Maxwell theory: \textit{Born-Infeld theory}. This leads to an action given by
\begin{equation}
    S_{\text{BI}}=-T_{Dp}\int d^{p+1}\xi\sqrt{-\det\left(\eta_{ab}+2\pi\alpha' F_{ab}\right)}~.
    \label{eq:BIaction}
\end{equation}
Here, the coordinates $\xi$ represent the $(p+1)$-dimensional worldvolume of the brane, and $T_{Dp}$ represents the tension of the brane, given by
\begin{equation}
    T_{Dp}=\frac{1}{(2\pi)^p g_s \alpha'^{\frac{p+1}{2}}}~.
    \label{eq:tensionDp}
\end{equation}
The Born-Infeld action, however, only describes the dynamics of the gauge field. We should also include the dynamics of the $d-p$ scalar fields that parametrize the fluctuations of the brane in the transverse directions. This should be the higher-dimensional analogous of the Nambu-Goto action. By inspection, the obvious candidate should be replacing the Minkowski metric by the induced metric on the worldvolume of the brane, $\eta_{ab}\to h_{ab}$, with $h_{ab}$ given in \eqref{eq:indmetricworldsheet}. The resulting action is the Dirac-Born-Infeld (DBI) action:
\begin{equation}
    S_{\text{DBI}}=-T_{Dp}\int d^{p+1}\xi\sqrt{-\det\left(h_{ab}+2\pi\alpha' F_{ab}\right)}
\end{equation}
The DBI action describes the dynamics of a D$p$-brane in flat space. The remaining step is to consider the situation in which the brane is located on the spacetime sourced by the closed string fields $G_{\mu\nu}$, $B_{\mu\nu}$ and $\phi$. The action is modified into
\begin{equation}
    S_{\text{DBI}}=-T_{Dp}\int d^{p+1}\xi~e^{-\tilde{\phi}} \sqrt{-\det\left(h_{ab}+ B_{ab}+2\pi \alpha'F_{ab}\right)}~.
    \label{eq:SDBI}
\end{equation}
The coupling to the non-trivial spacetime metric $G_{MN}$ is implicitly hidden in the induced metric $h_{ab}$. The requirement of gauge invariance explained in \eqref{eq:FplusB} makes it obvious that the Kalb-Ramond field has to appear in this way in the action. Finally, for the coupling to the dilaton it is customary to split its varying part $\phi$ from its asymptotic value at infinity, $\phi_0$, as $\phi=\phi_0+\tilde{\phi}$. The string coupling is determined by the asymptotic value of the dilaton, which is then absorbed into the brane tension, as in \eqref{eq:tensionDp}. This fact makes evident that D-branes are non-perturbative objects of string theory.

The full dynamics contains also the interactions of the $B$ field with the R-R forms $C_{(q+1)}$. This is given by the Wess-Zumino terms,
\begin{equation}
    S_{\text{WZ}}=T_{Dp}\int\sum_q P\left[C_{(q+1)}\right]\wedge e^{P[B]+2\pi\alpha' F}~,
    \label{eq:SWZ}
\end{equation}
where $P[...]$ denotes the pullback of a bulk field on the worldvolume of the brane. The total action is therefore the sum of the DBI plus the WZ terms,
\begin{equation}
    S_{Dp}=S_{\text{DBI}}+S_{\text{WZ}}~.
\end{equation}

We argued previously that when $N_c$ D$p$-branes are placed on top of each other, a $SU(N_c)$ gauge theory lives in the $(p+1)$-dimensional worldvolume of the brane. We are now in the position to know which particular gauge theory it is.

Considering for simplicity the case of a constant dilaton and vanishing Kalb-Ramond field, the low-energy action of a single D$p$-brane can be expanded and it correctly reduces to Maxwell electrodynamics, as can be seen by expanding the action \eqref{eq:BIaction} for sufficiently small $F_{\mu\nu}$ and its derivatives. The leading term in $\alpha'$ is given by
\begin{equation}
    S_{Dp}=-T_{Dp}(2\pi\alpha')^2\int d^{p+1}\xi\left[\frac{1}{4}F_{ab}F^{ab}+\frac{1}{2}\partial_a\phi^I\partial^a\phi^I+...\right]~,
    \label{eq:DBIexpand}
\end{equation}
with $\phi^I$ corresponding to the scalar fields in the tranverse directions to the brane. By comparing it with the usual form of the Yang-Mills action,
\begin{equation}
    S_{\text{YM}}=-\frac{1}{4g_{\text{YM}}^2}\int d^{p+1}\xi F_{ab}F^{ab}~,
    \label{eq:SYMabelian}
\end{equation}
we can obtain the relation between $g_{\text{YM}}$ and the string coupling $g_s$. A similar argument holds in the non-abelian case, arriving to the relation\footnote{Direct comparison between \eqref{eq:DBIexpand} and \eqref{eq:SYMabelian} leads to $g_{\text{YM}}^2=T_{Dp}^{-1}(2\pi\alpha')^{-2}$, without the factor of $2$ of \eqref{eq:gYMgs}. For a D$3$-brane, this gives $g_{\text{YM}}^2=2\pi g_s$. In the non-abelian case, there is an ambiguity in the normalization of the generators of the gauge group: $\Tr(T_aT_b)=d\h\delta_{ab}$. In D-brane physics $d=1$, leading again to $g_{\text{YM}}^2=2\pi g_s$. The conventional choice is $d=1/2$, which leads to $g_{\text{YM}}^2=4\pi g_s$. We adopt the latter, and hence \eqref{eq:gYMgs} acquires the extra factor of $2$ \cite{OBannon_2008}.}
\begin{equation}
    g_{\text{YM}}^2=2\h T_{Dp}^{-1}(2\pi\alpha')^{-2}=2(2\pi)^{p-2}\alpha'^{(p-3)/2}g_s~.
    \label{eq:gYMgs}
\end{equation}

In the case of a stack of $N_c$ coincident D$p$-branes, we argued earlier that at low energies the dynamics is governed by a non-abelian $SU(N_c)$ gauge theory. In this case, the leading term is given by
\begin{equation}
    S_{Dp}=-(2\pi\alpha')^2T_{Dp}\int d^{p+1}\xi \Tr\left(\frac{1}{4}F_{\mu\nu}F^{\mu\nu}+\frac{1}{2}D_\mu\phi^ID^\mu\phi^I-\frac{1}{4}\sum_{I\neq J}\left[\phi^I,\phi^J\right]^2\right)~,
    \label{eq:DBIexpansionnonabelian}
\end{equation} 
where the partial derivatives have been promoted to the standard covariant derivatives and the trace is over the gauge group indices.

Of particular interest is the case of $N_c$ D3-branes. In this case the worldvolume is (3+1)-dimensional and so is the gauge theory.
The same action can be obtained by counting supercharges: the introduction of D3 branes breaks half of the original supersymmetries of type IIB string theory, going from 32 supercharges to 16. This corresponds to the maximum number of supercharges in four dimensions. The theory with these properties is uniquely determined to be $\mathcal{N}=4$ SYM in four dimensions, a conformal field theory whose field content is a gauge field $A_\mu(x)$ ($\mu=0,...,3$), four Weyl fermions $\lambda_\alpha^a(x)$ ($a=1,...,4$ and $\alpha=1,2$) and six real scalars $\Phi^i(x)$ ($i=1,...,6$). They transform in the singlet, fundamental, and antisymmetric representation of $SU(4)_R$, respectively. Due to supersymmetry, all the fields transform in the adjoint representation of $SU(N_c)$. The bosonic part of the $\mathcal{N}=4$ SYM Lagrangian is precisely given by the action \eqref{eq:DBIexpansionnonabelian}.

\section{A tale of two perspectives: the AdS/CFT correspondence} \label{sec:AdSCFT}

\subsection{The decoupling limit}

Until now, we have only seen D-branes as hypersurfaces in ten-dimensional Minkowski space where open strings can end. However, an alternative picture of D$p$-branes exists. $p$-branes were originally conceived as classical solutions of the supergravity equations of motion. Concretely, they correspond to charged objects under the different R-R fields of the theory. The discovery that the D$p$-branes of string theory are charged under the same R-R fields made clear that the two objects are actually the same \cite{Polchinski_1995}. This alternative picture is fundamental in the formulation of the AdS/CFT correspondence. 

The supergravity solution corresponding to a stack of $N_c$ coincident D3-branes extended along the $(t,x_1,x_2,x_3)$ directions is given by
\begin{equation}
\begin{aligned}
    ds^2&=H^{-1/2}\left(-dt^2+dx_1^2+dx_2^2+dx_3^2\right)+H^{1/2}\left(dr^2+r^2d\Omega_5^2\right)~,\\
    C_{(4)}&=\left(H^{-1}-1\right)dt\wedge dx_1\wedge...\wedge dx_3~,\\
    B_{(2)}&=0~,
    \label{eq:metricD3branesIIB}
\end{aligned}
\end{equation}
where the function $H(r)$ is given by
\begin{equation}
    H(r)=1+\frac{L^4}{r^4}~,~~~~~L^4=4\pi g_sN_c\lss^4~
    \label{eq:H(r)}
\end{equation}
and $r^2=x_4^2+...+x_9^2$ represents the 6 directions transverse to the D3-branes. In this solution, the dilaton is a constant and determines the string coupling as $g_s=e^{\phi}$. This solution is a BPS solution of the type IIB supergravity equations of motion. The addition of D3-branes breaks the $SO(9,1)$ rotation symmetry of Minkowski spacetime into $SO(9,1)\to SO(3,1)\times SO(6)$. The first factor corresponds to Lorentz transformations along the worldvolume of the branes $(t,x_1,x_2,x_3)$, while the second factor corresponds to rotations in the remaining six transverse directions, $x_4,...,x_9$. This geometry has two very interesting limits. Far away from the D3-branes, at $r\gg L$, we can approximate $H(r)\simeq 1$ and the metric becomes the 10-dimensional flat Minkowski metric (with small corrections). In the opposite limit, $r\ll L$, we can neglect the "$1$" in \eqref{eq:H(r)} and the metric \eqref{eq:metricD3branesIIB} reduces to
\begin{equation}
    ds^2=\frac{r^2}{L^2}\left(-dt^2+dx_1^2+dx_2^2+dx_3^2\right)+\frac{L^2}{r^2}dr^2+L^2d\Omega_5^2~,
    \label{eq:AdS5xS5}
\end{equation}
where we identify an AdS$_5$ factor and a 5-sphere, with $L$ the curvature radius of AdS. Therefore, very close to the D3-branes, the geometry factorizes into AdS$_5\times S^5$.

We have established that D-branes admit two distinct descriptions. On one hand, they can be viewed as hypersurfaces in ten-dimensional Minkowski space where open strings can end\h\textemdash\h a viewpoint we shall refer to as the \textit{open-string perspective}. On the other hand, they also appear as solutions to the supergravity equations, with a geometry that is asymptotically flat at large distances but develops an AdS$_5\times S^5$ throat near the branes. We will refer to this as the \textit{closed-string perspective}. The choice of perspective depends on the value of the string coupling, $g_s$.

Suppose the case in which $g_sN_c=0$, and consider a stack of $N_c$ D3-branes in flat ten-dimensional spacetime. We take $N_c$ to be large but fixed. In this regime, the system consists of free open strings ending on the stack of D3-branes and free closed strings propagating in the ten-dimensional background, with no interactions. Now, if we introduce a small but non-zero coupling, $g_sN_c\ll 1$, the gravitational effects of the D3-branes remain negligible. This follows from \eqref{eq:H(r)}: since $g_sN_c\ll 1$ implies $L^4/\lss^4\ll 1$, it follows that at any distance greater than or comparable to the string length, we have $H(r)\simeq 1$ and the metric \eqref{eq:metricD3branesIIB} remains nearly flat. Therefore, the open-string perspective provides a valid and useful description in the regime $g_sN_c\ll 1$\footnote{For the argument above to work, it is essential that the tension of the brane scales as $g_s^{-1}$, as opposed to the $g_s^{-2}$ scaling that is typical in field theory solitons.}.

Now, consider the opposite limit, where $g_sN_c\gg 1$. In this regime, the backreaction of the branes becomes significant and must be taken into account. As a result, the branes are no longer treated as external objects but are instead replaced by the geometry \eqref{eq:metricD3branesIIB} that they generate. In this description, open strings are no longer present; the system consists solely of closed strings propagating in the curved background. Consequently, when $g_sN_c\gg 1$, the closed-string perspective becomes the most appropriate.

At this stage, we do not yet have a duality; we have merely described Type IIB string theory in two different limits, $g_sN_c\ll 1$ and $g_s N_c\gg 1$. The AdS/CFT correspondence emerges when we take the low-energy limit in both descriptions and compare the resulting theories. This procedure, known as the \textit{decoupling limit}, is what ultimately leads to the duality.

We now take the low-energy limit in the open-string perspective. As usual, in this limit, only the massless states survive for both open and closed strings, since we consider energies much smaller than the string energy scale $1/\lss$,
\begin{equation}
    E\ll \frac{1}{\lss}=\frac{1}{\sqrt{\alpha'}}~.
    \label{eq:lowEalpha}
\end{equation}
Equivalently, the low-energy limit can be understood as keeping energy finite while taking $\alpha'\to 0$. From the previous section, we know that the massless modes of the open strings give rise to $\mathcal{N}=4$ $SU(N)$ SYM theory in (3+1)-dimensional flat spacetime (the worldvolume of the branes), where, from \eqref{eq:gYMgs} for $p=3$ we identify $g_{\text{YM}}^2=4\pi g_s$. In the closed-string sector we have closed strings propagating in flat spacetime.

Next, let us examine the interactions between these modes. The interactions among closed strings are governed by the action of ten-dimensional supergravity, plus higher-derivative corrections. Since these corrections are suppressed by positive powers of $\alpha'$, they vanish in the $\alpha'\to 0$ limit, leaving behind free supergravity in ten-dimensional Minkowski space. The same reasoning applies to interactions between closed strings in the bulk with open strings on the brane: these interactions are also suppressed and disappear in the low-energy limit. As a result, open and closed strings decouple.

In contrast, the open strings on the brane remain interacting, the reason being that $g_{\text{YM}}$ is dimensionless in (3+1)-dimensions. The final result is a decoupled system consisting of free closed strings propagating in ten-dimensional flat spacetime and an interacting $\mathcal{N}=4$ $SU(N_c)$ SYM theory in (3+1)-dimensional flat spacetime.

We now analyze the low-energy limit from the closed-string perspective. In this description, there are no open strings, but only closed strings propagating in the ten-dimensional background \eqref{eq:metricD3branesIIB}. The low-energy limit corresponds to focusing on excitations with arbitrarily low energy as measured by a Minkowski observer at infinity. These excitations arise from two distinct contributions.

First, there are closed strings propagating in the asymptotic region. In the low-energy limit, only their massless modes survive, and their interactions vanish for the same reason as in the open-string perspective: interaction terms contain positive powers of $\alpha'$, which disappear as $\alpha'\to 0$.

The second contribution comes from finite-energy excitations in the throat region. Due to the infinite redshift they experience while climbing the gravitational potential, these excitations appear as low-energy modes to an observer at infinity. In fact, states with arbitrarily high proper energy are perceived as low-energy excitations if they remain sufficiently deep down the throat. As a result, the full tower of massive string states survives in this limit. 

There is evidence that these two sets of modes decouple. Massless, long-wavelength modes in the asymptotic region are unable to enter the throat, as their wavelengths far exceed its characteristic size. Conversely, modes localized deeper in the throat encounter an increasingly strong gravitational barrier, making it extremely difficult for them to escape to infinity.

Therefore, in the low-energy limit, the system once again decouples into two sectors: free closed strings propagating in ten-dimensional flat spacetime, and interacting type IIB closed strings in the throat region, given by AdS$_5\times S^5$.

Bringing everything together, we have seen that, in the low-energy limit, the two descriptions of the same system lead to the following decoupled effective theories:
\begin{itemize}
    \item Open-string perspective: Type IIB supergravity on $\mathbb{R}^{9,1}$ + $\mathcal{N}=4$ SYM theory on flat 4-dimensional spacetime.
    \item Closed-string perspective: Type IIB supergravity on $\mathbb{R}^{9,1}$ + Type IIB superstring theory on AdS$_5\times S^5$.
\end{itemize}
Type IIB supergravity on $\mathbb{R}^{9,1}$ is present in the two perspectives. Since the two descriptions should be equivalent, it is natural to conjecture that the remaining theories should be equivalent $(\Longleftrightarrow)$:

\begin{center}
    \Ovalbox{
    \centering
    \begin{tabular}{ccc}
        \parbox{0.4\textwidth}{
        \vspace{4pt}
        \centering $\mathcal{N}=4$ SYM theory \\ in  flat 4-dimensional spacetime
        \vspace{5pt}} & $\Longleftrightarrow$ & \parbox{0.4\textwidth}{
        \vspace{5pt}
        \centering Type IIB superstring theory\\ in AdS$_5\times S^5$.
        \vspace{5pt}}
    \end{tabular}
    }
\end{center}

This is, in summary, the statement of the AdS/CFT correspondence as proposed by Juan Maldacena in 1997 \cite{Maldacena_1997}. More detailed versions of the correspondence with precise relations between the parameters in the two sides will be provided in the following.

In the preceding discussion, the $\mathcal{N}=4$ SYM theory was considered at zero temperature. The corresponding supergravity solution, given by Eq.\eqref{eq:metricD3branesIIB}, is extremal (or BPS). However, supergravity also admits non-BPS solutions, among which the non-extremal black-brane solution is particularly relevant. The metric in this case takes the form
\begin{equation}
    ds^2=H^{-1/2}\left(-f(r)dt^2+dx_1^2+dx_2^2+dx_3^2\right)+H^{1/2}\left(\frac{dr^2}{f(r)}+r^2d\Omega_5^2\right)~,
    \label{eq:metricD3blackbranesIIB}
\end{equation}
where $H(r)$ and $L$ remain as in Eq.\eqref{eq:H(r)}, while $f(r)$ is the blackening factor, given by
\begin{equation}
    f(r)=1-\frac{r_h^4}{r^4}~.
    \label{eq:f(r)}
\end{equation}
The metric \eqref{eq:metricD3blackbranesIIB} describes a black brane with a horizon at $r = r_h$. In the near-horizon limit, the geometry reduces to
\begin{equation}
    ds^2=\frac{r^2}{L^2}\left(-f(r)dt^2+dx_1^2+dx_2^2+dx_3^2\right)+\frac{L^2}{r^2f(r)}dr^2+L^2d\Omega_5^2~.
    \label{eq:SchwAdS5xS5}
\end{equation}
The Hawking temperature associated with this horizon is
\begin{equation}
    T_H=\frac{r_h}{\pi L^2}~.
\end{equation}

As in the extremal case, taking the decoupling limit isolates the low-energy dynamics, leading once again to a duality between closed-string theory and a gauge theory. In this setup, however, the interpretation is that the gauge theory is at a finite temperature, which is identified with the Hawking temperature $T_H$ of the black brane. Therefore, we can conclude that $\mathcal{N}=4$ SYM theory in four dimensions at finite temperature is dual to type IIB string theory in an AdS black brane geometry.

\subsection{Matching of symmetries}
A first and important check of the correspondence is that the symmetries match on both sides. The symmetries of $\mathcal{N}=4$ SYM in 4 dimensions can be summarized as follows:
\begin{center}
    Conf(1,3)\quad$\times$\quad $SU(4)_R$\quad$\times$\quad 32 supercharges
\end{center}
Conf(1,3) refers to the conformal group of four-dimensional Minkowski spacetime. This contains dilatations and special conformal transformations in addition to the Poincaré group. $SU(4)_R$ refers to the R-symmetry group under the fields transform. The 32 supercharges correspond to the 16 supercharges introduced as an extension of the Poincaré algebra, and 16 more needed to close the algebra together with the conformal group (superconformal generators).

The symmetries of type IIB superstring theory in AdS$_5\times S^5$ can be summarized as follows:
\begin{center}
    $SO(2,4)$\quad$\times$\quad $SO(6)$\quad$\times$\quad 32 supercharges
\end{center}
Here, $SO(2,4)$ is the isometry group of AdS$_5$, which is isomorphic to the conformal group in 4 dimensions, $SO(2,4)\simeq \text{Conf}(1,3)$. The $SO(6)$ factor corresponds to the rotational symmetry of the 5-sphere, which is isomorphic to $SU(4)_R$. It is then evident that the bosonic symmetries match on both sides of the correspondence. For the fermionic symmetries, we will only mention that AdS$_5\times S^5$ is a maximally supersymmetric solution of type IIB superstring theory, and it possesses 32 Killing spinors. It can be seen that in both sides, the full set of symmetries group into a larger group, $SU(2,2|4)$, thereby showing complete agreement.

\subsection{Strong-weak duality}

The AdS/CFT correspondence states that the two theories involved are dynamically equivalent, meaning that all physical phenomena on one side have a well-defined counterpart on the other. This equivalence must hold across all possible parameter regimes of both theories.

This is known as the \textit{strongest form} of the duality. In its most general statement,

\begin{center}
    \Ovalbox{
    \centering
    \begin{tabular}{ccc}
        \parbox{0.4\textwidth}{
        \vspace{5pt}
        \centering $\mathcal{N}=4$ SYM theory \\ with gauge group $SU(N_c)$ and \\ coupling constant $g_{\text{YM}}$\\in flat 4-dimensional spacetime
        \vspace{5pt}} & $\Longleftrightarrow$ & \parbox{0.4\textwidth}{
        \vspace{5pt}
        \centering Type IIB superstring theory\\ in AdS$_5\times S^5$, with \\coupling constant $g_s$ and\\ $N_c$ units of $F_{(5)}$ flux on $S^5$.
        \vspace{5pt}}
    \end{tabular}
    }
\end{center}

The free parameters in the gauge theory side are the Yang-Mills coupling $\gYM$ and the rank of the gauge group $N_c$, while in the gravity side they are the string coupling $g_s$ and the ratio of the AdS radius $L$ to the string scale, $\frac{L}{\sqrt{\alpha'}}$. From equations \eqref{eq:gYMgs} and \eqref{eq:H(r)}, these parameters are related as
\begin{align}
    \gYM^2 & =4\pi g_s~, & \frac{L^4}{\alpha'^2} & =\gYM^2N_c=\lambda~,
    \label{eq:parametersAdSCFT}
\end{align}
where $\lambda$ is the 't Hooft coupling.

Despite the generality of the correspondence, explicit calculations are often difficult. In particular, string theory is only well-defined perturbatively, making it difficult to test the duality at large $g_s$. More manageable forms of the formulation can be achieved by working in some limits.

A useful regime is when string theory can be treated perturbatively, requiring $g_s\ll 1$ while keeping $L/\sqrt{\alpha'}$ fixed, but arbitrary. From \eqref{eq:parametersAdSCFT}, this corresponds to $\gYM\ll 1$. To ensure $L/\sqrt{\alpha'}$ remains finite, we must take $N_c\to \infty$ while holding the 't Hooft coupling $\lambda$ fixed. As previously discussed, this corresponds to the planar limit of the gauge theory, which is identified as classical string theory, where string loop corrections are suppressed.

This defines the \textit{strong} form of the duality:

\begin{center}
    \Ovalbox{
    \centering
    \begin{tabular}{ccc}
        \parbox{0.5\textwidth}{
        \vspace{5pt}
        \centering $\mathcal{N}=4$ SYM theory with gauge group\\ $SU(N_c)$ and coupling constant \\$g_{\text{YM}}\ll 1$, with $N_c\to\infty$,\\$\lambda=g_{\text{YM}}^2N_c$ fixed but arbitrary
        \vspace{5pt}} & $\Longleftrightarrow$ & \parbox{0.33\textwidth}{
        \vspace{5pt}
        \centering Classical ($g_s\ll 1$) type IIB \\superstring theory in AdS$_5\times S^5$.
        \vspace{5pt}}
    \end{tabular}
    }
\end{center}

Further insight is gained by considering the strong-coupling regime of the gauge theory, $\lambda\to\infty$. In this case, Eq. \eqref{eq:parametersAdSCFT} implies that $\frac{L}{\sqrt{\alpha'}}\gg 1$, meaning that in the gravity dual, the string length scale is much smaller than the AdS curvature scale. This is precisely the limit in which superstring theory can be effectively described by supergravity. This leads to the \textit{weak} form of the correspondence:

\begin{center}
    \Ovalbox{
    \centering
    \begin{tabular}{ccc}
        \parbox{0.5\textwidth}{
        \vspace{5pt}
        \centering $\mathcal{N}=4$ SYM theory with gauge group\\ $SU(N_c)$ and coupling constant \\$g_{\text{YM}}\ll 1$, with $N_c\to\infty$,\\$\lambda=g_{\text{YM}}^2N_c\to\infty$
        \vspace{5pt}
        } & $\Longleftrightarrow$ & \parbox{0.33\textwidth}{
        \vspace{5pt}
        \centering Classical ($g_s\ll 1$) type IIB \\supergravity in\\ weakly curved AdS$_5\times S^5$.
        \vspace{5pt}}
    \end{tabular}
    }
\end{center}

In this sense, the AdS/CFT correspondence can be seen as a strong/weak dualty: a strongly-coupled, large-$N_c$ supersymmetric quantum field theory is described by classical supergravity in a weakly curved spacetime.

\subsection{IR/UV duality and the holographic principle}
A fundamental aspect of the AdS/CFT correspondence is the IR/UV duality between the gravitational and gauge theory descriptions, which makes AdS/CFT a concrete realization of the holographic principle. To illustrate this, we start by considering the metric of AdS$_5$ in Poincaré coordinates,
\begin{equation}
    ds^2=\frac{r^2}{L^2}\left(-dt^2+dx_1^2+dx_2^2+dx_3^2\right)+\frac{L^2}{r^2}dr^2~.
    \label{eq:AdS5Poincare}
\end{equation}
These coordinates provide a natural geometric description of the system: at each fixed radial slice, the metric is conformally equivalent to Minkowski spacetime. Since these four coordinates correspond to directions parallel to the D3-branes, they are naturally interpreted as the gauge theory coordinates.

We now examine how the dilatation operator acts on both sides of the correspondence. Since $\mathcal{N}=4$ SYM is a conformal field theory, it is invariant under the transformation
\begin{equation}
    D~:\quad x^\mu\to\Lambda x^\mu~,
    \label{eq:dilatation}
\end{equation}
where $\Lambda$ is a constant. Given that the isometry group of AdS$_5$ is isomorphic to the conformal group in 4 dimensions, the metric \eqref{eq:AdS5Poincare} must also be invariant under this transformation. This is indeed the case if the radial coordinate transforms as $r\to r/\Lambda$.

This picture suggests a natural interpretation for the radial coordinate: as $r\to\infty$, we approach the "boundary" of AdS$_5$, while $r\to 0$ moves us deeper into the bulk. Since short-distance physics in the gauge theory corresponds to high-energy physics, and high-energy scales are associated with the UV regime of a quantum field theory, we can identify the radial coordinate $r$ as encoding the renormalization group (RG) flow of the field theory. For this reason, the boundary of AdS$_5$ is often referred to as the UV boundary. More generally, since a QFT is defined by a UV fixed point and its RG flow, it is common to say that the gauge theory "lives" at the boundary of AdS$_5$.

For a CFT, the RG flow is trivial, which is reflected geometrically by the fact that the dilatation transformation \eqref{eq:dilatation} is an exact isometry of AdS$_5$. However, in non-conformal theories, the RG flow becomes non-trivial, leading to a spacetime that is only asymptotically AdS. An example of this occurs in the black-brane geometry \eqref{eq:metricD3blackbranesIIB}, which describes a non-conformal field theory due to the presence of a finite temperature.

Beyond these observations, a key aspect of AdS/CFT is its realization of the holographic principle. While the gravitational theory is five-dimensional, the gauge theory is four-dimensional, indicating that all physical information in the bulk is encoded at the boundary. This aligns with the original proposals of 't Hooft \cite{tHooft_1993} and Susskind \cite{Susskind_1994}, which assert that in a gravitational theory, all information is encoded in a lower-dimensional theory defined at the boundary of the space. 

\subsection{Field-operator map}

In the original proposal of the AdS/CFT correspondence, Maldacena established the equivalence between $\mathcal{N}=4$ SYM theory and type IIB string theory on AdS$_5 \times S^5$. However, the proposal did not  provide a concrete prescription for computing observables and explicitly relating quantities between the two sides of the duality. This question was soon addressed in subsequent works by Witten, Gubser, Klebanov, and Polyakov \cite{Witten_1998, Gubser_1998}, leading to the formulation of the field/operator correspondence and the development of the holographic dictionary.

A crucial observation in this development is that the Yang-Mills coupling constant $g_{\text{YM}}$ is identified with the string coupling $g_s$, which in turn is determined by the boundary asymptotic value of the dilaton field. This first example provides the intuition that a gauge theory coupling is determined by the asymptotic behavior of a bulk field. This naturally suggests the following generalization: for a deformation in the field theory Lagrangian of the form
\begin{equation}
    \mathcal{L}\to\mathcal{L}+\phi_0(x)\mathcal{O}(x)~,
    \label{eq:Lsource}
\end{equation}
where $\mathcal{O}(x)$ is a local, gauge-invariant operator and $\phi_0(x)$ is a source, it is suggested that there has to exist a bulk field, $\Phi(z,x)$, such that it approaches $\phi_0(x)$ at the boundary. Schematically,
\begin{equation}
    \phi_0(x)\sim \lim_{z\to 0}\Phi(z,x)~,
    \label{eq:phi0bulkbdry}
\end{equation}
where the precise scaling relation will be determined later. This is the essence of the field/operator correspondence, which provides a direct mapping between bulk fields and boundary operators. Crucially, the bulk field and its corresponding boundary operator must share the same quantum numbers to ensure consistency in the identification. In particular, gauge-invariant operators of $\mathcal{N}=4$ SYM that transform under specific irreducible representations of the $SU(4)_R$ symmetry group are in one-to-one correspondence with states in type IIB string theory on AdS$_5 \times S^5$.

Two particularly significant examples illustrate this correspondence. The first is the case of a conserved current $J_\mu(x)$, whose source $A_\mu(x)$ is interpreted as an external background gauge field. In the holographic description, this source is identified with the asymptotic value of a bulk gauge field $A_\mu(z,x)$. The second example is the energy-momentum tensor $T_{\mu\nu}(x)$, which couples naturally to the boundary value of the bulk metric.

In general, fields will have an expansion near the boundary of the form
\begin{equation}
    f(z,x)\simeq A(x) z^{\alpha_1}+B(x) z^{\alpha_2}~,\qquad z\to 0~,
    \label{eq:UVexpansiongeneralfield}
\end{equation}
with $\alpha_1<\alpha_2$ and given by
\begin{equation}
    \alpha_1=d-\Delta-n~,\qquad \alpha_2=\Delta -n~,
    \label{eq:a1a2}
\end{equation}
where $n$ denotes the number of tensor indices of $f$, $d$ is the number of bulk spatial directions, and $\Delta$ is the conformal dimension of the dual operator, as we will see shortly. More generally, the conformal dimension $\Delta$ of a boundary operator is directly related to the mass of the corresponding bulk field.

The simplest illustrative case is that of a massive scalar field propagating in AdS$_{d+1}$ space.

We begin by considering the AdS$_{d+1}$ metric in Euclidean signature, given by
\begin{equation}
    ds^2=\frac{L^2}{z^2}\left[\delta_{\mu\nu}dx^\mu dx^\nu+dz^2\right]~,
    \label{eq:AdS5z}
\end{equation}
where we have defined the inverse radial coordinate $z$ as $z=L/r$. The action for a massive scalar field $\phi$ in a curved background is
\begin{equation}
    S=-\frac{1}{2}\int d^{d+1}x\sqrt{g}\left[g^{MN}\partial_M\phi\h\partial_N\phi+m^2\phi^2\right]~,
    \label{eq:actionscalarAdS}
\end{equation}
which leads to the equation of motion
\begin{equation}
    \frac{1}{\sqrt{g}}\partial_M\left(\sqrt{g}g^{MN}\partial_N\phi\right)-m^2\phi=0~.
\end{equation}
Using the explicit form of the AdS metric \eqref{eq:AdS5z}, this equation becomes
\begin{equation}
    z^{d+1}\partial_z\left(z^{1-d}\partial_z\phi\right)+z^2\delta^{\mu\nu}\partial_\mu\partial_\nu\phi-m^2L^2\phi=0~.
\end{equation}
By considering the field in Fourier space,
\begin{equation}
    \phi(z,x^\mu)=\int \frac{d^dk}{(2\pi)^d}e^{ik\cdot x}\hat{\phi}_k(z)~,
\end{equation}
its equation of motion becomes
\begin{equation}
    z^{d+1}\partial_z\left(z^{1-d}\partial_z\hat{\phi}_k\right)-k^2z^2\hat{\phi}_k-m^2L^2\hat{\phi}_k=0~.
    \label{eq:EomscalarFT}
\end{equation}

Near the boundary at $z=0$, it is not difficult to verify that an ansatz of the form $\hat{\phi}_k\sim z^\beta$ solves \eqref{eq:EomscalarFT}, provided that $\beta$ satisfies the characteristic equation
\begin{equation}
    \beta=\frac{d}{2}\pm\sqrt{\frac{d^2}{4}+m^2L^2}~.
\end{equation}

Returning to position space, the general near-boundary solution takes the form
\begin{equation}
    \phi(z,x)\approx A(x) z^{d-\Delta}+B(x)z^\Delta~,\qquad z\to 0~,
    \label{eq:expansionscalar}
\end{equation}
where $\Delta$ is given by
\begin{equation}
    \Delta=\frac{d}{2}+\nu~,\qquad \nu=\sqrt{\frac{d^2}{4}+m^2L^2}~.
\end{equation}

From this expression, we see that $\Delta$ remains real as long as the mass satisfies $m^2L^2\ge -d^2/4$. It has been shown \cite{Breitenlohner_1982,Breitenlohner_1982_2,Mezincescu_1984} that within this range, the theory remains stable, meaning that negative masses are allowed provided they do not violate this bound. This condition is known as the Breitenlohner-Freedman (BF) bound.

If the BF bound holds, we also find that
\begin{equation}
    d-\Delta\le \Delta~,\qquad \Longleftrightarrow \qquad \nu=2\Delta-d\ge 0~.
    \label{eq:relationexponents}
\end{equation}
This inequality implies that in the boundary expansion \eqref{eq:expansionscalar}, the first term dominates, while the second is subleading. Notably, if $m^2>0$, we have $d-\Delta<0$,  and this term diverges as $z\to 0$.

To correctly identify the boundary source $\phi_0(x)$ from the asymptotic behavior of the bulk field, we must rescale the latter to absorb this divergence. The appropriate identification is therefore
\begin{equation}
    \phi_0(x)=\lim_{z\to 0}z^{\Delta -d}\phi(z,x)~.
    \label{eq:phi0scaled}
\end{equation}
The same reasoning applies in the case of higher-rank fields \eqref{eq:UVexpansiongeneralfield}, \eqref{eq:a1a2}.

This discussion allows for a more precise interpretation of $\Delta$. If $\mathcal{O}$ is the operator dual to the bulk field $\phi(z,x)$, it appears in the CFT action as
\begin{equation}
    S_{CFT}\sim\int d^dx \sqrt{\gamma_\epsilon}\h \phi(\epsilon,x)\mathcal{O}(\epsilon,x)~,
\end{equation}
where $\gamma_\epsilon=\left(\frac{L}{\epsilon}\right)^{2d}$ is the determinant of the induced metric at the $z=\epsilon$ boundary. Using the asymptotic limit \eqref{eq:phi0scaled}, this action takes the form
\begin{equation}
    S_{CFT}\sim L^d \int d^dx\h\phi_0(x)\epsilon^{-\Delta}\mathcal{O}(\epsilon,x)~.
\end{equation}
For the action to remain finite as $\epsilon\to 0$ the boundary operator $\mathcal{O}(x)$ must be related to the bulk field $\mathcal{O}(z,x)$ as
\begin{equation}
    \mathcal{O}(\epsilon,x)=\epsilon^\Delta \mathcal{O}(x)~.
\end{equation}
This transformation corresponds precisely to a scale transformation in the CFT, showing that $\Delta$ is the scaling dimension of $\mathcal{O}(x)$. The same reasoning applies to \eqref{eq:phi0scaled}, implying that the bulk field has a scaling dimension of $d-\Delta$.

Interestingly, from the discussion following \eqref{eq:relationexponents}, we see that boundary operators can be classified according to the sign of the bulk mass squared: 
\begin{itemize}
    \item If $m^2>0$, then $\Delta>d$, and the operator is irrelevant.
    \item If $m^2=0$, then $\Delta=d$, and the operator is marginal. 
    \item If $m^2<0$, then $\Delta<d$, and the operator is relevant.
\end{itemize}

\subsection{Correlation functions}\label{subsec:correlationfunctions}
We are interested in computing Euclidean field theory correlation functions of the form
\begin{equation}
    \langle \mathcal{O}(x_1)~...~\mathcal{O}(x_n)\rangle
\end{equation}
from the gravity side of the correspondence. In field theory, these are computed from the generating functional, $\mathcal{Z}_{QFT}$, obtained by perturbing the Lagrangian with a source term:
\begin{equation}
    \mathcal{L}\to\mathcal{L}+\phi_0(x)\mathcal{O}(x)~.
\end{equation}
This yields the generating functional
\begin{equation}
    \mathcal{Z}_\text{QFT}\left[\phi_0\right]=\big\langle e^{\int d^dx\phi_0(x)\mathcal{O}(x)}\big\rangle_\text{QFT}~.
\end{equation}
According to the field/operator correspondence \eqref{eq:phi0scaled}, the partition functions on both theories should be equal, since the two theories are conjectured to be dynamically equivalent. This leads to the strong form of the AdS/CFT correspondence \cite{Witten_1998, Gubser_1998}:
\begin{equation}
    \big\langle e^{\int d^dx\phi_0(x)\mathcal{O}(x)}\big\rangle_\text{CFT}=\mathcal{Z}_\text{string}\left[\phi\to\phi_0\right]~,
    \label{eq:ZCFTstring}
\end{equation}
where $\mathcal{Z}_\text{string}\left[\phi\to\phi_0\right]$ is the full string theory partition function, integrated over all field configurations with asymptotic value $\phi \to \phi_0$. While formally elegant, direct computations using this expression are not possible since the full string partition function is unknown.

Things are more tractable, however, if we restrict oursevles to the weak form of the correspondence. In the weak form, we expect type IIB supergravity to be a saddle point of the full string theory partition function, allowing for the approximation
\begin{equation}
    \mathcal{Z}_\text{string}\left[\phi\to\phi_0\right]\approx e^{-S_\text{SUGRA}\left[\Tilde{\phi}\to\phi_0\right]}~,
\end{equation}
where now $\Tilde{\phi}$ denotes a solution of classical type IIB supergravity with asymptotic value $\phi_0$. Consequently, the holographic dictionary simplifies to
\begin{equation}
    \big\langle e^{\int d^dx\phi_0(x)\mathcal{O}(x)}\big\rangle_\text{CFT}=e^{-S_\text{SUGRA}\left[\Tilde{\phi}\to\phi_0\right]}~.
    \label{eq:ZCFTsugra}
\end{equation}
which states that the on-shell bulk supergravity action acts as the generating functional for connected correlation functions in the field theory side. The asymptotic values of the bulk fields act as sources for the field theory operators.

Therefore, to compute a field theory correlator, we proceed as follows:
\begin{itemize}
    \item Identify the bulk field $\phi$ dual to $\mathcal{O}$, typically via its dimension and symmetries.
    \item Solve the classical supergravity equations for $\phi$.
    \item Evaluate the on-shell supergravity action in terms of the boundary asymptotics $\phi_0$.
    \item Exponentiate the on-shell action and take functional derivatives with respect to $\phi_0$.
\end{itemize}

Actually, naïvely applying the procedure outlined above leads to another problem: both sides of \eqref{eq:ZCFTsugra} are divergent. The l.h.s. contains the usual UV divergences present in any field theory. The r.h.s. is divergent because of the infinite volume of AdS, giving rise to IR divergences. In the line of the IR/UV duality mentioned earlier, we observe it here again, since UV divergences in the field theory correspond to IR divergences in the gravitational theory. Therefore, there is the need of a consistent prescription to cancel these divergences and obtain finite answers. This procedure goes under the name of \textit{holographic renormalization} \cite{deHaro_2000,Skenderis_2002holoreno}. 

In holographic renormalization, following a similar logic to standard QFT, we regulate the bulk action by integrating only up to a cutoff surface at $z=\epsilon$, defining the \textit{regularized} on-shell action $S_{reg}$. Divergences are removed by adding local counterterms $S^{ct}_i$:
\begin{equation}
    S_{sub} = S_{reg} + \sum_i S^{ct}_i~.
    \label{eq:Ssub}
\end{equation}
Taking the limit $\epsilon \to 0$ yields the \textit{renormalized} action:
\begin{equation}
    S_{ren} = \lim_{\epsilon\to 0} S_{sub}~.
    \label{eq:Sren}
\end{equation}
This renormalized action is then used to compute correlation functions. As an illustration, we consider the one-point function of a scalar field $\phi$. The discussion will remain quite general, as the details depend on the specific setup under consideration. Explicit calculations are deferred to the cases of interest (see, in particular, Appendices \ref{app:holorenoD3D5} and \ref{app:holorenohelical}).

The one-point function for the operator $\mathcal{O}(x)$ dual to the scalar field $\phi(z,x)$ is given by
\begin{equation}
    \langle\mathcal{O}(x)\rangle_{\phi_0}=\frac{\delta S_{ren}\left[\phi\right]}{\delta \phi_0(x)}~.
\end{equation}
Taking into account the scaling relation between the boundary source and the bulk field, \eqref{eq:phi0scaled}, this expression can be rewritten as
\begin{equation}
    \langle\mathcal{O}(x)\rangle_{\phi_0}=\lim_{\epsilon\to 0}z^{d-\Delta}\frac{\delta S_{sub}\left[\phi\right]}{\delta \phi(z,x)}\Big\rvert_{z=\epsilon}~.
    \label{eq:rescaled1pfunction}
\end{equation}

To express the variations in a compact form, we write the action \eqref{eq:actionscalarAdS} as
\begin{equation}
    S=\int dz\h d^d x \mathcal{L}\left[\phi,\partial\phi\right]~.
\end{equation}
Variation of this action yields
\begin{equation}
    \delta S=\int dz\h d^dx\left[\frac{\partial\mathcal{L}}{\partial\phi}\delta\phi+\frac{\partial\mathcal{L}}{\partial(\partial_\mu\phi)}\delta(\partial_\mu\phi)\right]~.
\end{equation}
Using the standard identity $\delta(\partial_\mu\phi)=\partial_\mu(\delta\phi)$, integrating by parts, and imposing the Euler-Lagrange equations, leads to only a boundary term:
\begin{equation}
    \delta S^{on-shell}=\int_\epsilon^{\infty}dz\int d^d x\partial_z\left(\frac{\partial\mathcal{L}}{\partial(\partial_z\phi)}\delta\phi\right)=-\int d^d x\frac{\partial\mathcal{L}}{\partial(\partial_z\phi)}\delta\phi\Big\rvert_{z=\epsilon}~.
    \label{eq:deltaSonshell}
\end{equation}
Analogous to classical mechanics, where the conjugate momentum is defined with respect to time, we introduce the canonical momentum $\Pi(z,x)$ as
\begin{equation}
    \Pi(z,x)=-\frac{\partial\mathcal{L}}{\partial(\partial_z\phi(z,x))}~.
\end{equation}
Thus, \eqref{eq:deltaSonshell} can be rewritten as
\begin{equation}
    \frac{\delta S^{on-shell}}{\delta\phi(\epsilon,x)}=\Pi(\epsilon,x)~,
\end{equation}
which, as stated earlier, will be divergent in general. After defining the renormalized on-shell action in Eqs. \eqref{eq:Ssub}, \eqref{eq:Sren}, we introduce the renormalized momentum
\begin{equation}
    \Pi^{ren}(z,x)=\frac{\delta S_{ren}^{on-shell}}{\delta\phi(z,x)}~,
\end{equation}
which, near the boundary, takes the form
\begin{equation}
    \Pi^{ren}(\epsilon,x)=-\frac{\delta \mathcal{L}}{\partial(\partial_z\phi(z,x))}\Big\rvert_{z=\epsilon}+\frac{\delta S_{ct}}{\delta \phi(\epsilon,x)}~.
\end{equation}

Consequently, the one-point function for the scalar field follows from the near-boundary expansion as
\begin{equation}
    \langle\mathcal{O}(x)\rangle_{\phi_0}=\lim_{z\to 0}z^{d-\Delta}\Pi^{ren}(z,x)~.
\end{equation}

For the explicit case of a scalar field in AdS$_{d+1}$ with the action \eqref{eq:actionscalarAdS}, it can be shown that the one-point function is proportional to the subleading term in the field’s boundary expansion, Eq. \eqref{eq:expansionscalar}. Specifically, with the appropriate proportionality constant, the one-point function is given by
\begin{equation}
    \langle \mathcal{O}(x)\rangle_{\phi_0}=2\nu B(x)~.
\end{equation}
This result\h\textemdash\h namely, that the one-point function is determined by the subleading term in the near-boundary expansion\h\textemdash\h is a general feature of the AdS/CFT correspondence. We will encounter the same behavior throughout the following chapters, where the details for each case will be worked out explicitly.

%% file: Part1/Braneintersections/Braneintersections.tex
\renewcommand{\chapterquote}{\textit{%
“\textit{Three quarks for Muster Mark!\\
Sure he hasn’t got much of a bark.\\
And sure any he has it’s all beside the mark.}”
}\\[0.75em]
\normalfont— James Joyce, \textit{Finnegans Wake}.}
\chapter{Flavor brane physics}\label{chap:Branes}

In this chapter, we introduce the holographic models that will be employed throughout the Research sections of the thesis. We begin by reviewing the addition of flavor degrees of freedom within the AdS/CFT correspondence, focusing on two specific brane setups: the D3/D7 and D3/D5 intersections. We also review the framework of holographic renormalization and the holographic dictionary, which are essential tools to extract the field theory quantities from bulk gravitational data. The material in this first part is well established and can be found in standard references, including \cite{Karch_2001,DeWolfe_2001,KarchKatz_2002,Kruczenski_2003,Arean_2006,Kobayashi_2006,Mateos_2007,ErdmengerEvans_2007,CasalderreySolana_2011}.

In the second part of the chapter, we discuss the impact of introducing gauge fields on the worldvolume of the flavor branes. We review several aspects related to the presence of electric fields in Section \ref{sec:metallicAdSCFT}, and magnetic fields in Section \ref{sec:CSB}. These results provide the foundation for the analyses presented in Chapters \ref{chap:FloquetII} and \ref{chap:HelicalB}. The discussion closely follows and synthesizes results from \cite{Karch_2007,Filev_2007,Filev_2007_2,OBannon_2007,Kim_2011}.

\vspace{1cm}

\section{Adding flavor to AdS/CFT}

In the original formulation of the AdS/CFT correspondence, only fields in the adjoint representation of the gauge group were involved. These degrees of freedom arose from open strings with both endpoints attached to a stack of $N_c$ coincident D3-branes, sourcing a gauge field that gave rise to $\mathcal{N}=4$ SYM theory on their worldvolume. However, realistic theories like QCD contain not only fields in the adjoint representation (the gluons), but also quark degrees of freedom transforming in the fundamental representation of the gauge group. It is then evident that something needs to be modified in order to account for these degrees of freedom.

In string theory, the way of obtaining a field transforming in the fundamental representation is by considering strings with only one endpoint attached to the D3 branes. This means that the other endpoint has to be attached somewhere else. What "somewhere else" has to be was proposed in \cite{KarchKatz_2002}. Their method consists of adding $N_f$ new branes to the system, which for obvious reasons have been given the name \textit{flavor branes}. In this setup containing $N_c$ D3 branes and $N_f$ D$p$ branes, we identify three sectors of open strings: 3-3 strings, with both endpoints on the D3-branes, $p$-$p$ strings with both endpoints on the D$p$-branes, and 3-$p$, $p$-3 strings stretching between the D3's and the D7's.

In the low-energy limit, the 3-3 strings will continue to give rise to an interacting $\mathcal{N}=4$ SYM theory with gauge group $SU(N_c)$ on the worldvolume of the D3 branes. In this worldvolume, the endpoints of the 3-$p$ and $p$-3 strings give rise to pointlike charged particles transforming in the fundamental representation of $SU(N_c)$, as desired. As we will see, in the low-energy limit the remaining $p$-$p$ strings will not interact with the other open strings, and therefore the $U(N_f)$ gauge group on the D$p$-branes will act as a global flavour symmetry. Fluctuations of open strings on the D$p$-branes will give rise to mesonic states in the gauge theory.

\subsection{Decoupling limit with probe branes}

We now extend the arguments leading to the formulation of the AdS/CFT correspondence to the case in which, in addition to the $N_c$ D3-branes, we introduce $N_f$ flavor D$p$-branes.

As before, the key to the duality lies in analyzing the system in two different regimes of the effective coupling $g_s N_c$. When $g_s N_c \ll 1$, the appropriate description is the open-string perspective. In this picture, we have open and closed strings propagating in ten-dimensional flat space. However, now the open-string sector contains 3-3, 3-$p$, $p$-3 and $p$-$p$ strings, as mentioned above.

In the low-energy limit, the 3-3 strings give rise to the same interacting $\mathcal{N}=4$ $SU(N_c)$ SYM theory as before. The interactions among the open $p$-$p$ strings are governed by the corresponding 't Hooft coupling in the D$p$-branes. The interactions are controled by, respectively, 
\begin{equation}
    \lambda_{D3}=4\pi g_s N_c~,~~~~~~~\lambda_{Dp}=2(2\pi)^{p-2}\alpha'^{(p-3)/2}g_sN_f~.
    \label{eq:tHooftD3Dp}
\end{equation}

In the low-energy limit, $\lambda_{Dp}$ vanishes only for $p>3$. Since we want the $p$-$p$ strings to decouple, we will restrict ourselves to $p=5,7$. In these cases, $\lambda_{DP}$ vanishes as $\alpha'\to 0$, and these interactions become irrelevant at low energies. Closed strings also decouple, in the same way as in Section \ref{sec:AdSCFT}. The 3-$p$ strings couple to the 3-3 and $p$-$p$ open strings, with interaction strengths also given by the first and second expression in \eqref{eq:tHooftD3Dp}, respectively. Therefore, in the low-energy limit, their only remaining interactions are with the 3-3 strings. The spectrum of open strings stretching between D3- and D$p$-branes depends on the value of $p$ and the relative orientation of the branes. In the examples we will consider, the lightest states correspond to degrees of freedom transforming in the fundamental representation of the gauge group $SU(N_c)$, with a mass given by
\begin{equation}
    M_q=\frac{L}{2\pi\alpha'}~,
    \label{eq:Mqseparation}
\end{equation}
where $L$ is the separation between the D3- and the D$p$-branes. The subscript in $M_q$ refers to the fact that we will generically call these fermionic degrees of freedom \textit{quarks}, and \textit{squarks} their scalar superpartners.

Thus, in this limit, the system again decouples into: free closed strings and free open $p$-$p$ strings propagating in ten dimensional flat spacetime, and an interacting $\mathcal{N}=4$ SYM theory with gauge group $SU(N_c)$, coupled to matter fields transforming in the fundamental representation.

Consider now the opposite limit, $g_s N_c\gg 1$, where the closed-string perspective is appropriate. Again, in this regime the D3-branes and the 3-3 strings are not present anymore, but are replaced by the background geometry \eqref{eq:metricD3branesIIB}. 

A convenient choice is to work in the so-called \textit{probe limit}. In the probe limit, the number of flavor branes, $N_f$, is much smaller than $N_c$, $N_f\ll N_c$. We can assume also that $g_s N_f\ll 1$, which is consistent with $g_s N_c\gg 1$ for sufficiently small $N_f$. In this limit, the backreaction of the $Dp$-branes on the geometry can be neglected. Therefore, in addition to closed strings, we also have $p$-$p$ strings propagating in the background geometry sourced by the D3-branes. Once again, we distinguish between the asymptotically flat region and the near-horizon region. In the low-energy limit, these two regions decouple as before. In the asymptotically flat region, both closed and $p$-$p$ strings become free, as all interactions are controlled by the coupling $\lambda_{Dp}$, which vanishes when $\alpha'\to 0$. In the near-horizon region, however, the full interacting string theory survives, including the $p$-$p$ strings. As a result, the system again consists of two decoupled sectors: free closed strings and free $p$-$p$ strings propagating in flat ten dimensional spacetime, and type IIB closed strings in AdS$_5\times S^5$, coupled to the $p$-$p$ open strings. 

In the same way as before, the equivalence between the two descriptions leads us to conjecture the following version of the AdS/CFT correspondence:
\begin{center}
    \Ovalbox{
    \centering
    \begin{tabular}{ccc}
        \parbox{0.42\textwidth}{
        \vspace{5pt}
        \centering $\mathcal{N}=4$, $SU(N_c)$, SYM theory \\ coupled to $N_f$ flavors of \\fundamental degrees of freedom \\in the limit $N_f\ll N_c$
        \vspace{5pt}} & $\Longleftrightarrow$ & \parbox{0.42\textwidth}{
        \vspace{5pt}
        \centering Type IIB superstring theory\\ in AdS$_5\times S^5$, coupled to open\\ strings on the worldvolume of \\ $N_f$ D$p$-branes.
        \vspace{5pt}}
    \end{tabular}
    }
\end{center}

\section{The D3/D7 intersection}\label{sec:D3D7}

\subsection{Brane intersection and gauge theory}

As we mentioned before, in this thesis we will only consider $p=5,7$. We begin by constructing the D3/D7 intersection, from which we will also learn general features that will apply in the D3/D5 intersection as well. 

We introduce a set of $N_f$ D7 flavor branes, whose orientation relative to the $N_c$ D3-branes is detailed in Table \ref{tab:D3D7}. If the D3-branes and the D7-branes overlap, the original $SO(6)$ symmetry on the directions transverse to the D3-branes is broken to $SO(4)$ along the directions $X_4,...~,X_7$, and $SO(2)$ in the $(X_8,X_9)$ plane.

The field theory that arises from this intersection consists of the original $\mathcal{N}=4$ SYM theory coupled to $N_f$ $\mathcal{N}=2$ massless hypermultiplets transforming in the fundamental representation of $SU(N_c)$. The Lagrangian of this theory is written explicitly in \cite{Chesler_2006}. The matter content of the hypermultiplet consists of 2 Weyl spinors of opposite chirality, denoted as ($\psi$,$\bar{\psi}$) and 2 complex scalars, $(q,\Tilde{q})$. As mentioned before, we will call these hypermultiplet fields \textit{quarks} and \textit{squarks}, respectively.

By separating the D7-branes from the D3-branes, the $N_f$ hypermultiplets acquire a mass. If all the hypermultiplets have the same mass, the Lagrangian has a global $U(N_f)$ symmetry. In the brane picture, this corresponds to the situation in which the $N_f$ D7-branes are placed on top of each other. Different masses for each multiplet can be realized by placing each D7-brane at a different distance from the D3-branes, a situation we will not consider in this thesis.

\begin{table}
    \centering
    \begin{tabular}{ccccccccccc}
         & $X_0$ & $X_1$ & $X_2$ & $X_3$ & $X_4$ & $X_5$ & $X_6$ & $X_7$ & $X_8$ & $X_9$\\
        D3 & $\times$ & $\times$ & $\times$ & $\times$ & $-$ & $-$ & $-$ & $-$ & $-$ & $-$ \\
        D7 & $\times$ & $\times$ & $\times$ & $\times$ & $\times$ & $\times$ & $\times$ & $\times$ & $-$ & $-$ \\
    \end{tabular}
    \caption{D3/D7 brane intersection. Directions filled by the branes are denoted by $\times$. Directions orthogonal to them are denoted by $-$.}
    \label{tab:D3D7}
\end{table}

\subsection{Probe D7-branes at zero temperature}

The way in which the D7-branes are embedded in the geometry is determined by the sum of the DBI + WZ action, given in \eqref{eq:SDBI}, \eqref{eq:SWZ}. For a D7-brane,
\begin{equation}
    S_{D7}=-N_f T_{D7}\int d^8\xi \sqrt{-\det\left(h_{ab}+2\pi \alpha' F_{ab}\right)}+\frac{(2\pi\alpha')^2}{2}N_f T_{D7}\int P[C_{(4)}]\wedge F\wedge F~,
\end{equation}
where the tension of the brane, $T_{D7}$, is given by
\begin{equation}
    T_{D7}=\frac{1}{(2\pi)^7g_s\alpha'^4}~.
\end{equation}

The metric of AdS$_5\times S^5$ was written in \eqref{eq:AdS5xS5} as
\begin{equation}
    ds^2=\frac{r^2}{L^2}\left(-dt^2+dx_1^2+dx_2^2+dx_3^2\right)+\frac{L^2}{r^2}dr^2+L^2d\Omega_5^2~,
\end{equation}
where $r^2=x_4^2+...+x_9^2$ and $\Omega_5$ represent the 6 directions transverse to the D3-branes. It is useful to adapt this metric to the relevant coordinates in the presence of the two types of branes. The Minkowski directions $(t,x_1,x_2,x_3)$ are common to the D3- and D7-branes. Of the remaining 6 transverse directions, the D7-branes span the 4567-directions. We introduce spherical coordinates $(\rho,\Omega_3)$, with $\rho^2=x_4^2+...+x_7^2$ in these directions. The remaining two directions are transverse to both sets of branes, $(x_8,x_9)\equiv(w_1,w_2)$, which will parametrize the embedding of the D7-branes. The metric can therefore be written as
\begin{equation}
    ds^2=\frac{\rho^2+w_1^2+w_2^2}{L^2}\left(-dt^2+dx_1^2+dx_2^2+dx_3^2\right)+\frac{L^2}{\rho^2+w_1^2+w_2^2}\left(d\rho^2+\rho^2d\Omega_3^2+dw_1^2+dw_2^2\right)~.
    \label{eq:AdS5xS5D7brane}
\end{equation}

Lorentz invariance of the dual field theory forces the embedding to be independent of the Minkowski coordinates. We will also assume the embedding preserves the rotational symmetry of the $S^3$. Therefore, the only allowed dependence is on $\rho$. Moreover, the $SO(2)$ symmetry in the $(w_1,w_2)$ plane allows to set one of them to zero, $w_1=0$, $w_2\equiv w(\rho)$. Then, the induced metric of the D7-brane is
\begin{equation}
    ds^2=\frac{\rho^2+w^2}{L^2}\left(-dt^2+dx_1^2+dx_2^2+dx_3^2\right)+\frac{L^2}{\rho^2+w^2}\left((1+w'^2)d\rho^2+\rho^2d\Omega_3^2\right)~,
    \label{eq:D7induced}
\end{equation}
where primes denote derivatives with respect to the radial coordinate, $\rho$. We will set $L=1$ hereafter. The DBI action for this ansatz is
\begin{equation}
    S_{D7}=-N_f T_{D7}V_{\mathbb{R}^{3,1}}V_{S^3}\int d\rho~\rho^3\sqrt{1+w'^2}~.
\end{equation}
We have performed the integrals along the Minkowski directions and the 3-sphere. In the following we will always consider all quantities divided by the (infinte) volume of Minkowski space. We group the terms into a single overall constant, $\mathcal{N}$, defined as
\begin{equation}
    \mathcal{N}=N_f T_{D7}V_{S^3}=\frac{N_fN_c\h\lambda}{(2\pi)^4}~.
\end{equation}

The action is independent of the embedding function $w(\rho)$, which means there is a conserved quantity, determined by the equation of motion
\begin{equation}
    \frac{d}{d\rho}\left(\frac{\rho^3w'}{\sqrt{1+w'^2}}\right)=0~.
    \label{eq:eomw}
\end{equation}
The simple choice $w'(\rho)=0$ is an obvious solution of \eqref{eq:eomw}, giving a constant $w(\rho)\equiv M$. This solution is telling us that the D7-brane lies at a constant distance from the stack of D3-branes. In other words, it is not deformed by the presence of the D3-branes. This non-deformation of the D7-branes may seem strange at first sight, but it can be checked that the presence of supersymmetry allows for an exact cancellation of the forces on the D7-brane. This is the so-called "no-force condition" of D-branes \cite{Tseytlin_1996}. 

The constant $M$ determines the quark mass to be $M_q=M/2\pi\alpha'$. However, in more general situations with worldvolume gauge fields or non-zero temperature, the embedding function will not be a constant, and the D7-brane will develop a profile determined by $w(\rho)$, with asymptotic behavior
\begin{equation}
    w(\rho)=M+\frac{C}{\rho^2}+...
    \label{eq:wexpansion}
\end{equation}
We have just seen that the leading term in the expansion is related to the mass of the hypermultiplet. According to the dictionary, the boundary value of the embedding function acts as a source for a field theory operator. According to \eqref{eq:Lsource}, this boundary value couples linearly to the corresponding operator in the field theory Lagrangian. This means that this operator is obtained by $\partial_m \mathcal{L}$. This operator was obtained in \cite{Kobayashi_2006}, and it is given by
\begin{equation}
    \mathcal{O}_m=\Bar{\psi}\psi+\Tilde{q}\left(M_q+\sqrt{2}\Phi\right)\Tilde{q}^\dagger+q^\dagger\left(M_q+\sqrt{2}\Phi\right)q+h.c.
\end{equation}
Due to supersymmetry, it contains the mass term for the hypermultiplet fields, as well as an interaction term with a scalar field $\Phi$ of the $\mathcal{N}=4$ theory. We will not worry about this, and we will call $\mathcal{O}_m$ the mass operator. The subleading term, $C$, is related to the expectation value of the operator $\langle\mathcal{O}_m\rangle$, namely the quark condensate. The precise relation will be computed below.

Finally, notice that by introducing the solution into the induced metric \eqref{eq:D7induced}, a non-zero $w(\rho)$ gives to the radius of the 3-sphere a dependence on the radial direction of AdS$_5$, $r$. In particular, since $\rho^2=r^2-w^2$, we see that the radius of the 3-sphere shrinks to zero at a finite distance. For the case of a constant $w=w_0$, this happens at $r=w_0$. This effect is important: from the AdS$_5$ point of view, the D7-brane "ends" smoothly at the radial distance $r=w_0$.

\subsection{Probe D7-branes at non-zero temperature}\label{sec:D7probestemp}
In the case that a non-zero temperature is introduced, the system consists of a probe D7-brane in the black brane geometry \eqref{eq:SchwAdS5xS5}. This geometry now includes a black hole. At $T\neq 0$, all supersymmetry is broken, and therefore it is reasonable to expect that the forces on the D7-brane will not cancel each other anymore. This picture is consistent with the intuition that the gravitational attraction of the black hole will tend to bend the branes towards the interior of the bulk. Numerically solving the equation of motion for the embedding function shows that this is the case \cite{Mateos_2007}.

Qualitatively, the physics now is much richer than at zero temperature. We have seen that at zero temperature the D7-brane sits flat at a constant separation from the D3-branes. With a small but non-zero temperature (in terms of the dimensionless ratio $T/M_q$), the brane bends towards the bulk horizon but its tension is enough to compensate the gravitational attraction. The result is similar to what we described above: the brane "ends" smoothly at a finite radial distance. In this case, the induced metric on the D7-branes has no horizon and for this reason these solutions, or embeddings, have been given the name \textit{Minkowski embeddings}.

However, for sufficiently high temperatures, the branes can "fall" into the black hole horizon. In these cases, the worldvolume of the D7-branes intersect with the horizon of the background geometry, so they develop a \textit{worldvolume horizon}. These solutions are called \textit{black hole embeddings}.

\begin{figure}
    \centering
    \begin{subfigure}[t]{0.3\textwidth}
        \centering
        \vspace*{\fill}
        \includegraphics[width=\textwidth]{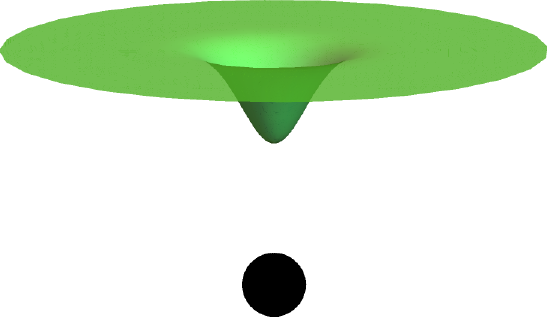}
    \end{subfigure}
    \hspace{0.02\textwidth}
    \begin{subfigure}[t]{0.3\textwidth}
        \centering
        \vspace*{\fill}
        \includegraphics[width=\textwidth]{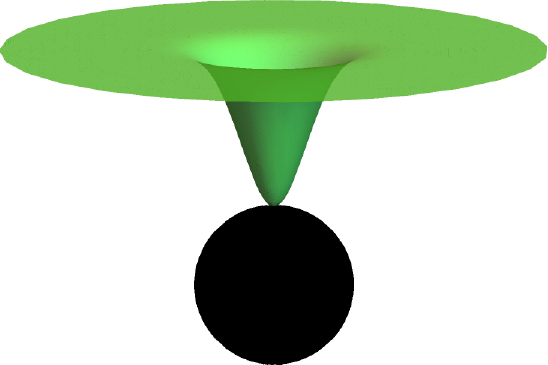}
    \end{subfigure}
    \hspace{0.02\textwidth}
    \begin{subfigure}[t]{0.3\textwidth}
        \centering
        \vspace*{\fill}
        \includegraphics[width=\textwidth]{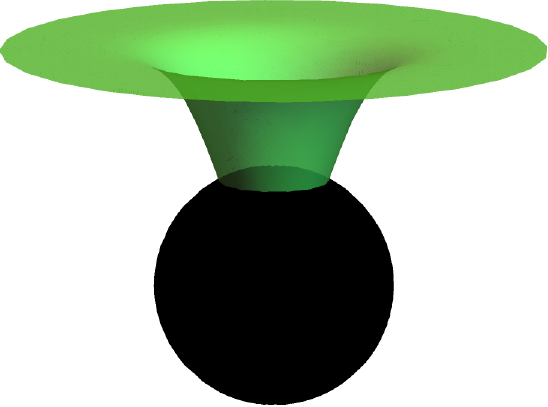}
    \end{subfigure}
    \caption{D7-brane embeddings for different background temperatures, increasing from left to right. At lower temperatures (left) we have Minkowski embeddings, wher the brane closes off smoothly outside the horizon. At higher temperatures the brane intersects the horizon, giving rise to black hole embeddings (right). The two behaviors are separated by the critical embedding (center). Inspired by Fig. 8.7 of \cite{CasalderreySolana_2011}.}
    \label{fig:3embeddings}
\end{figure}

The critical temperature that separates the two behaviors gives rise to the \textit{critical embedding}, in which the D7-brane remains outside the horizon but closes off smoothly precisely at $r=r_h$. The three behaviors are shown in Fig. \ref{fig:3embeddings}.

These behaviors have a direct field theory interpretation in terms of meson stability. At finite temperature, thermal effects generally weaken the quark-antiquark binding, leading to the expectation that mesons may dissociate. However, in the Minkowski phase, the fluctuations exhibit a discrete, normalizable spectrum, indicating the presence of stable mesons, as we will compute shortly. This corresponds to a phase where quark-antiquark bound states persist despite the finite temperature of the system.

In contrast, in the black hole phase, the spectrum is gapless and continuum. This corresponds to the melting of mesons, where the thermal energy is sufficient to overcome the binding, causing quark-antiquark pairs to dissociate.

However, this should not be confused with a confinement/deconfinement transition. The latter concerns whether quarks are confined into hadrons or can exist as free color charges. In contrast, meson melting is a bound-state phenomenon: even in a deconfined phase, some mesonic states may survive depending on their binding energy relative to the temperature. In the D3/D7 model, the gauge theory remains in a deconfined phase for all temperatures due to the presence of a black hole in the bulk.

\subsection{Holographic renormalization and quark condensate}\label{subsec:holorenoD3D7}

We now focus on how to extract field theory quantities from the D7-brane on-shell action. As we have mentioned above, the boundary value of the embedding function is related to the hypermultiplet mass. In this section we compute the expectation value of the mass operator, $\langle\mathcal{O}_m\rangle$, from the on-shell action of the D7-brane, and show that it is directly related to the subleading term in the UV expansion \eqref{eq:wexpansion}.

According to the holographic dictionary of Section \ref{subsec:correlationfunctions}, we need to solve the equation of motion for the brane embedding, evaluate the on-shell action, and take functional derivatives with respect to the boundary value of the field. As anticipated earlier, this quantity has IR divergences due to the infinite volume of AdS, corresponding to UV divergences of the field theory and therefore local counterterms need to be added.

In order to illustrate this, it is convenient to rewrite the AdS$_5\times S^5$ metric \eqref{eq:AdS5xS5D7brane} as
\begin{equation}
    ds^2=\frac{1}{z^2}\left(-dt^2+dx_1^2+dx_2^2+dx_3^2+dz^2\right)+d\theta^2+\sin^2\theta d\phi^2+\cos^2\theta d\Omega_3^2~,
    \label{eq:AdS5xS5sliceD7}
\end{equation}
where $z=1/r$, and therefore the boundary is now at $z=0$. The D7-brane spans the whole of AdS$_5$ as well as the $S^3$ inside the $S^5$. The size of the 3-sphere depends on $\theta$, being maximal at $\theta=0$ and shrinking to zero at $\theta=\pi/2$. The relation between these coordinates and the ones used in \eqref{eq:AdS5xS5D7brane} is
\begin{equation}
    \frac{1}{z^2}=r^2=\rho^2+w^2~,\quad \tan\theta=\frac{w}{\rho}~.
\end{equation}

Now the embedding function is $\theta(z)$. The reason for using this set of coordinates is that now the function $\theta(z)$ depends directly on the radial coordinate of AdS, as opposed to the cartesian embedding, which depends on $\rho$ rather than $r$ (or $z=1/r$). The induced metric is
\begin{equation}
    ds^2=\frac{1}{z^2}\left(-dt^2+dx_1^2+dx_2^2+dx_3^2\right)+\left(\frac{1}{z^2}+\theta'(z)^2\right)dz^2+\cos^2\theta d\Omega_3^2~.
    \label{eq:inducedD7thetaz}
\end{equation}
In these coordinates, we define the \textit{regularized} action as the DBI action with the radial direction integrated until a regulator surface $z=\epsilon$,
\begin{equation}
    S_{\text{D7}}^{reg}=-\mathcal{N}\int_\epsilon dz \frac{\cos^3\theta}{z^5}\sqrt{1+z^2\theta'^2}~.
    \label{eq:DBID7thetareg}
\end{equation}
The limit of integration in the bulk is irrelevant for the following discussion. We don't write it explicitly because it will depend on each particular setup.

From the equation of motion for $\theta(z)$ it can be seen that its behavior near $z=0$ is
\begin{equation}
    \theta(z)\sim z\left(\theta_0+\theta_2 z^2+ \mathcal{O}(z^4)\right)~.
    \label{eq:thetaUVexp}
\end{equation}
By imposing the equation of motion to be satisfied order by order in $z$, it can be seen that all higher-order coefficients $\theta_i$, with $i>2$ are fixed in terms of $\theta_0$ and $\theta_2$. This is consistent with the equation of motion being second order: we know the asymptotic value $\theta_0$ is related to the mass of the hypermultyplet. This is a constant we can choose as UV boundary condition for the integration. The second constant $\theta_2$, however, is not fixed at the UV, but at the other endpoint of the integration by requiring, for instance, regularity of the solution deep in the IR.\footnote{In the case of a finite temperature, this regularity condition will be imposed at the black hole horizon. In the case of electric fields present, regularity will be imposed at the singular shell. See Chapter \ref{chap:FloquetII} for details.} We will come back to this after reaching the final result for $\langle \mathcal{O}_m\rangle$.

Plugging the expansion \eqref{eq:thetaUVexp} into the action \eqref{eq:DBID7thetareg}, we find the divergent pieces to behave as
\begin{equation}
    S_{\text{D7}}^{reg}=-\mathcal{N}\int_\epsilon dz\left[\frac{1}{z^3}-\frac{\theta_0^2}{z^3}+\mathcal{O}(z)\right]=-\mathcal{N}\left[\frac{1}{4\epsilon^4}-\frac{\theta_0^2}{2\epsilon^2}+\mathcal{O}(\epsilon^2)\right]~.
\end{equation}
These divergences can be removed by adding local counterterms evaluated at $z=\epsilon$. These counterterms are obtained by the general method of holographic renormalization applied to the action of D$p$-branes. They are generically obtained in terms of the induced metric on the $z=\epsilon$ surface, denoted by $\gamma_{ij}$, and the Ricci scalar and tensor obtained from $\gamma_{ij}$. The details and the list of needed counterterms for the D3/D7 intersection can be found in \cite{Karch_2005_holorenobranes,Karch_2006}. In all cases we will consider, the induced metric at $z=\epsilon$ will be Ricci-flat, so some counterterms will be automatically zero. We write here the list of relevant countereterms that will be needed:
\begin{equation}
\begin{aligned}
    S^{ct}_1&=\frac{1}{4}\mathcal{N}\sqrt{-\gamma}~,\\
    S^{ct}_2&=-\frac{1}{2}\mathcal{N}\sqrt{-\gamma}\theta(\epsilon)^2~,\\
    S^{ct}_f&=\frac{5}{12}\mathcal{N}\sqrt{-\gamma}\theta(\epsilon)^4~.
    \label{eq:SctD3D7}
\end{aligned}
\end{equation}

Some comments are in order: the first counterterm is needed to regularize the infinite volume of AdS$_5$. $S^{ct}_2$ corresponds to the counterterm for a free scalar in AdS$_5$. Finally, $S^{ct}_f$ is a finite counterterm that can be added in the D3/D7 setup, with its only effect being a constant shift in the value of the on-shell action. Different choices for this finite term correspond to different renormalization schemes. In particular, when the background is supersymmetric, this term is fixed so that the on-shell action evaluated on the supersymmetric solution vanishes exactly \cite{Bianchi_2001, Karch_2005_holorenobranes}. A non-zero on-shell action in this case would imply a non-zero ground-state energy in the dual field theory, leading to the breaking of supersymmetry.

Armed with the necessary counterterms, we can now define the \textit{subtracted} on-shell action as
\begin{equation}
    S_{\text{D7}}^{sub}=S_{\text{D7}}^{reg}+\sum_iS^{ct}_i~,
\end{equation}
with $S^{ct}_i$ the counterterms listed above. This action is finite in the $\epsilon\to 0$ limit, defining the \textit{renormalized} on-shell action,
\begin{equation}
    S_{\text{D7}}^{ren}=\lim_{\epsilon\to 0}S_{\text{D7}}^{sub}~.
\end{equation}

According to the AdS/CFT dictionary, $-S_{\text{D7}}^{ren}$ is identified with the generating functional of the dual field theory. The renormalized one-point function for a scalar field $\phi(z)$ dual to an operator $\mathcal{O}$ of dimension $\Delta$ was given in \eqref{eq:rescaled1pfunction}. For the embedding function $\theta(z)$, $\Delta=3$, as follows from \eqref{eq:UVexpansiongeneralfield}, \eqref{eq:a1a2} and \eqref{eq:thetaUVexp}. Therefore, the renormalized $\langle \mathcal{O}_m\rangle$ is given by
\begin{equation}
    \langle \mathcal{O}_m\rangle =-(2\pi\alpha')\lim_{\epsilon\to 0}\epsilon\frac{\delta S_{\text{D7}}^{sub}}{\delta \theta(\epsilon)}~,
\end{equation}
where the normalization factor of $2\pi\alpha'$ arises because of \eqref{eq:Mqseparation}.

We can compute the variation of each term in $S_{\text{D7}}^{sub}$. For the bulk term $S_{\text{D7}}^{reg}$,
\begin{equation}
\begin{aligned}
    \delta S_{\text{D7}}^{reg}=\int_\epsilon dz\left[\frac{\partial \mathcal{L}}{\partial \theta'}\delta\theta'+\frac{\partial\mathcal{L}}{\partial\theta}\delta\theta\right]&=\frac{\partial \mathcal{L}}{\partial \theta'}\delta\theta\Big\rvert_\epsilon\\
    &=\mathcal{N}\left[\frac{\theta_0}{\epsilon^3}+\frac{3\theta_2-2\theta_0^3}{\epsilon}+\mathcal{O}(\epsilon)\right]\delta\theta(\epsilon)~,
\end{aligned}
\end{equation}
where in the first line we have integrated by parts and imposed the equation of motion for $\theta(z)$, and in the second line we have evaluated the remaining boundary term with the UV expansion \eqref{eq:thetaUVexp}.

For the counterterms, only $S^{ct}_2$ and $S^{ct}_f$ contribute, since $S^{ct}_1$ is independent of the embedding. Their contributions are
\begin{equation}
\begin{aligned}
    \delta S^{ct}_2&=-\mathcal{N}\sqrt{-\gamma}\theta(\epsilon)\delta \theta(\epsilon)=\mathcal{N}\left[-\frac{\theta_0}{\epsilon^3}-\frac{\theta_2}{\epsilon}+\mathcal{O}(\epsilon^0)\right]\delta \theta(\epsilon)\\
    \delta S^{ct}_f & = \frac{5}{3}\mathcal{N}\sqrt{-\gamma}\theta(\epsilon)^3\delta \theta(\epsilon)=\mathcal{N}\left[\frac{5\theta_0^3}{3\epsilon}+\mathcal{O}(\epsilon^0)\right]
\end{aligned}
\end{equation}

The total variation is
\begin{equation}
    \delta S^{sub}_{D7}=\mathcal{N}\left[\frac{2\theta_2}{\epsilon}-\frac{\theta_0^3}{3\epsilon}+\mathcal{O}(\epsilon^0)\right]~.
\end{equation}
Therefore, the one-point function $\langle \mathcal{O}_m\rangle$ will be given by
\begin{equation}
    \langle \mathcal{O}_m\rangle =-(2\pi\alpha')\lim_{\epsilon\to 0}\epsilon\frac{\delta S_{\text{D7}}^{sub}}{\delta \theta(\epsilon)}=-\mathcal{N}(2\pi\alpha')\left(2\theta_2-\frac{\theta_0^3}{3}\right)~,
    \label{eq:Om1pfunc}
\end{equation}

We can now return to the discussion above. We mentioned that fixing $\theta_0$ in the UV corresponds to fixing the mass in the Lagrangian. From \eqref{eq:Om1pfunc} we see that the subleading term, $\theta_2$, carries the information about $\langle \mathcal{O}_m\rangle$. We also mentioned that $\theta_2$ can be understood as the second boundary condition, in this case imposed deep in the IR. The field theory interpretation of this fact is that given a mass $m$, the dynamics of the system, including the IR physics, will determine $\langle \mathcal{O}_m\rangle$.

This behavior is not particular of the embedding function, but is a rather general feature in gauge/gravity duality, which we will use repeatedly in Chapters \ref{chap:FloquetII} and \ref{chap:HelicalB}: given a generic supergravity field, its leading, non-normalizable asymptotic value acts as a source for its dual field theory operator. The subleading, normalizable term determines the expectation value of the dual operator.

\subsection{Brane fluctuations and meson spectrum}\label{subsec:mesonspectrum}

The field theory dual to the D3/D7 brane intersection has a rich spectrum of mesons (quark-antiquark bound states). Remarkably, the masses of these mesons can be obtained from the gravitational dual. In particular, as suggested first in \cite{Karch_2002} and later computed in \cite{Kruczenski_2003}, the spectrum can be obtained by studying the fluctuations of the fields on the D7-brane. Later, the result was generalized to a larger variety of D$p$/D$q$-brane intersections \cite{Arean_2006}.

Here we consider only the fluctuations of the scalars. We begin by considering the metric in the form
\begin{equation}
    ds^2=\frac{\rho^2+w_1^2+w_2^2}{L^2}\left(-dt^2+dx_1^2+dx_2^2+dx_3^2\right)+\frac{L^2}{\rho^2+w_1^2+w_2^2}\left(d\rho^2+\rho^2d\Omega_3^2+dw_1^2+dw_2^2\right)~.
\end{equation}
We want to study fluctuations around the flat configuration $w_1=w_0$, $w_2=0$. In this case, the induced metric on the D7-brane is
\begin{equation}
    ds^2=\frac{\rho^2+w_0^2}{L^2}\left(-dt^2+dx_1^2+dx_2^2+dx_3^2\right)+\frac{L^2}{\rho^2+w_0^2}\left(d\rho^2+\rho^2d\Omega_3^2\right)~.
    \label{eq:D7inducedfluct}
\end{equation}
We can now study fluctuations of the type
\begin{equation}
    w_1=w_0+\Phi_1~,\qquad w_2=\Phi_2~.
\end{equation}
Expanding the DBI Lagrangian to second order in the fluctuations, we obtain
\begin{equation}
    \mathcal{L}_{D7}\simeq -\frac{1}{2}\rho^3\sqrt{\det(\Omega_3)}\frac{L^2}{\rho^2+w_0^2}h^{ab}\partial_a\Phi_i\partial_b\Phi_i~,
    \label{eq:LD7fluct}
\end{equation}
where $h^{ab}$ denotes the inverse of the induced metric \eqref{eq:D7inducedfluct} and $\det(\Omega_3)$ denotes the determinant of the unit 3-sphere metric. The equation of motion derived from the quadratic Lagrangian is
\begin{equation}
    \frac{L^4}{\left(\rho^2+w_0^2\right)^2}\partial^\mu\partial_\mu\Phi+\frac{1}{\rho^3}\partial_\rho(\rho^3\partial_\rho\Phi)+\frac{1}{\rho^2}\nabla^i\nabla_i\Phi=0~,
\end{equation}
where we have suppressed the indices $i$ in the fluctuations. The index $\mu$ denotes the directions $x^\mu=(t,...,x^3)$, and  $\nabla_i$ corresponds to the covariant derivative on $S^3$. The equation of motion can be solved with an ansatz of the form
\begin{equation}
    \Phi=\phi(\rho)e^{ik\cdot x}\mathcal{Y}^\ell(S^3)~,
\end{equation}
where $\mathcal{Y}^\ell(S^3)$ are the spherical harmonics on $S^3$, with
\begin{equation}
    \nabla_i\nabla^i\mathcal{Y}^\ell=-\ell(\ell+2)\mathcal{Y}^\ell~.
\end{equation}
The equation for $\phi(\rho)$ becomes
\begin{equation}
    \partial_\varrho^2\phi+\frac{3}{\varrho}\phi+\left(\frac{\bar{M}^2}{(1+\varrho^2)^2}-\frac{\ell(\ell+2)}{\varrho^2}\right)\phi=0~,
    \label{eq:eqphi}
\end{equation}
where we have defined the rescaled variables
\begin{equation}
    \varrho=\frac{\rho}{w_0}~,\qquad \bar{M}^2=-\frac{k^2L^4}{w_0^2}~.
\end{equation}
Eq. \eqref{eq:eqphi} can be solved in terms of hypergeometric functions. Imposing reality, normalizability, and regularity at the origin one obtains, after going back to the original $\rho$ coordinate \cite{Kruczenski_2003},
\begin{equation}
    \phi(\rho)=\frac{\rho^\ell}{(\rho^2+w_0^2)^{n+\ell+1}}F(-(n+\ell+1),-n;\ell+2;-\rho^2/w_0^2)~,
\end{equation}
provided that
\begin{equation}
    \bar{M}^2=4(n+\ell+1)(n+\ell+2)~,\quad \text{with}\quad n=0,1,2,...
\end{equation}
Identifying the masses of the mesons as $M^2=-k^2=\bar{M}^2w_0^2/L^4$, we obtain a spectrum given by
\begin{equation}
    M_s(n,\ell)=\frac{2w_0}{L^2}\sqrt{(n+\ell+1)(n+\ell+2)}~.
\end{equation}

In \cite{Arean_2006}, this result was generalized to other brane intersections. In particular, from there we can extract the same result for the D3/D5 intersection that will be used in Chapter \ref{chap:FloquetII}:
\begin{equation}
    M^{D3/D5}_s(n,\ell)=\frac{2w_0}{L^2}\sqrt{\left(n+\ell+\frac{1}{2}\right)\left(n+\ell+\frac{3}{2}\right)}~.
    \label{eq:mesonsD3D5}
\end{equation}

\section{The D3/D5 intersection}\label{sec:D3D5}

\subsection{Brane intersection and gauge theory}

We now consider the case in which $N_f$ D5-branes are added besides the $N_c$ D3-branes. We choose the D5-branes to span the $X_0,X_1,X_2$ directions, as well as the $X_4,X_5,X_6$ directions. Now, $X_3$ is parallel to the D3-branes but orthogonal to the D5-branes. We summarize the orientation of both sets of branes in Table \ref{tab:D3D5}. This intersection has an $SO(2,1)$ Lorentz symmetry along the directions $X_0,X_1,X_2$, and breaks the original $SO(6)$ of the $X_4,...,X_9$ coordinates into $SO(3)\times SO(3)$, along $(X_4,X_5,X_6)$ and $(X_7,X_8,X_9)$, respectively. The last $SO(3)$ symmetry will be further broken when the D5-branes are separated from the D3-branes, representing massive flavor fields.

\begin{table}
    \centering
    \begin{tabular}{ccccccccccc}
         & $X_0$ & $X_1$ & $X_2$ & $X_3$ & $X_4$ & $X_5$ & $X_6$ & $X_7$ & $X_8$ & $X_9$\\
        D3 & $\times$ & $\times$ & $\times$ & $\times$ & $-$ & $-$ & $-$ & $-$ & $-$ & $-$ \\
        D5 & $\times$ & $\times$ & $\times$ & $-$ & $\times$ & $\times$ & $\times$ & $-$ & $-$ & $-$ \\
    \end{tabular}
    \caption{D3/D5 brane intersection. Directions filled by the branes are denoted by $\times$. Directions orthogonal to them are denoted by $-$.}
    \label{tab:D3D5}
\end{table}

The decoupling limit in the open- and closed- string perspective when the low-energy limit is taken applies also in this case. However, the lower dimensionality of the intersection makes the flavor fields to live in $(2+1)$ dimensions. The resulting field theory corresponds to $\mathcal{N}=4$, $(3+1)$-dimensional SYM theory with gauge group $SU(N_c)$ coupled to $N_f$ $\mathcal{N}=4$ flavor hypermultiplets localized on the $(2+1)$-dimensional defect, giving rise to a \textit{defect field theory}. The Lagrangian and field content of the defect field theory is well known and it was constructed in detail in \cite{DeWolfe_2001}. Besides the adjoint fields of the $\mathcal{N}=4$ SYM theory, the $(2+1)$-dimensional flavor hypermultiplet contains two fermions (quarks) $\psi$ and two complex scalars (squarks) $q$.

It was shown in \cite{Karch_2001} that the probe D5-brane extends along an AdS$_4\times S^2$ of the ten-dimensional background. In this case, the holographic duality in the D5-brane worldvolume relates the fields at the boundary of AdS$_4$ to the operators of the dual (2+1)-dimensional defect theory.

\subsection{Probe D5-branes}

The action for $N_f$ D5-branes embedded in the background of $N_c$ D3-branes is
\begin{equation}
    S_{D5}=-N_f T_{D5}\int d^6\xi \sqrt{-\det\left(h_{ab}+2\pi \alpha' F_{ab}\right)}+2\pi\alpha' N_f T_{D5}\int P[C_{(4)}]\wedge F~,
\end{equation}
where $T_{\text{D5}}$ is the tension of the D5-brane, given by
\begin{equation}
    T_{\text{D5}}=\frac{1}{(2\pi)^5g_s\alpha'^3}~.
\end{equation}
Again, it is useful to adapt the coordinates in the AdS$_5\times S^5$ metric to the symmetries of the intersection. The Minkowski directions $(t,x_1,x_2)$ are common to the D3- and D5-branes. The $x_3$ coordinate is parallel to the D3-branes but orthogonal to the D5-branes. Of the remaining 6 transverse directions, the D5-branes span the 456-directions. We introduce spherical coordinates $(\rho,\Omega_2)$ with $\rho^2=x_4^2+x_5^2+x_6^2$ in these directions. The remaining three directions are transverse to both sets of branes, $(x_7,x_8,x_9)\equiv (w_1,w_2,w_3)$, which will parametrize the embedding of the D5-branes. The AdS$_5\times S^5$ metric can be written as
\begin{align}
\begin{split}
    ds^2=\frac{\rho^2+w_1^2+w_2^2+ w_3^2}{L^2}&\left(-dt^2+dx_1^2+dx_2^2+dx_3^2\right)\\
    +&\frac{L^2}{\rho^2+w_1^2+w_2^2+w_3^2}\left(d\rho^2+\rho^2d\Omega_2^2+dw_1^2+dw_2^2+dw_3^2\right)~.
    \label{eq:AdS5xS5D5brane}
\end{split}
\end{align}

The same symmetry arguments as in the case of a D7-brane allow us to consider only one of the three embedding functions, with its only dependence on $\rho$, $w_1=w_2=0$, $w_3\equiv w(\rho)$. Then, the induced metric on the D5-brane is
\begin{equation}
    ds^2=\frac{\rho^2+w^2}{L^2}\left(-dt^2+dx_1^2+dx_2^2\right)+\frac{L^2}{\rho^2+w^2}\left((1+w'^2)d\rho^2+\rho^2d\Omega_2^2\right)~,
\end{equation}
where primes again denote derivatives with respect to $\rho$ and we set $L=1$ from now on. The DBI action for this ansatz is
\begin{equation}
    S_{\text{D5}}=-\mathcal{N}\int d\rho~\rho^2\sqrt{1+w'^2}~,
\end{equation}
where again we have absorbed the prefactors into a constant $\mathcal{N}$, given by
\begin{equation}
    \mathcal{N}=4\pi N_f T_{D5}=\frac{N_fN_c\sqrt{\lambda}}{2\pi^3}~,
\end{equation}
having again divided by the infinite volume of $\mathbb{R}^{3,1}$. The profile $w(\rho)$ determines the way the D5-brane is embedded in the background geometry, with the 2-sphere shrinking to zero at a certain radial distance of the AdS factor. The difference with respect to the D3/D7 case is that now the subleading term of the asymptotic expansion has a different power of $\rho$,
\begin{equation}
    w(\rho)=M+\frac{C}{\rho}+...
\end{equation}

The qualitative picture in the presence of a non-zero temperature is the same as in the D3/D7 intersection: for low enough temperatures, the 2-sphere shrinks before intersecting the black hole horizon, giving rise to a Minkowski embedding. For higher temperatures, we will again obtain black hole embeddings, with both types of solutions separated by a critical embedding.

\subsection{Holographic renormalization and quark condensate}

We can repeat the same calculation done for the D3/D7 intersection, now for the D3/D5 case. Again, the boundary value of the embedding function $w(\rho)$ couples to the mass operator $\mathcal{O}_m$ in the boundary theory.
Again, it is convenient to rewrite the AdS$_5\times S^5$ metric in a similar form, in terms of $z=1/r$,
\begin{equation}
    ds^2=\frac{1}{z^2}\left(-dt^2+dx_1^2+dx_2^2+dx_3^2+dz^2\right)+d\theta^2+\sin^2\theta d\Omega_2^2+\cos^2\theta d\Omega_2^2~,
    \label{eq:AdS5xS5sliceD5}
\end{equation}
where the embedding function is $\theta(z)$, and the relation between the coordinates is the same as before,
\begin{equation}
    \frac{1}{z^2}=r^2=\rho^2+w^2~,\quad \tan\theta=\frac{w}{\rho}~.
\end{equation}
The induced metric on the D5-brane is
\begin{equation}
    ds^2=\frac{1}{z^2}\left(-dt^2+dx_1^2+dx_2^2\right)+\left(\frac{1}{z^2}+\theta'(z)^2\right)dz^2+\cos^2\theta d\Omega_2^2~.
    \label{eq:inducedD5thetaz}
\end{equation}

In these coordinates, the regularized action is given by
\begin{equation}
    S_{\text{D5}}^{reg}=-\mathcal{N}\int_\epsilon dz\frac{\cos^2\theta}{z^4}\sqrt{1+z^2\theta'^2}~,
\end{equation}
with $\theta(z)$ having an asymptotic expansion of the form
\begin{equation}
    \theta(z)\sim z\left(\theta_0+\theta_1 z+\mathcal{O}(z^2)\right)~.
\end{equation}
Again, the equation of motion determines all the higher-order coefficients in terms of $\theta_0$ and $\theta_1$. With this asymptotic behavior, the regularized action has a UV expansion of the form
\begin{equation}
    S_{\text{D5}}^{reg}=-\mathcal{N}\int_\epsilon dz\left[\frac{1}{z^4}-\frac{\theta_0^2}{2z^2}+\mathcal{O}(z^0)\right]=-\mathcal{N}\left[\frac{1}{3\epsilon^3}-\frac{\theta_0^2}{2\epsilon}+\mathcal{O}(\epsilon)\right]~.
\end{equation}
The list of necessary counterterms to cancel the divergences was also derived in \cite{Karch_2005_holorenobranes}. The counterterms for the case of a Ricci-flat metric at the regulator surface are
\begin{equation}
\begin{aligned}
    S^{ct}_1&=\frac{1}{3}\mathcal{N}\sqrt{-\gamma}~,&\quad \quad S^{ct}_2&=-\frac{1}{2}\mathcal{N}\sqrt{-\gamma}\theta(\epsilon)^2 ~,
\end{aligned}
\end{equation}
In this case, the renormalized $\langle \mathcal{O}_m\rangle$ is given by
\begin{equation}
    \langle \mathcal{O}_m\rangle =-(2\pi\alpha')\lim_{\epsilon\to 0}\epsilon\frac{\delta S_{\text{D5}}^{sub}}{\delta \theta(\epsilon)}~,
\end{equation}
with $S_{\text{D5}}^{sub}$ defined as before, $S_{\text{D5}}^{sub}=S_{\text{D5}}^{reg}+\sum_iS^{ct}_i$. The variation of each term now gives
\begin{equation}
\begin{aligned}
    \delta S^{reg}_{D5}&=\mathcal{N}\left[\frac{\theta_0}{\epsilon^2}+\frac{2\theta_1}{\epsilon}+\mathcal{O}(\epsilon^0)\right]\delta \theta(\epsilon)~,\\
    \delta S^{ct}_2 & = -\mathcal{N}\sqrt{-\gamma}\theta(\epsilon)=\mathcal{N}\left[-\frac{\theta_0}{\epsilon^2}-\frac{\theta_1}{\epsilon}+\mathcal{O}(\epsilon^0)\right]\delta \theta(\epsilon)~. 
\end{aligned}
\end{equation}

Therefore, the renormalized one point function $\langle \mathcal{O}_m\rangle$ in the D3/D5 model is given by
\begin{equation}
    \langle \mathcal{O}_m\rangle =-(2\pi\alpha')\lim_{\epsilon\to 0}\epsilon\frac{\delta S_{\text{D5}}^{sub}}{\delta \theta(\epsilon)}=-\mathcal{N}(2\pi\alpha')\h\theta_1~.
    \label{eq:Om1pfuncD3D5}
\end{equation}

We see again that the subleading term in the UV expansion of the field determines the one-point function of the dual field theory operator.

\section{Metallic AdS/CFT}\label{sec:metallicAdSCFT}

In the previous sections, we analyzed the D3/D7 and D3/D5 brane intersections without introducing gauge fields on their worldvolumes. In the next Chapter, we will examine in detail the field theory dual to the D3/D5 intersection in the presence of an external, periodic, time-dependent electric field. The dynamics of brane intersections under different gauge field configurations is significantly richer than the cases considered thus far. Therefore, before delving into the specific analysis, we outline here the general features that will be essential for the subsequent discussion. Once again, we will see that various phenomena in the gauge theory, arising when gauge fields are turned on, admit a natural geometric interpretation in the dual gravitational description.

We focus first on the qualitative features of the field theories dual to the D3/D7 and D3/D5 intersections. As previously mentioned, we work in the limit $N_f\ll N_c$, known in the field theory context as the \textit{quenched approximation}, which effectively neglects quark loop contributions.

As described in Section \ref{sec:D7probestemp}, at low temperatures and in the absence of an electric field, the quarks are bound into mesonic states with a discrete, gapped spectrum, computed in Section \ref{subsec:mesonspectrum}. In the gravitational picture, this corresponds to Minkowski embeddings of the probe branes. At higher temperatures, thermal fluctuations can dissociate the mesons, giving rise to the meson melting phenomenon. In this case, the probe branes intersect the black hole horizon, leading to black hole embeddings.

Introducing an external electric field produces similar qualitative effects. In particular, when both an electric field and a finite temperature are present, their combined influence governs the 
behavior of the system, as we will see in most of the results of Chapter \ref{chap:FloquetII}. As the electric field increases, the quark-antiquark pairs are pulled apart. Beyond a critical value, $E_{crit}$, mesons fully dissociate via the well-known Schwinger effect.

The charged quarks, once subject to an electric field, acquire a net velocity. Since the external field continuously injects energy into the system, the charges accelerate, resulting in a non-trivial time-dependent current. However, the quarks also interact with the adjoint degrees of freedom of the $\mathcal{N}=4$ SYM plasma, allowing energy and momentum to dissipate into the medium. The energy losses of the quarks scale as $N_fN_c$, whereas the energy density of the adjoint fields is of order $\mathcal{O}(N_c^2)$. In the limit $N_f\ll N_c$, the adjoint fields thus act as an effective heat bath that absorbs energy and momentum without significantly heating up. As a result, a balance between injected and dissipated energy can be achieved and the system evolves into a non-equilibrium steady state (NESS), which remains a valid approximation for times parametrically smaller than $N_c$. However, after a timescale of order $t\sim N_c$, the accumulated energy transfer to the bath becomes significant, and the NESS approximation ceases to be valid. In this thesis, we will restrict our analysis to the NESS regime.

Within this regime, we can define a conductivity tensor, $\sigma$, which characterizes the response current induced by the applied electric field
\begin{equation}
    \langle J_i\rangle=\sigma_{ij} E_j~,
\end{equation}
where $E_i$ are externally applied electric fields, and $\langle J_i\rangle$ denotes the induced currents. In general, determining the conductivity tensor is a highly non-trivial task, as it depends non-linearly on the electric field $E$, the temperature $T$, and the characteristic frequencies $\omega_n$ of possible time-dependent perturbations. In the regime where the electric field is a small perturbation to an equilibrium state, linear response theory applies, and the conductivity can be computed using the Kubo formula.

In the context of holography, the electric conductivity is typically extracted from the low-frequency behavior of bulk two-point functions of the gauge field. Pioneering works by Karch, O’Bannon, Sondhi, and Thompson \cite{Karch_2007, OBannon_2007, Karch_2008, Karch_2010} developed a method to compute the full nonlinear conductivity for arbitrary values of $E$ using the AdS/CFT correspondence in probe D-brane models. The method is much simpler in the sense that allows to extract the conductivity from a reality condition of the DBI action, by defining a special surface called the \textit{singular shell}. We will review this method below, where most of the material will follow these references, adapted to the specific cases relevant to this thesis.\footnote{For instance, their analysis includes a non-zero baryon density. In that case, a current is induced even for arbitrarily small electric fields, as the introduced charged quarks experience immediate acceleration. In the holographic dual description, this corresponds to the fact that only black hole embeddings exist at non-zero baryon density.}.

Another important effect takes place in the presence of electric fields: the charged quarks, when moving through the plasma, will continuously encounter a "hot wind" of oppositely charged quarks due to the Schwinger effect, moving in the opposite direction. Additionally, relativistic effects of the moving $\mathcal{N}=4$ plasma (in the reference frame of the quarks) affects their thermodynamic properties \cite{Liu_2006}. These effects give rise to an \textit{effective temperature}, which, in general, will be a function of the other parameters, $T_\text{eff}=T_\text{eff}(T,m,E)$. Notice that even at zero background temperature, we will generically have $T_\text{eff}\neq 0$ \cite{Kim_2011}.

In the brane picture, this effective temperature can be understood as the Hawking temperature of the open-string metric, describing the effective metric seen by the fluctuations on the probe brane. As we will see, the open-string metric has a horizon precisely at the singular shell.

\subsection{Nonlinear conductivity in AdS/CFT}\label{subsec:nonlinearsigma}

We consider the D3/D$q$-brane intersection at finite temperature, where later we will particularize to the cases $q=5$ and $q=7$.

The AdS$_5\times S^5$ metric can be written as
\begin{align}
\begin{split}
    ds^2=\frac{u^2}{L^2}&\left(-\frac{g(u)^2}{h(u)}dt^2+h(u)\left(dx^2+dy^2+dz^2\right)\right)+\frac{L^2}{u^2}du^2\\
    &\phantom{\left(-\frac{g(u)^2}{h(u)}dt^2+h(u)\left(dx^2\right.\right.}+L^2\left(\frac{d\psi^2}{1-\psi^2}+(1-\psi^2)d\Omega_n^2+\psi^2 d\Omega_{4-n}^2\right)~,
    \label{eq:metricD3Dqpsi}
\end{split}
\end{align}
where $u$ is an \textit{isotropic} coordinate which turns out to be useful at finite temperatures,
\begin{equation}
    u^2=\frac{r^2}{2}\left(1+\sqrt{1-\frac{r_h^4}{r^4}}\right)~,
    \label{eq:uconductivity}
\end{equation}
and the functions $g(u)$ and $h(u)$ are given by
\begin{equation}
    g(u)=1-\frac{u_h^4}{u^4}~~,\hspace{5mm}h(u)=1+\frac{u_h^4}{u^4}~.
\end{equation}
The black hole horizon is located at $u_h=r_h/\sqrt{2}$, and the corresponding Hawking temperature is given by
\begin{equation}
    T_H=\frac{r_h}{\pi L^2}=\frac{\sqrt{2}u_h}{\pi L^2}~,
\end{equation}
where we have written $T_H$ both in terms of the isotropic coordinate $u$ and the cartesian coordinate $r$. The 5-sphere has been parametrized as
\begin{equation}
    d\Omega_5^2=d\theta^2+\sin^2\theta d\Omega_n^2+\cos^2\theta d\Omega_{4-n}^2~,
\end{equation}
and we have defined $\psi\equiv\sin\theta$. For the D3/D7 and D3/D5 intersections, $n=3$ and $n=2$, respectively (see \eqref{eq:AdS5xS5sliceD7} and \eqref{eq:AdS5xS5sliceD5}, respectively). See Appendix \ref{app:coordinates} for details, where we provide the relations between the different coordinates used in the different chapters of the thesis.

Next, we introduce a worldvolume gauge field
\begin{equation}
    A_x(t,u)=-E\h t+A_x(u)~,
    \label{eq:gaugefieldE}
\end{equation}
which corresponds to a constant electric field $F_{xt}=E$ in the $x$-direction in the boundary theory, and we allow for a $u$ dependence in $A_x(u)$ in order to allow for a non-zero current. 

For an embedding of the form $\psi=\psi(u)$ and a fixed position in the $(4-n)$-sphere, the DBI action takes the form
\begin{equation}
    S_{Dq}=-\mathcal{N}_q\int g_{ii}^{(q-n-2)/2}g_{\Omega\Omega}^{n/2}\sqrt{g_{uu}\left(g_{ii}\abs{g_{tt}}-E^2\right)+g_{tt}A_x'^2}~,
    \label{eq:DBIactionDq}
\end{equation}
where the metric components are taken from \eqref{eq:metricD3Dqpsi}, with the modification $g_{uu} \to g_{uu} + \frac{\psi'^2}{1-\psi^2}$. The prefactor $\mathcal{N}_q$ is given by
\begin{equation}
    \mathcal{N}_q=N_fT_{Dq}V_n~,
\end{equation}
with $V_n$ the volume of the $n$-sphere, and the volume $\mathbb{R}^{p-n-1,1}$ has been factored out again.

Since the action \eqref{eq:DBIactionDq} depends only on $A_x'$ and not on $A_x$ explicitly, the system admits a conserved quantity, $J_x$, which allows us to express
\begin{equation}
    A_x'(u)=J_x\frac{\sqrt{g_{uu}}}{\sqrt{\abs{g_{tt}}}}\frac{\sqrt{g_{ii}\abs{g_{tt}}-E^2}}{\sqrt{g_{ii}^{q-n-2}\abs{g_{tt}}g_{\Omega\Omega}^n-J_x^2}}~.
    \label{eq:Axprime}
\end{equation}

The notation $J_x$ is not coincidental. From the equations of motion derived from \eqref{eq:DBIactionDq}, one can verify that $J_x$ corresponds precisely to the subleading term in the UV expansion of the gauge field. As is customary in the AdS/CFT correspondence, this quantity is identified with the boundary current $\langle J_x\rangle $. The identification is rigorously established by performing a careful variation of the renormalized on-shell action, taking into account the time-dependence of the gauge potential of \eqref{eq:gaugefieldE} \cite{Karch_2007, OBannon_2007}.

The system can be analyzed by deriving the equation of motion for the embedding function $\psi(u)$ from the action \eqref{eq:DBIactionDq}, and substituting the expression for $A_x'(u)$ from \eqref{eq:Axprime}. Alternatively, one can Legendre-transform the action with respect to $A_x'$, leading to
\begin{equation}
\begin{aligned}
    \bar{S}_{\text{Dq}}&=S_{\text{Dq}}-\int du~ A_x'\frac{\partial\mathcal{L}}{\partial A_x'}\\
    &=-\mathcal{N}_q\int du \frac{\sqrt{g_{uu}}}{\sqrt{g_{tt}}}\sqrt{\left(g_{ii}\abs{g_{tt}}-E^2\right)\left(\abs{g_{tt}}g_{ii}^{q-n-2}g_{\Omega\Omega}^n-J_x^2\right)}~.
\end{aligned}
\end{equation}

The observation in \cite{Karch_2007,OBannon_2007} is that, near the boundary, the two terms under the last square root are positive, since the factors $g_{ii}\abs{g_{tt}}$ and $\abs{g_{tt}}g_{ii}^{q-n-2}g_{\Omega\Omega}$ contain positive powers of $u$ in the limit $u\to \infty$. However, in the opposite limit, $u\to u_h$, these terms vanish, leading to two negative terms under the square root. Consequently, both terms under the square root change sign at some radial distance. To prevent an imaginary contribution to the action, both terms have to change sign at the same value of the radial coordinate, which we define as the \textit{critical radius}, denoted by $u=u_c$. The surface defined by $u=u_c$ is commonly referred to as the \textit{singular shell}, or \textit{pseudohorizon}, where the reason for the second name will become clear in the next section.

Remarkably, the condition that both terms under the square root vanish at the critical radius not only determines the location of the singular shell, but also fixes the conserved quantity $J_x$ in terms of the other parameters of the model, namely $E$ and $T$, as
\begin{equation}
    E=\sqrt{\abs{g_{tt}}g_{ii}}~\Big\rvert_{u=u_c}~,\qquad J_x=\sqrt{\abs{g_{tt}}g_{ii}^{q-n-2}g_{\Omega\Omega}^n}~\Big\rvert_{u=u_c}~.
\end{equation}
For the two cases of interest, this yields
\begin{equation}
\begin{aligned}
    E&=\frac{u_c^4-u_h^4}{u_c^2}~,\qquad &J_x&=(1-\psi_0^2)E~,& \qquad\text{D3/D5}\\
    E&=\frac{u_c^4-u_h^4}{u_c^2}~,\qquad &J_x&=\left(E^2+r_h^4\right)^{1/4}\left(1-\psi_0^2\right)^{3/2}E~,& \qquad\text{D3/D7}
    \label{eq:EJbothinters}
\end{aligned}
\end{equation}

This elegant mechanism for determining the response currents provides the aforementioned IR boundary condition that fully constrains the UV data. Notably, as previously mentioned, this result holds for arbitrary temperature and electric field, extending beyond the linear response regime, as long as the DBI action remains a valid approximation to the dynamics of the system.

Of particular interest is the case of the D3/D5 intersection, where, as seen from \eqref{eq:EJbothinters}, the conductivity is independent of both the applied electric field and the temperature. This striking feature motivated the authors of \cite{Karch_2010} to investigate the response of the system under arbitrary time-dependent electric fields. In this case, the conductivity can be generalized as
\begin{equation}
    J_x(t)=\int d\tau \sigma(\tau)E(t-\tau)~.
\end{equation}
Remarkably, their analysis demonstrated that, in the massless case, the conductivity remains constant even for arbitrary time-dependent electric fields and it is given by
\begin{equation}
    \sigma=\frac{2N_fN_c}{\pi\sqrt{\lambda}}~.
    \label{eq:sigmamassless}
\end{equation}
In particular, this relation must hold for the periodic electric field considered in Chapter \ref{chap:FloquetII}. The numerical results of Section \ref{sec:conductivitiesD3D5} confirm this behavior, with additional analytic proofs in different limits: in Appendix \ref{app:masslesssol}, we explicitly prove this result by solving the equations of motion analytically in the massless case. Furthermore, we extend this analytical proof to two additional scenarios: first, when considering small mass-deviations from the massless solution (Appendix \ref{app:smallmasssol}), and second, when studying AC conductivities by introducing a small, linearly polarized probe electric field on top of the rotating electric field (see Section \ref{subsec:opticalcond} and Appendix \ref{app:opticalcond} for details).

\subsection{Open-string metric and effective temperature}\label{subsec:Teff}

It turns out that the singular shell, introduced above as the surface where both terms under the square root in the action change sign simultaneously, is more than just a technical tool for obtaining response currents.

In the presence of background gauge fields, the open-string fluctuations on the probe branes do not couple to the background geometry $h_{ab}$, as in \eqref{eq:LD7fluct}, but instead to the so-called open-string metric (OSM) \cite{Seiberg_1999,Gibbons_2000}, given by
\begin{equation}
    \gamma_{ab}=h_{ab}+(2\pi\alpha')^2\mathcal{F}_{ac}\mathcal{F}_{bd}\h h^{cd}~,
    \label{eq:OSM}
\end{equation}
with $h_{ab}$ the induced metric on the brane and $\mathcal{F}$ the gauge-invariant field strength.

An interesting phenomenon arises when the OSM $\gamma_{ab}$ develops a horizon at a location distinct from that of the background metric, $h_{ab}$. This happens, for instance, in the presence of an electric field. In this case, the horizon of the OSM coincides precisely with the location of the singular shell \cite{Kim_2011}. As a result, for the open-string degrees of freedom on the probe brane, the OSM horizon acts as an event horizon, in the sense that it is a true causal boundary of spacetime. The associated Hawking temperature, distinct from the background temperature, is interpreted in the field theory as the effective temperature perceived by the quarks.

We now review the emergence of the OSM horizon and the effective temperature in the case of a constant electric field, which we will generalize in the next Chapter to rotating electric fields.

When a constant electric field is introduced, with the gauge field given in \eqref{eq:gaugefieldE}, the OSM \eqref{eq:OSM} takes the form
\begin{equation}
\begin{aligned}
    \gamma_{ab}d\xi^ad\xi^b=&-\left(\abs{g_{tt}}-\frac{E^2}{g_{xx}}\right)dt^2-\frac{2EA_x'}{g_{xx}}dt\h du+\left(g_{xx}-\frac{E^2}{\abs{g_{tt}}}+\frac{A_x'^2}{g_{uu}}\right)dx^2\\
    &+\left(g_{uu}+\frac{A_x'^2}{g_{xx}}\right)du^2+g_{xx}\left(dy^2+dz^2\right)+g_{\Omega\Omega}d\Omega_3^2~.
\end{aligned}
\end{equation}

To diagonalize the $(t,u)$ part of the OSM, we introduce a new time coordinate, $\tau$, via
\begin{equation}
    t=\tau-h(u)~,\qquad \text{where}\quad h'(u)=\frac{\gamma_{ut}}{\gamma_{tt}}~.
\end{equation}
For simplicity and clarity, we now focus on the massless case ($\psi=0$) with vanishing background temperature, $r_h=0$. The general case can be found in \cite{Kim_2011}. In the new variables, the metric reads
\begin{equation}
    \gamma_{ab}d\xi^ad\xi^b=-\frac{u^4-E^2}{u^2}d\tau^2+\frac{u^4}{u^6-E^3}du^2+\frac{u^4(E+u^2)}{u^4+E^2+E\h u^2}dx^2+u^2\left(dy^2+dz^2\right)+d\Omega_3^2~.
\end{equation}
This metric has an event horizon at $u=\sqrt{E}$, which precisely coincides with the location of the singular shell, as given by \eqref{eq:EJbothinters} for $u_h=0$. The result extends naturally to non-zero background temperature, $u_h\neq 0$. By expanding the Euclidean version of the metric around the horizon as $u=\sqrt{E}+\epsilon^2$ we obtain, for the $(\tau,\epsilon)$ space,
\begin{equation}
    \gamma_{ab}d\xi^ad\xi^b\big\rvert_{\tau,\epsilon}=4\sqrt{E}\h\epsilon^2\h d\tau^2+\frac{2}{3\sqrt{E}}d\epsilon^2~.
\end{equation}
Rescaling the coordinate $\epsilon$ as $\varepsilon^2\equiv\frac{2}{3\sqrt{E}}\epsilon^2$, the metric becomes
\begin{equation}
    \gamma_{ab}d\xi^ad\xi^b\big\rvert_{\tau,\varepsilon}=d\varepsilon^2+\varepsilon^26Ed\tau^2~.
\end{equation}
In order to avoid a conical singularity at the origin, the coordinate $\sqrt{6E}\h\tau$ needs to have periodicity $2\pi$, which means that $\tau$ needs to have periodicity $\beta=2\pi/\sqrt{6E}$. Its inverse defines the Hawking temperature of the OSM. For zero and non-zero background temperature, respectively, it is given by
\begin{equation}
\begin{aligned}
    T_\text{eff}&=\sqrt{\frac{3}{2}}\frac{\sqrt{E}}{\pi}~,\qquad &r_h=0~,\\
    T_\text{eff}&=\frac{1}{2\pi}\frac{(6E^2+4\h r_h^4)^{1/2}}{(E^2+r_h^4)^{1/4}}~,\qquad &r_h\neq 0~.
\end{aligned}
\end{equation}

For massive flavors, analogous expressions can be derived, although they become significantly more involved \cite{Kim_2011}. Notably, even when the background temperature is zero, the deconfined quarks experience a non-zero temperature due to the presence of the singular shell caused by the electric field. This temperature reflects the melting of sufficiently light mesons due to the Schwinger effect. Furthermore, we observe that $T_\text{eff}\ge T_H$, where the background Hawking temperature, $T_H=r_h/\pi$, is correctly recovered in the limit $E\to 0$.

In summary, we have seen that the singular shell, which initially appeared as a special surface where the IR boundary conditions can be imposed, acts as an event horizon for the open-string degrees of freedom propagating on the probe brane. For this reason, the singular shell will also be denoted as \textit{pseudohorizon} throughout the thesis.

\section{Chiral symmetry breaking in AdS/CFT}\label{sec:CSB}

We started this Chapter by discussing how flavor degrees of freedom can be incorporated into the AdS/CFT correspondence. The idea behind that was to bring holography closer to modeling real-world gauge theories, such as QCD. While a fully holographic description of QCD remains unknown, refined setups can capture some of its key features. A fundamental feature of low-energy QCD is the spontaneous breaking of chiral symmetry.

Various holographic models using probe branes exhibit chiral symmetry breaking. The simplest approach is to embed D7-branes in a non-supersymmetric gravity background. The first realization of this mechanism appeared in \cite{Babington_2003}, where D7-branes were placed in the Constable-Myers background \cite{Constable_1999}.

An alternative way to induce chiral symmetry breaking is by introducing a magnetic field $B$ in the original D3-brane geometry, a setup first explored in \cite{Filev_2007,Filev_2007_2}. This is the scenario we will focus on in this work.

Both constructions consider a relatively simple case, where the broken chiral symmetry is an abelian $U(1)_A$ symmetry. A more realistic realization of non-abelian chiral symmetry was proposed around the same time by Sakai and Sugimoto \cite{Sakai_2004, Sakai_2005}. Their model introduces $D8$ and $\overline{D8}$ probe branes in the background of a D4-brane stack.

We begin with a brief review of chiral symmetry breaking in QCD before turning to the holographic model of interest: a D3/D7 brane intersection with an external magnetic field on its worldvolume. This will be the main setup analyzed in Chapter \ref{chap:HelicalB}.

The Lagrangian of massless QCD is given by
\begin{equation}
    \mathcal{L}=-\frac{1}{4}F_{\mu\nu}^aF^{a\mu\nu}+\bar{\psi}_L\slashed{D}\psi_L+\bar{\psi}_R\slashed{D}\psi_R~,
\end{equation}
where $\psi_L$ and $\psi_R$ are the chiral components of the Dirac spinor $\Psi$,
\begin{equation}
    \Psi=\begin{pmatrix}
        \psi_L\\
        \psi_R
    \end{pmatrix}~.
\end{equation}
In the absence of a mass term, the QCD Lagrangian exhibits a global chiral symmetry $SU(N_f)\times SU(N_f)$, which acts independently on $\psi_L$ and $\psi_R$. Restricting to the three lightest quarks, $u$, $d$, $s$ ($N_f=3$), this symmetry takes the form
\begin{equation}
    \psi_L\rightarrow \exp\left(-i\theta_L\cdot \lambda\right)\psi_L~,\qquad \psi_R\rightarrow \exp\left(-i\theta_R\cdot \lambda\right)\psi_R~,
\end{equation}
where $\lambda^a$, with $a=1,...,8$ are the $SU(3)$ Gell-Mann matrices.

The introduction of a mass term,
\begin{equation}
    \mathcal{L}_m=-m\bar{\Psi}\Psi~,
\end{equation}
explicitly breaks this symmetry, reducing it to a single vector-like $SU(3)_V$. This explicit breaking is small for the light quarks, but more pronounced for heavier flavors.

Even in the absence of explicit breaking, QCD dynamics induce a spontaneous breaking of chiral symmetry due to strong interactions. Non-perturbative effects generate a non-vanishing expectation value for the quark bilinear operator
\begin{equation}
    \langle\bar{\Psi}\Psi\rangle=\langle\bar{\psi_L}\psi_R\rangle+h.c.\neq 0~,
\end{equation}
which acts as an order parameter for spontaneous chiral symmetry breaking. This non-zero condensate breaks again $SU(N_3)\times SU(N_3)$ down to $SU(3)_V$.

As a consequence of this symmetry breaking, the theory exhibits massless Goldstone bosons associated with the broken generators. In the case of three flavors, these correspond to the pions, kaons, and the eta meson, which acquire a small mass due to the explicit breaking by the quark mass term.

As mentioned earlier, the full non-abelian chiral symmetry breaking can be studied holographically using the Sakai-Sugimoto model. In the setup we will focus on, however, only a simpler abelian $U(1)_A$ symmetry is considered, under which $\psi_L$ and $\psi_R$ transform as
\begin{equation}
    \psi_L\to e^{i\alpha}\psi_L~,\qquad \psi_R\to e^{-i\alpha}\psi_R~.
\end{equation}
\subsection{A holographic model}

In Section \ref{sec:D3D7}, we analyzed the D3/D7 brane intersection at both zero and finite temperature. At zero temperature, supersymmetry prevents the formation of a quark condensate, and the embedding function is a constant, $w(\rho)=w_0$, meaning that the brane remains flat. At finite temperature, supersymmetry is completely broken, allowing for the emergence of a non-zero quark condensate. However, such a condensate only appears for non-zero quark mass: the only massless embedding is the flat configuration, which has a vanishing condensate.

To achieve chiral symmetry breaking, additional ingredients are required. The key observation of \cite{Filev_2007,Filev_2007_2} was that the introduction of a non-zero magnetic field $B$ induces a non-zero vacuum expectation value for the operator $\langle\bar{\Psi}\Psi\rangle$. Here we briefly review this effect, extending the discussion to finite temperature.

We consider $N_f$ probe D7-branes embedded in the geometry sourced by $N_c$ D3-branes, with $N_f\ll N_c$, so that we work in the probe limit. The field theory is taken at finite temperature, which means the background geometry contains an AdS$_5$ black hole. For this reason, the isotropic radial coordinate introduced in \eqref{eq:uconductivity} is the most suitable choice. Here, we adopt the cartesian embedding $(\rho,w)$, in which the metric takes the form
\begin{equation}
    ds^2=\frac{u^2}{L^2}\left(-\frac{g^2(u)}{h(u)}dt^2+h(u)(dx^2+dy^2+dz^2)\right)+\frac{L^2}{u^2}\Big(d\rho^2+dw^2+\rho^2d\Omega_3^2+w^2d\phi^2\Big)~,
\end{equation}
where
\begin{equation}
    g(u)=1-\frac{u_h^4}{u^4}~~,\hspace{5mm}h(u)=1+\frac{u_h^4}{u^4}~
\end{equation}
and $u^2=\rho^2+w^2$. The black hole horizon is located at $u_h=r_h/\sqrt{2}$. Again, we refer to Appendix \ref{app:coordinates} for details on the coordinate choice.

In this setup, we introduce a worldvolume gauge field
\begin{equation}
    2\pi\alpha'A=B\h x\h dy~,
\end{equation}
corresponding to a magnetic field in the $z$-direction. Assuming an embedding of the form $w=w(\rho)$, the DBI action for the D7-branes is
\begin{equation}
    S_{D7}=-\mathcal{N}\int d\rho ~\frac{\rho^3g}{\rho^2+w^2}\sqrt{\left(B^2+h\left(\rho^2+w^2\right)^2\right)\left(1+w'^2\right)},
    \label{eq:DBIBfield}
\end{equation}
where $\mathcal{N}=N_fT_{D7}$, and, as usual, we have implicitly divided both sides by the infinite volume of $\mathbb{R}^{3,1}$ and we work in units where $L=1$.

From the equation of motion derived from the action \eqref{eq:DBIBfield}, the asymptotic behavior of the embedding function $w(\rho)$ is found to be
\begin{equation}
    w(\rho)=M+\frac{C}{\rho^2}+...~,
\end{equation}
where the constant $C$ is related to the quark condensate via the holographic dictionary.

\begin{figure}
    \centering
    \begin{subfigure}[t]{0.48\textwidth}
        \centering
        \includegraphics[width=\textwidth]{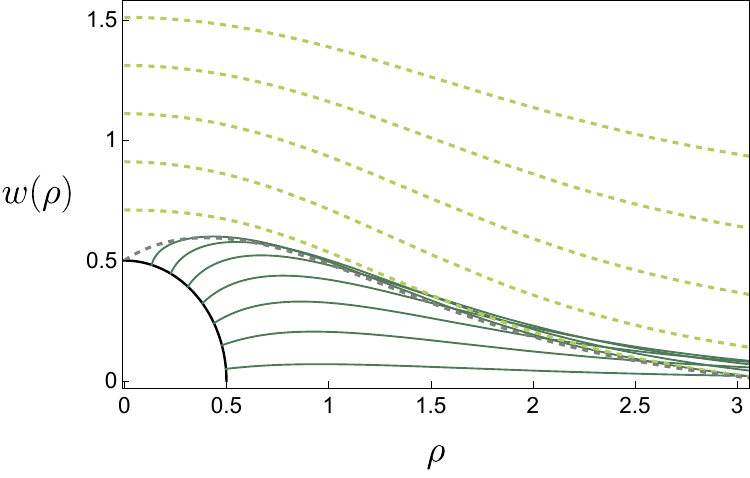}
    \end{subfigure}
    \hspace{0.02\textwidth}
    \begin{subfigure}[t]{0.48\textwidth}
        \centering
        \includegraphics[width=\textwidth]{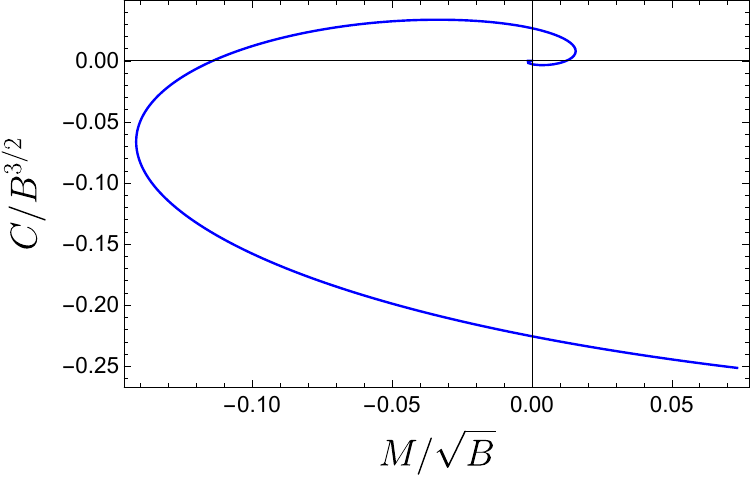}
    \end{subfigure}
    \caption{Left plot: Typical profiles for the embedding function $w(\rho)$ with $u_h=0.5$ and $B=5$. Non-flat embeddings with zero mass signal a non-vanishing quark condensate, thus realizing chiral symmetry breaking. Right plot: The spiral structure of the quark condensate.}
    \label{fig:CSBembeddingscondensate}
\end{figure}

As discussed earlier, at finite temperature, the system admits both black hole and Minkowski embeddings. By solving the equation for different values of $w_0\equiv w(0)$ or $w_0\equiv w(\rho_h)$ for Minkowski / black hole embeddings, respectively, we obtain the characteristic profiles of the embedding function $w(\rho)$, as shown in the left plot of Fig. \ref{fig:CSBembeddingscondensate}. The effects of the magnetic field and temperature act in opposite directions: while the black hole horizon pulls the branes towardsss it, the magnetic field induces a repulsive effect near the origin. This results in a bending of the brane, allowing for a zero quark mass in the ultraviolet while still generating a non-zero condensate, determined by $C$. The right plot of Fig. \ref{fig:CSBembeddingscondensate} illustrates the relation between $C$ and $m$, showing a finite condensate even when $M=0$. The spiral structure of the $C$-$M$ curve is a well-known feature of chiral symmetry breaking models. We will explore this behavior in more detail in the context of a helical magnetic field in Chapter \ref{chap:HelicalB}, where we will see that for sufficiently helical configurations, chiral symmetry is restored.

%% file: Part1/Braneintersections/BraneintersectionsAppendices.tex
\renewcommand{\chapterquote}{}
\chapterappendix{\thechapter}

\setcounter{equation}{0}
\setcounter{appendixsection}{\value{appendixsection}+1} 
\section{Coordinates}\label{app:coordinates}

In this appendix we provide the details between the coordinates used in the different chapters of the thesis. We will try to motivate the choice of coordinates in each case.

We begin by considering the Schwarzschild-AdS$_5\times S^5$ geometry,
\begin{equation}
    ds^2=\frac{r^2}{L^2}\left(-f(r)dt^2+dx^2+dy^2+dz^2\right)+\frac{L^2}{r^2}\left(\frac{dr^2}{f(r)}+r^2d\Omega_5^2\right)~,
    \label{eq:AdS5xS5app}
\end{equation}
where $f(r)$ is the blackening factor,
\begin{equation}
    f(r)=1-\frac{r_h^4}{r^4}~.
\end{equation}
At finite temperature, it is convenient to introduce an "isotropic" coordinate, $u$, defined as
\begin{equation}
    u^2=\frac{r^2}{2}\left(1+\sqrt{1-\frac{r_h^4}{r^4}}\right)~.
    \label{eq:u}
\end{equation}
The metric becomes
\begin{equation}
    ds^2=\frac{u^2}{L^2}\left(-\frac{g(u)^2}{h(u)}dt^2+h(u)\left(dx^2+dy^2+dz^2\right)\right)+\frac{L^2}{u^2}\left(du^2+u^2 d\Omega_5^2\right)~,
    \label{eq:metricisotropic}
\end{equation}
where the functions $g(u)$ and $h(u)$ are given by
\begin{equation}
    g(u)=1-\frac{u_h^4}{u^4}~~,\hspace{5mm}h(u)=1+\frac{u_h^4}{u^4}~.
\end{equation}
The black hole horizon is now at $u_h=r_h/\sqrt{2}$. Notice that the effect of this change of coordinates is to remove the blackening factor from the $g_{rr}$ component of the metric, making the presence of a flat 6-plane perpendicular to the horizon manifest. At zero temperature, where $r_h=0$ and $f(r)=1$, this is already manifest, and the two sets of coordinates coincide. Moreover, near the UV boundary, $u\sim r$.

We now embed a D$q$-brane in this geometry. A useful parametrization of the 5-sphere is
\begin{equation}
    d\Omega_5^2=d\theta^2+\cos^2\theta d\Omega_n^2+\sin^2\theta d\Omega_{4-n}^2~,
    \label{eq:metric5spheretheta}
\end{equation}
where $n$ denotes the number of coordinates parallel to both sets of branes. In the D3/D7 model, $n=3$, while in the D3/D5, $n=2$.

We define also $\psi\equiv \sin\theta$, in which case the metric becomes
\begin{equation}
    ds^2=\frac{u^2}{L^2}\left(-\frac{g(u)^2}{h(u)}dt^2+h(u)d\Vec{x}_3^2\right)+\frac{L^2}{u^2}du^2+L^2\left(\frac{d\psi^2}{1-\psi^2}+(1-\psi^2)d\Omega_n^2+\psi^2 d\Omega_{4-n}^2\right)~.
    \label{eq:AdS5xS5psi}
\end{equation}
with $\Vec{x}_3=(x,y,z)$. These will be the coordinates that we will generically use in Chapter \ref{chap:FloquetII}, except for the analytic solutions of Appendix \ref{app:analyticsols}, as we will detail below.

We will also consider \textit{cartesian} coordinates, $(\rho,w)$, related to $(u,\theta)$ as
\begin{equation}
    \rho=u\cos\theta~,\qquad w=u\sin\theta~,
    \label{eq:relationcartesianangular}
\end{equation}
which satisfy
\begin{equation}
    d\rho^2+dw^2=du^2+u^2d\theta^2~.
\end{equation}
In these coordinates, the metric \eqref{eq:AdS5xS5psi} becomes
\begin{equation}
    ds^2=\frac{u^2}{L^2}\left(-\frac{g(u)^2}{h(u)}dt^2+h(u)d\Vec{x}_3^2\right)+\frac{L^2}{u^2}\left(d\rho^2+dw^2+\rho^2d\Omega_n^2+w^2d\Omega_{4-n}^2\right)~,
    \label{eq:AdS5xS5wu}
\end{equation}
where $u$ has to be understood as written in terms of $\rho$ and $w$, \ie,
\begin{equation}
    u^2=\rho^2+w^2~.
\end{equation}

At zero temperature, where $g(u)=h(u)=1$, and $u=r$, the metric \eqref{eq:AdS5xS5wu} reduces to
\begin{equation}
    ds^2=\frac{\rho^2+w^2}{L^2}\left(-dt^2+d\Vec{x}_3^2\right)+\frac{L^2}{\rho^2+w^2}\left(d\rho^2+dw^2+\rho^2d\Omega_n^2+w^2d\Omega_{4-n}^2\right)~.
    \label{eq:AdS5xS5wrho}
\end{equation}

Notice that, although the name \textit{cartesian} is used, $w$ denotes the radius of the $\Omega_{4-n}$ space in spherical coordinates. This radius (squared) is the one written as $w_1^2+w_2^2$ in \eqref{eq:AdS5xS5D7brane} when we first introduced the D3/D7 intersection, and $w_1^2+w_2^2+w_3^2$ in \eqref{eq:AdS5xS5D5brane} for the D3/D5 case.

There is one case in which we will be interested in using the original radial coordinate, $r$, at finite temperature. This is the case of the analytic solutions of Appendix \ref{app:analyticsols}. In that case, we are going to use the original metric, \eqref{eq:AdS5xS5app}, with the parametrization of the 5-sphere in terms of $\theta$, \eqref{eq:metric5spheretheta}. The metric in this case is
\begin{equation}
    ds^2=\frac{r^2}{L^2}\left(-f(r)dt^2+dx^2+dy^2+dz^2\right)+\frac{L^2}{r^2}\frac{dr^2}{f(r)}+L^2\left(d\theta^2+\cos^2\theta d\Omega_n^2+\sin^2\theta d\Omega_{4-n}^2\right)~.
    \label{eq:AdS5xS5formassless}
\end{equation}

\restoredefaultnumbering
\restoredefaultsectioning

%% file: Part1/SYKwormholes/SYKwormholes.tex
\renewcommand{\chapterquote}{\textit{“Not only does God play dice, but...\\He sometimes throws them where they cannot be seen.”} \\[0.5em] \normalfont— Stephen Hawking, \textit{The Nature of Space and Time}.}
\chapter{The SYK model and traversable wormholes}\label{chap:SYKwormholes}

This Chapter departs from the topics of the previous two and shifts focus to traversable wormholes and the SYK model. We begin by reviewing the Gao-Jafferis-Wall mechanism, which shows how a direct coupling between asymptotic boundaries can render certain wormholes traversable \cite{Gao_2016}. We then present the Sachdev-Ye-Kitaev (SYK) model, highlighting its main features and its connection to nearly-AdS$_2$ gravity. Particular emphasis is given to the coupled SYK system proposed by Maldacena and Qi, which provides a concrete realization of a traversable wormhole \cite{Maldacena_2018}. Finally, we introduce the Schwinger-Keldysh formalism as a tool to study the real-time and out-of-equilibrium dynamics of the model, which will play a central role in the research chapters of this thesis.

The material presented here can be found in several comprehensive reviews and key original works, including \cite{Maldacena_Stanford_SYK,Gao_2016,Kitaev_2017,Sarosi_2017,Maldacena_2018,Rosenhaus_2018,Trunin_2020,Ramallo_SYK,Maldacena_2016,Mertens_2022,Turiaci_2024}.

\vspace{1cm}

\section{Motivation}

The SYK model is a quantum mechanical model involving $N$ Majorana fermions with random all-to-all quartic interactions. It was proposed in 2015 by Alexei Kitaev in a series of talks \cite{KitaevTalk1,KitaevTalk2} as a toy model for quantum gravity. Based on earlier work by Sachdev and Ye in the 1990s \cite{Sachdev_1992,Sachdev_2010}, the SYK model has recently attracted considerable attention for a wide range of reasons. Notably, the model is exactly solvable in the large-$N$ limit, even analytically in some cases. At low energies, the model exhibits an emergent (weakly broken) conformal symmetry, whose breaking is governed by a soft mode described by the Schwarzian action. The same action governs the dynamics of the only propagating degree of freedom in Jackiw-Teitelboim (JT) gravity \cite{Jackiw_1984,Teitelboim_1983}, a dilaton gravity theory in nearly-AdS$_2$ spacetime, which captures the near-horizon physics of nearly-extremal black holes \cite{Almheiri_2014,Nayak_2018,Kolekar_2018}.

In addition, the model provides a remarkable scenario to study quantum chaos and information scrambling \cite{Sonner_2017,LantagneHurtubise_2019,Kobrin_2020}. Quantum chaos, characterized by a sensitive dependence on initial conditions, manifests in complex many-body systems as the rapid delocalization of quantum information. In the large-$N$ regime, a key signature of this chaotic behavior is the exponential growth of out-of-time-ordered correlators (OTOCs), characterized by the Lyapunov exponent, $\lambda_L$ \cite{Shenker_2013,Bagrets_2017,Liao_2018,Anous_2019}. In the low-energy limit, the Lyapunov exponent in the SYK model saturates a universal bound \cite{Maldacena_2015,Maldacena_Stanford_SYK}. This mirrors the behavior observed in black hole scattering processes, which make them the fastest scramblers in nature \cite{Hayden_2007, Sekino_2008,Shenker_2013}. For these reasons, the SYK model is an exceptional toy model for exploring a wide range of phenomena, from aspects of condensed matter systems, such as strange metals, to fundamental questions in quantum chaos, information scrambling, and the dynamics of black holes.

Building on insights from semiclassical gravity, Gao, Jafferis, and Wall demonstrated that a wormhole can be rendered traversable by introducing a direct interaction between the two asymptotic boundaries of a maximally extended black hole spacetime \cite{Gao_2016}. Motivated by this mechanism and the deep analogies between SYK dynamics and AdS$_2$ gravity, Maldacena and Qi constructed a quantum mechanical model consisting of two coupled SYK systems \cite{Maldacena_2018}. This model provides a fully quantum realization of a traversable wormhole.

The traversability of this wormhole admits a clear interpretation within the quantum system \cite{Plugge_2020}. Consider the ground state of the model, $\ket{\Psi_0}$. In the absence of a coupling between the two SYK subsystems, an excitation created on the right (for instance, by acting with a Majorana operator, $\chi^j_R\ket{\Psi_0}$), quickly evolves into a highly scrambled state due to the chaotic nature of the SYK model. However, once a small coupling $\mu$ is turned on, a remarkable phenomenon takes place: after a characteristic time $t_{re}$, the excitation is unscrambled and reemerges on the left, with the state of the system approaching $\chi^j_L\ket{\Psi_0}$. In the gravitational picture, this process corresponds to the excitation having traversed the wormhole \cite{LantagneHurtubise_2019,Plugge_2020,Sahoo_2020,Haenel_2021}.

This striking mechanism has led to the development of proposals for "wormhole-inspired" teleportation protocols \cite{Gao_2019,Jafferis_2022}, some of which have been realized in experimental platforms designed to simulate aspects of SYK physics and quantum gravity \cite{Danshita_2016,Pikulin_2017,Chew_2017,Chen_2018,Franz_2018}. The ability of periodic drivings to modify quantum systems in a controlled way opens the possibility of improving these protocols by periodically driving certain parameters. Even if direct improvements are not achieved, studying the wormhole under different non-equilibrium perturbations can provide valuable insights into the underlying mechanisms that enable this teleportation.

\section{Traversable wormholes}\label{sec:traversableWHs}

\begin{figure}
\centering
\begin{tikzpicture}[baseline={(current bounding box.center)}]
    \def\L{2.5cm} 
    \draw[ultra thick,black]    (-\L,-\L) -- (-\L,\L);
    \draw[ultra thick,black]    (\L,-\L) -- (\L,\L);
    \draw[thick,dashed,black]   (-\L,-\L) -- (\L,\L);
    \draw[thick,dashed,black]   (\L,-\L) -- (-\L,\L);

    \draw[black] (-1.05*\L,0) node[left] {$L$};
    \draw[black] (1.05*\L,0) node[right] {$R$};

    \draw[decorate, decoration={snake, amplitude=2pt, segment length=10pt}, black, thick]
        (-\L,\L) -- (\L,\L);
    \draw[decorate, decoration={snake, amplitude=2pt, segment length=10pt}, black, thick]
        (-\L,-\L) -- (\L,-\L);
\end{tikzpicture}
\caption{Penrose diagram of AdS.}
\label{fig:PenroseAdS}
\end{figure}
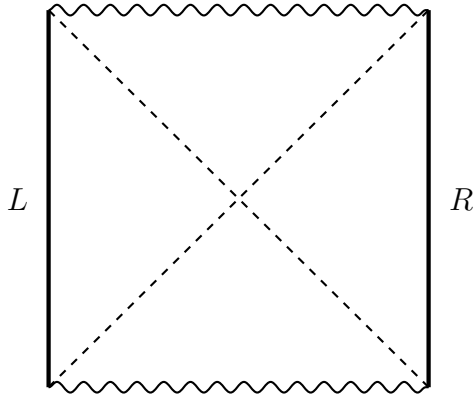

We begin by considering the maximally extended Schwarzschild-AdS geometry, whose Penrose diagram is depicted in Fig. \ref{fig:PenroseAdS}. It contains two asymptotically AdS regions, denoted by $L$ and $R$, connected in the bulk by an Einstein-Rosen bridge. For an observer in either region, the space appears as a black hole, with the corresponding white hole. The horizons are represented by dashed null lines, while curvy lines denote the future and past singularities.

In the context of AdS/CFT, each bulk geometry is dual to a particular state of the boundary field theory. The eternal Schwarzschild-AdS black hole is dual to the thermofield double (TFD) state \cite{Maldacena_2001}, defined in the Hilbert space of two copies of the CFT, each living on one of the asymptotic boundaries. The TFD state is given by
\begin{equation}
    \ket{\text{TFD}}=\frac{1}{\sqrt{Z(\beta)}}\sum_n e^{-\beta E_n/2}\ket{\smash{\bar{E}_n}}_L\ket{E_n}_R~,
\end{equation}
where $\ket{E_n}_{L,R}$ are the energy eigenstates of the left and right CFTs, respectively, and $\ket{\smash{\bar{E}_n}}_L=\Theta \ket{E_n}_L$, where $\Theta$ denotes the time-reversal operator. $Z(\beta)$ is the partition function of one copy of the CFT at inverse temperature $\beta$. The TFD is a maximally entangled state, and it provides a remarkable example of the ER$\h=\h$EPR conjecture, which postulates an equivalence between wormhole geometries with an Einstein-Rosen (ER) bride and quantum entanglement (EPR) \cite{VanRaamsdonk_2010,Maldacena_2013}. However, it is crucial to note that the two CFTs do not interact; the total Hamiltonian of the system is simply $H=H_L+H_R$.

\begin{figure}
\centering
\begin{subfigure}[b]{0.5\textwidth}
    \centering
    \raisebox{0.24\height}{ 
    \begin{tikzpicture}
    \def\radius{1} 
    \def\sep{3.5} 
    \def\npoints{200} 
    
    \draw[thick] (0,0) circle (\radius);
    \draw[black] (0,-1.3*\radius) node[below] {CFT $L$};
    \begin{scope}[opacity=0.3, transparency group]
    \foreach \i in {1,...,\npoints} {

    \pgfmathsetmacro{\x}{2*rnd*\radius - \radius}
    \pgfmathsetmacro{\y}{2*rnd*\radius - \radius}
    \pgfmathsetmacro{\dist}{\x*\x + \y*\y}
    \pgfmathsetmacro{\check}{\dist - 0.7*\radius*\radius}
        \ifdim \check pt < 0pt
            \fill[blue] (\x,\y) circle (1.3pt);
        \else
            \i=\i-1
        \fi}
    \end{scope}
    \draw[thick] (\sep,0) circle (\radius);
    \draw[black] (\sep,-1.3*\radius) node[below] {CFT $R$};
    \begin{scope}[opacity=0.3, transparency group]
    \foreach \i in {1,...,\npoints} {
    \pgfmathsetmacro{\x}{\sep+2*rnd*(\radius) - \radius}
    \pgfmathsetmacro{\y}{2*rnd*\radius - \radius}
    \pgfmathsetmacro{\dist}{(\x-\sep)*(\x-\sep) + \y*\y}
    \pgfmathsetmacro{\check}{\dist - 0.7*\radius*\radius}
        \ifdim \check pt < 0pt
            \fill[blue] (\x,\y) circle (1.3pt);
        \else
            \i=\i-1
        \fi
    }
    \end{scope}

    \fill[red] (\sep+0.5*\radius, 0.5*\radius) circle (2pt);
    \draw[-{Stealth[length=10pt, width=5pt]}, red, thick] (\sep+0.5*\radius, 0.5*\radius) -- (\sep-0.4*\radius,-0.6*\radius) -- (\sep+0.8*\radius,-0.1*\radius) -- (\sep+0*\radius,-0.8*\radius) -- (\sep-0.6*\radius,0.6*\radius) -- (\sep-0.4*\radius,-0.7*\radius) -- (\sep-0.1*\radius,0.7*\radius) -- (\sep+0.9*\radius,-0.2*\radius) -- (\sep-0.7*\radius,0.2*\radius) -- (\sep+0.5*\radius,0.7*\radius);

\end{tikzpicture}
}
    \caption{}
    \label{sfig:CFTpert}
\end{subfigure}
\begin{subfigure}[b]{0.45\textwidth}
    \centering
    \begin{tikzpicture}
    \def\L{2.5cm} 
    
    \draw[-{Stealth[scale=1]},thick,red]   (\L,-\L+0.1*\L) -- (-\L+0.1*\L+0.05*\L,\L-0.05*\L);
    
    \draw[ultra thick,black]    (-\L,-\L) -- (-\L,\L);
    \draw[ultra thick,black]    (\L,-\L) -- (\L,\L);
    \draw[thick,dashed,black]   (-\L,-\L) -- (\L,\L);
    \draw[thick,dashed,black]   (\L,-\L) -- (-\L,\L);

    \draw[black] (-1.05*\L,0) node[left] {$L$};
    \draw[black] (1.05*\L,0) node[right] {$R$};

    \draw[decorate, decoration={snake, amplitude=2pt, segment length=10pt}, black, thick]
        (-\L,\L) -- (\L,\L);
    \draw[decorate, decoration={snake, amplitude=2pt, segment length=10pt}, black, thick]
        (-\L,-\L) -- (\L,-\L);

    \fill[red] (\L,-\L+0.11*\L) circle (2.5pt);
    \draw[red] (1.05*\L,-\L+0.11*\L) node[right] {$\phi_R$};
    \end{tikzpicture}
    \caption{}
    \label{sfig:AdSpert}
\end{subfigure}
\caption{Left: Perturbation in the $R$ CFT. Since the two theories are decoupled, no signals can be transmitted between them. Right: Null perturbation in AdS. In this geometry, the perturbation falls into the black hole and reaches the singularity, not being able to reach the left asymptotic boundary.}
\end{figure}
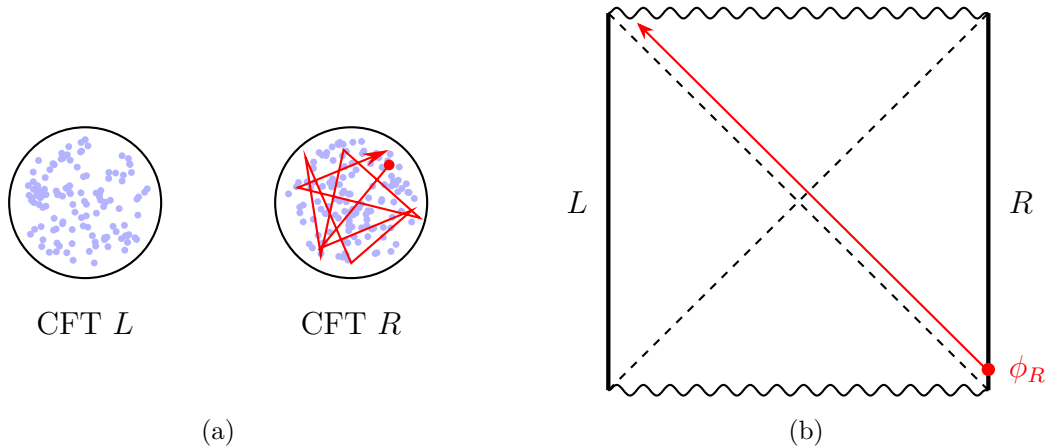

Because the two sides evolve independently, any perturbation added to one CFT remains confined to that side indefinitely (Fig. \ref{sfig:CFTpert}). This is mirrored in the gravitational picture: a small null perturbation in the bulk enters the black hole horizon and reaches the singularity without escaping to the opposite boundary, as shown in Fig. \ref{sfig:AdSpert}. Thus, the wormhole is non-traversable, consistently with the fact that EPR entangled pairs cannot be used to transmit information between the left and right systems.

The underlying reason for this non-traversability is the Averaged Null Energy Condition (ANEC), which states that for an infinite, null, achronal\footnote{A geodesic is achronal if no two of its points have a timelike separation. A well-known example of a chronal null geodesic where the ANEC does not apply is the case of Minkowski space compactified to a cylinder, leading to a negative Casimir energy.} geodesic with affine parameter $\lambda$, the integral of the null-null component of the energy-momentum tensor along the geodesic must be non-negative \cite{Graham_2007,Kontou_2012,Kontou_2015},
\begin{equation}
    \int_{-\infty}^\infty T_{\mu\nu}k^\mu k^\nu d\lambda \ge 0~.
    \label{eq:ANEC}
\end{equation}

Intuitively, in order for signals to be able to emerge in the other asymptotic region, we would need that a congruence of null geodesics, which focus towards each other on one end of the wormhole, would defocus when going out on the other end. According to the Raychaudhuri equation, this is only possible if the ANEC is violated\footnote{Catalan speakers may find this alarming, but no ducks will be harmed in this discussion.}.

While the impossibility of information transfer between the two sides appears robust from the CFT perspective, in the bulk one notices that a perturbation applied sufficiently far in the past seems "almost" capable of crossing to the opposite region. This naturally raises the question: could an additional perturbation modify the geometry in such a way as to allow a signal to traverse the wormhole?

\begin{figure}
\centering
\begin{subfigure}[b]{0.5\textwidth}
    \centering
    \begin{tikzpicture}
    \def\L{2.5cm} 
    \draw[ultra thick,black]    (-\L,-\L) -- (-1.75*\L,\L);
    \draw[ultra thick,black]    (\L,-\L) -- (\L,\L);
    \draw[thick,dashed,black]   (-\L,-\L) -- (\L,\L);
    \draw[thick,dashed,black]   (\L,-\L) -- (-\L,\L);
    \draw[thick,dashed,black]   (\L-0.75*\L,-\L) -- (-1.75*\L,\L);

    \draw[{Stealth[scale=1]}-{Stealth[scale=1]},black]   (-0.33*\L,0.27*\L) -- (-0.65*\L,-0.05*\L);
    \draw[black] (-0.35*\L,0.25*\L) node[left] {$\Delta X^+$};

    \draw[black] (-1.5*\L,0) node[left] {$L$};
    \draw[black] (1.05*\L,0) node[right] {$R$};

    \draw[decorate, decoration={snake, amplitude=2pt, segment length=10pt}, black, thick]
        (-1.75*\L,\L) -- (\L,\L);
    \draw[decorate, decoration={snake, amplitude=2pt, segment length=10pt}, black, thick]
        (-\L,-\L) -- (\L,-\L);

    \draw[-{Stealth[scale=1]},thick,red]   (\L,-\L+0.1*\L) -- (-\L+0.1*\L+0.05*\L,\L-0.05*\L);
    \fill[red] (\L,-\L+0.11*\L) circle (2.5pt);
    \draw[red] (1.05*\L,-\L+0.11*\L) node[right] {$\phi_R$};

    \draw[decorate, decoration={snake, amplitude=2pt, segment length=5pt}, green, very thick]
        (-\L,-\L+0.1*\L) -- (\L-0.1*\L-0.05*\L,\L-0.05*\L);
    \draw[green] (-1.05*\L,-\L) node[left] {$\boldsymbol{t_0}$};
    
    \end{tikzpicture}
    \caption{}
    \label{sfig:AdSdeformed}
\end{subfigure}
\hfill
\begin{subfigure}[b]{0.42\textwidth}
    \centering
    \begin{tikzpicture}
    \def\L{2.5cm} 
    \draw[ultra thick,black]    (-\L,-\L) -- (-\L,\L);
    \draw[ultra thick,black]    (\L,-\L) -- (\L,\L);
    \draw[thick,dashed,black]   (-\L,-\L) -- (\L,\L);
    \draw[thick,dashed,black]   (\L,-\L) -- (-\L,\L);

    \draw[black] (-1.05*\L,0) node[left] {$L$};
    \draw[black] (1.05*\L,0) node[right] {$R$};

    \draw[decorate, decoration={snake, amplitude=2pt, segment length=10pt}, black, thick]
        (-\L,\L) -- (\L,\L);
    \draw[decorate, decoration={snake, amplitude=2pt, segment length=10pt}, black, thick]
        (-\L,-\L) -- (\L,-\L);

    \draw[thick,red]   (\L,-\L+0.1*\L) -- (0*\L,0.1*\L);
    \draw[-{Stealth[scale=1]},thick,red] (0.3*\L,0.4*\L) -- (-0.25*\L,0.95*\L);
    \fill[red] (\L,-\L+0.11*\L) circle (2.5pt);
    \draw[red] (1.05*\L,-\L+0.11*\L) node[right] {$\phi_R$};

    \draw[{Stealth[scale=1]}-{Stealth[scale=1]},black]   (-0.27*\L,0.33*\L) -- (0.02*\L,0.62*\L);
    \draw[black] (0*\L,0.6*\L) node[left] {$\Delta X^+$};

    \draw[decorate, decoration={snake, amplitude=2pt, segment length=5pt}, green, very thick]
        (-\L+0.05*\L,-\L+0.1*\L+0.05*\L) -- (\L-0.1*\L-0.05*\L,\L-0.05*\L);
    \draw[green] (-1.05*\L,-0.9*\L) node[left] {$\boldsymbol{t_0}$};
    \end{tikzpicture}
    \caption{}
    \label{sfig:AdSshift}
\end{subfigure}
\caption{Left: The backreaction of a perturbation added at time $t_0$ on the left boundary causes the location of the horizon to shift an amount given by $\Delta X^+>0$, together with a deformation of the squared shape of the Penrose diagram. Right: This diagram represents exactly the same geometry as the left diagram, but in discontinuous coordinates in such a way that a continuous worldline suffers the corresponding shift in $\Delta X^+$ when it crosses the pulse of null energy.}
\label{fig:AdSperturbed}
\end{figure}
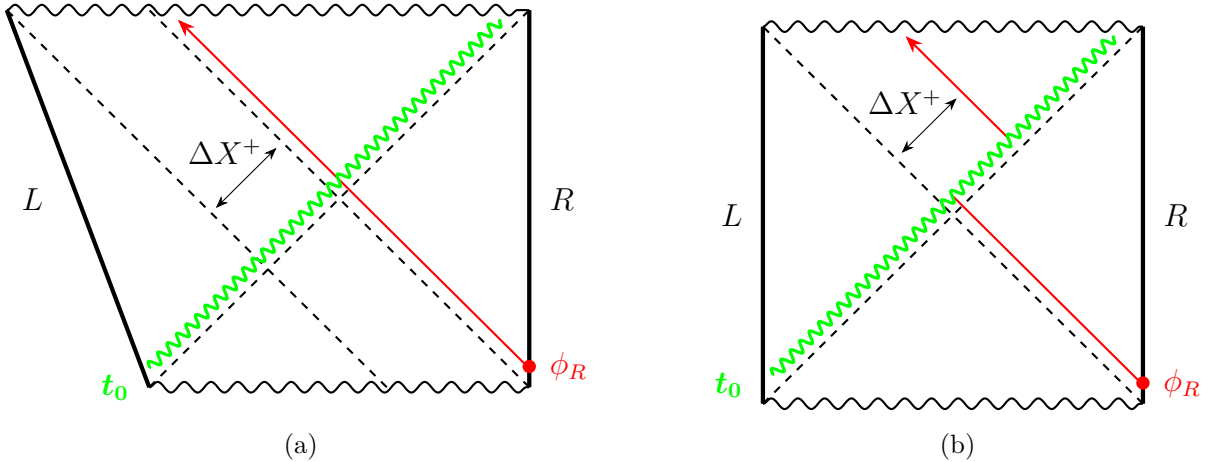

To analyze this, one can consider a small, spherically symmetric perturbation of the energy-momentum tensor, $T_{\mu\nu}\sim \mathcal{O}(\epsilon)$ introduced on the $R$ side at $t=t_0$. By solving the linearized Einstein's equations for the metric perturbation, it can be shown that the effect of this perturbation is a shift of the location of the horizon \cite{Gao_2016}. In terms of the null coordinates $X^+$, $X^-$, the shift is given by
\begin{equation}
    \Delta X^+=G_N\int_{-\infty}^\infty dX^-T_{--}~,
\end{equation}
where $G_N$ is Newton's constant, and we integrate the $T_{--}$ component of the energy-momentum tensor along the $X^-$ direction. If the ANEC holds, the shift in $X^+$ is necessarily positive, making it even harder for a signal to reach the opposite side. We show this configuration in Fig. \ref{fig:AdSperturbed}. This aligns with the intuition that adding energy to a black hole increases the size of its horizon, reducing the accessible region of spacetime for an external observer. Hence, the ANEC provides a robust argument for non-traversability in the gravitational description as well.

A natural question is whether one can modify the geometry in a controlled manner to enable signal propagation between the boundaries. As we have seen, a necessary condition for this is a violation of the ANEC, which would permit a negative shift in $X^+$.

An explicit mechanism for achieving this was proposed by Gao, Jafferis, and Wall \cite{Gao_2016}, who showed that a wormhole can be rendered traversable by introducing a direct interaction between the two boundaries of the eternal BTZ black hole in AdS$_3$ \cite{Banados_1992,Banados_1993}. Their proposal involves a double-trace deformation in the action of the form
\begin{equation}
    \delta S=\int dt\h d^{d-1}x\h h(t,x)\mathcal{O}_L(-t,x)\mathcal{O}_R(t,x)~,
    \label{eq:deltaSGJW}
\end{equation}
where $\mathcal{O}$ is a scalar operator of dimension less than $d/2$, dual to a scalar field $\phi$. The function $h(t,x)$ governs both the coupling strength and its time dependence. The null component of the energy-momentum tensor under this deformation was explicitly computed in \cite{Gao_2016}, and it was shown that, if this interaction is turned on at $t=0$, it generates negative null energy in the bulk in such a way that the ANEC is violated, rendering the wormhole traversable.

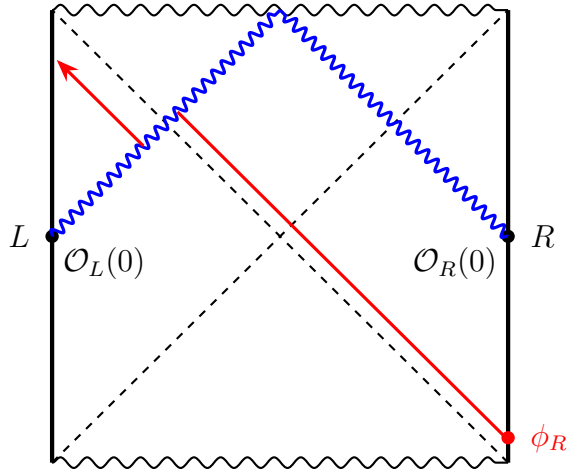
\begin{figure}
\centering
\begin{tikzpicture}[baseline={(current bounding box.center)}]
    \def\L{3cm} 
    \draw[ultra thick,black]    (-\L,-\L) -- (-\L,\L);
    \draw[ultra thick,black]    (\L,-\L) -- (\L,\L);
    \draw[black] (-1.05*\L,0) node[left] {$L$};
    \draw[black] (1.05*\L,0) node[right] {$R$};
    \draw[thick,dashed,black]   (-\L,-\L) -- (\L,\L);
    \draw[thick,dashed,black]   (\L,-\L) -- (-\L,\L);
    
    \draw[decorate, decoration={snake, amplitude=2pt, segment length=10pt}, black, thick]
        (-\L,\L) -- (\L,\L);
    \draw[decorate, decoration={snake, amplitude=2pt, segment length=10pt}, black, thick]
        (-\L,-\L) -- (\L,-\L);

    \fill[black] (-\L,0) circle (2.5pt);
    \fill[black] (\L,0) circle (2.5pt);
    \draw[black] (-\L,0) node[below right] {$\mathcal{O}_L(0)$};
    \draw[black] (\L,0) node[below left] {$\mathcal{O}_R(0)$};

    \draw[decorate, decoration={snake, amplitude=2pt, segment length=5pt}, blue, very thick]
        (-\L,0) -- (0,\L);
    \draw[decorate, decoration={snake, amplitude=2pt, segment length=5pt}, blue, very thick]
        (\L,0) -- (0,\L);

    \draw[very thick,red]   (\L,-\L+0.1*\L) -- (-0.45*\L,0.55*\L);
    \draw[-{Stealth[scale=1]},very thick,red] (-0.6*\L,0.4*\L) -- (-\L+0.02*\L,0.8*\L-0.02*\L);
    \fill[red] (\L,-\L+0.11*\L) circle (2.5pt);
    \draw[red] (1.05*\L,-\L+0.11*\L) node[right] {$\phi_R$};
\end{tikzpicture}
\caption{Summary of the traversable wormhole protocol of \cite{Gao_2016}.}
\label{fig:GJWprotocol}
\end{figure}

The precise protocol, depicted in Fig. \ref{fig:GJWprotocol}, proceeds as follows: the system begins in the unperturbed TFD state, corresponding to the AdS-Schwarzschild geometry. At an early time, a signal $\phi_R(t)$ is sent from the right boundary. In the unperturbed spacetime, this signal would be unable to reach the opposite boundary. However, at $t=0$, the evolution of the system is modified by the interaction \eqref{eq:deltaSGJW}, which at the boundary corresponds with an operator insertion $\sim e^{i h \mathcal{O}_L\mathcal{O}_R}$. This creates a pulse of negative energy in the bulk. When the signal later encounters this negative energy, the null shift it experiences is reversed, allowing it to emerge in the left asymptotic region, making the wormhole traversable.

An important clarification is in order: from the CFT perspective, the presence of an explicit interaction between the two CFTs makes it natural to expect that information can be transmitted from one side to the other. However, it is important to emphasize that the signal in this example is not simply transferred via the boundary interaction, but rather propagating through the wormhole.

Finally, it has been conjectured \cite{Gao_2016,Maldacena_2017,Susskind_2017} that the transmission of information through such a wormhole is the gravitational dual of quantum teleportation in the CFTs. A concrete realization of this mechanism in the SYK model has been described in \cite{Gao_2019}.

The next two sections are devoted to present the relation between the SYK model and the protocol described above. The impatient reader should know that the connection between the SYK model and wormhole teleportation arises from the fact that the IR limit of SYK is governed by the same action as JT gravity, a simple gravitational theory in AdS$_2$. While this double-trace deformation was first introduced in the context of AdS$_3$ \cite{Gao_2016}, the case of AdS$_2$ provides a simpler and more tractable setup \cite{Maldacena_2017}. For these reasons, we now move to describe the aspects of the SYK model relevant to our discussion, gravity in AdS$_2$, and the precise wormhole construction in the SYK model.

\section{The SYK model}

The SYK model describes a $(0+1)$-dimensional quantum mechanical system consisting of $N$ Majorana fermions $\chi_i$ with random, all-to-all, quartic interactions governed by the Hamiltonian
\begin{equation}
    H_{\text{SYK}}=\frac{1}{4!}\sum_{ijkl} J_{ijkl}\chi_i\chi_j\chi_k\chi_l~.
    \label{eq:HSYK}
\end{equation}
$\chi_i$ are Majorana fermions, $\chi_i=\chi_i^\dagger$, satisfying the usual anticommutation relations
\begin{equation}
    \left\{\chi_i,\chi_j\right\}=\delta_{ij}~,\qquad i,j=1,...,N~.
\end{equation}

In one dimension, Majorana fermions are dimensionless. The couplings $J_{ijkl}$ are random, independent Gaussian variables with mean and variance given by, respectively,
\begin{equation}
\overline{J_{ijkl}}=0~,\qquad \overline{J_{ijkl}^2}=\frac{3! J^2}{N^3} ~,
\label{eq:Jmeanvar}
\end{equation}
where $J$ is a constant with dimension of mass.

Of particular interest is the Euclidean time-ordered two-point function of the fermions, defined as
\begin{equation}
    G_{ij}(\tau)=\langle T\chi_i(\tau)\chi_j(0)\rangle\equiv \Theta(\tau)\langle \chi_i(\tau)\chi_j(0)\rangle-\Theta(-\tau)\langle\chi_j(0)\chi_i(\tau)\rangle~,
\end{equation}
where $\Theta$ is the Heaviside theta function, and the Majorana fields are written in the Heisenberg picture,
\begin{equation}
    \chi_i(\tau)=e^{\tau H_{\text{SYK}}}\chi_i\h e^{-\tau H_{\text{SYK}}}~.
\end{equation}
The normalized trace of the two-point function,
\begin{equation}
    G(\tau)=\frac{1}{N}\sum_{i=1}^N G_{ii}(\tau)~,
\end{equation}
plays a central role in the large-$N$ limit.

In the free case\footnote{The interest in the free case arises from the fact that, since the Majorana fields are dimensionless, the coupling in \eqref{eq:HSYK} is relevant. Therefore, in the UV the system will be well described by the free solution.}, $H_{\text{SYK}}=0$ and the time evolution of the fields is trivial. Using the anticommutation relations, this allows to obtain the free two-point function as
\begin{equation}
    G_{ij}^{0}(\tau)=\frac{1}{2}\delta_{ij}\sgn(\tau)~,\qquad G_0(\tau)=\frac{1}{N}\sum_{i=1}^N G_{ii}^0(\tau)=\frac{1}{2}\sgn(\tau)~.
\end{equation}

In the interacting model, the two-point functions can be computed using conventional perturbation theory in $J$. The theory contains a quartic vertex with four Majorana fields, proportional to $J_{ijkl}$. For each realization of the model, the Feynman diagrams are computed, and then the diagram has to be averaged over disorder.

The disorder average of the product of arbitrary couplings is given by
\begin{equation}
    \langle J_{i_1i_2i_3i_4}J_{j_1j_2j_3j_4}\rangle_J =\frac{3!J^2}{N^3}\delta_{i_1j_1}\delta_{i_2j_2}\delta_{i_3j_3}\delta_{i_4j_4}~.
    \label{eq:averageJ}
\end{equation}
For a product of a larger (even) number of couplings, Wick's theorem applies and the calculation reduces to combinations of \eqref{eq:averageJ}.

The diagrammatic expansion of the model simplifies greatly in the large-$N$ limit. It can be checked that, in this limit, the only dominant diagrams are melonic diagrams with disorder average pairing vertices inside a single melon\footnote{See the standard reviews \cite{Trunin_2020,Sarosi_2017} for details, and \cite{Bonzom_2018} for a more formal proof.}. Therefore, the interacting two-point function can be written diagramatically as
\begin{equation}
    G(\tau)=~
    \begin{tikzpicture}[baseline=-0.6ex]
    \def\L{1} 
    \draw[black, thick] (-\L,0) -- (\L,0);
    \end{tikzpicture}
    ~+~
    \begin{tikzpicture}[baseline=-0.6ex]
    \def\L{1} 
    \draw[black, thick] (-\L,0) -- (\L,0);
    \draw[black, thick] (0,0) circle (0.6*\L);
    \draw[domain=125:55, variable=\t, black, thick, dashed, dash pattern=on 2pt off 2pt] 
        plot ({\L*cos(\t)}, {-0.8*\L+\L*sin(\t)});
    \end{tikzpicture}
    ~+~
    \begin{tikzpicture}[baseline=-0.6ex]
    \def\L{1} 
    \draw[black, thick] (-\L,0) -- (\L,0);
    \draw[black, thick] (0,0) circle (0.6*\L);
    \draw[domain=125:55, variable=\t, black, thick, dashed, dash pattern=on 2pt off 2pt] 
        plot ({\L*cos(\t)}, {-0.8*\L+\L*sin(\t)});
    \draw[black, thick] (0,0.6*\L) circle (0.3*\L);
    \draw[domain=135:40, variable=\t, black, thick, dashed, dash pattern=on 2pt off 2pt] 
        plot ({0.35*\L*cos(\t)}, {0.35*\L+0.35*\L*sin(\t)});
    \end{tikzpicture}
    ~+~
    \begin{tikzpicture}[baseline=-0.6ex]
    \def\L{1} 
    \draw[black, thick] (-1.5*\L,0) -- (1.5*\L,0);
    \draw[black, thick] (-0.7*\L,0) circle (0.6*\L);
    \draw[black, thick] (0.7*\L,0) circle (0.6*\L);
    \draw[domain=125:55, variable=\t, black, thick, dashed, dash pattern=on 2pt off 2pt] 
        plot ({-0.7*\L+\L*cos(\t)}, {-0.8*\L+\L*sin(\t)});
    \draw[domain=125:55, variable=\t, black, thick, dashed, dash pattern=on 2pt off 2pt] 
        plot ({0.7*\L+\L*cos(\t)}, {-0.8*\L+\L*sin(\t)});
    \end{tikzpicture}~+~...~,
\end{equation}
which can be summarized in the following closed form
\begin{equation}
\begin{aligned}
    \begin{tikzpicture}[baseline=-0.6ex]
    \def\L{1} 
    \def\R{0.5*\L} 
    \draw[black, thick] (-\L,0) -- (-\R,0);
    \draw[black, thick] (\R,0) -- (\L,0);
    \draw[fill=green, opacity=0.1] (0,0) circle (\R);
    \draw[thick, black] (0,0) circle (\R);
    \node at (0,0) {\large $G$};
    \end{tikzpicture} & ~ = ~ 
    \begin{tikzpicture}[baseline=-0.6ex]
    \def\L{1} 
    \draw[black, thick] (-\L,0) -- (\L,0);
    \end{tikzpicture}
    ~+ ~
    \begin{tikzpicture}[baseline=-0.6ex]
    \def\L{1} 
    \def\R{0.5*\L} 
    \draw[black, thick] (-\L,0) -- (-\R,0);
    \draw[black, thick] (\R,0) -- (\L,0);
    \draw[fill=green, opacity=0.1] (0,0) circle (\R);
    \draw[thick, black] (0,0) circle (\R);
    \node at (0,0) {\large $\Sigma$};
    \end{tikzpicture}
    ~+ ~
    \begin{tikzpicture}[baseline=-0.6ex]
    \def\L{1} 
    \def\R{0.5*\L} 
    \draw[black, thick] (-1.5*\L,0) -- (-1.2*\L,0);
    \draw[black, thick] (-0.2*\L,0) -- (0.2*\L,0);
    \draw[black, thick] (1.2*\L,0) -- (1.5*\L,0);
    \draw[fill=green, opacity=0.1] (-0.7*\L,0) circle (\R);
    \draw[fill=green, opacity=0.1] (0.7*\L,0) circle (\R);
    \draw[thick, black] (-0.7*\L,0) circle (\R);
    \draw[thick, black] (0.7*\L,0) circle (\R);
    \node at (-0.7*\L,0) {\large $\Sigma$};
    \node at (0.7*\L,0) {\large $\Sigma$};
    \end{tikzpicture}
    ~+ ~...
    \\
    \begin{tikzpicture}[baseline=-0.6ex]
    \def\L{1} 
    \def\R{0.5*\L} 
    \draw[black, opacity=0] (-\L,0) -- (-\R,0);
    \draw[black, opacity=0] (\R,0) -- (\L,0);
    \draw[fill=green, opacity=0.1] (0,0) circle (\R);
    \draw[thick, black] (0,0) circle (\R);
    \node at (0,0) {\large $\Sigma$};
    \end{tikzpicture} & ~ = ~
    \begin{tikzpicture}[baseline=-0.6ex]
    \def\L{1} 
    \def\R{0.5*\L} 
    \draw[black, opacity=0] (-1.9*\L,0) -- (-1.4*\L,0);
    \draw[thick, black] (0,0) circle (1.4*\L);
    \draw[black, thick] (-1.4*\L,0) -- (1.4*\L,0);
    
    \draw[fill=white] (0,1.35*\L) circle (\R);
    \draw[fill=green, opacity=0.1] (0,1.35*\L) circle (\R);
    \draw[thick, black] (0,1.35*\L) circle (\R);
    \node at (0,1.4*\L) {\large $G$};

    \draw[fill=white] (0,0) circle (\R);
    \draw[fill=green, opacity=0.1] (0,0) circle (\R);
    \draw[thick, black] (0,0) circle (\R);
    \node at (0,0) {\large $G$};

    \draw[fill=white] (0,-1.35*\L) circle (\R);
    \draw[fill=green, opacity=0.1] (0,-1.35*\L) circle (\R);
    \draw[thick, black] (0,-1.35*\L) circle (\R);
    \node at (0,-1.4*\L) {\large $G$};
    \end{tikzpicture}
\end{aligned}
\end{equation}

These equations can be written as
\begin{equation}
\begin{aligned}
    G(\tau_1,\tau_2) & = G_0(\tau_1,\tau_2)+\int d\tau_3 d\tau_4 G_0(\tau_1,\tau_3)\Sigma(\tau_3,\tau_4)G(\tau_4,\tau_2)~,\\
    \Sigma(\tau_1,\tau_2) & \equiv J^2 G(\tau_1,\tau_2)^3~. 
    \label{eq:DysoneqSYK}
\end{aligned}
\end{equation}

Due to translation invariance, the two-point function depends only on time differences: $G(\tau_1,\tau_2)=G(\tau_1-\tau_2)\equiv G(\tau)$, $\Sigma(\tau_1,\tau_2)=\Sigma(\tau_1-\tau_2)\equiv\Sigma(\tau)$ and the equations \eqref{eq:DysoneqSYK} can be written as
\begin{equation}
\begin{aligned}
    G^{-1}(\omega) & = -i\omega -\Sigma(\omega)~,\\
    \Sigma(\tau) & = J^2 G(\tau)^3~,
    \label{eq:DysoneqSYKtauomega}
\end{aligned}
\end{equation}
where the first equation has been Fourier transformed.

These equations are valid both at zero and finite temperature. At finite temperature, the frequencies are discrete and are given by the Matsubara frequencies $\omega_n=\frac{2\pi}{\beta}(n+\frac{1}{2})$. The equations \eqref{eq:DysoneqSYKtauomega} can be easily solved numerically \cite{Maldacena_2016,Banerjee_2016}, starting from the free solution for the two-point function and transforming back and forth between the equations \eqref{eq:DysoneqSYKtauomega} in position and frequency space until the solution converges to the desired accuracy. We will use (the real time version of) this method in the following chapters, for two coupled SYK models. Many more technical details will be given for that setup.

Let us finally mention an interesting generalization of the model, consisting of allowing for $q$-body interactions, as opposed to the four-body interactions considered in the Hamiltonian \eqref{eq:HSYK}. Namely, one considers the Hamiltonian
\begin{equation}
    H_{\text{SYK,}q}=\frac{1}{q!}\sum_{i_1...i_q} J_{i_1...i_q}\chi_{i_1}\chi_{i_2}...\chi_{i_q}~,
    \label{eq:HSYKq}
\end{equation}
with the Gaussian distribution of the couplings generalized to
\begin{equation}
\overline{J_{ijkl}}=0~,\qquad \overline{J_{ijkl}^2}=\frac{2^{q-1}}{q}\frac{\mathcal{J}(q-1)!}{N^{q-1}} ~,
\label{eq:Jmeanvarq}
\end{equation}
where $\mathcal{J}^2\equiv J^2\frac{q}{2^{q-1}}$. In the large-$N$ limit, the Dyson equations can be generalized to
\begin{equation}
\begin{aligned}
    G^{-1}(\omega) & = -i\omega -\Sigma(\omega)~,\\
    \Sigma(\tau) & = J^2 G(\tau)^{q-1}~.
    \label{eq:DysoneqSYKtauomegaq}
\end{aligned}
\end{equation}
Remarkably, in the large-$q$ limit, these equations admit an analytic solution at all energy scales. This has to be contrasted with the case $q=4$, where one generically needs to resort to numerics, except in the large-$\beta J$ (IR) limit, as we will see below.

\subsection{Effective action}

We begin by considering the partition function, averaged over disorder. For a generic function $f(J_{ijkl})$, the disorder average is defined as
\begin{equation}
\begin{aligned}
    \overline{f}(J_{ijkl})&=\int D J_{ijkl} f(J_{ijkl})~,\qquad \text{where}\\
    DJ_{ijkl} & = \exp\left[-\frac{1}{4!}\frac{N^3}{12 J^2}\sum_{ijkl}J_{ijkl}^2\right]\Pi_{ijkl}\sqrt{\frac{N^3}{3!J^2}}\frac{dJ_{ijkl}}{\sqrt{2\pi}}~.
    \label{eq:disorder}
\end{aligned}
\end{equation}

The averaged partition function is\footnote{Here we adopt the annealed average, where one averages first the partition function and computes the expectation values from $\overline{Z}$. This is to be contrasted with the quenched average, where one computes the partition function with the replica method, and then averages the free energy $-\overline{\log Z}$. In order to obtain thermodynamic quantities, one should consider the latter. However, in the SYK model, both methods give the same result in the large-$N$ limit \cite{Sachdev_2015}.}
\begin{equation}
\begin{aligned}
    \overline{Z} & = \int DJ_{ijkl}D\chi_i \exp\left[-\frac{1}{2}\int d\tau\sum_{i=1}^N\chi_i\partial_\tau\chi_i+\frac{1}{4!}\sum_{i,j,k,l}J_{ijkl}\chi_i(\tau)\chi_j(\tau)\chi_k(\tau)\chi_l(\tau)\right]~.
    \label{eq:overlineZ}
\end{aligned}
\end{equation}

The integral over disorder is a Gaussian integral that can be directly evaluated. Introducing the Hubbard-Stratonovich bilocal fields
\begin{equation}
    O(\tau,\tau')\equiv\frac{1}{N}\sum_{i=1}^N\chi_i(\tau)\chi_i(\tau')~,
\end{equation}
the averaged partition function can be written as
\begin{equation}
    \overline{Z}=\int D\chi_i\exp\left[-\frac{1}{2}\int d\tau \sum_{i=1}^N\chi_i\partial_\tau\chi_i+\frac{NJ^2}{8}\int d\tau d\tau' O^4(\tau,\tau')\right]~,
\end{equation}
where the average \eqref{eq:averageJ} has been used.

It is conventional to insert the identity as
\begin{equation}
\begin{aligned}
    1 & = \int D G\h \delta \left[N\left(G(\tau,\tau')-O(\tau,\tau')\right)\right] \\
    & \sim \int DG D\Sigma \exp \left[-\frac{N}{2}\int d\tau d\tau' \Sigma(\tau,\tau')\left(G(\tau,\tau')-O(\tau,\tau')\right)\right]~,
    \label{eq:unitySYK}
\end{aligned}
\end{equation}
where $\Sigma(\tau,\tau')$ acts as a Lagrange multiplier, and will turn out to be the self-energy.

The $\delta$-function allows to exchange as many $O(\tau,\tau')$'s as desired by $G(\tau,\tau')$. After performing the functional integral over the fields $\chi_i$, the averaged partition function can be written in terms of an effective action,
\begin{equation}
    \overline{Z}=\int DG D\Sigma \exp\left[-S_\text{eff}\left(G,\Sigma\right)\right]~,
\end{equation} 
with $S_\text{eff}\left(G,\Sigma\right)$ given by
\begin{equation}
    \frac{S_\text{eff}}{N} = -\frac{1}{2}\log\det\left(\delta(\tau-\tau')\partial_{\tau'}-\Sigma(\tau,\tau')\right)+\frac{1}{2}\int d\tau d\tau'\left(\Sigma(\tau,\tau')G(\tau,\tau')+\frac{J^2}{4}G(\tau,\tau')^4\right)~.
    \label{eq:SeffSYK}
\end{equation}
Since $S_\text{eff}\propto N$, in the large-$N$ limit the path integral is dominated by the saddle in which
\begin{equation}
    \frac{\delta S_\text{eff}}{\delta G}=0~,\qquad \frac{\delta S_\text{eff}}{\delta \Sigma}=0~.
\end{equation}
It is not difficult to check that these variations lead precisely to the Schwinger-Dyson equations \eqref{eq:DysoneqSYK} (or \eqref{eq:DysoneqSYKtauomega}).

\subsection{IR limit and Schwarzian action}

A particularly interesting limit of the model is the low-energy (IR) limit. In the low-energy limit ($\omega\ll J$), we can neglect the $-i\omega$ term in \eqref{eq:DysoneqSYKtauomega}. Doing so reduces the equations (in position space) to
\begin{equation}
\begin{aligned}
    \int d\tau_3 G(\tau_1,\tau_3)\Sigma(\tau_3,\tau_2) & = -\delta(\tau_1-\tau_2)~,\\
    \Sigma(\tau_1,\tau_2) & = J^2 G(\tau_1,\tau_2)^3~,
    \label{eq:DysonSYKIR}
\end{aligned}
\end{equation}
where we have used that $G_0^{-1}(\tau,\tau')=\delta(\tau-\tau')\partial_\tau$.

These equations can be solved, at zero temperature, with a conformal ansatz of the form
\begin{equation}
    G(\tau)=\frac{b}{\abs{\tau}^{2\Delta}}\sgn(\tau)~,
    \label{eq:Gconfsolution}
\end{equation}
with $b$ and $\Delta$ two constants that can be determined by plugging the ansatz into the equations. Notice that $\Delta$ corresponds to the anomalous dimension of the fields in the IR. The result is
\begin{equation}
    J^2b^4=\frac{1}{4\pi}~,\qquad \Delta=\frac{1}{4}.
\end{equation}

It turns out that the IR solution just found is not unique. Actually,  there are infinite solutions. It can be checked that the equations \eqref{eq:DysonSYKIR} are invariant under reparametrizations of the time variable, $\tau\to f(\tau)$, with $f'(\tau)>0$, provided that the fields transform as
\begin{equation}
\begin{aligned}
    \Tilde{G}(\tau_1,\tau_2) & =\left[f'(\tau_1)f'(\tau_2)\right]^\Delta G(f(\tau_1),f(\tau_2))~,\\
    \Tilde{\Sigma}(\tau_1,\tau_2) & =\left[f'(\tau_1)f'(\tau_2)\right]^{3\Delta}\Sigma(f(\tau_1),f(\tau_2))~.
\end{aligned}
\end{equation}
It is important to recall that this reparametrization symmetry is only emergent in the IR, when we can drop the $-i\omega$ term in \eqref{eq:DysoneqSYKtauomega}. The symmetry is explicitly broken by this term away from the IR limit.

Thus, in the strict IR limit, we have an entire space of solutions, given by
\begin{equation}
    G(\tau_1,\tau_2)=\frac{1}{(4\pi)^{1/4}}\frac{\sgn(\tau_1-\tau_2)}{J^{2\Delta}}\frac{\left[f'(\tau_1)\right]^\Delta\left[f'(\tau_2)\right]^\Delta}{\abs{f(\tau_1)-f(\tau_2)}^{2\Delta}}~.
    \label{eq:Gconf}
\end{equation}

In particular, the two-point function at finite temperature can be obtained from the zero-temperature solution with the particular reparametrization $f(\tau)=\tan\frac{\pi\tau}{\beta}$:
\begin{equation}
    G(\tau)=b\left[\frac{\pi}{\beta\sin\frac{\pi\tau}{\beta}}\right]^{2\Delta}\sgn(\tau)~.
    \label{eq:confsolution}
\end{equation}

The breaking of the reparametrization invariance in the Schwinger-Dyson equations due to the $-i\omega$ term (or $\delta(\tau-\tau')\partial_\tau$ in time domain) can be studied in more detail. The change $\Sigma\to\Sigma-G_0^{-1}$ in the effective action \eqref{eq:SeffSYK} effectively separates the action into a conformally invariant part and a non-conformally invariant part:
\begin{equation}
\begin{aligned}
    \frac{S_{\text{CFT}}}{N}&=-\frac{1}{2}\log\det\left(-\Sigma(\tau,\tau')\right)+\frac{1}{2}\int d\tau d\tau'\left(\Sigma(\tau,\tau')G(\tau,\tau')-\frac{J^2}{4}G(\tau,\tau')^4\right)~,\\
    \frac{S_{\Sch}}{N}&=-\frac{1}{2}\int d\tau d\tau'G_0^{-1}(\tau,\tau')G(\tau,\tau')~.
\end{aligned}
\end{equation}

The action $S_{\text{CFT}}$ reproduces the IR limit of the Schwinger-Dyson equation, \eqref{eq:DysonSYKIR}. Since the change $\tau\to f(\tau)$ is a symmetry of $S_{\text{CFT}}$, this action does not depend on $f(\tau)$ (the change $\tau\to f(\tau)$ corresponds to moving within the manifold of conformal solutions). Moreover, for $S_{\Sch}$, the $\delta(\tau-\tau')$ term in $G_0^{-1}$ picks only contributions at small time differences, and therefore it can be neglected in the strict IR limit. However, one cannot simply throw away this non-invariant term. If one treats the $\delta(\tau-\tau')\partial_\tau$ term as a deviation from the exact IR limit, it can be seen that the fluctuations away from the conformal limit have a vanishing contribution to $S_{\text{CFT}}$ \cite{Sarosi_2017,Trunin_2020}. Therefore, the non-invariant part $S_{\Sch}$ has to be treated as a perturbation to the IR action, and compute its effect to leading order.

This deviation can be studied by expanding the conformal two-point function \eqref{eq:Gconf} in powers of $\tau_{12}\equiv \tau_1-\tau_2$. Defining also $\tau=\frac{\tau_1+\tau_2}{2}$, it can be seen that the conformal two-point function can be expanded as
\begin{equation}
    G(\tau_1,\tau_2)=\frac{1}{(4\pi)^{1/4}}\frac{\sgn(\tau_{12})}{\abs{J\tau_{12}}^{2\Delta}}\left(1+\frac{\Delta}{6}\tau_{12}^2\{f(\tau),\tau\}\right)~,
\end{equation}
where $\{f(\tau),\tau\}$ denotes the Schwarzian derivative,
\begin{equation}
    \{f(\tau),\tau\}=\frac{f'''}{f'}-\frac{3}{2}\left(\frac{f''}{f'}\right)^2~.
\end{equation}

Plugging this result for $G$ into $S_{\Sch}$ and subtracting the untransformed part leads, after some regularization\footnote{See \cite{Sarosi_2017,Kitaev_2017,Trunin_2020} for details.}, to
\begin{equation}
    \frac{S_{\Sch}}{N}=-\frac{C}{J}\int_{-\infty}^\infty\{f(\tau),\tau\}d\tau~,
    \label{eq:SchwSYK}
\end{equation}
where $C$ is some numerical constant that has to be determined numerically. The field $f(\tau)$ is usually referred to as the soft mode, the reparametrization mode, or the gravitational mode. 

\subsection{The chaotic nature of the SYK model}\label{subsec:LyapunovSYK}

As mentioned in the introduction, the SYK model has emerged as a powerful framework for exploring quantum chaos. Its features, which closely resemble those of black hole dynamics, have placed it at the center of current efforts to understand chaotic behavior in strongly interacting quantum systems. To appreciate its significance, it is instructive to begin by recalling the essential features of chaos in classical systems.

Classical chaos is characterized by an extreme sensitivity to initial conditions: trajectories in phase space that begin infinitesimally close to each other diverge exponentially over time. Let $q^i$, $p^i$, with $i=1,...,N$, denote the generalized coordinates and momenta of a classical system. This exponential sensitivity is commonly expressed using the Poisson bracket:
\begin{equation}
    \abs{\{q^i(t),p^j(0)\}}=\abs{\sum_{k=1}^N\frac{\partial q^i(t)}{\partial q^k(0)}\frac{\partial p^j(0)}{\partial p^k(0)}-\frac{\partial q^i(t)}{\partial p^k(0)}\frac{\partial p^j(0)}{\partial q^k(0)}}=\abs{\frac{\partial q^i(t)}{\partial q^j(0)}}\sim e^{\lambda_L t}~.
\end{equation}
The exponential growth is controled by $\lambda_L$, which is known as the classical Lyapunov exponent.

Extending this intuition to quantum systems is far from straightforward. In quantum mechanics, the very notion of a trajectory becomes ill-defined due to the uncertainty principle. As a result, defining quantum chaos in terms of trajectories is subtle, and one must instead identify quantities that capture its distinctive signatures in operator growth and correlation functions.

A widely used approach builds on the formal correspondence between Poisson brackets and quantum commutators in the semiclassical limit $\hbar\to 0$:
\begin{equation}
    \left[\hat{q}^i(t),\hat{p}^j(0)\right]\sim i\hbar\h \{q^i(t),p^j(0)\},\qquad \hbar\to 0~.
\end{equation}

For this purpose, a useful quantity is defined as
\begin{equation}
    C(t)=-\langle\left[V(t),W(0)\right]^2\rangle_\beta~,
    \label{eq:C(t)}
\end{equation}
where $V$ and $W$ are generic Hermitian operators with vanishing thermal expectation values, and the subscript $\beta$ denotes a thermal expectation value. A system is said to exhibit quantum chaos if $C(t)$ displays exponential growth for a wide class of operator pairs:
\begin{equation}
    C(t)\sim e^{2\lambda_L t}~,
\end{equation}
where $\lambda_L$ is referred to as the \textit{quantum} Lyapunov exponent.

To regulate potential divergences from operator products at coincident times, it is common to consider a regularized version of $C(t)$, in which the thermal density matrix is symmetrically distributed between the commutators:
\begin{equation}
    C(t)=-\Tr\big(\rho^{\frac{1}{2}}\left[V(t),W(0)\right]\rho^{\frac{1}{2}}\left[V(t),W(0)\right]\big)~,
\end{equation}
where $\rho=\frac{1}{Z}e^{-\beta H}$ is the thermal density matrix. Expanding this expression yields four terms: two time-ordered correlators (TOCs) and two out-of-time-ordered correlators (OTOCs),
\begin{equation}
    C(t) = 2\h \text{TOC}(t)-\text{OTOC}\left(t+\frac{i\beta}{4}\right)-\text{OTOC}\left(t-\frac{i\beta}{4}\right)~,
\end{equation}
with
\begin{equation}
\begin{aligned}
    \text{TOC}(t) & \equiv \Tr \big(V(t)\rho^{\frac{1}{2}}V(t)W(0)\rho^{\frac{1}{2}}W(0)\big)~,\\
    \text{OTOC}(t) & \equiv \Tr\big(\rho^{\frac{1}{4}}V(t)\rho^{\frac{1}{4}}W(0)\rho^{\frac{1}{4}}V(t)\rho^{\frac{1}{4}}W(0)\big)~.
\end{aligned}
\end{equation}

In chaotic systems, the typical behavior of $C(t)$ consists of three distinct regimes. Initially, $C(t)$ remains close to zero. After the dissipation time $t_d\sim\beta$, it enters a phase of exponential growth governed by the Lyapunov exponent. Finally, after the scrambling time $t_*\sim\beta\log N$, the growth saturates, and $C(t)$ approaches a late-time asymptotic value. It is important to notice that the region of exponential growth is only well-defined in systems where a large separation exists between $t_d$ and $t_*$.

The SYK model displays precisely this behavior at low temperatures, with a Lyapunov exponent saturating the universal upper bound $\lambda_L \leq 2\pi/\beta$ \cite{Maldacena_2015}. This makes the SYK model one of the few many-body quantum systems where such features of quantum chaos can be studied analytically.

We will consider the following four-point function, which can be written as a two-point function of the Euclidean correlators,
\begin{equation}
\begin{aligned}
    \frac{1}{N^2}\sum_{i,j}\langle \chi_i(\tau_1)\chi_i(\tau_2)\chi_j(\tau_3)\chi_j(\tau_4)\rangle & = \langle G(\tau_1,\tau_2) G(\tau_3,\tau_4)\rangle~\\
    & = \frac{\int \mathcal{D}G\h \mathcal{D}\Sigma\h G(\tau_1,\tau_2) G(\tau_3,\tau_4)e^{-S_\text{eff}(G,\Sigma)}}{\int \mathcal{D}G\h \mathcal{D}\Sigma e^{-S_\text{eff}(G,\Sigma)}}~.
\end{aligned}
\end{equation}
In the large-$N$ limit, this four-point function admits the expansion
\begin{equation}
    \langle G(\tau_1,\tau_2) G(\tau_3,\tau_4)\rangle=\overline{G(\tau_1,\tau_2)}\h  \overline{G(\tau_3,\tau_4)}\left[1+\frac{1}{N}\mathcal{F}(\tau_1,...,\tau_4)+...\right]~,
\end{equation}
and the connected piece can be obtained from the quadratic fluctuations of the effective action $S_\text{eff}$ around the saddle point solutions, denoted as  $\overline{G(\tau,\tau')}$ \cite{Maldacena_2016}.

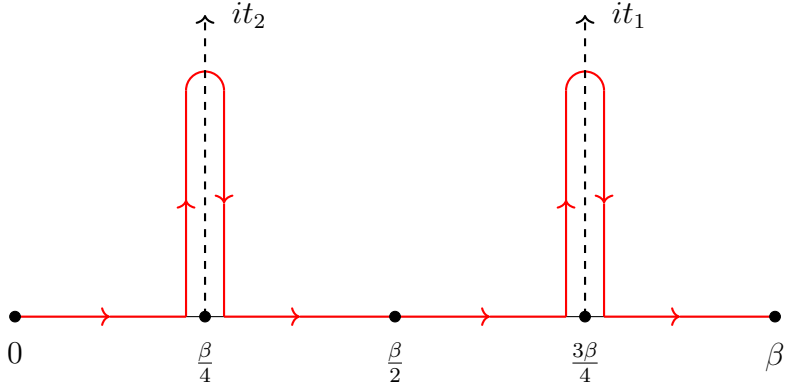
\begin{figure}
    \centering
    \begin{tikzpicture}
        \def\length{5cm}
        \def\eps{0.05*\length}
        \draw[-] (-\length, 0) -- (\length, 0);
        
        \node[below, yshift=-0.2cm] at (-\length, 0) {$0$};
        \node[below, yshift=-0.2cm] at (-\length/2, 0) {$\frac{\beta}{4}$};
        \node[below, yshift=-0.2cm] at (0, 0) {$\frac{\beta}{2}$};
        \node[below, yshift=-0.2cm] at (\length/2, 0) {$\frac{3\beta}{4}$};
        \node[below, yshift=-0.2cm] at (\length, 0) {$\beta$};
        
        \draw[thick,red, ->] (-\length, 0) -- (-3*\length/4, 0);
        \draw[thick,red] (-3*\length/4, 0) -- (-\length/2-\eps, 0);

        \draw[thick,red, ->] (-\length/2+\eps, 0) -- (-\length/4, 0);
        \draw[thick,red] (-\length/4, 0) -- (0, 0);

        \draw[thick,red, ->] (0, 0) -- (\length/4, 0);
        \draw[thick,red] (\length/4, 0) -- (\length/2-\eps, 0);

        \draw[thick,red, ->] (\length/2+\eps, 0) -- (3*\length/4, 0);
        \draw[thick,red] (3*\length/4, 0) -- (\length, 0);

        \draw[thick,red, ->] (-\length/2-\eps, 0) -- (-\length/2-\eps, 0.31*\length);
        \draw[thick,red,] (-\length/2-\eps, 0.3*\length) -- (-\length/2-\eps, 0.6*\length);

        \draw[thick,red, ->] (-\length/2+\eps,  0.6*\length) -- (-\length/2+\eps, 0.3*\length);
        \draw[thick,red] (-\length/2+\eps, 0.3*\length) -- (-\length/2+\eps, 0);

        \draw[thick,red, ->] (\length/2-\eps, 0) -- (\length/2-\eps, 0.31*\length);
        \draw[thick,red,] (\length/2-\eps, 0.3*\length) -- (\length/2-\eps, 0.6*\length);

        \draw[thick,red, ->] (\length/2+\eps,  0.6*\length) -- (\length/2+\eps, 0.3*\length);
        \draw[thick,red] (\length/2+\eps, 0.3*\length) -- (\length/2+\eps, 0);

        \filldraw[black] (-\length, 0) circle (2pt);
        \filldraw[black] (-\length/2, 0) circle (2pt);
        \filldraw[black] (0, 0) circle (2pt);
        \filldraw[black] (\length/2, 0) circle (2pt);
        \filldraw[black] (\length, 0) circle (2pt);
        
        
        \draw[thick, red] (-\length/2-\eps, 0.6*\length) arc[start angle=180,end angle=0,radius=\eps];
        \draw[thick, red] (\length/2-\eps, 0.6*\length) arc[start angle=180,end angle=0,radius=\eps];

        \draw[thick,dashed,black, ->] (-\length/2, 0) -- (-\length/2, 0.8*\length) node[right,xshift=0.2cm] {$it_2$};
        \draw[thick,dashed,black, ->] (\length/2, 0) -- (\length/2, 0.8*\length) node[right,xshift=0.2cm] {$it_1$};

    \end{tikzpicture}
    \caption{Usual contour used for the evaluation of the OTOC in the chaos regime.}
    \label{fig:contourLyapunov}
\end{figure}

The Lyapunov exponent is obtained by analytically continuing the time variable, and considering the following ordering of times,
\begin{equation}
    \tau_1=\frac{3\beta}{4}+it_1~,~~~~\tau_2=\frac{\beta}{4}+it_2~,~~~~\tau_3=\frac{\beta}{2}~,~~~~\tau_4=0~,
    \label{eq:timeorderings}
\end{equation}
with the integration along the contour depicted in Fig. \ref{fig:contourLyapunov}. The relevant fact is that the  connected piece $\mathcal{F}(\frac{3\beta}{4}+it_1,\frac{\beta}{4}+it_2,\frac{\beta}{2},0)\equiv \mathcal{F}(t_1,t_2)$ satisfies the following kernel equation
\begin{equation}
    \mathcal{F}(t_1,t_2)=\int dt_3 dt_4 K_R(t_1,...,t_4)\mathcal{F}(t_3,t_4)~,
    \label{eq:kerneleq}
\end{equation}
with the retarded kernel given by
\begin{equation}
    K_R(t_1,...,t_4)=-3J^2G_R(t_1,t_3)G_R(t_2,t_4)G_W(t_3,t_4)^2~.
    \label{eq:kernel}
\end{equation}
In \eqref{eq:kernel}, $G_R(t)$ is a retarded correlator, and $G_W(t)$ is a Wightman function obtained from the Euclidean correlator as $G_{W}(t)=G(\frac{\beta}{2}+it)$.

Eq. \eqref{eq:kerneleq} tells us that $\mathcal{F}(t_1,t_2)$ is an eigenfunction of the retarded kernel \eqref{eq:kernel} with eigenvalue 1. The Lyapunov exponent corresponds to the value of $\lambda_L$ for which the largest eigenvalue of $K_R$ crosses 1.

To solve this equation it is useful to introduce a {\em growing ansatz} of the form
\begin{equation}
    \mathcal{F}(t_1,t_2)=e^{\lambda_L(t_1+t_2)/2}f(t_1-t_2)~,
    \label{eq:F12ansatz}
\end{equation}
and Fourier transform \eqref{eq:kerneleq} to frequency space. Denoting by $f(\omega)$ the Fourier transform of $f(t_1-t_2)\equiv f(t)$, the equation can be written as\footnote{Here $G_W^2(\omega)$ refers to the Fourier transform of $G_W(t)^2$, and not the square of the Fourier transform of $G_W(t)$.}
\begin{equation}
    f(\omega)=3J^2|G_R(\omega+i\lambda_L/2)|^2\int_{-\infty}^\infty d\omega' G_W^2(\omega-\omega')f(\omega')~.
    \label{eq:fomega}
\end{equation}

This equation can be solved numerically in general, and analytically in the conformal limit. By analytically continuing the finite temperature conformal solution \eqref{eq:confsolution} and evaluating \eqref{eq:fomega}, one arrives to the equation
\begin{equation}
    f(\omega)=\frac{3}{4}\frac{\beta}{2\pi}\abs{\frac{\Gamma\left(\frac{1}{4}+\frac{\beta\lambda_L}{4\pi}-\frac{i\beta\omega}{2\pi}\right)}{\Gamma\left(\frac{3}{4}+\frac{\beta\lambda_L}{4\pi}-\frac{i\beta\omega}{2\pi}\right)}}^2\int_{-\infty}^{\infty}d\omega'\frac{f(\omega')}{\cosh\frac{\beta(\omega-\omega')}{2}}~.
    \label{eq:kerneleqomega}
\end{equation}

The solution to this equation is that the Lyapunov exponent is given by
\begin{equation}
    \lambda_L=\frac{2\pi}{\beta}~,
\end{equation}
which saturates the chaos bound. The details of the calculation can be found in Appendix \ref{app:Lyapunov}.

\begin{figure}
    \centering
    \includegraphics[width=0.6\textwidth]{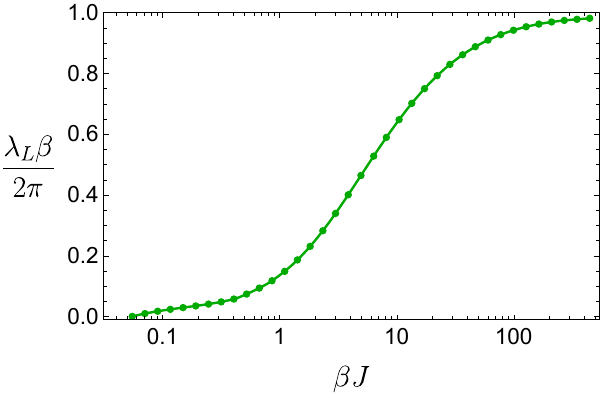}
    \caption{Lyapunov exponent of the SYK model. In the large-$\beta J$ limit, the Lyapunov exponent saturates the chaos bound \cite{Maldacena_2015}.}
    \label{fig:chaosq4}
\end{figure}

Away from the conformal limit, the numerical solution of this equation is shown in Fig. \ref{fig:chaosq4}.

\section{Eternal traversable wormhole}\label{sec:eternalWH}

Following the discussion in \cite{Maldacena_2018}, we first discuss gravity in nearly-AdS$_2$ spaces, and how a traversable wormhole can be realized in this context. We then construct a system of two coupled SYK models that, in the low-energy limit, will turn out to be dynamically equivalent to the traversable wormhole.

\subsection{Nearly-AdS\texorpdfstring{$_2$}{2} gravity}

The natural starting point to study gravity in two dimensions is to consider the Einstein-Hilbert action. However, in two dimensions, the integral of the Ricci scalar gives the Euler characteristic of the manifold, which is a topological invariant. As a result, the action does not lead to a well-defined variational principle since the notion of extremization makes no sense at all.

One possible extension would be to include matter fields. However, in the case of AdS$_2$ as the bulk spacetime, the standard approach to adding matter is not viable. This is because, in AdS$_2$, the combination $R_{\mu\nu}-\frac{1}{2}Rg_{\mu\nu}$ vanishes identically, implying through Einstein’s equations that $T_{\mu\nu}=0$.

A simple but non-trivial generalization is to consider a theory where the metric couples to a dilaton field, $\Phi$, with a certain potential for the dilaton, $V(\Phi)$. Such theories are broadly known as dilaton gravity theories.

A particularly important example in (1+1) dimensions is Jackiw-Teitelboim (JT) gravity \cite{Jackiw_1984,Teitelboim_1983}. Its Euclidean action is given by
\begin{equation}
    S_{\text{JT}}[g,\Phi]=-S_0\h\chi-\frac{1}{16\pi G_N}\left[\int\sqrt{g}\h\Phi\h (R+2)+2\oint \sqrt{h}\h\Phi_b\h K\right]+S_{\text{matter}}[\chi,g]~.
    \label{eq:actionJT}
\end{equation}
\noindent The first term is the topological term mentioned above. Its only effect is a shift of the action by a constant value and does not affect the dynamics. The most interesting physics comes from the central term. The theory can be coupled to matter fields, $\chi$, through the last term. However, it is assumed that the matter does not couple directly to the dilaton. This action appears naturally in the context of near-extremal black holes \cite{Almheiri_2014,Nayak_2018,Kolekar_2018}.

We can now derive the equations of motion and the classical solutions of the theory. Varying the action with respect to the dilaton yields
\begin{equation}
    R+2=0~,
    \label{eq:eomphi}
\end{equation}
which implies that the metric is locally AdS$_2$. This equation remains valid in the presence of matter since we assumed that the latter does not couple directly to the dilaton.

The equation of motion for $g_{\mu\nu}$ determines the bulk profile of the dilaton. After using \eqref{eq:eomphi}, it is given by
\begin{equation}
    \nabla_\mu\nabla_\nu\Phi-g_{\mu\nu}\Phi=0~.
    \label{eq:eomdilaton}
\end{equation}

It is instructive to assume first that the dilaton is a constant. In this case, we see from \eqref{eq:eomdilaton} that we need $\Phi=0$. Although the metric is locally AdS$_2$, its global structure remains undetermined. There are three relevant patches in AdS$_2$ (in its Lorentzian version), which we depict in Fig. \ref{fig:AdS2patches}:
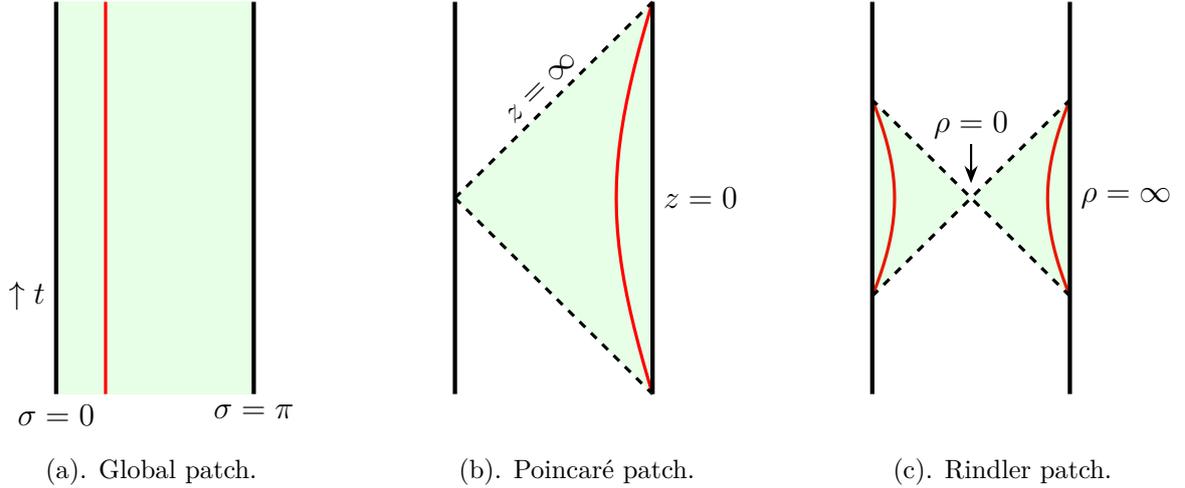
\begin{figure}
\centering
    \begin{subfigure}[b]{0.3\textwidth}
    \centering
    \begin{tikzpicture}
        \def\L{1.3} 
        \def\he{2*\L} 
        \fill[green, opacity=0.1] (-\L,-\he) rectangle (\L,\he);
        \draw[ultra thick,black]  (-\L,-\he) -- (-\L,\he) ;
        \draw[black] (-\L,-\he) node[below] {$\sigma=0$};
        \draw[ultra thick,black]  (\L,-\he) -- (\L,\he);
        \draw[black] (\L,-\he) node[below] {$\sigma=\pi$};
        \draw[black] (-\L,-\L) node[left] {$\uparrow t$};

        \draw[very thick,red]  (-0.5*\L,-\he) -- (-0.5*\L,\he) ;
    \end{tikzpicture}
    \caption{Global patch.}
    \label{sfig:AdS2global}
    \end{subfigure}
    \hfill
    \begin{subfigure}[b]{0.3\textwidth}
    \centering
    \begin{tikzpicture}
        \def\L{1.3} 
        \def\he{2*\L} 
        \fill[green, opacity=0.1] (-\L,0) -- (\L,2*\L) -- (\L,-2*\L) -- cycle;
        \draw[domain=-\he:\he,variable=\x,very thick,red] plot ({sqrt(0.15*\x*\x +0.1*\he*1*\he)},\x);
        \draw[ultra thick,black]  (-\L,-\he) -- (-\L,\he) ;
        \draw[black] (\L,0) node[right] {$z=0$};
        \draw[ultra thick,black]  (\L,-\he) -- (\L,\he);

        \draw[very thick,dashed,black]  (-\L,0) -- (\L,\he);
        \draw[very thick,dashed,black]  (-\L,0) -- (\L,-\he);
        \draw[black] (0,\L) node[above,rotate=45] {$z=\infty$};

        \draw[black,opacity=0] (-\L,-\he) node[below] {$\sigma=0$};
    \end{tikzpicture}
    \caption{Poincaré patch.}
    \label{sfig:AdS2Poincare}
    \end{subfigure}
    \hfill
    \begin{subfigure}[b]{0.3\textwidth}
    \centering
    \begin{tikzpicture}
        \def\L{1.3} 
        \def\he{2*\L} 

        \draw[domain=-\L:\L,variable=\x,very thick,red] plot ({sqrt(0.4*\x*\x +0.6*\L*\L)},\x);
        \draw[domain=-\L:\L,variable=\x,very thick,red] plot ({-sqrt(0.4*\x*\x +0.6*\L*\L)},\x);
        \fill[green, opacity=0.1] (-\L,-\L) -- (0,0) -- (-\L,\L) -- cycle;
        \fill[green, opacity=0.1] (\L,-\L) -- (0,0) -- (\L,\L) -- cycle;
        \draw[ultra thick,black]  (-\L,-\he) -- (-\L,\he) ;
        \draw[black] (\L,0) node[right] {$\rho=\infty$};
        \draw[ultra thick,black]  (\L,-\he) -- (\L,\he);

        \draw[very thick,dashed,black]  (-\L,-\L) -- (\L,\L);
        \draw[very thick,dashed,black]  (\L,-\L) -- (-\L,\L);
        \draw[black] (0,0.5*\L) node[above] {$\rho=0$};
        \draw[-{Stealth[scale=1]}, thick, black] (0,0.55*\L) -- (0,0.15*\L);
        \draw[black,opacity=0] (-\L,-\he) node[below] {$\sigma=0$};
    \end{tikzpicture}
    \caption{Rindler patch.}
    \label{sfig:AdS2Rindler}
    \end{subfigure}
    \caption{The three relevant patches of AdS$_2$. In each case, observers at fixed spatial coordinate follow the red trajectories.}
    \label{fig:AdS2patches}
\end{figure}

\begin{itemize}
    \item Global patch. This patch is parametrized by the coordinates $(t,\sigma)$, with $-\infty <t<\infty$ and $\sigma \in [0,\pi]$, with line element
    \begin{equation}
        ds^2=\frac{-dt^2+d\sigma^2}{\sin^2\sigma}~.
        \label{eq:AdS2global}
    \end{equation}
    The two boundaries are causally connected, behaving like a traversable wormhole. An observer at fixed $\sigma$ follows the vertical red trajectory in Fig. \ref{sfig:AdS2global}.
    \item Poincaré patch. In this patch, an observer sitting at a fixed $z$ follows the red trajectory of Fig. \ref{sfig:AdS2Poincare}. It is clear that the lines at a 45-degree angle represent a causal horizon for this observer. The line element is given by
    \begin{equation}
        ds^2=\frac{-dt_P^2+dz^2}{z^2}~.
    \end{equation}
    \item Rindler patch. The Rindler patch covers part of the Poincaré patch and is parametrized by the coordinates $(t_R,\rho)$, with line element
    \begin{equation}
        ds^2=-\sinh^2\rho\h dt_R^2+d\rho^2~.
        \label{eq:Ads2Rindler}
    \end{equation}
    Observers at fixed $\rho$ follow the red trajectories of Fig. \ref{sfig:AdS2Rindler}. The dashed lines represent the black hole horizon for these observers.
\end{itemize}

The different patches are related by changes of coordinates. An important relation is the following, relating the time coordinates in the different systems:
\begin{equation}
    t_P=\tan\frac{t_r}{2}=-\frac{1}{\tan\frac{t_l}{2}}=\tanh\frac{t_R}{2}~,
    \label{eq:relationtimes}
\end{equation}
where $t_l$, $t_r$ denote the global time coordinate at the left/right boundary, respectively. See Appendix \ref{app:AdS2} for details.

\begin{figure}
\centering
\begin{tikzpicture}[baseline={(current bounding box.center)}]
    \def\L{1.5} 
    \def\he{1.1*\L} 

    \filldraw[
        green!30, opacity=0.4,
        decorate,
        decoration={random steps, segment length=2pt, amplitude=1pt}
    ]
    (-\L,-\L)
    to[out=80, in=-80] (-\L,\L)
    -- (\L,\L)
    to[out=-100, in=100] (\L,-\L)
    -- cycle;
    \draw[ultra thick,black]  (-\L,-\he) -- (-\L,\he) ;
    \draw[ultra thick,black]  (\L,-\he) -- (\L,\he);

    \draw[very thick,dashed,black]  (-\L,-\L) -- (\L,\L);
    \draw[very thick,dashed,black]  (\L,-\L) -- (-\L,\L);

    \draw [black,very thick, decorate, decoration={random steps,segment length=2pt,amplitude=1pt}]
        (\L,-\L) to[out=100, in = -100] (\L,\L);
    \draw [black,very thick, decorate, decoration={random steps,segment length=2pt,amplitude=1pt}]
        (-\L,-\L) to[out=80, in = -80] (-\L,\L);

    \def\radius{1.5}
    \def\separ{5}
    \filldraw[fill=green!30, opacity=0.4, very thick, decorate, decoration={random steps,segment length=3pt,amplitude=1.5pt}] (\separ,0) circle (0.9*\radius);
    \draw[black, very thick] (\separ,0) circle (\radius);

\end{tikzpicture}
\caption{Lorentzian (left) and Euclidean (right) AdS$_2$ space with a cutoff surface near the boundary. Inspired by Fig. 2.17 of \cite{Turiaci_2024}.}
\label{fig:cutoffAdS}
\end{figure}
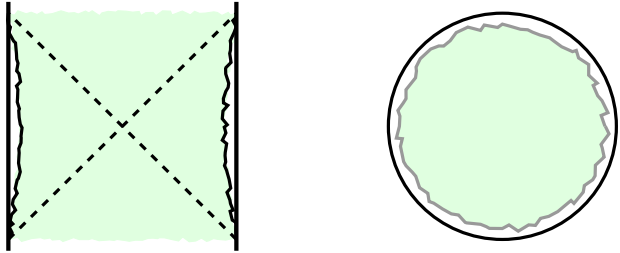

The physics in every patch is very different, and there has to be some mechanism to fix which choice we should consider. The issue here is that we have assumed a constant dilaton everywhere. The solution \cite{Maldacena_2016} is to allow for a non-trivial dilaton profile. Specifically, this requires imposing boundary conditions that break the conformal symmetry of AdS$_2$. One way to achieve this is by cutting off the spacetime with a non-trivial shape near the boundary (see Fig. \ref{fig:cutoffAdS}). Denoting by $u$ the proper time along the boundary trajectory, this trajectory can be parametrized in terms of the Poincaré coordinates as $t_P(u)$, $z(u)$. The location of the boundary trajectory is determined by means of two boundary conditions. The first one is to impose that the boundary trajectory has a fixed proper length, which in Euclidean signature reads
\begin{equation}
    g_{uu}du^2=\frac{t_P'^2+z'^2}{z^2}du^2=\frac{1}{\varepsilon^2}~,
    \label{eq:bdryproperlength}
\end{equation}
where primes denote derivatives with respect to $u$. Given an arbitrary $t_P(u)$, \eqref{eq:bdryproperlength} is satisfied by $z(u)=\varepsilon\h t_P'(u)$, at leading order in $\varepsilon$. Therefore, the boundary trajectory is parametrized solely in terms of the function $t_P(u)$.

The second boundary condition is imposed from the equation of motion \eqref{eq:eomdilaton}. Its most general solution is divergent near the boundary, with a $1/z$ divergence. The strength of this divergence gives a new dimensionful coupling, $\Phi_r$, defined as
\begin{equation}
    \Phi_b=\frac{\Phi_r}{\varepsilon}~,\qquad \varepsilon\to 0~,
\end{equation}
where $\Phi_b$ is the divergent value of the dilaton near the boundary, and $\Phi_r$ is a finite quantity.

It is not difficult to check that with these two conditions the boundary term in the action \eqref{eq:actionJT} can be written in terms of $\Phi_r$ and $t_P(u)$\footnote{We only need the boundary term because the Einstein-Hilbert term vanishes due to the equation of motion \eqref{eq:eomphi}.}. At leading order in $\varepsilon$, the extrinsic curvature is $K=1+\varepsilon^2\h\{t_P(u),u\}$, where $\{t_P(u),u\}$ denotes the Schwarzian derivative,
\begin{equation}
    \{f(z),z\}=\Sch(f(z),z)=\frac{f'''}{f'}-\frac{3}{2}\left(\frac{f''}{f'}\right)^2~.
\end{equation}

Therefore, the boundary action is
\begin{equation}
    S_{\text{bdy}}=-\frac{\Phi_r}{8\pi G_N}\int du\{t_P(u),u\}~,
    \label{eq:Schwactionbdry}
\end{equation}
where all the dynamics is captured by the single degree of freedom, $t_P(u)$. We can view $t_P(u)$ as a map between the physical time of the boundary trajectory and the Poincaré time in the interior.

This is a remarkable result: in nearly-AdS$_2$ gravity, all the dynamical information is encoded in the motion of the physical boundary, described by the trajectory $t_P(u)$. The action governing this degree of freedom is the Schwarzian action. Interestingly, the same Schwarzian action appeared in the low-energy limit of the SYK model, where we saw that it is the action governing the dynamics of the mode $f(\tau)$, which explicitly breaks the emergent reparametrization of the SYK model in the IR limit (see Eq. \eqref{eq:SchwSYK}).

\subsection{Making the AdS\texorpdfstring{$_2$}{2} wormhole traversable}

We have not answered yet which of the three patches of AdS$_2$ is a solution of the nearly-AdS$_2$ equations of motion.

The Schwarzian equation of motion arising from \eqref{eq:Schwactionbdry} is
\begin{equation}
    \phi_r\h\frac{\left[\Sch(t_P(u),u)\right]'}{t_P'(u)}=0~,
    \label{eq:eomT}
\end{equation}
with $\phi_r\equiv \frac{\Phi_r}{8\pi G_N}$. This equation is just the statement of energy conservation (the Noether charge corresponding to time-translation symmetry is precisely the Schwarzian).

A simple Lorentzian solution of \eqref{eq:eomT} is
\begin{equation}
    t_P(u)=\tanh\frac{\pi u}{\beta}~,\qquad \rightarrow \qquad t_R=\frac{2\pi}{\beta}u~,
    \label{eq:SchwarziansolutionRindler}
\end{equation}
where to go from the left to the right we have used \eqref{eq:relationtimes}. This result states that the boundary trajectory corresponds to a trajectory of fixed $\rho$ in the Rindler patch, \eqref{eq:Ads2Rindler}, where the two boundaries are causally disconnected. In the absence of matter, the global patch, in which the two boundaries are causally connected, is never realized. The same situation arises in the presence of matter, if the matter obeys the ANEC. Thus, the AdS$_2$ wormhole, much like its higher-dimensional counterpart discussed in Section \ref{sec:traversableWHs}, remains non-traversable. 

Following the original proposal of Gao, Jafferis, and Wall \cite{Gao_2016}, a direct coupling between the two asymptotic boundaries can make this wormhole traversable, as shown in \cite{Maldacena_2018}. The idea is to introduce a coupling in such a way that the solution to the Schwarzian equation of motion corresponds to lines of constant $\sigma$ in the global coordinates \eqref{eq:AdS2global}. The precise coupling is given by
\begin{equation}
    S_{int}=g\sum_{i=1}^N\int du\h \mathcal{O}_L^i(u)\mathcal{O}_R^i(u)~,
\end{equation}
where $\mathcal{O}^i$ are bulk fields evaluated at each boundary. We consider these boundary operators to have conformal dimension $\Delta$. For the same reason as in the higher-dimensional setup \cite{Gao_2016}, this interaction creates negative null energy in the bulk \cite{Maldacena_2018}. 

If $g$ is sufficiently small, the following replacement can be done in the path integral:
\begin{equation}
    \langle e^{i\h g \sum_i\int du\h \mathcal{O}_L^i(u)\mathcal{O}_R^i(u)}\rangle\simeq e^{i\h g \sum_i \int du\langle \mathcal{O}_L^i(u)\mathcal{O}_R^i(u) \rangle}
\end{equation}

The correlator of the exponential is given, in Poincaré coordinates, by
\begin{equation}
    \langle \mathcal{O}_L(t_{P,1})\mathcal{O}_R(t_{P,2})\rangle = \frac{1}{\abs{t_{P,1}-t_{P,2}}^{2\Delta}}~.
    \label{eq:correlatorPoincare}
\end{equation}

We can relate this correlator in Poincaré coordinates to the correlator in proper time $u$ by using the transformation of the primary fields under a change of coordinates $u\to t_P(u)$:
\begin{equation}
    \mathcal{O}(u)=\left(\frac{dt_P}{du}\right)^\Delta \mathcal{O}(t_P)~.
\end{equation}
Moreover, we change coordinates from $t_P$ to the global times at each boundary, $t_l$ and $t_r$, related to $t_P$ as in \eqref{eq:relationtimes}. The result is
\begin{equation}
    g\sum_{i=1}^N\langle \mathcal{O}_L^i(u)\mathcal{O}_R^i(u)\rangle =\frac{g N}{2^{2\Delta}}\left(\frac{t_l'(u)t_r'(u)}{\cos^2\left(\frac{t_l(u)-t_r(u)}{2}\right)}\right)^\Delta~.
    \label{eq:correlatortltr}
\end{equation}

After adding the coupling between the boundaries, the total Schwarzian action is given by
\begin{equation}
    S=\int du\left[-\phi_r \left\{\tan\frac{t_l(u)}{2},u\right\}-\phi_r \left\{\tan\frac{t_r(u)}{2},u\right\}+\frac{g N}{2^{2\Delta}}\left(\frac{t_l'(u)t_r'(u)}{\cos^2\left(\frac{t_l(u)-t_r(u)}{2}\right)}\right)^\Delta\right]~.
    \label{eq:2coupledSchw}
\end{equation}
Notice that we have also written the boundary Schwarzian factors, \eqref{eq:Schwactionbdry}, in terms of the time on each boundary, using \eqref{eq:relationtimes}\footnote{Although the relations between $t_P$ and $t_l$, $t_r$ are different, the Schwarzians are the same on both sides. This is because the Schwarzian derivative is invariant under Möbius transformations, $\Sch\left(\frac{a\h f+b}{c\h f+d},z\right)=\Sch(f,z)$, with $a,b,c,d\in \mathbb{R}$ such that $ab-cd=1$. From \eqref{eq:relationtimes} we see that the transformation from $t_P$ to $t_l$ is obtained from the transformation from $t_P$ to $t_r$ by a Möbius transformation with $a,d=0$, $b=1$, $c=-1$.}.

In Appendix \ref{app:Schwarzianeqs} we analyze in detail this action and its solutions. Let us just mention here its most relevant aspects for the present discussion. First of all, in order to compare with the same action that will appear in the SYK construction, we re-scale the boundary time $u$ as $u\to \frac{N}{\phi_r}u$, which leads to the action
\begin{equation}
    S=N\int du\left[-\left\{\tan\frac{t_l(u)}{2},u\right\}-\left\{\tan\frac{t_r(u)}{2},u\right\}+\eta\left(\frac{t_l'(u)t_r'(u)}{\cos^2\left(\frac{t_l(u)-t_r(u)}{2}\right)}\right)^\Delta\right]~,
    \label{eq:2coupledSchwresc}
\end{equation}
with $\eta\equiv \frac{g}{2^{2\Delta}}\left(\frac{N}{\phi_r}\right)^{2\Delta -1}$.

As we discuss in the appendix, this action is invariant under the $SL(2,\mathbb{R})$ infinitesimal transformations
\begin{equation}
\begin{aligned}
    \delta t_l & = \varepsilon^0+\varepsilon^+e^{it_l}+\varepsilon^-e^{-it_l}~,\\
    \delta t_r & = \varepsilon^0-\varepsilon^+e^{it_r}-\varepsilon^-e^{-it_r}~.
\end{aligned}
\end{equation}

Choosing a gauge in which $t_l(u)=t_r(u)\equiv t(u)$ and imposing the gauge constraint that the Noether charges associated to this symmetry vanish \cite{Maldacena_2016,Maldacena_2017,Maldacena_2018}, one is able to reduce the problem to the following equation (see Appendix \ref{app:Schwarzianeqs}),
\begin{equation}
    \varphi''=-e^{2\varphi}+\eta\Delta e^{2\Delta \varphi}~,
\end{equation}
where $\varphi\equiv \log t'$. This corresponds to the equation of motion of a particle in a potential
\begin{equation}
    V(\varphi)=e^{2\varphi}-\eta e^{2\Delta \varphi}~.
\end{equation}

There is a solution for which $\varphi$ is a constant, which corresponds to the stable point at the minimum of the potential. In terms of $t'$, this solution is given by
\begin{equation}
    (t')^{2(1-\Delta)}=\eta\Delta~.
    \label{eq:tpsolution}
\end{equation}

A constant derivative $t'$ means that the global times at the boundaries are given by
\begin{equation}
    t_l(u)=t_r(u)=t' u~,
\end{equation}
which means that now the boundary trajectories are straight lines of constant $\sigma$ in the global coordinates. Therefore, the two boundary trajectories are in causal contact at all times, which creates an eternal traversable wormhole \cite{Maldacena_2018}.

The value of $t'$ in \eqref{eq:tpsolution} determines the energy of bulk excitations, which by $SL(2,\mathbb{R})$ symmetry are fixed to be
\begin{equation}
    E_n^\text{conf}=t'(\Delta+n)~,\qquad n=0,1,2,...
    \label{eq:conformalspectrum}
\end{equation}

Finally, expanding the potential around the minimum, we obtain a harmonic oscillator of frequency
\begin{equation}
    \omega_0=\sqrt{2(1-\Delta)}t'~.
\end{equation}
This corresponds a quantum harmonic oscillator with energy levels
\begin{equation}
    E_n^\text{bg}=t'\sqrt{2(1-\Delta)}\left(n+\frac{1}{2}\right)~.
    \label{eq:bdrygravspectrum}
\end{equation}
These are the excitations of the "boundary graviton" degree of freedom.

As noted in \cite{Maldacena_2018}, the same action \eqref{eq:2coupledSchwresc} governs the low-energy limit of two coupled SYK models, which we now review. This identification suggests that by coupling two SYK models, we can study wormhole physics from a purely quantum mechanical point of view.

\subsection{Two coupled SYK models}\label{subsec:2coupledSYK}

We consider now the model proposed in \cite{Maldacena_2018}, which consists of two identical SYK systems, each containing $N$ Majorana fermions with all-to-all random interactions. The two systems are coupled through a bilinear interaction that pairs fermions from each system. The Hamiltonian is given by
\begin{equation}   
    H=\frac{1}{4!}\sum_{a=L,R}\sum_{ijkl}J_{ijkl}\chi_a^i\chi_a^j\chi_a^k\chi_a^l+i\mu\sum_j \chi_L^j\chi_R^j~,
    \label{eq:Hamilt2coupledSYK}
\end{equation}
$a=L,R$ denotes the left and right SYK systems, respectively, and the coupling between the two SYK factors is denoted by $\mu$. The random couplings $J_{ijkl}$ are drawn from a Gaussian distribution with zero mean and variance given by \eqref{eq:Jmeanvar}. Importantly, we consider the couplings to be the same on each side.

Following the same analysis as in the single SYK case, the averaged partition function can be written in terms of an effective action in terms of the fields $G_{ab}$ and $\Sigma_{ab}$. $G_{ab}$ is defined as
\begin{equation}
    G_{ab}(\tau,\tau')=\frac{1}{N}\sum_j \chi_a^j(\tau)\chi_b^j(\tau')~,
\end{equation}
where now we have four types of correlators: $G_{LL}$, $G_{LR}$, $G_{RL}$, and $G_{RR}$. $\Sigma_{ab}$ will turn out to correspond to the self-energies.

After averaging over the random couplings and integrating out the Majorana fields, we can write the averaged partition function as
\begin{equation}
    \overline{Z}=\int\mathcal{D}G_{ab}\mathcal{D}\Sigma_{ab}~e^{-S_\text{eff}[G,\Sigma]}~,
\end{equation}
with
\begin{align}
\begin{split}
    \frac{S_\text{eff}[G,\Sigma]}{N}=&~-\frac{1}{2}\log\det \left(\delta(\tau-\tau')\left(\delta_{ab}\partial_\tau-\mu \sigma_{ab}^y\right)-\Sigma_{ab}(\tau,\tau')\right)\\
    &+\frac{1}{2}\int d\tau d\tau'\sum_{a,b}\Sigma_{ab}(\tau,\tau')G_{ab}(\tau,\tau')-\frac{J^2}{8}\sum_{a,b}\int d\tau d\tau' \left[G_{ab}(\tau,\tau')\right]^4~.
\label{eq:Seffeuclidean}
\end{split}
\end{align}
The matrix $\sigma^y$ is given by $\sigma_{LL}^y=\sigma_{RR}^y=0$, $\sigma_{LR}^y=-\sigma_{RL}^y=-i$.

In the large-$N$ limit, the dynamics of the system are governed by the saddle point equations 
\begin{align}
\begin{split}
    G_{LL}(i\omega_n)&=-\frac{i\omega_n+\Sigma_{LL}}{(i\omega_n+\Sigma_{LL})^2+(i\mu-\Sigma_{LR})^2}\\[10pt]
    G_{LR}(i\omega_n)&=-\frac{i\mu-\Sigma_{LR}}{(i\omega_n+\Sigma_{LL})^2+(i\mu-\Sigma_{LR})^2}~,
\label{eq:SDequilibrium}
\end{split}
\end{align}
where the self-energies $\Sigma_{ab}$ are given by
\begin{equation}
    \Sigma_{ab}(\tau)=J^2 G_{ab}(\tau)^3~.    
    \label{eq:Sigmaeq}
\end{equation}

For small $\mu$, the low-energy regime of the system is expected to behave approximately as two copies of the decoupled SYK models, plus a relevant deformation due to the $\mu$ coupling. It was shown in \cite{Maldacena_2018} that, for small $\mu$, the ground state of the coupled system is very close to the TFD state of the two decoupled systems, with a particular inverse temperature determined by the left-right coupling, $\beta(\mu)$.

In this limit, the approximate conformal symmetry allows us to obtain the left-right correlators from reparametrizations of the single side conformal correlators, in the same way as in \eqref{eq:correlatorPoincare}-\eqref{eq:correlatortltr}. From the two-point function in the IR limit, given in \eqref{eq:Gconfsolution}, one obtains
\begin{equation}
    \langle \chi_L(t_l)\chi_R(t_r)\rangle=c_\Delta\frac{i}{\abs{2\mathcal{J}\cos\frac{t_l-t_r}{2}}^{2\Delta}}~,
    \label{eq:lrcorrelator}
\end{equation}
where $2\mathcal{J}^2=J^2$, and $c_\Delta=\frac{1}{2}\left[(1-2\Delta)\frac{\tan \pi\Delta}{\pi\Delta}\right]^\Delta$. In order to determine whether these correlators are close to the real solution, we need to allow for arbitrary reparametrizations $t_l(u)$ and $t_r(u)$ and study the resulting effective action. In the same way as in the previous section, each SYK system leads to a Schwarzian action, while the interaction term, after exponentiating and considering the small-$\mu$ limit, contributes with the correlator \eqref{eq:lrcorrelator}. The resulting effective action is
\begin{equation}
    S=N\int du \left[-\frac{\alpha_S}{\mathcal{J}}\left(\left\{\tan\frac{t_l(u)}{2},u\right\}+\left\{\tan\frac{t_r(u)}{2},u\right\}\right)+\mu\frac{c_\Delta}{(2\mathcal{J})^{2\Delta}}\left(\frac{t'_l(u)t'_r(u)}{\cos^2{\frac{t_l(u)-t_r(u)}{2}}}\right)^\Delta\right]~,
    \label{eq:2coupledSYKSchw}
\end{equation}
which is the same action (up to identification of constants and a rescaling of the boundary time, see Appendix \ref{app:Schwarzianeqs}) as \eqref{eq:2coupledSchw}. Therefore, the analysis we performed in that section, as well as in the appendix, also applies to the two-coupled SYK system. In particular, let us emphasize here the results obtained in equations \eqref{eq:conformalspectrum}, \eqref{eq:bdrygravspectrum}: if the identification of the AdS$_2$ wormhole and two coupled SYK models is correct, in the low-$\mu$ regime we should expect the quantum mechanical system to exhibit the same two towers of states, with energies given by
\begin{align}
    E_n^\text{conf} & = t'(\Delta+n)~,\label{eq:2towersconf}\\
    E_n^\text{bg} & = t'\sqrt{2(1-\Delta)}\left(n+\frac{1}{2}\right)~,\label{eq:2towersbg}
\end{align}
corresponding to conformal excitations and the boundary graviton, respectively. These two towers\footnote{Tolkien fans might complain that any union of Two Towers should be named \textit{Mordor} and \textit{Isengard}, but we will stick with \textit{conformal} and \textit{boundary graviton}.} will be important in the research part of the thesis.

Away from the low-energy limit, the Dyson equations \eqref{eq:SDequilibrium} can be solved numerically with the same method as in the single SYK case. The physics of the coupled model is characterized by two stable phases in the $(T, \mu)$ parameter space, in units of $J = 1$.

At low temperatures and small coupling $\mu$, the system exhibits a gapped phase identified with the traversable wormhole geometry in which the two boundaries are smoothly connected through the interior. We are going to refer to this phase as the wormhole (WH) phase from now on.

At high temperatures or sufficiently large coupling $\mu$, the system transitions into a phase where the two SYK models behave approximately as decoupled systems. From a gravitational perspective, this phase corresponds to two disconnected AdS$_2$ black holes. For this reason, this phase is commonly referred to as the black hole (BH) phase. The entanglement between the two boundaries is significantly reduced in this phase, and the wormhole geometry breaks down. This transition is analogous to the Hawking-Page transition between thermal AdS and AdS black hole geometries in higher dimensions.

\begin{figure}
    \centering
    \begin{subfigure}[c]{0.55\textwidth}
        \centering
        \includegraphics[width=\textwidth]{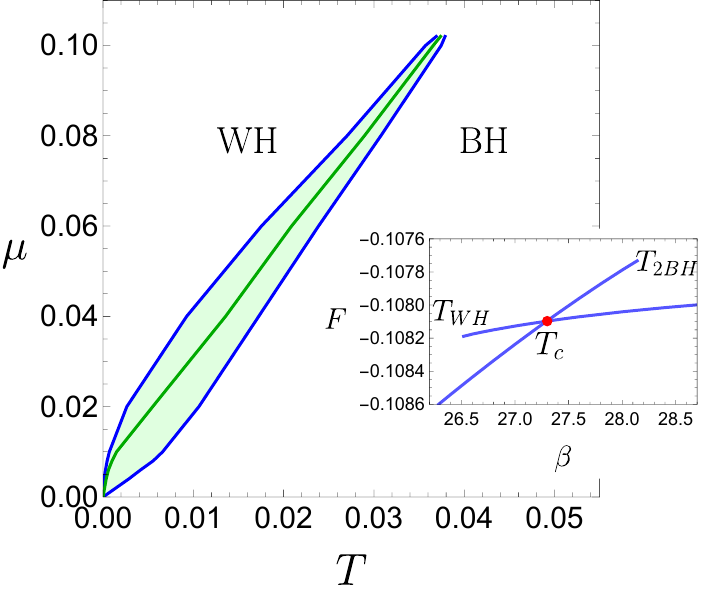}
    \end{subfigure}
    \hspace{0.02\textwidth}
    \begin{subfigure}[c]{0.41\textwidth}
        \centering
        \includegraphics[width=\textwidth]{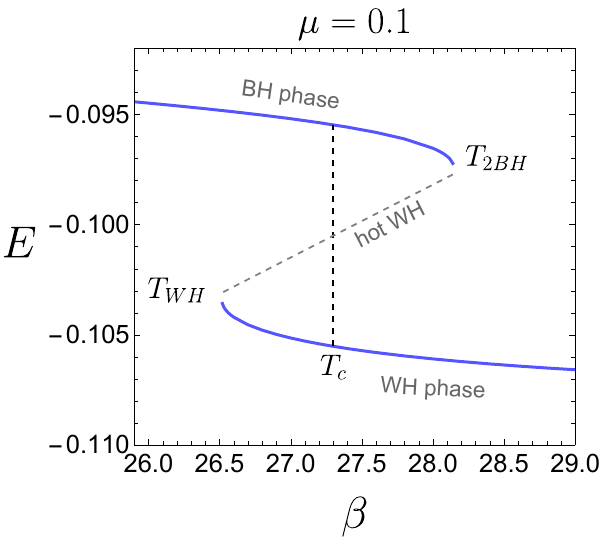}
    \end{subfigure}
    \caption{Phase diagram of the coupled model. Left: the two phases in the $(T,\mu)$ plane. The green area denotes a region of coexistence between the two phases. The green line denotes the critical temperature at which the phases switch dominance, as dictated by the minimum value of the free energy, as shown in the inset for $\mu=0.1$. Right: phase diagram in the energy vs inverse temperature plane.}
    \label{fig:phasediagram}
\end{figure}

Connecting these two phases is an unstable intermediate phase, referred to as the "hot wormhole" \cite{Maldacena_2018, Maldacena_2019}. While this phase cannot be observed in equilibrium canonical simulations due to its instability, it has been conjectured to be stable in the microcanonical ensemble \cite{Maldacena_2018}. The phase diagram for $\mu=0.1$ is shown in Fig. \ref{fig:phasediagram}. The unstable phase can be analyzed analytically in the large-$q$ limit. In this limit, as the system transitions between phases, the energy evolves smoothly, suggesting that, although the geometry undergoes a discrete change, the underlying quantum dynamics remain continuous.

This observation is particularly relevant for cooling processes, where the system can dynamically relax from a high-energy state with decoupled black holes into the low-energy wormhole ground state. As demonstrated in \cite{Maldacena_2019}, coupling the system to a cold bath allows it to explore the unstable phase during the transition. Notably, for $q=4$, the smooth transition between the black hole and wormhole phases persists, indicating that the system effectively traverses the unstable phase as part of the relaxation process. We will explore this in detail in Chapter \ref{chap:FloquetSYK}.

\section{Real-time formalism and revival dynamics}

The work of Maldacena and Qi provided compelling evidence that the system consisting of two coupled SYK models with Hamiltonian \eqref{eq:Hamilt2coupledSYK} has a ground state that can be identified with a traversable wormhole geometry in AdS$_2$.

As was argued in Section \ref{sec:traversableWHs}, this traversability of the wormhole should be seen in the SYK side of the correspondence: if we consider first the decoupled case ($\mu=0$), we can imagine creating an excitation on the right system, $\chi_R^i\ket{G}$, with $\ket{G}$ being the ground state of the model. Due to the maximally chaotic nature of the SYK model, this excitation will rapidly dissipate and its quantum information will be \textit{scrambled} among the degrees of freedom of the right system.

However, when the coupling between the two sides is turned on, $\mu\neq 0$, the wormhole-like behavior of the system should be able to allow the scrambled excitation to emerge on the left side, where it should be \textit{unscrambled} to its original form: the state of the system is approximately $\chi_L^i\ket{G}$.

This \textit{revival} of the excitations between the two sides of the wormhole was not seen in the exact numerical solution of the model in Euclidean simulations. The pioneering work by Plugge, Lantagne-Hurtubise and Franz \cite{Plugge_2020} successfully observed these revivals by studying the real-time version of the model, which we briefly review here since this will be our starting point in Chapter \ref{chap:FloquetSYK}.

The idea is to analytically continue the Dyson equations \eqref{eq:SDequilibrium}-\eqref{eq:Sigmaeq} to real time and perform a numerical iterative method to solve them. These analytically continued equations will be extensively used in the Research part of the thesis, and for this reason we devote an appendix (in the Research part) to derive them; see Appendix \ref{app:realtimeeqs}. The real-time version of Eq. \eqref{eq:SDequilibrium} is
\begin{align}
\begin{split}
    G_{LL}^R(\omega)&=-\frac{\omega+i\delta +\Sigma_{LL}}{(\omega+i\delta +\Sigma_{LL})^2+(i\mu-\Sigma_{LR})^2}\\[10pt]
    G_{LR}^R(\omega)&=-\frac{i\mu-\Sigma_{LR}}{(\omega+i\delta +\Sigma_{LL})^2+(i\mu-\Sigma_{LR})^2}~,
\label{eq:SDeqsreal}
\end{split}
\end{align}
where the imaginary Matsubara frequencies have been analytically continued to real time by the standard procedure $i\omega_n\rightarrow \omega +i\delta$ \cite{Bruus_2004}, and $G_{ab}^R$ are retarded propagators. The real-time version of \eqref{eq:Sigmaeq} is
\begin{align}
\begin{split}
    \Sigma_{LL}^R(\omega)&=2iJ^2\int_{0}^{\infty}dt e^{i(\omega+i\delta)t}\Re\left[n_{LL}^3(t)\right]\\
    \Sigma_{LR}^R(\omega)&=2iJ^2\int_{0}^{\infty}dt e^{i(\omega+i\delta)t}\Im\left[n_{LR}^3(t)\right]~,
\label{eq:SigmaReIm}
\end{split}
\end{align}
where the functions $n_{LL}(t)$, $n_{LR}(t)$ are defined as
\begin{equation}
    n_{ab}(t)\equiv\int_{-\infty}^{\infty}d\omega\rho_{ab}(\omega)n_F(\omega)e^{-i\omega t}~,
\label{eq:occupations}
\end{equation}
with $n_F(\omega)$ the Fermi distribution function, and $\rho_{ab}(\omega)$ the spectral functions, see Appendix \ref{app:realtimeeqs} for details.

The numerical method allows to obtain the spectral functions, $\rho_{ab}(\omega)$, which in turn provide the greater and lesser components of the real-time two-point functions,
\begin{align}
\begin{split}
    G_{ab}^>(t_1,t_2)&=-\frac{i}{N}\sum_j\langle\chi_a^j(t_1)\chi_b^j(t_2)\rangle~,\\
    G_{ab}^<(t_1,t_2)&=-\frac{i}{N}\sum_j\langle\chi_b^j(t_2)\chi_a^j(t_1)\rangle~
    \label{eq:greaterlesser}
\end{split}
\end{align}
(see Eq. \eqref{eq:greaterlesserrho}).

\begin{figure}
    \centering
    \begin{subfigure}[c]{0.48\textwidth}
        \centering
        \includegraphics[width=\textwidth]{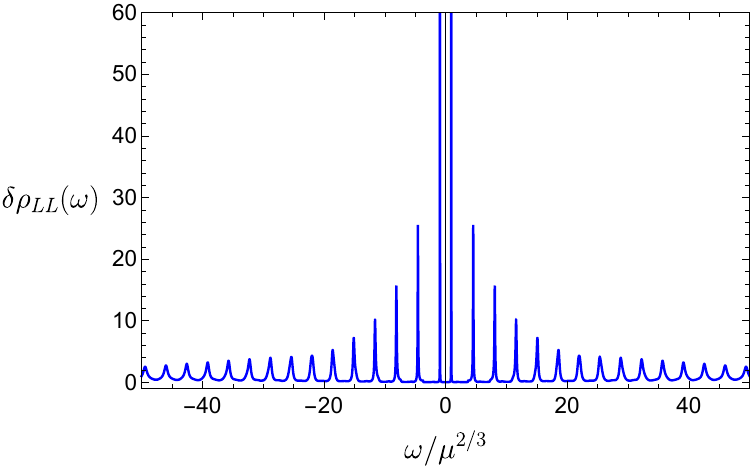}
    \end{subfigure}
    \hspace{0.02\textwidth}
    \begin{subfigure}[c]{0.48\textwidth}
        \centering
        \includegraphics[width=\textwidth]{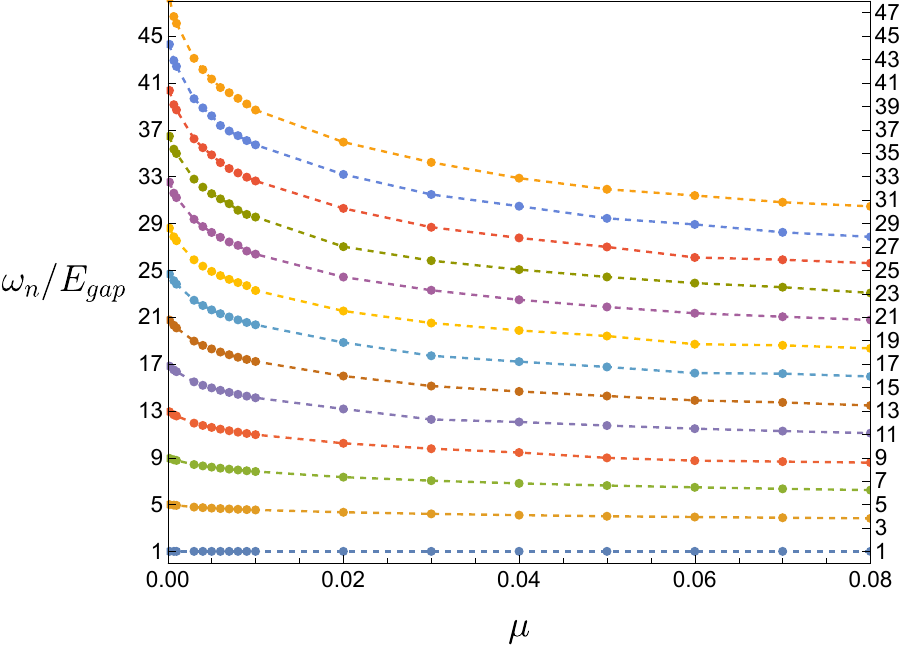}
    \end{subfigure}
    \caption{Reproduction of Fig. 1 of \cite{Plugge_2020} using our codes. Left: relative spectral function, $\delta \rho_{LL}=\rho_{LL}/\rho_\text{SYK}$, showing the peak spectrum for $\mu=0.001$, $\beta=20000$. Right: spectrum of the coupled model for different values of $\mu$. At low $\mu$ the spectrum approaches the conformal tower \eqref{eq:2towersconf} with $\Delta=1/4$.}
    \label{fig:spectrallevels}
\end{figure}

The first result of \cite{Plugge_2020} was to notice that the spectral function differs a lot between the two phases of the model. The black hole phase has a continuum spectrum, while in the wormhole phase the spectral function contains a series of evenly-spaced spectral peaks at discrete frequencies, $\omega_n$. In the low-$\mu$ regime, these peaks agree with the conformal tower \eqref{eq:2towersconf} predicted by holography, with $\Delta=1/4$ and an energy gap $E_{\text{gap}}\sim \Delta\mu^{2/3}$ \footnote{The energy gap is determined by the lowest level among the conformal tower and the boundary graviton spectrum; see Eqs.~\eqref{eq:2towersconf}--\eqref{eq:2towersbg}. For $\Delta<1/2$, it is set by the lowest conformal level, whereas at $\Delta=1/2$ the two lowest levels cross, and the gap is instead set by the lowest boundary graviton state. Since we will always work with $\Delta=1/4$, $E_{gap}$ will always be given by the lowest energy in the conformal tower.}. We show these results of \cite{Plugge_2020} in Fig. \ref{fig:spectrallevels}. The boundary graviton states remained unobserved in their analysis.

\begin{figure}
    \centering
    \includegraphics[width=0.6\linewidth]{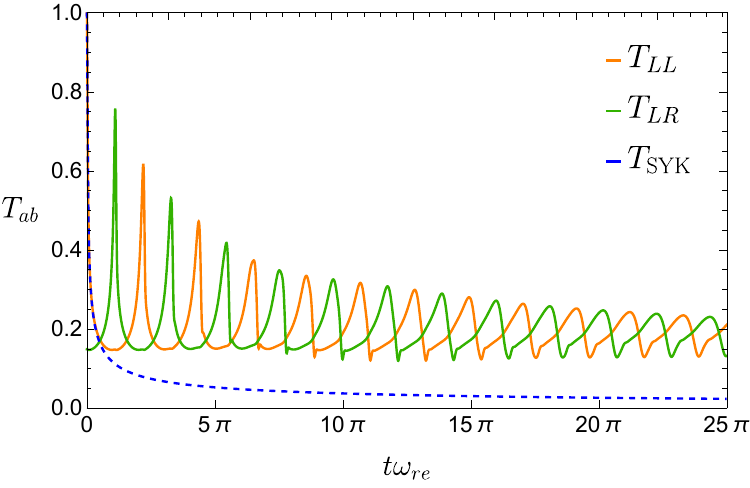}
    \caption{Reproduction of Fig. 2 of \cite{Plugge_2020} using our codes. Transmission coefficients in the wormhole phase (orange and green). The phase opposition of the revivals in the wormhole phase is consistent with having a traversable wormhole. In the black hole phase the correlators decay exponentially with time, with a behavior much closer to the case of two decoupled SYK models (dashed blue). Also the frequency scaling as $t_{re} \sim \mu^{2/3}$ is, for small $\mu$, much higher than the natural coupling frequency $\sim\mu$.}
    \label{fig:LHrevivals}
\end{figure}

The other result of \cite{Plugge_2020} that strengthens the traversable wormhole behavior of the two-coupled model is the observation of revivals. These revivals are observed in the time evolution of the transmission amplitudes between the two systems, defined as
\begin{equation}
    T_{ab}(t)=2\abs{G^>_{ab}(t,0)}~.
\end{equation}
The transmission amplitudes encode the probability of recovering $\chi^j_a$ at time $t$  after having inserted $\chi^j_b$ at time $t=0$. Fig. \ref{fig:LHrevivals} shows the different behaviors in the two phases. On the left plot, the peaked out-of-phase revivals  are interpreted as being duals of a probe particle traversing from one side of the wormhole to the other and bouncing back and forth. The period of the oscillations scales as $\mu^{2/3}$ \cite{Plugge_2020, Haenel_2021, Zhang_2020}, in agreement with the holographic prediction \cite{Maldacena_2018}. The slow decay in the overall envelope indicates that the model allows for particles to thermalize. This feature still deserves an explanation from the dual gravitational point of view.

\section{The Schwinger-Keldysh formalism, \textit{or There and Back Again}}\label{sec:SchwingerKeldysh}

The real-time formulation of the two-coupled SYK models, as discussed so far, captures only the behavior of the system in thermal equilibrium. In Chapter \ref{chap:FloquetSYK}, we will extend our analysis to include various periodic drivings of the parameters.

To address such scenarios, one requires a formalism explicitly designed to handle non-equilibrium dynamics. A powerful and widely used approach is the Schwinger-Keldysh (or closed-time-path) formalism, which provides a consistent framework for computing real-time correlation functions in quantum systems out of equilibrium. In what follows, we motivate this formalism and outline how the SYK model can be embedded within it. Our presentation largely follows the exposition in \cite{Maciejko_2007}. Other useful reviews on the topic include \cite{Jauho_2006,Kamenev_2011,Haehl_2016,Haehl_2024}.

In standard equilibrium perturbation theory, the central object of interest is the time-ordered Green’s function, defined as
\begin{equation}
    i\h G(t,\Vec{x};t',\Vec{x}')=\bra{\Omega}T[\psi(t,\Vec{x})\psi^\dagger (t',\Vec{x}')]\ket{\Omega}~,
    \label{eq:GFgstate}
\end{equation}
where $\ket{\Omega}$ denotes the ground state of an interacting theory governed by a Hamiltonian $H$\footnote{In general one considers the expectation value \begin{equation}
    i G(t,\Vec{x};t',\Vec{x}')=\langle T[\psi(t,\Vec{x})\psi^\dagger (t',\Vec{x}')]\rangle=\Tr\left[\rho T[\psi(t,\Vec{x})\psi^\dagger (t',\Vec{x}')]\right]~,
    \label{eq:GFgenericrho}
\end{equation}
where $\rho$ is an equilibrium density matrix. At zero temperature, this reduces to a projector onto the ground state, $\rho=\ket{\Omega}\bra{\Omega}$ recovering \eqref{eq:GFgstate}. At finite temperature, the perturbative expansion is formally equivalent but formulated in terms of Euclidean time evolution.}. The field operator $\psi(t,\Vec{x})$ is defined in the Heisenberg picture,
\begin{equation}
    \psi(t,\Vec{x})=U^\dagger(t,0)\psi(\Vec{x})U(t,0)~,
    \label{eq:fieldsHeisen}
\end{equation}
with the time evolution operator $U(t,t')$ given by
\begin{equation}
    U(t,t')=T\exp\left[-i\int_{t'}^tdt_1H(t_1)\right]~.
\end{equation}

To develop a perturbative expansion, one usually moves to the interaction picture by decomposing the Hamiltonian as $H=H_0+H_{int}$, $H_0$ is a solvable (usually free) Hamiltonian, and $H_{int}$ contains the interaction terms. Denoting with a subscript "$0$" the interaction picture fields and states with respect to $H_0$, we have
\begin{equation}
\begin{aligned}
    \hat{\psi}_0(t,\Vec{x}) & =e^{iH_0t}\psi(\Vec{x})e^{-iH_0t}~,\\
    \ket{\Omega(t)}_0 & = S(t,0)\ket{\Omega(0)}_0=S(t,0)\ket{\Omega},
\end{aligned}
\end{equation}
where in the last equality we have used that the Schrödinger and interaction pictures are defined to coincide at $t=0$. The S-matrix operator is given by
\begin{equation}
    S(\infty,-\infty)=T\exp\left[-i\int_{-\infty}^\infty dt_1\hat{H}_{int}(t_1)\right]~.
\end{equation}

Letting $\ket{\phi_0}$ to denote the ground state of $H_0$, the time-ordered Green’s function takes the form
\begin{equation}
    i\h G(x,x')=\frac{\bra{\phi_0}T\left[S(\infty,-\infty)\h\hat{\psi_0}(t,\Vec{x})\hat{\psi_0}(t',\Vec{x}')\right]\ket{\phi_0}}{\bra{\phi_0}S(\infty,-\infty)\ket{\phi_0}}~,
    \label{eq:GDysonpert}
\end{equation}
which is the conventional starting point for perturbative expansions in equilibrium quantum field theory.

These standard techniques break down when the system is driven out of equilibrium, as we now discuss. Consider a generic time-dependent Hamiltonian of the form
\begin{equation}
    \mathcal{H}(t)=H+V(t)~,
    \label{eq:H0pV}
\end{equation}
where $H$ denotes the unperturbed Hamiltonian, which may already include interactions, and $V(t)$ is an arbitrary time-dependent perturbation. In this setting, we are interested in computing the time-ordered Green's function
\begin{equation}
    i\h G(x,x')=\bra{\Phi}T[\psi(x)\psi^\dagger(x')]\ket{\Phi}~,
    \label{eq:GFPhi}
\end{equation}
where $\ket{\Phi}$ is, for now, an arbitrary state. Here and in what follows, we use the shorthand $x\equiv(t,\Vec{x})$.

To proceed, we move to the interaction picture, now defined with respect to the unperturbed part $H$:
\begin{equation}
    \hat{\psi}(t,\Vec{x})=e^{iHt}\psi(\Vec{x})e^{-iHt}~,
    \label{eq:fieldsinter}
\end{equation}
where $\psi(\Vec{x})$ is the Schrödinger picture field.

The evolution operator $\mathcal{U}(t,t')$ and the corresponding S-matrix $\mathcal{S}(t,t')$ for the full Hamiltonian $\mathcal{H}(t)$ are given by\footnote{Following the notation in \cite{Maciejko_2007}, we denote by $U$ and $S$ the evolution operator and S-matrix associated to the unperturbed Hamiltonian, $H$, while $\mathcal{U}$ and $\mathcal{S}$ refer to the analogous quantities for the full time-dependent Hamiltonian $\mathcal{H}(t)$.}
\begin{equation}
\begin{aligned}
    \mathcal{U}(t,t') & = T\exp\left[-i\int_{t'}^tdt_1\mathcal{H}(t_1)\right]~,\\
    \mathcal{S}(t,t') & = T\exp\left[-i\int_{-\infty}^\infty dt_1\hat{V}(t_1)\right]~,
\end{aligned}
\end{equation}
where $\hat{V}(t)$ denotes the perturbation in the interaction picture.

The states in the interaction picture evolve according to
\begin{equation}
    \ket{\Phi(t)}_I=\mathcal{S}(t,0)\ket{\Phi(0)}_I=\mathcal{S}(t,0)\ket{\Phi}~,
    \label{eq:statesinter}
\end{equation}
where again we used that the interaction and Schrödinger pictures coincide at $t=0$.

Using the relations between the Heisenberg and interaction-picture fields from \eqref{eq:fieldsHeisen} and \eqref{eq:fieldsinter}, together with the time-evolved interaction-picture states from \eqref{eq:statesinter}, the general non-equilibrium Green's function \eqref{eq:GFPhi} can be expressed as
\begin{equation}
    i\h G(x,x')=\prescript{}{I}{\bra{\Phi(\infty)}}T\left[\mathcal{S}(\infty,-\infty)\h\hat{\psi}(t,\Vec{x})\hat{\psi}^\dagger(t'\Vec{x}')\right]\ket{\Phi(-\infty)}_I~.
    \label{eq:generalG}
\end{equation}

This expression is completely general. To recover the standard form used in perturbation theory, \eqref{eq:GDysonpert}, an important assumption must be made: namely, that the interaction can be switched on and off adiabatically. A typical example is the Hamiltonian
\begin{equation}
    \mathcal{H}_\epsilon(t)=H+e^{-\epsilon\abs{t}}V~,\qquad \text{with}~\epsilon\to 0^+~.
\end{equation}

We choose $\ket{\Phi}$ to be the ground state of the Hamiltonian $\mathcal{H}_0=H+V$, which we denote by $\ket{\Psi_0}$. According to the adiabatic theorem, the state $\ket{\Phi(\pm \infty)}_I$ must be an eigenstate of $H$.  Assuming the ground state is non-degenerate, the asymptotic states must coincide up to a phase:
\begin{equation}
    \ket{\Phi(\infty)}_I=e^{iL}\ket{\Phi(-\infty)}_I=e^{iL}\ket{\Phi_0}~,
\end{equation}
where we define $\ket{\Phi_0} \equiv \ket{\Phi(-\infty)}_I$. Since $\ket{\Phi(\infty)}_I=\mathcal{S}(\infty,-\infty)\ket{\Phi_0}$, this implies that the phase is given by
\begin{equation}
    e^{iL}=\bra{\Phi_0}\mathcal{S}(\infty,-\infty)\ket{\Phi_0}~,
\end{equation}
Substituting this into \eqref{eq:generalG}, one immediately recovers \eqref{eq:GDysonpert}.

The discussion above breaks down in non-equilibrium situations, where the evolution is non-adiabatic for a general time-dependent Hamiltonian, and the initial and final states no longer differ by a phase.

A formalism to handle such cases was developed long ago by Schwinger \cite{Schwinger_1960}, Keldysh \cite{Keldysh_1964}, and Feynman and Vernon \cite{Feynman_1963}. The central idea is to eliminate the dependence on the asymptotic future state $\ket{\Phi(\infty)}_I$ by expressing all quantities solely in terms of the asymptotic past state, $\ket{\Phi(-\infty)}_I$. In particular, one substitutes
\begin{equation}
    \ket{\Phi(\infty)}_I=\mathcal{S}(\infty,-\infty)\ket{\Phi(-\infty)}_I~
\end{equation}
in \eqref{eq:generalG}:
\begin{equation}
    i\h G(x,x')=\prescript{}{I}{\bra{\Phi(-\infty)}}\mathcal{S}(-\infty,\infty)T\left[\mathcal{S}(\infty,-\infty)\h\hat{\psi}(t,\Vec{x})\hat{\psi}^\dagger(t'\Vec{x}')\right]\ket{\Phi(-\infty)}_I~.
    \label{eq:generalGstatepast}
\end{equation}

\begin{figure}
    \centering
    \begin{tikzpicture}
        \def\Lx{5.5}
        \draw[-{Stealth[length=5pt, width=5pt]}] (0, 0.2) -- (0.515*\Lx, 0.2);
        \draw[-] (0.515*\Lx, 0.2) -- (\Lx, 0.2);
        \draw[-{Stealth[length=5pt, width=5pt]}] (\Lx, -0.2) -- (0.475*\Lx, -0.2);
        \draw[-] (0.475*\Lx, -0.2) -- (0, -0.2);

        \node[above] at (0.3*\Lx, 0.2) {$C_+$};
        \node[below] at (0.3*\Lx, -0.2) {$C_-$};
        \node[left] at (0, 0) {$-\infty$};
        \node[right] at (\Lx, 0) {$+\infty$};
    \end{tikzpicture}
    \caption{Schwinger-Keldysh contour, with two branches that extend from $-\infty$ to $+\infty$.}
    \label{fig:2branches}
\end{figure}
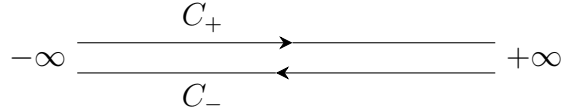

However, this introduces a factor of $\mathcal{S}(-\infty,\infty)$ outside the time-ordering operation. To bring it inside, one must extend the notion of time ordering to a closed-time path, or Schwinger-Keldysh contour, $\mathcal{C} = C_+ \cup C_-$, which includes both forward and backward time evolution. This is illustrated in Fig. \ref{fig:2branches}.

On the Schwinger-Keldysh contour, the usual notion of time ordering is replaced by contour ordering, denoted by $T_\mathcal{C}$, where the upper branch, $C_+$, is considered to be earlier than the lower branch, $C_-$ (in the contour sense). Accordingly, the time-ordered Green’s function \eqref{eq:generalGstatepast} is replaced by its contour-ordered counterpart. Extending the time arguments $t$, $t'$ to variables defined on the whole contour $\mathcal{C}$, the Green's function becomes
\begin{equation}
    i G(t,\Vec{x};t',\Vec{x}')= \prescript{}{I}{\bra{\Phi(-\infty)}}T_\mathcal{C}\left[\mathcal{S}_\mathcal{C}(-\infty,-\infty)\hat{\psi}(t,\Vec{x})\hat{\psi}^\dagger(t',\Vec{x}')\right]\ket{\Phi(-\infty)}_I~,
\end{equation}
where the contour evolution operator $\mathcal{S}_\mathcal{C}$ is defined as
\begin{equation}
    \mathcal{S}_\mathcal{C}(-\infty,\infty)=T_\mathcal{C}\exp\left[-i\oint_\mathcal{C}dt_1\hat{V}(t_1)\right]~.
\end{equation}

\begin{figure}
    \centering
    \begin{tikzpicture}
        \begin{scope}
            \def\Lx{5.5}
            \draw[->] (0, 0) -- (\Lx, 0) node[right] {$\mathrm{Re}\,t$};
            \draw[->] (0, -1) -- (0, 1) node[above] {$\mathrm{Im}\,t$};

            \draw[->, thick, red] (0, 0.2) -- (0.8*\Lx, 0.2);
            \draw[->, thick, red] (0.8*\Lx, -0.2) -- (0, -0.2);
            \draw[thick, red] (0.8*\Lx, 0.2) arc[start angle=90,end angle=-90,radius=0.2];
            \draw[->, thick, red] (0.8*\Lx, -0.2) -- (0.78*\Lx, -0.2);

            \filldraw[blue] (0.6*\Lx, 0.2) circle (2pt);
            \filldraw[blue] (0.4*\Lx, -0.2) circle (2pt);
            
            \node[left] at (0, 0) {$t_0$};
            \node[above, blue] at (0.6*\Lx, 0.3) {$\psi(t_+)$};
            \node[below, blue] at (0.4*\Lx, -0.3) {$\psi(t_-)$};
            \node[above] at (0.2*\Lx, 0.2) {$\textcolor{red}{\mathcal{C}}$};

        \end{scope}
        \begin{scope}[xshift=8.5cm] 
            \def\Lx{5.5}
            \draw[->] (0, 0) -- (\Lx, 0) node[right] {$\mathrm{Re}\,t$};
            \draw[->] (0, -1) -- (0, 1) node[above] {$\mathrm{Im}\,t$};

            \draw[->, thick, red] (0, 0.2) -- (0.8*\Lx, 0.2);
            \draw[-, thick, red] (0.8*\Lx, -0.2) -- (0, -0.2);
            \draw[thick, red] (0.8*\Lx, 0.2) arc[start angle=90,end angle=-90,radius=0.2];
            \draw[->, thick, red] (0.8*\Lx, -0.2) -- (0.78*\Lx, -0.2);
            \draw[->, thick, red] (0, -0.2) -- (0, -1);

            \filldraw[blue] (0.6*\Lx, 0.2) circle (2pt);
            \filldraw[blue] (0.4*\Lx, -0.2) circle (2pt);
            
            \node[left] at (0, 0) {$t_0$};
            \node[above, blue] at (0.6*\Lx, 0.3) {$\psi(t_+)$};
            \node[below, blue] at (0.4*\Lx, -0.3) {$\psi(t_-)$};
            \node[left] at (0, -1) {$t_0-i\beta$};
            \node[above] at (0.2*\Lx, 0.2) {$\textcolor{red}{\mathcal{C}}$};
        \end{scope}
    \end{tikzpicture}
    \caption{Schwinger-Keldysh contour for zero-temperature field theory (left) and its finite-temperature version (right) when the density matrix is known at a time $t_0$. The time coordinate is promoted to a complex variable.}
    \label{fig:Keldyshcontourbckgr}
\end{figure}
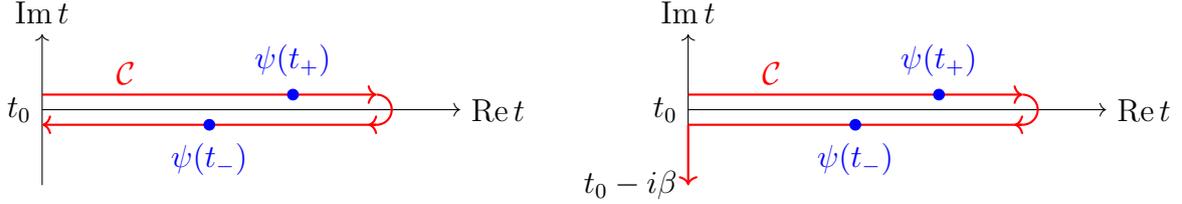

Actually, if the equilibrium state of the system at a given time $t_0$ is known, the contour does not need to extend from $-\infty$ to $+\infty$ and back. Instead, it suffices to consider the contour from $t_0$  until the latest time of operator insertion, either $t$ or $t'$, and then back to $t_0$. In the case where the density matrix at $t=t_0$ corresponds to a thermal equilibrium state, a small excursion in imaginary time must be added. Both types of contours are depicted in Fig. \ref{fig:Keldyshcontourbckgr}\footnote{The forward-and-return path of the Schwinger-Keldysh contour is reminiscent of \textit{The Hobbit, or There and Back Again} (J.R.R. Tolkien, 1937), which inspired the title of the section.}.

The power of this formalism lies in allowing the use of standard perturbation theory techniques, such as Feynman diagrams and the Dyson equation. However, a price must be paid: we now need to account for the Keldysh contour, and the doubling of the contour introduces four possible configurations for the contour-ordered Green's function, depending on which branch of the contour the times are inserted. Denoting $+$, $-$ for the forward and backward branches, respectively, the Green’s function can be written as a matrix
\begin{equation}
    G(x,x')=\begin{pmatrix} G_{++} & G_{+-}\\ G_{-+} & G_{--} \end{pmatrix}=\begin{pmatrix} G^T & G^< \\ G^> & G^{\overline{T}} \end{pmatrix}~,
\end{equation}
where in the second equality, we identify the usual time-ordered, anti-time-ordered, greater, and lesser components of the Green's function. These are defined as
\begin{equation}
\begin{aligned}
    G^T(x,x') & = -i\langle T\left[\psi(x)\psi^\dagger(x')\right]\rangle~,\\
    G^{\tilde{T}}(x,x') & = -i\langle\overline{T}\left[\psi(x)\psi^\dagger(x')\right]\rangle~,\\
    G^>(x,x') & = -i\langle\psi(x)\psi^\dagger(x')\rangle~,\\
    G^<(x,x') & = -i\langle \psi^\dagger(x')\psi(x)\rangle~.
\end{aligned}
\end{equation}

Actually, not all four components are independent. From their definitions, it is not difficult to check that the following relation holds:
\begin{equation}
    G^T+G^{\overline{T}}=G^>+G^<~.
\end{equation}
Thus, only three Green's functions are independent. It is common to define the Keldysh, retarded, and advanced Green's functions as
\begin{equation}
\begin{aligned}
    G^K(x,x') & = G^>(x,x')+G^<(x,x') ~,\\
    G^R(x,x') & = \theta(t-t')\left[G^>(x,x')-G^<(x,x')\right]~,\\
    G^A(x,x') & = -\theta(t'-t)\left[G^>(x,x')-G^<(x,x')\right]~.
\end{aligned}
\end{equation}

In terms of the contour-ordered Green's function, the Dyson equation can be written as
\begin{equation}
\begin{aligned}
    G & = G_0+G_0\h\Sigma\h G~,\\
    G & = G_0+G\h\Sigma\h G_0~,
    \label{eq:DysonSchwKel}
\end{aligned}
\end{equation}
where $G$ denotes the exact Green's function, $G_0$ is the unperturbed Green's function, and $\Sigma$ is the irreducible self-energy. In the terms $G_0\h\Sigma\h G$ and $G\h\Sigma\h G_0$, a double convolution along the Keldysh contour is implicit.

In practice, it is useful to rewrite these integrals in terms of a single ordinary time variable, using the real-time Green's functions $G^>$, $G^<$, $G^R$, $G^A$ (and similarly for the self-energies).  The "dictionary" that translates between contour integrals and ordinary time integrals is known as the \textit{Langreth rules} \cite{Langreth_1972,Langreth_1976} which can be summarized as follows:
\begin{equation}
\begin{aligned}
    (AB)^{\stackrel{\scriptstyle >}{\scriptstyle <}} & = A^RB^{\stackrel{\scriptstyle >}{\scriptstyle <}}+A^{\stackrel{\scriptstyle >}{\scriptstyle <}}B^A~,\\
    (AB)^{R,A} & = A^{R,A}B^{R,A}~,\\
    (ABC)^{\stackrel{\scriptstyle >}{\scriptstyle <}} & = A^RB^RC^{\stackrel{\scriptstyle >}{\scriptstyle <}}+A^RB^{\stackrel{\scriptstyle >}{\scriptstyle <}}C^A+A^{\stackrel{\scriptstyle >}{\scriptstyle <}}B^AC^A~,\\
    (ABC)^{R,A} & = A^{R,A}B^{R,A}C^{R,A}~,
    \label{eq:Langreth}
\end{aligned}
\end{equation}
where the products $AB$ are understood as convolutions along the Keldysh contour. For example, taking $C\equiv A B$, the first identity in \eqref{eq:Langreth} is
\begin{equation}
\begin{aligned}
    C^<(t,t') & = \int_\mathcal{C}dt_1 A(t,t_1)B(t_1,t') \\
    & = \int_{-\infty}^\infty \left[A^R(t,t_1)B^<(t_1,t')+A^<(t,t_1)B^A(t_1,t')\right]~,
\end{aligned}
\end{equation}
with similar expressions for the greater component and the other identities in \eqref{eq:Langreth}.

\subsection{Non-equilibrium SYK}\label{subsec:noneqSYK}

We conclude this Chapter by presenting the real-time formulation of the SYK model within the Schwinger–Keldysh formalism, mainly following the results of \cite{Eberlein_2017}. This framework will later be extended to the case of two coupled SYK models in the research part of the thesis (see Chapter \ref{chap:FloquetSYK} for the main discussion, with technical details in Appendices \ref{app:KBequations} and \ref{app:numerics}).

The action of the SYK model on the Keldysh contour takes the form
\begin{equation}
    S[\chi]=\int_\mathcal{C}dt\left[\frac{i}{2}\sum_i\chi_i(t)\partial_t\chi_i(t)-H(t)\right]~,
\end{equation}
where the Hamiltonian corresponds to the real-time version of the standard SYK Hamiltonian introduced in\eqref{eq:HSYK}:
\begin{equation}
    H(t)=\frac{1}{4!}\sum_{i,j,k,l} f(t)J_{ijkl}\chi_i\chi_j\chi_k\chi_l~.
\end{equation}
We have included an arbitrary time-dependent function, $f(t)$, in front of the random couplings. In \cite{Eberlein_2017}, this time dependence is taken to model a quantum quench that drives the system out of equilibrium\footnote{In our case, we will consider both a time-dependent coupling between the two SYK copies and a time-dependent interaction strength within each copy, see equation \ref{eq:H2SYKtdep}.}. The distribution of couplings remains the same as in the Euclidean formulation: a Gaussian distribution with vanishing mean and variance given by
\begin{equation}
    \overline{J_{ijkl}^2}=\frac{3! J^2}{N^3} ~.
\label{eq:Jmeanvarrealt}
\end{equation}

Therefore, the action is
\begin{equation}
    S[\chi]=\int_\mathcal{C} dt\left[\frac{i}{2}\sum_i\chi_i(t)\partial_t\chi_i(t)+\frac{1}{4!}\sum_{i,j,k,l}f(t)J_{ijkl}\chi_i(t)\chi_j(t)\chi_k(t)\chi_l(t)\right]~,
\end{equation}
where the integral is taken along the Keldysh contour, $\mathcal{C}$. The disorder-averaged partition function then reads
\begin{equation}
    \overline{Z}=\int DJ_{ijkl}D\chi\exp\big[i\h S[\chi]\big]~,
\end{equation}
with $DJ_{ijkl}$ defined as in \eqref{eq:disorder}. As in the Euclidean case, the integral over disorder is Gaussian and can be performed exactly. With the introduction of bilocal fields $O(t,t')$, defined as
\begin{equation}
    O(t,t')=-\frac{i}{N}\sum_j\chi_j(t)\chi_j(t')~,
\end{equation}
the partition function can be written as
\begin{equation}
    \overline{Z}=\int D\chi\exp\left[-\frac{1}{2}\int_\mathcal{C} dt\sum_{i=1}^N\chi_i\partial_t\chi_i-\frac{N}{8}\int_\mathcal{C} dt dt' J(t)J(t')O^4(t,t')\right]~,
\end{equation}
where we have defined $J(t)\equiv J f(t)$.

As in the Euclidean formulation, it is customary to insert a resolution of the identity, analogous to~\eqref{eq:unitySYK}. This allows one to integrate out the Majorana fermions explicitly. Defining the inverse free propagator as
\begin{equation}
    G_0^{-1}(t_1,t_2)=i\partial_{t_1}\delta_\mathcal{C}(t_1-t_2)~,
\end{equation}
the disorder-averaged partition function can be expressed in terms of an effective action as
\begin{equation}
    \overline{Z}=\int DG D\Sigma \exp\left[iS_\text{eff}\left(G,\Sigma\right)\right]~,
\end{equation} 
with the effective action given by
\begin{equation}
    \frac{S_\text{eff}}{N} = -\frac{i}{2}\log\det\left(-i\h G_0^{-1}+\Sigma(t,t')\right)+\frac{i}{2}\int_\mathcal{C} dt dt'\left(\Sigma(t,t')G(t,t')+\frac{J(t)J(t')}{4}G(t,t')^4\right)~.
    \label{eq:SeffSYKrealt}
\end{equation}

In the large-$N$ limit, the dynamics is governed by the saddle-point equations
\begin{equation}
    \frac{\delta S_\text{eff}}{\delta G}=0~,\qquad \frac{\delta S_\text{eff}}{\delta \Sigma}=0~,
\end{equation}
which yield the Schwinger-Dyson equations for the bilocal fields $G(t,t')$ and $\Sigma(t,t')$. These are given by\footnote{It can be shown that this Schwinger-Keldysh equation is equivalent to \eqref{eq:DysonSchwKel}.}
\begin{equation}
\begin{aligned}
    G_0^{-1}(t_1,t_2) & =G^{-1}(t_1,t_2)+\Sigma(t_1,t_2)~,\\
    \Sigma(t,t') & =-J(t)J(t')G(t,t')^3~,
    \label{eq:Dysoneqsrealtime}
\end{aligned}
\end{equation}
where the second equation defines the self-energy in terms of the full two-point function.

By taking the greater and lesser components of the self-energy, one obtains
\begin{equation}
    \Sigma^{>,<}(t,t')=-J(t)J(t')G^{>,<}(t,t')^3~.
\end{equation}

To derive the equations that are ultimately used, one may convolute the first line of \eqref{eq:Dysoneqsrealtime} with $G$ from the left and the right, yielding the Kadanoff-Baym equations:
\begin{equation}
\begin{aligned}
    \int_\mathcal{C}dt_3\h G_0^{-1}(t_1,t_3)G(t_3,t_2)=\delta_\mathcal{C}(t_1,t_2)+\int_\mathcal{C}dt_3\h\Sigma(t_1,t_3)G(t_3,t_2)~,\\
    \int_\mathcal{C}dt_3\h G(t_1,t_3)G_0^{-1}(t_3,t_2)=\delta_\mathcal{C}(t_1,t_2)+\int_\mathcal{C}dt_3\h G(t_1,t_3)\Sigma(t_3,t_2)~.
\end{aligned}
\end{equation}

To obtain the equations of motion for the greater Green's function $G^>(t_1,t_2)$, we consider the case where $t_1\in C_-$, $t_2\in C_+$. Applying the Langreth rules introduced in~\eqref{eq:Langreth}, we find
\begin{align}
    i\partial_{t_1}G^>(t_1,t_2) & =\int_{-\infty}^\infty\h dt_3\left[\Sigma^R(t_1,t_3)G^>(t_3,t_2)+\Sigma^>(t_1,t_3)G^A(t_3,t_2)\right]~,\label{eq:KBSYKa}\\
    -i\partial_{t_2}G^>(t_1,t_2) & = \int_{-\infty}^\infty\h dt_3\left[G^R(t_1,t_3)\Sigma^>(t_3,t_2)+G^>(t_1,t_3)\Sigma^A(t_3,t_2)\right]~,
    \label{eq:KBSYKb}
\end{align}
where the retarded and advanced self-energies are defined as
\begin{equation}
\begin{aligned}
    \Sigma^R(t_1,t_2) & = \theta(t_1-t_2)\left[\Sigma^>(t_1,t_2)-\Sigma^<(t_1,t_2)\right] ~,\\
    \Sigma^A(t_1,t_2) & = -\theta(t_2-t_1)\left[G^>(t_1,t_2)-G^<(t_1,t_2)\right]~.
\end{aligned}
\end{equation}

Although analogous equations can be derived for the lesser component $G^<(t_1,t_2)$, they are not needed here, since, for Majorana fermions, the greater and lesser components are related by
\begin{equation}
    G^>(t_1,t_2)=-G^<(t_2,t_1)~,
\end{equation}
which follows directly from the definitions.

These equations govern the real-time dynamics of the SYK model in the Schwinger-Keldysh formalism and serve as the starting point for the out-of-equilibrium analysis developed in later chapters. Appendix~\ref{app:numerics} provides further details on the numerical algorithm employed to solve the Kadanoff–Baym equations~\eqref{eq:KBSYKa}–\eqref{eq:KBSYKb}.

%% file: Part1/SYKwormholes/SYKwormholesAppendices.tex
\renewcommand{\chapterquote}{}
\chapterappendix{\thechapter}

\setcounter{equation}{0}
\setcounter{appendixsection}{\value{appendixsection}+1} 
\section{Details on the Lyapunov exponent}\label{app:Lyapunov}

We consider the kernel equation
\begin{equation}
    \mathcal{F}(t_1,t_2)=\int dt_3 dt_4 K_R(t_1,...,t_4)\mathcal{F}(t_3,t_4)~,
    \label{eq:kerneleqApp}
\end{equation}
where the kernel is given by
\begin{equation}
    K_R(t_1,...,t_4)=-3J^2G_R(t_1,t_3)G_R(t_2,t_4)G_W(t_3,t_4)^2~,
    \label{eq:kernelApp}
\end{equation}
with $G_R(t)$ a retarded correlator, and $G_W(t)$ a Wightman function defined as $G_W(t)=G(\frac{\beta}{2}+it)$.

With the growing ansatz
\begin{equation}
    \mathcal{F}(t_1,t_2)=e^{\lambda_L(t_1+t_2)/2}f(t_1-t_2)~,
    \label{eq:F12ansatzApp}
\end{equation}
Eq. \eqref{eq:kerneleqApp} becomes
\begin{equation}
     e^{\lambda_L(t_1+t_2)/2}f(t_1-t_2)=-3J^2\int dt_3 dt_4 G_R(t_1-t_3)G_R(t_2-t_4)G_W(t_3-t_4)^2e^{\lambda_L(t_3+t_4)/2}f(t_3-t_4)~,
     \label{eq:expandkerneleq}
\end{equation}
where we used the fact that in equilibrium the correlators depend only on the difference of times.

Using the convolution theorem for a product of Fourier transforms, we can write
\begin{equation}
\begin{aligned}
    G_W(t_3-t_4)^2f(t_3-t_4)&=\frac{1}{2\pi}\int_{-\infty}^{\infty}d\omega (G_W^2*f)(\omega)e^{-i\omega(t_3-t_4)}\\
    &=\frac{1}{2\pi}\int_{-\infty}^{\infty}d\omega~e^{-i\omega(t_3-t_4)}\int_{-\infty}^{\infty}d\omega'G_W^2(\omega-\omega')f(\omega')~,
    \label{eq:convoGf}
\end{aligned}
\end{equation}
where $G_W^2$ here denotes the Fourier transform of the square of $G_W$, not the square of the Fourier transform. Introducing \eqref{eq:convoGf} into \eqref{eq:expandkerneleq} and rearranging terms,
\begin{equation}
\begin{aligned}
    e^{\lambda_L(t_1+t_2)/2}f(t_1-t_2)=-&\frac{3J^2}{2\pi}\int_{-\infty}^{\infty}d\omega\int_{-\infty}^{\infty}d\omega' G_W^2(\omega-\omega')f(\omega')\\ \times &\int_{-\infty}^{\infty}dt_3G_R(t_1-t_3)e^{-i(\omega +i\lambda_L/2)t_3}
     \int_{-\infty}^{\infty}dt_4G_R(t_2-t_4)e^{i(\omega -i\lambda_L/2)t_4}
\end{aligned}
\end{equation}
Defining $x\equiv t_1-t_3$, we have
\begin{equation}
    \int_{-\infty}^{\infty}dx G_R(x)e^{i(\omega+i\lambda_L/2)x}e^{-i(\omega+i\lambda_L/2)t_1}=G_R(\omega+i\lambda_L/2)e^{-i(\omega+i\lambda_L/2)t_1}~,
\end{equation}
and similarly for $y\equiv t_2-t_4$:
\begin{equation}
    \int_{-\infty}^{\infty}dy G_R(y)e^{i(-\omega+i\lambda_L/2)y}e^{i(\omega-i\lambda_L/2)t_2}=G_R(-\omega+i\lambda_L/2)e^{i(\omega-i\lambda_L/2)t_2}~.
\end{equation}

We then have
\begin{equation}
     f(t_1-t_2)=\frac{3J^2}{2\pi}\int_{-\infty}^{\infty}d\omega~e^{-i\omega(t_1-t_2)}\abs{G_R(\omega+i\lambda_L/2)}^2 \int_{-\infty}^{\infty} d\omega'G_W^2(\omega-\omega')f(\omega')
\end{equation}
where we have used that $G_R(\omega+i\lambda_L/2)G_R(-\omega+i\lambda_L/2)=-\abs{G_R(\omega+i\lambda_L/2)}^2$. Fourier transforming also the lhs, we find
\begin{equation}
    f(\omega)=3J^2|G_R(\omega+i\lambda_L/2)|^2\int_{-\infty}^\infty d\omega' G_W^2(\omega-\omega')f(\omega')~.
    \label{eq:fomegaApp}
\end{equation}

\subsection{Conformal limit}

In the conformal limit, we have
\begin{equation}
    G(\tau)=b\left[\frac{\pi}{\beta \sin \frac{\pi \tau}{\beta}}\right]^{2\Delta}~,~~~~J^2b^4=\frac{1}{4\pi}~,~~~~\Delta=1/4
\end{equation}
Analytic continuation to real time yields \cite{Maldacena_2016}
\begin{align}
    \langle \chi(t)\chi(0)\rangle&=be^{-i\pi \Delta}\left[\frac{\pi}{\beta \sinh \frac{\pi t}{\beta}}\right]^{2\Delta}=iG^>(t)~,\\
    \langle \chi(0)\chi(t)\rangle&=-be^{i\pi \Delta}\left[\frac{\pi}{\beta \sinh \frac{\pi t}{\beta}}\right]^{2\Delta}=iG^<(t)~.
\end{align}
The retarded correlator is (for $\Delta=1/4$)
\begin{equation}
    G^R(t)=\theta(t)\left(G^>(t)-G^<(t)\right)=-\sqrt{2}ib\left[\frac{\pi}{\beta \sinh \frac{\pi t}{\beta}}\right]^{1/2}\theta(t)~.
    \label{eq:Gretconf}
\end{equation}
The Wightman correlator $G_W(t)\equiv  G(it+\beta/2)$ is
\begin{equation}
    G_W^2(t)=\frac{\pi b^2}{\beta \cosh \frac{\pi t}{\beta}}~.
    \label{eq:GWconf}
\end{equation}

Their Fourier transforms are, respectively,
\begin{equation}
\begin{aligned}
    G^R(\omega)&=\int_{-\infty}^{\infty}dt G^R(t) e^{i\omega t}=-ib\sqrt{\beta}\frac{\Gamma\left(\frac{1}{4}-i\frac{\beta \omega}{2\pi}\right)}{\Gamma\left(\frac{3}{4}-i\frac{\beta \omega}{2\pi}\right)}~,\\
    G_{lr}^2(\omega)&=\frac{1}{2\pi}\int_{-\infty}^{\infty}dt G_{lr}^2(t) e^{i\omega t}=\frac{b^2}{2\cosh\frac{\beta \omega}{2}}~.
\end{aligned}
\end{equation}

We can now introduce these expressions into \eqref{eq:fomegaApp}, which leads to 
\begin{equation}
    f(\omega)=\frac{3}{4}\frac{\beta}{2\pi}\abs{\frac{\Gamma\left(\frac{1}{4}+\frac{\beta\lambda_L}{4\pi}-\frac{i\beta\omega}{2\pi}\right)}{\Gamma\left(\frac{3}{4}+\frac{\beta\lambda_L}{4\pi}-\frac{i\beta\omega}{2\pi}\right)}}^2\int_{-\infty}^{\infty}d\omega'\frac{f(\omega')}{\cosh\frac{\beta(\omega-\omega')}{2}}~.
    \label{eq:kerneleqomegaApp}
\end{equation}

It is convenient to define $h\equiv \frac{\beta \lambda_L}{2\pi}$, and change variables to $u\equiv \frac{\beta \omega}{2\pi}$:
\begin{equation}
    f(u)=\frac{3}{4}\abs{\frac{\Gamma\left(\frac{1}{4}+\frac{h}{2}-iu\right)}{\Gamma\left(\frac{3}{4}+\frac{h}{2}-iu\right)}}^2\int_{-\infty}^{\infty}du'\frac{f(u')}{\cosh(\pi(u-u'))}~.
    \label{eq:kernelequ}
\end{equation}
Using the following property of the Gamma function,
\begin{equation}
    \abs{\Gamma\left(1/2+ib\right)}^2=\frac{\pi}{\cosh \pi b}~,
\end{equation}
we can rewrite \eqref{eq:kernelequ} as
\begin{equation}
    f(u)=\frac{3}{4\pi}\abs{\frac{\Gamma\left(\frac{1}{4}+\frac{h}{2}-iu\right)}{\Gamma\left(\frac{3}{4}+\frac{h}{2}-iu\right)}}^2\int_{-\infty}^{\infty}du'\abs{\Gamma\left(\frac{1}{2}+i(u-u')\right)}^2f(u')~.
    \label{eq:fu}
\end{equation}

Now, an ansatz is proposed for $f(u)$ \cite{Banerjee_2016}:
\begin{equation}
    f(u)=\abs{\Gamma\left(\frac{1}{4}+\frac{h}{2}+iu\right)}^2~,
\end{equation}
and the integral over $u'$ becomes
\begin{align}
\begin{split}
    &\int_{-\infty}^{\infty}du'\abs{\Gamma\left(\frac{1}{2}+i(u-u')\right)}^2\abs{\Gamma\left(\frac{1}{4}+\frac{h}{2}+iu'\right)}^2\\
    =&\int_{-\infty}^{\infty}du'\Gamma\left(\frac{1}{2}+i(u-u')\right)\Gamma\left(\frac{1}{2}-i(u-u')\right)\Gamma\left(\frac{1}{4}+\frac{h}{2}+iu'\right)\Gamma\left(\frac{1}{4}+\frac{h}{2}-iu'\right)
\end{split}
\end{align}
This integral can be evaluated using the following integral of the product of four $\Gamma$-functions \cite{Gradshteyn_1943}
\begin{equation}
    \int_{-i\infty}^{i\infty}\Gamma(\alpha+s)\Gamma(\beta+s)\Gamma(\gamma-s)\Gamma(\delta-s)ds=2\pi i\frac{\Gamma(\alpha+\gamma)\Gamma(\alpha+\delta)\Gamma(\beta+\gamma)\Gamma(\beta+\delta)}{\Gamma(\alpha+\beta+\gamma+\delta)}~,
\end{equation}
for Re $\alpha$, Re $\beta$, Re $\gamma$, Re $\delta>0$. This corresponds to our integral if we identify $s=iu'$, $\alpha=\frac{1}{2}-iu$, $\beta=\delta=\frac{1}{4}+\frac{h}{2}$, $\gamma=\frac{1}{2}+iu$. Evaluating the integral and simplifying the gamma functions we get
\begin{equation}
    \int_{-\infty}^{\infty}du'\abs{\Gamma\left(\frac{1}{2}+i(u-u')\right)}^2\abs{\Gamma\left(\frac{1}{4}+\frac{h}{2}+iu'\right)}^2=2\pi \frac{\Gamma\left(\frac{1}{2}+h\right)}{\Gamma\left(\frac{3}{2}+h\right)}\abs{\Gamma\left(\frac{3}{4}+\frac{h}{2}-iu\right)}^2~,
\end{equation}
and now it is immediate to see that \eqref{eq:fu} is satisfied, provided that
\begin{equation}
    \frac{\Gamma\left(\frac{1}{2}+h\right)}{\Gamma\left(\frac{3}{2}+h\right)}=\frac{2}{3}~,
\end{equation}
whose solution is $h=1$, meaning that, from the definition of $h$,
\begin{equation}
    \lambda_L=\frac{2\pi}{\beta}~.
\end{equation}

\setcounter{equation}{0}
\setcounter{appendixsection}{\value{appendixsection}+1} 
\section{AdS\texorpdfstring{$_2$}{2} space}\label{app:AdS2}

The two-dimensional anti-de Sitter space can be represented as the hyperboloid
\begin{equation}
    -(Y^{-1})^2-(Y^{0})^2+(Y^1)^2=-1~,
    \label{eq:AdS2hyperboloid}
\end{equation}
with the metric
\begin{equation}
    ds^2=-(dY^{-1})^2-(dY^0)^2+(dY^1)^2~.
\end{equation}

\begin{figure}
\centering
    \begin{subfigure}[b]{0.3\textwidth}
    \centering
    \begin{tikzpicture}
    \centering
        \def\L{1.8} 
        \def\he{1.7*\L} 
        \fill[green, opacity=0.1] (-\L,-\he) rectangle (\L,\he);
        \draw[ultra thick,black]  (-\L,-\he) -- (-\L,\he) ;
        \draw[ultra thick,black]  (\L,-\he) -- (\L,\he);
        \draw[black] (0.8*\L,-1.5*\L) node[left] {$t$};
        \draw[-{Stealth[scale=1]}, thick, black] (0.8*\L,-1.5*\L) -- (0.8*\L,-0.8*\L);
        \draw[-{Stealth[scale=1]}, thick, black] (-0.8*\L,-1.5*\L) -- (-0.1*\L,-1.5*\L);
        \draw[black] (-0.8*\L,-1.5*\L) node[above] {$\sigma$};
    \end{tikzpicture}
    \caption{Global patch of AdS$_2$.}
    \label{sfig:AdS2globalApp}
    \end{subfigure}
    \hfill
    \begin{subfigure}[b]{0.3\textwidth}
    \centering
    \begin{tikzpicture}
    \centering
        \def\L{1.53} 
        \def\he{2*\L} 
        \fill[green, opacity=0.1] (-\L,0) -- (\L,2*\L) -- (\L,-2*\L) -- cycle;
        \draw[ultra thick,black]  (-\L,-\he) -- (-\L,\he) ;
        \draw[black] (\L,0.2*\L) node[left] {$\abs{z}$};
         \draw[-{Stealth[scale=1]}, thick, black] (0.9*\L,0) -- (0.2*\L,0);
         \draw[-{Stealth[scale=1]},thick, black] (-0.15*\L,0) arc[start angle=180, end angle=130, radius=\L];
         \draw[thick, black] (-0.15*\L,0) arc[start angle=180, end angle=230, radius=0.9*\L];
         \draw[black] (-0.15*\L,0) node[left] {$t_P$};
        \draw[ultra thick,black]  (\L,-\he) -- (\L,\he);

        \draw[very thick,dashed,black]  (-\L,0) -- (\L,\he);
        \draw[very thick,dashed,black]  (-\L,0) -- (\L,-\he);
    \end{tikzpicture}
    \caption{Poincaré patch of AdS$_2$.}
    \label{sfig:AdS2PoincareApp}
    \end{subfigure}
    \hfill
    \begin{subfigure}[b]{0.3\textwidth}
    \centering
    \begin{tikzpicture}
        \def\L{1.8} 
        \def\he{1.7*\L} 

        \fill[green, opacity=0.1] (-\L,-\L) -- (0,0) -- (-\L,\L) -- cycle;
        \fill[green, opacity=0.1] (\L,-\L) -- (0,0) -- (\L,\L) -- cycle;
        \draw[ultra thick,black]  (-\L,-\he) -- (-\L,\he) ;
        \draw[ultra thick,black]  (\L,-\he) -- (\L,\he);

        \draw[black] (0.6*\L,-0.2*\L) node[right] {$\rho$};
        \draw[-{Stealth[scale=1]}, thick, black] (0.5*\L,0) -- (0.9*\L,0);

        \draw[-{Stealth[scale=1]},thick, black] (0.4*\L,0) arc[start angle=180, end angle=130, radius=0.5*\L];
        \draw[thick, black] (0.4*\L,0) arc[start angle=180, end angle=230, radius=0.5*\L];
        \draw[black] (0.45*\L,0) node[left] {$t_R$};

        \draw[very thick,dashed,black]  (-\L,-\L) -- (\L,\L);
        \draw[very thick,dashed,black]  (\L,-\L) -- (-\L,\L);
    \end{tikzpicture}
    \caption{Rindler patch of AdS$_2$.}
    \label{sfig:AdS2RindlerApp}
    \end{subfigure}
    \caption{Global, Poincaré and Rindler patches of AdS$_2$.}
    \label{fig:AdS2patchesApp}
\end{figure}
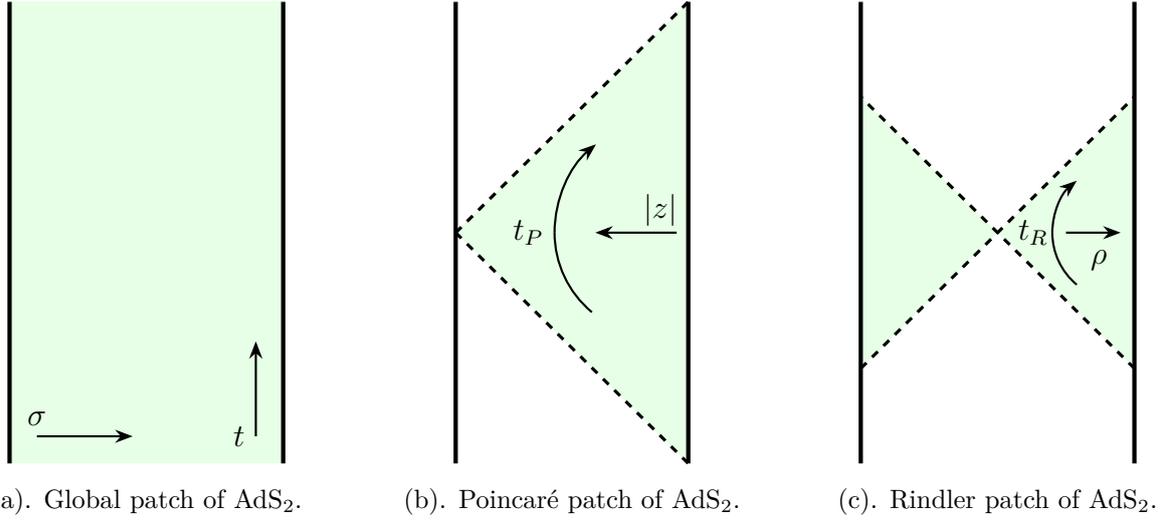

A particular parametrization satisfying the hyperboloid Eq. \eqref{eq:AdS2hyperboloid} can be written in terms of two coordinates $(t,\sigma)$, defined as
\begin{equation}
    Y^{-1}=\frac{\cos t}{\sin\sigma}~,\qquad Y^0=\frac{\sin t}{\sin\sigma}~,\qquad Y^{1}=-\frac{1}{\tan\sigma}~,
    \label{eq:Ycoordsglobal}
\end{equation}
with $\sigma\in [0,\pi]$, and $-\infty<t<\infty$. The metric in these coordinates is given by
\begin{equation}
    ds^2=\frac{-dt^2+d\sigma^2}{\sin^2\sigma}~.
\end{equation}
$(t,\sigma)$ is a system of global coordinates, which covers the full AdS$_2$ space. The AdS$_2$ space can be represented by a Penrose diagram consisting of a strip of width $\pi$, as we show in Fig. \ref{sfig:AdS2globalApp}. It has two boundaries, at $\sigma=(0,\pi)$, where the metric is divergent due to the $\sin^2\sigma$ factor. To describe the boundaries, we introduce projective coordinates $X^M$ by removing the divergent factor in the parametrization of the $Y^M$ coordinates. In these projective coordinates, the two boundaries are located at
\begin{equation}
\begin{aligned}
    X^1 & = -1~,&\qquad (\text{left boundary, at}~\sigma=0)~,\\
    X^1 & = 1~,&\qquad (\text{right boundary, at}~\sigma=\pi)~.
\end{aligned}
\end{equation}

The boundaries are parametrized by a time coordinate. Denoting by $t_l$ and $t_r$ the time coordinate at each of the boundaries, we have
\begin{equation}
\begin{aligned}
    \text{left bdry:} & \qquad X^{-1}=\cos t_l~,~~X^0=\sin t_l~,~~X^1=-1~, \\
    \text{right bdry:} & \qquad X^{-1}=\cos t_r~,~~X^0=\sin t_r~,~~X^1=1~.
\end{aligned}
\end{equation}

Notice that, on each boundary, we have the combinations
\begin{equation}
\begin{aligned}
    \frac{X^0}{X^{-1}+X^1}\Big\rvert_l=-\frac{1}{\tan\frac{t_r}{2}}~,\qquad \frac{X^0}{X^{-1}+X^1}\Big\rvert_r=\tan\frac{t_r}{2}~.
    \label{eq:combtltr}
\end{aligned}
\end{equation}

We can introduce the Poincaré coordinates $(t_P,z)$, related to the global coordinates $(t,\sigma)$ as
\begin{equation}
\begin{aligned}
    t_P & = -\frac{1}{2}\left[\cot\left(\frac{t+\sigma}{2}\right)+\cot\left(\frac{t-\sigma}{2}\right)\right]~,\\
    z & = \frac{1}{2}\left[\cot\left(\frac{t+\sigma}{2}\right)-\cot\left(\frac{t-\sigma}{2}\right)\right]~.
    \label{eq:Poincareglobal}
\end{aligned}
\end{equation}

In Poincaré coordinates, the metric is given by
\begin{equation}
    ds^2=\frac{-dt_P^2+dz^2}{z^2}~.
\end{equation}

From the relation between Poincaré and global coordinates \eqref{eq:Poincareglobal} and the relation to the $Y^M$ coordinates \eqref{eq:Ycoordsglobal}, we can relate the Poincaré time to the $Y^M$ coordinates (and therefore, the projective coordinates $X^M$). The result is
\begin{equation}
    t_P=\frac{X^0}{X^1+X^{-1}}~,
    \label{eq:combPoinc}
\end{equation}
which allows us to relate the Poincaré time to the global times $t_l$ and $t_r$ at the boundaries:
\begin{equation}
    t_P=\tan\frac{t_r}{2}=-\frac{1}{\tan\frac{t_l}{2}}~.
\end{equation}

From \eqref{eq:Poincareglobal} we notice that the Poincaré coordinates do not cover the full spacetime, but just a triangular region that collapses to a point at $(t=0,\sigma=0)$ (see Fig. \ref{sfig:AdS2PoincareApp}).

Finally, we can introduce Rindler coordinates, $(t_R,\rho)$, from a different parametrization of the original $Y^M$ coordinates which also satisfies the hyperboloid Eq. \eqref{eq:AdS2hyperboloid},
\begin{equation}
    Y^{-1}=\cosh\rho~,\qquad Y^0=\sinh\rho\sinh t_R~,\qquad Y^1=\sinh\rho\cosh t_R~.
\end{equation}
with $-\infty< t_R,\rho < \infty$. The metric in these coordinates is given by
\begin{equation}
    ds^2=-\sinh^2\rho\h dt_R^2+d\rho^2~.
\end{equation}

The boundaries are located at $\rho\to\pm \infty$. If we focus on the right boundary, $\rho\to\infty$, we have
\begin{equation}
    Y^{-1}\sim \frac{1}{2}e^{\rho}~,\qquad Y^0\sim \frac{1}{2}e^\rho\sinh t_R~,\qquad Y^1\sim \frac{1}{2}e^\rho\cosh t_R~.
\end{equation}
Defining the projective coordinates $X^M$ by absorbing the prefactor $\frac{1}{2}e^\rho$, we can find again the combination
\begin{equation}
    \frac{X^0}{X^{-1}+X^1}\Big\rvert_r=\tanh\frac{t_R}{2}~.
    \label{eq:combRindler}
\end{equation}

Combining equations \eqref{eq:combtltr}, \eqref{eq:combPoinc} and \eqref{eq:combRindler} we obtain the relation of the Poincaré time with the other time coordinates.
\begin{equation}
    t_P=\tan\frac{t_r}{2}=-\frac{1}{\tan\frac{t_l}{2}}=\tanh\frac{t_R}{2}~.
\end{equation}

The Rindler coordinates cover part of the Poincaré patch, as we show in Fig. \ref{sfig:AdS2RindlerApp}.

\setcounter{equation}{0}
\setcounter{appendixsection}{\value{appendixsection}+1} 
\section{Two coupled Schwarzian theories}\label{app:Schwarzianeqs}

\subsection{Single Schwarzian theory}
Before studying the full left-right coupled theory, let us go back to the theory with a single Schwarzian, with Lagrangian
\begin{equation}
    L=-N\h \Sch\left(\tan\frac{t(u)}{2},u\right)~,
\end{equation}
with $N$ being a constant. The equation of motion of the Schwarzian theory can be written as
\begin{equation}
    \frac{\left[\Sch(t(u),u)\right]'}{t'(u)}=0~.
\end{equation}

Given a function $f(z)$, we can change to a new function $g(z)$, with $f$ and $g$ related as $f=f(g)$, by using the following property of the Schwarzian derivative,
\begin{equation}
    \Sch(f,z)=\Sch(g,z)+\Sch(f,g)\left(\frac{dg}{dz}\right)^2~.
\end{equation}
In our case, with $f(u)=\tan\frac{t(u)}{2}$, we obtain
\begin{equation}
    L=-N\left[\Sch\left(t(u),u\right)+\frac{1}{2}t'^2\right]~,
\end{equation}
where primes denote derivatives with respect to $u$. Making the change of variables
\begin{equation}
    t'=e^\varphi~,
    \label{eq:tpvarphi}
\end{equation}
the Lagrangian becomes
\begin{equation}
    L=-N\left[\varphi''-\frac{1}{2}\varphi'^2+\frac{1}{2}e^{2\varphi}\right]~.
\end{equation}

We can now study the symmetries of $L$. Since the Lagrangian depends on the derivatives of $t$, it is clear that it is invariant under the infinitesimal transformation
\begin{equation}
    t\to t+\delta t~,\qquad \text{with}\qquad \delta t=\varepsilon^0~\text{a constant}~.
\end{equation}

It can be checked that the Lagrangian is also invariant under the infinitesimal transformations
\begin{equation}
    t\to t+\delta t~,\qquad \delta t=\varepsilon^\pm e^{\pm i t}~,
\end{equation}
with $\varepsilon^\pm$ constants. Therefore, $L$ is invariant under the infinitesimal $SL(2,\mathbb{R})$ transformations
\begin{equation}
    \delta t=\varepsilon^0+\varepsilon^+e^{it}+\varepsilon^-e^{-it}~.
    \label{eq:SL2transf}
\end{equation}

We can obtain the Noether charges associated to these symmetries. For the Schwarzian Lagrangian with the transformation \eqref{eq:SL2transf}, these are given by
\begin{equation}
\begin{aligned}
    \frac{Q_0^S}{N} & =-t'+\frac{t''^2}{t'^3}-\frac{t'''}{t'^2}~,\\
    \frac{Q_\pm^S}{N} & =e^{\pm it}\left[\pm i\frac{t''}{t'}+\frac{t''^2}{t'^3}-\frac{t'''}{t'^2}\right]~.
    \label{eq:charges1Schw}
\end{aligned}
\end{equation}

It is not difficult to check that conservation of $Q_0$ is already implied by the equation of motion, \ie,
\begin{equation}
    \frac{dQ_0^S}{du}=-\frac{\left[\Sch(t(u),u)\right]'}{t'(u)}=0~.
\end{equation}

\subsection{Two coupled boundaries}
We can perform a similar analysis for the coupled Schwarzian theory. We consider a coupled Schwarzian theory with action
\begin{equation}
    S=N\int d\Tilde{u}\left[-\left\{\tan\frac{t_l(\Tilde{u})}{2},\Tilde{u}\right\}-\left\{\tan\frac{t_r(\Tilde{u})}{2},\Tilde{u}\right\}+\eta\left(\frac{t_l'(\Tilde{u})t_r'(\Tilde{u})}{\cos^2\left(\frac{t_l(\Tilde{u})-t_r(\Tilde{u})}{2}\right)}\right)^\Delta\right]~.
    \label{eq:2coupledSchwApp}
\end{equation}
We denote by $\Tilde{u}$ the boundary time because it is a re-scaled boundary time compared to the actions that appear in the AdS$_2$ context, \eqref{eq:2coupledSchw}, and the two-coupled SYK system, \eqref{eq:2coupledSYKSchw}. The precise relation between parameters is
\begin{equation}
    \Tilde{u}=\frac{N}{\phi_r}u=\frac{\mathcal{J}}{\alpha_S}u~,\qquad \eta=\frac{g}{2^{2\Delta}}\left(\frac{N}{\phi_r}\right)^{2\Delta -1}=\frac{\mu \alpha_S}{\mathcal{J}}\frac{c_\Delta}{(2\alpha_S)^{2\Delta}}~.
    \label{eq:rescaledbdrytime}
\end{equation}

Clearly, the two Schwarzian terms are invariant under separate $SL(2,\mathbb{R})$ transformations of the form
\begin{equation}
\begin{aligned}
    \delta t_l & = \varepsilon^0+\varepsilon^+e^{it_l}+\varepsilon^-e^{-it_l}~,\\
    \delta t_r & = \overline{\varepsilon}^0+\overline{\varepsilon}^+e^{it_r}+\overline{\varepsilon}^-e^{-it_r}~.
\end{aligned}
\end{equation}

However, due to the coupling between the boundaries, there must exist a relation between the left and right parameters of the transformation for the total action to be invariant. For instance, when $\delta t_{l,r}$ are constant, the interaction term is only invariant if the cosine in the denominator does not change, which happens if $\delta t_l=\delta t_r$. Thus, we must have
\begin{equation}
    \varepsilon^0=\overline{\varepsilon}^0~.
\end{equation}
Explicit calculation shows that the interaction term is invariant if, in addition,
\begin{equation}
    \overline{\varepsilon}^\pm=-\varepsilon^\pm~.
\end{equation}

Therefore, the total action is invariant under the infinitesimal transformation
\begin{equation}
\begin{aligned}
    \delta t_l & = \varepsilon^0+\varepsilon^+e^{it_l}+\varepsilon^-e^{-it_l}~,\\
    \delta t_r & = \varepsilon^0-\varepsilon^+e^{it_r}-\varepsilon^-e^{-it_r}~.
\end{aligned}
\end{equation}

We can now proceed as before and obtain the Noether charges associated to this symmetry. The contributions from the left and the right boundaries are given by \eqref{eq:charges1Schw}. The interaction term gives an extra contribution to the total charges, which now are given by
\begin{equation}
\begin{aligned}
    \frac{Q_0}{N} & = Q_0^S(t_l)+Q_0^S(t_r)+\Delta\eta \left(\frac{1}{t_l'}+\frac{1}{t_r'}\right)\left[\frac{t_l'\h t_r'}{\cos^2\left(\frac{t_l-t_r}{2}\right)}\right]^\Delta~,\\
    \frac{Q_\pm}{N} & = Q_\pm^S(t_l)-Q_\pm^S(t_r)+\Delta\eta \left(\frac{e^{\pm i t_l}}{t_l'}-\frac{e^{\pm i t_r}}{t_r'}\right)\left[\frac{t_l'\h t_r'}{\cos^2\left(\frac{t_l-t_r}{2}\right)}\right]^\Delta~.
\end{aligned}
\end{equation}

Again, it is not difficult to check that the equations of motion derived from the action \eqref{eq:2coupledSchwApp} are equivalent to the statement of $Q_0$ charge conservation.

This can be used to solve the equations of motion. The idea is to consider the $SL(2,\mathbb{R})$ symmetry and to impose the vanishing of the charges. The vanishing of $Q_+$ and $Q_-$ is achieved automatically by taking $t_l=t_r$. Since any solution can be gauge transformed to a solution with $t_l=t_r$ (see Appendix B of \cite{Maldacena_2018}), we can safely set $t_l=t_r$ from now on. 

It only remains to require that $Q_0=0$. For $t_l=t_r\equiv t$, we have
\begin{equation}
    \frac{Q_0}{N}=2\left[-t'+\frac{t''^2}{t'^3}-\frac{t'''}{t'^2}\right]+2\Delta \eta (t')^{2\Delta-1}~.
\end{equation}

With the same change as in \eqref{eq:tpvarphi}, we get
\begin{equation}
    \frac{Q_0}{N}=2e^{-\varphi}\left[-\varphi''-e^{2\varphi}+\eta\Delta e^{2\Delta \varphi}\right]~.
\end{equation}

Thus, $Q_0=0$ is achieved if
\begin{equation}
    \varphi''=-e^{2\varphi}+\eta\Delta e^{2\Delta \varphi}~,
\end{equation}
which corresponds to the equation of motion of a particle in a potential
\begin{equation}
    V(\varphi)=e^{2\varphi}-\eta e^{2\Delta \varphi}~.
\end{equation}
There is a solution for which $\varphi$ is a constant, which corresponds to the stable point at the minimum of the potential. Going back to the $t$ coordinate (Eq. \eqref{eq:tpvarphi}), this solution is given by
\begin{equation}
    (t')^{2(1-\Delta)}=\eta\Delta~.
    \label{eq:tpsolutionApp}
\end{equation}

A constant derivative $t'$ means that the global times at the boundaries are given by
\begin{equation}
    t_l(\Tilde{u})=t_r(\Tilde{u})=t' \Tilde{u}~.
\end{equation}

\restoredefaultnumbering
\restoredefaultsectioning

%% file: Part2/FloquetII/FloquetII.tex
\renewcommand{\chapterquote}{
\raggedleft
\begin{minipage}{0.4\textwidth}
\color[gray]{0.6}
\textit{If you pull hard from here,}\\
\textit{and I pull hard from there}\\
\textit{surely it will fall, fall fall}\\
\textit{and we can free ourselves.}
\end{minipage}
\begin{minipage}{0.35\textwidth}
\raggedleft
\textit{Si tu l'estires fort per aquí,}\\
\textit{i jo l'estiro fort per allà,}\\
\textit{segur que tomba, tomba tomba}\\
\textit{i ens podrem alliberar.}
\end{minipage}

\vspace{2em}

\raggedleft\normalfont— Lluís Llach, \textit{L'Estaca}.}

\chapter{Holographic Floquet states in low dimensions}\label{chap:FloquetII}

We arrive now at the first Chapter of research in this thesis. Almost all the material in this Chapter is based on, and adapted from, \textit{Holographic Floquet states in low dimensions (II)} \cite{Berenguer_2022}.

In this Chapter, we study the D3/D5 brane intersection at finite temperature under a periodic, time-dependent electric field. The system has a non-equilibrium phase diagram with conductive and insulator phases. The external driving induces a rotating current due to vacuum polarization (in the insulator phase) and to Schwinger effect (in the conductive phase). For some particular values of the driving frequency the external field resonates with the vector mesons of the model and a rotating current can be produced even in the limit of vanishing driving field, defining the \textit{vector meson Floquet condensates}. For all temperatures, at given intercalated frequencies, we find new dual states that we name \textit{Floquet suppression points}, where the vacuum polarization vanishes even in the presence of an electric field. From the data we infer that these states exist both in the conductive and insulating phases. In the massless limit we find a linear and instantaneous conductivity law, recovering the results reviewed in Section \ref{subsec:nonlinearsigma}. We also examine the photovoltaic AC and DC current as the response to an oscillating probe electric field and see that rising the temperature suppresses the photovoltaic Hall current.

\vspace{1cm}

\section{Introduction}

Periodically driven systems form a separate chapter in the book of non-equilibrium dynamics. Much progress has been achieved both at theoretical and experimental levels in the path to control their effective long time dynamics (see references in the Introduction). When the rates of energy injection and dissipation are able to balance, the system can evolve into a non-equilibrium steady state (NESS).

In a series of previous papers, the existence of a Floquet NESS has been studied in the context of the AdS/CFT correspondence both in the case of a D3/D7 system \cite{Hashimoto_2016,Kinoshita_2017}, and of a D3/D5 system \cite{Garbayo_2020}, providing controlled examples of driven, dissipative systems in holography. In the D3/D7 system, both massless and massive flavour branes were considered in the presence of external rotating electric fields. In the massless case, the authors identified an induced Hall conductivity, while the analysis for massive flavours revealed a rich phase structure. This analysis was extended to the D3/D5 intersection in \cite{Garbayo_2020}, where the dual field theory lives on a two-dimensional defect embedded in four-dimensional $\mathcal{N}=4$ SYM theory \cite{DeWolfe_2001, Karch_2002, Erdmenger_2002}.

As explained in Section \ref{sec:metallicAdSCFT}, when a medium is subjected to an external electric field, a vacuum polarization due to virtual charged particles (quarks) is produced. As a consequence, an oscillating polarization current is induced. For weak fields, the system remains in a gapped, insulating phase where the induced current is purely transverse to the driving field, and energy dissipation is absent. However, beyond a critical field strength, real pair production via the Schwinger effect occurs, and the system transitions into a conducting phase characterized by dissipative currents and a Joule heating.

When the external field is periodic in time, the critical value $E_c$ at which this transition occurs becomes a function of the driving frequency $\Omega$, \ie, $E_c=E_c(\Omega)$ \cite{Takayoshi_2020}. Remarkably, there exist resonant frequencies $\Omega=\Omega_c$ for which the transition occurs even at vanishing field, $E_c(\Omega_c)=0$. At these frequencies, the external driving resonates with vector meson excitations in the dual field theory, leading to the formation of a non-trivial rotating current even in the absence of an applied electric field. Following \cite{Kinoshita_2017,Garbayo_2020}, we refer to this phase as a \textit{vector meson Floquet condensate}.

In this work, we revisit the D3/D5 setup, now at finite temperature, which deconfines the adjoint degrees of freedom and breaks supersymmetry. This adds charged carriers to the ones previously formed by Schwinger pair production. These are naturally melted mesons that are present at finite temperature for low enough quark mass. Therefore, the phenomenology is expected to yield a continuous deformation of the case at zero temperature.

We will work in the probe approximation and will hence neglect the backreaction of the D5-branes on the background geometry. The fluctuation of the fields living on the brane are dual to the mesonic excitations of the gauge theory. In \cite{Arean_2006} the complete analysis of these fluctations was performed for the D3-D5 system and the exact spectrum of mesons was found, as we also reviewed in Section \ref{subsec:mesonspectrum}.

A crucial simplification arises from moving to a rotating frame. In this frame, the action becomes time-independent, and the resulting equations of motion are ordinary differential equations. This allows us to apply the techniques developed in \cite{Karch_2007}, where one-point functions are obtained from the condition that the on-shell action remains real (see Section \ref{sec:metallicAdSCFT}). While this IR regularity mechanism has been widely used for static configurations, its application to time-dependent sources remains limited. Notably, in two spatial dimensions and for massless flavors, the current is known to respond linearly and instantaneously to the electric field: $J(t) = \sigma E(t)$ \cite{Karch_2010} (see Eq. \eqref{eq:sigmamassless}). Our results are consistent with this observation in the massless and small-mass regime.

Another key feature of the system is the presence of an effective temperature, $T_\text{eff}$, perceived by the open string degrees of freedom on the brane. This arises from the so-called open string metric (see Section \ref{subsec:Teff}) and has been argued to behave as a genuine temperature, satisfying fluctuation-dissipation relations \cite{Sonner_2012}. In most known examples, $T_\text{eff} > T$, where $T$ is the temperature of the bulk plasma. While exceptions to this inequality have been found \cite{Nakamura_2013}, our analysis across parameter space reveals no such anomalies in the D3/D5 model.

The plan of the Chapter is the following. Section \ref{sec:D3D5temp} sets up the holographic model, introduces the relevant field theory quantities, and discusses the classification of brane embeddings. We also analyze the behavior of the effective temperature. In Section \ref{sec:phasespace}, we examine the phase structure of the system in detail. Section \ref{sec:conductivitiesD3D5} is devoted to transport phenomena: we study both the rotating current induced by the periodic driving and the photovoltaic current, which characterizes the response to a secondary probe field on top of the rotating electric field. In Section \ref{sec:mesons}, we analyze the mesonic spectrum via linearized Minkowski embeddings. In Section \ref{sec:problemsnoneq} we comment on the difficulties of interpreting the on-shell action thermodynamically. Our findings and open questions are summarized in Section \ref{sec:FloquetIIconclusions}.

Finally, the Chapter includes several appendices with technical details and supplementary results. Appendix \ref{app:analyticsols} presents analytic solutions for zero and small quark mass. Appendix \ref{app:holorenoD3D5} contains the holographic renormalization and dictionary. Appendix \ref{app:opticalcond} elaborates on the calculation of optical conductivities, including the analytic massless case. For completeness, Appendix \ref{app:D3D7} presents the corresponding phase structure for the D3/D7 system.

\section{D3/D5 system at finite temperature}\label{sec:D3D5temp}

We work again in the coordinates of Section \ref{subsec:nonlinearsigma}. In these coordinates, the AdS$_5\times S^5$ metric is given by
\begin{equation}
\begin{aligned}
    ds^2=\frac{u^2}{L^2}&\left(-\frac{g(u)^2}{h(u)}dt^2+h(u)\left(dx^2+dy^2+dz^2\right)\right)+\frac{L^2}{u^2}du^2\\
    &\phantom{\left(-\frac{g(u)^2}{h(u)}dt^2+h(u)\left(dx^2\right.\right.}+L^2\left(\frac{d\psi^2}{1-\psi^2}+(1-\psi^2)d\Omega_2^2+\psi^2 d\Omega_2^2\right)~,
    \label{eq:AdS5xS5FloquetII}
\end{aligned}
\end{equation}
with the functions $g(u)$ and $h(u)$ given by
\begin{equation}
    g(u)=1-\frac{u_h^4}{u^4}~~,\hspace{5mm}h(u)=1+\frac{u_h^4}{u^4}~,
    \label{eq:ghfunctions}
\end{equation}
The black hole horizon is located at $u_h=r_h/\sqrt{2}$, with the Hawking temperature given by
\begin{equation}
    T_H=\frac{r_h}{\pi L^2}=\frac{\sqrt{2}u_h}{\pi L^2}~.
    \label{eq:TBH}
\end{equation}
In the following, we set $L=1$. We will parametrize background temperatures in terms of $r_h/m$.

We want to study the response of the system to an external circularly polarized electric field,
\begin{equation}
    \begin{pmatrix} \mathcal{E}_x(t) \\ \mathcal{E}_y(t) \end{pmatrix}=\begin{pmatrix} \cos\Omega t & -\sin\Omega t\\ \sin\Omega t & \cos\Omega t \end{pmatrix}\begin{pmatrix} E_x \\ E_y\end{pmatrix}\equiv \mathcal{O}(t)\vec{E}~,
    \label{eq:rotatingE}
\end{equation}
with $\vec E=(E_x,E_y)$ denoting the electric field at $t=0$. It is convenient to introduce the complexified electric field
\begin{equation}
    \mathcal{E}_x+i\mathcal{E}_y=Ee^{i\Omega t}~,
    \label{eq:complexE}
\end{equation}
with $E=E_x+iE_y$. This electric field can be derived from a vector potential
\begin{equation}
    A_x+iA_y=Ae^{i\Omega t}~,
    \label{eq:complexA}
\end{equation}
with $A=iE/\Omega$. Holographically, this corresponds to turning on a worldvolume gauge field $A(t,u)$ that approaches \eqref{eq:complexA} at the boundary of AdS$_4$. We introduce it as
\begin{equation}
    2\pi\alpha'A(t,u)=A_x(t,u)~dx+A_y(t,u)~dy~,
\end{equation}
with field strength given by
\begin{equation}
    2\pi\alpha'F=\dot{A}_x~dt\wedge dx+A_x'~du\wedge dx+\dot{A}_y~dt\wedge dy+A_y'~du\wedge dy~.
\end{equation}
The complexifications above motivate to do the same at the level of the bulk fields,
\begin{equation}
\begin{aligned}
    2\pi\alpha'\h A(t,u)=A_x(t,u)+i A_y(t,u)&\equiv c(t,u)e^{i\Omega t}\\
    &\equiv b(t,u)e^{i\left(\Omega t+\chi(t,u)\right)}~,
    \label{eq:defctu}
\end{aligned}
\end{equation}
where in the second line we have defined two new real variables, $b(t,u)$ and $\chi(t,u)$, representing the magnitude and the phase of $c(t,u)$, respectively. 

The D5-branes will be located at $z=0$. For the embedding function we are going to consider a general dependence on the radial coordinate, and time, $\psi=\psi(t,u)$. However, as we will see shortly, all time-dependence can be eliminated and the resulting problem becomes one-dimensional, with dependence only on the holographic coordinate. The induced metric on the D5-branes is
\begin{equation}
\begin{aligned}
    ds^2=&-\left(u^2\frac{g(u)^2}{h(u)}-\frac{\dot{\psi}^2}{1-\psi^2}\right)dt^2+u^2h(u)\left(dx^2+dy^2\right)+\left(\frac{1}{u^2}+\frac{\psi'^2}{1-\psi^2}\right)du^2\\    
    &+\frac{2\dot{\psi}\psi'}{1-\psi^2}dt\hspace{0.5mm}du+(1-\psi^2)d\Omega_2^2~.
\end{aligned}
\end{equation}

In terms of these variables, the DBI action for the $N_f$ probe D5-branes is given by
\begin{equation}
\begin{aligned}
    S_{D5}=&-\mathcal{N}\int dt~du\sqrt{1-\psi^2}\Bigg[h\left(1-\psi^2+u^2\psi'^2\right)\left(u^4g^2-\dot{b}^2-\left(\Omega+\dot{\chi}\right)^2b^2\right)\Bigg.\\
    &\phantom{-N_f T_{D5}}-u^2h~\dot{\psi}^2\left(h+b'^2+b^2\chi'^2\right)+2u^2h~\dot{\psi}~\psi'\left(\dot{b}~b'+\left(\Omega+\dot{\chi}\right)b^2\chi'^2\right)\\
    &\phantom{-N_f T_{D5}}\Bigg.+\left(1-\psi^2\right)\left[u^4g^2\left(b'^2+b^2\chi'^2\right)-b^2\left(\dot{b}\chi'-\left(\Omega+\dot{\chi}\right)b'\right)^2\right]\Bigg]^{1/2}~,
    \label{eq:actionD5tdep}
\end{aligned}
\end{equation}
with $\mathcal{N}=4\pi N_fT_{\text{D5}} V_{\mathbb{R}^{2,1}}$. The factors $V_{\mathbb{R}^{2,1}}$ and $4\pi$ arise from the integration along the $(x,y)$-directions and the two-sphere, respectively. It turns out that a consistent ansatz corresponds to taking the fields $\psi(t,u)$, $b(t,u)$ and $\chi(t,u)$ to be time-independent:
\begin{equation}
    \psi=\psi(u)~,\quad\quad b=b(u)~,\quad\quad\chi=\chi(u)~.
\end{equation} 
With this truncation, the action \eqref{eq:actionD5tdep} becomes
\begin{align}
\begin{split}
    S_{D5}=-\mathcal{N}\int du \sqrt{1-\psi^2}\Bigg[\left(u^4g^2-\Omega^2b^2\right)\left[\left(1-\psi^2\right)\left(h+b'^2\right)+u^2h\psi'^2\right]\Bigg.\\
    \Bigg.+u^4g^2b^2\left(1-\psi^2\right)\chi'^2\Bigg]^{1/2}~,
    \label{eq:actionD5}
\end{split}
\end{align}
where we have re-defined the overall constant to $\mathcal{N}=4\pi N_fT_{\text{D5}}$. On both sides we have implicitly divided by the volume of $\mathbb{R}^{2,1}$. The prefactor $\mathcal{N}$ may be written in terms of field theory quantities as
\begin{equation}
    \mathcal{N}=4\pi N_fT_{\text{D5}}=\frac{N_fN_c\sqrt{\lambda}}{2\pi^3}~.
    \label{eq:ND3D5}
\end{equation}

From this action it is obvious that all time-dependence has been removed, and the system has effectively become one-dimensional, with only dependence on the holographic coordinate, $u$. The fact that the electric field is time-dependent and rotating is seen through the $\Omega$-dependence in the action\footnote{This is a notorious difference with respect to the case of a constant electric field, that is at the heart of many differences among the two cases. In the constant field case, $A_x = -Et +...$ \cite{Karch_2007}, the action depends explicitly on $E$, whereas here the electric field will be extracted from the asymptotic behavior of the $b$ and $\chi$ fields. See \eqref{eq:UVexpansions}.}. This is the action we will use to derive the equations of motion for the fields $\psi$, $b$ and $\chi$. These equations are long and unilluminating, so they won't be reproduced here.

As usual, by imposing that the equations must be satisfied order by order in $u$ near the UV boundary, we can find the asymptotic behavior of the fields. By expanding $\psi(u)=\sum_{n=0}^\infty \psi_n u^{-n}$, and similarly for $b$ and $\chi$, we find the following behavior:
\begin{align}
    \psi(u)&=\frac{m}{u}+\frac{C}{u^2}+...~,\\
    b(u)e^{i\chi(u)}&=\frac{i E}{\Omega}+\frac{J}{u}+...~,
    \label{eq:UVexpansions}
\end{align}
where we have denoted $\psi_1$ and $\psi_2$ as $m$ and $C$, respectively, and in the second line we have
\begin{equation}
\begin{aligned}
    \frac{iE}{\Omega} &= \lim_{u\to\infty}b(u)e^{i\chi(u)} = b_0\cos\chi_0+i\h b_0\sin\chi_0~,\\
    J & = -\lim_{u\to\infty} u^2 \left[b'(u)+ib(u)\chi'(u)\right]e^{i\chi(u)} =  e^{i\chi_0}\left(b_1+ib_0\chi_1\right)~,
    \label{eq:UVexpansionsEJ}
\end{aligned}
\end{equation}
with $E$ the complex electric field, defined in \eqref{eq:complexE}-\eqref{eq:complexA}. The same complexified notation is used for $J$, defined as $J=J_x+iJ_y$. From here we can read off the one-point functions for the quark condensate $\langle O_m\rangle$ and the electric current $\mathcal{J}_\text{YM}$ in the boundary theory. The precise relation between $E$, $J$, $m$ and $C$ and the electric field ${\cal E}_\text{YM}(t)$, the electric current ${\cal J}_\text{YM}(t)$, quark mass $m_q$ and quark condensate $\langle O_m\rangle $ in the boundary theory is derived in Appendix \ref{app:holorenoD3D5}:
\begin{equation}
\begin{aligned}
    \mathcal{E}_\text{YM}(t) &= \frac{\sqrt{\lambda}}{2\pi}E e^{i\Omega t}~, & \qquad \mathcal{J}_\text{YM}(t) &= \frac{N_fN_c}{\pi^2}Je^{i\Omega t}~, \\
    m_q~ &= \frac{\sqrt{\lambda}}{2\pi}m~, & \qquad \langle O_m\rangle &= -\frac{N_fN_c}{\pi^2}C~,
    \label{eq:dictionaryD3D5}
\end{aligned}
\end{equation}
where $\lambda=g^2_\text{YM}N_c$ is the 't Hooft coupling of the $\mathcal{N}=4$ theory.

It is worth noting that the action \eqref{eq:actionD5} depends only on the derivative of $\chi(u)$, namely $\chi'(u)$, and not on $\chi(u)$ itself. Therefore, there exists a conserved quantity defined as
\begin{equation}
    q\equiv\Omega \frac{\partial \mathcal{L}}{\partial \chi'}\sim J_xE_x+J_yE_y=\vec{J}\cdot \vec{E}~,
    \label{eq:qconserved}
\end{equation}
with $\mathcal{L}$ the DBI Lagrangian. With $\sim$ we denote the UV expansion of this quantity, using \eqref{eq:UVexpansions}. Therefore we see that $q$ acquires the meaning of a Joule heating. Only black hole embeddings have a non-zero value for this quantity. The stationarity of the background metric upon this energy injection can only be understood as a transient effect due to the imbalance $N_f/N_c \sim 0$ that is present in the probe limit. 
In the presence of a black hole in the bulk, the long time effect of a non-negligible backreaction  would be a slow increase in the horizon radius. 

We can use the conserved quantity to obtain $\chi'$ as
\begin{equation}
    \chi'^2=q^2\frac{\left(u^4g^2-\Omega^2b^2\right)}{u^4g^2b^2\left(1-\psi^2\right)\left(u^4g^2\Omega^2b^2\left(1-\psi^2\right)^2-q^2\right)}\left(\left(1-\psi^2\right)b'^2+h\left(1-\psi^2+u^2\psi'^2\right)\right)~.
\end{equation}
Using this, we can Legendre transform the action to get rid of $\chi'$, and write an action in terms of $q$,
\begin{equation}
\begin{aligned}
    \bar{S}_{\text{D5}}&=S_{\text{D5}}-\int du~ \chi'\frac{\partial\mathcal{L}}{\partial \chi'}\\
    &=-\mathcal{N}\int du \frac{\sqrt{b'^2+h\left(1+\frac{u^2\psi'^2}{1-\psi^2}\right)}}{u^2g\Omega\abs{b}}\sqrt{\left[u^4g^2-\Omega^2b^2\right]\left[u^4g^2\Omega^2b^2\left(1-\psi^2\right)^2-q^2\right]}~.
    \label{eq:actionLTtransf}
\end{aligned}
\end{equation}

\begin{figure}
    \centering
    \includegraphics[width=0.34\textwidth]{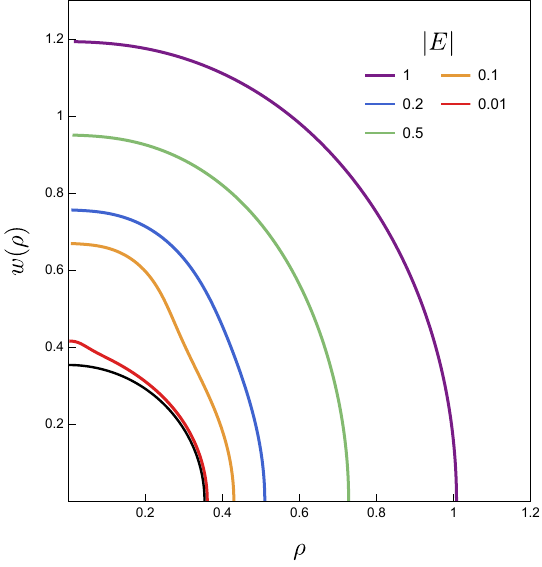}
    \hspace{0.02\textwidth} 
    \includegraphics[width=0.45\textwidth]{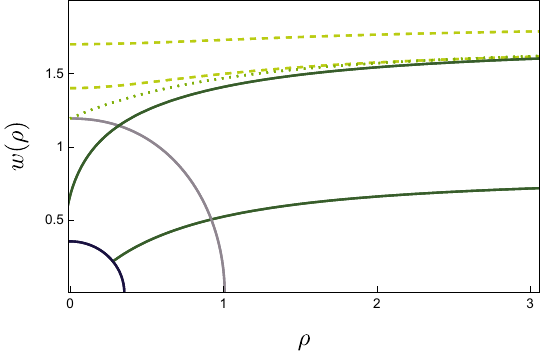}
    \caption{Left: shape of the singular shell for different applied electric fields. The axes are $\rho =u\sqrt{1-\psi^2}$ and $w=u\psi$ with $\psi=\sin\theta$. The curves are the points $\rho^2+w^2=u_c(E,\psi_0)$ with $\psi_0=\sin\theta_0$ the embedding angle at the singular shell. Outside the limit $E\to 0$ the shell shape is non spherical. Hence, unlike the case of a constant electric field, in the rotating situation we have a non-trivial dependence of $u_c$ on the mass $m$ of the D5-brane at fixed $|E|$. Right: profiles of different embeddings for $\Omega=|E|=1$. The dashed lines represent black hole embeddings. They can be either regular (thermal), ending on the horizon, or singular, ending in a conical singularity. The dotted line represents a critical embedding. The solid lines represent Minkowski embeddings. The black hole and pseudo-horizon are shown by the black (inner) and gray (outer) lines.}
    \label{fig:rcEmbeddings}
\end{figure}

As has become the usual case when there is an electric field switched on \cite{Karch_2007}, demanding reality of the Legendre-transformed action \eqref{eq:actionLTtransf} imposes that the two terms under the second square root must change sign (and therefore, vanish) at the same point $u=u_c$, which locates the singular shell. At once, this condition fixes both  the value of $u_c$ and that of the conserved quantity $q$
\begin{equation}
    \Omega\h b_0=\frac{u_c^4-u_h^4}{u_c^2}~, \qquad q=\frac{(u_c^4-u_h^4)^2}{u_c^4}(1-\psi_0^2)~,
    \label{eq:b0andq}
\end{equation}
where $b_0=b(u_c)$, $\psi_0=\psi(u_c)$. As $\Omega b_0\geq 0$, we see that  $u_c\geq u_h$, and the singular shell sits always at a larger radius than the black hole horizon. Moreover, here it is the IR data, $b_0$, what controls the position of the singular shell, $u_c(u_h, b_0)$. This is another important difference with respect to the case of constant $E$  \cite{Karch_2007}, where the position of the singular shell depends exclusively on the UV value $\abs{E}$. Therefore, in this case the shape of the critical surface at constant $|E|$ is non-spherical, as can be seen in the left plot of Fig. \ref{fig:rcEmbeddings}.

An important remark for later use is the fact that both the worldvolume electric field and the black hole horizon add up their effects of bending the brane in the IR towards the origin. This will mean that as we increase the temperature we will find black hole embeddings with milder electric fields.

The following  is a  scaling symmetry of the lagrangian  $\mathcal{L} \to \alpha^2\mathcal{L}$ and the boundary conditions
\begin{equation}
\begin{aligned}
    t &\to t/\alpha~, & u &\to \alpha\h u~, & w & \to \alpha\h w~, & b & \to \alpha\h b~, & \chi & \to\chi~,\\
    \Omega &\to \alpha\h \Omega~, & E &\to \alpha^2\h E~, & J & \to \alpha^2\h J~, & \psi & \to \psi~, & m & \to\alpha\h m~,\\
    C &\to \alpha^2C, & q &\to \alpha^4q~, & T & \to \alpha\h T~, 
    \label{eq:scaling}
\end{aligned}
\end{equation}
By choosing $\alpha=1/m$ in \eqref{eq:scaling} we can take $m=1$ and deal with the remaining quantities in units of (the  appropriate powers of) $m$. 

\subsection{Types of embeddings}

There are three types of embeddings in place now. First of all, we find the \textit{Minkowski} embeddings, which do not intersect the singular shell. They end up closing  smoothly at a value of $u=u_0>u_c$, where $\psi=1$. With \textit{black hole} (BH) embeddings we will generically denote solutions that intersect the singular shell. This accounts for the fact that for a worldvolume observer the singular shell acts as an event  horizon, inducing thermal effects through Schwinger pair production. For this reason, we will term interchangeably singular shell and effective horizon. Finally, the \textit{conical} embeddings will denote the solutions ending outside but just above the singular shell. Notice that this differs from the usual definition of critical embeddings when the electric field is not present, where they are defined as those ending just above the black hole horizon. We depict the different embeddings in the right plot of Fig. \ref{fig:rcEmbeddings}.

\subsubsection{Black hole embeddings}
Black hole embeddings  can be further subdivided into two classes,  \textit{thermal} and \textit{conical}.  The first ones hit the bulk black hole horizon, while the second ones close up at $\psi=1$ with a conical singularity, most likely a reflection of the sink of energy pumped by the electric field in a conducting, albeit non-dissipative system. These conical black hole embeddings are the remnant of the ones that were studied in \cite{Garbayo_2020} at zero temperature. For a constant electric field they were analyzed in \cite{Mas_2009,Kim_2011}. The two lower embeddings in the right plot of Fig. \ref{fig:rcEmbeddings} are examples of each class.

As a technical  remark, notice that, in order to solve numerically for the  black hole embeddings,  boundary conditions must be placed at the singular shell. This can be done by expanding the fields as
\begin{equation}
\begin{aligned}
    \psi(u) & = \psi_0+\psi_1(u-u_c)+...\\
    b(u) & = b_0+b_1(u-u_c)+...\\
    \chi(u) & = \chi_0 +\chi_1(u-u_c)+...~,
    \label{eq:coefsBHemb}
\end{aligned}
\end{equation}
with $\psi_0=\psi(u_c)$, and similarly for $b_0$ and $\chi_0$. These parameters can be chosen freely and give rise to different solutions with different UV asymptotics. However, the equations of motion derived from the action \eqref{eq:actionD5} are naïvely divergent at $u=u_c$. Therefore, the three derivatives $\psi_1$, $b_1$ and $\chi_1$ are not free but commanded by regularity. The precise expressions are lengthy and provide no further insight, so they are omitted here.

In contrast, we could choose boundary conditions at the background horizon for the embedding function $\psi$ and the module $b$, but not for the phase  $\chi$, since this function diverges logarithmically as $u\to u_h$. Starting from the effective horizon, then, one can integrate either outwards or inwards and this is how the thermal and/or conical embeddings are distinguished.

\subsubsection{Minkowski and critical embeddings}
Minkowski embeddings do not intersect the singular shell. They close off smoothly at $u=u_0>u_c$, where $\psi=1$. The values $b_0=b(u_0)$ and $\chi_0=\chi(u_0)$ are free parameters and, again, regularity of the equations at $u=u_c$ determines the fields to behave as
\begin{equation}
\begin{aligned}
    \psi(u)&=1-\frac{3\left(u_0^4+u_h^4\right)\left(u_0^8+u_h^8-u_0^4\left(2u_h^4+\Omega^2b_0^2\right)\right)}{u_0\left(u_0^4-u_h^4\right)\left(3\left(u_0^8+u_h^8\right)+u_0^4\left(2u_h^4-\Omega^2b_0^2\right)\right)}(u-u_0)+...\\
    b(u)&=b_0-\frac{u_0b_0\Omega^2\left(u_0^4+u_h^4\right)^2}{\left(u_0^4-u_h^4\right)\left(3\left(u_0^8+u_h^8\right)+u_0^4\left(2u_h^4-\Omega^2b_0^2\right)\right)}(u-u_0)+...
\end{aligned}
\end{equation}
We haven't written an expansion for $\chi$ because the equations of motion impose all the terms $\chi_n$, with $n\ge 1$, to vanish. Therefore, for Minkowski embeddings, $\chi(u)=\chi_0$ is a constant along the holographic direction, only determining the initial direction at $t=0$ of the electric field in the boundary theory. For this reason, as mentioned earlier, the Joule heating $q$ vanishes for Minkowski embeddings (recall that $q\propto \chi'$).

Critical embeddings can be understood as Minkowski embeddings, in the limit where $\Omega b_0\to\frac{u_c^4-u_h^4}{u_c^2}$, as dictated by the position of the singular shell, Eq. \eqref{eq:b0andq}. This leads to the following expansion for the critical embeddings:
\begin{equation}
\begin{aligned}
    \psi(u)&=1-\frac{4\left(u_c^4+u_h^4\right)+\Omega^2u_c^2}{2u_c^2\left(u_c^4+u_h^4\right)}(u-u_0)^2+...\\
    b(u)&=\frac{u_c^4-u_h^4}{\Omega u_c^2}-\frac{\Omega}{2u_c}(u-u_0)+...
\end{aligned}
\end{equation}

\subsection{Effective metric and effective temperature}\label{subsec:Teffrotating}
As mentioned earlier, the critical radius $u_c$ signals the position of an event horizon in the induced open string metric, which governs the dynamics of the worldsheet fluctuations. The effective temperature $T_\text{eff}$ was defined in Section \ref{subsec:Teff} as the Hawking temperature associated with the black hole horizon of this metric. We now extend the results to the present case of a rotating electric field.

Let $h_{ab}$ be the induced six-dimensional metric and $\mathcal{F}_{ab}$ the worldvolume gauge field. The effective open string metric $\gamma_{ab}$ is defined as
\begin{equation}
    \gamma_{ab}=h_{ab}+(2\pi\alpha')^2\mathcal{F}_{ac}\mathcal{F}_{bd}~h^{cd}~.
\end{equation}
In order to write the form of this metric for our ansatz when the embedding is parameterized as a function $\psi=\psi(u)$, let us define the function $F(u)$ as
\begin{equation}
    F(u)\equiv\frac{1}{h}\left(u^2g^2-\frac{\Omega^2\abs{c}^2}{u^2}\right)~,
\end{equation}
and the complex one-forms $e_{\pm}$ as
\begin{equation}
    e_{\pm}=e^{\mp i\Omega t}(dx\pm idy)~.
\end{equation}
Then, we have
\begin{align}
\begin{split}
    \gamma_{ab}d\xi^a d\xi^b=-F(u) dt^2+\left[\frac{1}{u^2}+\frac{\psi'^2}{1-\psi^2}+\frac{\abs{c'}^2}{u^2h}\right]du^2-\frac{2\h \Omega}{u^2h}\Im(c\bar{c}')dt\h du\\
    +\frac{1}{4}\left[\frac{1}{u^2}+\frac{\psi'^2}{1-\psi^2}\right]^{-1}(c'e_-+\bar{c}'e_+)^2+\frac{\Omega^2h}{4u^2g^2}(c\h e_-+\bar{c}\h e_+)^2+\frac{h^2F(u)}{g^2}e_+e_-~,
    \label{eq:eff_metric_non_diagonal}
\end{split}
\end{align}
where $c(u)=b(u)\h e^{i\chi(u)}$ is the complexified field potential in the rotating frame. We can diagonalize the $(t,u)$ part of the metric following the same procedure as in Section \ref{subsec:Teff}, with the change
\begin{equation}
    t=\tau-h(u)~,\qquad \text{where}\quad h'(u)=\frac{\gamma_{ut}}{\gamma_{tt}}~.
\end{equation}
In the new variables, the $(\tau,u)$ part of the metric reads
\begin{equation}
\begin{aligned}
    \gamma_{ab}d\xi^a d\xi^b\big\rvert_{\tau,u}&=\gamma_{tt}d\tau^2+\left(\gamma_{uu}-\frac{\gamma_{tu}^2}{\gamma_{tt}}\right)du^2\\
    &=-F(u)d\tau^2+\left[\frac{1}{u^2}+\frac{\psi'^2}{1-\psi^2}+\frac{\abs{c'}^2}{u^2h}+\frac{\Omega^2}{u^2h^2F(u)}\Im(c\h \bar{c}')\right]du^2
\end{aligned}
\end{equation}

\begin{figure}
    \centering
    \begin{subfigure}[t]{0.48\textwidth}
        \centering
        \includegraphics[width=\textwidth]{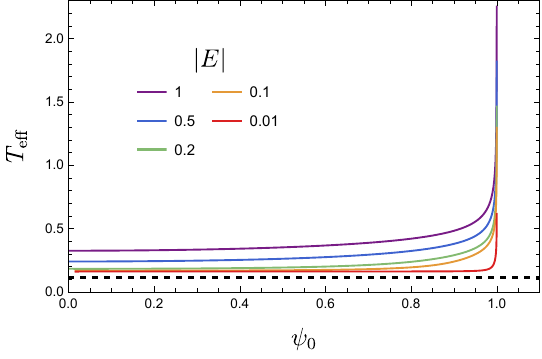}
        \label{sfig:TPlot}
    \end{subfigure}
    \hspace{0.02\textwidth} 
    \begin{subfigure}[t]{0.48\textwidth}
        \centering
        \vspace{-5.05cm} 
        \includegraphics[width=\textwidth]{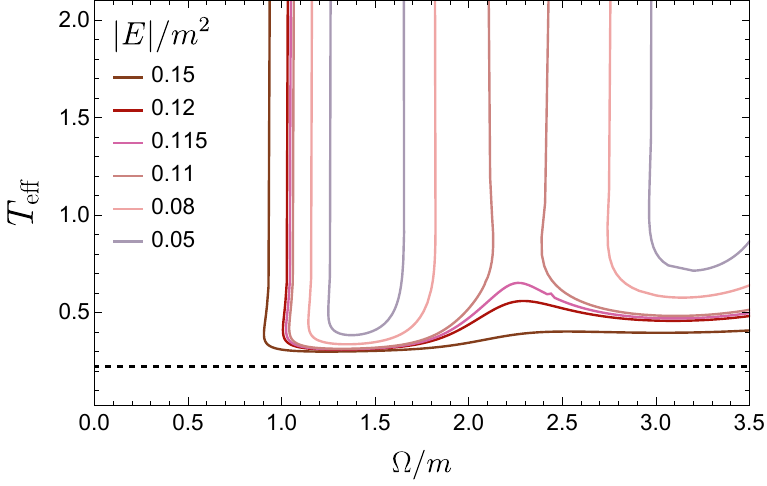}
        \label{sfig:EffectiveTlobes}
    \end{subfigure}
    \caption{Left: effective temperature for different BH embeddings, labeled by the insertion angle at the effective horizon, $\psi_0=\psi(u_c)$. Right: effective temperature as we vary $\Omega/m$ at fixed $|E|/m^2$. In both cases, the divergences in $T_{\text{eff}}$ arise as the curves come close to the critical embeddings. In both plots, the lowest dashed line in black signals the background temperature.}
    \label{fig:Teff}
\end{figure}

This metric has an event horizon at the singular shell, $u=u_c$, where the function $F(u)$ vanishes. Using the field expansions around $u=u_h$ defined in \eqref{eq:coefsBHemb} the Euclidean metric can be expanded near the horizon. Following the same procedure as in Section \ref{subsec:Teff}, but with more involved expressions, enforcing the correct periodicity of the time coordinate to eliminate the conical singularity at the origin leads to the effective temperature of the OSM,
\begin{equation}
    T_{\text{eff}}=\frac{2u_ch(u_c)-\Omega b_1}{2\pi b_0\chi_1}~,
    \label{eq:TeffD5}
\end{equation}
where $b_0$, $b_1$ and $\chi_1$ are the coefficients defined in \eqref{eq:coefsBHemb} for black hole embeddings. As in the vast majority of the situations encountered in the literature \cite{Kim_2011,Kundu_2013,Nakamura_2013,Kundu_2019}, here we also find that $T_{\text{eff}} > T_H$, as far as we have been able to  scan, as shown in Fig. \ref{fig:Teff}.

\section{Phase space}\label{sec:phasespace}

As just discussed, the different types of IR boundary conditions correspond to different behaviors for the probe branes. This behavior has to have an effect on the field theory phase space. The standard lore in flavor branes is that Minkowski (black hole) embeddings are dual to insulating (conducting) phases of the dual field theory. In the case of a rotating electric field, we must be more careful. Actually, two types of currents emerge. Black hole embeddings carry dissipating currents because of the presence of fundamental carriers. The external driving has to supply energy in order to maintain the stationary rotating current. 

For Minkowski embeddings, $J$ is a polarization current. In analogy with the case of the D3/D7 intersection \cite{Kinoshita_2017}, we interpret this polarization as a coherent alignement of the vector meson vacuum fluctuations parallel to the electric field. The polarization current is the time derivative of the polarization and rotates at right angles with the electric field, signalling zero Joule heating, hence not dissipating any energy. This conservative aspect allows for the  possibility to have non-zero persistent current even in the limit of vanishing driving field. This effect happens at discrete driving frequencies. As we will see, there exists a dual possibility, in which the polarizability is dynamically suppressed, even at finite driving field.

Using the different IR boundary conditions, we can integrate the equations of motion up to the UV boundary. For black hole embeddings, this is done by specifying the values $\psi_0$, $b_0$, $\chi_0$ at the singular shell $u=u_c$, with the derivatives of the fields dictated by regularity. For Minkowski embeddings, the IR boundary conditions are imposed at the point where the brane ends, $u=u_0$. In both cases, the solution is unique under these boundary conditions. From the asymptotic behavior of the fields, Eq. \eqref{eq:UVexpansions}, we can read off the values of $m$, $C$, $E$, $J$. Due to the scaling symmetry \eqref{eq:scaling}, these quantities are not all independent. Therefore, we will work with the dimensionless quantities $T/m$, $r_h/m$, $\Omega/m$, $E/m^2$, $J/m^2$ and ${\cal C}/m^2$.

\begin{figure}
    \centering
    \begin{subfigure}[t]{0.48\textwidth}
        \centering
        \includegraphics[width=\textwidth]{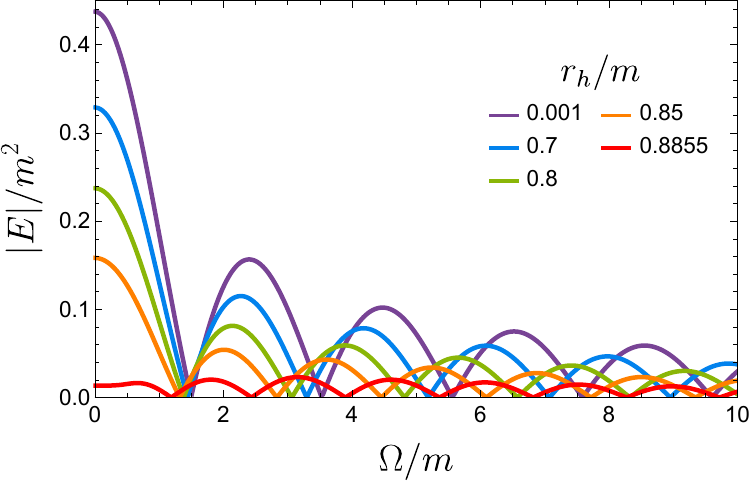}
    \end{subfigure}
    \hspace{0.02\textwidth}
    \begin{subfigure}[t]{0.48\textwidth}
        \centering
        \includegraphics[width=\textwidth]{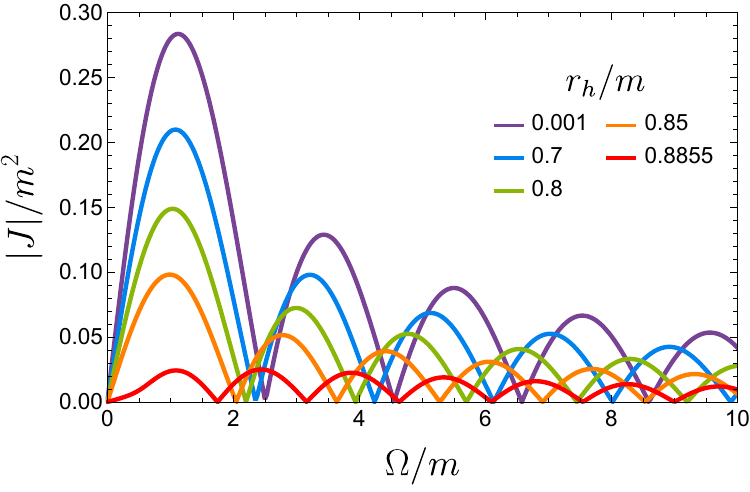}
    \end{subfigure}
    \caption{Electric field and current of the critical embeddings versus driving frequency, for different values of $r_h/m$. The frequencies for which $|E|/m^2$ vanishes are the critical frequencies $\Omega_c/m$ of the vector meson Floquet condensates. The regions of black hole embeddings between lobes, above such critical points, are termed \textit{wedges} in the text. In the right plot we show the value of the current along the curve of critical embeddings. We observe points with $|J|/m^2=0$ somewhere close to the maxima of the lobes in the left plot. We will term these points \textit{Floquet suppression points}, and will study them in detail in Section \ref{subsec:suprpoints}.}
    \label{fig:LobesEJD3D5}
\end{figure}

In Fig. \ref{fig:LobesEJD3D5} we show the phase diagram in the $(E/m^2,\Omega/m)$ parameter space. Different lines correspond to different temperatures, parametrized by $r_h/m$. The limit $r_h/m\to 0$ (purple line) recovers the results of \cite{Garbayo_2020}. For each temperature, the solid line represents the electric field and frequency of the critical embeddings. Roughly speaking, points above such curves correspond to black hole embeddings, whereas those below are Minkowski\footnote{However, close to the critical embeddings, the boundary quantities are usually multivalued (see \cite{Mateos_2006,Mateos_2007} and Fig. \ref{fig:JCvsE} below) and we can find both types of embeddings in the near vicinity.}.

This lobed structure was also present at zero temperature \cite{Garbayo_2020}, and in the D3/D7 model \cite{Kinoshita_2017}. We notice that $E/m^2$ has a series of maxima whose height decreases as $\Omega/m$ increases. Of particular interest are the points in which, for particular values of the driving frequency, the critical electric field vanishes. We denote these frequencies as $\Omega_c$. For these frequencies, however, the electric current is non-zero, as we see in the right plot of Fig. \ref{fig:LobesEJD3D5}. This means that, for these particular frequencies, the insulator-conductor transition can be triggered with vanishing external electric field. Physically, for these frequencies the driving field enters in resonance with the vector meson excitations of the gauge theory. We will refer to these states as \textit{vector meson Floquet condensates}, following \cite{Hashimoto_2016,Kinoshita_2017}. Actually, these $E=0$ Floquet states exist also for Minkowski embeddings in a finite range of frequencies, $\Omega_c\le\Omega\le \Omega_{meson}$, where $\Omega_{meson}$ denotes the mass of a vector meson in the supersymmetric defect theory. These masses, whose calculation was reviewed in Section \ref{subsec:mesonspectrum} were computed in \cite{Arean_2006} and are given by
\begin{equation}
    \Omega_{meson}/m=2\sqrt{\left(n+\frac{1}{2}\right)\left(n+\frac{3}{2}\right)}~,\qquad\qquad n=0,1,2,...
    \label{eq:mesonfreqs}
\end{equation}
For the zero-temperature D3/D5 intersection, these frequencies can be obtained analytically by considering the linearized fluctuations of the brane, and imposing that the critical electric field vanishes. This was done in \cite{Garbayo_2020}, and in Section \ref{sec:mesons} we review this calculation and extend it to non-zero temperature, which has to be solved numerically. For comparison, at zero temperature the critical frequencies and the meson frequencies are given by
\begin{equation}
\begin{aligned}
    \Omega_c/m & = 1.497,~3.531,~5.568,~7.585,~...\\
    \Omega_{meson}/m & = 1.732,~3.873,~5.916,~7.937,~...
\end{aligned}
\end{equation}

From Fig. \ref{fig:LobesEJD3D5} we can observe that the effect of increasing the temperature is a depletion of the  height of the lobes with the rising of $r_h/m$, until they fully disappear beyond some temperature. Let us pause to describe the origin of this damping effect.  We choose to measure  dimensionful quantities in units of the quark mass. In particular, the curves above are drawn each one for a fixed value of $r_h/m$. Remember that both the electric field and the temperature tend to bend the probe brane towards the origin in the IR. Let us fix a mass  $m=1$ for concreteness. Then, for a small value of $r_h$, we can still switch on and fine tune the electric field to make the embedding bend  enough so as to touch the critical surface. As $r_h$ grows, this supplemental field needed becomes less and less, which accounts for the drop in the lobe structure to be seen on the  plots in  Fig. \ref{fig:LobesEJD3D5}. Finally, there is a maximum value for $r_h/m=0.8897$ beyond which all the embeddings are of black hole type for any value of $|E|$ and $\Omega$.

Fig. \ref{fig:JCvsE} unfolds the fine structure in the vicinity of the first vector meson Floquet condensate. The left (right) plots show the values of the current $|J|/m^2$ (condensate $C/m^2$) as a function of the applied electric field, $|E|/m^2$, for different values of the frequency $\Omega/m$, close to the first resonant frequency $\Omega_c$. From top to bottom, the temperature increases parametrized by $r_h/m$. The upper case, with $r_h/m=0.5$ is almost indistinguishable from the case with $r_h/m=0$ studied in \cite{Kinoshita_2017,Garbayo_2020}. Points on the continuous (dashed) lines are for  black hole (Minkowski) embeddings. From the lower left corner, all curves start at the  Minkowski solution with $|J|=|E|=0$ (no singular shell). For $\Omega<\Omega_c$, the behavior is as shown in the green curves with $\Omega/m=1.2$. As the electric field increases, a non-dissipative polarization current builds up along the dashed portion of the green curves. At some point, the curve becomes multivalued, the prelude of a presumably discontinuous phase transition to a black hole configuration (a point on the continuous curve segment upwards) where a dissipative conduction current is allowed. The nature and exact occurrence of this transition is beyond the reach of equilibrium thermodynamics where free energy evaluation is enough. We will comment more about these issues in Section \ref{sec:problemsnoneq}.

\begin{figure}
    \centering
    \begin{subfigure}[t]{0.48\textwidth}
        \centering
        \includegraphics[width=\textwidth]{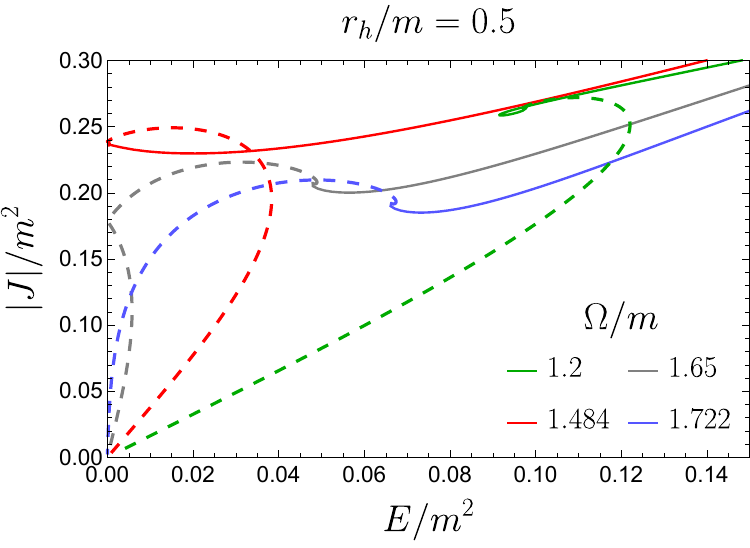}
        \label{sfig:JErh0p5}
    \end{subfigure}
    \hspace{0.02\textwidth}
    \begin{subfigure}[t]{0.48\textwidth}
        \centering
        \includegraphics[width=\textwidth]{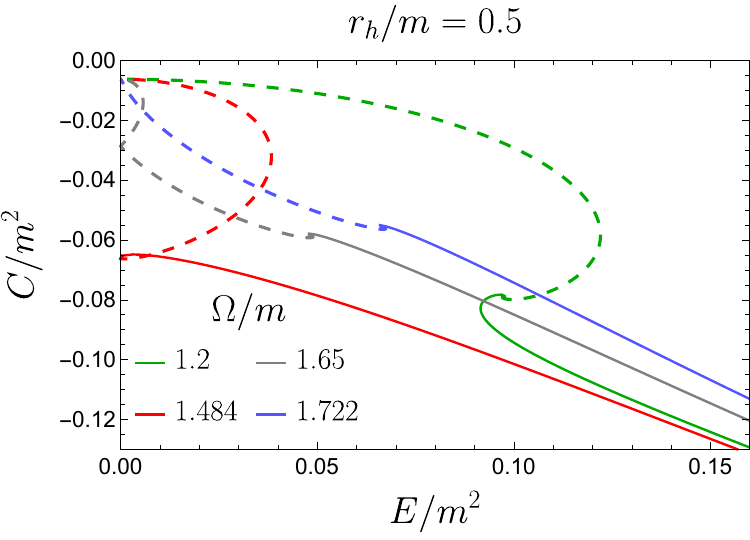}
        \label{sfig:cErh0p5}
    \end{subfigure}
    
    \centering
    \begin{subfigure}[t]{0.48\textwidth}
        \centering
        \includegraphics[width=\textwidth]{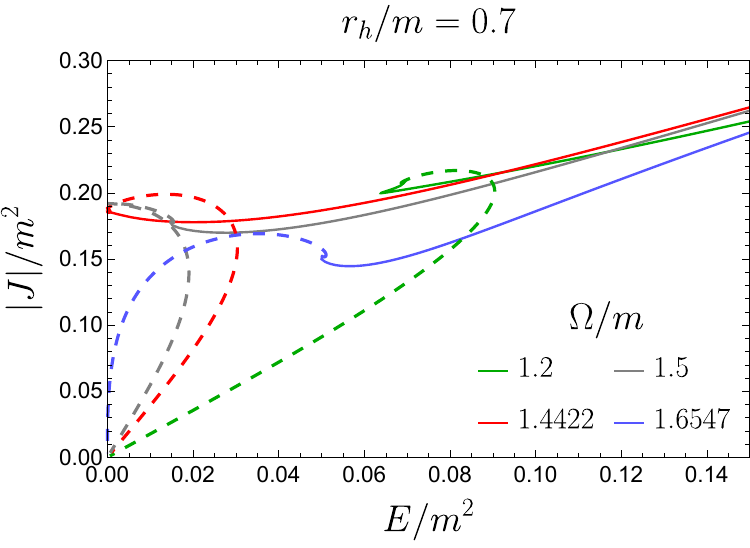}
        \label{sfig:JErh0p7}
    \end{subfigure}
    \hspace{0.02\textwidth}
    \begin{subfigure}[t]{0.48\textwidth}
        \centering
        \includegraphics[width=\textwidth]{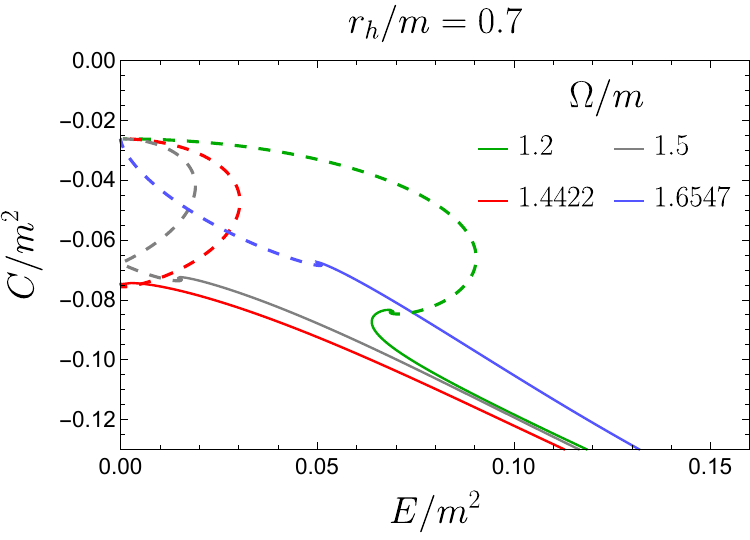}
        \label{sfig:cErh0p7}
    \end{subfigure}

    \centering
    \begin{subfigure}[t]{0.48\textwidth}
        \centering
        \includegraphics[width=\textwidth]{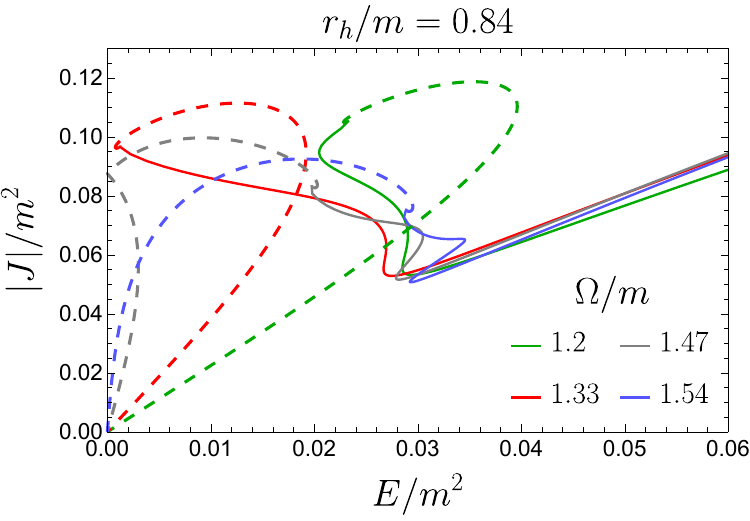}
        \label{sfig:JErh0p84}
    \end{subfigure}
    \hspace{0.02\textwidth}
    \begin{subfigure}[t]{0.48\textwidth}
        \centering
        \includegraphics[width=\textwidth]{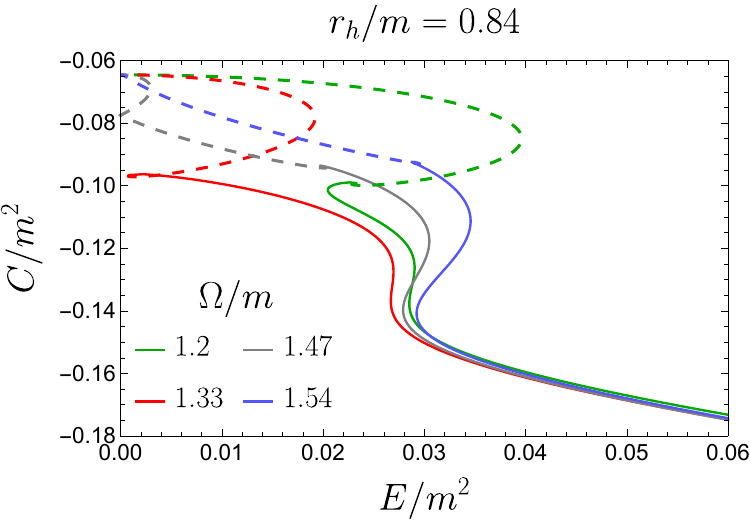}
        \label{sfig:cErh0p84}
    \end{subfigure}
    \caption{We show the electric current $|J|/m^2$ (left) and condensate $C/m^2$ (right) versus electric field for $r_h/m=0.5$, $0.7$ and $0.84$, from top to bottom. The dashed curves represent the insulator (Minkowski) phase and the solid curves the conductive (BH) phase (beware the opposite code with respect to \cite{Garbayo_2020}). In all figures, we show curves for a driving frequency $\Omega<\Omega_c$ (green), $\Omega=\Omega_c$ (red), $\Omega_c<\Omega<\Omega_{meson}$ (gray) and $\Omega=\Omega_{meson}$ (blue). We show that $E/m^2$ can only vanish for $J/m^2\neq0$ for Minkowski embeddings with $\Omega_c\leq \Omega\leq \Omega_{meson}$.}
    \label{fig:JCvsE}
\end{figure}

\begin{figure}[H]
    \centering
    \begin{subfigure}[t]{0.48\textwidth}
        \centering
        \includegraphics[width=\textwidth]{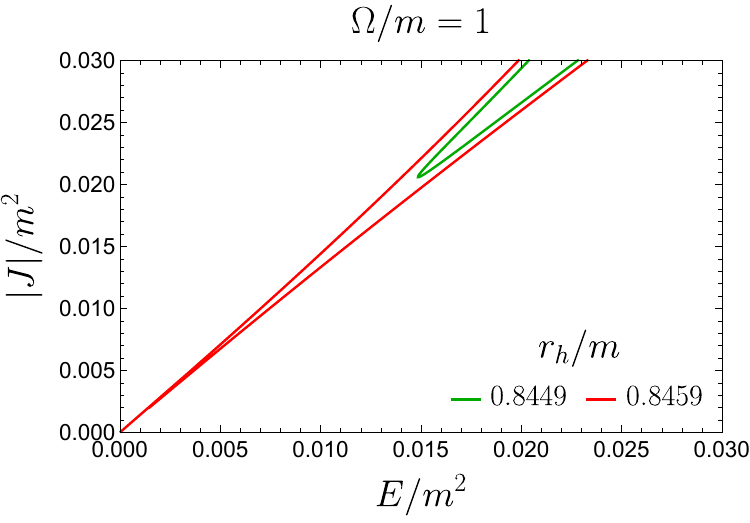}
    \end{subfigure}
    \hspace{0.02\textwidth}
    \begin{subfigure}[t]{0.48\textwidth}
        \centering
        \includegraphics[width=\textwidth]{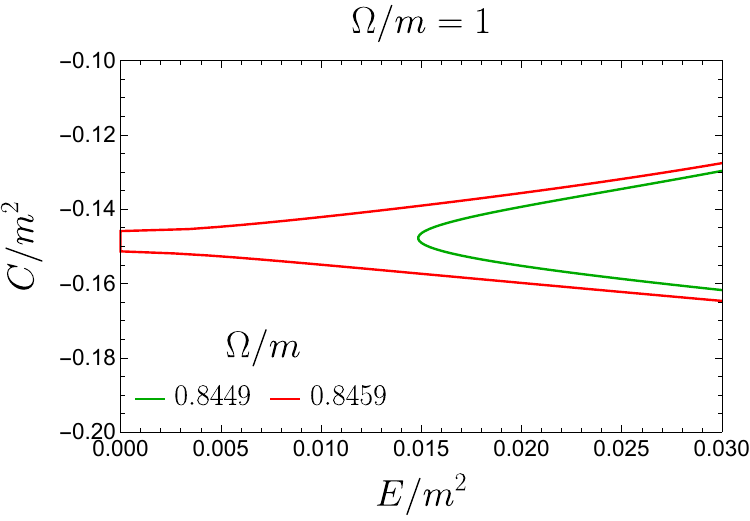}
    \end{subfigure}
    
    \centering
    \begin{subfigure}[t]{0.48\textwidth}
        \centering
        \includegraphics[width=\textwidth]{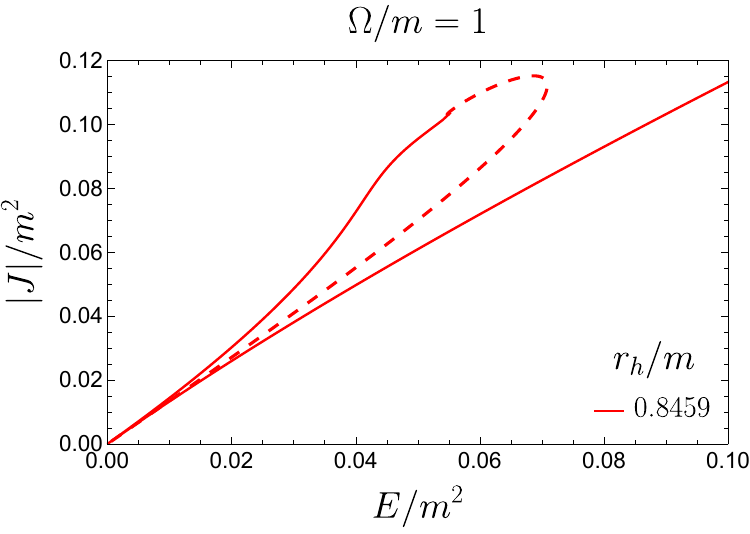}
    \end{subfigure}
    \hspace{0.02\textwidth}
    \begin{subfigure}[t]{0.48\textwidth}
        \centering
        \includegraphics[width=\textwidth]{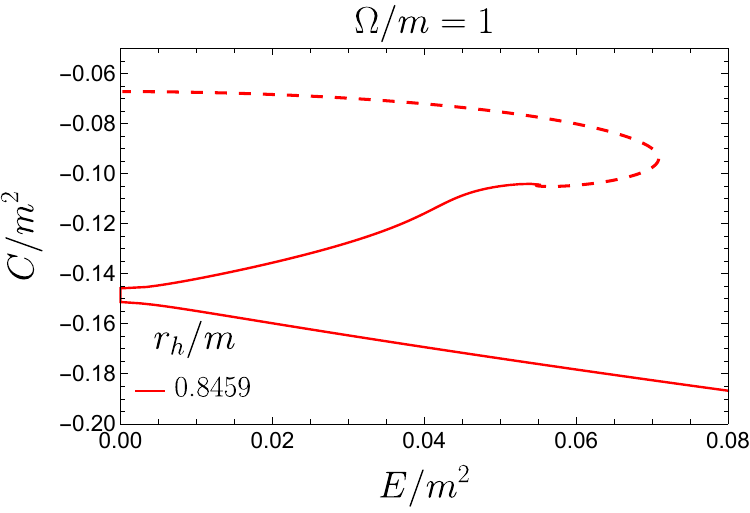}
    \end{subfigure}
    \caption{Electric current (left) and condensate (right) versus $|E|/m^2$ for $\Omega/m=1$. The continuous lines represent BH embeddings, while the dashed ones are the Minkowski embeddings. The upper plots show that for $r_h/m$ around 0.845, we start getting $E/m^2=0$ for BH embeddings. The lower plots are the full curves for the $r_h/m=0.8459$ case. Notice that the electric field increases again, to recover the limit $|J| = \sigma |E|$.}
    \label{fig:JFullCurves}
\end{figure}

When increasing the frequency $\Omega$, all the curves get displaced towards the left and, at the value $ \Omega_c$, which depends on $r_h/m$, they contact the axis $E/m^2=0$ at a non-zero value for both $|J|/m^2$ and $C/m^2$ (red curves). Precise computation reveals the contact point to correspond to a critical embedding, as it is evident from the change dashed $\to$ solid at the contact point. This is the first zero in Fig. \ref{fig:LobesEJD3D5}. Further increase in $\Omega$ makes this contact point slide down the vertical axis, now inside the Minkowski branch. Eventually, it reaches zero, merging again with the trivial Minkowski embedding with  $|E|=|J| = 0$. All the embeddings having $|J|/m^2\neq 0$ with $|E|/m^2=0$ build the manifold of \textit{vector meson  Floquet condensates} \cite{Kinoshita_2017}, which exist for $\Omega_c\le\Omega\le \Omega_{meson}$, as already mentioned before.

When $r_h/m\simeq 0.84$ we start seeing an interference between the two horizons as they come close together. The effect is a multivaluedness in the curve of black hole embeddings (solid lines) that precludes the monotonic growth of $|J|/m^2$ and $|C|/m^2$ with $|E|/m^2$ that is seen at lower temperatures. In all cases, in the large-field regime, the Ohmic behavior $|J| = \sigma |E|$, with constant $\sigma$, is reached. We prove this in Appendix \ref{app:analyticsols}, where we find analytic solutions for the fields in this limit. However, in the small interval  $r_h/m \in (0.84, 0.8897)$, this regime is not approached monotonically, and we find a multivaluedness of $|J|/m^2$ and $C/m^2$ as functions of $|E|/m^2$. This looks similar to the multivaluedness encountered in \cite{Ishii_2018} within the superconducting phase. We, however, encounter this multivaluedness in the conducting phase (the normal phase there). In Fig. \ref{fig:JFullCurves}, we see the interference effect between the two nearby horizons is so effective that a trivial configuration with  $|J|=|E|=0$  is again attained, but now inside the branch of black hole embeddings.

\begin{figure}[H]
    \centering
    \begin{subfigure}[t]{0.48\textwidth}
        \centering
        \includegraphics[width=\textwidth]{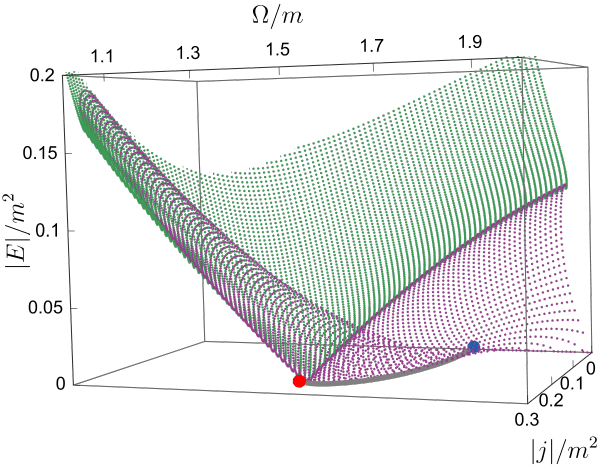}
    \end{subfigure}
    \hspace{0.02\textwidth}
    \begin{subfigure}[t]{0.48\textwidth}
        \centering
        \includegraphics[width=\textwidth]{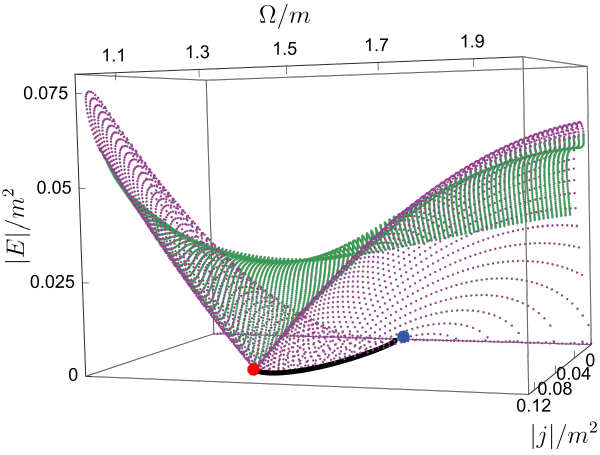}
    \end{subfigure}
    
    \centering
    \begin{subfigure}[t]{0.48\textwidth}
        \centering
        \includegraphics[width=\textwidth]{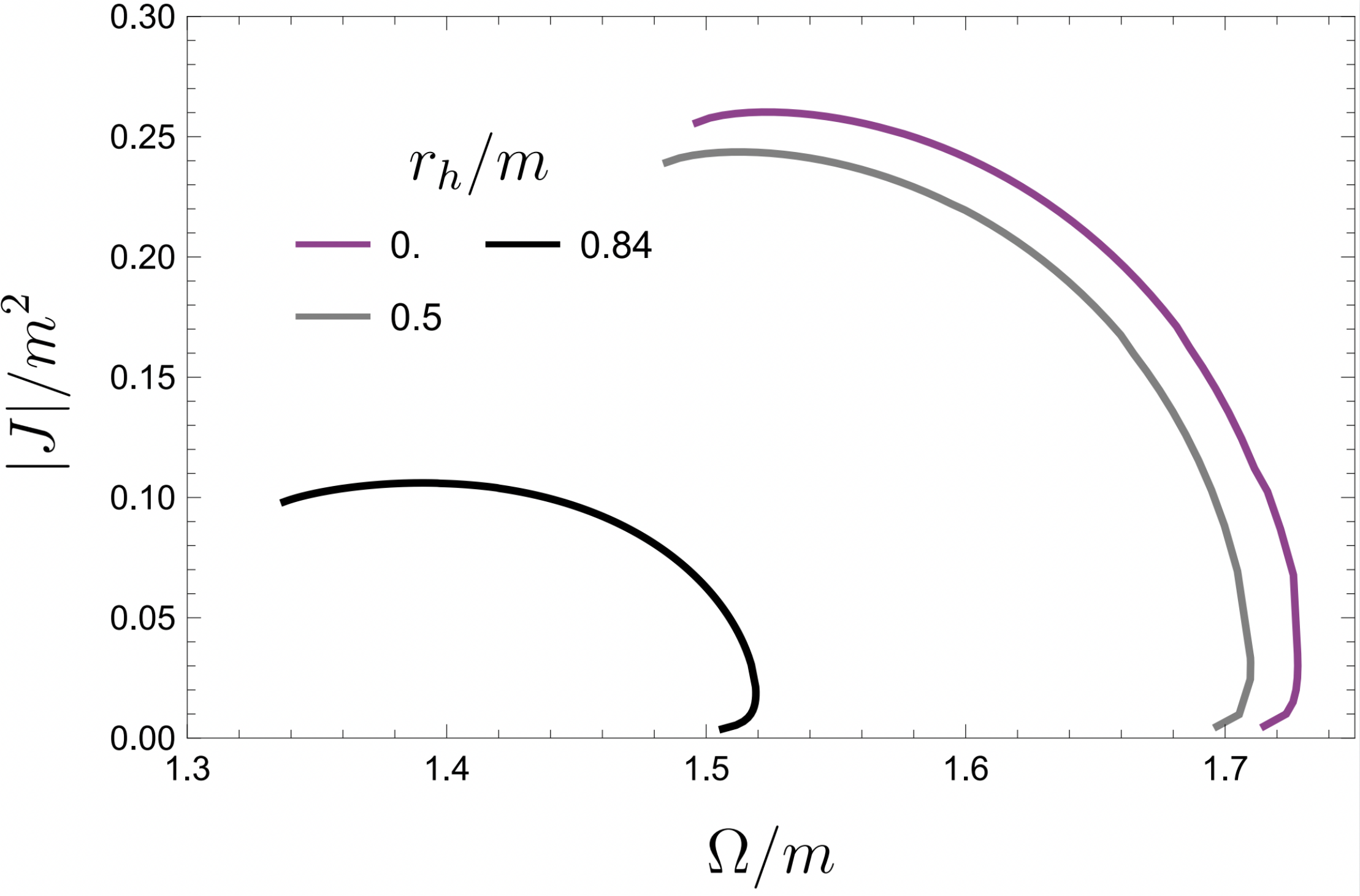}
    \end{subfigure}
    \hspace{0.02\textwidth}
    \begin{subfigure}[t]{0.48\textwidth}
        \centering
        \includegraphics[width=\textwidth]{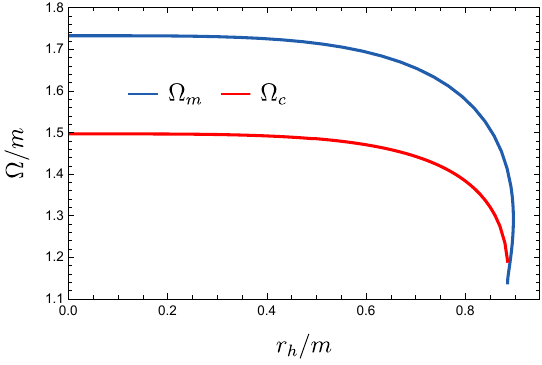}
    \end{subfigure}
    \caption{\textit{Top:} 3D development of the lobed curves in Fig. \ref{fig:LobesEJD3D5} in the vicinity of the first critical point, with an extra $\Omega$ axis, for $r_h/m=0.5$ (left) and $r_h/m=0.84$ (right). Purple (green) surfaces belong to the Minkowski (BH) phase.  Beware the difference in vertical scales. The gray thick lines represent the $|E|=0$ vector meson Floquet condensates, from $\Omega_c$ (red dot) to $\Omega_{meson}$ (blue dot). \textit{Lower left:} vertical view of the upper left plot, where the gray lines in that plot have been graphed for three different temperatures. \textit{Lower right:}  movement of the two extreme points in the curve as a function of $r_h$. They are the critical embedding and the first meson mass frequency respectively. We see that the effect of the background temperature is  mild (less than 10 \%) until the value $r_h/m \sim 0.8$ is reached.}
    \label{fig:3D}
\end{figure}

In Fig. \ref{fig:3D} we zoom again in the region near the first resonance. We have promoted the frequencies in each of the plots in Fig. \ref{fig:JCvsE} to a third $\Omega$ axis, where the associated curves of \ref{fig:JCvsE} are sections of a 3D surface. The surface "touches" the bottom plane $|E|/m^2=0$ in a curve which is the full manifold of vector meson Floquet condensates. This curve interpolates between two endpoints. On one end, $|J|/m^2\neq0$ (red dot), corresponding to $\Omega_c$. On the other end (blue dot), $|J|/m^2\neq0$ and we find the frequency corresponding to the mass of the first vector meson condensate, \ie, the fluctuations in the probe brane worldvolume gauge field \cite{Arean_2006}. As mentioned before, at non-zero temperature these masses have to be found numerically, which we do in Section \ref{sec:mesons} and check they correspond to this point (see Fig. \ref{fig:mesons} for the first three meson masses). Turning on $r_h$ causes an overall shift of these curves ($\Omega_c$ and $\Omega_{meson}$) towards lower values of $\Omega$, which can be inferred from the movement of the extreme points as shown on the lower right plot in Fig. \ref{fig:3D}.

On general grounds, the influence of the temperature on the results at $r_h/m=0$ is small until we approach the maximum temperature $r_h/m = 0.8897$, when the lobes in Fig. \ref{fig:LobesEJD3D5} are very small. Regarding the upper plots in Fig. \ref{fig:3D}, notice how the manifold of black hole embeddings folds down for the right plot $r_h/m=0.85$, in contrast with the monotonic growth on the left one, at $r_h/m=0.5$. This is precisely the multivaluedness remarked for higher temperatures inside the branch of BH embeddings, also observed in Fig. \ref{fig:JCvsE}.

To finish this section, let us comment on the possibility of accurately locating the first-order phase transitions where the phase space curves become multivalued. The usual prescription of comparing the free energies is valid in thermodynamical equilibrium. The usual holographic prescription that proposes the euclidean gravitational action for such a construct is not working properly in the present context of a non-equilibrium steady state (see Section \ref{sec:problemsnoneq} for details, and also \cite{Ishii_2018} for similar concerns and \cite{Kundu_2019} for a review on the topic). A more sophisticated approach using techniques tailored for non-equilibrium open systems as applied to the holographic context is an interesting project to carry out also here. Eventually, an exact dynamical simulation with a slowly varying $|E|/m^2$ should be the right thing to do.

\subsection{Floquet supression points}\label{subsec:suprpoints}
In Fig. \ref{fig:LobesEJD3D5}, on the right plot, we already mentioned the presence of points within the line of critical embeddings where the current vanishes $\abs{J}=0$, even in the presence of a non-zero electric field. They roughly coincide with the points where the electric field becomes maximal within the same family. We will term these points \textit{Floquet suppression points} and the corresponding states \textit{Floquet-suppressed states}. As we will show, the existence of these points extends to the Minkowski embeddings and, in a sense to be explained in the next section, also to the black hole embeddings. 
Focusing on critical and Minkowksi embeddings, we have already mentioned that the current has its origin in the polarizability, $\tilde\pi$, of the vacuum, $\mathcal{P} = \tilde\pi \mathcal{E}$, with $\mathcal{P}= \langle \bar{\psi} (\gamma_x +i\gamma_y) \psi \rangle$ \cite{Kinoshita_2017}. Hence $\mathcal{J} = \dot{\mathcal{P}} =i\Omega \tilde\pi \mathcal{E}$ and a vanishing value of $\mathcal{J} = 0$ implies $\tilde\pi = 0$, i.e.  the polarizability is \textit{dynamically} suppressed.

This suppression of the vacuum polarizability for certain frequencies is similar to  well known (and searched for) dynamical effects in other examples of Floquet engineering. For example, in periodically driven lattices, hopping between neighbouring sites, although present in the bare hamiltonian, can be completely suppressed by tuning the ratio of frequency to amplitude, leading to induced dynamical localization (see \cite{Bukov_2014,Oka_2018,Giovannini_2019} for references).

\begin{figure}[H]
    \centering
    \begin{subfigure}[t]{0.48\textwidth}
        \centering
        \includegraphics[width=\textwidth]{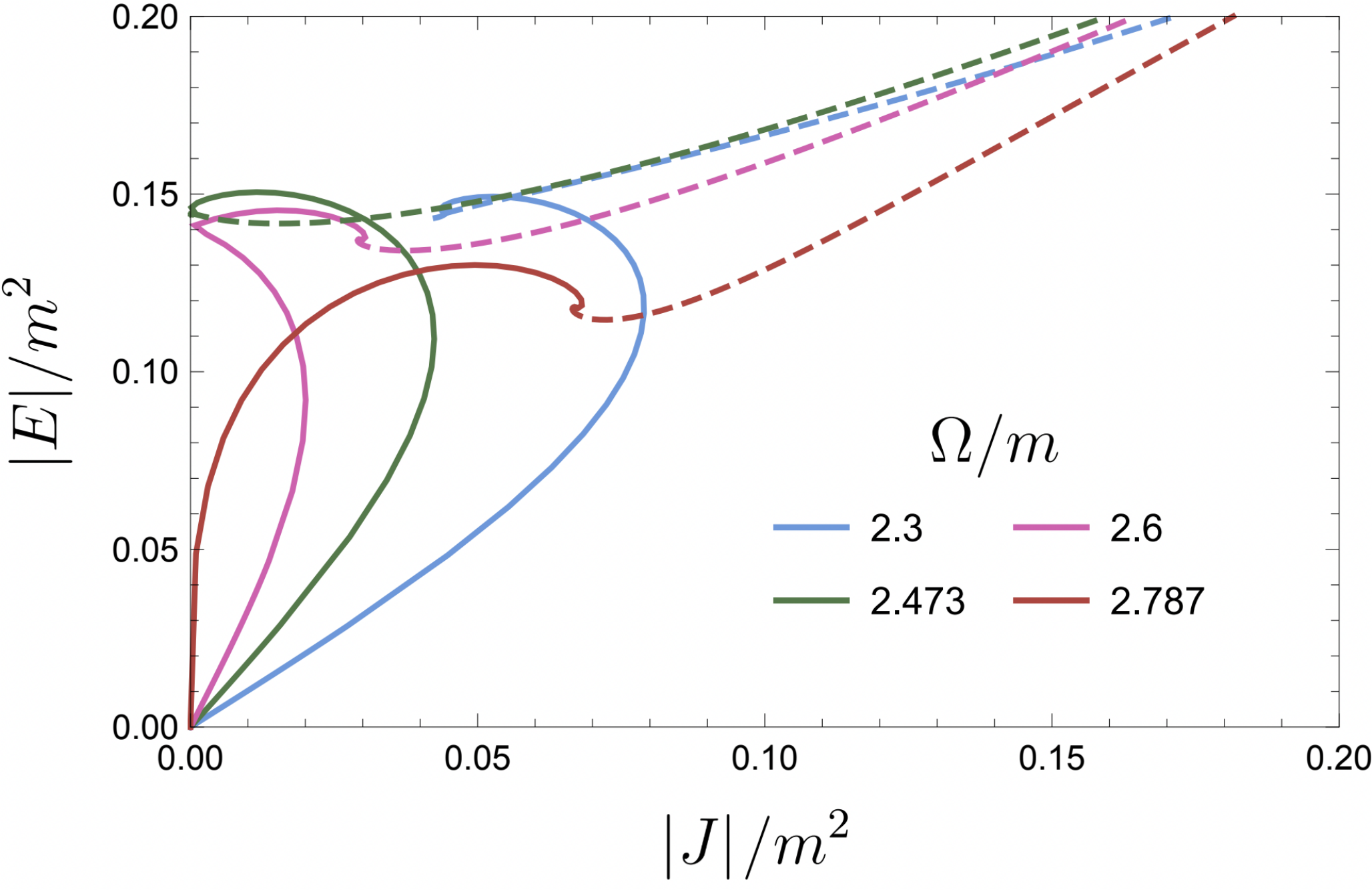}
    \end{subfigure}
    \hspace{0.02\textwidth}
    \begin{subfigure}[t]{0.48\textwidth}
        \centering
        \includegraphics[width=\textwidth]{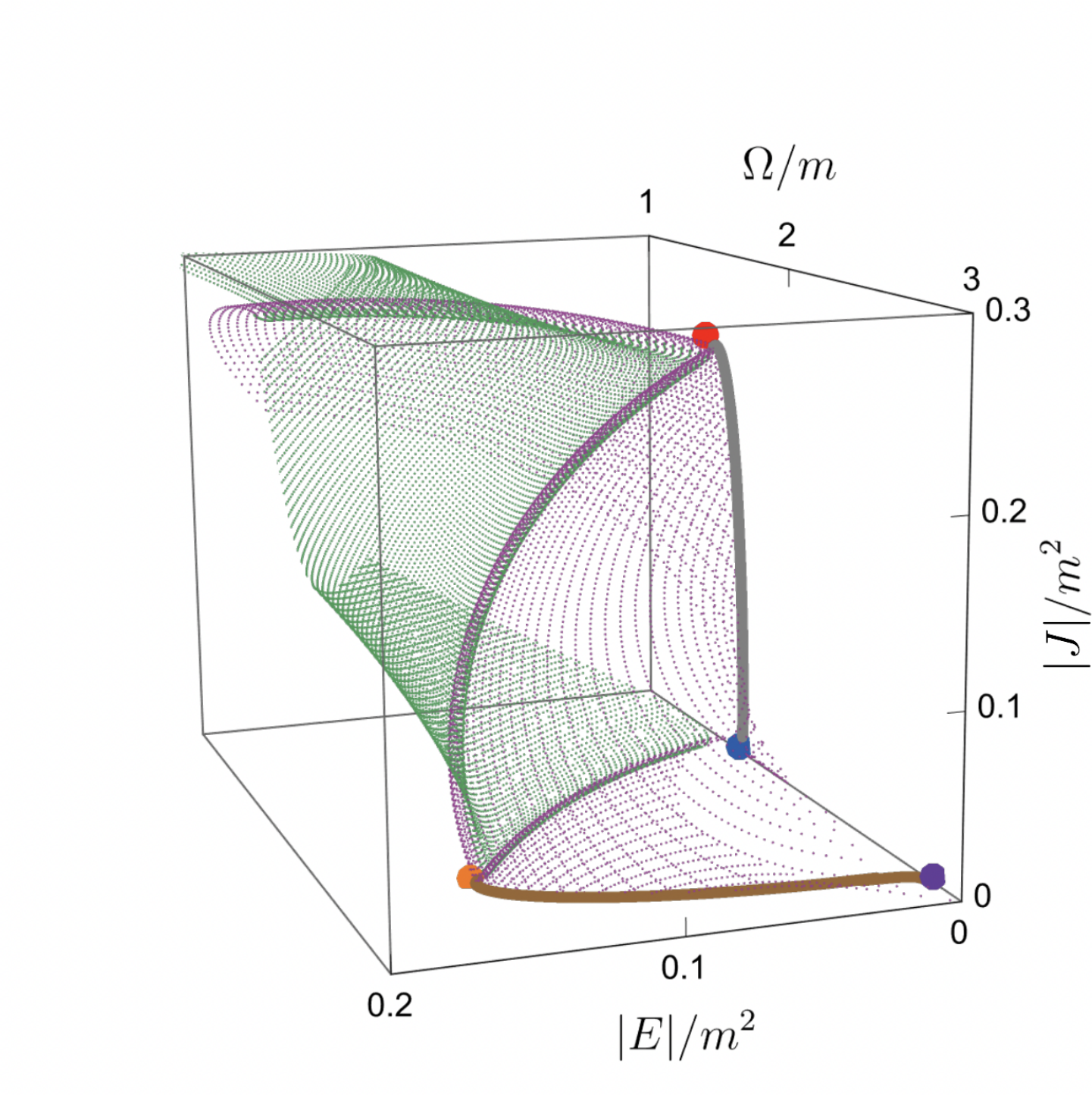}
    \end{subfigure}
    
    \centering
    \begin{subfigure}[t]{0.48\textwidth}
        \centering
        \includegraphics[width=\textwidth]{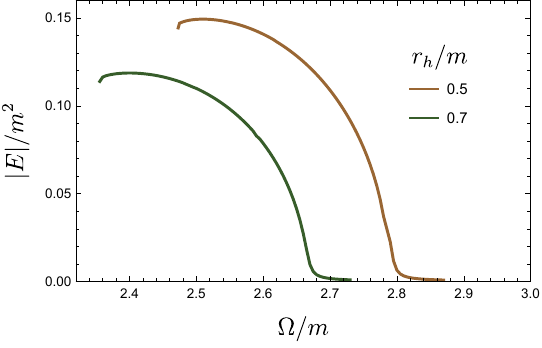}
    \end{subfigure}
    \hspace{0.02\textwidth}
    \begin{subfigure}[t]{0.48\textwidth}
        \centering
        \includegraphics[width=\textwidth]{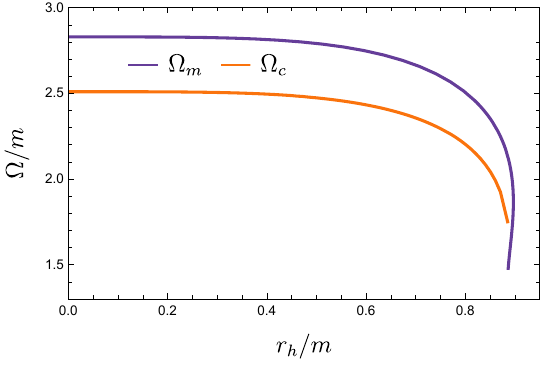}
    \end{subfigure}
    \caption{\textit{Top:} On the left, plots of $|J|/m^2$ vs. $|E|/m^2$, where we exchanged the axes to make apparent the striking similarity with the plots in Fig. \ref{fig:JCvsE}. On the right, 3D plot where the range of $\Omega$ has been extended to cover the first vector meson Floquet condensate (grey curve at $|E|=0$, now vertical), as well as the first Floquet suppressed condensate (brown curve at $|J|=0$, horizontal). \textit{Bottom:} The left plot shows view of the $|J|=0$ plane of the upper plots. The suppressing effect of the temperature is apparent. The right plot is the downshift in $\Omega$ of the two extreme points (orange and violet) in the plot above this, as a function of $r_h/m$.}
    \label{fig:3Dj}
\end{figure}

Fig. \ref{fig:JCvsE} was built by scanning frequencies $\Omega \in (1.2, 1.7)$, \ie, around the point of the first vector meson Floquet condensate. In Fig. \ref{fig:3Dj} we show, on the upper left plot, the similar curves setting instead $\Omega \in (2.3,2.7)$, that is, in a small interval around the first Floquet suppression point. The result exhibits a remarkable similarity, but with $|J|/m^2$ and $|E|/m^2$ axes exchanged. Indeed the symmetry is not exact, as can be seen by comparing the curves in the lower left plots in Figs. \ref{fig:3D}  and \ref{fig:3Dj}. These are, respectively, the curves of vector meson Floquet condensates and Floquet suppression points. In the same way as for the vector meson condensates, the frequencies at the endpoints corresponding to the Minkowski branch can be obtained by studying the linearized fluctuations of the worldvolume gauge field, subject to the boundary condition $|J|=0$, in this case. At zero temperature this calculation can be performed analytically giving (see \cite{Garbayo_2020} eqs. (C.2) and (C.3), and Section \ref{sec:mesons} here)
\begin{equation}
    \abs{J} = 0 \quad \to \quad\Omega_k/m  = 2\sqrt{k(k+1)}=2.828,~4.899,~...
\end{equation}
This is the "dual" result of \eqref{eq:mesonfreqs}.

Upon rising the temperature, $r_h>0$, these quantities get shifted downwards, as shown on the lower right plots in Figs. \ref{fig:3D} and \ref{fig:3Dj}. The (almost) symmetry between vector meson Floquet condensates and  Floquet suppression points is highlighted on the top right plot of Fig. \ref{fig:3Dj}, where both manifolds have been included within the same 3D development.

It is worth mentioning that the existence of these Floquet suppression points is not restricted to the D3/D5 system, hence is not apparently linked to the dimensionality. For completeness, we devote Appendix \ref{app:D3D7} to the twin version of this section in the context of a D3/D7 scenario. Apart from the discrepancy in the precise numerical values, the global picture is the same. For example, in Fig. \ref{fig:LobesEJD3D7} we reproduce the lobe structure for the D3/D7 system, which is analogous to the one found in Fig. \ref{fig:LobesEJD3D5} for D3/D5. Also the $(|J|/m^2,|E|/m^2)$
curves in Fig. \ref{fig:JvsED3D7} are very similar counterparts of the ones in Figs. \ref{fig:JCvsE} and \ref{fig:3Dj}.

\section{Conductivities}\label{sec:conductivitiesD3D5}

In this section we analyze the response of the system under the externally applied electric fields, focusing on two disctinct but related notions of conductivity. The first, which we refer to as \textit{non-linear conductivity}, has been already encountered in Section \ref{subsec:nonlinearsigma} and characterizes the non-linear relation between the induced current and the applied electric field. It was mentioned there that, in the massless case, the conductivity is a constant for any time-dependent electric field \cite{Karch_2010}. Our results below agree with this behavior.

The second part of our analysis involves optical conductivities, where we introduce a small linearly polarized probe electric field on top of the existing rotating background field. From the response of the system to this small perturbation we are able to extract AC and DC conductivities, following the proposal of \cite{Oka_2010} and mimicking the strategy in \cite{Hashimoto_2016,Garbayo_2020}.

\subsection{Non-linear conductivity}\label{subsec:nonlinearcond}

The relation between the current vector and the electric field vector defines a \textit{rotating current (RC) conductivity}\footnote{Notice  that  in the rotating frame we write $\sigma_{RC}$ as we are dealing here with a single  Fourier component  $\Omega$  of the rotational time dependence. In general, in the lab frame, we would write instead $\mathcal{J}(t)=\int d\tau \sigma_{RC}(\tau) \mathcal{E}(t-\tau)$. Also the use of complex instead of vector notation is implicit.}

\begin{equation}
    \begin{pmatrix} J_x\\ J_y \end{pmatrix}=\begin{pmatrix} \sigma_{xx} & \sigma_{xy} \\ -\sigma_{xy} & \sigma_{xx} \end{pmatrix}\begin{pmatrix} E_x \\ E_y\end{pmatrix}~,
    \label{eq:sigmaRC}
\end{equation}
where the form of the matrix is dictated by rotating symmetry in the $(x,y)$-plane. Inverting these relation yields
\begin{equation}
    \sigma_{xx}=\frac{E_xJ_x+E_yJ_y}{E_x^2+E_y^2}=\frac{\abs{J}}{\abs{E}}\cos\delta~,\qquad\qquad \sigma_{xy}=\frac{E_yJ_x-E_xJ_y}{E_x^2+E_y^2}=\frac{\abs{J}}{\abs{E}}\sin\delta~,
\end{equation}
where $\delta$ is the angle formed by the vectors $(J_x,J_y)$ and $(E_x,E_y)$ in the $(x,y)$-plane. Therefore, in complex notation we can write it as
\begin{equation}
    J=\sigma_{RC}E~,
\end{equation}
where $\sigma_{RC}=\sigma_{xx}-i\sigma_{xy}$ is a complex number which is, itself, a non-linear function of $|E|$ (by rotational symmetry) and $\Omega$.

\begin{figure}
    \centering
    \begin{subfigure}[t]{0.48\textwidth}
        \centering
        \includegraphics[width=\textwidth]{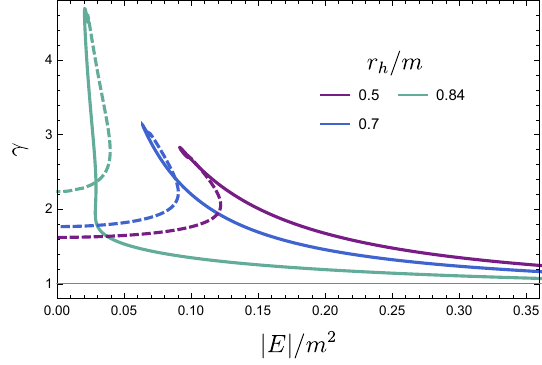}
    \end{subfigure}
    \hspace{0.02\textwidth}
    \begin{subfigure}[t]{0.48\textwidth}
        \centering
        \includegraphics[width=\textwidth]{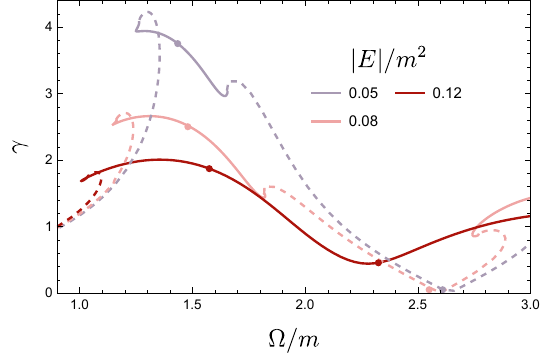}
    \end{subfigure}
    \caption{Modulus of the non-linear conductivity $\gamma = \abs{\tilde\sigma_{RC}}$ as a function of $|E|/m^2$  for fixed $\Omega/m =1.2$ at different temperatures (left plot) or as a function of $\Omega/m$ for fixed values of $|E|/m^2$ at $r_h/m=0.7$ (right plot). The dots are related to the analysis of Fig. \ref{fig:deltaBHrh0p7}. For large $|E|/m^2$ the curves asymptote to the value $\gamma = 1$.}
    \label{fig:NLconduct}
\end{figure}

Writing
\begin{equation}
    \sigma_{RC} = \gamma e^{-i\delta}, 
\end{equation}
the modulus $\gamma$ is the non-linear conductivity, whose value is plotted in Fig. \ref{fig:NLconduct}. The phase $\delta$ encodes the angle between the instantaneous vectors $\vec{J}=(J_x,J_y)$ and $\vec{E}=(E_x,E_y)$. This is why we will refer to $\delta$ as the angle, even if we use complex instead of vector notation. For a given $|E|$ this relative angle controls the Joule heating,
\begin{equation}
    q = \gamma |E|^2 \cos \delta~.
\end{equation}
Its microscopic origin is unclear although we will make an attempt to put forward a consistent picture after we have collected all the bits and pieces.

For Minkowski embeddings, $\vec{J}$ and $\vec{E}$ are perpendicular and $q=0$. This is consistent with the picture of the polarization of the meson condensate into dipoles aligned with the electric field. It leaves two possibilities for $\delta$: $\delta = \pm \pi/2$. In \cite{Kinoshita_2017}, only the positive sign was considered, as it is natural to think that the polarization and the electric field are parallel vectors. We will show here that the existence of both signs is a natural consequence of the presence of Floquet suppression points.

\begin{figure}
    \centering
    \begin{subfigure}[t]{0.48\textwidth}
        \centering
        \includegraphics[width=\textwidth]{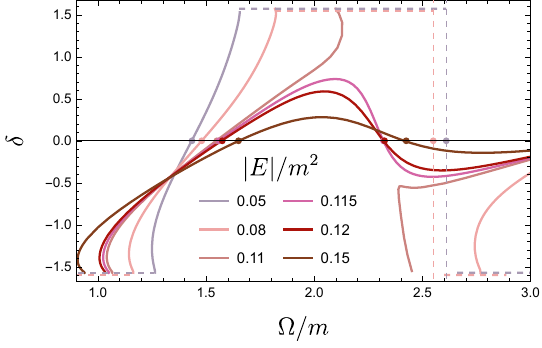}
    \end{subfigure}
    \hspace{0.02\textwidth}
    \begin{subfigure}[t]{0.48\textwidth}
        \centering
        \includegraphics[width=\textwidth]{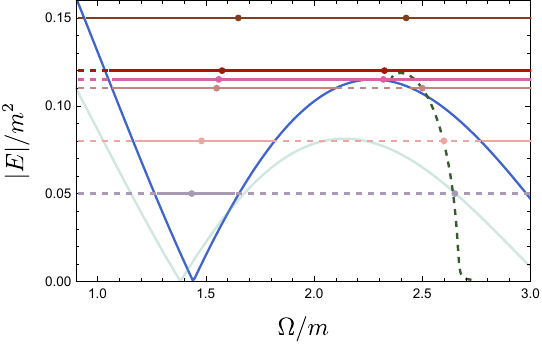}
    \end{subfigure}
    \caption{Continuous lines are black hole embeddings, whereas dashed lines are Minkowski embeddings (with a slight vertical offset for clarity). \textit{Left:} relative angle $\delta$ for different embeddings at various fixed values of the electric field and with $r_h/m=0.7$ for varying $\Omega/m$. \textit{Right:} a zoom of the first lobe region in Fig. \ref{fig:LobesEJD3D5}, where the color code for constant $|E|/m^2$ lines corresponds to the ones on the left plot. The dots indicate the frequencies where the angle $\delta$ either becomes zero in the BH segments, or flips sign in the Minkowski segments. In this later case, joining all the points gives the dashed green curve. The fact that this curve of Minkowski embeddings exits the lower lobe is related to the spiralling multivaluedness of the phase space curves in the vicinity of the critical embeddings. We have added another (dimmed green) lobed curve with higher temperature $r_h/m=0.8$ to show that the effect of rising the temperature is similar to that produced by increasing the electric field $|E|/m^2$.}
    \label{fig:deltaBHrh0p7}
\end{figure}

In Fig. \ref{fig:deltaBHrh0p7} we observe the behavior of the relative angle $\delta$ as we scan embeddings along the horizontal lines of constant $|E|$ while increasing $\Omega$, as shown in the right plot. As usual, solid (dashed) lines correspond to BH (Minkowski) embeddings. On the left plot, using the same color coding, we can see the value of the angle, $\delta$, as we move along these sets of solutions. Notice the jumps $\delta = \pi/2 \to - \pi/2$ that occur within the dashed segment, i.e. for Minkowski embeddings. They seem to reflect a discontinuous transition but this is not the case. Indeed, looking at the right plot in Fig. \ref{fig:NLconduct} we see that, precisely at those points, we find a vanishing value for the module of the polarization current $\gamma=0\to |J|=0$. The corresponding  $\Omega$ frequencies have been signalled with a dot on the right plot in Fig. \ref{fig:deltaBHrh0p7}. Joining all such Floquet suppression points yields the almost vertical green dashed curve of Fig. \ref{fig:deltaBHrh0p7} which is, precisely, the same curve represented in the lower left plot of Fig. \ref{fig:3Dj}, also in (solid) green.

In summary, the transition $\delta = \pi/2 \to - \pi/2$ occurs through a Floquet-suppressed state where the polarizability $\tilde\pi$ vanishes and transits smoothly  from positive to negative. This is remarkable as it states that, for ample intervals in the range of driving frequencies $\Omega$, the polarization of the meson condensate is \textit{antiparallel} to the applied electric field!

Looking back to the left plot in Fig. \ref{fig:deltaBHrh0p7} we notice that the opposite transition $\delta =-\pi/2\to +\pi/2$ is \textit{not} discontinuous. It occurs through a sequence of black hole embeddings that interpolate between those values along a curve that crosses smoothly the axis  $\delta = 0$ with finite slope. A look at the right plot in Fig. \ref{fig:NLconduct} reveals that, in contrast, at those points $\gamma$ stays strictly positive.

Putting  all the information together, the interpretation we find most plausible is as follows: in general, the total current will be a mixture $J = J_{con} + J_{pol}$ of conduction (dissipative) and a polarization (conservative) currents \cite{Hartnoll_2007,Karch_2008}. The precise contribution of each component is controlled by the driving frequency $\Omega$ and by $|E|$. The conduction component $J_{con}$, embodied by deconfined charged carriers, is parallel to the applied electric field. The polarizarion component $J_{pol}$, as explained above, is perpendicular. The vector sum of these two components gives $J$ and $E$ a relative phase angle $\delta$.

Changing $\Omega$ at fixed $|E|/m^2$, like on the left plot in Fig. \ref{fig:deltaBHrh0p7}, we find that $J_{pol}$ vanishes at given frequencies $\Omega$, thereby flipping $\delta = \pm \pi/2\to \mp \pi/2$. In the gapped (Minkowski phase) only this component of the current is present. In the gapless (BH)  phase, both components generically contribute. We interpret  the points where $\delta=0$ as precisely signaling that, there also, $J_{pol}=0$. Thereby the total current becomes parallel to the electric field. This is the reason why the transition in $\delta$ is continuous in the BH phase (solid segments). If this picture makes sense, the conclusion is that \textit{we also have Floquet suppression points within the BH phase}. It is just that in the BH phase this vanishing is masked by the conduction component $J = J_{con}+0$. In summary, all the dots in Fig. \ref{fig:NLconduct} and \ref{fig:deltaBHrh0p7} correspond to Floquet suppressed states. As we approach the boundaries of these segments, the conduction component disappears $J_{con}\to 0$, and the polarization component survives, making $J= J_{pol}$ and $E$ mutually perpendicular again.

For large enough $|E|/m^2$ we always stay within the phase of BH embeddings, and the (solid) curves smoothly relax down to the asymptotic regime where $\delta=0$. We interpret this as the vanishing of the $J_{pol}$ component in this limit. This is the same effect we get for large temperature $T/m\gg 1$ as both are indistinguishable from the limit of small mass $m\to 0$.

In Appendix \ref{app:masslesssol} we prove exactly this fact, $\sigma_{RC}=1$, for massless flavors. This implies that, in this case, the response is both instantaneous and linear. We make contact and fully agree here with the results  in \cite{Karch_2010}. In a sense, the claim  there is stronger as it applies to \textit{any} time dependence of a linearly polarized electric field $\mathcal{E}_x(t)$ at the boundary. Here, on one side, we go to a rotational polarization ansatz and, moreover, in Appendix \ref{app:smallmasssol} we prove this result to hold also at linearized order in a small mass $\delta m$. Linearity of the response entails that it should also extend to arbitrary two-dimensional time dependent electric fields $\vec{\mathcal{E}}(t)$ at linear order in small masses.

\subsection{Photovoltaic optical conductivities}\label{subsec:opticalcond}

The Floquet engineering of an induced Hall effect is termed usually \textit{photovoltaic Hall effect} \cite{Oka_2008}. In \cite{Hashimoto_2016}, following the proposal in \cite{Oka_2010}, the photovoltaic optical response  was obtained for massless charge carriers in the D3/D7 model and an optical Hall current was found. This study was extended in \cite{Garbayo_2020} to massive flavors in the D3/D5 model,  and observed an intricate behavior in the wedge region between the lobes in Fig. \ref{fig:LobesEJD3D5}, with multiple resonance peaks present. The physics in this wedge is presumably controlled by the vector meson Floquet condensate at zero temperature, that signals the presence of a quantum phase transition. In the present work, first, we would like to see how the presence of a temperature affects those results.

In order to study the photovoltaic current of the model, the proposal in  \cite{Oka_2010} is to analyze the response of our system to an additional linearly polarized electric field on top of the circularly driven background \eqref{eq:rotatingE}. In vector cartesian notation, the total electric field is now
\begin{equation}
    \vec{\mathcal{E}}(t)=O(t)\vec{E}+\vec{\epsilon}(t)=O(t)\vec{E}+\vec{\epsilon}\h e^{-i\omega t}~,
\end{equation}
where $\vec\epsilon$ is a constant vector such that $\abs{\vec{\epsilon}}\ll|\vec{E}|$. We want to extract the effective conductivities that arise from the effect of this perturbation on the current. In particular, we expect a change of the form $\vec{\mathcal{J}}(t)\to O(t)\vec J+\delta \vec J(t)$, giving rise to an effective conductivity defined by
\begin{equation}
    \delta\vec{J}(t)=\boldsymbol{\sigma}\cdot\vec{\epsilon}(t)~.
\end{equation}
The goal is to determine $\boldsymbol{\sigma}$.

The perturbation of the electric field will also mean a change in the gauge potential $\vec{a}+\delta \vec{a}$, so that  $c(r)$ develops a time dependent perturbation
\begin{equation}
    \vec{a}(t,r)+\delta \vec{a}(t,r)=O(t)\Big(\vec{c}(r)+\delta\vec{c}(t,r)\Big)~.
\end{equation}
Now the bulk gauge potential $\vec{a}(t,r)+\delta\vec{a}(t,r)$ has to match the full electric field at the boundary,
\begin{equation}
    \vec{a}(t,r=\infty)+\delta \vec{a}(t,r=\infty)=-\frac{1}{\Omega}O(t)\epsilon \vec{E}-\frac{i}{\omega}\vec{\epsilon}\h e^{-i\omega t}~.
    \label{eq:pert}
\end{equation}
Since the fluctuations of the gauge field will also couple to the embedding functions,  $\theta(r)\to\theta(r) + \delta \theta(t,r)$, and we will be dealing with a 3 component vector of fluctuations, $\delta \vec \xi(t,\rho) = (\delta c_x,\delta c_y,\delta \theta)$. The general formalism to study these fluctuations was developed in \cite{Garbayo_2020}, following the analysis of \cite{Hashimoto_2016} for the massless D3-D7 system. We devote Appendix \ref{app:opticalcond}  for the details of the calculation and here we only outline the main results.

The equations for the perturbations can be solved in terms of the matrices $\mathbf{M}_\pm$, defined as
\begin{equation}
    \textbf{M}_\pm = \frac{1}{2}\begin{pmatrix} 1 & \pm i\\ \mp i & 1 \end{pmatrix}~,
\end{equation}
as well as two other matrices, $\boldsymbol{X}_{\pm}$ which, in general, have to be determined numerically. It can be shown that $\delta \vec{\mathcal{J}}$ contains three modes, which oscillate with frequencies $\omega$ and $\omega\pm2\Omega$ (the heterodyning mixing modes \cite{Oka_2016}),
\begin{equation}
    \delta \vec{\mathcal{J}}=\Big[\boldsymbol{\sigma}(\omega)e^{-i\omega t}+\boldsymbol{\sigma}^{+}(\omega)e^{-i(\omega+2\Omega)t}+\boldsymbol{\sigma}^{-}(\omega)e^{-i(\omega-2\Omega)t}\Big]\vec{\epsilon}~,
    \label{eq:sigmasgeneral}
\end{equation}
where $\boldsymbol{\sigma}(\omega)$, $\boldsymbol{\sigma^+}(\omega)$ and $\boldsymbol{\sigma}^-(\omega)$ are the conductiviy matrices corresponding to the frequencies $\omega$, $\omega+2\Omega$ and $\omega-2\Omega$, and are given by
\begin{equation}
\begin{aligned}
    \boldsymbol{\sigma}(\omega) & = -\frac{i}{\omega}\left(\textbf{M}_+\textbf{X}_+\textbf{M}_++\textbf{M}_-\textbf{X}_-\textbf{M}_-\right)~,\\
    \boldsymbol{\sigma}^+(\omega) & = -\frac{i}{\omega}\textbf{M}_-\textbf{X}_-\textbf{M}_+~,\\
    \boldsymbol{\sigma}^-(\omega) & = -\frac{i}{\omega}\textbf{M}_+\textbf{X}_+\textbf{M}_-~.
\end{aligned}
\end{equation}

\begin{figure}
    \centering
    \begin{subfigure}[t]{0.48\textwidth}
        \centering
        \includegraphics[width=\textwidth]{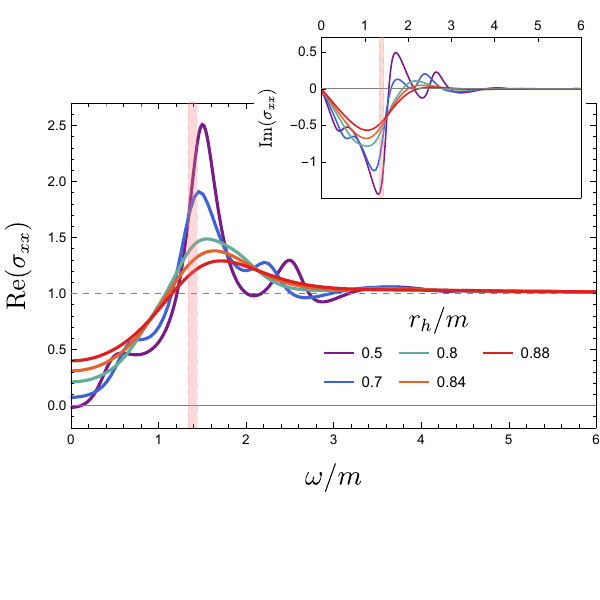}
    \end{subfigure}
    \hspace{0.02\textwidth}
    \begin{subfigure}[t]{0.48\textwidth}
        \centering
        \includegraphics[width=\textwidth]{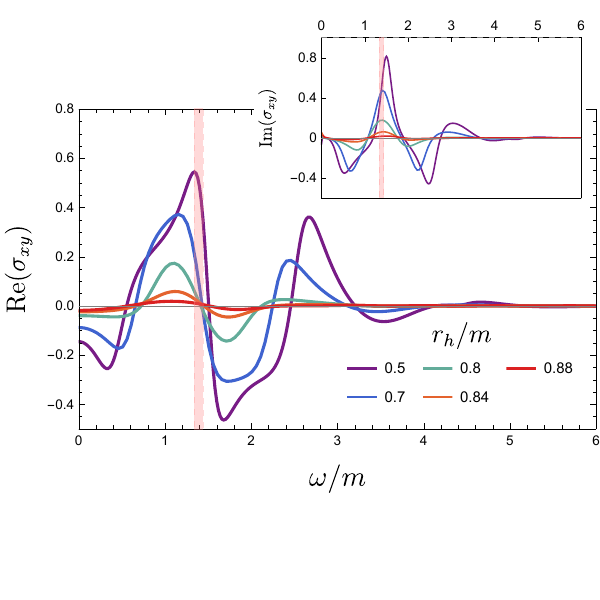}
    \end{subfigure}
    
    \centering
    \begin{subfigure}[t]{0.48\textwidth}
        \centering
        \includegraphics[width=\textwidth]{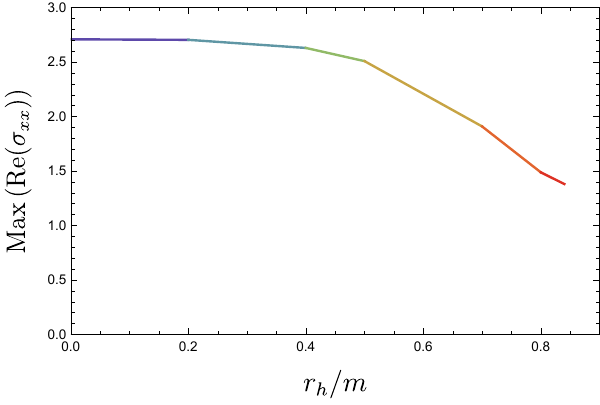}
    \end{subfigure}
    \hspace{0.02\textwidth}
    \begin{subfigure}[t]{0.48\textwidth}
        \centering
        \includegraphics[width=\textwidth]{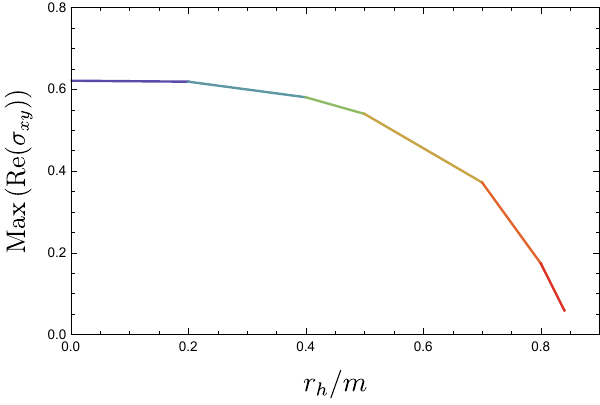}
    \end{subfigure}
    \caption{\textit{Top:} AC conductivities for four values of $r_h/m$ at their corresponding $\Omega_c(r_h)$, for $E/m^2=0.1$. The pink band marks the range where the four values of $\Omega_c(r_h)$ belong. The main peak is close to this region. \textit{Bottom:} Variation with the temperature of the maximum value of the real part of $\boldsymbol{\sigma}_{xx}$ and $\boldsymbol{\sigma}_{xy}$ for $E/m^2=0.1$.}
    \label{fig:sig_rh_Omegac}
\end{figure}

See Appendix \ref{app:opticalcond} for details. The results are contained in in Fig. \ref{fig:sig_rh_Omegac}. The curves represent the absortion spectrum $\boldsymbol{\sigma}_{xx}(\omega)$ and the Hall conductivity $\boldsymbol{\sigma}_{xy}(\omega)$. The background rotating electric field has been fixed to $|E|/m^2=0.1$. Its frequency has been set to the first critical frequency $\Omega_c(r_h)$, which decreases with the temperature as seen in the lower right plot of Fig. \ref{fig:3D}. The band in which these values lie for the chosen temperatures has been signaled by a vertical band in pink. The curves for $\boldsymbol{\sigma}_{xx}$ and $\boldsymbol{\sigma}_{xy}$ show a smooth deformation of the ones in \cite{Garbayo_2020} (Fig. 10) for the same value of $|E|/m^2$. 

Succinctly stated, the temperature in the gluon bath in general destroys the AC Hall optical conductivity, and hence, also the DC Hall conductivity. Again, the effect of the temperature is similar to the one caused by an increase of the electric field. Namely, the conductivity peaks get roughened and lowered. In a sense, both agents, the temperature and the electric field, act in parallel by enhancing the amount of deconfined charge carriers in the medium.

\begin{figure}
    \centering
    \begin{subfigure}[t]{0.48\textwidth}
        \centering
        \includegraphics[width=\textwidth]{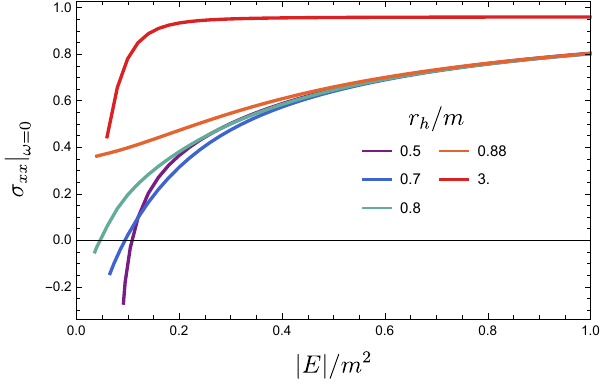}
    \end{subfigure}
    \hspace{0.02\textwidth}
    \begin{subfigure}[t]{0.48\textwidth}
        \centering
        \includegraphics[width=\textwidth]{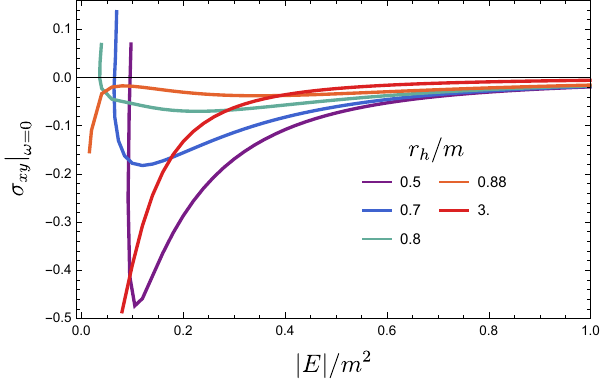}
    \end{subfigure}
    \caption{DC conductivities as functions of $|E|/m^2$ for different values of $r_h/m$ at $\Omega/m=1$. At high temperatures the conductivity tensor tends towards the identity $\boldsymbol{\sigma}_{xx} =1$ and $\boldsymbol{\sigma}_{xy} = 0$.}
    \label{fig:sigDC_rh_Omegac}
\end{figure}

The effect becomes more pronounced beyond some temperature $r_h/m \sim 0.5$, as shown in the lower plots of Fig. \ref{fig:sig_rh_Omegac}. In the large-temperature limit, $r_h \to \infty$, all conductivities, both AC and DC  tend towards $\boldsymbol{\sigma}_{xx} = 1~,~\boldsymbol{\sigma}_{xy} = 0$ (see also Fig. \ref{fig:sigDC_rh_Omegac}). This result is the same we obtained for the rotating current conductivity $\sigma_{RC}$ in the massless limit. Since in this case the electric field is linearly polarized, rather than circularly, we see this as a further evidence in favor of the fact that the response will be Ohmic and instantaneous for an arbitrary time dependence of the electric field in the plane, $J(t) = \sigma E(t)$.

A peculiar observation is that the frequencies $\omega$ of the highest peaks in the absorption spectrum $\Re(\boldsymbol{\sigma}_{xx})$ slightly deviate above the one of the driving $\omega \gtrapprox \Omega$ (within the vertical band in pink). This was also observed in \cite{Garbayo_2020} (Fig. 10), where the drift is seen to be enhanced with increasing $|E|$. We could not offer any explanation to this. Here we can see that there is a very similar shift in the driving frequencies $\Omega$ of the Floquet suppression points inside the BH phase (solid segments in Fig. \ref{fig:deltaBHrh0p7}), with increasing $|E|$ towards the right. We have tried to make sense of this qualitative coincidence but could not find an exact numerical agreement.

\subsubsection{Massless limit}

It was shown in \cite{Garbayo_2020} that, in the massless case, the perturbed system can be solved analytically. Remarkably, the same happens at finite temperature.

In the massless case, the fluctuations of the embedding function decouple from those of the gauge field $\delta \vec{c}$. The unperturbed configuration corresponds to the massless analytic solution obtained in Appendix \ref{app:masslesssol}. Surprisingly, the coupled equations for $\delta c_x$ and $\delta c_y$ can be solved analytically, obtaining the matrices
\begin{equation}
    \textbf{X}_\pm=i\omega\h \mathbf{1}~.
\end{equation}
Plugging these $\mathbf{X}_\pm$ matrices in \eqref{eq:sigmasgeneral}, we obtain the conductivity matrices in the massless case, namely
\begin{equation}
    \boldsymbol{\sigma}(\omega)  =\mathbf{1}~,\qquad\qquad  \boldsymbol{\sigma}^+(\omega)  = \boldsymbol{\sigma}^-(\omega)=0~,
\end{equation}
which is exactly the same result as the one found in \cite{Garbayo_2020} at zero temperature. As in the case of the non-linear current, after correctly normalizing the electric field and current according to \eqref{eq:dictionaryD3D5}, we obtain that the conductivity is given by
\begin{equation}
    \boldsymbol{\sigma}(\Omega)=\frac{2N_f N_c}{\pi\sqrt{\lambda}}\mathbf{1}~,
\end{equation}
where here also the result confirms the expectations put forward in \cite{Karch_2010}. The details of this analytic solution are relegated to Appendix \ref{app:masslessconduc}.

\section{Linearized Minkowski embeddings and meson spectrum}\label{sec:mesons}

We have seen that two different values of the driving frequency play an important role around each resonance. On one hand, the critical frequency $\Omega_c$, at one endpoint of the line of vector meson Floquet condensates, and $\Omega_{meson}$, at the other endpoint (see Fig. \ref{fig:3D}). We have claimed that this endpoint, $\Omega_{meson}$ corresponds to the frequencies of the vector mesons of the theory, obtained from the quadratic scalar fluctuations on the brane and written in \eqref{eq:mesonfreqs}. In this section we want to prove this claim.

This observation was proven in \cite{Kinoshita_2017} for the D3/D7 model, and later proven in \cite{Garbayo_2020} for the D3/D5 model at zero temperature. The zero-temperature result is analytic and we begin this section by reviewing it. Later on we confirm that this is still true at finite temperature by numerically computing the meson spectrum $\Omega_{meson}$ at different temperatures and comparing with the endpoints of the obtained vector meson condensates.

\subsection{Zero temperature: analytic results}

For this analysis it is convenient to switch to the cartesian coordinates $(\rho,w)$. The analysis of \cite{Garbayo_2020} considered first the linearized equations for the gauge field $c(\rho)$ around $c=0$\footnote{This is indeed the adequate regime to consider because the line of vector meson condensates has $E=0$, and moreover its endpoint has $J=0$.}, and for the embedding around $w=1$. The linearized equation for the gauge field turns out to be
\begin{equation}
    c''+\frac{2}{\rho}c'+\frac{\Omega^2}{(1+\rho^2)^2}\h c=0~,
\end{equation}
whose solution is
\begin{equation}
    c(\rho)=c_1g_1(\rho)+c_2g_2(\rho)~,
\end{equation}
where $c_1$ and $c_2$ are constants and $g_1(\rho)$ and $g_2(\rho)$ are the functions
\begin{equation}
\begin{aligned}
    g_1(\rho)&=\frac{\sqrt{1+\rho^2}}{\rho}\sin\left(\sqrt{1+\Omega^2}\arctan\rho\right)~,\\
    g_2(\rho)&=\frac{\sqrt{1+\rho^2}}{\rho}\cos\left(\sqrt{1+\Omega^2}\arctan\rho\right)~.
\end{aligned}
\end{equation}

The UV expansions of the functions $g_1(\rho)$ and $g_2(\rho)$ are, respectively,
\begin{equation}
\begin{aligned}
    g_1(\rho) & = \sin\left(\frac{\pi}{2}\sqrt{1+\Omega^2}\right)-\sqrt{1+\Omega^2}\cos\left(\frac{\pi}{2}\sqrt{1+\Omega^2}\right)\frac{1}{\rho}+...~,\\
    g_2(\rho) & = \cos\left(\frac{\pi}{2}\sqrt{1+\Omega^2}\right)+\sqrt{1+\Omega^2}\sin\left(\frac{\pi}{2}\sqrt{1+\Omega^2}\right)\frac{1}{\rho}+...~,
    \label{eq:UVlinearized}
\end{aligned}
\end{equation}
while near the axis $\rho=0$ they behave as
\begin{equation}
\begin{aligned}
    g_1(\rho) & = \sqrt{1+\Omega^2}-\frac{\Omega^2\sqrt{1+\Omega^2}}{6}\rho^2+...\\
    g_2(\rho) & = \frac{1}{\rho}-\frac{\Omega^2}{2}\rho+...~.
\end{aligned}
\end{equation}

Since Minkowski embeddings are regular around $\rho=0$, regularity forces $g_2(\rho)$ to vanish. In this case, the solution is simply
\begin{equation}
    c(\rho)=c_1\frac{\sqrt{1+\rho^2}}{\rho}\sin\left(\sqrt{1+\Omega^2}\arctan\rho\right)~.
\end{equation}
From the UV expansion \eqref{eq:UVlinearized} we can read the electric field and current as
\begin{equation}
\begin{aligned}
    E&=-i\h c_1\Omega\sin\left(\frac{\pi}{2}\sqrt{1+\Omega^2}\right)~,\\
    J&=-c_1\sqrt{1+\Omega^2}\cos\left(\frac{\pi}{2}\sqrt{1+\Omega^2}\right)~.
\end{aligned}
\end{equation}
To find the resonant frequencies of the Floquet condensates we impose that $E=0$, which means that
\begin{equation}
    \sqrt{1+\Omega_n^2}=2(n+1)~,\qquad \text{for}\quad n=0,1,2,...~,
\end{equation}
which corresponds precisely to the masses of the vector mesons in the D3/D5 model, \cite{Arean_2006}, written in \eqref{eq:mesonsD3D5} and \eqref{eq:mesonfreqs}.

\subsection{Finite temperature: numerical results}

At finite temperature we need to find the frequencies numerically. We begin by writing the DBI action in the coordinates $(\rho,w)$ for the fields $w$ and $c$:
\begin{equation}
\begin{aligned}
    S_{D5}&=-\mathcal{N}\int d\rho \frac{\rho^3}{(\rho^2+w^2)^3}\Bigg[(\rho^2+w^2)^2\left((\rho^2+w^2)^2-u_h^4\right)^2\abs{c'}^2-\Omega^2(\rho^2+w^2)^4\Re(c\bar{c}')^2+\Bigg.\\
    &\phantom{-}\Bigg.+\left((\rho^2+w^2)^2+u_h^4\right)\left(\left((\rho^2+w^2)^2-u_h^4\right)^2-\Omega^2(\rho^2+w^2)^2\abs{c}^2\right)(1+w'^2)\Bigg]^{1/2}~.
    \label{eq:SD5cartesian}
\end{aligned}
\end{equation}

Let us consider the case in which the gauge field $c(\rho)$ vanishes, and let $\tilde w=\tilde w(\rho)$ denote the Minkowski embedding function in such case. It can be obtained by solving the equation of motion of the action $\tilde{S}_{D5}=S_{D5}(c=0)$,
\begin{equation}
    \tilde{S}_{D5}=-\mathcal{N}\int d\rho\frac{\rho^2\left(u_h^4-(\rho^2+\tilde{w}^2\right))}{(\rho^2+\tilde{w}^2)^3}\sqrt{\left[(\rho^2+\tilde{w}^2)^2+u_h^4\right](1+\tilde{w}'^2)}
\end{equation}

Let us  next suppose that we perturb around the $c=0$ solution by making $c\to\delta c$ and $w\to \tilde{w}+\delta w$ in the equations of motion derived from the Lagrangian density \eqref{eq:SD5cartesian}. It is easy to see that, at first order in the variations, the equation for $\delta c$ reads
\begin{align}
\begin{split}
    \partial_\rho\Bigg(&\frac{\rho^2\left(u_h^4-(\rho^2+\tilde{w}^2)^2\right)}{(\rho^2+\tilde{w}^2)\sqrt{\left((\rho^2+\tilde{w}^2)^2+u_h^4\right)(1+\tilde{w}'^2)}} \delta c' \Bigg)\\
    & \hspace{0.25\textwidth}+ \frac{\rho^2 \Omega^2\sqrt{(\rho^2+\tilde{w}^2)^2+u_h^4}}{(\rho^2+\tilde{w}^2)\left(u_h^4-(\rho^2+\tilde{w}^2)^2\right)\sqrt{1+\tilde{w}'^2}}\delta c = 0~.
    \label{eq:eomdeltac}
\end{split}
\end{align}

\begin{figure}
    \centering
    \begin{subfigure}[t]{0.48\textwidth}
        \centering
        \includegraphics[width=\textwidth]{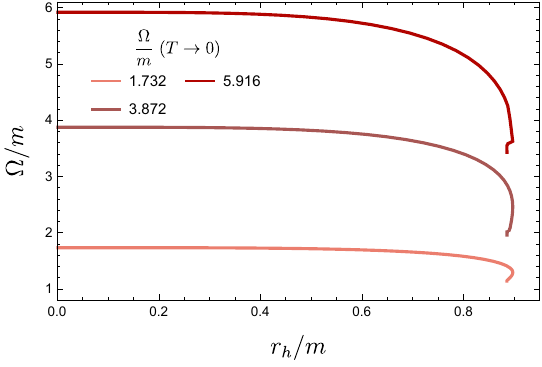}
    \end{subfigure}
    \hspace{0.02\textwidth}
    \begin{subfigure}[t]{0.48\textwidth}
        \centering
        \includegraphics[width=\textwidth]{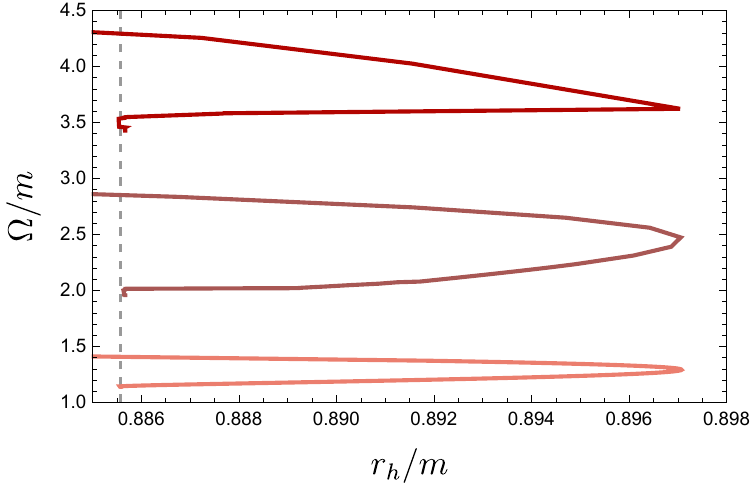}
    \end{subfigure}
    \caption{$\Omega/m$ variation with $r_h/m$. The right figure shows the endpoint of the curves. Both of them are bivaluated between $r_h/m=0.8855$ to $r_h=0.8897$, and stop for $r_h/m=0.8855$.}
    \label{fig:mesons}
\end{figure}

To obtain the resonant frequencies of the mesonic Floquet condensates we integrate \eqref{eq:eomdeltac} and find the solutions that are regular at $\rho=0$ and such that $E=0$ when we reach $J=0$ after each critical driving frequency $\Omega_c/m$ (see the lower plot in Fig. \ref{fig:3D}). This last condition is only possible when the frequency $\Omega/m$ takes values in a discrete set (which depends on the horizon radius $r_h$). For 
$r_h/m=0$, we recover the analytic results reviewed just above. For $r_h/m\neq 0$, the masses of these mesonic states decrease as $r_h/m$ grows. At some value of $r_h/m$ they cease to exist (see Fig. \ref{fig:mesons}), as the mere Minkowski embedding themselves do (see Section \ref{sec:phasespace}). This behavior is dual to the meson melting phenomenon, as discussed in \cite{Mateos_2007}.

\section{Non-equilibrium thermodynamics}\label{sec:problemsnoneq}

Before concluding this Chapter, we briefly comment on the (unsuccessful) application of standard thermodynamic arguments in analyzing the phase structure of our system. A common method for identifying phase transitions involves computing the free energy in both competing phases. This quantity typically develops a characteristic swallow-tail structure, with a crossing point signaling a change in the dominant phase. Our goal is to apply a similar strategy to the insulator/conductor transition in the present model.

In the holographic context, the free energy is given by (minus) the on-shell Euclidean action. For our setup, this is obtained by Wick rotating the original DBI action~\eqref{eq:actionD5}, resulting in
\begin{align}
\begin{split}
    S^E_{D5}=\mathcal{N}\int du \sqrt{1-\psi^2}\Bigg[\left(u^4g^2-\Omega^2b^2\right)\left[\left(1-\psi^2\right)\left(h+b'^2\right)+u^2h\psi'^2\right]\Bigg.\\
    \Bigg.+u^4g^2b^2\left(1-\psi^2\right)\chi'^2\Bigg]^{1/2}~,
    \label{eq:actionD5Eucl}
\end{split}
\end{align}
where the functions $g$ and $h$ are defined in \eqref{eq:ghfunctions}.

To render the action finite, we introduce appropriate counterterms, which are discussed in detail in Appendix~\ref{app:holorenoD3D5}.\footnote{The overall sign differs from that in the appendix due to the Wick rotation to Euclidean signature.} These take the form
\begin{equation}
\begin{aligned}
    S^{ct}_1 & = -\frac{1}{3}\mathcal{N}\sqrt{-\gamma}~,\\
    S^{ct}_2 & = \frac{1}{2}\mathcal{N}\sqrt{-\gamma}\theta(\epsilon)^2~,
\end{aligned}
\end{equation}
where $\gamma$ denotes the determinant of the induced metric on the regulator surface, $z=\epsilon$.

\begin{figure}
    \centering
    \includegraphics[width=0.65\textwidth]{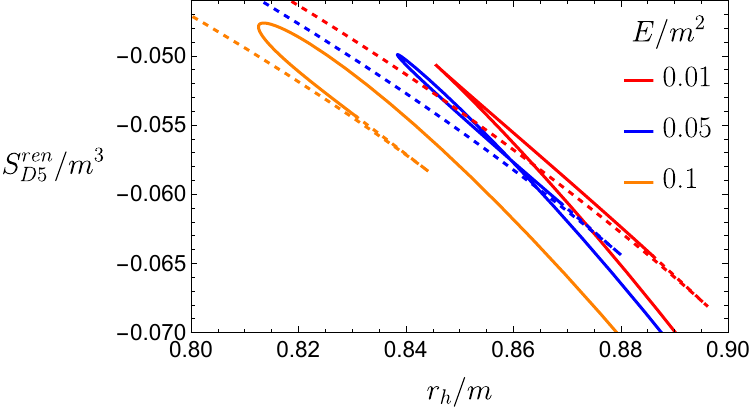}
    \caption{On-shell Euclidean action for different values of the electric field. Solid (dashed) lines correspond to BH (Minkowski) embeddings.}
    \label{fig:OnshellD5}
\end{figure}

The time-independence of the Lorentzian action suggests that the free energy can again be defined as the negative of the renormalized on-shell Euclidean action~\eqref{eq:actionD5Eucl}. Under this definition, one might expect that evaluating the free energy for both black hole and Minkowski embeddings would reveal a swallow-tail structure, characteristic of a first-order phase transition. However, as shown in Fig. \ref{fig:OnshellD5}, this expectation is not borne out.

Several comments are in order. First, the on-shell action involves an integration over the holographic radial coordinate. In the literature, two choices for the infrared cutoff are commonly considered: $u_{min} =u_h$, corresponding to the background horizon, and $u_{min} = u_c$, the critical surface of the brane embedding. This ambiguity is deeply tied to the non-equilibrium nature of the system: when the electric field is absent, the only consistent IR boundary is the background horizon.

The first choice is problematic because the phase $\chi$ diverges logarithmically as $\log (u-u_h)$. This mirrors the well-known divergence in $A_x$ on the case of a constant electric field in the $x$-direction \cite{Karch_2008}. In contrast, the second option $u_{min}=u_c$ provides the infrared regulator adopted throughout this work. This choice is widely regarded as the appropriate one~\cite{Kim_2011,Kundu_2013,Banerjee_2015}, as the singular shell at $u_c$ effectively functions as an event horizon for the open string degrees of freedom (see Sections~\ref{subsec:Teff} and~\ref{subsec:Teffrotating}). For this reason, we adhere to the second prescription.

We see that only for very small electric fields does the free energy curve resemble that of a conventional first-order phase transition. As $|E|/m^2$ increases, however, the behavior deviates significantly from the expected swallow-tail structure. Extending the integration domain to $u_{min} = u_h$ does not cure this problem.

The existence of the singular shell also complicates the interpretation of the on-shell action as a generator of response functions. This can be seen by considering general variations of the Euclidean action. On-shell, the variation of the bulk term~\eqref{eq:actionD5Eucl} yields only boundary contributions:
\begin{equation}
\begin{aligned}
	\delta S^E_{D5}&=\left[\frac{\partial \mathcal{L}}{\partial \psi'}\delta\psi+\frac{\partial\mathcal{L}}{\partial b'}\delta b+\frac{\partial \mathcal{L}}{\partial \chi'}\delta \chi\right]\Bigg\rvert^{z_c}_{z=\epsilon}\\
&=-\mathcal{N}\left[\frac{m\h\delta m}{\epsilon}+m\h\delta C+2C\delta m+\vec{J}\cdot\delta\vec{A}-\frac{q}{\Omega}\delta\chi\big\rvert_{z_c}\right]~,
\end{aligned}
\end{equation}
where the first four terms originate from the boundary at $z=\epsilon$, while the last term arises from the singular shell. Notably, this final contribution vanishes when the electric field is absent.

The variation of the counterterms yields
\begin{equation}
	\delta S^{ct}=\delta S^{ct}_1+\delta S^{ct}_2=\mathcal{N}\left[\frac{m\h\delta m}{\epsilon}+m\h\delta C+C\delta m\right]~.
\end{equation}

Adding up, the total variation of the renormalized action becomes
\begin{equation}
    \delta S_{D5}^{ren}=\delta S^E_{D5}+\delta S^{ct}=-\mathcal{N}\left[C\delta m+\vec{J}\cdot\delta\vec{A}-\frac{q}{\Omega}\delta\chi\big\rvert_{z_c}\right]~.
\end{equation}

The appearance of the term evaluated at the singular shell may underlie the atypical behavior of the free energy. As observed in~\cite{Ishii_2018}, this issue is related to the real-time holography prescription~\cite{Son_2002}, which instructs one to retain only boundary contributions and disregard those at the horizon. In our context, neglecting the contribution at the singular shell implies that the current and condensate defined in the main text follow this prescription. That is, they are extracted from the variation
\begin{equation}
    \delta S_{D5}^{ren}=\delta S^E_{D5}+\delta S^{ct}=-\mathcal{N}\left[C\delta m+\vec{J}\cdot\delta\vec{A}\right]~.
    \label{eq:dFnohorizon}
\end{equation}

This confirms the internal consistency of the holographic dictionary for our setup: the one-point functions $\langle\mathcal{O}_m\rangle$ and $\mathcal{J}_{\text{YM}}$ are correctly obtained from functional derivatives of the renormalized on-shell action with respect to the sources $m$ and $\vec{A}$, provided that contributions from the singular shell are excluded. 

To further verify this, we construct a family of solutions $\{\psi(u),b(u),\chi(u)\}$ parameterized by the insertion angle $\psi_0$ at the singular shell. By adjusting $b(u_c)$ and $\chi(u_c)$ so as to keep $|E|/m^2$ fixed at the boundary, we can evaluate
\begin{equation}
	S_{D5}^{ren}(\psi_0)=-\mathcal{N}\int_0^{\psi_0}\left[C(x)m'(x)+\vec{J}(x)\cdot\vec{A}'(x)\right]dx
\end{equation}
and confirm that the resulting curves coincide with the direct computation of the on-shell action shown in Fig. \ref{fig:OnshellD5}.

This implies that a naïve identification of minus the renormalized on-shell action with a free energy leads to the relation
\begin{equation}
    dF_{\text{naive}}=C\h dm+ \vec{J}\cdot d\vec{A}~,
\end{equation}
where, for simplicity, we have absorbed the overall normalization factor $\mathcal{N}$. For this expression to define a well-posed differential (\ie, for $dF_{\text{naive}}$ to be integrable), the following integrability conditions on the mixed partial derivatives must hold:
\begin{equation}
    \frac{\partial C}{\partial A_{x}}\Big\rvert_{m,A_{y}}=\frac{\partial  J_x}{\partial m}\Big\rvert_{A_{x},A_{y}}~,\qquad
    \frac{\partial C}{\partial A_{y}}\Big\rvert_{m,A_{x}}=\frac{\partial  J_y}{\partial m}\Big\rvert_{A_{x},A_{y}}~,\qquad
    \frac{\partial}{\partial \vec{A}}\times\vec{J}~\Big\rvert_m=0~,
    \label{eq:CrossDer}
\end{equation}
where $A_{i}$ denotes the boundary value of the $i$-th component of the gauge field.

\begin{figure}
    \centering
    \begin{subfigure}[t]{0.4\textwidth}
        \centering
        \includegraphics[width=\textwidth]{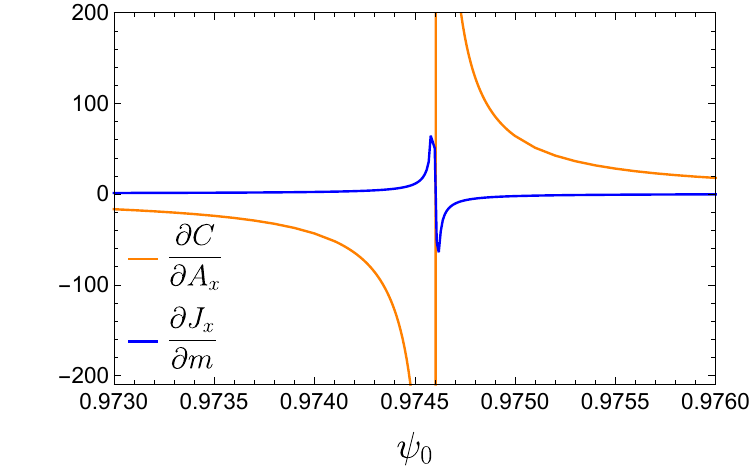}
    \end{subfigure}
    \hspace{0.02\textwidth}
    \begin{subfigure}[t]{0.4\textwidth}
        \centering
        \includegraphics[width=\textwidth]{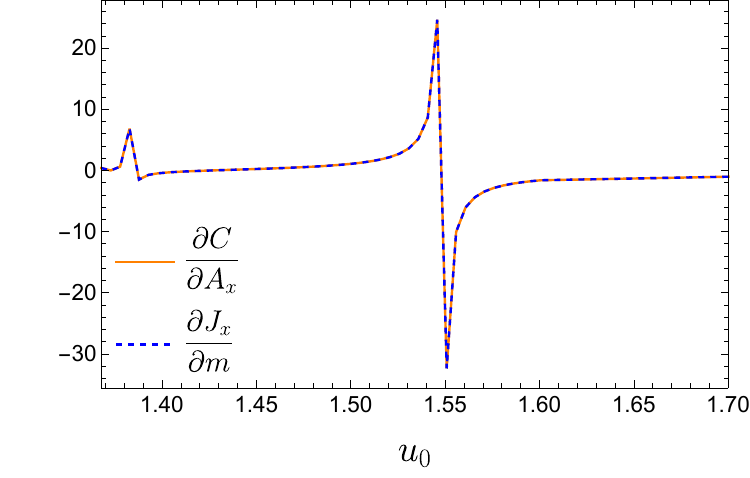}
    \end{subfigure}\\
    \centering
    \begin{subfigure}[t]{0.4\textwidth}
        \centering
        \includegraphics[width=\textwidth]{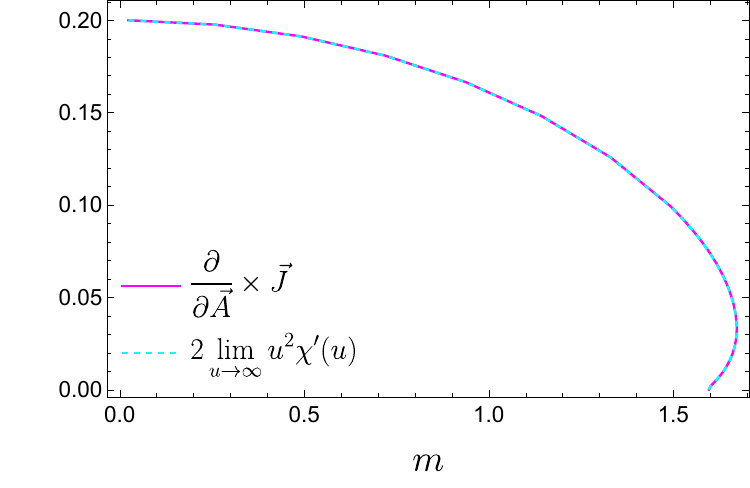}
    \end{subfigure}
    \hspace{0.02\textwidth}
    \begin{subfigure}[t]{0.4\textwidth}
        \centering
        \includegraphics[width=\textwidth]{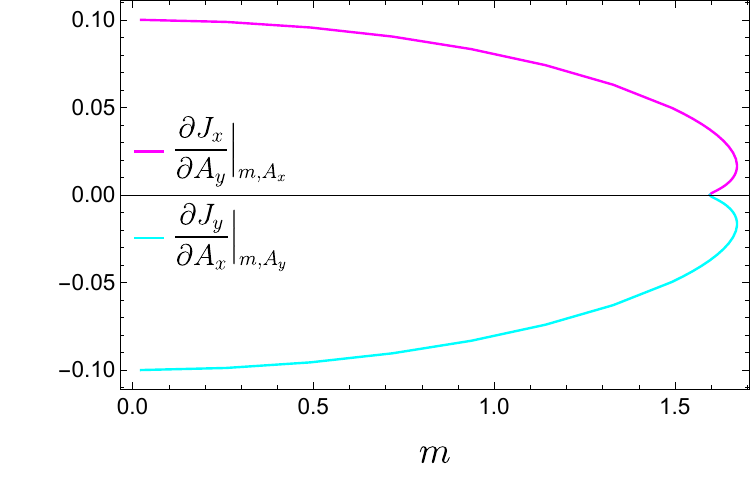}
    \end{subfigure}
    \caption{Top: First equality in Eq. \eqref{eq:CrossDer} for $\Omega=1$. On the left we show it for black hole embeddings, where it is violated, while on the right we observe that Minkowski embeddings correctly satisfy it. Bottom: on the left, numerical check of Eq. \eqref{eq:rotderlimit}. On the right, third equality in \eqref{eq:CrossDer} for $\Omega=0.1$. In the massless limit, the difference between the two derivatives approaches $2\Omega$, in agreement with the analytic result \eqref{eq:CrossDer2Omega}.}
    \label{fig:CrossDer}
\end{figure}

Evaluating the derivatives, we find that the integrability conditions are not satisfied, as illustrated in Fig. \ref{fig:CrossDer}. In fact, the violation of the last equality can be explicitly demonstrated in the massless case, where analytic solutions are known (see Appendix \ref{app:masslesssol}). Returning to the original holographic coordinate $u$, from the UV expansions \eqref{eq:UVexpansionsEJ} we identify the following expressions for the boundary values of the gauge field:
\begin{equation}
\begin{aligned}
    A_{x}&\equiv\lim_{z\rightarrow0}A_x(z)=b_0\cos\chi_0~,\\
    A_{y}&\equiv\lim_{z\rightarrow 0}A_y(z)=b_0\sin\chi_0~
\end{aligned}
\end{equation}
and also for the components of the current:
\begin{equation}
\begin{aligned}
      J_x&=b_1\cos\chi_0-b_0\chi_1\sin\chi_0~,\\
      J_y&=b_1\sin\chi_0+b_0\chi_1\cos\chi_0~,
\end{aligned}
\end{equation}
from where we immediately see that
\begin{equation}
    \frac{\partial J_x}{\partial A_{y}}=-\chi_1~,\qquad \frac{\partial J_y}{\partial A_{x}}=\chi_1\qquad \Rightarrow\qquad \frac{\partial}{\partial \vec{A}}\times\vec{J}=2\lim_{u\rightarrow\infty}u^2\chi'(u),
    \label{eq:rotderlimit}
\end{equation}
where we have used that $\displaystyle \chi_1=-\lim_{u\to\infty}u^2\chi'(u)$.

Using the analytic solution for $\chi$ in the massless case (see Appendix \eqref{app:masslesssol}), this yields
\begin{equation}
    \frac{\partial}{\partial \vec{A}}\times\vec{J}=2\Omega~,
    \label{eq:CrossDer2Omega}
\end{equation}
which matches the result reported in \cite{Ishii_2018} for a holographic superconductor under a rotating electric field. This behavior is further confirmed by the numerical results shown in Fig. \ref{fig:CrossDer}.

These results cast doubt on the interpretation of the Euclidean on-shell action as a free energy. In particular, the disappearance of the cusp would imply a negative entropy, as discussed in Appendix C of \cite{Mateos_2007}.

In conclusion, although the rotating ansatz \eqref{eq:defctu} removes the explicit time dependence from the action, the non-equilibrium character of the system remains manifest in features such as those discussed above. It is important to emphasize, however, that the extraction of response functions from the subleading terms in the near-boundary expansion of bulk fields remains valid within the framework of real-time holography \cite{Herzog_2002,Skenderis_2008_short,Skenderis_2008_long}. Nonetheless, other aspects, such as the nature and precise location of the phase transition, lie beyond the scope of equilibrium thermodynamics. This issue has been addressed in several works \cite{Sasa_2004,Albash_2007,Bergman_2008,Nakamura_2012,Nakamura_2013,Kundu_2013,Banerjee_2015,Kundu_2019}, but to our knowledge, no unambiguous resolution has been established.

This limitation of standard equilibrium AdS/CFT motivates the shift in focus in the second part of this thesis, which explores the non-equilibrium dynamics of the SYK model. It is therefore a suitable moment to justify this seemingly abrupt change of topic.

To study generic non-equilibrium configurations within holography, the most widely accepted approach is the one developed by Skenderis and van Rees \cite{Skenderis_2008_short,Skenderis_2008_long,vanRees_2009}. Their prescription requires choosing a suitable contour in the complex time plane, reflecting the field theory setup, and constructing a corresponding bulk geometry of mixed signature. This is done by gluing Euclidean segments to describe the imaginary parts of the contour and Lorentzian ones for the real-time evolution.

Despite its conceptual clarity, this prescription is technically demanding. It requires solving a coupled system of bulk equations, subject to boundary conditions that ensure the continuity of both the fields and their derivatives at the gluing surfaces between the Euclidean and Lorentzian segments. In most cases, these equations can only be tackled numerically.

To the best of our knowledge, no comprehensive studies have applied this formalism to D3/D5 or D3/D7 brane intersections driven out of equilibrium. Pursuing such an analysis remains an open and compelling direction for future work.

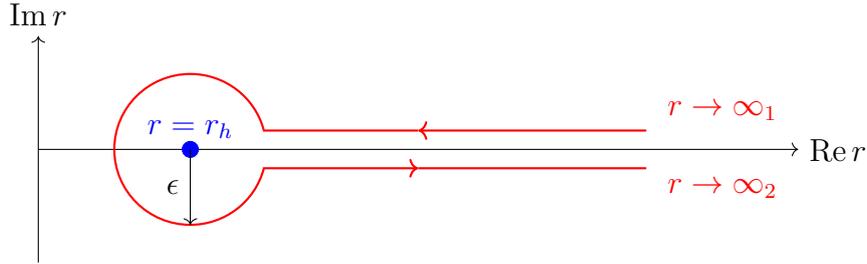
\begin{figure}
    \centering
    \begin{tikzpicture}

        \def\L{10}
        \draw[->] (0, 0) -- (\L, 0) node[right] {$\mathrm{Re}\,r$};
        \draw[->] (0, -0.15*\L) -- (0, 0.15*\L) node[above] {$\mathrm{Im}\,r$};

        \filldraw[blue] (0.2*\L, 0) circle (3pt) node[above] {$r=r_h$};

        \draw[->, thick, red] (0.8*\L, 0.025*\L) -- (0.5*\L, 0.025*\L);
        \draw[thick, red] (0.5*\L, 0.025*\L) -- (0.2955*\L, 0.025*\L);
        \node[above, red] at (0.9*\L, 0.025*\L) {$r\to\infty_1$};

        \draw[->, thick, red] (0.2955*\L, -0.025*\L) -- (0.5*\L, -0.025*\L);
        \draw[thick, red] (0.5*\L, -0.025*\L) -- (0.8*\L, -0.025*\L);
        \node[below, red] at (0.9*\L, -0.025*\L) {$r\to\infty_2$};

        \draw[thick, red] (0.1*\L, 0) arc[start angle=180,end angle=15,radius=0.1*\L];
        \draw[thick, red] (0.1*\L, 0) arc[start angle=-180,end angle=-15,radius=0.1*\L];

        \draw[->] (0.2*\L, 0) -- (0.2*\L, -0.1*\L);
        \node[left, black] at (0.2*\L, -0.05*\L) {$\epsilon$};
    \end{tikzpicture}
    \caption{Gravitational dual of the Schwinger-Keldysh contour proposed in \cite{Glorioso_2018}.}
    \label{fig:Glorioso}
\end{figure}

An alternative approach to out-of-equilibrium holography was proposed by Glorioso, Crossley, and Liu \cite{Glorioso_2018}. Their method involves an analytic continuation of the radial coordinate in the (generally time-dependent) bulk geometry. In this construction, the gravitational dual of the Schwinger-Keldysh contour is a path that starts at the UV boundary, extends into the bulk, and encircles the point $r=r_h$ along a small circle in the complex plane (see Fig. \ref{fig:Glorioso}). This prescription enables the computation of scalar two-point functions in both equilibrium states and in slowly varying non-equilibrium backgrounds with a horizon.

The desire to understand these holographic prescriptions for constructing Schwinger-Keldysh propagators prompted a broader study of the Schwinger-Keldysh formalism in generic out-of-equilibrium systems, with the aim of eventually applying it to our brane models. Along the way, we took a detour when the necessary tools became available to investigate non-equilibrium dynamics in the SYK model, an area we had been exploring in parallel due to its growing relevance.

This will be the subject of Chapter \ref{chap:FloquetSYK}. Before turning to that, we conclude this part with a summary and final remarks, and then proceed to study chiral symmetry breaking and restoration under helical magnetic fields in the D3/D7 system.

\section{Conclusions}\label{sec:FloquetIIconclusions}

This Chapter has pursued, along the line of previous works \cite{Kinoshita_2017,Garbayo_2020}, the study of holographic Floquet flavor systems driven by an external rotating electric field. We focused on the D3/D5 system but we found that most of the results are qualitatively robust and shared by the D3/D7 setup. We have sharpened the findings in \cite{Garbayo_2020} in several directions.

First, we studied the effects of having the system heated at some non-zero temperature. In this  case, the dual geometry has two types of horizons: the usual event horizon of the closed string geometry in the bulk, and the effective horizon of the open string metric on the brane. Respectively, we can associate two temperatures to them: the Hawking temperature $T_H$ of the  background \eqref{eq:TBH} and the effective temperature $T_\text{eff}$ written in \eqref{eq:TeffD5} experienced by the worldsheet degrees of freedom. We have scanned throughout all our phase space and checked that $T_\text{eff}>T_H$.

The main effect of the background temperature is the addition of deconfined charged carriers to the system. Such carriers add to the ones produced by the electric field through the Schwinger mechanism of dielectric breakdown. We show that one of the main results in previous works, namely, the presence of vector meson Floquet condensates, persists at finite temperature, signaling the robustness of this non-perturbative effect. The lobbed structure of the line of critical embeddings is also found here, but it gets depleted in height and is completely washed away for high enough values of the temperature, namely, for a radius horizon $r_h/m \geq 0.8897$ in units of the quark mass (see Fig. \ref{fig:LobesEJD3D5}).

At high temperatures, some interesting effects occur when the background and effective horizons come close together. These include a multi valuedness that resembles a second-order phase transition within the conductive black hole phase (Fig. \ref{fig:JCvsE}). Also new solutions with vanishing $|E|=|J|=0$ appear within this phase (Fig. \ref{fig:JFullCurves}).

Secondly, we have remarked the relevance of the so  called {\em Floquet suppression points}. These states were missed in previous analysis but are common to both D3/D5 and D3/D7 systems both at zero and finite temperature. We have shown that the phase portrait very close to these points is strikingly similar to the one in the vicinity of the vector meson Floquet condensates, up to an exchange of $J$ with $E$. This calls for a deeper study in search for a sounder duality. From the physical point of view, these new points exhibit a dynamical suppression of the vacuum polarizability. It could be attributed to a dynamical screening of the effective dipole charge of the meson fluctuations at strong coupling and for precise frequencies. It bears resemblance to similar effects in the realm of Floquet condensed matter systems where, for example, hopping terms can be seen to vanish at fined tuned frequencies of the driving. This is the type of effects that make Floquet driving an appealing paradigm in the search for mechanisms that could help in suppressing quantum decoherence.

Thirdly, we have also pursued the analysis initiated in \cite{Garbayo_2020} concerning two different notions of conductivities in Section \ref{sec:conductivitiesD3D5}. The first one is the non-linear rotating conductivity of Section \ref{subsec:nonlinearcond}. The relative phase (angle) between $J$ and $E$ has an interesting information that we interpret in terms of a possible variable admixture of two types of currents: rotating polarization and charge flow. The polarization current is the only one present in the Minkowski phase while both are present in the BH states. The global picture that emerges from the analysis shows that the Floquet suppressed states are points where the polarizability of the vacuum switches smoothly from positive to negative. This is remarkable as it states that, for ample intervals in the range of driving frequencies, $\Omega$, the meson condensate is  polarized  {\em antiparallel} to the applied electric field! Again here, this result is amongst the class of remarkable effects that one can find in the context of Floquet engineering \cite{Giovannini_2019} of condensed matter systems. For example, it is worth citing ref. \cite{Schmidt_2019}, where paramagnetism can be turned into  diamagnetism under a strong driving in the Rabi model coupled to a heat bath.

In the limit of large electric field, and/or large temperature, we agree with the results in \cite{Karch_2010}. In this limit the polarization rotating current gets suppressed, $J_{pol}(t)\to 0$,  whereas the conduction currect satisfies an Ohmic instantaneous response for an arbitrary frequency of the rotating driving $J(t)\sim  J_{con}(t) = \sigma  E(t)$ with $\sigma$ a real constant.  Linearity suggests the possibility of this being also true for any electric field time dependence in 2+1 dimensions. We also find agreement with the predictions in \cite{Karch_2010} in the limit of small mass. 

The other type of conductivities involves the optical AC and DC conductivities in the presence of a driving. The interesting pattern with peaks found in \cite{Garbayo_2020} deep inside the wedges between the lobes in phase space still exists for low to moderate temperatures, but gets dissolved as soon as the height of the lobes is depleted at high  temperature. The highest peaks shift with growing $|E|$ and stay close to the position of the (also drifting) Floquet suppression points.
 
Our work could be continued along several directions. One clear option would be to add chemical potential and/or magnetic components to the gauge field in order to explore the complete phase space of the D3-D5 model. To verify the universality of our results we could consider the ABJM model \cite{Aharony_2008} driven by a rotating electric field. This last model has a rich topological structure and is dual to a $(2+1)$-dimensional conformal field theory. The flavor branes in this case are D6-branes \cite{Hohenegger_2009, Gaiotto_2009} (the thermodynamics of these flavor branes has been studied in \cite{Jokela_2012}). Another direction worth pursuing is the analysis of the effects of backreaction for a large number of  flavor branes. Interestingly, backreacted D3-D5 backgrounds have been constructed in \cite{Conde_2016, Penin_2017, Jokela_2019, Garbayo_2022} using the smearing approximation reviewed in \cite{Nunez_2010}. One could also study the driving generated by moving the brane periodically in time (i.e. oscillating or rotating). This type of configurations were considered in \cite{Das_2010, Hoyos_2011}, and are the natural setup to find resonances that could be interpreted as Floquet condensates of other type of mesons, like scalar mesons.
 
Finally, another aspect that deserves further attention is the actual nature of the phase transition. It can be triggered by an admixture of both temperature and/or electric field. While the multi-valuedness of the state curves in Fig. \ref{fig:JCvsE} suggest an area law for the transition point, the actual location it is not consistent with the $(|J|/m^2,|E|/m^2)$ and the $(C/m^2,|E|/m^2)$ curves.

This points to the difficulties, out of equilibrium, in defining a {\em bona fide} free energy from which the one-point functions are derived \cite{Ishii_2018}. Techniques developed specifically for open non-equilibrium steady states should be used instead. In particular, it would be very interesting to apply the Schwinger-Keldysh approach of \cite{Glorioso_2018,Jana_2020} to these systems in order to calculate properties of the corresponding steady states like equilibrium of phases and phase transitions.

\clearpage

%% file: Part2/FloquetII/FloquetIIAppendices.tex
\renewcommand{\chapterquote}{}
\chapterappendix{\thechapter}

\setcounter{appendixsection}{\value{appendixsection}+1} 
\section{Analytic solutions}\label{app:analyticsols}

The equations of motion admit analytic solutions in certain limits. In particular, we can find analytic expressions for the gauge field in the massless case. By considering small masses as a deviation of this analytic solution, the equations for the deviations can also be solved analytically. This is a remarkable feature of the D3/D5 model that is not present in the D3/D7 case \cite{Hashimoto_2016,Kinoshita_2017}. There, even in the massless case, the equations have to be solved numerically. In this appendix we obtain the two analytic solutions mentioned above. The procedure follows the same steps as in \cite{Garbayo_2020}, providing a non-zero temperature generalization to the solutions found there.

In both cases, it is convenient to use the coordinates \eqref{eq:AdS5xS5formassless}, with the embedding parametrized as $\theta(r)$. The DBI action \eqref{eq:actionD5} in these coordinates is given by
\begin{equation}
\begin{aligned}
    S_{\text{D5}}=-\mathcal{N}\int dr\frac{\cos^2\theta}{\sqrt{r^4-r_h^4}}&\Big(\big[r^4-r_h^4-\Omega^2b^2\big]\big[\left(r^4-r_h^4\right)\left(b'^2+r^2\theta'^2\right)+r^4\big]\Big.\\
    &\phantom{\Big[\big[r^4-r_h^4-\Omega^2b^2\big]\big[\left(r^4-r_h^4\right)\big.\Big.}\Big.+\left(r^4-r_h^4\right)^2b^2\chi'^2\Big)^{1/2}~,
    \label{eq:SD5app}
\end{aligned}
\end{equation}

where again we have consistently chosen the fields to be time-independent, and the constant $\mathcal{N}$ is given in \eqref{eq:ND3D5}. The cyclic coordinate $\chi(r)$ is now related to $\theta(r)$ and $b(r)$ as
\begin{equation}
    \chi'(r)=\frac{q~\sqrt{r^4-r_h^4-\Omega^2 b^2}\sqrt{\left(r^4-r_h^4\right) \left(b'^2+r^2\theta'^2\right)+r^4}}{\abs{b}\left(r^4-r_h^4\right)\sqrt{\left(r^4-r_h^4\right)\Omega^2 b^2\cos^4\theta-q^2}}~.
\label{eq:chiprime}
\end{equation}
Legendre-transforming the action to eliminate $\chi'$, we obtain
\begin{equation}
\begin{aligned}
    \bar{S}_{\text{D5}}&=S_{\text{D5}}-\int dr~ \chi'\frac{\partial\mathcal{L}}{\partial \chi'}\\
    &=-\mathcal{N}\int dr \frac{\sqrt{r^4+\left(r^4-r_h^4\right)\left(b'^2+r^2\theta'^2\right)}}{\left(r^4-r_h^4\right)\Omega\abs{b}}\sqrt{\left[r^4-r_h^4-\Omega^2b^2\right]\left[\left(r^4-r_h^4\right)\Omega^2b^2\cos^4\theta-q^2\right]}~.
    \label{eq:actionLTtransftheta}
\end{aligned}
\end{equation}

The tortoise coordinates $(\tau, r_*)$ in this parameterization are defined as
\begin{align}
    d\tau &= dt-A_{\theta}(r)~dr~, & dr_* &= B_{\theta}(r)~dr~,
    \label{eq:tortoise}
\end{align}
where $A_\theta(r)$ and $B_\theta(r)$ are the functions
\begin{align}
    A_{\theta}(r) &= \frac{\Omega b^2\chi'}{r^4-r_h^4-b^2~\Omega^2}~, & B_{\theta}(r) &= \frac{\mathcal{L}}{\cos^2\theta \left(r^4-r_h^4-b^2~\Omega^2\right)}~,
\end{align}
with ${\cal L}$ being the lagrangian density in \eqref{eq:SD5app}. In terms of $(\tau, r_*)$ the $(t,r)$ part of the effective metric takes the form
\begin{equation}
    \frac{r^4-r_h^4-b^2\Omega^2}{r^2}\left(-d\tau^2+dr_*^2\right)~,
\end{equation}
which means that, indeed, these new coordinates are tortoise coordinates for the effective open string metric.

\subsection{Massless solution}\label{app:masslesssol}
Let us now consider a massless embedding with $\theta=0$. The action \eqref{eq:actionLTtransftheta} for this case takes the form
\begin{equation}
    \bar{S}_{\text{D5}}=-\mathcal{N}\int dr\frac{\sqrt{r^4+\left(r^4-r_h^4\right)b'^2}}{\left(r^4-r_h^4\right)\Omega\abs{b}}\sqrt{\left[r^4-r_h^4-\Omega^2b^2\right]\left[\left(r^4-r_h^4\right)\Omega^2b^2-q^2\right]}~.
    \label{eq:actionLTtransfmassless}
\end{equation}
Both factors in the second square root must vanish simultaneously at the singular shell, $r=r_c$, in order to keep the action real. From this condition, we get
\begin{equation}
    b_0^2=\frac{r_c^4-r_h^4}{\Omega^2}\quad, \quad\quad q^2=(r_c^4-r_h^4)\Omega^2b_0^2~.
    \label{eq:b0andqmassless}
\end{equation}
Combining both equations we obtain
\begin{equation}
    b_0^2=\frac{r_c^4-r_h^4}{\Omega^2}=\frac{q}{\Omega^2}~.
    \label{eq:b0massless}
\end{equation}
Actually, one can verify from the equation of motion of $\bar{S}_{\text{D5}}$ that a constant $b(r)=b_0$ is a solution. Moreover, plugging $b(r)=b_0$ into the right-hand side of \eqref{eq:chiprime} (with $\theta=0$), we get that the phase $\chi(r)$ in the massless case satisfies
\begin{equation}
    \chi'(r)=\Omega \frac{r^2}{r^4-r_h^4}=\frac{\Omega}{r^2f(r)}~.
    \label{eq:chi_0_prime}
\end{equation}
This equation can be integrated directly. By defining a new function $\Lambda(x)$ as
\begin{equation}
    \Lambda(x) \equiv \log\frac{x-1}{x+1}-2\arccot x~, \qquad\frac{d\Lambda}{dx} = \frac{4x^2}{x^4-1}~,
    \label{eq:Lambdadef}
\end{equation}
one has
\begin{equation}
    \chi(r)=\frac{\Omega}{4r_h}\Lambda(r/r_h)~,
    \label{eq:chimassless}
\end{equation}
where we have fixed the integration constant in such a way that $\Lambda(x)\to 0$ for $x\to\infty$ ($\Lambda(x)\approx -4/x$ for large $x$). Therefore, the complexified gauge potential $c(r)=b(r)e^{i\chi(r)}$ is given by
\begin{equation}
    c(r)=\frac{\sqrt{r_c^4-r_h^4}}{\Omega}e^{\frac{i\Omega}{4r_h}\Lambda(r/r_h)}~.
    \label{eq:c_massless}
\end{equation}

By expanding the right-hand side of \eqref{eq:c_massless} for large $r$ and comparing the result with \eqref{eq:UVexpansions}, we get the electric field $E$ and the current $J$ to be given by
\begin{equation}
    E=J=-i\sqrt{r_c^4-r_h^4}=-i\sqrt{q}~.
    \label{eq:EJmassless}
\end{equation}
Thus, $E$ and $J$ are parallel and equal, which corresponds to $\sigma_{xx}=1$ and $\sigma_{xy}=0$ in the conductivity tensor \eqref{eq:sigmaRC}. Using \eqref{eq:dictionaryD3D5} it is easy to derive the correctly normalized physical conductivity, relating the the physical electric field and the physical current, as
\begin{equation}
    \sigma(\Omega)=\frac{2N_f N_c}{\pi\sqrt{\lambda}}~,
\end{equation}
matching the results of \cite{Karch_2010}, which we reviewed in Section \ref{subsec:nonlinearsigma}. Remarkably, the response of the system is not only linear, but instantaneous.

Finally, let's consider the effective temperature, which we found to be in general
\begin{equation}
    T_{\text{eff}}=\frac{2u_ch(u_c)-\Omega b_1}{2\pi b_0\chi_1}~.
    \label{eq:TeffApp}
\end{equation}

For the massless solution found above, $b_1=0$, and the value of $\chi_1$ can be read from \eqref{eq:chi_0_prime} by taking $r=r_c$. We obtain\footnote{The coefficients $b_1$ and $\chi_1$ in \eqref{eq:TeffApp} were defined from the series expansion in the $u$ coordinate. Therefore the change of coordinates \eqref{eq:u} needs to be used.}
\begin{equation}
    T_{\text{eff}}=\frac{r_c}{\pi}=\frac{(q+r_h^4)^{1/4}}{\pi}\ge T_H=\frac{r_h}{\pi}~.
\end{equation}

\subsection{Small-mass solution}\label{app:smallmasssol}
Let us now consider small-mass solutions, in which $b(r)$ and $\theta(r)$ are given by
\begin{equation}
    b(r)=b_0+\beta(r)~,\qquad\theta(r)= \lambda(r)~,
\end{equation}
where the functions $\beta(r)$ and $\lambda(r)$ are treated as small perturbations. At first order in these functions, they satisfy the following linear equations of motion
\begin{equation}
\begin{aligned}
    \frac{d}{dr}\left[(r^4f-q)\lambda'\right]+2r^2\lambda&=0~,\\
    \frac{d}{dr}\left[\frac{r^4f-1}{r^2}\beta'\right]+\frac{4r^2\Omega^2}{r^4f-q}\beta&=0~,
\end{aligned}
\end{equation}
which can be written as
\begin{equation}
\begin{aligned}
    (r^4-r_c^4)\lambda''+4r^3\lambda'+2r^2\lambda&=0~,\\
    r^2(r^4-r_c^4)\beta''+2r(r^4+r_c^4)\beta'+4\Omega^2\frac{r^6}{r^4-r_c^4}\beta&=0~.
    \label{eq:smallmasseqs}
\end{aligned}
\end{equation}
These equations are equivalent to those solved in Appendix B of \cite{Garbayo_2020}, with the only difference being the value of the critical radius, $r_c$. Therefore, we can just adapt their results to our case. Let us start by writing the general solution for the differential equation satisfied by $\lambda(r)$:
\begin{equation}
    \lambda(r)=c_1\h_2F_1\left(\frac{1}{4},\frac{1}{2};\frac{3}{4};\frac{r^4}{r_c^4}\right)+c_2r\h_2F_1\left(\frac{1}{2},\frac{3}{4};\frac{5}{4};\frac{r^4}{r_c^4}\right)~,
    \label{eq:lambdasol}
\end{equation}
where $_2F_1$ is the ordinary hypergeometric function, and $c_1$ and $c_2$ are constants. The hypergeometric functions in \eqref{eq:lambdasol} have a logarithmic divergence as we approach the singular shell, $r\to r_c$. Imposing the solution to be regular at $r=r_c$, we find that the integration constants $c_1$ and $c_2$ must satisfy the relation
\begin{equation}
    \frac{c_1}{c_2}=-\frac{r_c}{4}\left[\frac{\Gamma\left(\frac{1}{4}\right)}{\Gamma\left(\frac{3}{4}\right)}\right]^2
    \label{eq:c1c2}
\end{equation}
We can expand the solution \eqref{eq:lambdasol} near the boundary and, by comparing with the known expansion for $\theta(r)$,
\begin{equation}
    \theta(r)\approx \frac{m}{r}+\frac{C}{r^2}+...~, \qquad r \to \infty~,
\end{equation}
we can relate the parameters $m$ and $C$ to the integration constants. Eliminating $c_1$ using \eqref{eq:c1c2}, we find the relations
\begin{align}
    m & = r_c\h c_1 =  -\frac{r_c^2}{4}\left[\frac{\Gamma\left(\frac{1}{4}\right)}{\Gamma\left(\frac{3}{4}\right)}\right]^2c_2~,\\
    C & = r_c^3\h c_2~.
\end{align}
It follows that $C$ and $m$ are linearly related in this small-mass solutions
\begin{equation}
    C=-r_c\left[\frac{2\Gamma\left(\frac{3}{4}\right)}{\Gamma\left(\frac{1}{4}\right)}\right]^2m~,
\end{equation}
in agreement with the asymptotic behavior in the right plots of Fig. \ref{fig:JCvsE}.

In summary, in the small-mass regime, the embedding function $\theta(r)$ can be written as
\begin{equation}
    \theta(r)=\frac{m}{r_c}F\left(\frac{1}{4},\frac{1}{2};\frac{3}{4};\frac{r^4}{r_c^4}\right)+\frac{C}{r_c^3}\h r\h F\left(\frac{1}{2},\frac{3}{4};\frac{5}{4};\frac{r^4}{r_c^4}\right)~.
\end{equation}

Let us next look at the equation satisfied by the gauge field perturbation $\beta(r)$ in the system \eqref{eq:smallmasseqs}. Its general solution is
\begin{equation}
    \beta(r)=d\h\exp\left[-\frac{i\Omega}{2r_c}\Lambda(r/r_c)\right]+d^*\exp\left[\frac{i\Omega}{2r_c}\Lambda(r/r_c)\right]~,
    \label{eq:betasol}
\end{equation}
where $d$ is a complex constant and $\Lambda$ is defined in \eqref{eq:Lambdadef}. It is interesting to find the solution for the complex potential $c(r)=b(r)e^{i\chi(r)}$. Let us denote
\begin{equation}
    \delta c(r)\equiv c(r)-c_0(r)~,
\end{equation}
where $c_0(r)$ is the solution for the massless case obtained in \eqref{eq:c_massless}. $\delta c$ satisfies the following differential equation at linear order
\begin{align}
\begin{split}
    r(r^4-r_c^4) (r^4-r_h^4)^2\delta c''+ 2(r^4-r_h^4)\big[r^8+(r_c^4-r_h^4)r^3(r+i\Omega)-r_c^4 r_h^4\big]\delta c'\,+\\
    +r^5\Omega\big[\Omega~r^4+(r_c^4-2r_h^4)\Omega-4i(r_c^4-r_h^4)r\big]\delta c=0~.
\end{split}
\end{align}
Instead of trying to solve directly this equation we notice that $\delta c$, at first order, can be written as
\begin{equation}
    \delta c(r)=e^{i\chi_0(r)}\left(\beta(r)+i\h b_0\h \delta \chi(r)\right)~,
    \label{eq:deltac}
\end{equation}
where $\chi_0(r)$ and $b_0$ denote the massless solutions in \eqref{eq:b0massless} and \eqref{eq:chimassless}, respectively, and $\delta \chi(r)=\chi(r)-\chi_0(r)$. The equation satisfied by $\delta \chi(r)$ is obtained by expanding \eqref{eq:chiprime} at linear order in the perturbations, giving
\begin{equation}
    \delta \chi'=-\frac{2r^2\Omega}{b_0(r^4-r_c^4)}\beta~,
\end{equation}
which can be readily integrated:
\begin{equation}
    \delta \chi(r)=-\frac{i\h d}{b_0}\exp\left[-\frac{i\Omega}{2r_c}\Lambda(r/r_c)\right]+\frac{i\h d^*}{b_0}\exp\left[\frac{i\Omega}{2r_c}\Lambda(r/r_c)\right]+\varphi~,
    \label{eq:deltachisol}
\end{equation}
where $\varphi$ is a constant. Plugging \eqref{eq:betasol} and \eqref{eq:deltachisol} into \eqref{eq:deltac}, we obtain:
\begin{equation}
    \delta c(r)=A\exp\left[i\frac{\Omega}{4}\left(\frac{1}{r_h}\Lambda(r/r_h)-\frac{2}{r_c}\Lambda(r/r_c)\right)\right]+B\exp\left[\frac{i\Omega}{4r_h}\Lambda(r/r_h)\right]~,
    \label{eq:deltacsol}
\end{equation}
where $A=2 d$ and $B=i\frac{\varphi}{\Omega}\sqrt{r_c^4-r_h^4}$.

To proceed further we have to impose a regularity condition to the general solution 
\eqref{eq:deltacsol}. With this purpose, let us write $\delta c(r)$ in terms of the tortoise coordinates \eqref{eq:tortoise} for the open string metric. In the small-mass case it is easy to demonstrate that \eqref{eq:tortoise} can be integrated to give the following relation between the tortoise coordinates $(\tau, r_*)$ and our original coordinates $(t,r)$,
\begin{align}
    \tau & = t+\frac{1}{4r_h}\Lambda(r/r_h)-r_*~, & r_* & = \frac{1}{4r_c}\Lambda(r/r_c)~.
\end{align}
The new radial coordinate $r_*$ varies from $r_*=-\infty$ at the pseudohorizon to $r_*=0$ at the UV boundary. Actually, one can prove that in these regions it can be related to $r$ as
\begin{equation}
\begin{aligned}
    r_* & = -\frac{1}{r}+\mathcal{O}(r^{-5})~, & (r & \to \infty)~,\\
    r_* & \sim \frac{1}{4r_c}\log(r-r_c)~, & (r & \to r_c)~.
    \label{eq:tortoisesolution}
\end{aligned}
\end{equation}

Let us next consider the time-dependent gauge potential $\delta a(r,t)$, given by
\begin{equation}
    \delta a(r,t)=\delta c(r) e^{i\Omega t}~.
\end{equation}
By combining \eqref{eq:deltacsol} and \eqref{eq:tortoisesolution}) it straightforward to prove that $\delta a$ can be written in terms of $(\tau, r_*)$ simply as
\begin{equation}
    \delta a(\tau,r_*)=Ae^{i\Omega (\tau-r_*)}+Be^{i\Omega (\tau+r_*}~.
    \label{eq:deltaatortoise}
\end{equation}
We next impose an infalling boundary condition, which amounts to select the solutions with $A=0$ in \eqref{eq:deltaatortoise}. Therefore, the regular solutions we are looking for are
\begin{equation}
    \delta a(\tau,r_*)=Be^{i\Omega (\tau+r_*)}~.
\end{equation}
Equivalently, $\delta c(r)$ is given by
\begin{equation}
    \delta c(r)=B\exp\left[\frac{i\Omega}{4r_h}\Lambda(r/r_h)\right]~.
\end{equation}

We can  now read off the electric field and current from the asymptotic behavior of $\delta c$, namely
\begin{equation}
    \delta c(r) \approx B-\frac{i\h\Omega B}{r}+...~, \qquad (r  \to \infty)~.
\end{equation}

Thus, we have
\begin{equation}
    \delta E=\delta J = -i\h\Omega B~,
\end{equation}
which means that the equality of $E$ and $J$ of the massless solution is mantained at first order in the small-mass regime, in agreement with the asymptotic behavior in the left plots of Fig. \ref{fig:JCvsE}.

\setcounter{equation}{0}
\setcounter{appendixsection}{\value{appendixsection}+1} 
\section{Holographic renormalization and dictionary}\label{app:holorenoD3D5}

We begin with the metric of AdS$_5\times S^5$ written in the $\theta$-parametrization \eqref{eq:metricisotropic}-\eqref{eq:metric5spheretheta}, and we introduce the inverse radial coordinate $z=1/u$. The AdS$_5\times S^5$ metric becomes
\begin{equation}
    ds^2=\frac{1}{z^2}\left[-\frac{g(z)^2}{h(z)}dt^2+h(z)d\vec{x}_3^2+dz^2\right]+d\theta^2+\cos^2\theta d\Omega_2^2+\sin^2\theta d\tilde{\Omega}_2^2~,
\end{equation}
with
\begin{equation}
    g(z)=1-\frac{z^4}{z_h^4}~,\qquad h(z)=1+\frac{z^4}{z_h^4}~,
\end{equation}
and the horizon is located at $z_h=1/u_h$.

The ansatz for the brane embedding and the gauge field \eqref{eq:defctu} translates in these coordinates into
\begin{equation}
    (2\pi\alpha')(A_x+iA_y)=c(z)e^{i\Omega t}~,\qquad \theta=\theta(z)~.
\end{equation}

In these coordinates, the DBI action is given by
\begin{equation}
    S_{D5}=-\mathcal{N}\int_{0}^{z_0}dz\frac{\cos^2\theta}{z^4}\sqrt{z^4g^2\abs{c'}^2-z^8\Omega^2\Re(c\h c'^*)^2+h\left(g^2-z^4\Omega^2\abs{c}\right)\left(1+z^2\theta'^2\right)}~,
\end{equation}
with $\mathcal{N}=4\pi N_fT_{D5}$, and primes denote derivatives with respect to $z$. The integral goes from the boundary $z=0$ to the bulk interior, denoted by $z_0$\footnote{For black hole embeddings, $z_0$ is the location of the singular shell. For Minkowski embeddings, $z_0$ is the value of $z$ at which the brane ends.}. The asymptotic expansions \eqref{eq:UVexpansions} become
\begin{equation}
\begin{aligned}
    \theta(z) & = Mz+Cz^2+\mathcal{O}(z^4)~,\\
    c(z) & = c_\infty+Jz-\frac{\Omega^2c_\infty}{2}z^2+\mathcal{O}(z^3)~.\\
\end{aligned}
\end{equation}

Integrating the DBI action up to a UV cutoff at $z=\epsilon$, we find the asymptotic behavior
\begin{equation}
\begin{aligned}
    S_{D5}&=-\mathcal{N}\int_\epsilon dz\left[\frac{1}{z^4}-\frac{M^2}{2z^2}+\mathcal{O}(\epsilon^0)\right]\\
    &=-\mathcal{N}\left[\frac{1}{3\epsilon^3}-\frac{M^2}{2\epsilon}+\mathcal{O}(\epsilon)\right]~.
\end{aligned}
\end{equation}

The local counterterms needed to regularize the divergences were are given by \cite{Karch_2005_holorenobranes}
\begin{equation}
\begin{aligned}
    S^{ct}_1 & = \frac{1}{3}\mathcal{N}\sqrt{-\gamma}~,\\
    S^{ct}_2 & = -\frac{1}{2}\mathcal{N}\sqrt{-\gamma}\theta(\epsilon)^2~.
\end{aligned}
\end{equation}

The contribution of the counterterms is
\begin{equation}
    S^{ct}=-\mathcal{N}\left[-\frac{1}{3\epsilon^3}+\frac{M^2}{2\epsilon}+MC\right]~,
\end{equation}
where the first term is the contribution of $S^{ct}_1$, and the rest comes from $S^{ct}_2$. These contributions correctly cancel the divergences when $\epsilon\to 0$.

\subsection{Quark condensate and electric current}

We now consider the one-point functions. We begin by the quark condensate $\langle\mathcal{O}_m\rangle$,
\begin{equation}
    \langle \mathcal{O}_m\rangle =-(2\pi\alpha')\lim_{\epsilon\to 0}\epsilon\frac{\delta S_{\text{D5}}^{sub}}{\delta \theta(\epsilon)}~,
    \label{eq:condensateD3D5}
\end{equation}
with $S^{sub}_{D5}=S^{reg}_{D5}+\sum_iS^{ct}_i$. The contribution from the regularized action on-shell can be written as a boundary term,
\begin{equation}
    \delta S^{reg}_{D5}=\frac{\partial \mathcal{L}}{\partial \theta'}\delta \theta\big\rvert_\epsilon=\mathcal{N}\left[\frac{M}{\epsilon^2}+\frac{2C}{\epsilon}+\mathcal{O}(\epsilon^0)\right]\delta\theta(\epsilon)~.
\end{equation}
The counterterms contribute as
\begin{equation}
    \delta S^{ct}_2=\mathcal{N}\left[-\frac{M}{\epsilon^2}-\frac{C}{\epsilon}+\mathcal{O}(\epsilon^0)\right]\delta\theta(\epsilon)~.
\end{equation}

Adding the two contributions, we readily get from \eqref{eq:condensateD3D5}
\begin{equation}
    \langle\mathcal{O}_m\rangle=-\frac{N_fN_c}{\pi^2}C~,
\end{equation}
where we have written $\mathcal{N}$ in terms of field theory quantities using \eqref{eq:ND3D5}.

We now move on to the current,
\begin{equation}
    \mathcal{J}_\text{YM}(t) =(2\pi\alpha')\lim_{\epsilon\to 0}\left[\frac{\delta S^{sub}_{D5}}{\delta c_x(\epsilon)}+i\frac{\delta S^{sub}_{D5}}{\delta c_y(\epsilon)}\right]e^{i\Omega t}~,
    \label{eq:currentD3D5}
\end{equation}
with $c_x$ and $c_y$ defined as $c(z)=c_x(z)+ic_y(z)$.

The contribution from $S^{reg}_{D5}$ is again a boundary term,
\begin{equation}
    \delta S^{reg}_{D5}=\frac{\partial \mathcal{L}}{\partial c_x'}\delta c_x\big\rvert_\epsilon+i\frac{\partial \mathcal{L}}{\partial c_y'}\delta c_y\big\rvert_\epsilon=\mathcal{N}\left[J_x \delta c_x(\epsilon)+iJ_y\delta c_y(\epsilon)+\mathcal{O}(\epsilon)\right]~,
\end{equation}
which is the only contribution since the counterterms do not involve the gauge field\footnote{This is not the case in the D3/D7 model, where a logarithmic counterterm has to be added, leading to an ambiguity in the definition of the current. See Chapter \ref{chap:HelicalB}.}.

Therefore, we readily get from \eqref{eq:currentD3D5}
\begin{equation}
    \mathcal{J}_{\text{YM}}(t) =\frac{N_fN_c}{\pi^2}Je^{i\Omega t}~,
\end{equation}
where $J=J_x+iJ_y$.

\setcounter{equation}{0}
\setcounter{appendixsection}{\value{appendixsection}+1} 
\section{Details on optical conductivities}\label{app:opticalcond}

In this appendix we provide the details concerning the photovoltaic optical conductivities of Section \ref{subsec:opticalcond}. We analyze the response of our system to an additional linearly polarized electric field on top of the circularly driven background \eqref{eq:rotatingE}. In vector cartesian notation, the total electric field is now
\begin{equation}
    \vec{\mathcal{E}}(t)=O(t)\vec{E}+\vec{\epsilon}(t)=O(t)\vec{E}+\vec{\epsilon}\h e^{-i\omega t}~,
\end{equation}
where $\vec\epsilon$ is a constant vector such that $\abs{\vec{\epsilon}}\ll|\vec{E}|$. The change in the current, $\vec{\mathcal{J}}(t)\to O(t)\vec J+\delta \vec J(t)$, defines the conductivity $\boldsymbol{\sigma}$ as $\delta\vec{J}(t)=\boldsymbol{\sigma}\cdot\vec{\epsilon}(t)$. We define the vector of fluctuations, $\delta \vec \xi(t,\rho) = (\delta c_x,\delta c_y,\delta \theta)$.

First of all, let us write the perturbed equations in terms of the tortoise coordinates $(\tau, r_*)$ defined in \eqref{eq:tortoise}. We get
\begin{equation}
    \Big( \partial_\tau^2-\partial_{r_*}^2+\textbf{A}(r)\partial_\tau+\textbf{B}(r)\partial_{r_*}+\textbf{C}(r)\Big)\delta \vec{\xi}=0~,
    \label{eq:perteq}
\end{equation}
where $\textbf{A}$, $\textbf{B}$ and $\textbf{C}$ are $3\times 3$ matrices which at the pseudohorizon $r=r_c$ satisfy
\begin{equation}
    \textbf{A}(r=r_c) =-\textbf{B}(r=r_c)\equiv \textbf{A}_c~, \qquad \textbf{C}(r=r_c)  =0~.
\end{equation}
Thus, in this limit, which corresponds to $r_*\to -\infty$, the fluctuation equation
\eqref{eq:perteq} becomes
\begin{equation}
    \left(\partial_\tau^2-\partial_{r_*}^2\right)\delta\vec{\xi}+\textbf{A}_c\left(\partial_\tau-\partial_{r_*}\right)\partial \vec{\xi}=0~,
\end{equation}
whose general solution takes the form
\begin{equation}
    \delta \vec{\xi}=\vec{f}(\tau+r_*)+e^{-\textbf{A}_cr_*}\vec{g}(\tau-r_*)~.
\end{equation}
We will impose that $\vec{g}=0$, which selects the ingoing wave boundary condition at the pseudohorizon. Let us next look at the UV boundary condition \eqref{eq:pert}. First of all, we rewrite the rotation matrix $O(t)$ as
\begin{equation}
    O(t)=\textbf{M}_+e^{i\Omega t}+\textbf{M}_-e^{-i\Omega t}~,
\end{equation}
where
\begin{equation}
    \textbf{M}_\pm = \frac{1}{2}\begin{pmatrix} 1 & \pm i\\ \mp i & 1 \end{pmatrix}~.
\end{equation}
Then, defining the frequencies $\omega_\pm=\omega\pm\Omega$, the boundary UV condition of $\delta\vec{c}$ can be written as
\begin{equation}
    \delta \vec{c}(t,r=\infty)=-\frac{i}{\omega}\Big(\textbf{M}_+e^{-i\omega_+ t}+\textbf{M}_- e^{-i\omega_- t}\Big)\vec{\epsilon}~.
    \label{eq:deltacbdry}
\end{equation}
Let us assume that $\delta\vec{c}(t,r)$ and $\delta\theta(t,r)$ oscillate with frequencies  $\omega_{\pm}$
\begin{equation}
    \delta \vec{c}(t,r)  =\vec{\beta}_+(r)e^{-i\omega_+ t}+\vec{\beta_-}(r)e^{-i\omega_- t}~, \qquad \delta \theta(t,r)  = \gamma_+(r) e^{-i\omega_+ t}+\gamma_-(r) e^{-i\omega_- t}~.
    \label{eq:deltactheta}
\end{equation}
Then, our system of equations \eqref{eq:perteq} for the fluctuations becomes
\begin{equation}
    \left[\frac{d^2}{dr_*^2}-\textbf{B}(r)\frac{d}{dr_*}+\omega_\pm^2+i\omega_\pm\textbf{A}(r)-\textbf{C}(r)\right]\vec{\xi}_{\pm}=0~.
\end{equation}
Let us write the  boundary expansions for the fields in the form
\begin{equation}
    \delta\vec{c}(t,r)  \approx \delta \vec{c}^{(0)}(t)+\frac{\delta \vec{c}^{(1}(t)}{r}+...~, \qquad \vec{\beta}_\pm(r)  = \vec{\beta}_\pm^{(0)}+\frac{\vec{\beta}_\pm^{(1)}}{r}+...~.
\end{equation}
Plugging these expansions in \eqref{eq:deltactheta} we get
\begin{equation}
    \delta \vec{c}^{(0)}(t)  = \vec{\beta}_+^{(0)}e^{-i\omega_+ t}+\vec{\beta}_-^{(0)}e^{-i\omega_- t}~, \qquad \delta \vec{c}^{(1)}(t)  = \vec{\beta}_+^{(1)}e^{-i\omega_+ t} + \vec{\beta}_-^{(1)}e^{-i\omega_- t}~.
\end{equation}
Comparing the first of these equations with \eqref{eq:deltacbdry} we conclude that
\begin{equation}
    \vec{\beta}_\pm^{(0)}=-\frac{i}{\omega}\textbf{M}_\pm\vec{\epsilon}~.
    \label{eq:betapm0}
\end{equation}
The subleading vectors $\vec{\beta}_\pm^{(1)}$ determine the variation of the current $\delta \vec{\mathcal{J}}(t)=O(t)\delta \vec{c}^{(1)}(t)$, namely
\begin{equation}
    \delta \vec{\mathcal{J}}(t)=e^{-i\omega t}\Big(\textbf{M}_+\vec{\beta}_+^{(1)}+\textbf{M}_-\vec{\beta}_-^{(1)}+\textbf{M}_+\vec{\beta}_-^{(1)}e^{2i\Omega t}+\textbf{M}_- \vec{\beta}_+^{(1)}e^{-2i\Omega t}\Big)~.
    \label{eq:deltacJ}
\end{equation}
For a regular solution the vectors $\vec{\beta}_\pm^{(1)}$ and $\vec{\beta}_\pm^{(0)}$ are related. Let us write this relation as
\begin{equation}
    \vec{\beta}_\pm^{(1)}=\mathbf{X}_\pm\vec{\beta}_\pm^{(0)}=-\frac{i}{\omega}\boldsymbol{X}_\pm\boldsymbol{M}_\pm\vec{\epsilon}~,
    \label{eq:beta1and0}
\end{equation}
where ${\textbf{X}_{\pm}}$ are $2\times 2$ matrices that, in general, must be determined numerically (see \cite{Garbayo_2020} for details). Plugging \eqref{eq:beta1and0} into \eqref{eq:deltacJ} we get a relation between the current $\delta{\vec{\cal{J}}}$ and the applied electric field $\vec{\epsilon}$
\begin{equation}
    \delta \vec{\mathcal{J}}=\Big[\boldsymbol{\sigma}(\omega)e^{-i\omega t}+\boldsymbol{\sigma}^{+}(\omega)e^{-i(\omega+2\Omega)t}+\boldsymbol{\sigma}^{-}(\omega)e^{-i(\omega-2\Omega)t}\Big]\vec{\epsilon}~,
\end{equation}
where $\boldsymbol{\sigma}(\omega)$, $\boldsymbol{\sigma^+}(\omega)$ and $\boldsymbol{\sigma}^-(\omega)$ are the conductiviy matrices corresponding to the frequencies $\omega$, $\omega+2\Omega$ and $\omega-2\Omega$, given by
\begin{equation}
\begin{aligned}
    \boldsymbol{\sigma}(\omega) & = -\frac{i}{\omega}\left(\textbf{M}_+\textbf{X}_+\textbf{M}_++\textbf{M}_-\textbf{X}_-\textbf{M}_-\right)~,\\
    \boldsymbol{\sigma}^+(\omega) & = -\frac{i}{\omega}\textbf{M}_-\textbf{X}_-\textbf{M}_+~,\\
    \boldsymbol{\sigma}^-(\omega) & = -\frac{i}{\omega}\textbf{M}_+\textbf{X}_+\textbf{M}_-~.
    \label{eq:sigmasgeneralapp}
\end{aligned}
\end{equation}

\subsection{Massless case}\label{app:masslessconduc}

In the massless case the fluctuations of the embedding function decouple from those of the gauge field $\delta \vec{c}$. Therefore, since we are interested in computing conductivities, we can concentrate in studying the equations for $\delta c_x$ and $\delta c_y$. In order to write these  equations in a more convenient form, let us define the following differential operators

\begin{equation}
\begin{aligned}
    \mathcal{O}_1 & \equiv \partial_t^2 + \frac{(r^4-r_h^4)(r_c^4-r^4)}{\rho^4(r_c^4+r^4-2r_h^4)}\partial_r^2 
    - \frac{2(r^4-r_h^4)(r_c^4-r_h^4)}{r^2(r_c^4+r^4-2r_h^4)}\partial_t\partial_r  + \frac{4r(r_c^4-r_h^4)}{(r_c^4+r^4-2r_h^4)}\partial_t
    \\
    & \hspace{0.45\textwidth}- \frac{2(r^4-r_h^4)^2(r_c^4+r^4)}{r^5(r_c^4+r^4-2r_h^4)}\partial_r~, \\
    \mathcal{O}_2 & \equiv -2\partial_t + \frac{2(r^4-r_h^4)(r_c^4-r_h^4)}{\rho^2(r_c^4+r^4-2r_h^4)}\partial_r 
    + \frac{4r(r^4-r_h^4)}{r_c^4+r^4-2r_h^4}~.
\end{aligned}
\end{equation}
Then, one can show that $\delta c_x$ and $\delta c_y$ satisfy the following system of coupled second-order differential equations
\begin{align}
    \left(\mathcal{O}_1-\Omega^2\right)\delta c_x+\Omega\h\mathcal{O}_2\h\delta c_y & =0~, & \left(\mathcal{O}_1-\Omega^2\right)\delta c_y+\Omega\h\mathcal{O}_2\h\delta c_x & =0~.
\end{align}
To decouple these equations, let us consider the following  complex combinations of $\delta c_x$ and $\delta c_y$
\begin{align}
    \eta(t,r) & \equiv \delta c_x(t,r)+i\delta c_y(t,r)~, & \tilde{\eta}(t,r) & \equiv \delta c_x(t,r)-i\delta c_y(t,r)~.
\end{align}
Notice that $\tilde{\eta}$ is not the complex conjugate of $\eta$ since $\delta c_x$ and $\delta c_y$ are not necessarily real. It is straightforward to verify that the equations for $\eta$ and $\tilde{\eta}$ are indeed decoupled and given by
\begin{align}
    \left(\mathcal{O}_1-\Omega^2\right)\eta-i\Omega\h\mathcal{O}_2\h\eta & = 0~, & \left(\mathcal{O}_1-\Omega^2\right)\tilde{\eta}-i\Omega\h\mathcal{O}_2\h\tilde{\eta} & = 0~.
\end{align}
Let us now separate variables as
\begin{align}
    \eta(t,r) & = \beta(r)\h e^{-i\omega t}~, & \tilde{\eta}(t,r) & = \tilde{\beta}(r)\h e^{-i\omega t}~,
\end{align}
for some frequency $\omega$. Then, we obtain the following differential equation for $\beta$
\begin{align}
\begin{split}
    r(r^4-r_c^4)(r^4-r_h^4)\beta''  =& -2(r^4-r_h^4)\left[r^8-r_c^4r_h^4+r^4(r_c^4-r_h^4)-ir^3(r_c^4-r_h^4)(\omega-\Omega)\right]\beta'\\
    &-r^5(\omega-\Omega)\left[4ir(r_c^4-r_h^4)+r^4(\omega-\Omega)+(r_c^4-2r_h^4)(\omega-\Omega)\right]\beta ~.
\end{split}
\end{align}
The equation for $\tilde{\beta}$ is the same, but with $(\omega+\Omega)$ instead of  $(\omega-\Omega)$. Then, remarkably, one can find the following general solutions 
\begin{equation}
\begin{aligned}
    \beta(r)&=e^{i\frac{\Omega-\omega}{4r_h}\Lambda(r/r_h)}\left[A+Be^{-i\frac{\Omega-\omega}{2r_c}\Lambda(r/r_c)}\right]~,\\
    \tilde{\beta}(r)&=e^{-i\frac{\Omega+\omega}{4r_h}\Lambda(r/r_h)}\left[\tilde{A}+\tilde{B}e^{i\frac{\Omega+\omega}{2r_c}\Lambda(r/r_c)}\right]~,
    \label{eq:betassolution}
\end{aligned}
\end{equation}
where $\Lambda$ is the function defined in \eqref{eq:Lambdadef} and $A$, $B$, $\tilde A$ and $\tilde B$ are complex constants which are determined by imposing boundary conditions both at the IR and UV. First of all, we write the solutions we found in terms of the tortoise coordinates $(\tau, r_*)$ of \eqref{eq:tortoisesolution}. Actually, by inspecting the expression of $\eta$ obtained from \eqref{eq:betassolution} one easily demonstrates that, in terms of the tortoise variables, it can be simply written as
\begin{equation}
    \eta(\tau,r_*)=e^{\frac{\Omega}{4r_h}\Lambda(r/r_h)}\left[A~e^{-i\omega (\tau+r_*)}+B~e^{-2i\Omega r_*}~e^{-i\omega(\tau-r_*)}\right]~.
\end{equation}
It is now clear that $\eta(\tau, r_*)$ is the superposition of ingoing and outgoing waves at the pseudohorizon. The infalling regularity condition requires that $B$ vanishes. Then, writing $\eta$ in our original $(t,r)$  coordinates, we have
\begin{equation}
    \eta(t,r)=A~e^{-i\frac{\omega-\Omega}{4r_h}\Lambda(r/r_h)}~e^{-i\omega t}~.
\end{equation}
We can proceed similarly with $\tilde{\eta}$ and conclude that we should require that $\tilde B=0$. Therefore,
\begin{equation}
    \tilde{\eta}(t,r)=\tilde{A}~e^{-i\frac{\omega+\Omega}{4r_h}\Lambda(r/r_h)}~e^{-i\omega t}~.
\end{equation}
Therefore, we obtain that the fluctuations $\delta c_x$ and $\delta c_y$ regular at the pseudohorizon are
\begin{equation}
\begin{aligned}
    \delta c_x(t,r)&=\frac{1}{2}\left[A~e^{-i\frac{\omega-\Omega}{4r_h}\Lambda(r/r_h)}+\tilde{A}~e^{-i\frac{\omega+\Omega}{4r_h}\Lambda(r/r_h)}\right]e^{-i\omega t}~,\\
    \delta c_y(t,r)&=\frac{1}{2i}\left[A~e^{-i\frac{\omega-\Omega}{4r_h}\Lambda(r/r_h)}-\tilde{A}~e^{-i\frac{\omega+\Omega}{4r_h}\Lambda(r/r_h)}\right]e^{-i\omega t}~.
    \label{eq:deltacxcy}
\end{aligned}
\end{equation}
Let us now impose the boundary conditions at the UV. To fulfill the UV boundary condition \eqref{eq:deltacbdry} we sum two solutions of the form \eqref{eq:deltacxcy} with frequencies $\omega_+=\omega+\Omega$ and $\omega_-=\omega-\Omega$. Let $A_{\pm}$ and $\tilde{A}_{\pm}$ denote the constants in \eqref{eq:deltacxcy} with frequency $\omega_{\pm}$. From the leading UV terms we get that, in order to satisfy the boundary condition \eqref{eq:betapm0}, the constants $A_{\pm}$ and $\tilde{A}_{\pm}$ must be related to $\epsilon_x$ and $\epsilon_y$ as
\begin{align}
    A_+ & = -\frac{i}{\omega}(\epsilon_x+i\epsilon_y)~, & \tilde{A}_- & = -\frac{i}{\omega}(\epsilon_x-i\epsilon_y)~, & \tilde{A}_+ & = A_-=0~.
\end{align}
Moreover, from the analysis of the subleading UV terms we conclude that
\begin{equation}
    \vec{\beta}_\pm^{(1)}=\textbf{M}_\pm\vec{\epsilon}~.
    \label{eq:betapm1}
\end{equation}
By comparing \eqref{eq:betapm1} and \eqref{eq:beta1and0}, we get that the matrices $\mathbf{X}_\pm$ are given by
\begin{equation}
    \textbf{X}_\pm=i\omega \mathbf{1}~.
\end{equation}
Plugging these $\mathbf{X}_\pm$ matrices in \eqref{eq:sigmasgeneralapp}, we obtain the conductivity matrices in the massless case, namely
\begin{align}
    \boldsymbol{\sigma}(\omega) & =\boldsymbol{1}~, & \boldsymbol{\sigma}^+(\omega) & = \boldsymbol{\sigma}^-(\omega)=0~,
\end{align}
which is exactly the same result as the one found in \cite{Garbayo_2020} at zero temperature. As in the case of the non-linear current, here also the result confirms the expectations put forward in \cite{Karch_2010}.

\setcounter{equation}{0}
\setcounter{appendixsection}{\value{appendixsection}+1} 
\section{Phase space structure for the D3/D7 system}\label{app:D3D7}
\begin{figure}
    \centering
    \begin{subfigure}[t]{0.48\textwidth}
        \centering
        \includegraphics[width=\textwidth]{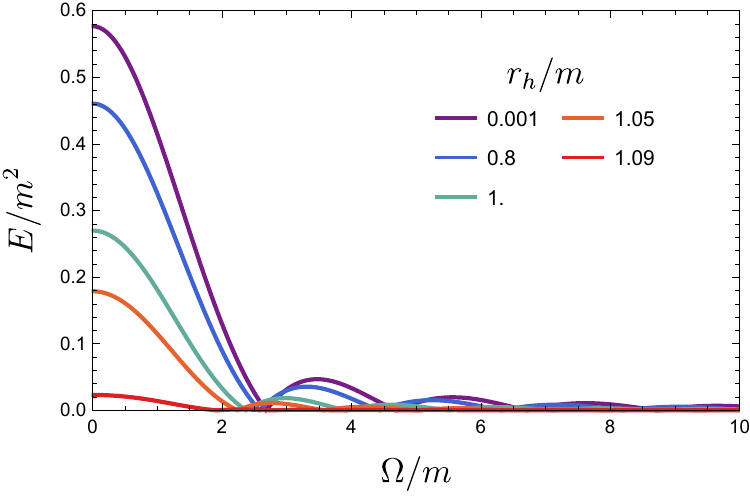}
    \end{subfigure}
    \hspace{0.02\textwidth}
    \begin{subfigure}[t]{0.48\textwidth}
        \centering
        \includegraphics[width=\textwidth]{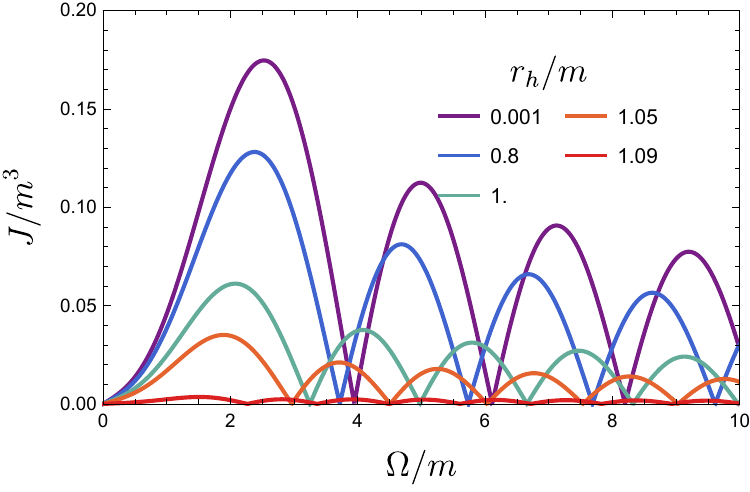}
    \end{subfigure}
    \caption{Electric field and current of the critical embeddings versus driving frequency, for different values of $r_h$ in the D3/D7 system. The structure is mainly the same as the one found for the D3/D5 case. Here the maximum temperature for the Minkowski embeddings is slightly higher, while the height of the lobes is suppressed faster as one increases the driving frequency.}
    \label{fig:LobesEJD3D7}
\end{figure}

It is worth mentioning that the existence of these Floquet suppression points is not restricted to the D3/D5 system. In Fig. \ref{fig:LobesEJD3D7} we reproduce the lobe structure for the D3/D7 system, which is analogous to the one found in Fig. \ref{fig:LobesEJD3D5} for D3/D5. The $r_h\to0$ limit for the electric field plot coincides, of course, with the one studied in \cite{Kinoshita_2017}. 

We find again a set of points where the induced current vanishes while the electric field is close to its maximum value within the lobe. This seems to indicate the presence of a range of frequencies for which, in Minkowski embeddings, $\abs{j}=0$ with $\abs{E}\neq0$, ranging from the critical frequency of the current $\Omega_{c,j}$ ($c$ stands for critical) up to some maximum value $\Omega_{m,j}$ ($m$ stands form meson). 

\begin{figure}
    \centering
    \begin{subfigure}[t]{0.48\textwidth}
        \centering
        \includegraphics[width=\textwidth]{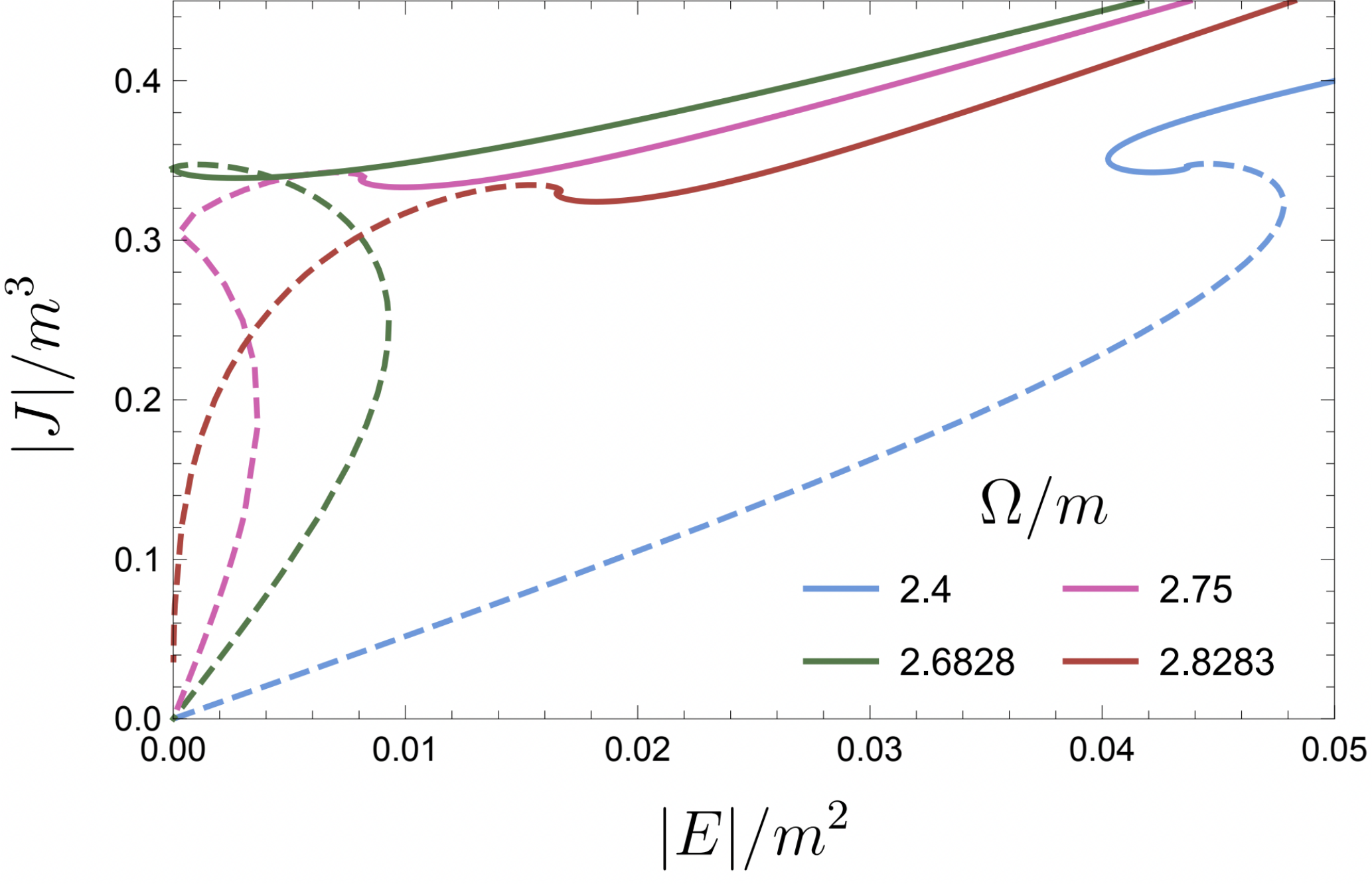}
    \end{subfigure}
    \hspace{0.02\textwidth}
    \begin{subfigure}[t]{0.48\textwidth}
        \centering
        \includegraphics[width=\textwidth]{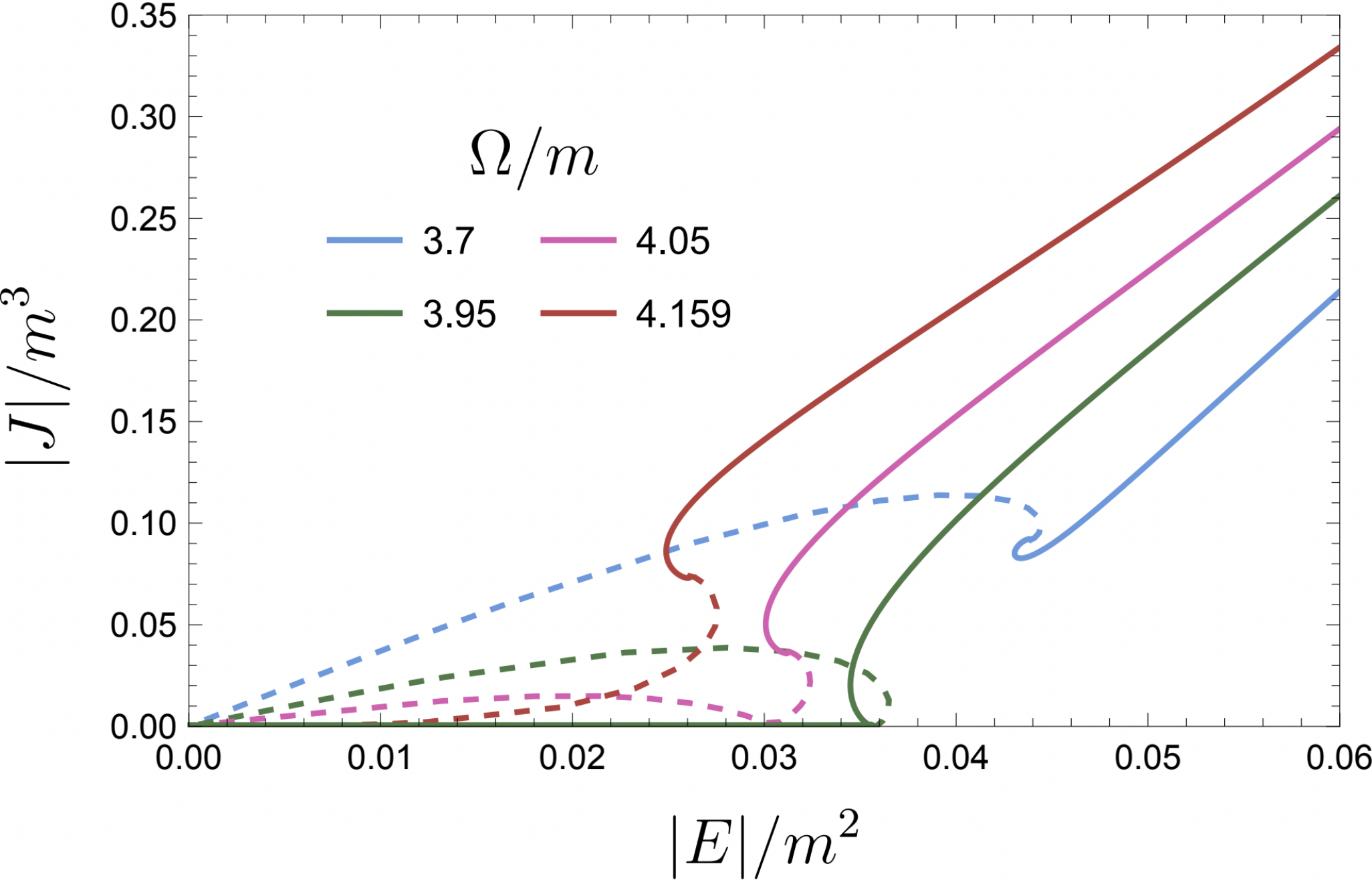}
    \end{subfigure}
    \caption{Electric current $|J|/m^3$ versus electric field $|E|/m^2$ for $r_h/m=0.001$. The solid curves represent the insulator (Minkowski) phase and the dashed curves the conductive (BH) phase.  The driving frequency is fixed to some $\Omega/m<\Omega_c/m$ (blue), $\Omega/m=\Omega_{c}/m$ (green), $\Omega_{c}/m<\Omega/m<\Omega_{meson}/m$ (pink) and $\Omega/m=\Omega_{meson}/m$ (red). The curves on the left, with $\Omega/m \sim 2.68$  scan the region close to the Floquet vector meson condensates and were obtained in \cite{Kinoshita_2017}. On the right, for $\Omega/m \sim 3.95$ we do the same around the first Floquet suppression point.}
    \label{fig:JvsED3D7}
\end{figure}

The analytic computation of $\Omega_{m,j}$ in the $T=0$ limit of the D3/D7 system is more complicated than the one corresponding to the D3/D5 intersection but it can, however, be obtained numerically. We have found that the Floquet suppression points first appear for $\Omega/m\in\left(3.950,4.159\right)$, and plotted in Fig. \ref{fig:JvsED3D7} the modulus of the current and the electric field for frequencies close or within that range. As anticipated, the similarity with the structure studied in \cite{Kinoshita_2017} is quite obvious, and thus we conclude that a 3D diagram as that of Fig. \ref{fig:3Dj} is also found when studying D3/D7 branes.

\restoredefaultnumbering
\restoredefaultsectioning

%% file: Part2/HelicalB/HelicalB.tex
\renewcommand{\chapterquote}{\begin{flushright}
\begin{minipage}{0.45\textwidth}
\color[gray]{0.6}
\textit{They want my heart in the right place}\\
\textit{But when I look down}\\
\textit{It beats on the left.}
\end{minipage}
\begin{minipage}{0.47\textwidth}
\raggedleft
\textit{Sie wollen mein Herz am rechten Fleck}\\
\textit{Doch seh ich dann nach unten weg}\\
\textit{Da schl\"{a}gt es links.}
\end{minipage}

\vspace{2em}

\normalfont— Rammstein, \textit{Links 2, 3, 4}.
\end{flushright}}

\chapter{Chiral symmetry breaking and restoration by helical magnetic fields}\label{chap:HelicalB}

Almost all the material in this Chapter is based on, and adapted from, \textit{Chiral symmetry breaking and restoration by helical magnetic fields} \cite{Berenguer_2025}.

We study the effects of helical magnetic fields on chiral symmetry breaking within the D3/D7-brane model. By analyzing the brane embeddings, we obtain three types of massless solutions, corresponding to three phases with different behavior in the dual field theory. From the study of quark condensates, free energy, and electric currents, we find that helical magnetic fields can counteract uniform-field-induced symmetry breaking, driving the system towards symmetry restoration. We also find an effect analogous to the chiral magnetic effect whereby the current is parallel to the magnetic field. We further study the massive case, and find that the helical configuration is less effective in erasing the first-order phase transition that is present in the case of a constant magnetic field.

This Chapter deviates slightly from the main focus of this thesis, in the sense that it is the only Chapter in which we are not going to deal with a non-equilibrium system. However, it was the rotating ansatz for the gauge field in Chapter \ref{chap:FloquetII} what motivated the helical magnetic field that we introduce here.

\vspace{1cm}

\section{Introduction}

Among the various holographic models developed during this thesis, the D3/D7-brane intersection \cite{KarchKatz_2002} (see Section \ref{sec:D3D7}) has emerged as a prominent tool for studying fundamental aspects of QCD, a framework commonly referred to as AdS/QCD. In this model, a probe D7-brane is embedded into an asymptotically AdS spacetime, providing a powerful framework for non-perturbative analyses of gauge theory dynamics, particularly in the presence of external fields and finite temperature.

A key feature of low-energy QCD is the spontaneous breaking of chiral symmetry \cite{Bergman_2012}. Within the AdS/QCD framework, this phenomenon can be robustly modeled. For instance, it is well known that a uniform magnetic field induces chiral symmetry breaking \cite{Filev_2007,Filev_2007_2}.

However, heavy ion collision experiments suggest that the magnetic fields generated in such environments are neither static nor uniform. In particular, helical magnetic fields have been identified as a relevant configuration \cite{Skokov_2009,Voronyuk_2011,Bzdak_2011,Deng_2012}. The interplay between such non-uniform magnetic fields and the dynamics of chiral symmetry breaking and restoration remains poorly understood. Therefore, developing methods to incorporate such spatially varying magnetic fields within AdS/QCD is crucial for gaining deeper insight into real-world QCD dynamics.

In this work, we investigate the effects of helical magnetic fields on chiral symmetry dynamics within the AdS/QCD framework. The theory of interest is the $\mathcal{N}=4$ SYM in $(3+1)$-dimensions with gauge group $SU(N_c)$, coupled to a number $N_f\ll N_c$ of $\mathcal{N}=2$ flavor hypermultiplets in the fundamental representation of $SU(N_c)$. Holographically, this system is modeled as $N_f$ probe D7-branes embedded in the AdS$_5\times S^5$ geometry sourced by a stack of $N_c$ coincident D3-branes \cite{KarchKatz_2002} (see Chapter \ref{chap:Branes}). Additionally, we include an axial $U(1)_A$ field $A_j^5=b/2\h\delta_{jz}$, which describes a Weyl semimetal. The holographic realization of this setup was developed in \cite{Fadafan_2020}. 

Employing the model described above, we introduce a spatially varying magnetic field with non-zero helicity to explore its influence on chiral symmetry breaking and restoration. Our results show that the helical structure of the magnetic field counteracts the symmetry-breaking effects of a constant magnetic field, ultimately driving the system towards chiral symmetry restoration. We are also interested in the possibility that a helical magnetic field could induce the helical structure of the brane, determined by the parameter $b$. We find that this effect is not induced by the helical magnetic field.

The Chapter is organized as follows: In Section \ref{sec:HelicalB} we introduce the model and relate the bulk fields to the gauge theory quantities at the boundary. We also outline the different types of solutions encountered. Section \ref{sec:CSBrestoration} focuses on the massless solutions, classifying them according to their chiral symmetry breaking pattern. We compare their free energies, chiral condensates and response currents, demonstrating that sufficiently helical magnetic fields restore chiral symmetry. We extend the analysis to the massive case in Section \ref{sec:CSBmassive}. We conclude in Section \ref{sec:HelicalBconclusions} with an outlook and future directions. Additional details of the derivations and analysis are relegated to Appendices \ref{app:holorenohelical} and \ref{app:discretescale}.

\section{Helical magnetic fields in the D3/D7 model}\label{sec:HelicalB}

We consider the AdS$_5\times S^5$ geometry generated by a stack of $N_c$ D3-branes. Since we will be working at zero temperature, we do not need to introduce the isotropic coordinate $u$ and the most suitable coordinates are the cartesian coordinates defined in \eqref{eq:AdS5xS5wrho}:
\begin{equation}
    ds^2=\frac{\rho^2+w^2}{L^2}[-dt^2+dx^2+dy^2+dz^2]+\frac{L^2}{\rho^2+w^2}[d\rho^2 + dw^2 +\rho^2 d\Omega_3^2 +w^2d\phi^2]~.
  \label{eq:AdS5xS5helical}
\end{equation}

The 4-form potential is given by
\begin{equation}
    C_4=\frac{(\rho^2+w^2)^2}{L^4}dt\wedge dx \wedge dy \wedge dz - \frac{L^4\rho^4}{(\rho^2+w^2)^2} d\phi \wedge 
 d\Omega_3~.
\end{equation}
We use the same orientation between the D3 and the D7-branes as the one explained in Section \ref{sec:D3D7}, which we summarize again in Table \ref{tab:D3D7helical} for convenience.

To introduce a helical magnetic field, we consider the following ansatz\footnote{Notice that this ansatz is motivated by the rotating electric field ansatz considered in the previous Chapter, $A_x+iA_y\propto e^{i\Omega t}$, based on the previous works \cite{Hashimoto_2016,Kinoshita_2017,Ishii_2018,Garbayo_2020,Berenguer_2022}. We will see that the equations of motion of the brane are given by ordinary differential equations even though the gauge field is $z$-dependent.}
\begin{equation}
    2\pi\alpha'(A_x+iA_y)=a(r)e^{ikz}~,\qquad w=w(r)~,\qquad \phi=b\h z
    \label{eq:ansatzhelical}
\end{equation}

This choice explicitly breaks translational symmetry along the $z$-direction. Although we could include an $r$-dependence for $\phi$ as $\phi = bz + \Phi(r)$, any such dependence leads to a non-zero imaginary part in the action, signalling a tachyonic instability \cite{Fadafan_2020}, which ultimately forces $\Phi(r)$ to be constant. Under the above ansatz, we obtain
\begin{equation}
\begin{aligned}
    (2\pi \alpha')^2 F \wedge F &= -2aa'k\, d\rho \wedge dx \wedge dy \wedge dz~, \\
    P[C_4] &= (\rho^2+w^2)^2\, dt \wedge dx \wedge dy \wedge dz - \frac{\rho^4}{(\rho^2+w^2)^2} b\, dz \wedge d\Omega_3
\end{aligned}
\end{equation}
Therefore, the Wess-Zumino term vanishes in the setup under consideration. 

Substituting the above ansatz into the DBI action, we obtain
\begin{equation}
    S_{D7}=-\mathcal{N}\int d\rho \frac{\rho^3}{\rho^2+w^2}\sqrt{\left(k^2a^2+b^2w^2+\left(\rho^2+w^2\right)^2\right)\left(1+a'^2+w'^2\right)}~,
    \label{eq:DBIhelical}
\end{equation}
where
\begin{equation}
    \mathcal{N}=2\pi^2N_fT_{D7}=\frac{N_fN_c\h\lambda}{(2\pi)^4}~
    \label{eq:ND3D7}
\end{equation}
and we set $L=1$. In the second equality we have related it to the field theory quantities, and again we have implictly divided by the volume of $\mathbb{R}^{3,1}$. The Wess-Zumino term vanishes in the setup under consideration.

\begin{table}
    \centering
    \begin{tabular}{ccccccccccc}
         & $X_0$ & $X_1$ & $X_2$ & $X_3$ & $X_4$ & $X_5$ & $X_6$ & $X_7$ & $X_8$ & $X_9$\\
        D3 & $\times$ & $\times$ & $\times$ & $\times$ & $-$ & $-$ & $-$ & $-$ & $-$ & $-$ \\
        D7 & $\times$ & $\times$ & $\times$ & $\times$ & $\times$ & $\times$ & $\times$ & $\times$ & $-$ & $-$ \\
    \end{tabular}
    \caption{D3/D7 brane intersection. Directions filled by the branes are denoted by $\times$. Directions orthogonal to them are denoted by $-$.}
    \label{tab:D3D7helical}
\end{table}

The equations of motion derived from this action are
\begin{equation}
\begin{aligned}
    a''&=-\frac{1+a'^2+w'^2}{\rho(\rho^2+w^2)\left(k^2a^2+b^2w^2+(\rho^2+w^2)^2\right)}\Big[a'(\rho^2+3w^2)(k^2a^2+b^2w^2)\Big.\\
    &\phantom{-\frac{1+a'^2+w'^2}{\rho(\rho^2+w^2)\left(k^2a^2+b^2w^2+(\rho^2+w^2)^2\right)}\Big[a'}\Big.+3a'^2(\rho^2`w^2)^3-k^2\rho\h a(\rho^2+w^2)\Big]~,\\
    w''&=-\frac{1+a'^2+w'^2}{\rho(\rho^2+w^2)\left(k^2a^2+b^2w^2+(\rho^2+w^2)^2\right)}\Big[w'(\rho^2+3w^2)(k^2a^2+b^2w^2)\Big.\\
    &\phantom{------------.}\Big.+2k^2\rho\h a^2w+b^2\rho\h w(w^2-\rho^2)+3w'(\rho^2+w^2)^2\Big]~.
    \label{eq:Eomhelical}
\end{aligned}
\end{equation}

In the same lines as before, we will solve numerically the equations of motion \eqref{eq:Eomhelical} to determine the profiles $w(\rho)$ and $a(\rho)$.

\subsection{Physical quantities in the boundary theory}
By using the equations of motion near the AdS boundary, $\rho\to\infty$, the fields $w(\rho)$ and $a(\rho)$ can be expanded as
\begin{equation}
\begin{aligned}
    w(\rho) &=M+\frac{C}{\rho^2}-\frac{b^2M}{2\rho^2}\log\left(\frac{\rho}{M}\right)+...\\
    a(\rho) & = a_\infty+\frac{J}{2\rho^2}-\frac{k^2a_\infty}{2\rho^2}\log\left(\frac{\rho}{\sqrt{ka_\infty}}\right)+...
    \label{eq:UVexpansionhelical}
\end{aligned}
\end{equation}
Logarithmic terms appear in the asymptotic expansion. In their arguments, we have normalized the radial coordinate $\rho$ by the constants $\sqrt{k a_\infty}$ and $M$. This normalization can always be done by redefining $J$ and $C$. As we will see shortly, with the above definition of $J$ and $C$, these quantities remain covariant under scaling transformations.

The coefficients in the expansion are related to the physical quantities in the boundary theory. For example, from \eqref{eq:ansatzhelical} and \eqref{eq:UVexpansionhelical}, the asymptotic value of the gauge field, which corresponds to the external gauge field in the boundary theory, is given by $(A_x + i A_y)\rvert_{\rho\to\infty} = a_\infty e^{ikz}/(2\pi\alpha')$. The magnetic field in the boundary theory is given by $\Vec{\mathcal{B}} = \nabla \times (A_x, A_y, A_z)\rvert_{\rho\to\infty}$. Its explicit expression is 
\begin{equation}
    \mathcal{B}_x+i\mathcal{B}_y=-\frac{\sqrt{\lambda}}{2\pi}ka_\infty e^{ikz}~,\qquad \mathcal{B}_z=0~,
    \label{eq:BxByBz}
\end{equation}
where $\lambda$ is the 't Hooft coupling. The gauge field above describes a helical magnetic field: for each $(x,y)$-plane, the magnetic field is homogeneous, but its direction rotates as we vary the coordinate $z$. We define
\begin{equation}
    B=k\h a_\infty~,
    \label{eq:Bfielddef}
\end{equation}
which characterizes the amplitude of the helical magnetic field in the boundary theory. Since the DBI action is invariant under $a(\rho) \to -a(\rho)$, we can assume $B \geq 0$ without loss of generality. A uniform magnetic field, independent of $z$, is recovered in the double-scaling limit $k \to 0$, while keeping $B$ finite. This limit corresponds to the setup considered in \cite{Evans_2024}.

It was shown in \cite{Fadafan_2020} that the parameter $b$ corresponds to a component of the non-dynamical $U(1)_A$ gauge field, $A_\mu^5\rvert$. The other parameters, $(M, J, C)$, are related to the quark mass $M_q$, the electric current $\mathcal{J}$, and the quark condensate $\langle \mathcal{O}_m \rangle$ in the boundary theory, respectively. The correspondence is summarized as follows:
\begin{equation}
\begin{aligned}
    M_q & = \frac{\sqrt{\lambda}}{2\pi} M~, & \quad A^5_\mu & =\frac{1}{2}\partial_\mu\phi=\frac{b}{2}\delta_\mu^z\\
    \langle \mathcal{O}_m \rangle & = \frac{N_c N_f \sqrt{\lambda}}{4\pi^3} \left(-C + \frac{b^2 M}{4}\right)~, & & \\
    \mathcal{J}_x + i \mathcal{J}_y & = \frac{N_f N_c \sqrt{\lambda}}{(2\pi)^3} \left(J + \frac{k B}{4}\right) e^{ikz}~, & \quad \mathcal{J}_z\rvert_\text{QFT} & = 0
    \label{eq:PhysicalQ}
\end{aligned}
\end{equation}
A detailed derivation of $\mathcal{J}$ and $\langle \mathcal{O}_m \rangle$ is provided in Appendix \ref{app:holorenohelical}. In the massless case $M=0$, the parameter $b$ can be removed by a field redefinition in the boundary theory and becomes irrelevant.
We also introduce the free energy as
\begin{equation}
 F =-\left(S_{D7}^{on-shell} + S^{ct}\right)~,
 \label{fdef}
\end{equation}
where $S_{D7}^{on-shell} $ is the on-shell DBI action \eqref{eq:DBIhelical}, and $S^{ct}$ contains the counterterms, whose explicit expressions are given in Appendix \ref{app:holorenohelical}. The free energy will be used to determine the most stable solution.

The equations of motion are invariant under the following scaling transformations:
\begin{equation}
\begin{aligned}
    (t, x, y, z) & \to \lambda^{-1} (t, x, y, z)~,  & \rho & \to \lambda\h \rho ~, & (k, b) & \to\lambda\h (k, b)~,\\
    w(\rho) & \to \lambda\h w(\rho)~, & a(\rho) & \to \lambda\h a(\rho)~, & &
\end{aligned}
\end{equation}
where $\lambda$ is an arbitrary positive constant. Additionally, the transformation laws for quantities in the asymptotic expansion are given by
\begin{equation}
\begin{aligned}
    M & \to \lambda M~,  & C & \to \lambda^3 C ~, & a_\infty & \to \lambda\h a_\infty~,\\
    J & \to \lambda^3 J~, & B & \to \lambda^2 B~, & F & \to \lambda^4 F~.
    \label{scalsym}
\end{aligned}
\end{equation}
We will consider scale-invariant combinations of parameters. We will mostly normalize physical quantities by the magnetic field $B$, since we are interested in the physics of a non-zero helical magnetic field.

\subsection{Types of embeddings}\label{sec:Solutions}
In this section, we numerically obtain the solutions for the fields in the brane worldvolume. The gravitational background is pure AdS$_5\times S^5$ and there is no event horizon.

\subsubsection{Minkowski embeddings}
In this case there is no singular shell. In fact, the equations of motion \eqref{eq:Eomhelical} are regular everywhere, meaning that all solutions extend to the axis $\rho\to 0$. 

Here, we focus on the case where $a_0\equiv a(0)\neq 0$ and $w_0\equiv w(0)\neq 0$. 
For $\rho\sim 0$, the solutions behave as
\begin{equation}
\begin{split}
    a(\rho) & =a_0+\frac{k^2 a_0}{8(k^2 a_0^2+ b^2 w_0^2 + w_0^4)}\rho^2+...~, \\
    w(\rho) & =w_0-\frac{(2 k^2 a_0^2 + b^2 w_0^2)}{8 w_0 (k^2 a_0^2 + b^2 w_0^2 + w_0^4)}\rho^2+...~.
    \label{eq:Minkexpansion}
\end{split}
\end{equation}

\begin{figure}
    \centering
    \begin{subfigure}[t]{0.48\textwidth}
        \centering
        \includegraphics[width=\textwidth]{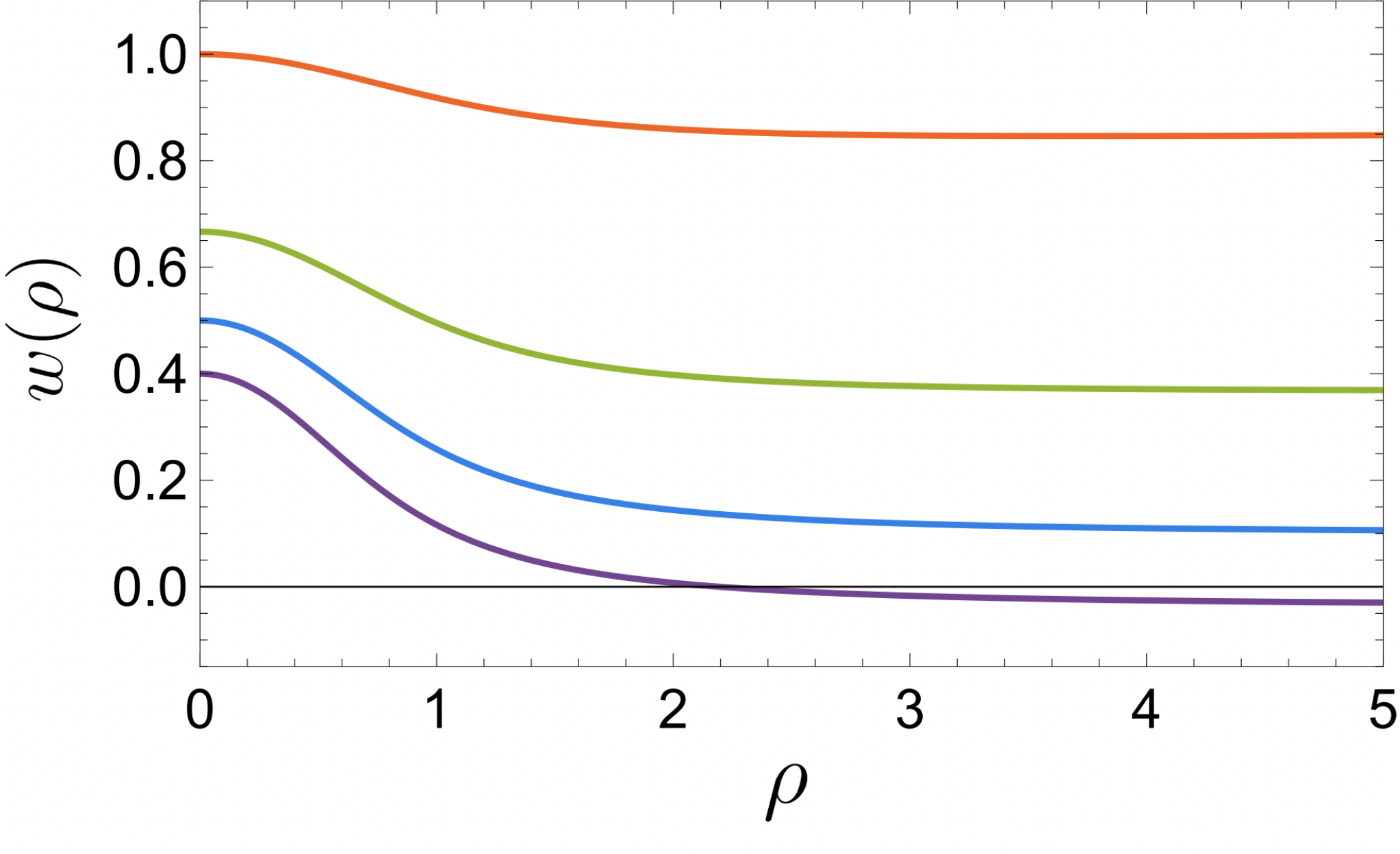}
    \end{subfigure}
    \hspace{0.02\textwidth}
    \begin{subfigure}[t]{0.48\textwidth}
        \centering
        \includegraphics[width=\textwidth]{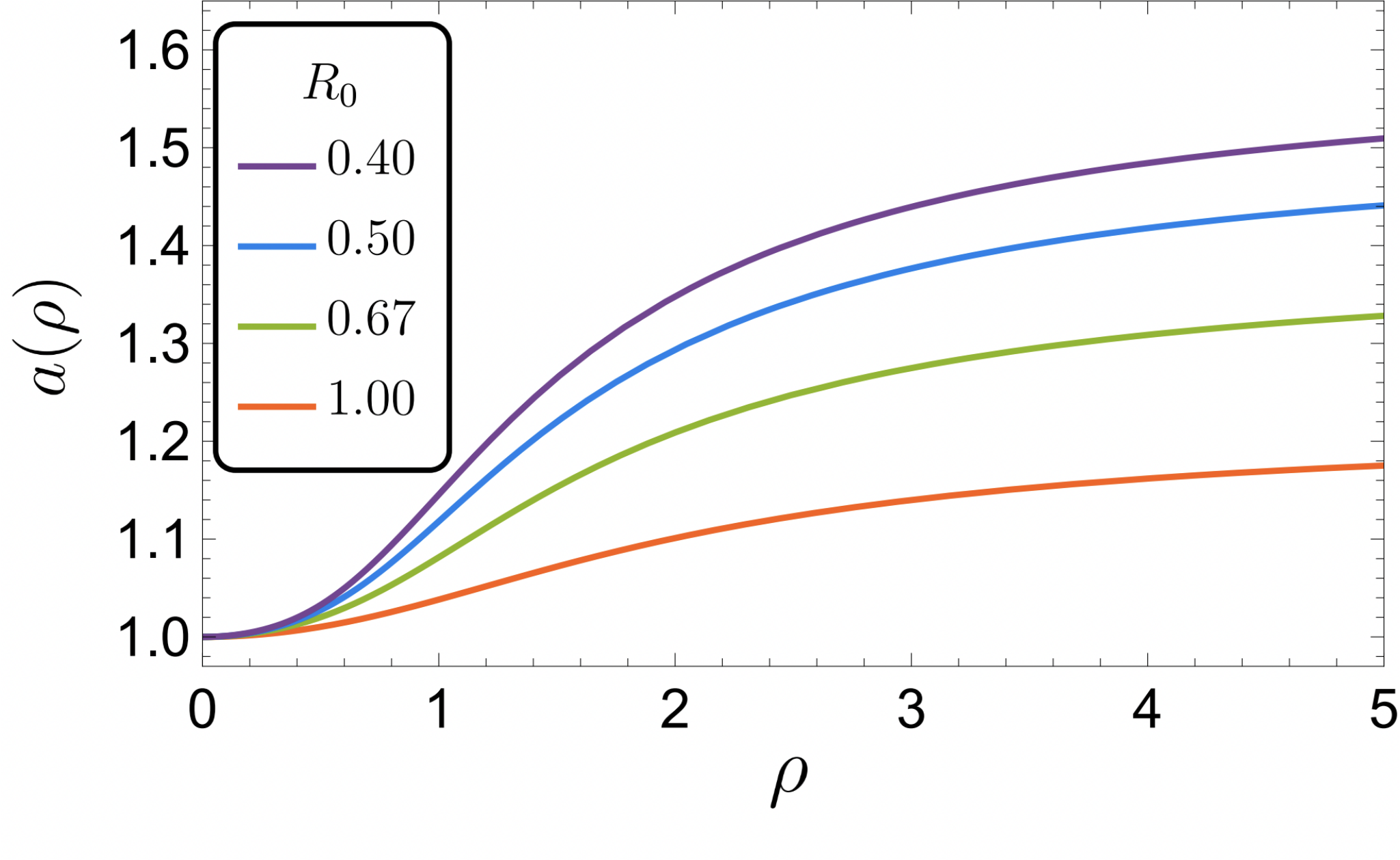}
    \end{subfigure}
    \caption{Typical solutions of $w(\rho)$ (left) and $a(\rho)$ (right) for the Minkowski embedding. We show the solutions for several values of $w_{0}$ while the other parameters are fixed as $k=b=a_{0}=1$.}
    \label{fig:Minksols}
\end{figure}

Fig. \ref{fig:Minksols} shows typical solutions under the above boundary conditions. Here, we vary $w_0$ as $w_0= \{0.40, 0.50, 0.67, 1.00\}$ while the other parameters are fixed as $k=b=a_0=1$. Following the standard nomenclature, we refer to this type of solutions as "Minkowski embeddings".

\subsubsection{Exponential embeddings}
\begin{figure}
    \centering
    \begin{subfigure}[t]{0.48\textwidth}
        \centering
        \includegraphics[width=\textwidth]{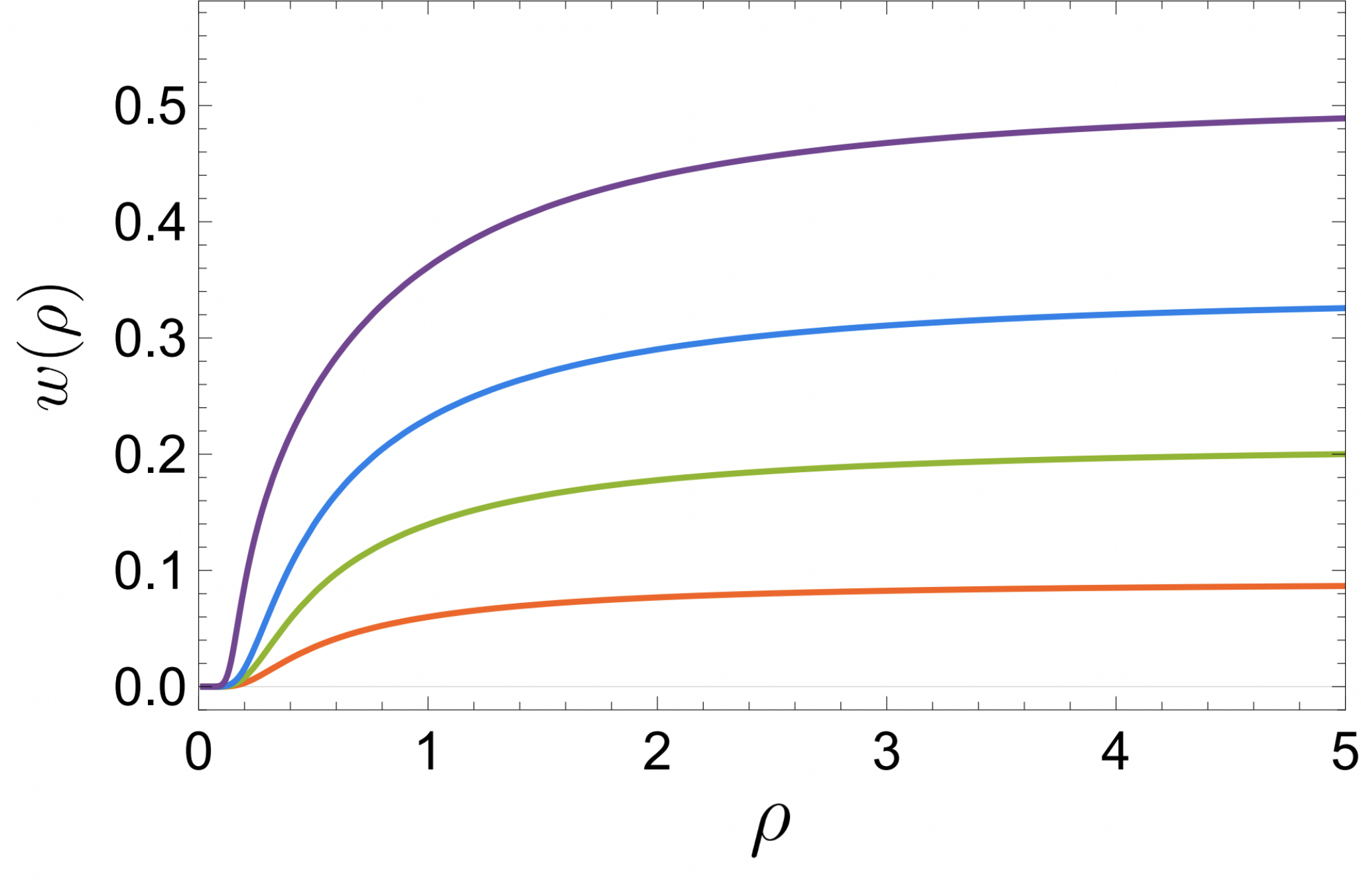}
    \end{subfigure}
    \hspace{0.02\textwidth}
    \begin{subfigure}[t]{0.48\textwidth}
        \centering
        \includegraphics[width=\textwidth]{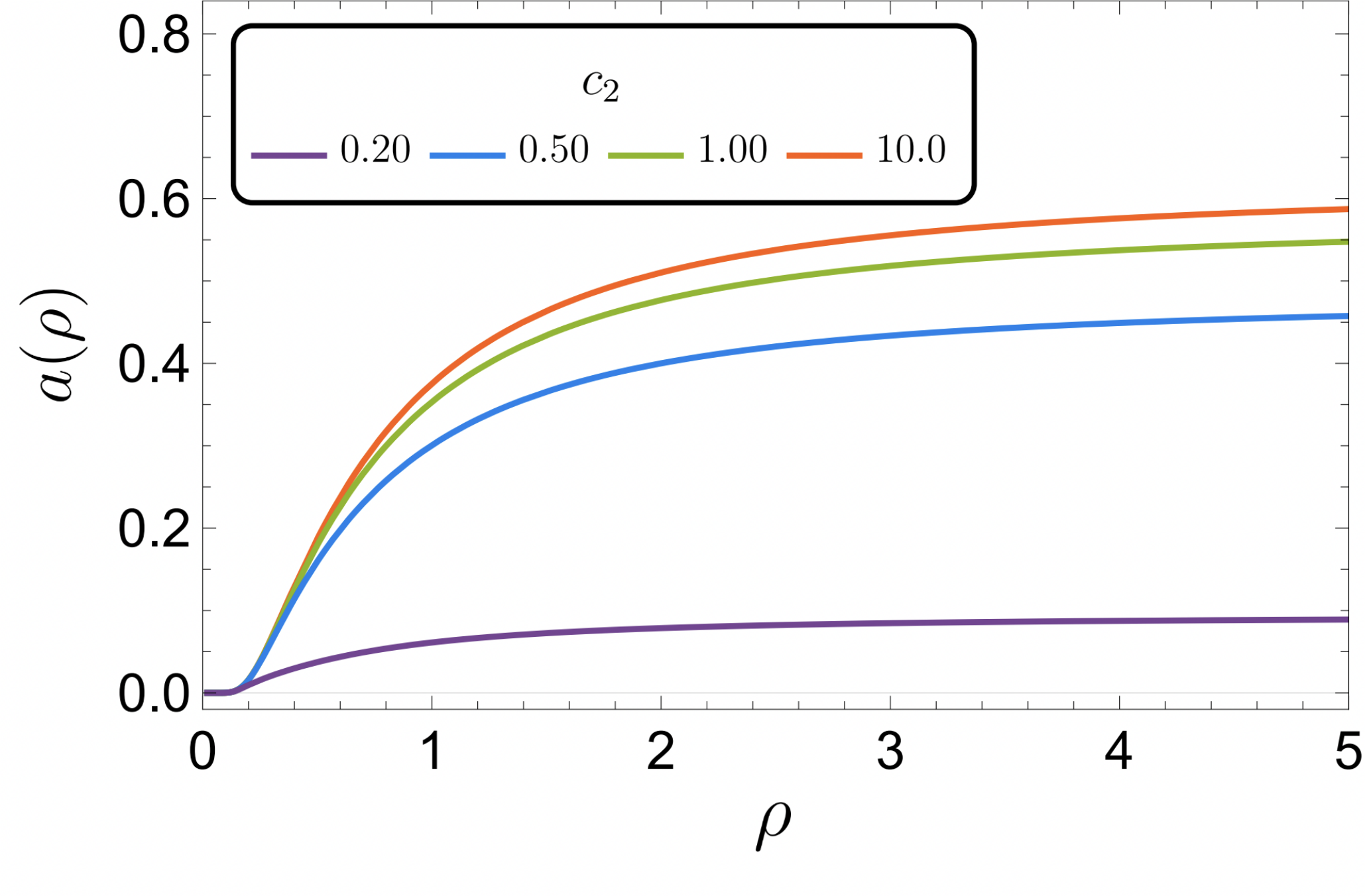}
    \end{subfigure}
    \caption{Typical solutions of $w(\rho)$ (left) and $a(\rho)$ (right) for the exponential embedding. We show the solutions for several values of $c_{2}$ while the other parameters are fixed as $k=b=c_{1}=1$.}
    \label{fig:Expsols}
\end{figure}

For $a_0=w_0=0$, we easily find the trivial solution $a(\rho)=w(\rho)=0$. However, as shown in \cite{Fadafan_2020}, there exists a family of non-trivial solutions for which both $a(\rho)$ and $w(\rho)$ approach $0$ as $\rho\to 0$. To find these solutions, we assume that $a(\rho)$ and $w(\rho)$ are sufficiently small near the axis $\rho=0$ and we consider the linearized equations of motion. Substituting $a(\rho)\to \epsilon\h a(\rho)$ and $w(\rho)\to \epsilon\h w(\rho)$ into the equations of motion \eqref{eq:Eomhelical}, and expanding to first order in $\epsilon$, we obtain the following decoupled equations
\begin{equation}
    a''+\frac{3}{\rho}a'-\frac{k^2}{\rho^4}a=0\ ,\qquad
    w''+\frac{3}{\rho}w'-\frac{b^2}{\rho^4}w=0\ .
\end{equation}
The regular solutions at $\rho=0$ are given by $a\propto K_1(k/\rho)/\rho$ and $w\propto K_1(b/\rho)/\rho$, where $K_\alpha(x)$ is the modified Bessel function of the first kind. In the vicinity of $\rho = 0$, these solutions can be approximated as
\begin{equation}
    a(\rho)\simeq  \frac{c_1}{\sqrt{\rho}}e^{-k/\rho}~,\qquad 
    w(\rho)\simeq \frac{c_2}{\sqrt{\rho}}e^{-b/\rho}~.
    \label{eq:Expexpansion}
\end{equation}
Solutions satisfying these boundary conditions are referred to as "exponential embeddings". Typical profiles of these solutions are shown in Fig. \ref{fig:Expsols}. 

Notice that for the exponential embeddings, the solutions with $M=0$ are possible only when $w$ is the trivial embedding with $c_2=0$, which means there are no chiral symmetry breaking solutions of this type.

\section{Chiral symmetry breaking and its restoration}\label{sec:CSBrestoration}

As we have reviewed in Section \ref{sec:CSB}, an effect of introducing a worldvolume magnetic field is the possibility of triggering spontaneous chiral symmetry breaking. We are going to see that we find a similar behavior here; however, we must now consider the two types of solutions we have found so far.

\subsection{Classifying solutions for \texorpdfstring{$M=0$}{M=0}}
In this section, we only focus on the massless case, $M=0$, and study spontaneous chiral symmetry breaking in the presence of a helical magnetic field. We classify the solutions into three types:
\begin{description}
\item[I] Chiral symmetry preserving Minkowski embeddings:
$$a(\rho)\in F_\text{Mink}\ ,\quad  R(\rho)\equiv 0~.$$
\item[II] Chiral symmetry breaking Minkowski embeddings:
$$a(\rho)\in F_\text{Mink}\ ,\quad  R(\rho)\in F_\text{Mink}~.$$
\item[III] Chiral symmetry preserving exponential embeddings: 
$$a(\rho)\in F_\text{Exp}\ ,\quad  R(\rho)\equiv 0~.$$
\end{description}

\begin{figure}
    \centering
    \begin{subfigure}[t]{0.3\textwidth}
        \centering
        \includegraphics[width=\textwidth]{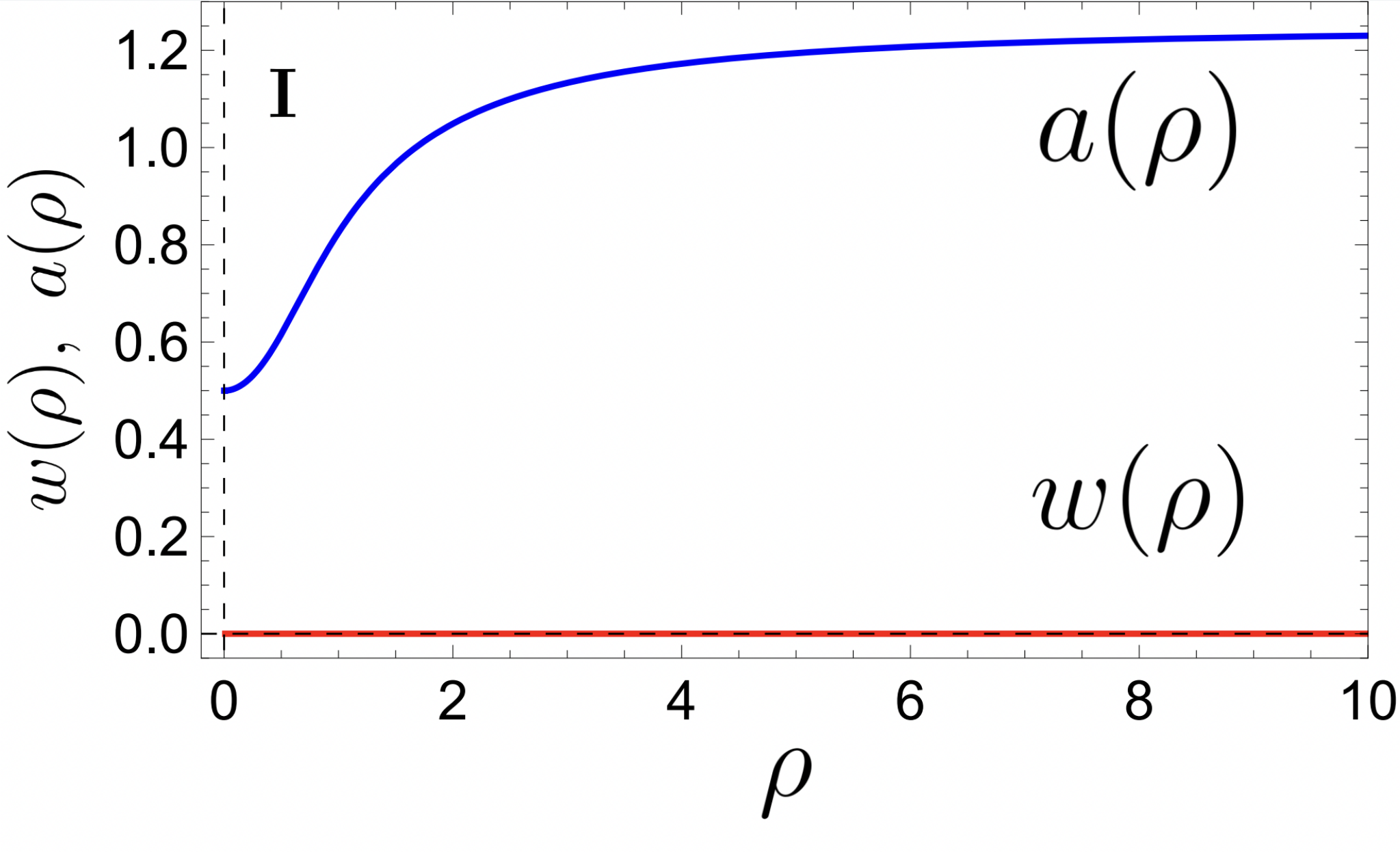}
    \end{subfigure}
    \hspace{0.02\textwidth}
    \begin{subfigure}[t]{0.3\textwidth}
        \centering
        \includegraphics[width=\textwidth]{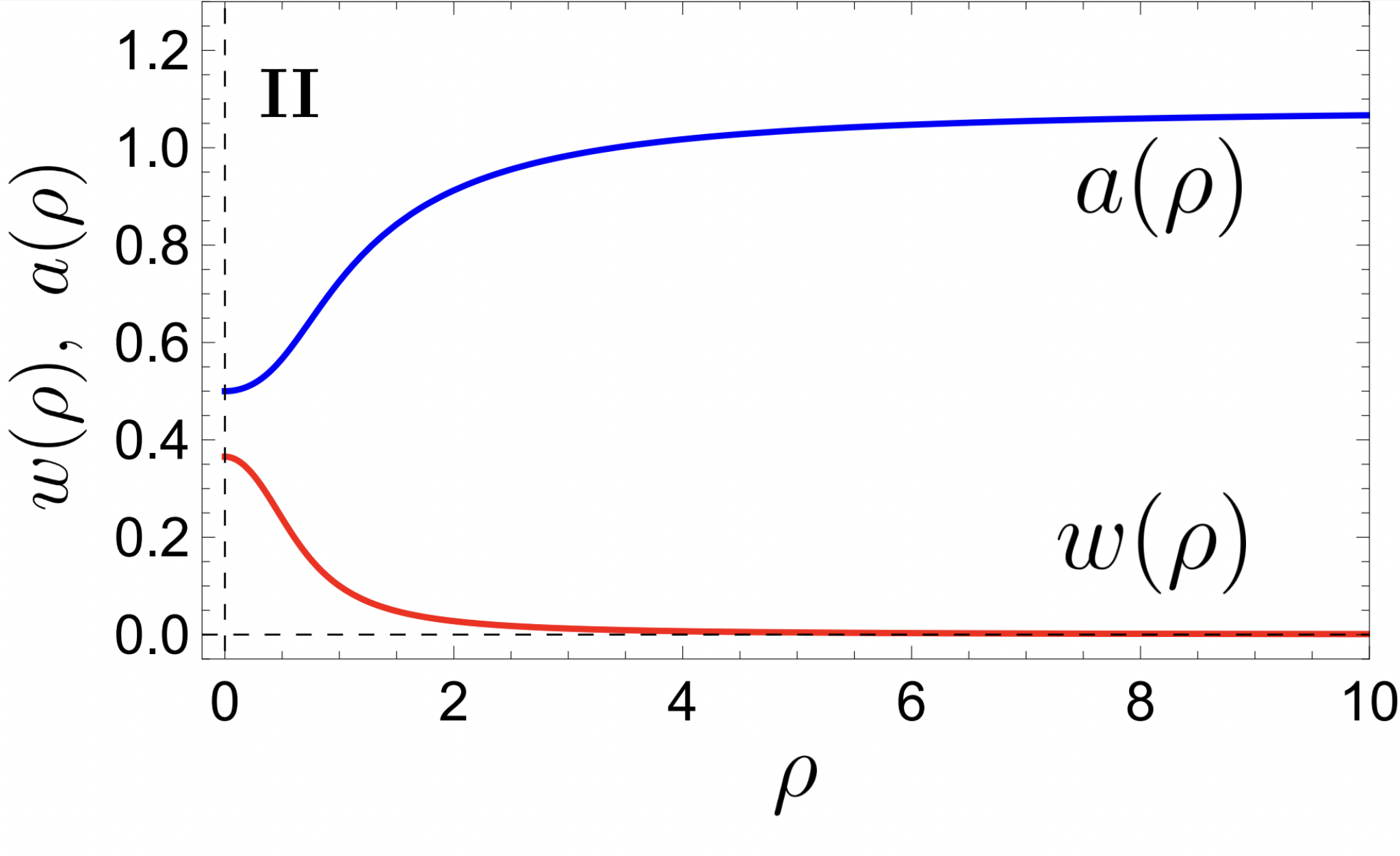}
    \end{subfigure}
    \hspace{0.02\textwidth}
    \begin{subfigure}[t]{0.3\textwidth}
        \centering
        \includegraphics[width=\textwidth]{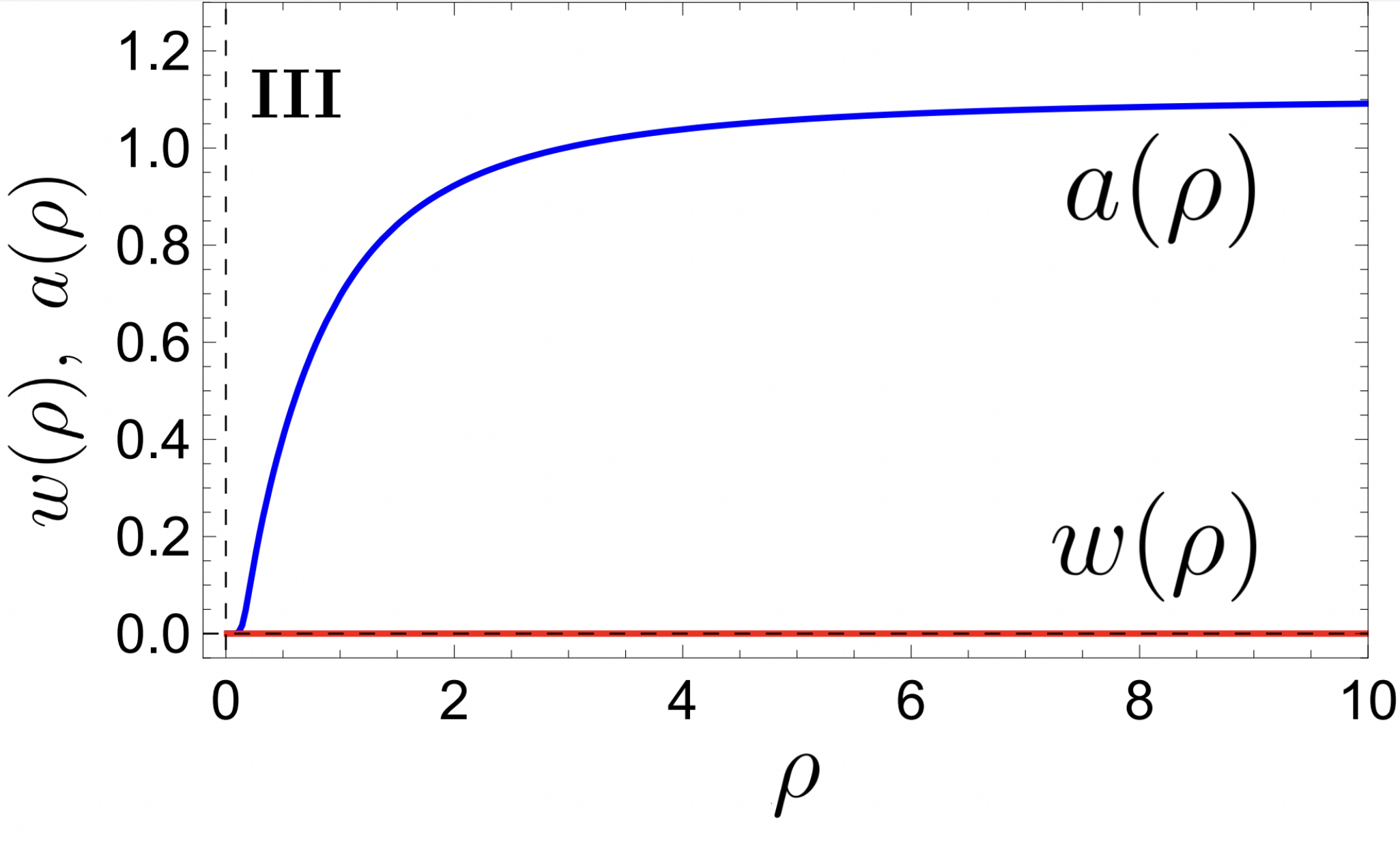}
    \end{subfigure}
    \caption{Typical solutions of $a(\rho)$ and $w(\rho)$ for type I (left), II (middle), and III (right) with $b=0$.}
    \label{fig:3types}
\end{figure}

Here, $F_\text{Mink}$ and $F_\text{Exp}$ are sets of non-trivial functions which behave as described by \eqref{eq:Minkexpansion} and \eqref{eq:Expexpansion} near the axis, respectively. Typical profiles of types I, II and III solutions are shown in Fig. \ref{fig:3types}.

For type II solutions, the brane profile $w(\rho)$ is non-trivial but asymptotically approaches zero as $\rho\to\infty$. In this case, the quark condensate is non-zero and chiral symmetry is spontaneously broken. 

In contrast, for type III solutions, where $a(\rho)\in F_\text{Exp}$ and $w(\rho)\in F_\text{Exp}$, there is no massless solution. Therefore, for type III, the brane profile $w(\rho)$ must correspond to the trivial solution. 

\begin{figure}
    \centering
    \begin{subfigure}[t]{0.48\textwidth}
        \centering
        \includegraphics[width=\textwidth]{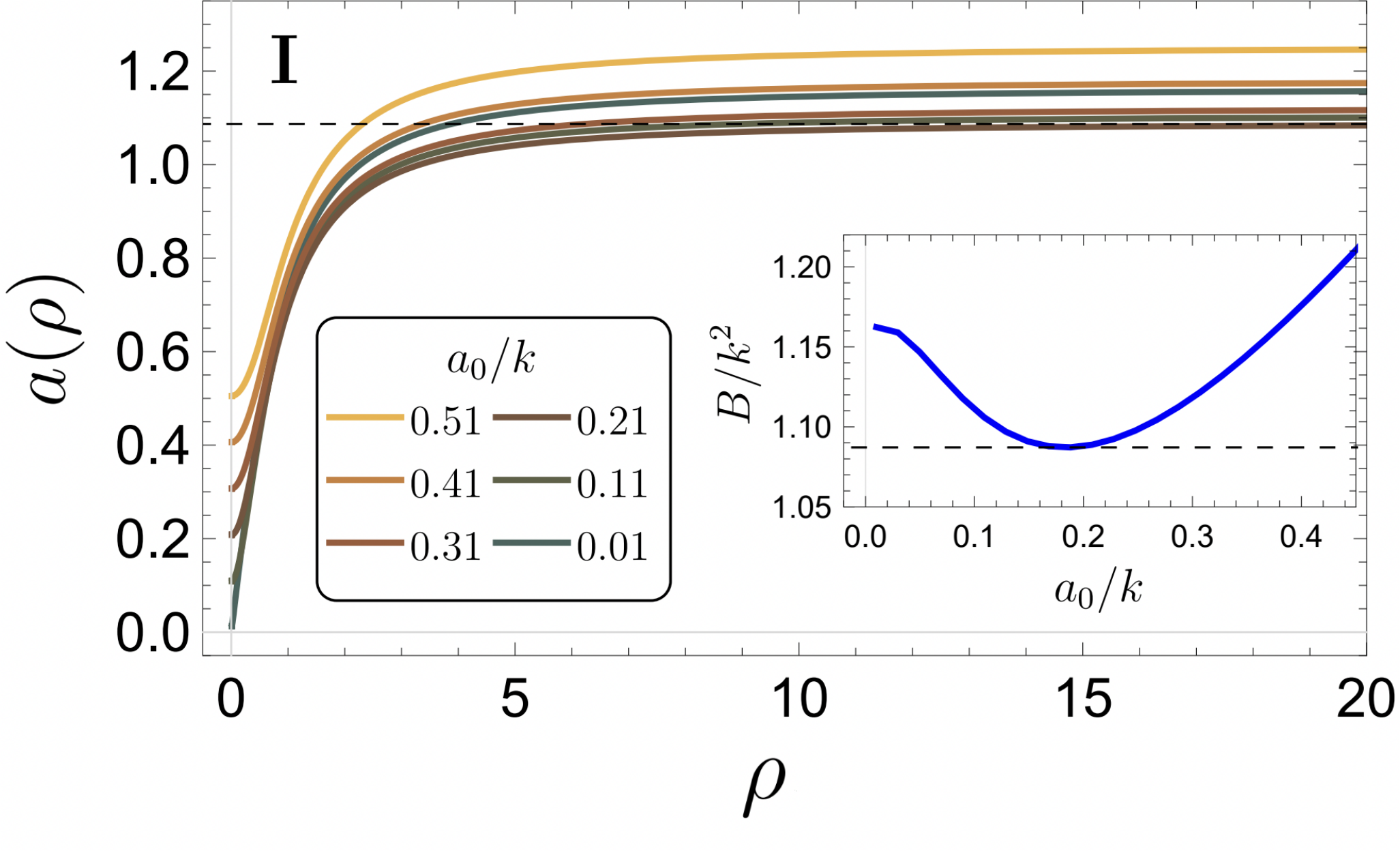}
    \end{subfigure}
    \hspace{0.02\textwidth}
    \begin{subfigure}[t]{0.48\textwidth}
        \centering
        \includegraphics[width=\textwidth]{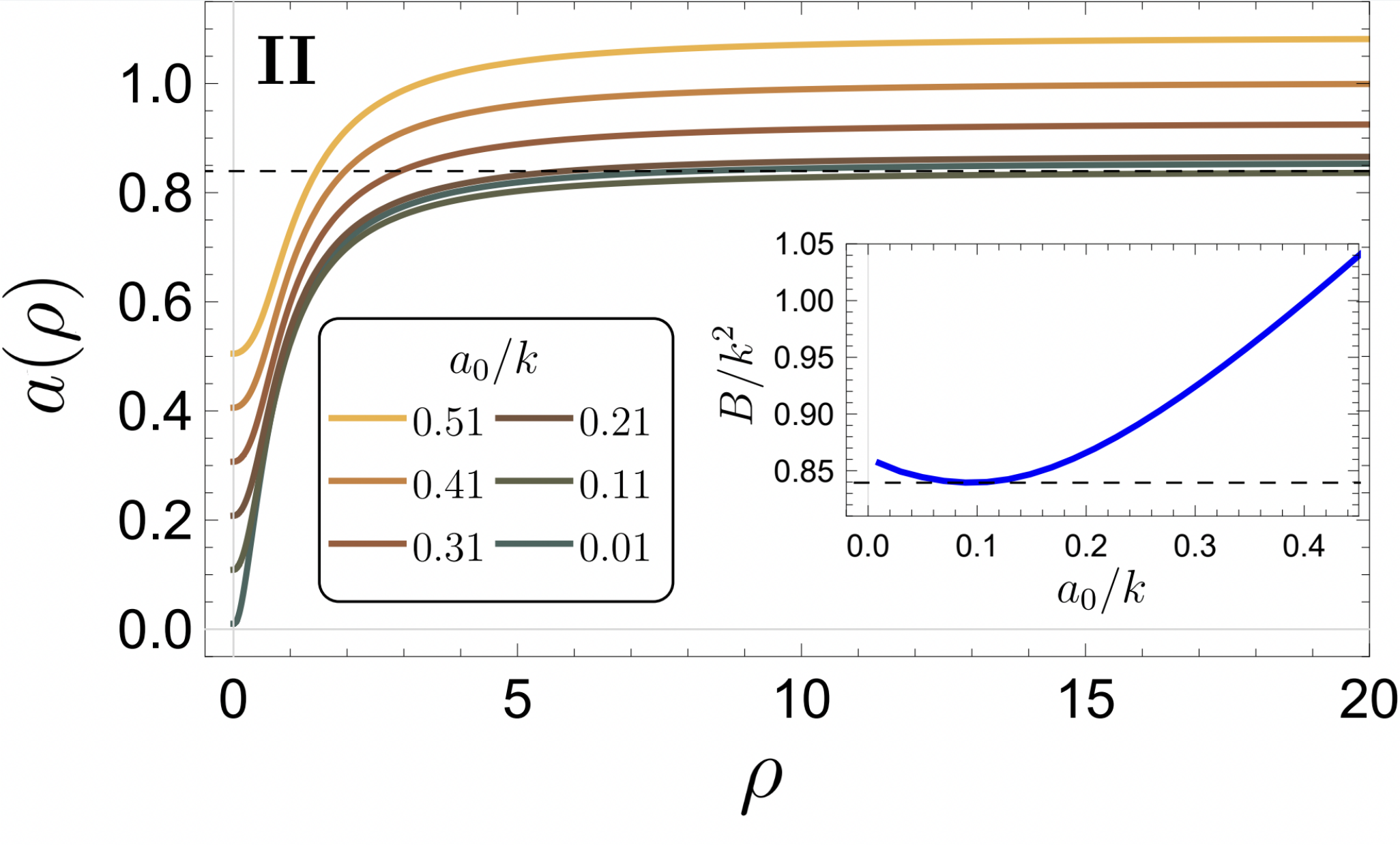}
    \end{subfigure}
    \hspace{0.02\textwidth}
    \begin{subfigure}[t]{0.48\textwidth}
        \centering
        \includegraphics[width=\textwidth]{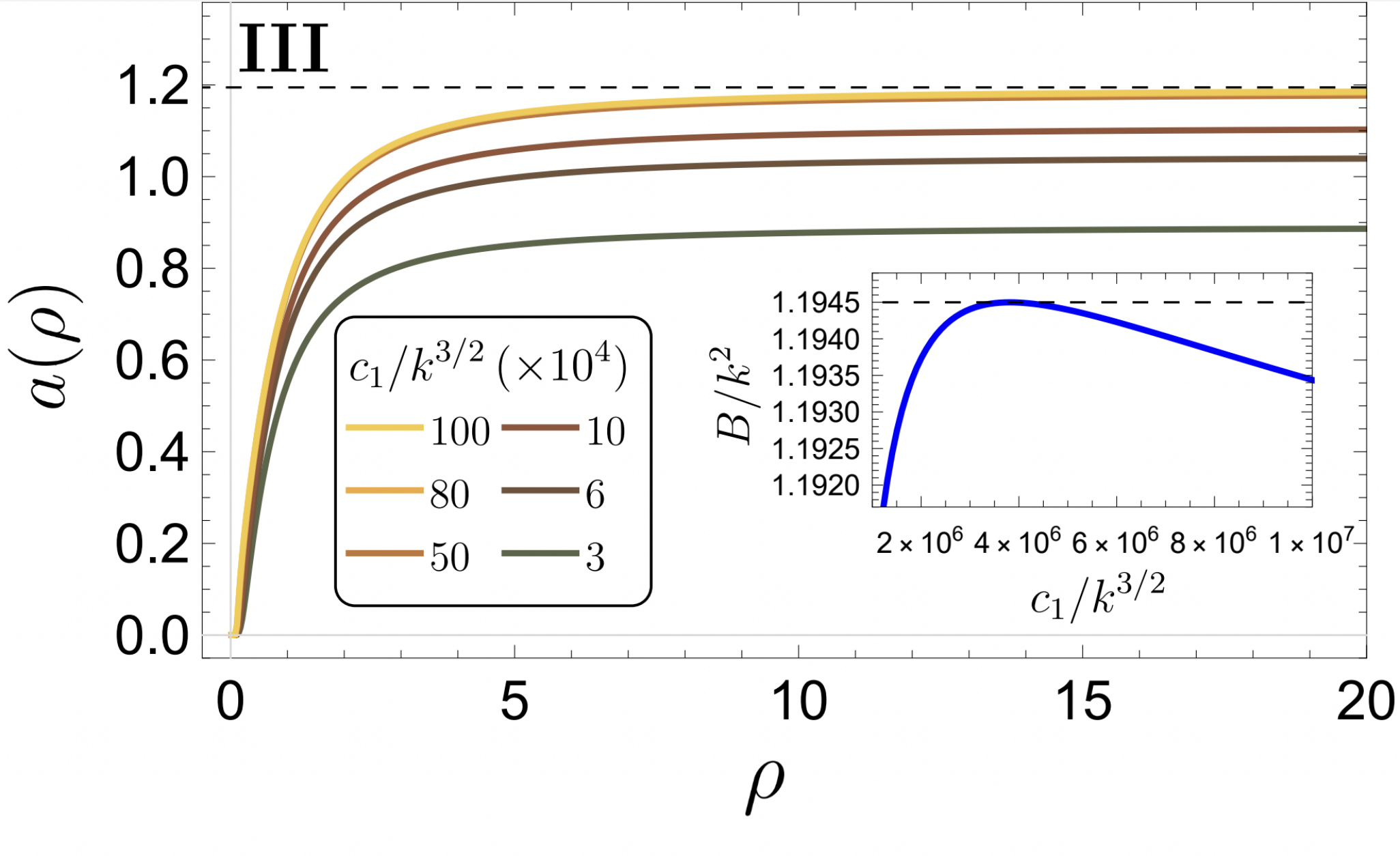}
    \end{subfigure}
    \caption{Typical solutions of $a(\rho)$ for type I (left), II (middle), and III (right) with $b=0$. The inset shows the value of the magnetic field as a function of the parameters, $a_0$ for type I and II, and $c_1$ for type III. The dashed line denotes the lower bound of $B/k^{2}$ for type I and II, and the upper bound for type III.}
    \label{fig:3types2}
\end{figure}

Fig. \ref{fig:3types2} shows the profiles of $a(\rho)$ for type I (left), II (middle), and III (right), with the parameters normalized by $k$. In each panel, we vary the value of the parameter $a_{0}/k$ for types I and II, and $c_{1}$ for type III. The inset shows the magnetic field $B/k^{2}$ as a function of these parameters. 

As indicated by the black dashed line, we observe that there is a lower bound of $B/k^{2}\gtrsim 1.09$ for type I, and of $B/k^{2}\gtrsim 0.84$ for type II, and an upper bound $B/k^{2}\lesssim 1.1945$ for type III. In other words, for a fixed magnetic field $B$, the wavenumber $k$ is bounded by $k/\sqrt{B}\lesssim 0.958$ and $1.091$ for type I and II, respectively, and $k/\sqrt{B}\gtrsim0.915$ for type III. These bounds for each type of solutions can be visually confirmed by the plots of free energy, quark condensate, and electric current in the following section.

\begin{figure}
    \centering
    \includegraphics[width=0.7\textwidth]{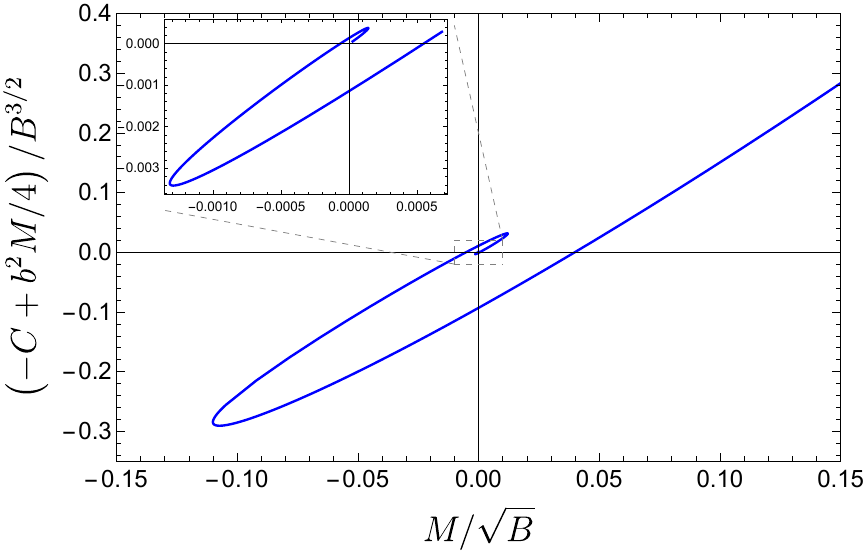}
    \caption{The quark condensate $-C+b^{2}M/4$ is plotted as a function of the quark mass $M$, normalized by the magnetic field $B$. The value of $w_{0}$ decreases from the outer to the inner regions of the spiral, with $k=b=a_{0}=1$ fixed. The inset provides a zoom-in on the region near the origin.}
    \label{fig:C_vs_M}
\end{figure}

Fig. \ref{fig:C_vs_M} shows the quark condensate as a function of the quark mass for type II solutions, normalized by the magnetic field $B$. The value of $w_{0}$ decreases from the outer to the inner regions of the spiral in the plot, with the other parameters fixed to $k=b=a_{0}=1$. The inset reveals a self-similar spiral behavior near the origin, indicating the instability of the embedding around $w_{0}=0$. Note that only the probe brane profile approaches $w=0$ as $w_{0}\to 0$; the profile of $a(\rho)$ is still of Minkowski type, with $a_{0}=1$ fixed. This spiral shows the same qualitative behavior as in the case of a homogeneous magnetic field \cite{Filev_2007,Filev_2007_2}.

This spiral behavior implies the existence of an infinite number of non-trivial solutions, even in the massless case. These solutions generally have non-zero quark condensate and correspond to chiral symmetry-breaking phases in the dual field theory. A more detailed analysis of this spiral behavior can be found in Appendix \ref{app:discretescale}.

\subsection{Comparing free energies}
For the massless case $M=0$, solutions are parameterized by two wavenumbers, $k$ and $b$, with $B$ used for normalization. Notice that in the massless case, the value of $b$ is not determined by the boundary condition at infinity, \ie, it cannot be considered a source in the boundary theory. By introducing cartesian coordinates $(x_8, x_9) = (R \cos \phi, R \sin \phi)$, the boundary condition for the brane can be written as $(x_8, x_9)\rvert_{\rho \to \infty} = (M \cos(bz), M \sin(bz))$. 

For $M \neq 0$, we can control the parameter $b$ by imposing the appropriate boundary condition at infinity. However, when $M = 0$, the boundary condition becomes trivial, and we can no longer use it to control the value of $b$. Consequently, $b$ will be determined dynamically\footnote{The parameter $b$ can be removed by a chiral rotation of the Dirac field in the boundary theory when $M=0$~\cite{Fadafan_2020}.}. We fix $k/\sqrt{B}$ and search for the value of $b/\sqrt{B}$ that minimizes the free energy.

\begin{figure}[h!]
   \centering
    \includegraphics[width=0.6\textwidth]{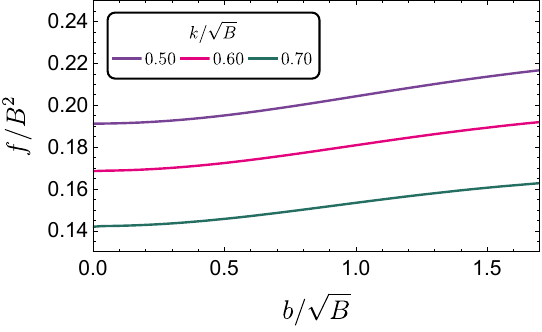}
    \caption{Free energy $f/B^2$ as a function of $b/\sqrt{B}$ for type II solutions. We show three different values for the wavenumber of the helical magnetic field.}
    \label{fig:f_vs_k_b}
\end{figure}

Here, we compute the free energy for type II solutions as a function of these two parameters. As mentioned earlier, there are infinitely many type II solutions that share the same wavenumbers $(k/\sqrt{B}, b/\sqrt{B})$. Here, we focus only on the zero-node solution, as it is found to have the minimum free energy among type II solutions. For type I and type III solutions, the brane profile $w(\rho)$ is trivial, and their free energies are independent of $b$. We will consider their free energies shortly. Fig. \ref{fig:f_vs_k_b} shows the free energy density $f/B^2$ for type II solutions as a function of $b/\sqrt{B}$ for $k/\sqrt{B}=0.5, 0.6$ and $0.7$. For a fixed value of $k/\sqrt{B}$, the brane achieves minimum free energy when $b/\sqrt{B}=0$. This indicates that solutions with $b=0$ are "thermodynamically" favored. In other words, even in the presence of a helical magnetic field, the helical structure of the brane embedding is not induced.

\begin{figure}
   \centering
    \includegraphics[width=0.6\textwidth]{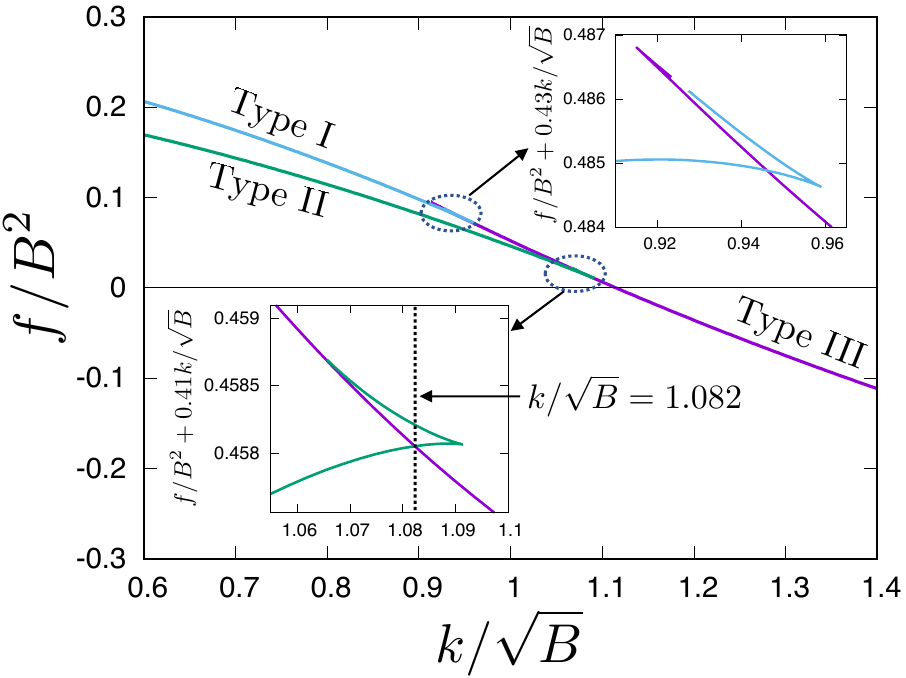}
    \caption{Free energy $f/B^2$ as a function of $k/\sqrt{B}$ for type I, II and III solutions. For type II solutions, we set $b/\sqrt{B}=0$. The inset shows a close-up view of the region around the connection points between type I and type III, as well as between type II and type III. To enhance visibility, the vertical axes in the insets are adjusted to $f/B^2 + \alpha k/\sqrt{B}$ with $\alpha = 0.41$ and $0.43$.}
    \label{fig:f_vs_k}
\end{figure}

We now examine the free energy of solutions with $b=0$. Fig. \ref{fig:f_vs_k} shows the free energy $f/B^2$ as a function of $k/\sqrt{B}$ for type I, II, and III solutions. The insets show magnified views of the vicinity of the connection points between type I and type III, as well as between type II and type III. To enhance visibility, the vertical axes in the insets are rotated to $f/B^2 + 0.41k/\sqrt{B}$ and $f/B^2 + 0.43k/\sqrt{B}$, respectively. All curves for the free energy exhibit a folding behavior at the connection points.

Type I solutions always have a larger free energy than type II solutions, and thus they cannot be the favored solutions.

As observed earlier, type II and type III solutions exist for $k/\sqrt{B} \lesssim 1.09$ and $k/\sqrt{B} \gtrsim 0.915$, respectively. Type II and III solutions have the smallest free energy for $k/\sqrt{B}<1.082$ and $k/\sqrt{B}>1.082$, respectively, indicating a phase transition between type II and type III at $k/\sqrt{B}=1.082$.

\begin{figure}
   \centering
    \includegraphics[width=0.6\textwidth]{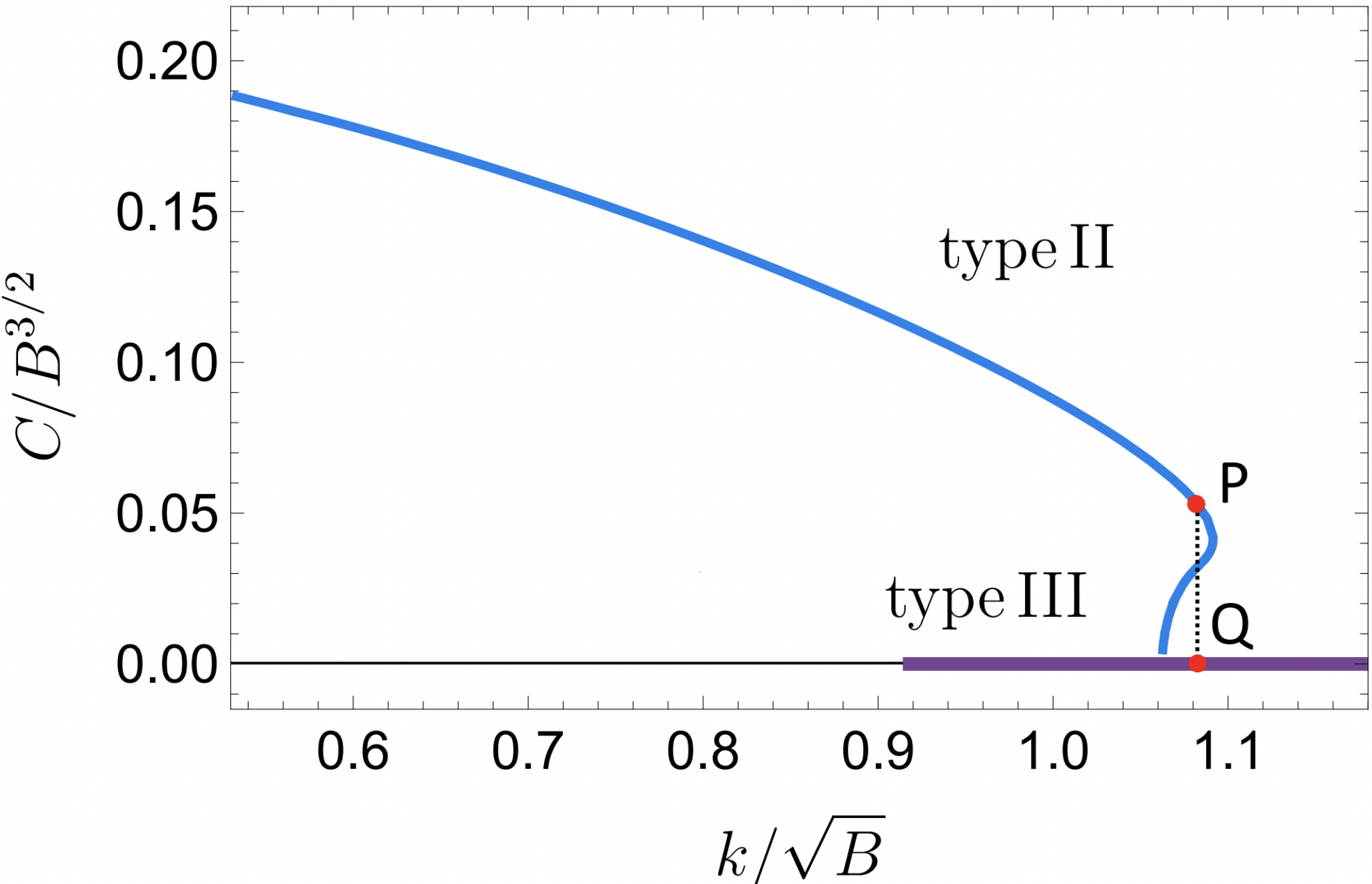}
    \caption{The quark condensate for the type II and III solutions. Points P and Q are the transition points. There is a discontinuous change of the quark condensate due to the transition.}
    \label{fig:kcplot}
\end{figure}

Fig. \ref{fig:kcplot} shows the quark condensate for type II and type III solutions. Points P and Q mark the transition points. The phase transition causes a discontinuous change in the quark condensate. Since type III solutions have zero quark condensate, the restoration of chiral symmetry for $k/\sqrt{B}>1.082$ is evident in this figure.

\subsection{On the helical electric current}
From equations \eqref{eq:BxByBz} and \eqref{eq:PhysicalQ}, we obtain an electric current proportional to the helical magnetic field:
\begin{equation}
    \mathcal{\vec{J}}=-\frac{N_c N_f}{(2\pi)^2}\left(\frac{J}{B} + \frac{k}{4}\right) \mathcal{\vec{B}}~.
\end{equation}

\begin{figure}
   \centering
    \includegraphics[width=0.6\textwidth]{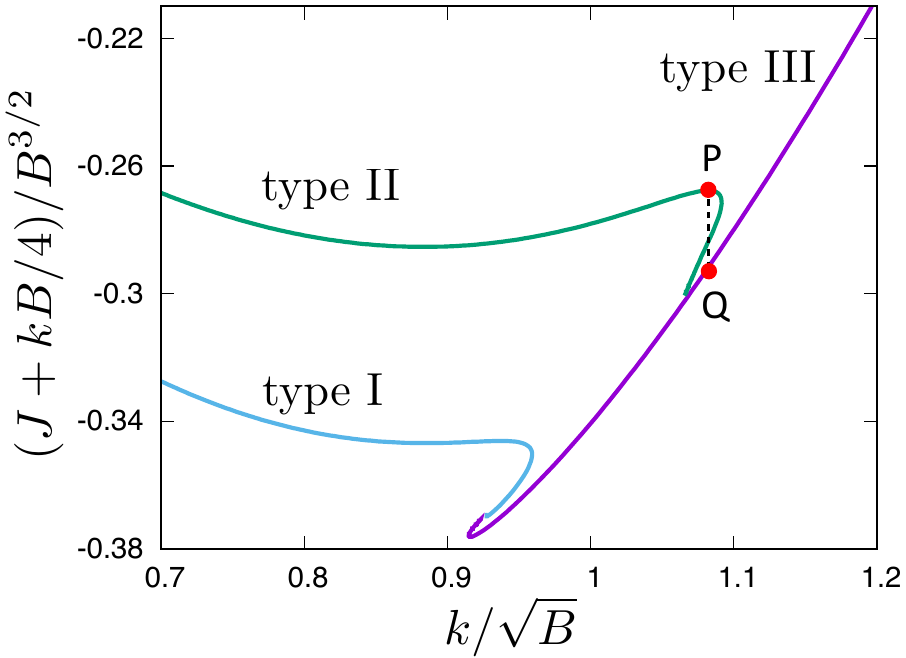}
    \caption{The electric current $(J+kB/4)/B^{3/2}$ as a function of the wavenumber $k/\sqrt{B}$. Points P and Q mark the transition points. The phase transition induces a discontinuous change in the current.}
    \label{fig:kJplot}
\end{figure}

This is a somewhat surprising prediction of the holographic model: there is a non-vanishing electric current parallel to the magnetic field in each plane. Fig. \ref{fig:kJplot} shows the electric current $(J+k B/4)/B^{3/2}$ as a function of the wavenumber $k/\sqrt{B}$. Again, we focus on the massless case: $M=0$. The dashed vertical line represents the critical wavenumber of the phase transition: $k/\sqrt{B}=1.082$. Points P and Q are the transition points, where the current undergoes a discontinous change, similar to the quark condensate.

The microscopic origin of this electric current remains an open question, since we are here in the confined phase and mesons are expected to be electrically neutral. A more detailed analysis of this effect will be adressed in the future. 

\section{Massive case}\label{sec:CSBmassive}

In the absence of a magnetic field, the model exhibits a first-order phase transition from a Weyl semimetal to a trivial insulator as the ratio $|m/b|$ increases \cite{Fadafan_2020}. When a constant magnetic field $B_x$ is applied in the $x$-direction \cite{Evans_2024}, the first-order phase transition persists for small values of $B_x$. However, as $B_x$ increases, the transition region gradually shrinks and eventually vanishes, indicating the existence of a critical point along the line of first-order transitions. In this section we explore the impact of a helical magnetic field on this scenario. In the limit $k\rightarrow 0$, with $k a_{\infty}=B$ fixed, our results reduce to those of \cite{Evans_2024}.

To facilitate comparison with \cite{Evans_2024}, we normalize all quantities by $b$, setting $b=1$ throughout this section. In these units, the critical point occurs at $B_x=B_c\simeq 0.06$.

\begin{figure}
    \centering
    \includegraphics[width=0.46\textwidth]{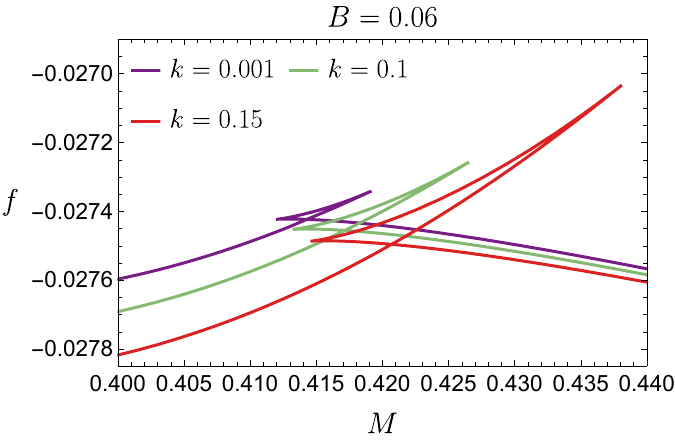}
    \includegraphics[width=0.51\textwidth]{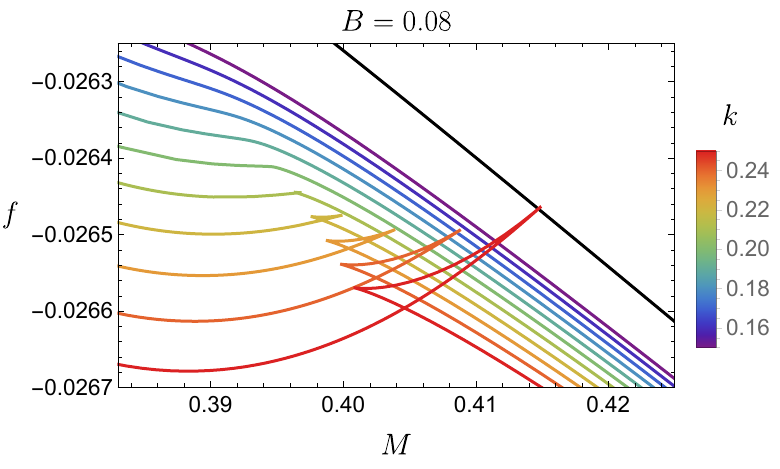}
    \caption{Free energy for two values of the magnetic field. On the left, for $B\lesssim B_c$, we observe that as the phase $k$ increases, the swallow tail structure becomes more pronounced. On the right, for $B=0.08>B_c$, the swallow tail structure emerges when the magnetic field is sufficiently helical, around $k\sim 0.21$. The black line corresponds to the constant $B$ field limit of \cite{Evans_2024}.}
    \label{fig:fmassive}
\end{figure}

\begin{figure}
    \centering
    \includegraphics[width=0.49\textwidth]{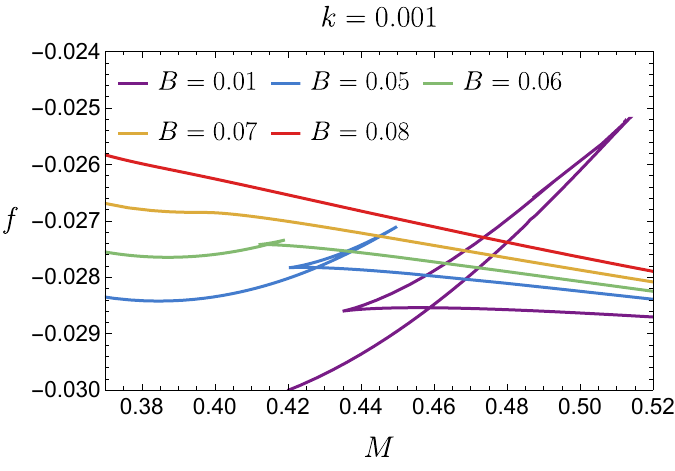}
    \includegraphics[width=0.49\textwidth]{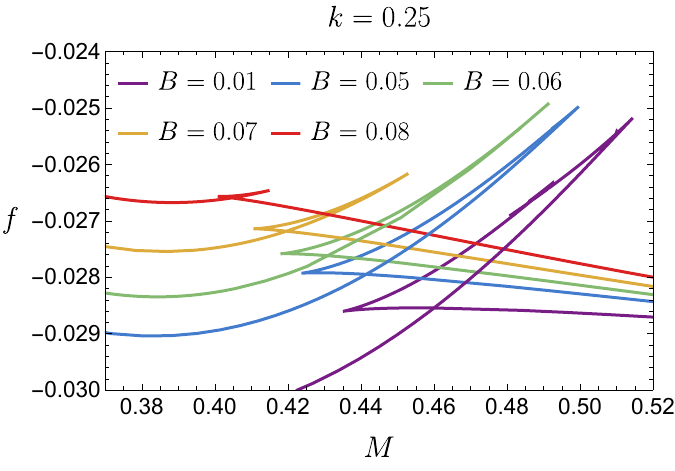}
    \caption{Alternatively, we show the free energy for two values of $k$. On the left, we show the limit of a constant magnetic field from \cite{Evans_2024}. Around $B_c\simeq 0.06$ the phase transition disappears. On the right, we show the free energy for the same values of $B$, but now with a helical structure ($k=0.25$). We observe that the swallow-tail structure of the first-order phase transition is enhanced, even for $B>B_c$.}
    \label{fig:fmassivek}
\end{figure}

Our findings show that, above $B_c$, the helical structure of the magnetic field can reintroduce the phase transition. This behavior is illustrated in Figs. \ref{fig:fmassive} and \ref{fig:fmassivek}.

\section{Conclusions}\label{sec:HelicalBconclusions}

In this Chapter we have explored the effects of introducing a helical magnetic field in a D3/D7-brane model holographically dual to a Weyl semimetal. We first examined the case of massless flavor branes, where constant magnetic fields are known to induce chiral symmetry breaking. Our results show that helical magnetic fields counteract the symmetry-breaking effects and drive the system towards chiral symmetry restoration. Initially, we expected the brane embedding to undergo an unstable transition to a helicoidal structure, similar to that of the magnetic field itself. However, our thermodynamic analysis reveals the opposite: the stable configuration is that of a non-helical brane. Thus, from the perspective of the brane embeddings, there are only two such solutions: symmetry-preserving and symmetry-breaking solutions, similar to the case of constant magnetic fields. Notably, the gauge potential admits a class of new solutions, akin to those found in \cite{Fadafan_2020}, which we also term "exponential embeddings" with a little abuse of language. Together, these solutions combine to form three distinct phases, with first-order phase transitions between them, characterized by the familiar self-similar spiraling behavior seen in many flavor-brane systems. The helical magnetic field also induces a parallel electric current on the flavor degrees of freedom, whose microscopic origin is not fully understood.

In the final part of this Chapter, we have examined  the massive case. Here, the helical structure of the brane embeddings can be enforced as a boundary condition rather than being dynamically chosen. Our work extends the findings of \cite{Evans_2024}, where a constant magnetic field was shown to erase the first-order phase transition between exponential embeddings and Minkowski embeddings.

Future investigations could extend this analysis to finite-temperature effects, which are crucial for understanding environments such as the early-universe or heavy-ion collisions. Time-dependent helical magnetic fields also present an exciting avenue for study, as they could model dynamically evolving systems and offer insights into transient phenomena in non-equilibrium settings like heavy ion collisions.

Generalizing the findings to other holographic models or incorporating more realistic QCD-like dynamics would further establish the universality of these phenomena. Exploring the coupling of helical magnetic fields with additional degrees of freedom, such as scalar or pseudoscalar fields, could uncover deeper connections between field configurations and symmetry-breaking patterns.

In summary, our study highlights the rich phenomenology associated with spatially tunable external fields in holographic models, demonstrating their relevance in exploring the phase space of strongly coupled quantum field theories.

%% file: Part2/HelicalB/HelicalBAppendices.tex
\renewcommand{\chapterquote}{}
\chapterappendix{\thechapter}

\setcounter{equation}{0}
\setcounter{appendixsection}{\value{appendixsection}+1} 
\section{Holographic renormalization and dictionary}\label{app:holorenohelical}

We begin with the metric of AdS$_5\times S^5$ written in \eqref{eq:AdS5xS5helical} and we change to the angular coordinates $(r,\theta)$ using \eqref{eq:relationcartesianangular}. Moreover, we introduce the inverse radial coordinate\footnote{In the rest of the thesis, the inverse radial coordinate is denoted as $z$. Here, and only in this appendix, we call it $u$ in order to distinguish it from the $z$-spatial direction. $u$ must not be confused with the isotropic coordinate \eqref{eq:u}, which plays no role in this Chapter since we always work at zero temperature.} $u=1/r$. The AdS$_5\times S^5$ metric becomes
\begin{equation}
    ds^2=\frac{1}{u^2}\left(-dt^2+d\Vec{x}_3^2+du^2\right)+d\theta^2+\cos^2\theta d\Omega_3^2+\sin^2\theta d\phi^2~.
\end{equation}

The ansatz for the brane embedding and the gauge field \eqref{eq:ansatzhelical} translates in these coordinates into
\begin{equation}
    2\pi\alpha'(A_x+iA_y)=a(u)e^{ikz}~,\qquad \theta=\theta(u)~,\qquad \phi=bz~.
\end{equation}

In these coordinates, the DBI action is given by
\begin{equation}
    S_{D7}=-\mathcal{N}\int_{0}^{u_0} du \frac{\cos^3\theta}{u^5}\sqrt{\left(1+k^2u^4a^2+b^2u^2\sin^2\theta\right)\left(1+u^4a'^2+u^2\theta'^2\right)}~,
    \label{eq:DBIhelicalapp}
\end{equation}
with $\mathcal{N}=2\pi^2N_fT_{D7}$, and primes denote derivatives with respect to $u$. The integral goes from the boundary $u=0$ to the bulk interior, denoted by $u_0$\footnote{In the present case, $u_0=\infty$ since we are working at zero temperature. At finite temperature, the integral would be until the black hole horizon.}. The asymptotic expansions \eqref{eq:UVexpansionhelical} become
\begin{equation}
\begin{aligned}
    \theta(u) &=M u+\left(C+\frac{M^3}{6}+\frac{b^2M}{2}\log(Mu)\right)u^3+...\\
    a(u) & = a_\infty+ \left(\frac{J}{2}+\frac{b^2a_\infty}{2}\log(\sqrt{B} u)\right)u^2 +...
    \label{eq:UVexpansionatheta}
\end{aligned}
\end{equation}
We have chosen the constants in the logarithms such that their arguments become scale-invariant. This can always be done by a redefinition of $C$ and $J$.

Integrating the DBI action up to a UV cutoff at $u=\epsilon$, we find the asymptotic behavior
\begin{equation}
\begin{aligned}
    S_{D7}&=-\mathcal{N}\int_\epsilon du\left[\frac{1}{u^5}-\frac{M^2}{u^3}+\frac{a_\infty^2 k^2}{2u}+\frac{b^2 M^2}{u}+\mathcal{O}(u)\right]\\
    &=-\mathcal{N}\left[\frac{1}{4\epsilon^4}-\frac{M^2}{2\epsilon^2}-\frac{a_\infty ^2k^2}{2}\log\epsilon-b^2M^2\log\epsilon+\mathcal{O}(\epsilon^2)\right]~.
\end{aligned}
\end{equation}

The local counterterms needed to regularize these divergences are the usual ones \cite{Karch_2005_holorenobranes,Karch_2006}, but modified to account for the presence of the phase $\phi(z)$. The most natural thing is to group the embedding function $\theta(u)$ and its phase $\phi$ into a single field $\Theta(z,u)=\theta(u)e^{i\phi(z)}$. The counterterms become \cite{Hoyos_2011,Fadafan_2020}
\begin{equation}
\begin{aligned}
    S^{ct}_1&=\frac{1}{4}\mathcal{N}\sqrt{-\gamma}~,&\quad S^{ct}_2&=-\frac{1}{2}\mathcal{N}\sqrt{-\gamma}\abs{\Theta(\epsilon)}^2~, \\
    S^{ct}_3 & = \frac{1}{2}\mathcal{N}\sqrt{-\gamma}\log\abs{\Theta}\Theta^*\square_{\gamma}\Theta ~,&\quad S^{ct}_4&=\frac{1}{4}\mathcal{N}\sqrt{-\gamma}\Theta^*\Box_\gamma\Theta~,\\
    S^{ct}_5 & = -\frac{1}{4}\mathcal{N}\sqrt{-\gamma}F_{\mu\nu}F^{\mu\nu}\log(\epsilon/u_0)~, & S^{ct}_f&=\frac{5}{12}\mathcal{N}\sqrt{-\gamma}\abs{\Theta(\epsilon)}^4~,
\end{aligned}
\end{equation}

\noindent with $\square_{\gamma}\Theta=\frac{1}{\sqrt{-\gamma}}\partial_{\mu}\left(\sqrt{-\gamma}\gamma^{\mu\nu}\partial_{\nu}\Theta\right)$. A constant $u_0$ is introduced to make the argument of the logarithm scale-invariant. 

Their contributions are
\begin{equation}
\begin{aligned}
    S^{ct}_1&=\mathcal{N}\frac{1}{4\epsilon^4}~,&\quad~ S^{ct}_2&=-\mathcal{N}\left[M\h C+\frac{M^4}{6}+\frac{M^2}{2\epsilon^2}+\frac{b^2M^2}{2}\log(M\epsilon)\right]~, \\
    S^{ct}_3 & =-\mathcal{N}\frac{b^2M^2}{2}\log(M\epsilon) ~,&\quad~ S^{ct}_4&=-\mathcal{N}\frac{b^2M^2}{4}~,\\
    S^{ct}_5 & = -\mathcal{N}\frac{k^2a_\infty^2}{2}\log(\epsilon/u_0)~,&\quad~ S^{ct}_f&=\mathcal{N}\frac{5}{12}M^4~,
\end{aligned}
\end{equation}
giving a total contribution of
\begin{equation}
    S^{ct}=-\mathcal{N}\left[-\frac{1}{4\epsilon^4}+\frac{M^2}{2\epsilon^2}+b^2M^2\log(M\epsilon)+MC-\frac{M^2}{4}(M^2-b^2)+\frac{k^2a_\infty^2}{2}\log (\epsilon/u_0)\right]~,
    \label{Sct}
\end{equation}
which correctly cancels the divergences when $\epsilon\to 0$. 

\subsection{Quark condensate and electric current}

We now consider the one-point functions. We begin with the quark condensate $\langle\mathcal{O}_m\rangle$
\begin{equation}
    \langle \mathcal{O}_m\rangle =-(2\pi\alpha')\lim_{\epsilon\to 0}\epsilon\frac{\delta S_{\text{D7}}^{sub}}{\delta \theta(\epsilon)}~,
    \label{eq:condensate}
\end{equation}
with $S^{sub}_{D7}=S^{reg}_{D7}+\sum_iS^{ct}_i$. The contribution from the regularized action on-shell can be written as a boundary term,
\begin{equation}
    \delta S^{reg}_{D7}=\frac{\partial \mathcal{L}}{\partial \theta'}\delta \theta\big\rvert_\epsilon=\mathcal{N}\left[\frac{M}{\epsilon^3}+\frac{3C}{\epsilon}+\frac{b^2M}{2\epsilon}-\frac{3M^3}{2\epsilon}+\frac{3b^2M}{2\epsilon}\log(M\epsilon)+\mathcal{O}(\epsilon)\right]\delta\theta(\epsilon)~.
\end{equation}

The counterterms contribute as
\begin{equation}
\begin{aligned}
    \delta S^{ct}_2&=\mathcal{N}\left[-\frac{M}{\epsilon^3}-\frac{C}{\epsilon}-\frac{M^3}{6\epsilon}-\frac{b^2M}{2\epsilon}\log(M\epsilon)\right]\delta\theta(\epsilon)~,\\
    \delta S^{ct}_3&=\mathcal{N}\left[-\frac{b^2M}{2\epsilon}-\frac{b^2M}{\epsilon}\log(M\epsilon)\right]\delta\theta(\epsilon)~,\\
    \delta S^{ct}_4&=\mathcal{N}\left[-\frac{b^2M}{2\epsilon}\right]\delta\theta(\epsilon)~,\\
    \delta S^{ct}_f&=\mathcal{N}\left[\frac{5M^3}{3\epsilon}\right]\delta\theta(\epsilon)~.
\end{aligned}
\end{equation}

Adding all contributions, we readily get from \eqref{eq:condensate}
\begin{equation}
    \langle\mathcal{O}_m\rangle=\frac{N_fN_c\sqrt{\lambda}}{4\pi^3}\left[-C+\frac{b^2M}{4}\right]~,
\end{equation}
where we have written $\mathcal{N}$ in terms of field theory quantities using \eqref{eq:ND3D7}.

We now move on to the current,
\begin{equation}
    \mathcal{J}_x + i \mathcal{J}_y =(2\pi\alpha')\lim_{\epsilon\to 0}\frac{\delta S^{sub}_{D7}}{\delta a(\epsilon)}e^{ikz}~.
    \label{eq:current}
\end{equation}

The contribution from $S^{reg}_{D7}$ is again a boundary term,
\begin{equation}
    \delta S^{reg}_{D7}=\frac{\partial \mathcal{L}}{\partial a'}\delta a\big\rvert_\epsilon=\mathcal{N}\left[J+\frac{k a_\infty^2}{2}+k^2 a_\infty\log(\sqrt{k a_\infty}\epsilon)+\mathcal{O}(\epsilon)\right]\delta a(\epsilon)~.
\end{equation}

The contribution from the counterterms is (only $S^{ct}_5$ contributes)
\begin{equation}
    \delta S^{ct}_5=\mathcal{N}\left[-\frac{k^2a_\infty}{4}-k^2a_\infty\log\left(\frac{\sqrt{ka_\infty}\epsilon}{\gamma}\right)\right]\delta a(\epsilon)~,
\end{equation}
where have set $u_0=\gamma/\sqrt{ka_\infty}$, with $\gamma$ an arbitrary constant. Adding both contributions, we obtain
\begin{equation}
    \mathcal{J}_x + i \mathcal{J}_y=\frac{N_fN_c\sqrt{\lambda}}{(2\pi)^3}\left[J+k^2a_\infty\left(\frac{1}{4}+\log\gamma\right)\right]e^{ikz}~,
\end{equation}
where we have used \eqref{eq:ND3D7} again. In the main text, we have set $\gamma=1$.

\setcounter{equation}{0}
\setcounter{appendixsection}{\value{appendixsection}+1} 
\section{Discrete scale invariance}\label{app:discretescale}

\begin{figure}
   \centering
    \includegraphics[width=0.6\textwidth]{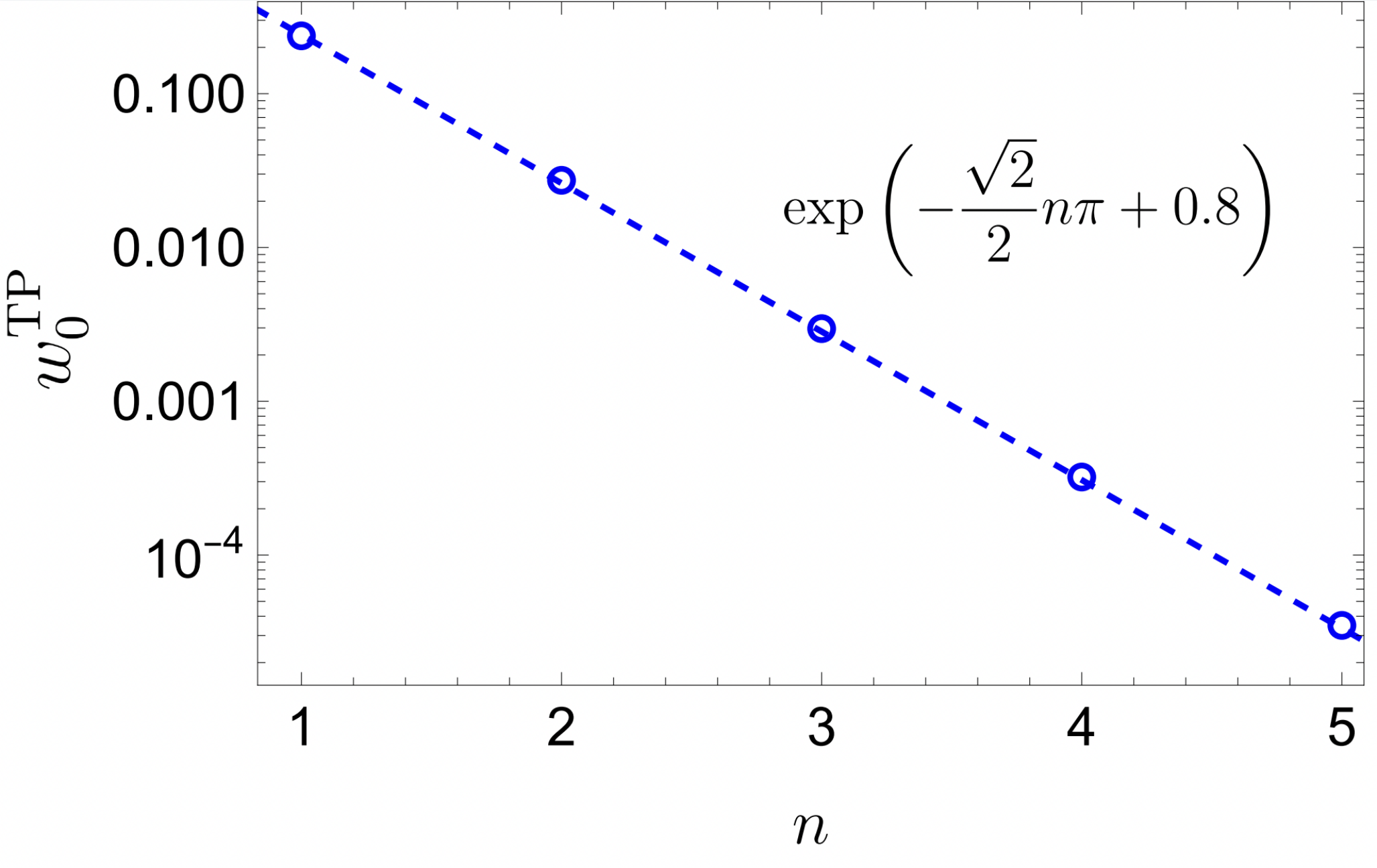}
    \caption{The electric current $(J+kB/4)/B^{3/2}$ as a function of the wavenumber $k/\sqrt{B}$. Points P and Q mark the transition points. The phase transition induces a discontinuous change in the current.}
    \label{fig:w0TP}
\end{figure}

In this appendix we study in more detail the spiral structure of the condensate observed in Fig. \ref{fig:C_vs_M}. Let $w_{0n}$ ($n = 0,1,2,\cdots$) represent the values of $w_0$ where the quark mass $M$ vanishes, arranged in descending order. The brane profile $w = w(\rho)$ corresponding to $w_0 = w_{0n}$ intersects the horizontal axis $n$ times (see Fig. \ref{fig:C_vs_M}). We will refer to such a solution as the $n$-node solution.

Now, let's examine the self-similarity structure near the origin in more detail. As $w_{0}\to 0$, $w(\rho)$ becomes sufficiently small, while $a(\rho)$ remains finite. We linearize the equations by substituting $w(\rho) \to \epsilon\h w(\rho)$ and considering the first-order term in $\epsilon$. This leads to the linearized equation for $w(\rho)$ near $\rho=0$,
\begin{equation}
    w'' + \frac{w'}{\rho} + \frac{2w}{\rho^2}=0.
    \label{eq:smallw}
\end{equation}
One can easily check that the equation has the scaling symmetry $w(\rho)\to w(\mu \rho)/\mu$, with $\mu$ a real positive constant. The solution of \eqref{eq:smallw} is given by $w \sim \rho^{\nu_{\pm}}$, with a purely imaginary exponent $\nu_{\pm }=\pm\sqrt{2} i$, or equivalently,
\begin{equation}
    w(r) = C_{1} \sin \left( \sqrt{2}\log \rho \right) + C_{2} \cos\left( \sqrt{2}\log \rho \right),
\end{equation}
where $C_1$ and $C_{2}$ are constants. The scaling symmetry of the equation leads to the coefficients transformation
\begin{equation}
    \begin{pmatrix}
    C_{1} \\
    C_{2} 
    \end{pmatrix}
    \to
    \frac{1}{\mu}
    \begin{pmatrix}
    \cos(\sqrt{2} \log \mu) & -\sin(\sqrt{2} \log \mu) \\
    \sin(\sqrt{2} \log \mu) & \cos(\sqrt{2} \log \mu)
    \end{pmatrix}
    \begin{pmatrix}
    C_{1} \\
    C_{2} 
    \end{pmatrix}.
\end{equation}

This scaling symmetry and transformation imply that the brane embedding shows discrete self-similarity with a period $\mu_{2\pi} = \exp (\sqrt{2}\pi)$.
Although this result comes from the analysis near $\rho=0$, it naturally explains the oscillatory behavior of the quantities at the boundary, such as the spiral behaviour in Fig. \ref{fig:C_vs_M}. 

To illustrate this, we plot the $n$-th turning point in the spiral behavior of Fig. \ref{fig:C_vs_M}, and the corresponding value of $w_{0}\equiv w_{0}^{\rm TP}$ in Fig.~\ref{fig:w0TP}.

Since $w_{0}$ scales as $w_{0} \to w_{0}/\mu$, each turning point is expected to appear at a half-period:~$\mu_{\pi}=\exp(\sqrt{2}\pi/2)$. Hence, we obtain $w_{0}^{\rm TP} \propto \exp(-\sqrt{2}n\pi/2)$. As shown in the dashed line in Fig. \ref{fig:w0TP}, we confirm that the turning point of the spiral in the quark mass and condensate is well described by the self-similarity of the brane embedding.

A similar analysis has been performed near the critical embedding at finite temperature \cite{Frolov_2006,Mateos_2006,Mateos_2007}, or in the presence of a finite electric field \cite{Ishigaki_2023}. However, in our case, the self-similarity appears near the trivial solution $w=0$, rather than near the critical embedding with the conical singularity at the black hole horizon or effective horizon created by an external electric field.

\restoredefaultnumbering
\restoredefaultsectioning

%% file: Part2/FloquetSYK/FloquetSYK.tex
\renewcommand{\chapterquote}{\textit{
"But what do we have left, once we abandon the lie?\\
Chaos, a gaping pit waiting to swallow us all."}\\[0.5em]
\textit{"Chaos isn't a pit. Chaos is a ladder."
}\\[0.5em]
\normalfont— Game of Thrones.}
\chapter{Floquet SYK wormholes}\label{chap:FloquetSYK}

Almost all the material in this Chapter is based on, and adapted from, \textit{Floquet SYK wormholes} \cite{Berenguer_2024} and \textit{Hot wormholes and chaos dynamics in a two-coupled SYK model} \cite{Berenguer_2025_SYK}.

We study the non-equilibrium dynamics of two coupled SYK models, conjectured to be holographically dual to an eternal traversable wormhole in AdS$_2$. We consider different periodic drivings of the parameters of the system as well as a sudden coupling to a cold bath. We analyze the energy flows in the wormhole and black hole phases of the model as a function of the driving frequency. Our numerical results show a series of resonant frequencies in which the energy absorption and heating are enhanced significantly and the transmission coefficients drop, signalling a closure of the wormhole. These frequencies correspond to part of the conformal tower of states and to the boundary graviton of the dual gravitational theory. When driving the strength of the separate SYK terms we find that the transmission can be enhanced by suitably tuning the driving.

We then use two of these driving protocols to access the unstable region of the phase diagram, the hot wormhole, which is inaccessible through equilibrium simulations. One protocol involves cooling the system via a coupling to a thermal bath, while in the other we periodically drive the coupling parameter between the two sides. We numerically compute the Lyapunov exponents of the hot wormhole for the two cases. Our results uncover a rich structure within this phase, including both thermal and non-thermal solutions. These behaviors are analyzed in detail, with partial insights provided by the Schwarzian approximation, which captures certain but not all aspects of the observed dynamics.


\section{Introduction}

As introduced in Chapter \ref{chap:SYKwormholes}, the SYK model admits a low-energy dual description in terms of JT gravity on AdS$_2$. The thermal AdS$_2$ space has two causally disconnected boundaries. As shown in \cite{Gao_2016} (see also Section \ref{sec:traversableWHs}), a suitable double-trace deformation can render these boundaries causally connected. Building on this idea, Maldacena and Qi proposed in \cite{Maldacena_2018} an extended construction involving two coupled SYK models, providing a concrete realization of an eternal traversable wormhole in AdS$_2$. 

This setup offers a tractable framework for exploring phases of matter with distinct gravitational duals. The model exhibits a Hawking-Page-like transition between a gapless, high-temperature phase and a gapped, low-temperature phase. It has been argued that these phases correspond to duals of a two-sided AdS$_2$ black hole and an eternal traversable wormhole geometry, respectively. The traversability of the wormhole is supported by the negative null energy generated through the coupling between the two boundaries \cite{Gao_2016, Maldacena_2017}.

In addition to these two stable phases, the model also features an intermediate unstable phase, commonly referred to as the "hot wormhole" phase, which is inaccessible through standard equilibrium simulations in the canonical ensemble \cite{Maldacena_2018}. This unstable phase is also present when the Majorana fermions are replaced by complex fermions \cite{GarciaGarcia_2020_08}.

This traversable wormhole serves as a laboratory for studying quantum teleportation protocols: it has been demonstrated that a small perturbation applied to one SYK system can effectively teleport quantum information to the other, exploiting the geometric connectivity of the wormhole \cite{Gao_2019, Brown_2019, Nezami_2021}. This remarkable phenomenon has significantly advanced our understanding of the interplay between quantum information theory and gravitational physics.

The SYK model and its generalizations have been widely studied both in the high-energy and condensed matter contexts. To our knowledge, only a handful of studies involve time-dependent dynamics. In \cite{Maldacena_2019}, the authors address the dynamical formation of the wormhole. Other works instead deal with the time evolution after a sudden quench \cite{Eberlein_2017,Bhattacharya_2018,Larzul_2021, Zhou_2020,Zhang_2020}. In this Chapter, following the lines of this thesis, we are interested mainly in periodic drivings.


In this Chapter we study the out-of-equilibrium dynamics of the two-coupled model under periodic drivings. Specifically, we drive the relevant coupling, $\mu$, sinusoidally in time. We also explore drivings of the interaction strengths $J_{L,R}$ of each SYK model independently. Although the latter are less physically motivated, as they imply a time-dependent variance of the coupling distribution, they provide useful comparative insight. 

For the driving in $\mu$, the overall behavior is broadly consistent with the expectations. However, within the wormhole phase, a significant heating amplification appears when driving at specific subsets of the eigenfrequencies of the undriven system. While the resonant behavior seems intuitive, understanding why the driving couples only to half of the spectrum requires careful consideration. Of particular interest is the second resonance, which appears to correspond to exciting a boundary graviton mode.

Our numerical results suggest that the system can be driven into the unstable hot wormhole regime dynamically. Very little is known about this phase, and one of our goals is to extract new information about its physical properties. For instance, while the Lyapunov exponents of the two stable phases have been computed in \cite{Nosaka_2020}, those of the hot wormhole phase remained unknown. Using non-equilibrium protocols, we access this regime and compute its chaos exponents.

This is achieved through two distinct out-of-equilibrium protocols. The first, introduced in \cite{Maldacena_2019}, involves cooling a high-temperature black hole solution by coupling the system to a cold bath. This cooling process enables a quasi-static evolution that transitions through the unstable phase. The second method involves the driving in $\mu$ mentioned above, injecting energy into the system in a low-temperature wormhole solution and reaching the hot wormhole phase. This is advantageous in the sense that it preserves the intrinsic properties of the original system since it does not require the action of an external bath. By driving the coupling parameter, we uncover a richer structure of potential evolution endpoints within the unstable phase. These findings are confirmed by a third non-equilibrium protocol that combines the two previous ones. We also provide a qualitative understanding of the observed phenomena using the Schwarzian effective theory, shedding light on the dynamics of this exotic phase.

The Chapter is organized as follows. Section \ref{sec:2SYKrealtime} reviews the non-equilibrium formalism and defines the observables used to track the evolution of the system under the drivings. In Section \ref{sec:numresults1}, we present numerical results for various driving protocols, finding an interesting pattern of resonances and interpreting them using the Schwarzian action. Section \ref{sec:divingHW} revisits the cooling protocol of \cite{Maldacena_2019}, deriving the equations required to compute the Lyapunov exponents in the different phases. The results of this computation are presented in Section \ref{sec:numresults2}. In Section \ref{sec:HotWHColdBH}, we undertake a more detailed study of the hot wormhole phase. Finally, we summarize our findings in Section \ref{sec:FloquetSYKconclusions}.

\section{The model}\label{sec:2SYKrealtime}

We consider the real-time version of the two coupled SYK model we reviewed in Section \ref{subsec:2coupledSYK}. The relevant dynamics is contained in the following time-dependent Hamiltonian 
\begin{equation}
H(t)=\sum_{a=L,R}\frac{1}{4!}\sum_{ijkl}f_a(t)J_{ijkl}\chi_a^i\chi_a^j\chi_a^k\chi_a^l+i\mu(t)\sum_j \chi_L^j\chi_R^j ~.
\label{eq:H2SYKtdep}
\end{equation}
Again, $\chi_a^i$ are Majorana fermions  with $i=1,...,N$ and $a = L,R$, satisfying the usual anticommutation relations $\left\{\chi_a^i,\chi_b^j\right\}=\delta^{ij}\delta_{ab}$. $J_{ijkl}$ are real constants drawn from a Gaussian distribution with mean and variance given by
\begin{equation}
\overline{J_{ijkl}}=0~,~~~~~\overline{J_{ijkl}^2}=\frac{3! J^2}{N^3} ~.
\label{eq:Jmeanvar2SYKtdep}
\end{equation}
Setting $\mu(t) = \mu$ to a constant value and $f_{L/R}(t)\rightarrow 1$ recovers the model dual to a traversable wormhole in \cite{Maldacena_2018}.

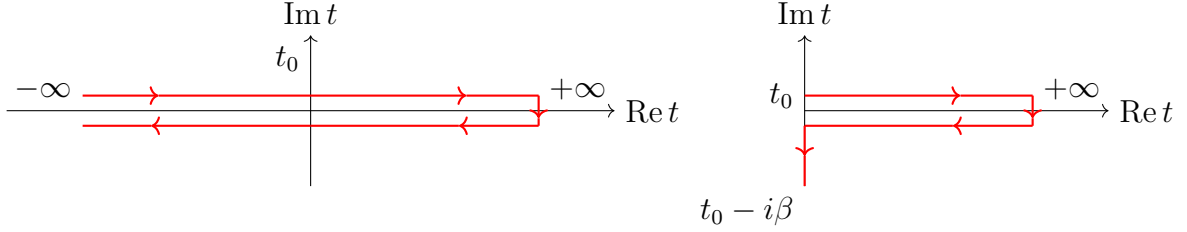
\begin{figure}
    \centering
    \begin{tikzpicture}
        \begin{scope}
            \draw[->] (-4, 0) -- (4, 0) node[right] {$\mathrm{Re}\,t$};
            \draw[->] (0, -1) -- (0, 1) node[above] {$\mathrm{Im}\,t$};
            
            \node[below left] at (0, 1) {$t_0$};
            \node[above right] at (3, 0) {$+\infty$};
            \node[above left] at (-3, 0) {$-\infty$};
            
            \draw[thick,red, ->] (-3, 0.2) -- (-2, 0.2);
            \draw[thick,red, ->] (-2, 0.2) -- (2.05, 0.2);
            \draw[thick,red] (2.05, 0.2) -- (3, 0.2); 
            \draw[thick,red, ->] (3, 0.2) -- (3, -0.1);
            \draw[thick,red] (3, -0.1) -- (3, -0.2); 
            \draw[thick,red, ->] (3, -0.2) -- (1.95, -0.2);
            \draw[thick,red] (1.95, -0.2) -- (0, -0.2);
            \draw[thick,red,->] (0, -0.2) -- (-2.1, -0.2); 
            \draw[thick,red] (-2.1, -0.2) -- (-3, -0.2);
        \end{scope}
        
        \begin{scope}[xshift=6.5cm] 
            \draw[->] (0, 0) -- (4, 0) node[right] {$\mathrm{Re}\,t$};
            \draw[->] (0, -1) -- (0, 1) node[above] {$\mathrm{Im}\,t$};
            
            \node[below left] at (0, 0.5) {$t_0$};
            \node[below left] at (0, -1) {$t_0 - i\beta$};
            \node[above right] at (3, 0) {$+\infty$};
            
            \draw[thick,red, ->] (0, 0.2) -- (2.05, 0.2);
            \draw[thick,red] (2.05, 0.2) -- (3, 0.2); 
            \draw[thick,red, ->] (3, 0.2) -- (3, -0.1);
            \draw[thick,red] (3, -0.1) -- (3, -0.2); 
            \draw[thick,red, ->] (3, -0.2) -- (1.95, -0.2);
            \draw[thick,red] (1.95, -0.2) -- (0, -0.2); 
            \draw[thick,red, ->] (0, -0.2) -- (0, -0.6);
            \draw[thick,red] (0, -0.6) -- (0, -1); 
        \end{scope}
    \end{tikzpicture}
    \caption{Contours used in the numerical integration. At $t=t_0$,  the drivings are turned on. Prior to that, the system is at equilibrium. When using the standard predictor-corrector method the initial condition must be explicitly provided (left). When using the NESSi package, the L-shaped contour (right) calculates the initial thermal equilibrium state directly in the Matsubara formalism.}
    \label{fig:Keldyshcontour}
\end{figure}

In order to study the non-equilibrium dynamics of the model proposed above, the Schwinger-Keldysh formalism is instrumental. In this formalism, observables are computed by evolving the system forwards and backwards in time along the  Keldysh contour ${\cal C}$ (see Fig. \ref{fig:Keldyshcontour} and Section \ref{sec:SchwingerKeldysh}). The action is therefore
\begin{equation}
    S=\int_\mathcal{C}dt\Biggl[\frac{i}{2}\sum_{a=L,R}\sum_j\chi_a^j\partial_t\chi_a^j-\frac{1}{4!}\sum_{a=L,R}\sum_{i,j,k,l}f_a(t) J_{ijkl}~\chi_a^i\chi_a^j\chi_a^k\chi_a^l-\frac{i\mu(t)}{2}\sum_j\left(\chi_L^j\chi_R^j-\chi_R^j\chi_L^j\right)\Biggr]~.
\end{equation}

After performing the disorder average, the partition function can be written in terms of an effective action $S_\text{eff}[G,\Sigma]$, whose arguments are the contour-ordered Green's functions
\begin{equation}
    i G_{ab}(t_1,t_2)=\frac{1}{N}\sum_j\langle \mathcal{T}_\mathcal{C}\chi_a^j(t_1)\chi_b^j(t_2)\rangle
    \label{eq:contourGF}
\end{equation}
and the associated self-energy  $\Sigma(t_1,t_2)$. That is,
\begin{equation}
\overline{Z}=\int\mathcal{D}G_{ab}\mathcal{D}\Sigma_{ab}\exp\left[\frac{iS_\text{eff}[G,\Sigma]}{N}\right]
\end{equation}
with
\begin{equation}
\begin{aligned}
    \frac{iS_\text{eff}[G,\Sigma]}{N} & =\frac{1}{2}\log\det \left(-i\left[G_0^{-1}\right]_{ab}(t_1,t_2)-\mu_{ab}(t_1)\delta(t_1-t_2)+i\Sigma_{ab}(t_1,t_2)\right)\\
    &-\frac{1}{2}\int_\mathcal{C}dt_1dt_2\sum_{a,b}\Big(\Sigma_{ab}(t_1,t_2)G_{ab}(t_1,t_2)+\frac{J_L(t)J_R(t)}{4}G_{ab}(t_1,t_2)^4\Big)~,
\end{aligned}
\end{equation}
where $J_a(t)\equiv J f_a(t)$ and
\begin{equation}
    \mu_{ab}(t)=\begin{pmatrix}
0 & \mu(t) \\
-\mu(t) & 0
\end{pmatrix}~.
\end{equation}

The large-$N$ saddle point equations of motion are $\delta S_\text{eff}/\delta G_{ab}=0$ and $\delta S_\text{eff}/\delta 
\Sigma_{ab}
=0$.
In terms of the greater and lesser components, defined as follows
\begin{align}
\begin{split}
    G_{ab}^>(t_1,t_2)&=-\frac{i}{N}\sum_j\langle\chi_a^j(t_1)\chi_b^j(t_2)\rangle~,\\
    G_{ab}^<(t_1,t_2)&=-\frac{i}{N}\sum_j\langle\chi_b^j(t_2)\chi_a^j(t_1)\rangle~,
\end{split}
\end{align}
we can obtain the Kadanoff-Baym equations for the two-point functions $G_{ab}^>(t_1,t_2)$ (see  Appendix \ref{app:KBequations} for details)
\begin{equation}
\begin{aligned}
    i\partial_{t_1}G_{ab}^>(t_1,t_2) & = i\mu_{ac}(t_1)G_{cb}^>(t_1,t_2)+\int_{-\infty}^{\infty}dt~ \Sigma_{ac}^R(t_1,t)G_{cb}^>(t,t_2)+\int_{-\infty}^{\infty}dt~ \Sigma_{ac}^>(t_1,t)G_{cb}^A(t,t_2)~,\\
    -i\partial_{t_2}G_{ab}^>(t_1,t_2) & = i G_{ac}^>(t_1,t_2)\mu_{cb}(t_2)+\int_{-\infty}^{\infty}dt~ G_{ac}^R(t_1,t)\Sigma_{cb}^>(t,t_2)+\int_{-\infty}^{\infty}dt ~G_{ac}^>(t_1,t)\Sigma_{cb}^A(t,t_2)~.
    \label{eq:KBeqs}
\end{aligned}
\end{equation}

For the self-energies we find 
\begin{equation}
    \Sigma^>_{ab}(t_1,t_2) =-J_a(t_1)J_b(t_2)G^>_{ab}(t_1,t_2)^3~.
    \label{eq:Sigmarealt}
\end{equation}

Following the literature \cite{Haldar_2019,Bhattacharya_2018,Eberlein_2017,Kuhlenkamp_2019,Larzul_2021,Maldacena_2019,Zhang_2020,Hosseinabadi_2023,Berenguer_2024,Guo_2024,Jaramillo_2024}, the prevailing approach to integrate these equations numerically involves a two-times grid and employs a predictor-corrector method. The reason a two-times grid is needed is because, out of equilibrium, the Green's functions depend in general on two times, instead of  only on time differences, as it happens in equilibrium. This makes the non-equilibrium  numerical solving much more resource consuming than the equilibrium one. In Appendix \ref{app:numerics} we provide details on the numerical method to solve these equations.

As explained in Section \ref{subsec:2coupledSYK}, the phase diagram of the model shows two phases, separated by a first-order phase transition \cite{Maldacena_2018}. The low temperature phase, containing the ground state, which is identified with a traversable wormhole geometry, and the high temperature phase, more similar to two disconnected thermal SYK systems, which is identified with 2 AdS black holes (see Fig. \ref{fig:phasediagramdrivings}). In the canonical ensemble both phases are connected by an unstable phase, the so-called hot wormhole. This phase is presumably  stable in the microcanonical ensemble \cite{Maldacena_2019}. 

A key feature of the wormhole phase is the regularly peaked structure of the spectral function, as opposed to the continuum shown in the SYK case. 
As shown in \cite{Plugge_2020}, this spectrum neatly approaches the conformal spectrum $E_n^\text{conf} = t'(\Delta+n)$, with $\Delta = 1/4$ and $t' \propto \mu^{2/3}$, in the limit $\mu\to 0$. Such regularity is behind the revival phenomenon that shows up as the dual counterpart to the traversing of the wormhole by a probe particle \cite{Maldacena_2017,Bak_2018}. These revivals are neatly seen in the time evolution of the transmission coefficients
\begin{equation}
    T_{ab}(t)=2\abs{G^>_{ab}(t,0)}~.
\end{equation}

\begin{figure}
    \centering
    \includegraphics[width=0.49\linewidth]{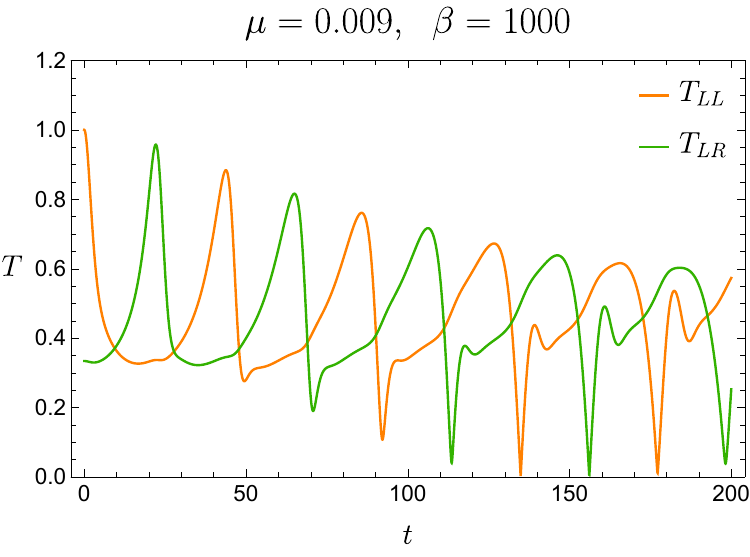}
    \hfill
    \includegraphics[width=0.49\linewidth]{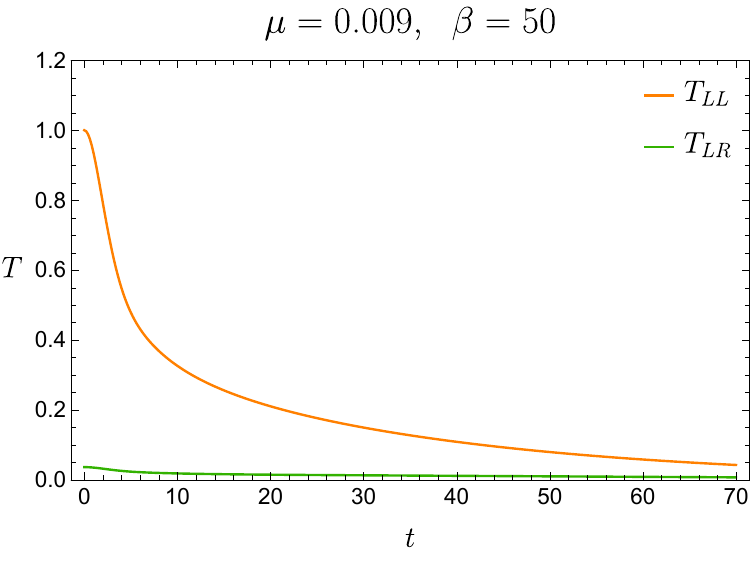}
    \caption{Left (right): transmission coefficients in the wormhole (black hole) phase. The phase opposition of the revivals in the wormhole phase is consistent with having a traversable wormhole. In the black hole phase the correlators decay exponentially with time. Also the frequency scaling as $\sim \mu^{2/3}$ is, for small $\mu\ll J$, much higher than the natural coupling frequency $\sim \mu$.}
    \label{fig:RevivalsWHBH}
\end{figure}

\begin{figure}[H]
    \centering
    \includegraphics[width=0.47\textwidth]{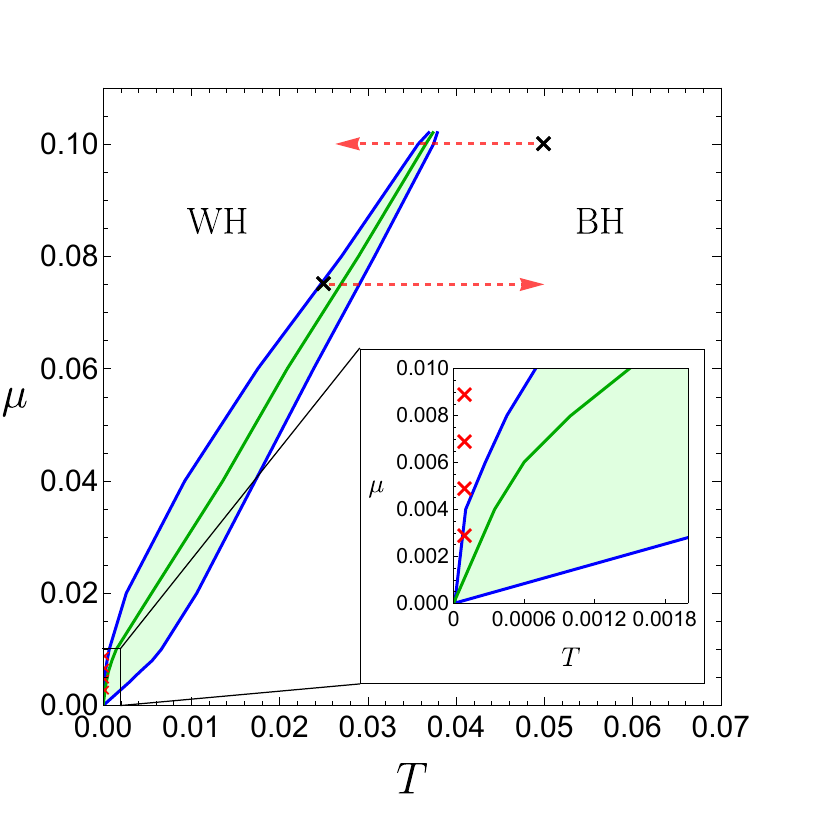}
    \caption[]{$\mu$-$T$ phase diagram separating the two phases. The colored region shows where the two phases coexist. On the left (right) of the green line, the wormhole (black hole) phase is more stable. We show in black crosses the areas of the phase diagram that have been explored in other non-equilibrium analysis of the model \cite{Maldacena_2019,Zhang_2020}, with arrows showing the trajectories of their processes\protect \footnotemark. We show in red crosses the areas we are able to reach with NESSi, much deeper inside the wormhole regime. Right: phase diagram for $\mu=0.01$. The existence of a canonically unstable but microcanonically stable phase can be seen by computing the energy of each phase.}
    \label{fig:phasediagramdrivings}
\end{figure}

The transmission coefficients, $T_{ab}$, encode the probability of recovering $\chi^j_a$ at time $t$  after having inserted $\chi^j_b$ at time $t=0$. Fig. \ref{fig:RevivalsWHBH} shows the different behaviors in the two phases. On the left plot the peaked out-of-phase revivals are interpreted as being duals of a probe particle traversing from one side of the wormhole to the other and bouncing forever back and forth. The slow decay in the overall envelope still deserves an explanation from the dual gravitational point of view.

\footnotetext{Strictly speaking, the phase diagrams of \cite{Maldacena_2019,Zhang_2020} are not exactly equal as the one showed here because their setups are different: in \cite{Maldacena_2019} they couple the system to a cold bath, and in \cite{Zhang_2020} they consider the complex version of the model. However, as far as numerical resources are concerned, these distinctions are not important.}
In our protocol, initially the system is set to a thermal equilibrium state characterized by a temperature parameter, $\beta$, and the drivings are turned on at $t_0=0$. Specifics regarding the derivation of the equilibrium initial conditions can be found in Appendix \ref{app:realtimeeqs}.

The predictor-corrector method hinges upon the exponential decay of equilibrium Green's functions, a typical behavior in the black hole phase. However, in the wormhole phase, the revival dynamics significantly slows down the decay, especially at high $\beta$ values which renders it impractical. To circumvent these limitations, leveraging the NESSi (Non-Equilibrium Systems Simulation \cite{Schuler_2019}) library has proven crucial for exploring the most interesting region of the phase diagram $-$specifically, the low-$\mu$, low-$T$ area within the wormhole phase. In Fig. \ref{fig:phasediagramdrivings} we show the phase diagram in the $(T,\mu)$ plane (as we did in Fig. \ref{fig:phasediagram}), showing the regions we are able to explore with NESSi, compared to other similar works. NESSi adopts the L-shaped Kadanoff-Baym contour shown on the right plot in Fig. \ref{fig:Keldyshcontour}, hence obtaining the equilibrium state directly from the Matsubara formalism. Additionally, NESSi incorporates advanced high-order integration routines, vital for achieving high-fidelity results. In order to verify the results shown in this thesis, we have used both integration schemes when possible.

Finally, let us recall that, in the low-energy limit, $\mu\ll J$, the model is effectively governed by the following Schwarzian action (see Section \ref{subsec:2coupledSYK} and Appendix \ref{app:Schwarzianeqs}) \cite{Maldacena_2018}
\begin{equation}
    S=N\int du \left[-\frac{\alpha_S}{\mathcal{J}}\left(\left\{\tan\frac{t_l(u)}{2},u\right\}+\left\{\tan\frac{t_r(u)}{2},u\right\}\right)+\mu\frac{c_\Delta}{(2\mathcal{J})^{2\Delta}}\left(\frac{t'_l(u)t'_r(u)}{\cos^2{\frac{t_l(u)-t_r(u)}{2}}}\right)^\Delta\right]~,
    \label{eq:Schwaction2SYK}
\end{equation}
where $2\mathcal{J}^2=J^2$, $c_\Delta=\frac{1}{2}\left[(1-2\Delta)\frac{\tan \pi\Delta}{\pi\Delta}\right]^\Delta$, and $\alpha_S$ is a numerical constant \cite{Maldacena_Stanford_SYK}.

It is also a good point to recall that the model has two integerly spaced spectrums: the "conformal tower" and the boundary graviton. Since these two towers will be important, we just rewrite them here (see Section \ref{subsec:2coupledSYK}):
\begin{align}
    E_n^\text{conf} & = t'(\Delta+n)~,\label{eq:2towersconfresearch}\\
    E_n^\text{bg} & = \omega_0\left(n+\frac{1}{2}\right)~,\qquad \omega_0=t'\sqrt{2(1-\Delta)}~.\label{eq:2towersbgresearch}
\end{align}
with $\Delta=1/4$ in our setup, and $t'\sim \mu^{2/3}$ in the low-$\mu$ limit.

\subsection{Observables of interest}

In order to monitor the response of the system to the driving, we will analyze the time evolution of the total energy, $E_{tot}(t)=\langle H\rangle =\langle H_L\rangle+\langle H_R\rangle+\langle H_{int}\rangle$ in the hamiltonian \eqref{eq:H2SYKtdep} for different frequencies. The energy can be written in terms of the Green's functions, and in the case of arbitrary time-dependent couplings $\mu(t)$ and $J_{L/R}(t)$ it acquires the following form
\begin{equation}
\begin{aligned}
    \frac{1}{N}\langle H_L(t)\rangle=-\frac{iJ_L(t)}{4}\int_{-\infty}^{t}dt'\Big[&J_L(t')\left(G_{LL}^>(t,t')^4-G_{LL}^<(t,t')^4\right)\Big.\\
    \Big.+&J_R(t')\left(G_{LR}^>(t,t')^4-G_{LR}^<(t,t')^4\right)\Big]~,
    \label{eq:E_L}
\end{aligned}
\end{equation}
with the corresponding expression for $\langle H_R(t)\rangle$ upon doing $L\leftrightarrow R$. The total energy is given by
\begin{equation}
    E_{tot}(t)=E_L(t)+E_R(t)+E_{int}(t)~,
    \label{eq:Etot}
\end{equation}
where $E_{int}(t)=-N\mu(t)G^>_{LR}(t,t)$. Details are provided in Appendix \ref{app:energytdep}. 
However, it is evident from the expressions that the time-dependent modulations $J_{L/R}(t)$ as well as $\mu(t)$ will dominate and will "hide" the part of the evolution of the energy that comes from the response of the system to the drivings, encoded in the time evolution of the Green's functions. For that reason, we choose to evaluate these expressions taking the equilibrium values of the couplings $J_{L/R}$ and $\mu$. This corresponds to the expectation values of the undriven hamiltonian, $\langle H\rangle (t)$.\footnote{We thank C. Kuhlenkamp for clarification on this point. }

We would also like to study the heating dynamics. To track the temperature evolution during the process, we use a variant of the fluctuation-dissipation (FD) relation, which, in equilibrium, states that
\begin{equation}
    \frac{i G^K(\omega)}{\rho(\omega)}=\tanh\frac{\beta\omega}{2}~,
    \label{eq:FDTequil}
\end{equation}
where $G^K(\omega)$ is the Fourier transform of the Keldysh Green's function, defined in \eqref{eq:retadvkel}, and $\rho(\omega)=-2\Im G^R(\omega)$ is the spectral function. This relation follows from the KMS condition
\begin{equation}
    G^>(\omega)=-e^{-\beta\omega}G^<(\omega)~.
\end{equation}
Eq. \eqref{eq:FDTequil} allows to extract the temperature of the system from the Green's functions. 

Out of equilibrium, the notion of temperature is not well-defined. However, if the process is slow enough, the correlation functions are expected to exhibit near-thermal behavior at a particular effective temperature. One way of defining such temperature is by rotating the time variables into relative time $t$ and average time $\mathcal{T}$, defined as
\begin{equation}
    t=t_1-t_2,~~~~~\mathcal{T}=\frac{t_1+t_2}{2}~,
\end{equation}
and computing the Fourier transform with respect to $t$. For a general function $G(t_1,t_2)\rightarrow G(\mathcal{T},t)$, this yields the Wigner transform $G(\mathcal{T},\omega)$,
\begin{equation}
    G(\mathcal{T},\omega)=\int dt~e^{i\omega t}G(\mathcal{T},t)~.
    \label{eq:Wignertransf}
\end{equation}

Defined this way, the time-translation symmetry breaking due to the driving is contained in the dependence on $\mathcal{T}$. Now one can define the $\mathcal{T}$-dependent quantities $G^K(\mathcal{T},\omega)$, $\rho(\mathcal{T},\omega)$, and fit
\begin{equation}
    \frac{i G^K_{LL}(\mathcal{T},\omega)}{\rho_{LL}(\mathcal{T},\omega)}=\tanh\frac{\beta_\text{eff}(\mathcal{T})\omega}{2}
    \label{eq:betaeff}
\end{equation}
at each average time $\mathcal{T}$ to obtain a time-dependent effective temperature.

The caveat is that this quotient will only approach a hyperbolic tangent when the system is not too far from an equilibrium configuration. We will use this $\beta_\text{eff}(\mathcal{T})$ as an indicator of the heating. At a numerical level, the Wigner transform cannot be computed in a finite domain of times
unless the decay in relative time $t = t_1-t_2$ is strong enough. In the wormhole phase the oscillations decay at a very slow rate, so this strategy cannot be used in general and we will face strong limitations when computing the effective temperature for large $\beta$ (small $\mu$) wormholes. However, the energy is numerically easier to compute in all cases.

\section{Numerical results (I): Resonant drivings}\label{sec:numresults1}

This section provides a first batch of numerical results. We focus on periodic drivings of the couplings and study the response of the system. We first study the black hole and wormhole phases under periodic drivings of the coupling $\mu$, gaining some insight from the Schwarzian action at the end. We also drive the couplings $J_L(t)$, $J_R(t)$.

\subsection{Driving the \texorpdfstring{$L$}{L}-\texorpdfstring{$R$}{R} coupling \texorpdfstring{$\mu$}{mu}}\label{subsec:mudriving}

We begin our numerical study by perturbing sinusoidally  $\mu$, that is the coupling between the $L$ and $R$ sides appearing in \eqref{eq:H2SYKtdep}
\begin{equation}
        \mu(t)=\mu_0(1+a\sin\Omega t)~,
        \label{eq:drivingmu}
\end{equation}
with $\mu_0$ its equilibrium value.

The most interesting effects arise in the wormhole phase (low $\mu$, large $\beta$), but we begin by considering the black hole phase for later comparison.

\begin{figure}[t]
    \centering
    \includegraphics[width=0.49\textwidth]{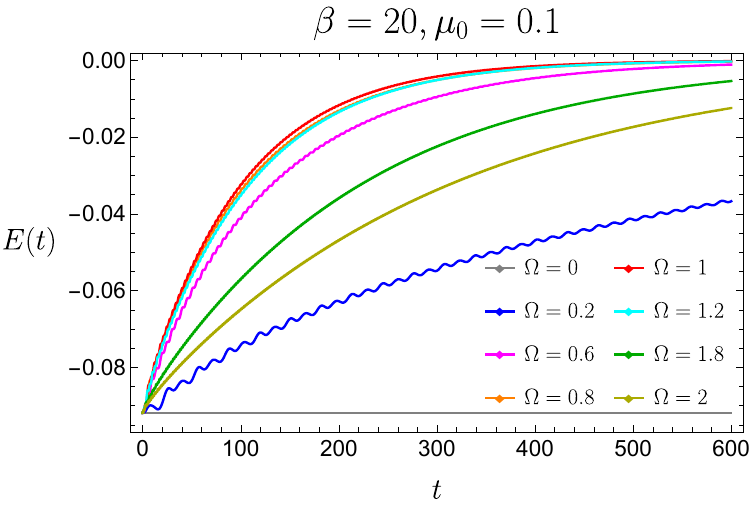}
    \includegraphics[width=0.49\textwidth]{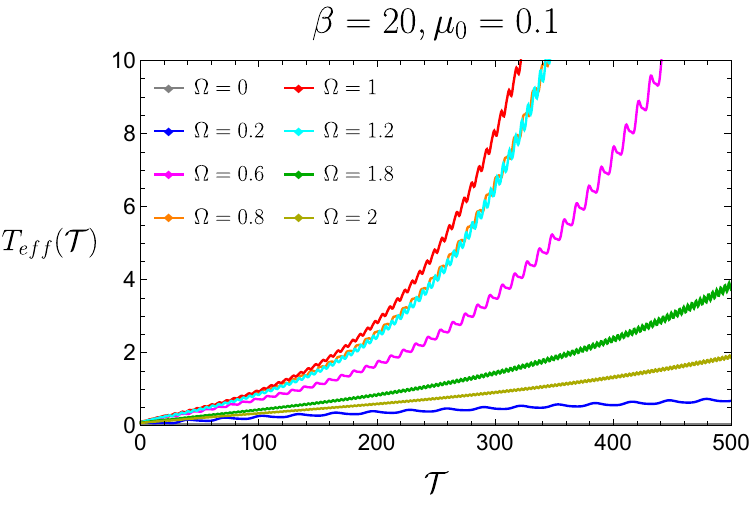}
    \caption{Time dependence of the energy and the effective temperature for different frequencies of the driving in the black hole phase. The amplitude is taken to be $a=0.7$. We observe that the absorption and the heating reach a maximum value for $\Omega\sim 1$ and then they decrease again.}
    \label{fig:BHenergyTeff}
\end{figure}

Some examples of the $t$-dependent energy, $E_{tot}(t)$,  computed using expressions \eqref{eq:E_L}-\eqref{eq:Etot}, and the ${\mathcal T}$-dependent effective temperature, $T_\text{eff}({\mathcal T})$, from \eqref{eq:betaeff}, are shown in Fig. \ref{fig:BHenergyTeff}.

\begin{figure}[t]
     \centering
     \begin{subfigure}[t]{0.47\textwidth}
         \centering
         \includegraphics[width=\textwidth]{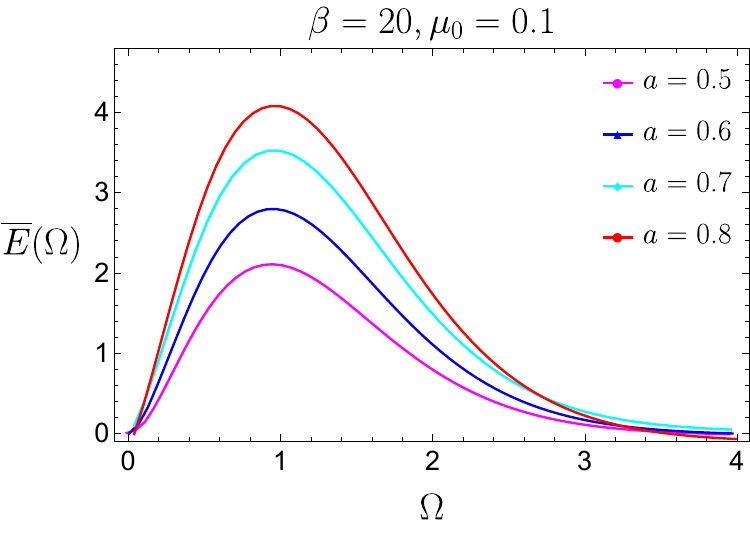}
         \caption{}
         \label{sfig:BHintegr}
     \end{subfigure}
     \begin{subfigure}[t]{0.49\textwidth}
         \centering
         \includegraphics[width=\textwidth]{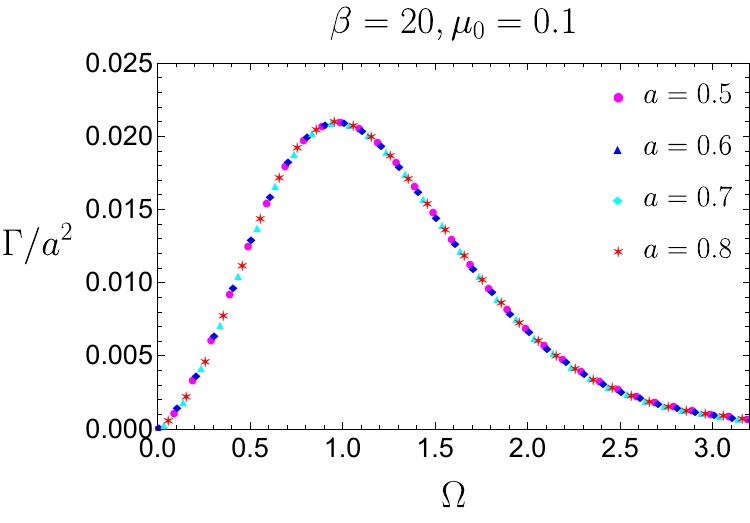}
         \caption{}
         \label{sfig:BHfit}
     \end{subfigure}
        \caption{Left: integrated energy $\overline{E}(\Omega)$ for different amplitudes in the black hole phase. We observe a maximum around $\Omega\sim 1$. Right: heating rate $\Gamma$. We find that the rate has a universal behavior $\Gamma(a,\Omega)=a^2f(\Omega)$, as it was already observed in \cite{Kuhlenkamp_2019}.}
        \label{fig:BHintegrfit}
\end{figure}

In order to characterize and study the net absorption of energy as a function of the driving frequency $\Omega$, we will make use of the integrated energy 
\begin{equation}
    \overline{E}=\int_{0}^{100}\left[E_{tot}(t)-E_{tot}(0)\right]dt~.
\end{equation}
for different driving frequencies. The choice of upper value $t\leq 100$  will become clear in the analysis of the wormhole phase.  Fig. \ref{sfig:BHintegr} shows the value of  $\overline{E}$ as a function of $\Omega$ for four amplitudes $a$. In all cases the plots  reach maximum around $\Omega\sim 1$.  We fitted the energy to an exponential,
\begin{equation}
    \abs{E_{tot}(t)}\sim e^{-\Gamma(a,\Omega) t}~,
\end{equation}
with a driving-dependent heating rate $\Gamma(a,\Omega)$. Fig. \ref{sfig:BHfit} shows this rate, where we observe a universal behavior, $\Gamma(a,\Omega)=a^2f(\Omega)$, that matches the one  observed in \cite{Kuhlenkamp_2019} in the strange-metal phase.

New physics appears if the initial state belongs, instead, to the wormhole phase. Consider for instance an initial equilibrium solution with $\beta=1000$, $\mu=0.009$. At $t=0$ we turn on the driving \eqref{eq:drivingmu} and see how the transmission amplitudes between the two sides are affected by the injection of energy. For high amplitudes of the driving ($a\sim \mathcal{O}(1)$) the injection of energy is so violent that the system transitions to the black hole phase almost immediately. To remain in the wormhole phase for longer times it is necessary to choose smaller perturbation amplitudes.

\begin{figure}
     \centering
     \begin{subfigure}[t]{0.49\textwidth}
         \centering
         \includegraphics[width=\textwidth]{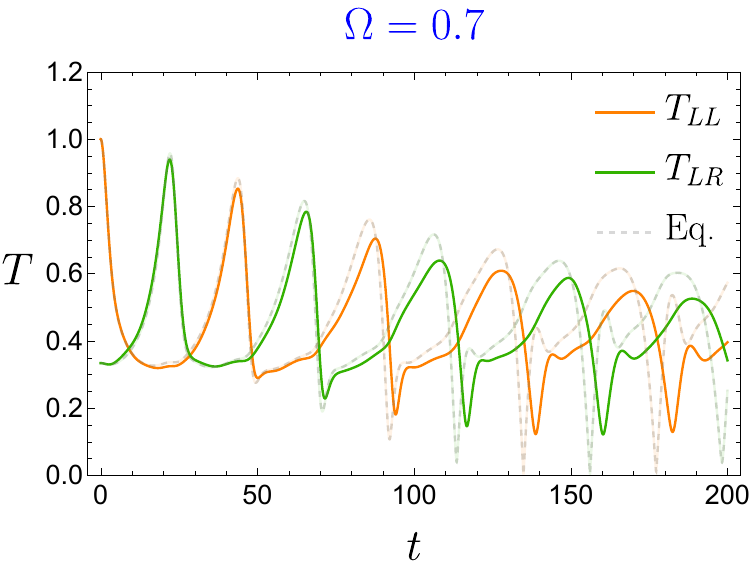}
         \caption{}
         \label{sfig:WHrevivalsnores}
     \end{subfigure}
     \begin{subfigure}[t]{0.49\textwidth}
         \centering
         \includegraphics[width=\textwidth]{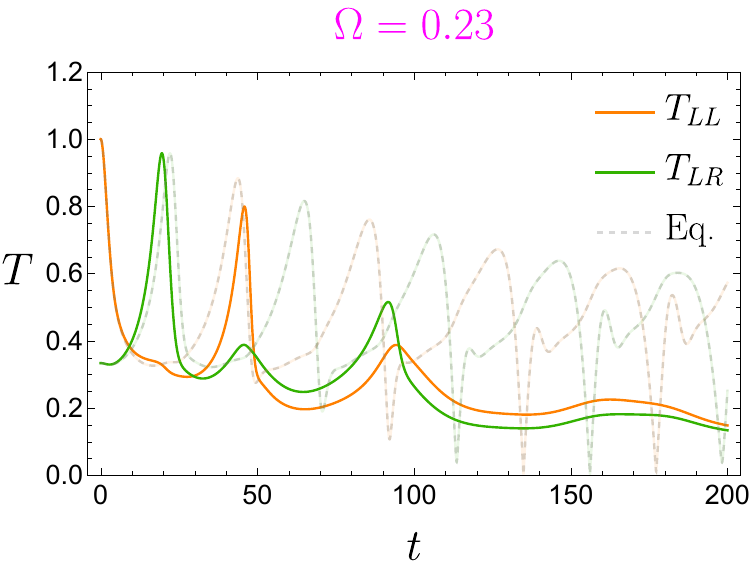}
         \caption{}
         \label{sfig:WHrevivalsres}
     \end{subfigure}
     \begin{subfigure}[t]{0.55\textwidth}
         \centering
         \includegraphics[width=\textwidth]{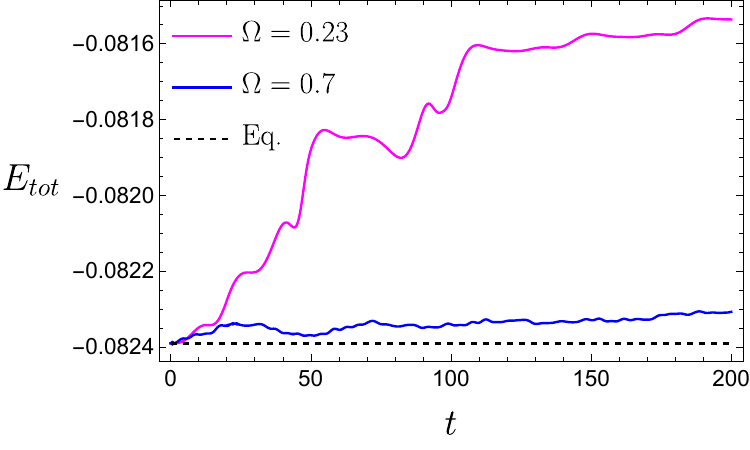}
         \caption{}
         \label{sfig:WHenergy}
     \end{subfigure}
        \caption{The two different behaviors we observe. In \ref{sfig:WHrevivalsnores} for non-resonant frequencies the revivals retain mostly their equilibrium form. In \ref{sfig:WHrevivalsres}, for resonant frequencies we observe a rapid decay of the transmission amplitudes while they cease to be in phase opposition, signalling that the wormhole has closed. For comparison,  we plot the shape of the revivals in equilibrium in light dashed. \ref{sfig:WHenergy}: we show the behavior of the energy for the two cases. We observe how, when driving at the resonant frequencies, the early time absorption becomes exponential. Between $t=50$ and $t=100$, the behavior changes. Looking at \ref{sfig:WHrevivalsres}, this corresponds to the moment where the transition to the black hole phase is happening. After $t=100$, the behavior is essentially the same as in the non-resonant case. }
        \label{fig:WHrevivals}
\end{figure}

When $a$ is small enough, we find two types of behavior upon varying the driving frequency $\Omega$. In most of the frequency range, the revivals are almost unaffected (Fig. \ref{sfig:WHrevivalsnores}). However, for some discrete set of frequencies, the transmission amplitudes decay and become similar to those of the black hole phase (Fig. \ref{sfig:WHrevivalsres}). We interpret this effect as a resonance that triggers a transition to the high-temperature phase.

One would like to understand the nature of the resonant frequencies. As we show in Fig. \ref{sfig:WHenergy}, the absorption of energy is very different between resonant and non-resonant frequencies: while in the non-resonant case the energy increases very slowly, in the resonant frequencies the heating is exponential at early times, triggering the transition to the high-temperature (black hole) phase. In section \ref{subsec:forminghotWH} we analyze this transition in more detail.

\begin{figure}[!ht]
     \centering
     \begin{subfigure}[t]{0.49\textwidth}
         \centering
         \includegraphics[width=\textwidth]{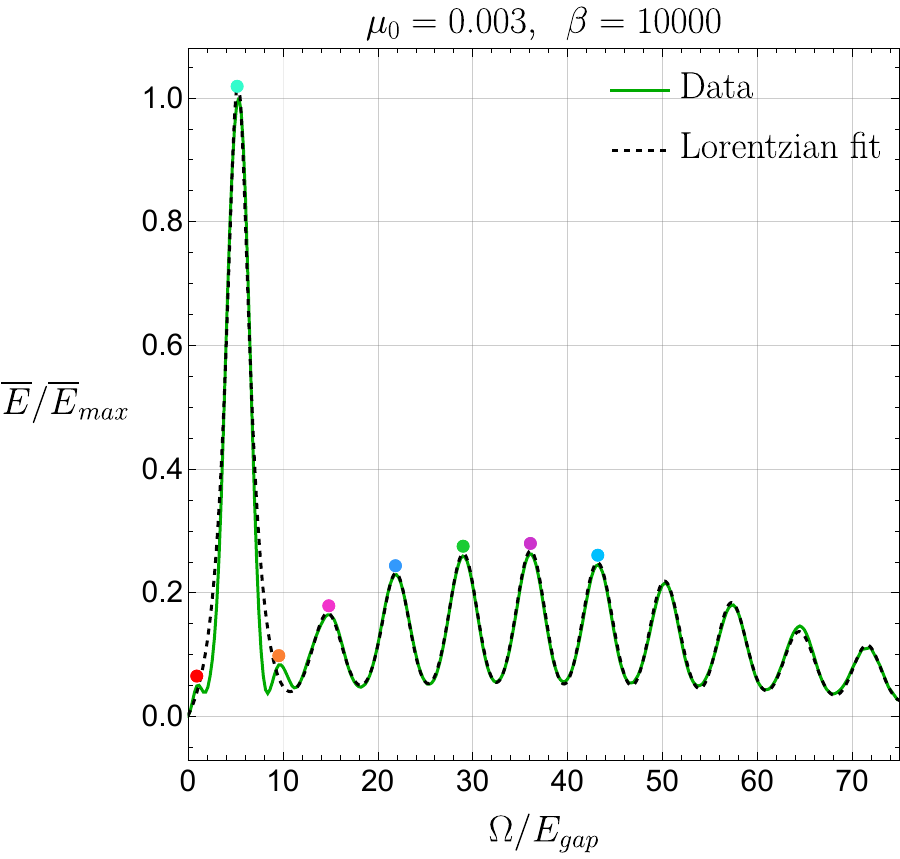}
         \caption{}
         \label{sfig:WHintegrenergy}
     \end{subfigure}
     \begin{subfigure}[t]{0.49\textwidth}
         \centering
         \includegraphics[width=\textwidth]{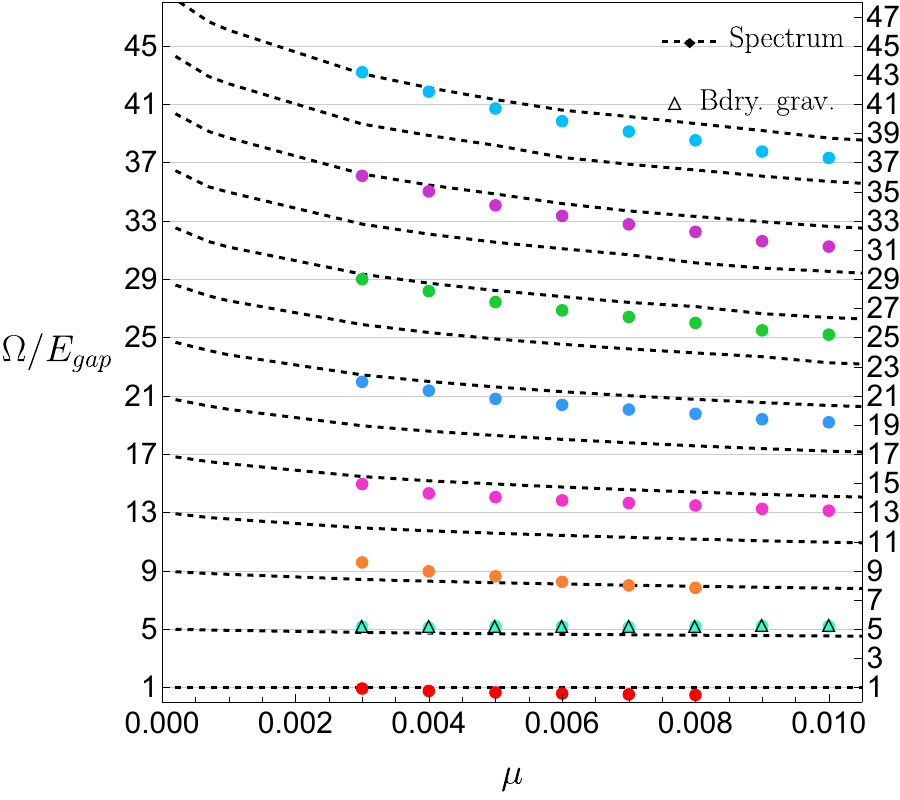}
         \caption{}
         \label{sfig:Peaksspectrum}
     \end{subfigure}
        \caption{Left:~(normalized) integrated energy for an amplitude $a=1/50$. We observe a series of absorption peaks at the frequencies where the revivals are suppressed. The peaks are well approximated by a sum of Lorentzians. Right: Solid dots mark absorption peaks for varying $\mu$,  with the same color coding as in the left plot. In dashed, a fit to the  spectrum of the undriven model computed numerically in \cite{Plugge_2020}. Notice that for small $\mu$ the spectrum approaches the conformal tower \eqref{eq:2towersconfresearch}, whose lowest level defines $E_{gap}$. We also show the frequency $\omega_0(\mu)=t'(\mu)\sqrt{2(1-\Delta)}$ of the boundary graviton (see Eq. \eqref{eq:2towersbgresearch}), which coincides with our second (and highest) absorption peak on the left.}
        \label{fig:WHresonances}
\end{figure}

The dependence of the  energy absorption,  $\overline E$, with the driving frequency, $\Omega$, departs significantly from the one in the black hole phase shown in Fig. \ref{sfig:BHintegr}. Now, Fig. \ref{sfig:WHintegrenergy} exhibits the integrated energy as one varies the frequency\footnote{For this, we choose an amplitude small enough to remain in the wormhole phase during all the time used for the integrated energy ($t=100$). In this way we capture the intrinsic behavior of the wormhole phase, and we don't get contamination from the transition to the black hole phase, which will be reached inevitably at some later time.}.    The peaks correspond precisely to the frequencies where the revivals get maximally suppressed, and the overall curve is well fitted to a sum of Lorentzians
\begin{equation}
    \overline{E}(\Omega)\sim \sum_{i=\text{peaks}}\frac{1}{\pi}\frac{A_i}{(\Omega-\Omega_i)^2+\gamma_i^2}~,
\end{equation}
with $A_i$, $\gamma_i$, $\Omega_i$ being characteristic of each resonant peak.\footnote{The global fit works better if we don't include the two small peaks before and after the highest one. However, we consider them to be resonances.}

The resonant frequencies are easily extracted from Fig. \ref{sfig:WHintegrenergy}, but the question of their meaning remains open. For that, we repeated the analysis for lower values of $\mu$. In each case, we compute the integrated energy and identify the resonant frequencies as the frequencies where the absorption has a maximum and the corresponding transmission coefficients are highly suppressed. After all the resonances are collected, the natural thing to do is to compare them with the spectrum of the undriven system, which was computed numerically in \cite{Plugge_2020}.

From the results, shown in Fig. \ref{sfig:Peaksspectrum}, one observes that the resonances indeed come very close to the levels of the spectrum. Remarkably, this only happens in about half of the spectrum. This is a very intriguing feature for which we don't have a clear explanation. To study this phenomenon from the dual gravity side, a fully backreacted gravitational calculation in the dual geometry seems to be needed. Following the calculations in \cite{Gao_2016, Bak_2018}, one could study how the violation of the averaged null energy condition is affected by the driving.

The second resonance stands out prominently. Although it is close to the second level ($n=1$) of the conformal spectrum, it would be the only peak that violates the evenly spaced structure found for the others. 

In fact, it seems to be closer to the {\it natural frequency} of the boundary graviton, $\omega_0$ given in \eqref{eq:2towersbgresearch}. This is surprising at first sight, since graviton states don't seem to show up as features of the spectral function \cite{Plugge_2020}. This fact was justified from being expected only as a $1/N$ order effect beyond the saddle point calculation involved here.

In contrast, the fact that we can see it could come from driving, precisely, the parameter that gives rise to this spectrum. We will give more support to this claim in Section \ref{subsec:drivingSchwarzian}, analytically in the limit where the Schwarzian approximation is reliable.

\subsection{Forming hot wormholes?}\label{subsec:forminghotWH}

It has been  argued  that the black hole  and  wormhole phases are connected by a set of metastable states referred to as "hot wormholes" \cite{Maldacena_2018,Maldacena_2019}. Being unstable  in the canonical ensemble, one does not expect to find them  when solving the equilibrium Schwinger-Dyson equations at a given temperature. By putting the system in thermal contact with a cooler bath, evidence of the presence of those states is obtained from the behaviour of the temperature with energy \cite{Maldacena_2019}.

In Fig. \ref{fig:WHrevivals}, we showed that by driving periodically $\mu$ in a resonant frequency it is possible to make the wormhole transition into a black hole. In this section we look for a better characterization of that transition, in the seek for signals that might correspond to such hot wormhole states.

\begin{figure}
     \centering
     \begin{subfigure}[t]{0.55\textwidth}
         \centering
         \includegraphics[width=\textwidth]{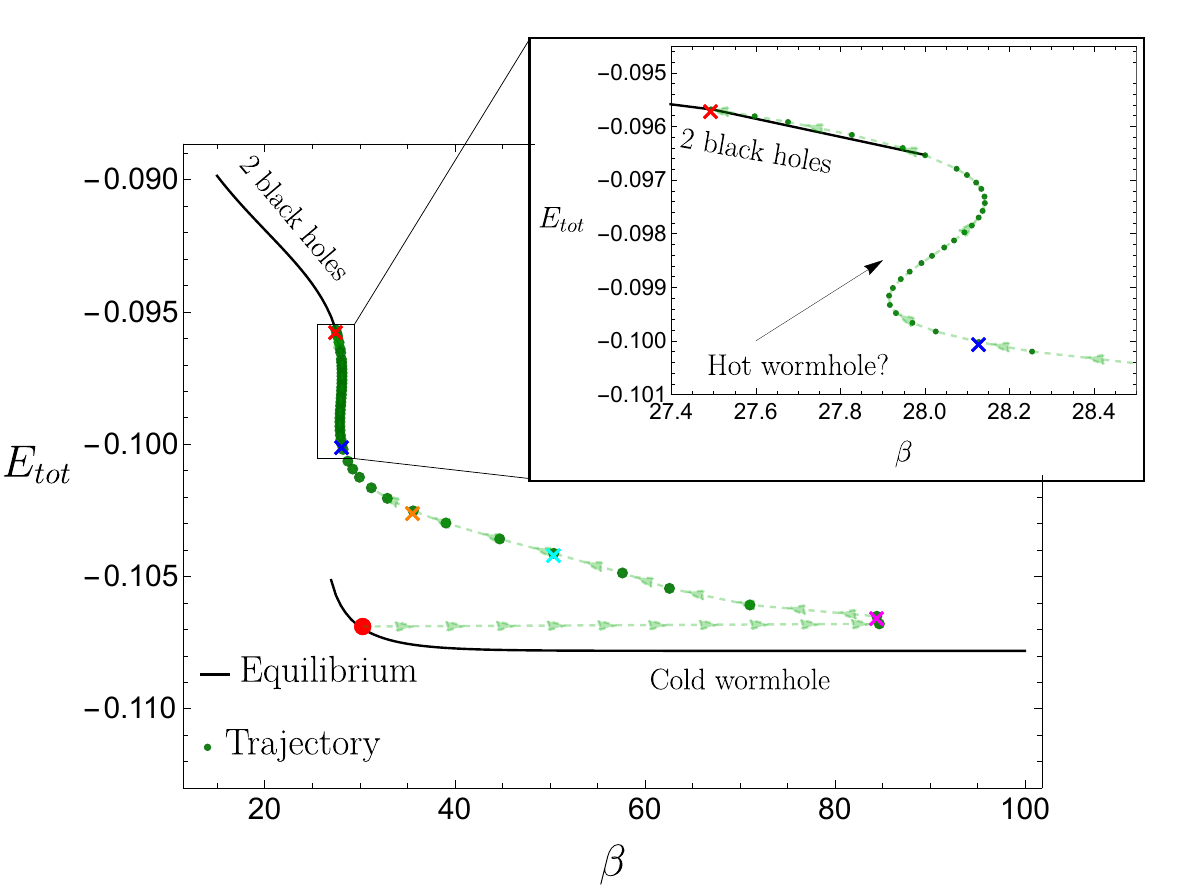}
         \caption{}
         \label{sfig:hotWHAll}
     \end{subfigure}
     \begin{subfigure}[t]{0.44\textwidth}
         \centering
         \includegraphics[width=\textwidth]{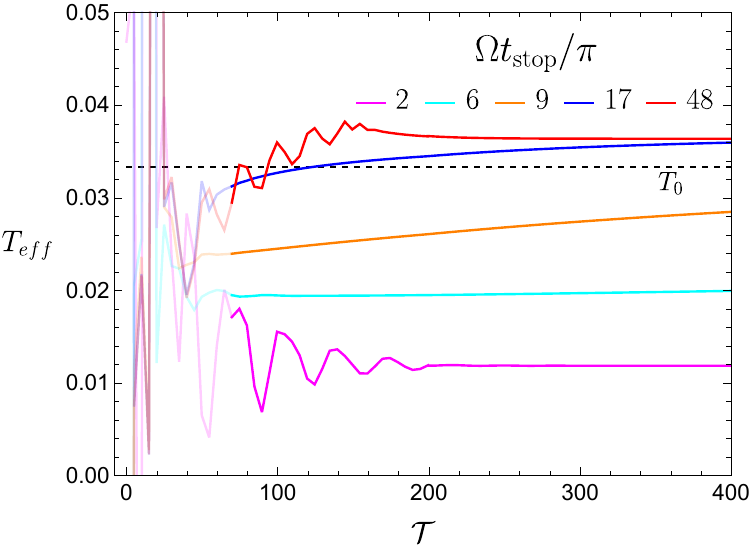}
         \caption{}
         \label{sfig:hotWHTeff}
     \end{subfigure}
        \caption{Left: asymptotic equilibrium trajectory in the phase diagram. The initial equilibrium solution (red dot) is a wormhole solution of $\mu=0.1$, $\beta=30$, very close to the transition point. We take the drivings to have amplitude $a=0.1$ and frequency $\Omega=1$, and each green dot represents the final equilibrium temperature and energy for growing $t^n_{stop}=\frac{\pi}{\Omega}n$ along the direction of the green arrows. We see that the system follows a trajectory that joins smoothly with the black hole phase. Zooming at the points with larger $t_{stop}$, we see that the trajectory has a shape that is characteristic of a phase transition between the phases, as expected. Right: time evolution of the effective temperature for the highlighted points with colored crosses of Fig. \ref{sfig:hotWHAll}.}
        \label{fig:hotWH}
\end{figure}

Our protocol starts by initializing a wormhole state at equilibrium near the threshold. Then we activate the driving as in \eqref{eq:drivingmu}, and sustain it for a certain number of half cycles, $t_{stop}=\frac{\pi}{\Omega}n$. After this, the driving is quenched without discontinuities, and the system returns smoothly to isolation. As a consequence of the driving, the energy of the system will have risen in a way proportional to $n$, and remain as a constant of motion after switch-off. Effectively we are in the microcanonical ensemble and we may continue the free evolution to look for late time thermalization. After equilibrium has settled, the effective temperature of the state can be determined using the aforementioned method \eqref{eq:betaeff}. Each end state gives a point in an $E$-$\beta$ plot.  By repeating this procedure with increasing values of $t_{stop}$, the obtained points trace out the green curve on the left plot in Fig. \ref{fig:hotWH}.

Monitoring the evolution of the effective temperature down to equilibration requires that the Green's functions decay rapidly enough at the boundaries of the time domain. This is numerically demanding in the wormhole phase where the Green's functions decay slowly for small $\mu$ and large $\beta$.  This puts restrictions in these parameters in order to make  the numerics fit within our resources: we will chose for our initial equilibrium state a wormhole with $\beta=30$, $\mu=0.1$. This is close to the phase transition temperature $\beta\approx27.5$ but far from the small $\mu$, large $\beta$ limit considered in previous sections.

In Fig. \ref{fig:hotWH} we present the results of this driving protocol. We show in black the equilibrium phase diagram (see Fig. \ref{fig:phasediagram} right). Each green dot represents a different simulation, with a different value of $t_{stop}$, all starting in the same initial configuration in the wormhole phase represented by the red dot. The dashed line with arrows shows the direction of increasing $t_{stop}$. We highlight five points in the trajectory with color crosses for which we show, on the right plot, the evolution of the effective temperature as computed using \eqref{eq:betaeff}. We observe that for small values of $t_{stop}$, the system  decreases notably its final  temperature. For larger values of $t_{stop}$, the system follows a trajectory that joins smoothly with the equilibrium curve of the black hole phase. 

In summary, whereas driving the system periodically triggers a  transition into the black hole phase (see previous section), a quenched version of the protocol followed by later free evolution brings the system into some stationary state that we propose to correspond to the conjectured hot wormhole state. 
The states with high $\beta$ obtained for small values of $t_{stop}$ (magenta, light blue and orange points) are equilibrium states that deserve some further study. We will return to them in Section \ref{subsec:chaosFloquet} and \ref{sec:HotWHColdBH}.

\subsection{Schwarzian analysis}\label{subsec:drivingSchwarzian}
We want to study the Schwarzian limit under our periodic driving with the goal of understanding the nature of the dominant peak observed in Section \ref{subsec:mudriving}. In order to do this, we need to make the  replacement $\mu\rightarrow\mu(u)$ in the action \eqref{eq:Schwaction2SYK}, which, in terms of the rescaled boundary time $\tilde{u}$ (see Eq. \eqref{eq:rescaledbdrytime}), translates into the same replacement in the equation of motion:
\begin{equation}
    -e^{2\varphi}-\varphi''+\Delta\eta(\tilde{u}) e^{2\Delta \varphi}=0~,
    \label{eq:schweq}
\end{equation}
where $\eta(\tilde{u})\equiv \frac{ \alpha_S}{\mathcal{J}}\frac{c_\Delta}{(2\alpha_S)^{2\Delta}}\mu(\tilde{u})$ and $\varphi=\log t'$.

The general solution has to be found numerically, but we can predict the existence of at least one resonance in the limit that the driving is a small perturbation. We consider a driving of the form
\begin{equation}
    \eta~\rightarrow~\eta(\tilde{u})=\eta_0(1+a f(\tilde{u}))~,
\end{equation}
where $\eta_0$ is the equilibrium value at the minimum of the potential, and we take $a f(\tilde{u})\ll 1$ (at the end we will consider $f(\tilde{u})=\sin\Omega \tilde{u}$, so we will only need $a\ll1$). If we now assume that the deviation from the equilibrium value $\varphi_0$ will also be small, $\varphi(\tilde{u})\approx \varphi_0+\phi(\tilde{u})$, with $\abs{\phi(\tilde{u})}\ll 1$, at first order in $a f(\tilde{u})$ and $\phi(\tilde{u})$, Eq. \eqref{eq:schweq} becomes that of a forced harmonic oscillator
\begin{equation}
    \phi''(\tilde{u})+\omega_0^2\phi(\tilde{u})=\frac{\omega_0^2}{2(1-\Delta)}af(\tilde{u})~,
    \label{eq:forcedoscillator}
\end{equation}
with $\omega_0$ the frequency of the oscillations that give rise to the boundary graviton excitations, \eqref{eq:2towersbgresearch}, and the driving of $\eta$, $a f(\tilde{u})$, acting as the external force. The solution of \eqref{eq:forcedoscillator} for  general $f(\tilde{u})$ is well known. In the particular case of a sinusoidal driving, $f(\tilde{u})=\sin\Omega \tilde{u}$, it takes the form
\begin{equation}
    \phi(\tilde{u})=\frac{a\h\omega_0}{2(1-\Delta)}\frac{1}{\Omega^2-\omega_0^2}\Big[\Omega\sin\omega_0\tilde{u}-\omega_0\sin\Omega\tilde{u}\Big]~,
\end{equation}
which shows a resonance when $\Omega\rightarrow\omega_0$.\footnote{The limit $\Omega\rightarrow\omega_0$ leads to the usual oscillation with a lineraly growing term, $$\lim_{\Omega\rightarrow\omega_0}\phi(\tilde{u})=\frac{a}{4(1-\Delta)}\Big[\sin\omega_0\tilde{u}-\omega_0\tilde{u}\cos\omega_0\tilde{u}\Big]~.$$} Of course, in that limit the assumption that $\abs{\phi(\tilde{u})}\ll 1$ breaks down and the solution is not valid anymore, but the existence of a resonance in $\phi(\tilde{u})$ (and therefore in $\varphi(\tilde{u})$) will translate into a noticeable change of behavior in the propagators (recall that $\varphi(\tilde{u})=\log t'(\tilde{u})$ and $t'(\tilde{u})$ determines the correlation functions via \eqref{eq:correlatortltr}). This analysis confirms that the highest resonant peak we observe corresponds to an excitation of the boundary graviton degrees of freedom.

\subsection{Driving the source couplings \texorpdfstring{$J_{L,R}(t)$}{J(t)}}\label{subsec:drivingJLR}

Next we shall report on our findings for a periodic driving of the other couplings in the time-dependent Hamiltonian  \eqref{eq:H2SYKtdep}, namely the source couplings $J_{L,R}(t) =J f_{L,R}(t)$. This driving is less motivated physically from an operational point of view as it implies a periodic variation of the variance of the Gaussian distribution of couplings in each SYK.  
Our analysis will be more qualitative than the one carried out for the driving in $\mu(t)$. A full and thorough study will be reported elsewhere.

\begin{figure}
     \centering
     \begin{subfigure}[t]{0.49\textwidth}
         \centering
         \includegraphics[width=\textwidth]{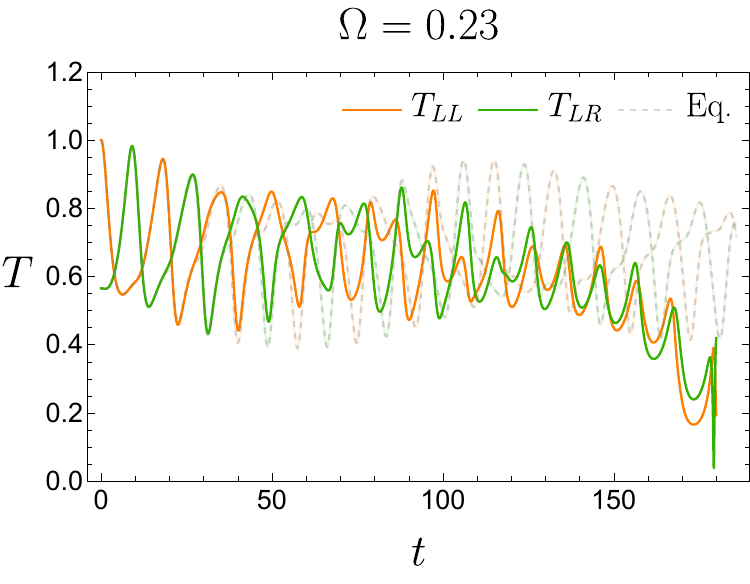}
         \caption{}
         \label{sfig:WHJLdepl}
     \end{subfigure}
     \begin{subfigure}[t]{0.49\textwidth}
         \centering
         \includegraphics[width=\textwidth]{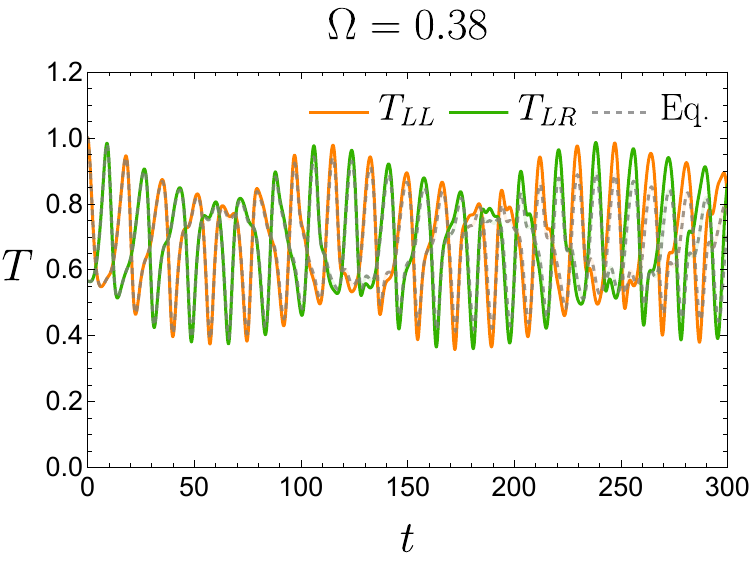}
         \caption{}
         \label{sfig:WHJLJRenh}
     \end{subfigure}
        \caption{Left: Depletion of the transmission coefficients under the asymmetric driving $J_L(t) = J+a\sin \omega t$ while $J_R$ stays constant.  Right: enhancement of the transmission under the driving $J_L(t) = J+a\sin \omega t$ and $J_R(t) = J-a\sin \omega t$. In both cases, the initial equilibrium solution has $\mu=0.05$, $\beta=100$.}
        \label{fig:WHdriveJL}
\end{figure}

We solved the Kadanoff-Baym equations for two different cases. In both of them $J_L(t) = J  + a\sin \omega t$ for the $L$ sector after $t=0$. For the $R$ sector, in one case we will leave  $J_R$ constant. In the other we will impose   $J_R(t) = J  - a\sin \omega t$, thus a time periodic generalization of the imbalanced interactions case studied in \cite{Haenel_2021}. The main features we have encountered are essentially two, both of them showing up in both types of driving. 

\begin{figure}
    \centering
    \includegraphics[width=0.6\textwidth]{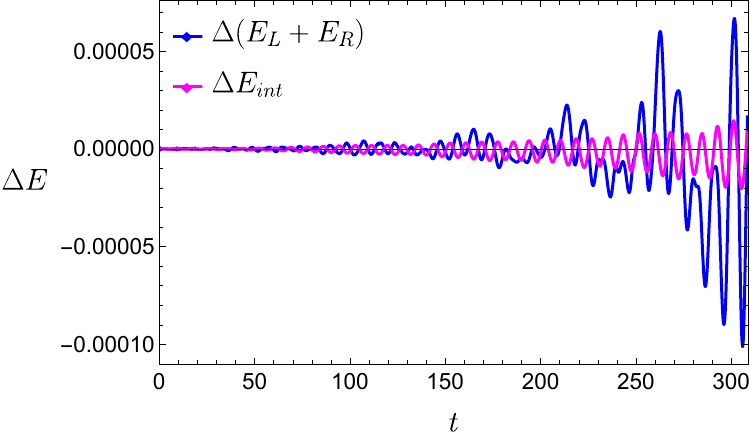}
    \caption{Here we plot the gain in energy, reflecting that we are close to a resonant frequency. The gain occurs in all the components of the total energy, including the interaction part. This is a physical instability that drives the system to another phase which we cannot track up to the end in our simulations because the numerical convergence is lost. Hence we only show the simulation in the reliable interval. }
    \label{fig:WHJLenergy}
\end{figure}

On one hand, at late times the evolution is always unstable. The total energy grows unboundedly (see Fig. \ref{fig:WHJLenergy}) and the numerics break down before we can reach the  endpoint of this instability. This instability gets enhanced by either increasing the amplitude of the perturbation, or by driving at certain resonant frequencies. In Fig. \ref{sfig:WHJLdepl} we see a  behavior similar to the one found for the driving in $\mu$ for a sinusoidal perturbation of just $J_L(t)$ at a very particular value of its frequency,  leaving $J_R$ constant. As mentioned, the difference here is that also numerical convergence is lost and simulations cannot be continued to late times. This was essential in previous sections to establish the nature of the effect as a transition to the black hole phase.

For sufficiently small perturbations $a\ll 1$, we have found an intriguing  effect at early times, much before the simulations stop being reliable: an enhancement in the transmission. To get a grasp on this effect it is necessary to recall that, in the undriven case, there is a monotonic drop in the maximum of the oscillations of the transmission coefficients that eventually would lead to a thermalization of the excitation a  late times, exponential in the inverse temperature $\beta$ \cite{Plugge_2020}. The nature of this dissipative mechanism is obscure from the dual gravity point of view and deserves a study.

In Fig. \ref{sfig:WHJLJRenh} we have compared the free evolution with that of a simultaneous driving of $J_L$ and $J_R$ with $a/J = 0.0015$, hence, extremely tiny. For times where the simulation is still reliable way before the instability sets in, $t<300$, a transient period can be seen where an enhancement of the transmission coefficients $T_{LR}$ occurs.

\section{Diving into (hot) wormholes}\label{sec:divingHW}

The results of Section \ref{subsec:forminghotWH} suggest that the hot wormhole phase can be accessed by injecting energy into the system in a controlled manner. In \cite{Maldacena_2019}, this phase was instead reached by coupling the system to a cold bath and allowing it to cool. In this section, we review the latter approach and argue that the same set of equations used in \cite{Nosaka_2020} to compute the Lyapunov exponents of the two stable phases also apply to the hot wormhole. These results, together with our earlier analysis of periodic drivings, will serve as the basis for a detailed study of this phase in Sections \ref{sec:numresults2} and \ref{sec:HotWHColdBH}.

\subsection{Wormhole formation in real time}

We discuss the real-time dynamics of the two-coupled SYK model after it is suddenly coupled to a cold bath, along the lines given in \cite{Maldacena_2019}.

\begin{figure}
    \centering
    \begin{subfigure}[t]{0.5\textwidth}
        \centering
        \includegraphics[width=\textwidth]{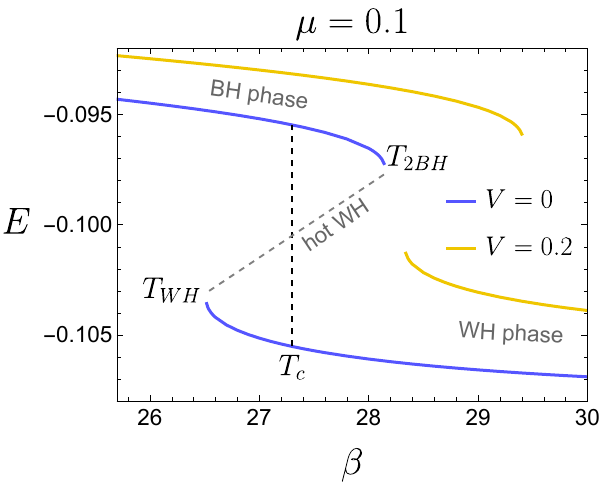}
        \label{sfig:phasediagE}
    \end{subfigure}
    \hspace{0.02\textwidth} 
    \begin{subfigure}[t]{0.36\textwidth}
        \centering
        \vspace{-5cm} 
        \includegraphics[width=\textwidth]{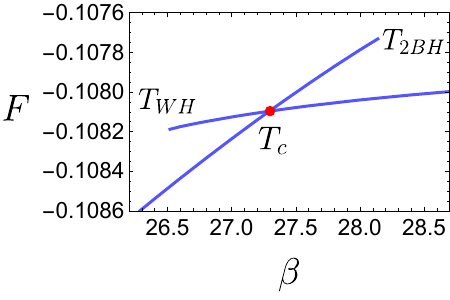}
        \label{sfig:phasediagF}
    \end{subfigure}
    \caption{Left: Phase diagram of the coupled model for $\mu=0.1$ (blue lines). In yellow we show the phase diagram of the model coupled to a cold bath, with coupling strength $V$. Right: The transition inverse temperature $\beta_c\sim 27.3$ is obtained by comparing the free energies in the Euclidean formalism, computed as the on-shell effective action \eqref{eq:SeffeuclideanApp}. The red dot corresponds to the vertical line in the left figure.}
    \label{fig:phasediagrambath}
\end{figure}

The system begins in a high-temperature (BH) state at $T>T_{2BH}$, where the dynamics of the two SYK models are mostly decoupled. At $t=0$, we couple the system to a cold thermal bath with temperature $T_B < T_{WH}$, so that the system evolves towards an equilibrium wormhole configuration at $T=T_B$.
The numerical results suggest that the system approximately follows the microcanonical equilibrium curve during the evolution, thereby passing through the unstable hot wormhole phase of Fig. \ref{fig:phasediagrambath}.

To model the bath, we consider two independent SYK systems, each composed of $M$ fermions, denoted by $\psi^i$ and $\tilde{\psi}^i$. The distribution of random couplings is gaussian as in \eqref{eq:Jmeanvar2SYKtdep} with the replacement $N\rightarrow M$, and its variance denoted by $J_B$. Finally, each  SYK-bath term is coupled to a corresponding SYK factor of the original model. The interaction Hamiltonian between the bath and the system is taken as follows
\begin{equation}
    H_{B,S}=\frac{1}{3!}\sum_{ijk}\sum_{\alpha}V_{ijk\alpha}\chi_L^\alpha\psi^i\psi^j\psi^k+\frac{1}{3!}\sum_{ijk}\sum_{\alpha}\tilde{V}_{ijk\alpha}\chi_R^\alpha\tilde{\psi}^i\tilde{\psi}^j\tilde{\psi}^k~.
    \label{eq:bathHamiltonian}
\end{equation}

Again, the couplings $V_{ijk\alpha}$ and $\tilde{V}_{ijk\alpha}$ are taken to be random, with a Gaussian distribution of zero mean, and variance given by
\begin{equation}
    \overline{V_{ijk\alpha}^2}=\overline{\tilde{V}_{ijk\alpha}^2}=\frac{3! V^2}{M^3}~.
\end{equation}
To ensure the system-bath interaction is marginal, each system fermion interacts with three fermions from the bath \cite{Almheiri_2019,Maldacena_2019}. By choosing $M \gg N$, the backreaction of the system on the bath can be neglected.

Performing the  disorder average over the couplings $V_{ijk\alpha}$ and $\tilde{V}_{ijk\alpha}$ gives an extra contribution to the effective action \eqref{eq:Seffeuclidean} of the form 
\begin{equation}
    \frac{S[G,\Sigma]_{bath}}{N}= -\frac{V^2}{2}\int d\tau_1 d\tau_2 \left(G_{LL}\left(\tau_1, \tau_2\right)+G_{RR}\left(\tau_1, \tau_2\right)\right)G_B^3\left(\tau_1, \tau_2\right)~.
    \label{eq:bathcontribution}
\end{equation}

In the protocol described above, the coupling $V$ between the system and the bath is turned on at $t=0$, pushing the system out of equilibrium. The sudden coupling breaks time translation invariance, and the evolution of the system must be analyzed using the non-equilibrium techniques we have been using so far.

The Kadanoff-Baym equations for the coupled model, \eqref{eq:KBeqs} are still valid in this setup, with the only difference appearing in the self-energies, which now include the contribution of the baths,
\begin{equation}
    \Sigma_{ab}(t_1,t_2)=-J_a(t_1)J_b(t_2)G_{ab}(t_1,t_2)^3~-V(t_1)V(t_2)G_B(t_1,t_2)^3\delta_{ab}.
\label{eq:SigmaKB}
\end{equation}

It was shown in \cite{Maldacena_2019} that, when this protocol is implemented, tracking $\beta_\text{eff}(\mathcal{T})$ during the process reveals that the system follows the microcanonical equilibrium curve, going through the region of the phase diagram with negative specific heat, associated with the hot wormhole.

If the whole process happens slowly enough, it is reasonable to expect that the configurations of the system during the evolution will correspond to the true (albeit unstable) solutions of the equilibrium Schwinger-Dyson equations \eqref{eq:SDequilibrium}-\eqref{eq:Sigmaeq}, with the replacement $\Sigma_{LL}\rightarrow \Sigma_{LL}+V^2 G_B^3$ (and similarly for $\Sigma_{RR}$), which accounts for the contributions of the bath. We have checked numerically that this is the case. This means that by extracting the $\mathcal{T}$-dependent spectral functions $\rho(\mathcal{T},\omega)$ we have access to all the information regarding these unstable solutions. In particular, we are going to use this approach to extract the chaos exponents. In this way we go beyond the work in \cite{Nosaka_2020}, where these exponents were obtained for the two stable phases.

However, it is important to notice, as mentioned before, that in order to study the system around the unstable points, we must bring it out of equilibrium and let it relax. For example, the strategy of coupling to a cold bath modifies the equilibrium solutions (see Fig. \ref{fig:phasediagrambath}), meaning the chaos exponents extracted using this method correspond to a different setup than the one originally considered.

For this reason, we are going to compare the results with the ones obtained using the method of Section \ref{subsec:forminghotWH}, consisting of injecting energy to the system through a periodic driving of $\mu$, which does not require to modify the initial system. The energy is again computed according to \eqref{eq:E_L}-\eqref{eq:Etot}, where now the couplings $J_L$ and $J_R$ are constant in time.

\subsection{Chaos exponents}

We will also be interested in computing the chaos exponents of the model. These were already obtained for the two stable phases in \cite{Nosaka_2020}. As we will see later, with our driving protocols we are able to access the unstable region of the phase diagram, allowing us to extract several features of these previously unobserved region of the phase diagram. Among these features, we will be interested in the Lyapunov exponent.

Following the same procedure explained in Section \ref{subsec:LyapunovSYK}, we can obtain the quantum chaotic behavior of the model from the connected part of the following four-point function,
\begin{align}
\begin{split}
    \frac{1}{N^2}\sum_{i,j}\langle \chi_a^i(\tau_1)\chi_b^i(\tau_2)\chi_c^j(\tau_3)\chi_d^j(\tau_4)\rangle&=\langle G_{ab}(\tau_1,\tau_2) G_{cd}(\tau_3,\tau_4)\rangle\\
    &=\frac{\int \mathcal{D}G\mathcal{D}\Sigma~ G_{ab}(\tau_1,\tau_2)G_{cd}(\tau_3,\tau_4)e^{-S_\text{eff}\left[G,\Sigma\right]}}{\int \mathcal{D}G\mathcal{D}\Sigma~e^{-S_\text{eff}\left[G,\Sigma\right]}}~,
\end{split}
\end{align}
which in the large-$N$ limit admits the expansion
\begin{equation}
    \langle G_{ab}(\tau_1,\tau_2) G_{cd}(\tau_3,\tau_4)\rangle=\overline{G_{ab}(\tau_1,\tau_2)} ~\overline{G_{cd}(\tau_3,\tau_4)}\left[1+\frac{1}{N}\mathcal{F}_{abcd}(\tau_1,\tau_2,\tau_3,\tau_4)+...\right]~.
\end{equation}
Then, the connected piece can be obtained from the quadratic fluctuations of the effective action $S_\text{eff}$ around the saddle point solutions, denoted as  $\overline{G_{ab}(\tau_1,\tau_2)}$ \cite{Maldacena_Stanford_SYK}.

This was done in \cite{Nosaka_2020} for the model under consideration. After choosing the time orderings as in \eqref{eq:timeorderings} and Fig. \ref{fig:contourLyapunov}, the kernel equation can be generalized to (see \cite{Nosaka_2020} for details)
\begin{equation}
   \mathcal{F}_{abcd}(t_1,t_2)=\sum_{e,f}\int dt_3 dt_4 \, K_{abef}^R(t_1,...,t_4)\mathcal{F}_{efcd}(t_3,t_4)
    \label{eq:kerneleq2c}~
\end{equation}
with the retarded kernel given by
\begin{equation}
    K_{abcd}^R(t_1,...,t_4)=-3J^2G_{ac}^R(t_1-t_3)G_{bd}^R(t_2-t_4)G_{cd}^{W}(t_3-t_4)^2~.
    \label{eq:kernel2c}
\end{equation}

In \eqref{eq:kernel2c} $G_{ab}^R(t)$ is a retarded correlator, and $G_{ab}^{W}(t)$ is a Wightman function obtained from the Euclidean correlator as $G_{ab}^{W}(t)=G_{ab}(\frac{\beta}{2}+it)$. Remarkably, Eq. \eqref{eq:kernel2c} remains valid even in the presence of the bath. The reason is that the way in which we chose to couple each SYK to the bath (Eq. \eqref{eq:bathHamiltonian}) leads to a contribution in the large-$N$ effective action that is linear in the two-point functions of the system (see Eq. \eqref{eq:bathcontribution}). Since \eqref{eq:kernel2c} is obtained from the quadratic fluctuations of the action, the new terms do not contribute to this order\footnote{This was noticed in \cite{GarciaGarcia_2024} for a similar setup.}.

As in the single SYK case, to solve this equation it is useful to introduce a {\em growing ansatz} of the form
\begin{equation}
    \mathcal{F}_{abcd}(t_1,t_2)=e^{\lambda_L(t_1+t_2)}f_{abcd}(t_1-t_2)~.
    \label{eq:ansatzF}
\end{equation}

Although \cite{Nosaka_2020} uses the equations in time domain, we find it convenient to rewrite Eq. \eqref{eq:kerneleq2c}, with the ansatz \eqref{eq:ansatzF}, in frequency space\footnote{Here $(G_{ef}^{W})^2(\omega)$ refers to the Fourier transform of $G_{ab}^{W}(t)^2$, and not the square of the Fourier transform of $G_{ab}^{W}(t)$.}:
\begin{equation}
    f_{abcd}(\omega)=-3J^2\sum_{e,f}G_{ae}^R(\omega+i\lambda_L/2)G_{bf}^R(-\omega+i\lambda_L/2)\int_{-\infty}^{\infty}d\omega'(G_{ef}^{W})^2(\omega-\omega')f_{efcd}(\omega')~~.
    \label{eq:kerneleqfreq}
\end{equation}
As noted in \cite{Nosaka_2020}, the indices $cd$ do not mix in the equation, so we can take $cd=LL$ and omit this dependence. From now on, $f_{ab}\equiv f_{abLL}$. It is also convenient to express the different correlators in terms of the spectral functions $\rho_{ab}(\omega)$, since these are the outputs of the real-time equilibrium simulations \cite{Berenguer_2024,LantagneHurtubise_2019,Plugge_2020}. With our conventions, the precise relations are
\begin{equation}
    G_{ab}^R(\omega)=-\sigma_{ab}\int d\omega' \frac{\rho_{ab}(\omega')}{\omega-\omega'}~~~~,~~~~~~\sigma=\mathbbm{1}+i\sigma_x=\begin{pmatrix}
    1 & i \\
    i & 1
    \end{pmatrix}~
\end{equation}
for the retarded correlator in frequency space, and
\begin{equation}
    G_{ab}(\tau)=\tilde{\sigma}_{ab}\int_{-\infty}^{\infty}d\omega\frac{e^{-\omega\tau}\rho_{ab}(\omega)}{1+e^{-\beta\omega}}~~\rightarrow~~G_{ab}^W(t)=G_{ab}(\beta/2+it)=\tilde{\sigma}_{ab}\int_{-\infty}^{\infty}d\omega\frac{e^{-i \omega t}\rho_{ab}(\omega)}{2\cosh{\frac{\beta\omega}{2}}}~,
\end{equation}
with $\tilde{\sigma}=\mathbbm{1}-\sigma_x=\begin{pmatrix}
    1 & -1 \\
    -1 & 1
    \end{pmatrix}$, for the Wightman function. Plugging everything back into \eqref{eq:kerneleqfreq} gives
\begin{equation}
f_{ab}(\omega) =\int d\omega' \sum_{cd}K_{abcd}(\omega,\omega') f_{cd}(\omega')~, \label{eq:fabomega}
\end{equation}
with
\begin{align}
\begin{split}
    K_{abcd}(\omega,\omega')=-\frac{3J^2}{4} \sigma_{ac}\sigma_{bd}\int d\omega_1\frac{\rho_{ac}(\omega_1)}{\omega-\omega_1+i\lambda_L/2}&\int d\omega_2\frac{\rho_{bd}(\omega_2)}{-\omega-\omega_2+i\lambda_L/2}\\ \rule{0mm}{7mm}
    \times&  \int d\omega_3\, \frac{\rho_{cd}(\omega_3)\rho_{cd}(\omega-\omega'-\omega_3)}{\cosh\displaystyle\frac{\beta\omega_3}{2}\cosh\frac{\beta(\omega-\omega'-\omega_3)}{2}}~.
    \label{eq:fabomega2}
\end{split}
\end{align}

Eq. \eqref{eq:fabomega} tells us that $f_{ab}(\omega)$ is an eigenfunction of the kernel matrix $K$ given in \eqref{eq:fabomega2} with eigenvalue 1. The Lyapunov exponent corresponds to the value of $\lambda_L$ for which the largest eigenvalue of $K$ crosses 1.

In the following we compute the chaos exponents of the hot wormhole phase using the two aforementioned approaches.

\section{Numerical results (II): chaos exponents}\label{sec:numresults2}

\subsection{Chaos exponents: wormhole formation}

\begin{figure}[htbp!]
    \centering
    \begin{minipage}[t]{0.3\textwidth}
        \centering
        \begin{subfigure}[t]{\textwidth}
            \centering
            \includegraphics[width=\textwidth]{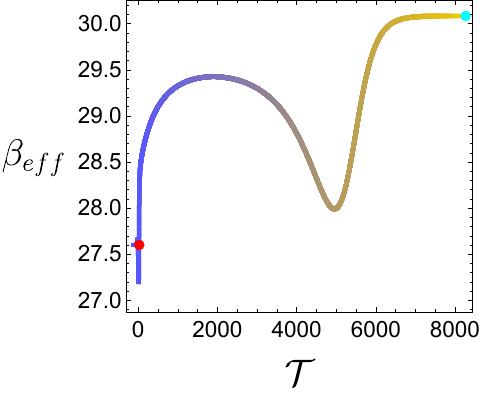}
        \end{subfigure}
        \vspace{1cm}
        \begin{subfigure}[b]{\textwidth}
            \centering
            \includegraphics[width=\textwidth]{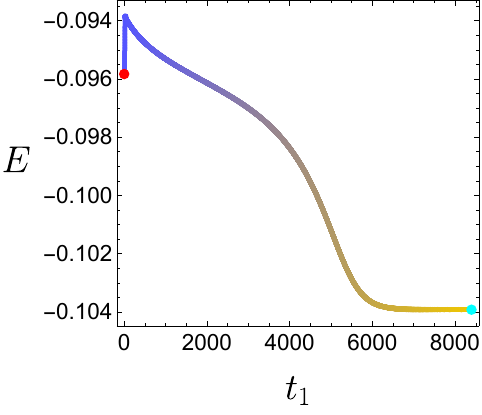}
        \end{subfigure}
    \end{minipage}
    \hspace{0.02\textwidth} 
    \begin{minipage}[c]{0.45\textwidth}
        \centering
        \begin{subfigure}[t]{\textwidth}
            \centering
            \includegraphics[width=\textwidth]{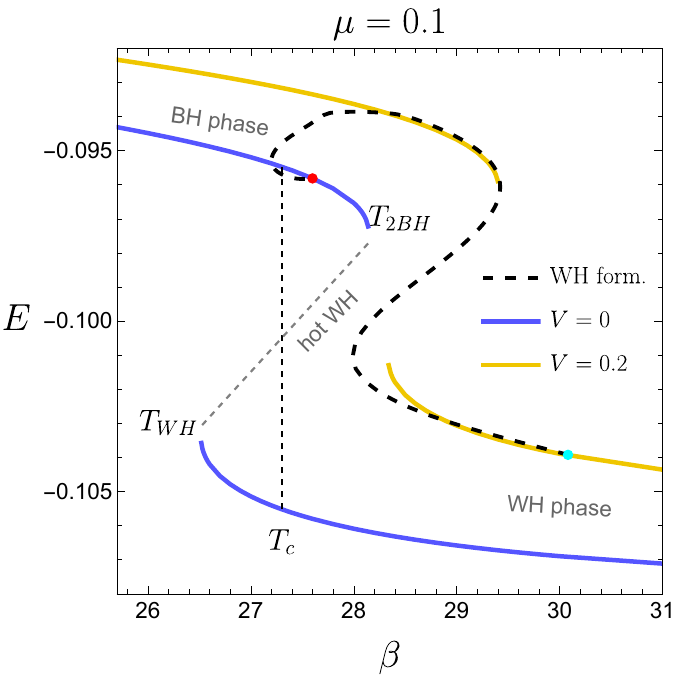}
        \end{subfigure}
    \end{minipage}
    \caption{Wormhole formation in real time \cite{Maldacena_2019}. We show the cooling process that allows us to transition from the black hole to the wormhole phase by turning on a coupling to a cold bath. The system explores the hot wormhole solutions of the new phase diagram. On the left we show the inverse effective temperature and the energy during the process. The two quantities are combined into the black dashed line in the right plot. The color gradient on the left plots stresses the fact that we begin in the original closed system, but we end up in the system containing the bath.}
    \label{fig:WHformation}
\end{figure}

We analyze first the protocol in which the cooling process is done by means of a cold bath. Throughout the simulations, we fix the coupling strengths of the SYK factors to $J=J_B=1$ and the interaction between the two sides to $\mu=0.1$. In this protocol, we initialize the system in the black hole phase, at a temperature slightly higher than $T_{2BH}$. We have chosen $\beta_i=27.6$. At $t=0$, a coupling to a cold bath at $\beta_B=30$, with interaction strength $V=0.2$, is turned on. We turn it on linearly between $t=0$ and $t=20$ to avoid abrupt changes in the parameters, which would render additional, undesired non-equilibrium effects. The numerical results in Fig. \ref{fig:WHformation} show that the equilibration process brings the system from the black hole to the wormhole solutions, going through a region where the temperature increases. The latter is identified with the negative specific heat solutions corresponding to the hot wormhole, in agreement with \cite{Maldacena_2019}. We also overlap the time-dependent energy and temperature with the equilibrium phase diagram.

\begin{figure}
    \centering
    \includegraphics[width=0.45\textwidth]{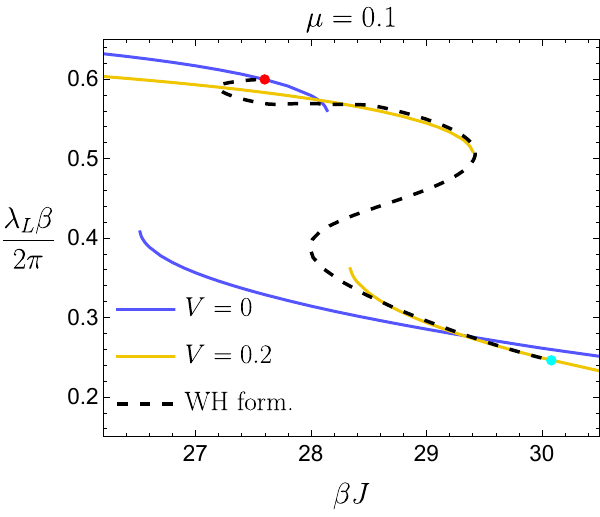}
    \caption{Chaos exponents of the hot wormhole phase obtained at each average time $\mathcal{T}$ during the cooling process.}
    \label{fig:WHformchaos}
\end{figure}

From Fig. \ref{fig:WHformation} we observe that, as expected, introducing the bath alters the hot wormhole solutions making them differ from those initially sought. In Section \ref{subsec:chaosFloquet}, we will revisit this issue using a different non-equilibrium protocol that avoids modifying the original problem.

We can ignore this for now, and proceed with the determination of the chaos exponents, using \eqref{eq:fabomega}. Fig. \ref{fig:WHformchaos} shows, in blue,  the chaos exponents of the BH and WH phases of the original model, which were already obtained in \cite{Nosaka_2020}. In yellow, the ones corresponding to the system coupled to the baths in equilibrium. In dashed we show the chaos exponents obtained from our non-equilibrium simulation, where the spectral function and temperature at each average time $\mathcal{T}$ are obtained from the non-equilibrium Green's functions as in \eqref{eq:Wignertransf}-\eqref{eq:betaeff}. We observe how, similarly as it happened with the energy, the Lyapunov exponent of the hot wormhole solutions interpolates quite smoothly between the two phases. This  matching was already conjectured in Fig. 18 of \cite{Nosaka_2020}. As mentioned above, the caveat of this protocol is that the system for which we are exploring the unstable phase is not the original one, but rather the one in contact with the cold baths.

\begin{figure}
    \centering
    \includegraphics[width=0.49\textwidth]{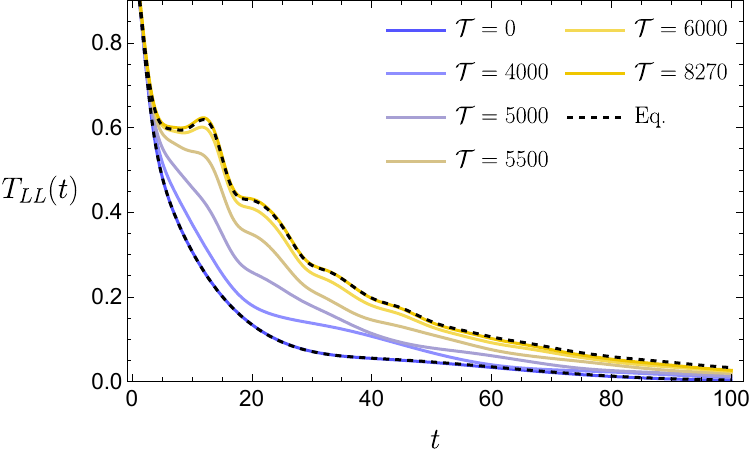}
    \hspace{0.05cm}
    \includegraphics[width=0.49\textwidth]{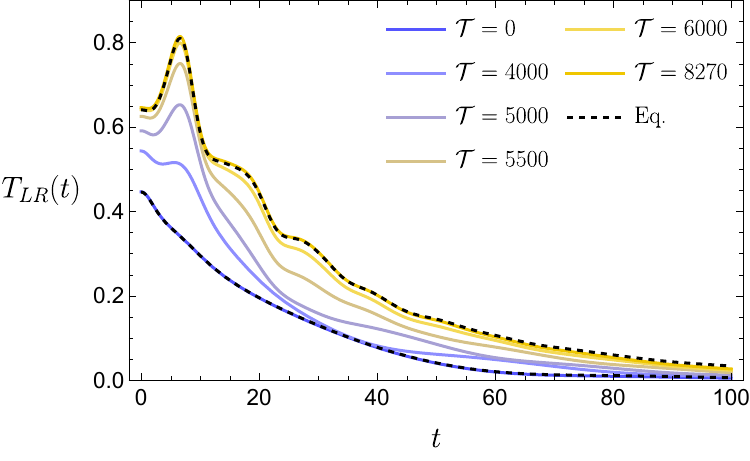}
    \caption{Wormhole formation in real time. By coupling the system, initially in a black hole solution ($\mathcal{T}=0$), to a cold bath, a wormhole solution can be reached once the equilibrium with the bath is achieved ($\mathcal{T}=8270$ in our simulations). The dashed lines show the transmission coefficients of the equilibrium solutions for $\beta=27.6$ and $\beta=30$ (initial and end points in Fig. \ref{fig:WHformation}). The coupling to  the baths reduces the observed revivals in the wormhole phase in comparison with the isolated case (see \cite{LantagneHurtubise_2019,Berenguer_2024} and Fig. \ref{fig:TLLLRFloquet} below).}
    \label{fig:TLLLRWHformation}
\end{figure}

Once $\beta_\text{eff}(\mathcal{T})$ and $\rho_{ab}(\omega,\mathcal{T})$ have been computed, as demanded for the calculation of the chaos exponents, they can further be also used to extract the pseudo-equilibrium Green's functions $G_{ab}^>(\mathcal{T},t)$, $G_{ab}^<(\mathcal{T},t)$, and from them, the transmission amplitudes, $T_{ab}(\mathcal{T},t)=2\abs{G_{ab}^>(\mathcal{T},t)}$ \cite{Berenguer_2024}. Plotting these quantities allows to see visually the process of wormhole formation studied in \cite{Maldacena_2019}, as we show in Fig.~\ref{fig:TLLLRWHformation}.

\subsection{Chaos exponents: Floquet SYK wormholes} \label{subsec:chaosFloquet}

We consider again the initial system, namely, two coupled SYK models, with the hamiltonian given in \eqref{eq:H2SYKtdep}. In this part of the analysis we effectively explore the microcanonical ensemble by injecting controlled amounts of energy. We do so by periodically driving the parameter $\mu$, following the protocol described in Section \ref{subsec:forminghotWH}. The initial equilibrium state is a wormhole with $\mu=0.1$ and $\beta_i=30$, which is very close to the transition temperature, $\beta_c\sim 27.3$. At $t=0$, we turn on the driving as $\mu(t)=\mu\left(1+a\sin\left(\Omega t\right)\right)$, with $a=0.1$ and $\Omega=1.0$, which corresponds to a resonant frequency of the wormhole phase (see Section \eqref{subsec:mudriving}), thereby injecting energy into the system at an exponential rate. After $n$ half cycles, the driving is stopped at $t_{\text{stop}}^n=\frac{\pi}{\Omega}n$, and $\mu$ returns to its static initial value, $\mu=0.1$. As a result, the energy of the system increases for $t<t_{\text{stop}}^n$ and stabilizes at a constant value $E_f$ beyond that point. Asymptotically, the system is then expected to reach a state, which may or may not be a thermal equilibrium state. 

\begin{figure}[H]
    \begin{minipage}[t]{0.35\textwidth}
        \centering
        \begin{subfigure}[t]{\textwidth}
            \centering
            \includegraphics[width=\textwidth]{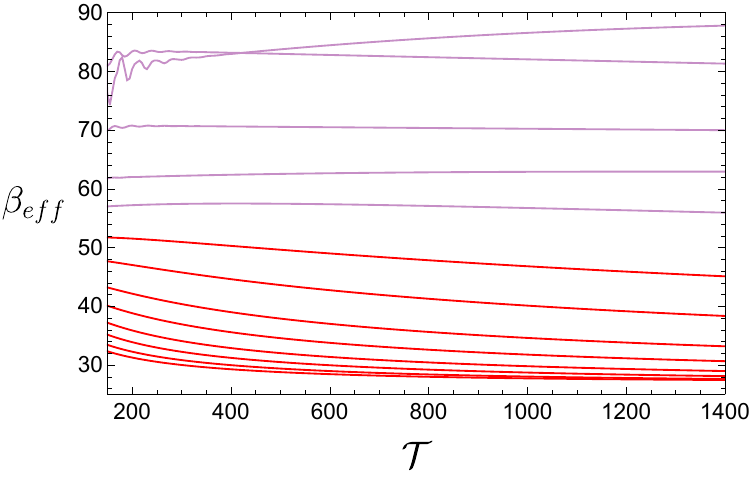}
            \caption{}
            \label{sfig:betaeffnonthermal}
            \vspace{0.1cm}
        \end{subfigure}
        \begin{subfigure}[b]{\textwidth}
            \centering
            \includegraphics[width=\textwidth]{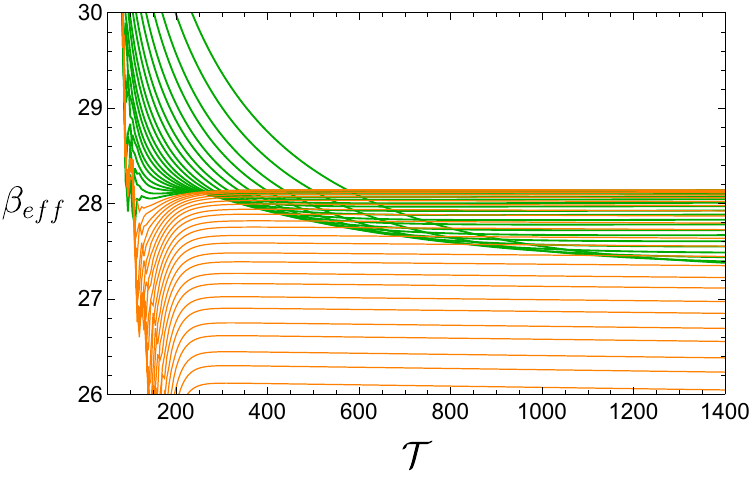}
            \caption{}
            \label{sfig:betaeffthermal}
        \end{subfigure}
    \end{minipage}
    \hspace{0.01\textwidth}
    \centering
    \vspace{0.3cm}
    \begin{minipage}[c]{0.55\textwidth}
        \vspace{0.7cm}
        \centering
        \begin{subfigure}[t]{\textwidth}
            \centering
            \vspace{1cm}
            \includegraphics[width=\textwidth]{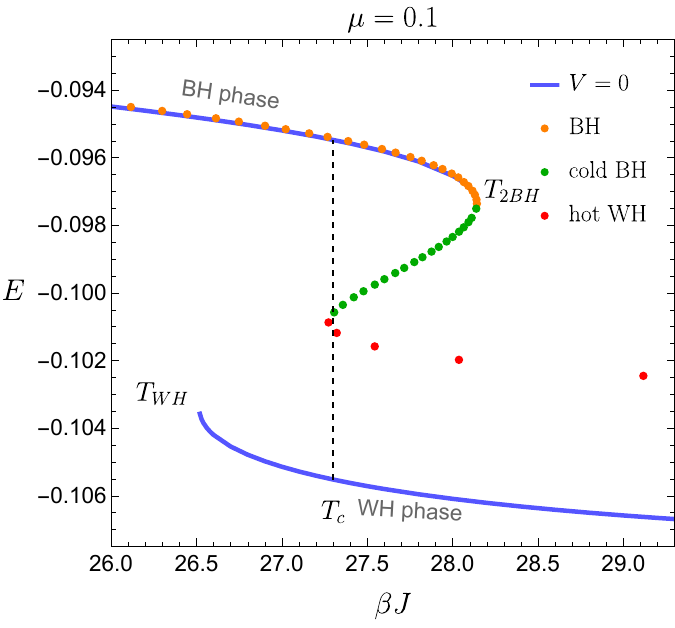}
            \caption{}
            \label{sfig:Floquetphasediagram}
        \end{subfigure}
    \end{minipage}
    \begin{subfigure}[c]{0.49\textwidth}
    \includegraphics[width=0.49\textwidth]{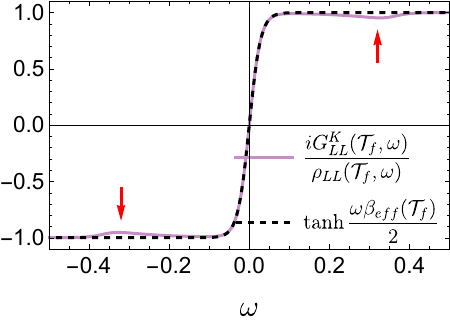}
    \includegraphics[width=0.49\textwidth]{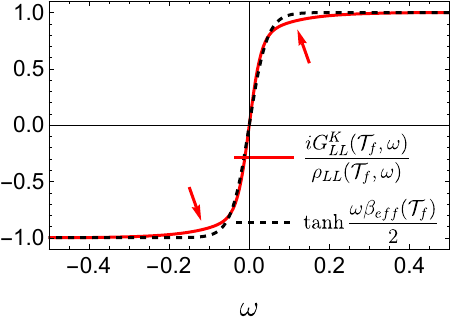}
    \caption{}
    \label{sfig:tanhnonthermal}
    \end{subfigure}
    \begin{subfigure}[c]{0.49\textwidth}
    \includegraphics[width=0.49\textwidth]{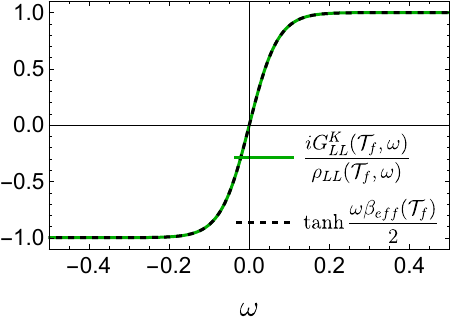}
    \includegraphics[width=0.49\textwidth]{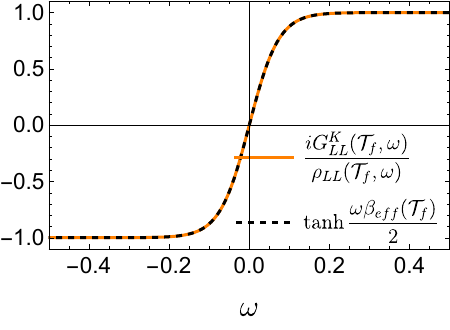}
    \caption{}
    \label{sfig:tanhthermal}
    \end{subfigure}
    \caption{Figs. \ref{sfig:betaeffnonthermal}, \ref{sfig:betaeffthermal}: Effective inverse temperature for the non-thermal ($n\in \left[1,13\right]$ in Fig. \ref{sfig:betaeffnonthermal}) and thermal ($n\in \left[14,69\right]$ in Fig. \ref{sfig:betaeffthermal}) solutions. Only the asymptotic values in \ref{sfig:betaeffthermal} can be interpreted as proper temperatures since only there the FD is satisfied exactly. Fig. \ref{sfig:Floquetphasediagram}: we locate these solutions in the phase diagram and observe that the green thermal ones belong to the hot wormhole phase. The presence of the (red) non-thermal solutions in the diagram is explained in Section \ref{sec:HotWHColdBH}. Fig. \ref{sfig:tanhnonthermal}: FD relation at large $\mathcal{T}$ for $n=3, 6$ (non-thermal). Fig. \ref{sfig:tanhthermal}: FD relation at large $\mathcal{T}$ for $n=25, 50$ (thermal).}
    \label{fig:Floquetbetasdiagram}
\end{figure}

We repeat the simulation for increasing values of $n$, which allows to characterize the thermalization under different energy injections. In each simulation, we analyze the asymptotic behavior of the system. Our numerical capabilities are limited to $t_{max}=1500$, which is enough to characterize the thermalization in most cases. The results are shown in Fig. \ref{fig:Floquetbetasdiagram}, where two main scenarios have to be distinguished: for $n<14$, the system does not fully thermalize within our simulation time. By this we mean that the fluctuation-dissipation relation \eqref{eq:betaeff} does not hold exactly. As a result, the values of $\beta_\text{eff}(\mathcal{T})$ obtained from the fits are not meaningful, and therefore we cannot use \eqref{eq:fabomega} to determine the chaos exponents for these solutions. We show the evolution of the (meaningless) effective inverse  \noindent temperature of the non-thermal solutions in Fig. \ref{sfig:betaeffnonthermal}. Each line corresponds to a different $t_{\text{stop}}^n$ ($n=1,2,3...$). The fact that the results of the fit to \eqref{eq:betaeff} become $\mathcal{T}$-independent for the purple lines could suggest that the effect of the driving has been a cooling of the system to a much larger $\beta$ than the initial one. We want to stress that this is not exactly the case, since the FD relation (although $\mathcal{T}$-independent), is not satisfied, as we show in Fig. \ref{sfig:tanhnonthermal} for two particular choices of $n$. The colors purple and red denote two distinct behaviors within the non-thermal solutions, whose details are relegated to Section \ref{sec:HotWHColdBH}, where we will also explore whether they could thermalize over longer timescales.

\begin{figure}
    \centering
    \includegraphics[width=0.45\textwidth]{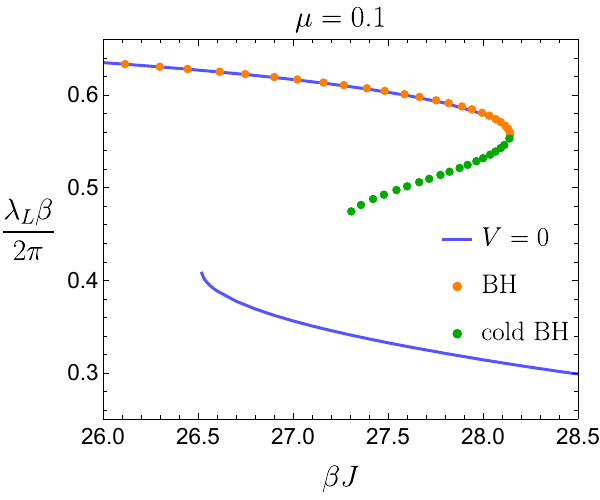}
    \caption{Chaos exponents of the final equilibrium solutions that thermalize within our simulation time. With our protocol we are able to explore approximately half of the unstable region of the phase diagram.}
    \label{fig:Floquetdiagramchaos}
\end{figure}

Conversely, for $n\geq14$, the system relaxes to a thermal state after a finite time. Coincidentally, when this change of behavior occurs, $\beta_\text{eff}\left(\mathcal{T} \to \infty\right)\approx \beta_{c}$. We show the thermal solutions in Fig. \ref{sfig:betaeffthermal}. Again, in Fig. \ref{sfig:tanhthermal} we check the FD relation, which holds for these solutions and allows us to obtain the final inverse temperature of the system. In \ref{sfig:Floquetphasediagram} we locate the thermal solutions in the equilibrium phase diagram, where we find that these solutions belong to the hot wormhole phase. Again, the two colors in the set of thermal solutions signal a change of behavior in the way the effective inverse temperature approaches its final value. Comparing Figs. \ref{sfig:betaeffthermal} and \ref{sfig:Floquetphasediagram} we notice this change in concavity/convexity coincides with the moment the system ceases to stabilize in a hot wormhole solution, and reaches the black hole phase. 

In general, we observe that our driving protocol fails to capture all possible states within the coexistence region, instead it only converges to points along the upper half segment of the hot wormhole phase. In Section \ref{subsec:Schwarzian} we will use the Schwarzian approximation to show that this is what we should expect. In other words, we will show that for small energy injections, the system does not transition to the hot wormhole phase, but instead it explores excited states of the cold wormhole.

In all these cases (thermal solutions of the hot wormhole and black hole solutions), it is possible to compute the chaos exponents. Remarkably, these correspond to the ones we were seeking at first, which do not require the addition of a bath. The results are shown in Fig. \ref{fig:Floquetdiagramchaos}.

\subsection{The last dance: a mixture of both approaches}\label{subsec:bothtogether}

A natural question is whether the results obtained above reflect intrinsic properties of the model or are merely artifacts of the chosen driving protocol. To address this, we consider an alternative protocol that incorporates a combination of the two methods above. In this case, starting from a black hole solution, the system is coupled to a cold bath for a finite time before being decoupled as it transitions through the (modified) hot wormhole phase. The goal is to determine whether energy extraction, rather than injection, allows the system to stabilize within the previously unobserved region of the hot wormhole branch. Once the bath is turned off, the system goes back to its original form, and we expect the final equilibrium states to align with those of the unperturbed model, in the lines of the analysis above.

\begin{figure}[H]
    \begin{minipage}[t]{0.45\textwidth}
        \centering
        \begin{subfigure}[t]{\textwidth}
            \centering
            \includegraphics[width=\textwidth]{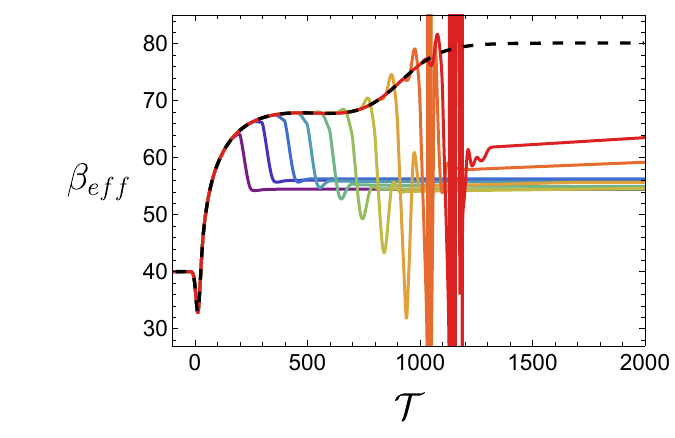}
            \vspace{-0.4cm}
        \end{subfigure}
        \begin{subfigure}[b]{\textwidth}
            \centering
            \includegraphics[width=\textwidth]{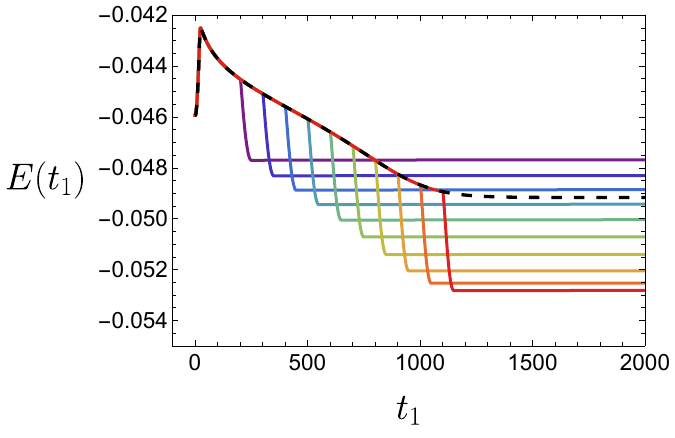}
        \end{subfigure}
    \end{minipage}
    \hspace{0.01\textwidth}
    \centering
    \vspace{0.3cm}
    \begin{minipage}[c]{0.5\textwidth}
        \vspace{0.7cm}
        \centering
        \begin{subfigure}[t]{\textwidth}
            \centering
            \vspace{1cm}
            \includegraphics[width=\textwidth]{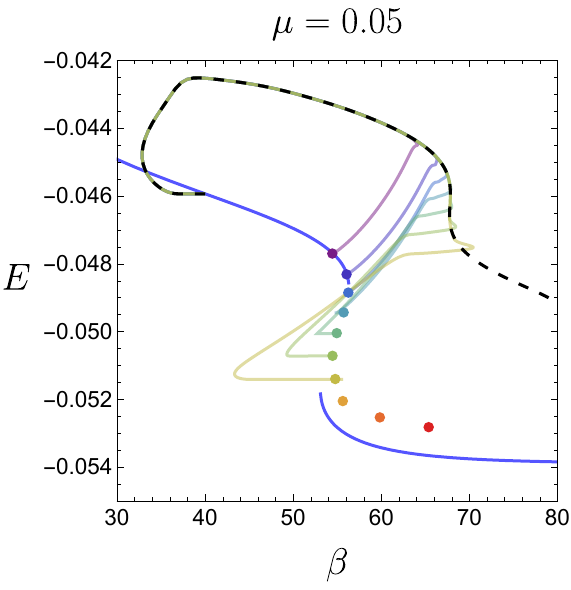}
        \end{subfigure}
    \end{minipage}
    \begin{subfigure}[c]{\textwidth}
    \centering
    \includegraphics[width=0.3\textwidth]{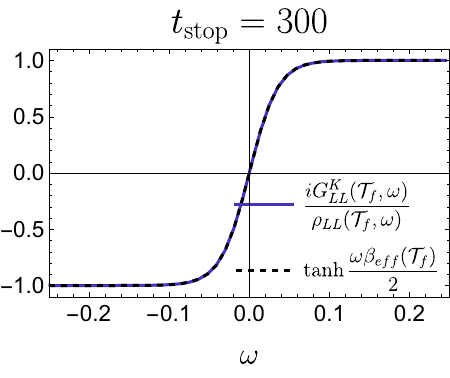}
    \includegraphics[width=0.3\textwidth]{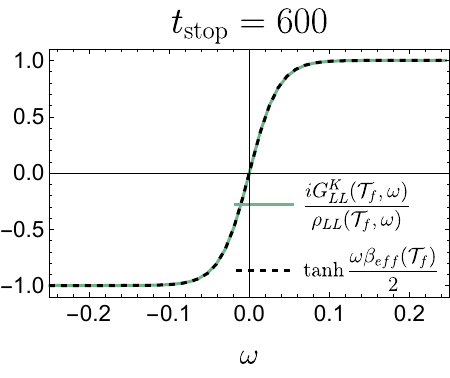}
    \includegraphics[width=0.3\textwidth]{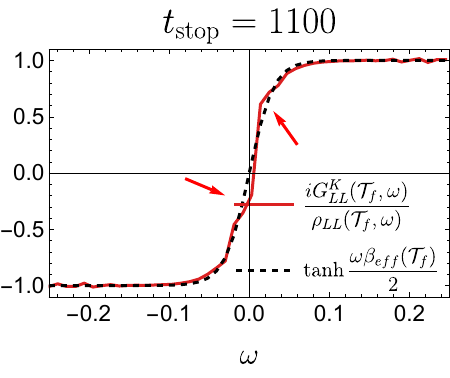}
    \label{sfig:tanhthermalstopbath}
    \end{subfigure}
    \caption{The figure displays the time evolution of the inverse effective temperature and energy, along with their trajectory in the $E$-$\beta$ phase diagram (right plot). For visual clarity, the trajectory lines for the last three points, where $\beta$ exhibits strong oscillations, have been omitted. $t_{\text{stop}}$ increases from purple to red, taking values $t_{\text{stop}}=200,300,...,1100$. The lower panels present the fluctuation-dissipation relation for three representative cases, illustrating the distinct behaviors observed.}
    \label{fig:stopbathresults}
\end{figure}

By turning off the coupling to the bath at different $t_{\text{stop}}$, we can extract controlled amounts of energy. The results\footnote{For this analysis we are using different parameters than in the previous section. In particular, we are using the same parameters as in \cite{Maldacena_2019}, where $J=J_B=0.5$, $\mu=0.05$, and $V=0.2$. The initial inverse temperature is $\beta_i=40$, and the bath is at $\beta_B=80$.}, shown in Fig. \ref{fig:stopbathresults}, reveal the same qualitative behavior as in previous cases—namely, this protocol captures only the equilibrium solutions corresponding to the upper segment of the hot wormhole branch. In these solutions, the fluctuation-dissipation relation holds, and equilibration to the final state occurs almost instantaneously. However, for larger energy extractions, the system fails to thermalize, leading to a time-dependent effective temperature that lacks physical meaning.

This provides further evidence that the structure of the unstable region of the phase diagram is more intricate than initially thought, and obtaining the complete set of hot wormhole solutions is non-trivial. Motivated by this, the next section examines the structure of this phase in greater detail.

\section{Hot wormholes vs cold black holes}\label{sec:HotWHColdBH}

\begin{figure}
    \centering
    \includegraphics[width=0.49\textwidth]{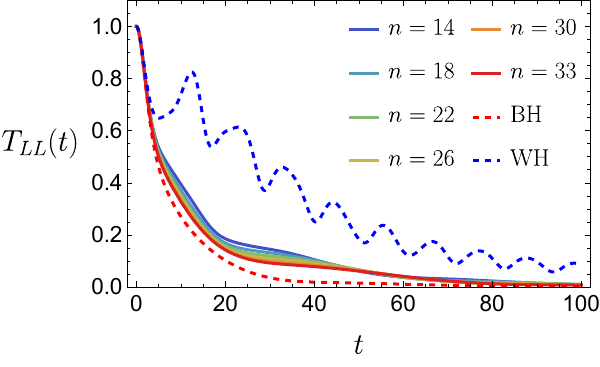}
    \hspace{0.05cm}
    \includegraphics[width=0.49\textwidth]{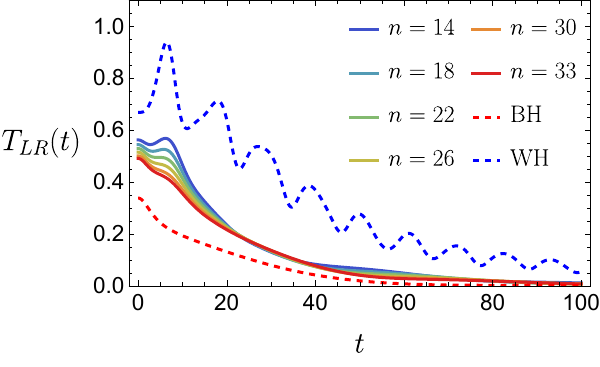}
    \caption{Transmission amplitudes of four different final states within the unstable phase. We show typical profiles of the amplitudes for the black hole phase (dashed red for $\beta=20$), and the wormhole phase (dashed blue for $\beta=30$).}
    \label{fig:TLLLRFloquet}
\end{figure}

In this section we put the focus on the structure of the unstable region of the phase diagram. We begin by showing the behavior of the thermal solutions from the previous section corresponding to the unstable branch (green solutions in Figs. \ref{fig:Floquetbetasdiagram} and \ref{fig:Floquetdiagramchaos}, with $n \in \left[14,33\right]$). In Fig. \ref{fig:TLLLRFloquet} we show the transmission amplitudes $T_{LL}(t)$ and $T_{LR}(t)$ for some of them. We also show, for comparison, the typical profiles of solutions belonging to the black hole phase (dashed red lines), and the wormhole phase (dashed blue lines). The transmission amplitudes of the new solutions show that as we go deeper in the unstable phase (lower $n$), a peak in the transmission amplitude $T_{LR}$ emerges. However, their overall  \noindent behavior is closer to a black hole than to a wormhole.

For the non-thermal solutions, \ie, purple and red curves in Fig. \ref{sfig:betaeffnonthermal}, with $n\in [1,13]$, we can distinguish two classes. For the purple curves,  the results of the fits to the (not precisely satisfied) FD relation are approximately $\mathcal{T}$-independent. In the next subsection we interpret these as excited states of the stable wormhole phase. 

The other solutions (in red in  Fig. \ref{sfig:betaeffnonthermal}) pose greater challenges for interpretation. They show a slowly varying (although meaningless) effective temperature. The fits to the FD relation improve as $\mathcal{T}$ increases, indicating a potential thermalization at late times. In order to extract the endpoint of this thermalization, we fit the red solutions to an exponential of the form
\begin{equation}
    \beta_\text{eff}(\mathcal{T})=Ae^{-\gamma \mathcal{T}}+\beta_{\infty}~,
\end{equation}
with $\beta_{\infty}=\lim_{\mathcal{T}\rightarrow \infty}\beta_\text{eff}(\mathcal{T})$. The reasonable expectation is that these solutions would fill the interpolating segment in the hot wormhole phase between from $\beta_c\sim 27.3$ to the wormhole phase. Surprisingly, the results of the exponential fits give $\beta_\infty>\beta_c$, \ie, the red dots in Fig. \ref{sfig:Floquetphasediagram}. These {\em almost} thermalized solutions are suspicious from the point of view of the phase diagram. Their true existence as thermal equilibrium solutions  would imply   that the system suffers a decrease in temperature after a very small injection of energy. If the fits to the FD relation improve with time but never become exact, these solutions have a nature similar to the purple non-thermal solutions and the temperature is not well-defined. If, on the contrary, the FD relation is satisfied at asymptotically late times, the mechanism that governs this thermalization is unknown. This effect is very similar to the observation that the sharp revival oscillations in the wormhole phase have an envelope that decays as a power-law in time \cite{Plugge_2020}, suggesting the possibility that the two phenomena could be related.

In addition to this, we observe that the turning point that signals the transition between thermal (green) and non-thermal (red) solutions is aligned with the temperature of the first-order Hawking-Page like  transition between the black hole phase and the wormhole phase, at $\beta_c\sim 27.3$. These observations, together with the behavior of the transition amplitudes shown in  Fig. \ref{fig:TLLLRFloquet} suggest that the thermalized purple solutions should be rather called "cold black holes". The  {\em true} hot wormholes, that would eventually fill the rest of the segment joining the two stable phases, remain inaccessible to our driving protocol.

\subsection{Schwarzian analysis}\label{subsec:Schwarzian}

In this section we want to provide qualitative arguments for why we can only access a sub-region of the unstable phase with our protocol. In other words, why the system only thermalizes for energy injections above a certain threshold. Although for the values of the parameters used in the simulations the Schwarzian approximation does not allow for precise quantitative comparisons\footnote{For smaller values of $\mu$, where the Schwarzian approximation is valid, the phase transition occurs at higher $\beta$. In that regime, the characteristic decay times of the Green's functions drastically increases and so do the computational resources needed for non-equilibrium simulations.}, the qualitative behavior is similar \cite{Maldacena_2019} and it allows us to gain some intuition.

\begin{figure}[ht]
    \centering
    \includegraphics[width=0.5\textwidth]{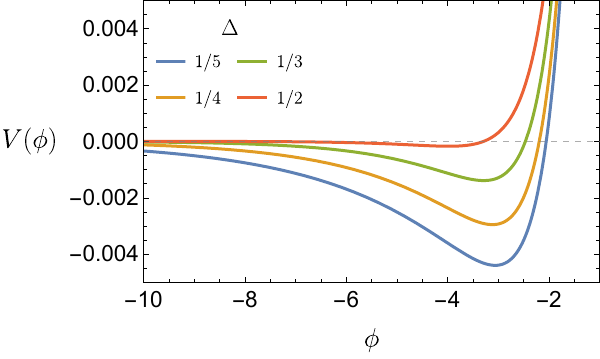}
    \caption{Shape of the potential \eqref{eq:potential} for different values of $\Delta$. The case $\Delta=1/2$ can be solved analytically.}
    \label{fig:potential}
\end{figure}

We consider the coupled Schwarzian action \eqref{eq:Schwaction2SYK}, which, after choosing $t_l(u)=t_r(u)\equiv t(u)$, it reduces to
\begin{equation}
    S=-N~\frac{2\alpha_S}{\mathcal{J}}\int du \left[\left\{\tan\frac{t(u)}{2},u\right\}-\tilde{\mu}\h t'(u)^{2\Delta}\right],
    \label{eq:Schwactiontprime}
\end{equation}
with $\tilde{\mu}\equiv \mu\frac{\mathcal{J}c_\Delta}{2\alpha_S(2\mathcal{J})^{2\Delta}}$. In terms of a new field $\phi(u)$, defined as $t'(u)=e^{\phi(u)}$, the equations of motion of \eqref{eq:Schwactiontprime} are
\begin{equation}
    \phi''(u)+e^{2\phi(u)}-2\Delta\tilde{\mu}e^{2\Delta\phi(u)}=0~,
    \label{eq:EoMphi}
\end{equation}
which correspond to the motion of a particle under the influence of the potential
\begin{equation}
    V(\phi)=\frac{1}{2}e^{2\phi}-\tilde{\mu}e^{2\Delta\phi}~,
    \label{eq:potential}
\end{equation}
which we show in Fig. \ref{fig:potential}.

The simplest solution to \eqref{eq:EoMphi} has the linear form $t(u)=t'u$, where $t'$ is a constant. This solution corresponds to the configuration in which the particle sits at the minimum of the potential. We want to consider all the other solutions to the equation of motion \eqref{eq:EoMphi}. By looking at the potential we see that these will be bounded and oscillating solutions when $E<0$, but they are unbounded when $E>0$, with the energy given by
\begin{equation}
    E=\frac{2\alpha_S}{\mathcal{J}}\left(\frac{1}{2}\phi'^2+\frac{1}{2}e^{2\phi}-\tilde{\mu}e^{2\Delta\phi}\right).
    \label{eq:energySchwarz}
\end{equation}

\begin{figure}[ht]
    \centering
    \includegraphics[width=0.5\textwidth]{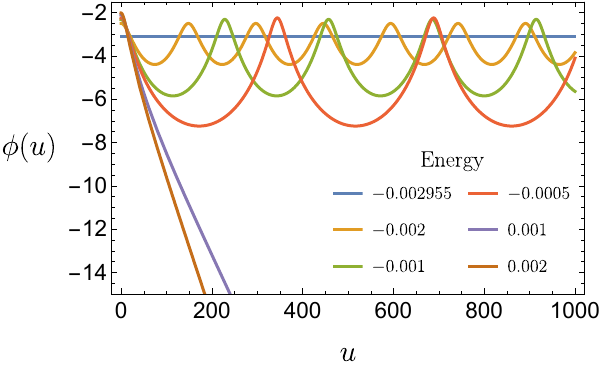}
    \caption{Typical bounded ($E<0$) and unbounded ($E\ge 0$) solutions of the potential \eqref{eq:potential}. The constant blue line corresponds to the solution sitting at the minimum.}
    \label{fig:solutionsE}
\end{figure}

The solutions for arbitrary values of the energy can be found analytically when $\Delta=1/2$ \cite{Dhar_2018}. For the present case ($\Delta=1/4$) they have to be obtained numerically. Typical profiles of the solutions are shown in Fig. \ref{fig:solutionsE}.

Let's focus on the constant solution at the minimum of the potential. We aim to demonstrate that we can bring it to an oscillating solution with a higher energy by turning on the driving of the coupling $\tilde{\mu}$. For this, we change $\tilde{\mu}\rightarrow\tilde{\mu}(u)$ in the action \eqref{eq:Schwaction2SYK}. The same change translates into the equation of motion \eqref{eq:EoMphi}.

\begin{figure}[ht]
    \centering
    \includegraphics[width=0.6\textwidth]{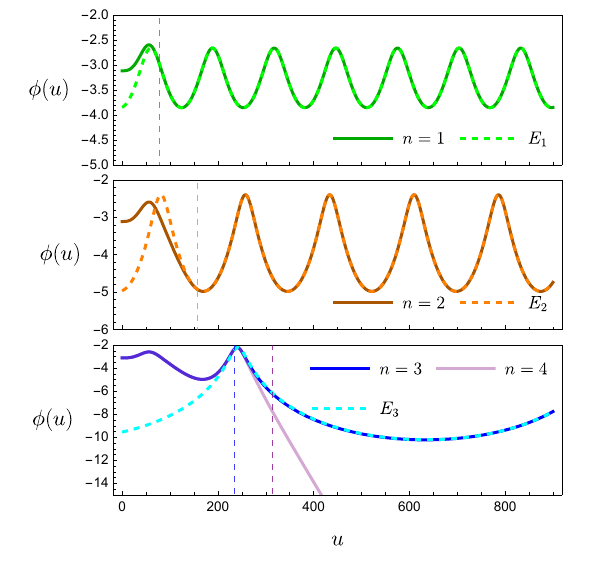}
    \caption{Solid lines: numerical solutions of the equation of motion \eqref{eq:EoMphi} for different $u_{\text{stop}}$ (marked as vertical lines). In particular, $n=1,2,3,4$. For $n=1,2,3$, the injection of energy turns out to be small enough to remain inside the potential. The final energy is evaluated as \eqref{eq:energySchwarz} and the solution of the equation of motion with the corresponding final energy is overlaid in dashed lines. The case $n=4$ corresponds to a driving that has increased the energy to a positive value, leading to an unbounded solution.}
    \label{fig:solutionsdriving}
\end{figure}

We can again solve numerically this equation for the different drivings $\tilde{\mu}(u)=\tilde{\mu}(1+a\sin\Omega u)$, where we turn on the driving at $u=0$, but we stop it at $u_{\text{stop}}=\frac{\pi}{\Omega}n$. We solve it for different $u_{\text{stop}}$, with $n=1,2,3,...$. In all the cases, we choose as initial conditions, $\phi(0)=\phi_0$, $\phi'(0)=0$, which correspond to the solution at the minimum. The results (Fig. \ref{fig:solutionsdriving}) show that once the driving is turned on, $\phi(u)$ begins to oscillate. As soon as we turn off the driving (marked as dashed, vertical lines) two distinct behaviors appear: in all cases, the energy has increased, and depending on the magnitude of this increase, the solution remains bounded, or becomes unbounded. We evaluate the final energy of these solutions and check that they correspond to the oscillating solutions of Fig. \ref{fig:solutionsE} with the corresponding energy (dashed lines).

This proves that, by periodically driving the coupling $\tilde{\mu}$, we excite the solutions with higher energy in this potential. These are the non-thermal solutions that we find for small injections of energy, which we understand as excited solutions of the stable wormhole phase. For higher energies, the particle escapes from the potential, leading to the thermal solutions.

\subsection{Matter contribution}
The previous Schwarzian analysis offers a simplified yet insightful framework for understanding the dynamics of the system, revealing two classes of behavior: oscillating solutions that remain perpetually trapped within the potential, corresponding to the purple lines in Fig. \ref{sfig:betaeffnonthermal}, and unbounded solutions that escape to infinity, transitioning the system into the black hole phase (orange lines in Fig. \ref{sfig:betaeffthermal}). However, the Schwarzian picture fails to capture the full range of observed behaviors, leaving two critical types of solutions unexplained.

First, there are numerical solutions that reach the hot wormhole phase (green lines in Fig. \ref{sfig:betaeffthermal}). In order to capture the existence of this unstable phase in the Schwarzian picture, one needs to include in the free energy the contributions of the matter fields in the throat. At low enough temperature this can be approximated by the contribution of the lightest excitation of the theory, consisting of a fermion with mass $\Delta=1/4$, with $F\approx -\frac{1}{\beta}e^{-t'\beta/4}$  \cite{Maldacena_2019}. Such contribution introduces a correction to the Schwarzian potential \eqref{eq:potential}, resulting in the following change
\begin{equation}
    V(\phi)\rightarrow \tilde{V}(\phi,\beta)=\frac{1}{2}e^{2\phi}-\tilde{\mu}e^{\phi/2}-\frac{\mathcal{J}}{2\alpha_S\beta}e^{-\frac{\beta}{4}e^{\phi}}~.
\label{eq:effectivepotential}
\end{equation}

This potential retains the minimum of the original potential, but for low enough $\beta$ (high enough $T$) it develops a local maximum, which corresponds precisely to the unstable hot wormhole solution. For a sufficiently high energy injection, the particle may overcome the barrier and roll down to $\phi\to -\infty$, to meet the black hole solution. According to this picture, there should be a single (and very fine-tuned) value of the injected energy for which the particle stays at the maximum. However, this feature does not seem to occur for a concrete energy injection. The existence of the green solutions in Figs. \ref{sfig:betaeffthermal} and \ref{sfig:Floquetphasediagram} indicates that there is a intermediate range of energy injections for which the effective particle will end up sitting for a long time on top of the local maximum, in agreement with the metastable character of the cold black hole solutions. This precise fine-tuning may come from the fact that a consistent treatment of \eqref{eq:effectivepotential} should involve a $\beta_\text{eff}(\mu(t))$ dependence, since the driving induces a change of the effective temperature. This dynamical backreaction of the potential in response to the driving may give rise to an attractor mechanism that, for a range of such energy injections, makes the maximum of the potential and the rest point of the particle to meet with high precision. While this fine-tuning might appear suspicious, it is ultimately dictated by the Schwinger-Dyson equations, which admit a single solution for a given energy.

\begin{figure}[ht]
    \centering
    \includegraphics[width=0.49\textwidth]{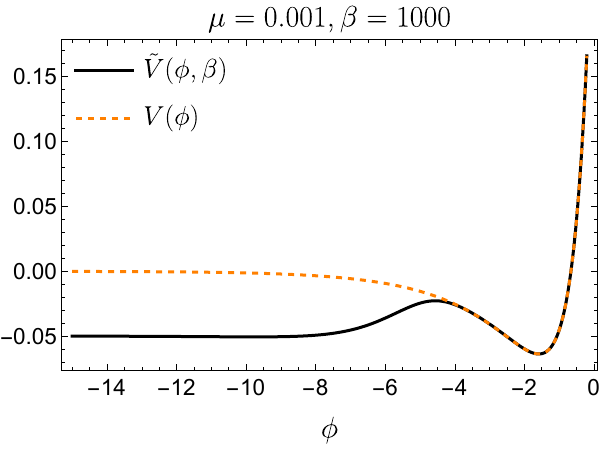}
    \includegraphics[width=0.49\textwidth]{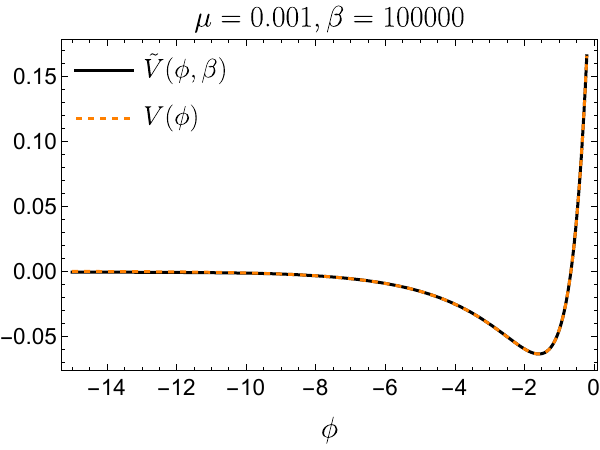}
    \caption{Original (orange, dashed) and modified (black) potential for two different temperatures.}
    \label{fig:newpotential}
\end{figure}

As we have seen, the Schwarzian description is not devoid of  significant challenges.
It also does not fully capture  the dynamics of the red solutions in Fig. \ref{sfig:betaeffnonthermal}. As explained in the previous section, these solutions show a slowly varying (albeit not well defined) effective temperature, with an improvement of the fits to the FD relation as $\mathcal{T}$ increases, suggesting the possibility of an eventual thermalization at late times. The picture of the Schwarzian potential does not provide a mechanism for this as, for trapped solutions, the oscillations should persist indefinitely.

\section{Conclusions}\label{sec:FloquetSYKconclusions}

In this Chapter we studied the out-of-equilibrium dynamics of a system of two coupled SYK models which is  holographically dual to a traversable wormhole in AdS$_2$. In our setup we consider a time-dependent coupling between sides, $\mu(t)$, as well as time-dependent interaction strengths on each of the SYK's, $J_{L,R}(t)$. We also consider the case in which the system is suddenly coupled to a cold bath. Our main tool are the Kadanoff-Baym equations (\ref{eq:KBeqs}) for the non-equilibrium Green's functions, which we integrate numerically.

Under the action of the drivings the system absorbs energy and gets heated exponentially. In the case of the driving $\mu(t)$, our numerical results show a clear enhancement of the heating and a depletion of the transmission coefficients for some discrete set of resonant frequencies of the driving. We identified these resonant frequencies with half of the conformal tower of excitations. We also find a pronounced peak for which we provide numerical and analytical evidence as being a resonance coupling the driving to the boundary graviton of the dual theory. This is a remarkable result, since the boundary graviton states are not observed in the large-$N$ equilibrium analysis of the model.

We have also integrated the non-equilibrium equations for drivings involving the couplings $J_L(t)$ and $J_R(t)$. The late time simulations become unstable both physically and numerically. For small enough amplitudes of the perturbations, the early time evolution indicates a transient enhancement of the signal. 

We then moved on to study the unstable “hot wormhole” phase of the model, which is inaccessible through equilibrium simulations. Using two distinct non-equilibrium protocols, we explored this phase dynamically. The first protocol involves coupling the system to a cold bath, enabling a quasi-static cooling process that transitions the system from a black hole to a wormhole configuration. In the second protocol we use again a  driving of the $LR$ coupling, $\mu$, periodically in time, injecting energy directly into the system.

The chaotic dynamics of the hot wormhole phase is characterized in both cases through the computation of the Lyapunov exponent. While the first method modifies the equilibrium properties due to the coupling to the bath, the second protocol probes the hot wormhole phase of the original model. Remarkably, driving $\mu$ reveals previously unobserved non-equilibrium behaviors, including thermal and non-thermal states. The emergence of these states is confirmed again with a protocol that combines the two previous ones, in which we couple the system to a cold bath, but the coupling is turned off after a certain time. 

A partial understanding of these phenomena is provided by a Schwarzian analysis, though it does not capture the full range of observed behaviors. This suggests a richer structure within the unstable phase, which warrants further investigation. A complete reformulation of the problem in the microcanonical ensemble, where the negative heat phase is stable, would be the most appropriate framework to answer these open questions.

There are many possible new directions worth exploring in the near future. The asymmetric driving $J_L, J_R$ investigated in this work is not particularly appealing from the physical point of view, as it involves manipulating the variance of the gaussian distribution of couplings. A different option would amount to inserting a modulated operator on one side and monitoring the effect it causes on the other. One can, for example, add a periodic mass deformation as the one introduced in \cite{Kourkoulou_2017} on one of the sides of the wormhole.

The two-coupled SYK model is far from being maximally chaotic, and tighter chaos bounds proposed in the literature \cite{Avdoshkin_2019, Gu_2021} merit comparison with our numerical results. The numerical techniques developed for this work are easily adaptable to other setups with equilibrium and non-equilibrium protocols. A related system was proposed in \cite{GarciaGarcia_2020_10,GarciaGarcia_2022}, introducing complex couplings in the SYK factors, which led to a Euclidean-to-Lorentzian wormhole transition.

A related problem is the study of the transport properties of driven SYK islands connected to thermal reservoirs \cite{Kruchkov_2019,Larzul_2022}. Also to understand the connection of our periodic driving in $\mu$ to the repeated measurement mechanism proposed in \cite{Milekhin_2022} would be very interesting.

A detailed gravity description of the time-dependent processes studied here is clearly desirable. At low energy, our setup should be described by an AdS$_2$ JT gravity theory with a suitable dilaton field. In particular, it would be interesting to determine how the driving changes the violation of the average null energy condition with respect to values calculated in \cite{Bak_2018, Gao_2016}.

It is also interesting to analyze the entanglement entropy between the two SYK systems when their coupling is periodically driven. One could use for this study the Schwarzian model as in \cite{Chen_2019}. A natural question is whether or not there are entanglement resonances occurring at particular frequencies, as those observed in \cite{Sauer_2011} for multipartite quantum systems.

Finally, a large-$q$ analysis of the model is also promising, given the existence of analytical results in this limit \cite{Maldacena_Stanford_SYK, Maldacena_2018}. This could provide greater control over the transition region and a deeper understanding of the non-thermal solutions. Additionally, broader studies of quantum dynamics and scrambling, including Krylov complexity, may offer new insights into the interplay between chaos, complexity, and gravity.

%% file: Part2/FloquetSYK/FloquetSYKAppendices.tex
\renewcommand{\chapterquote}{}
\chapterappendix{\thechapter}

\setcounter{equation}{0}
\setcounter{appendixsection}{\value{appendixsection}+1} 
\section{Effective action and Kadanoff-Baym equations}\label{app:KBequations}

We consider the following time-dependent version of the hamiltonian proposed in \cite{Maldacena_2018},
\begin{equation}   
    H(t)=\sum_{a=L,R}\frac{1}{4!}\sum_{ijkl}f_a(t)J_{ijkl}\chi_a^i\chi_a^j\chi_a^k\chi_a^l+i\mu(t)\sum_j \chi_L^j\chi_R^j~,
\end{equation}
 which couples two identical SYK models with a bilinear term and relevant coupling $\mu$, which now we take to be time-dependent: $\mu(t)$. We also allow for a time dependence in the interaction term of each SYK. $\chi_a^i$ are Majorana operators satisfying the usual anticommutation relations $\left\{\chi_a^i,\chi_b^j\right\}=\delta^{ij}\delta_{ab}$ and $J_{ijkl}$ are real constants drawn from a gaussian distribution with mean and variance given by
\begin{equation}
    \overline{J_{ijkl}}=0,~~~~~\overline{J_{ijkl}^2}=\frac{3! J^2}{N^3}~.
\label{eq:Jmeanvarapp}
\end{equation}

For a general function $f(J_{ijkl})$, the disorder average over the couplings is defined as
\begin{equation}
    \overline{f}(J_{ijkl})=\int\mathcal{D}J_{ijkl} f(J_{ijkl})~,
\end{equation}
where in the case of a gaussian distribution, $\mathcal{D}J_{ijkl}$ is
\begin{equation}
    \mathcal{D}J_{ijkl}=\left(\prod_{ijkl}\frac{d J_{ijkl}}{\sqrt{2\pi\alpha}}\right)\exp\left[-\frac{1}{2\alpha}\frac{1}{4!}\sum_{ijkl}J_{ijkl}^2\right]
\label{eq:DJijkl}
\end{equation}
with $\alpha=\frac{3! J^2}{N^3}$. The averaged partition function is written as\footnote{Again, we adopt the annealed average, see footnote before Eq. \eqref{eq:overlineZ}.}
\begin{equation}
    \overline{Z}=\int\mathcal{D}\chi_L\mathcal{D}\chi_R\mathcal{D}J_{ijkl}~e^{i S[\chi_L,\chi_R]}~,
\end{equation}
with the action
\begin{equation}
    S=\int_\mathcal{C}dt\Biggl[\frac{i}{2}\sum_{a=L,R}\sum_j\chi_a^j\partial_t\chi_a^j-\frac{1}{4!}\sum_{a=L,R}\sum_{i,j,k,l}f_a(t) J_{ijkl}~\chi_a^i\chi_a^j\chi_a^k\chi_a^l-\frac{i\mu(t)}{2}\sum_j\left(\chi_L^j\chi_R^j-\chi_R^j\chi_L^j\right)\Biggr]
\end{equation}
integrated along the Keldysh contour $\mathcal{C}$ (Fig. \ref{fig:Keldyshcontour}).
The gaussian integral over the couplings can be directly evaluated to be

\begin{align}
\begin{split}
    \int \mathcal{D}J_{ijkl} \exp\Biggl[-\frac{1}{4!}\sum_{a=L,R}\sum_{i,j,k,l}\int_{\mathcal{C}}&  dt f_a(t)J_{ijkl} ~\chi_a^i\chi_a^j\chi_a^k\chi_a^l\Biggr]\\
    &=\exp\left[-\frac{N}{2}\frac{J^4}{4}\sum_{a,b}\int_{\mathcal{C}} dt_1~dt_2 f_a(t_1)f_b(t_2)\left[O_{ab}(t_1,t_2)\right]^4\right]~,
\end{split}
\end{align}
where we have defined the bilocal field $O_{ab}(t_1,t_2)$ as
\begin{equation}
    O_{ab}(t_1,t_2)=-\frac{i}{N}\sum_j\chi_a^j(t_1)\chi_b^j(t_2)~.
\end{equation}
In terms of this bilocal field the averaged partition function is
\begin{align}
\begin{split}
    \overline{Z}=\int \mathcal{D}\chi_L\mathcal{D}\chi_R\exp\Biggl[  &-\frac{1}{2}\sum_{a,b}\int_\mathcal{C} dt_1~dt_2 \sum_j\chi_a^j(t_1)\Big[(\delta_{ab}\partial_{t_1}-\mu_{ab}(t_1))\delta(t_1-t_2)\Big]\chi_b^j(t_2)\\
    &-\frac{N}{2}\frac{J^4}{4}\sum_{a,b}\int_\mathcal{C} dt_1 dt_2f_a(t_1)f_b(t_2)\left[O_{ab}(t_1,t_2)\right]^4\Biggr]~,
\end{split}
\end{align}
where the matrix $\mu_{ab}$ is
\begin{equation}
    \mu_{ab}(t)=\begin{pmatrix}
0 & \mu(t) \\
-\mu(t) & 0
\end{pmatrix}~.
\end{equation}
We can insert the identity as
\begin{align}
\begin{split}
    1&=\int \mathcal{D}G_{ab}~\delta\left[N\left(G_{ab}(t_1,t_2)-O_{ab}(t_1,t_2)\right)\right]\\
    &\sim \int \mathcal{D}G_{ab}\mathcal{D}\Sigma_{ab}\exp\left[-\frac{N}{2}\int_\mathcal{C} dt_1 dt_2 \Sigma_{ab}(t_1,t_2)\left(G_{ab}(t_1,t_2)-O_{ab}(t_1,t_2)\right)\right]~.
\end{split}
\end{align}
Therefore, the averaged partition function is
\begin{align}
\begin{split}
    \overline{Z}=&\int\mathcal{D}\chi_L\mathcal{D}\chi_R\mathcal{D}G_{ab}\mathcal{D}\Sigma_{ab}\\
    &\times\exp\left[-\frac{1}{2}\sum_{a,b}\int_\mathcal{C} dt_1 dt_2\sum_j\chi_a^j(t_1)\Big[ (\delta_{ab}\partial_{t_1}-\mu_{ab}(t_1))\delta_{\mathcal{C}}(t_1-t_2)+i\Sigma_{ab}(t_1,t_2) \Big]\chi_b ^j(t_2)\right]\\
    &\times \exp\left[-\frac{N}{2}\int_\mathcal{C} dt_1dt_2\sum_{a,b}\Big(\Sigma_{ab}(t_1,t_2)G_{ab}(t_1,t_2)+\frac{J^2}{4}f_a(t_1)f_b(t_2)G_{ab}(t_1,t_2)^4\Big)\right]~.
    \label{eq:AvgZrealt}
\end{split}
\end{align}
Now we want to integrate over the fermions. For that, let's define $\left[G_0^{-1}\right]_{ab}\equiv i\delta_{ab}\partial_{t_1}\delta_\mathcal{C}(t_1-t_2)$. Then, the integral to do is
\begin{align}
\begin{split}
    \int\mathcal{D}\chi_L\mathcal{D}\chi_R&\exp\left[-\frac{1}{2}\sum_{a,b}\sum_j\int_\mathcal{C}dt_1dt_2\chi_a^j(t_1)\Big[-i\left[G_0^{-1}\right]_{ab}-\mu_{ab}(t_1)\delta(t_1-t_2)+i\Sigma_{ab}(t_1,t_2)\Big]\chi_b^j(t_2) \right]\\
    &=\Big[\det\left(-i\left[G_0^{-1}\right]_{ab}(t_1,t_2)-\mu_{ab}(t_1)\delta(t_1-t_2)+i\Sigma_{ab}(t_1,t_2)\right)\Big]^{N/2}\\
    &=\exp\left[\frac{N}{2}\log\det\left(-i\left[G_0^{-1}\right]_{ab}(t_1,t_2)-\mu_{ab}(t_1)\delta(t_1-t_2)+i\Sigma_{ab}(t_1,t_2)\right)\right]~.
\end{split}
\end{align}
Therefore, the averaged partition function can be written in terms of an effective action $S_\text{eff}[G,\Sigma]$,
\begin{equation}
    \overline{Z}=\int\mathcal{D}G_{ab}\mathcal{D}\Sigma_{ab}~e^{iS_\text{eff}[G,\Sigma]}~,
\end{equation}
with
\begin{align}
\begin{split}
    \frac{1}{N}iS_\text{eff}[G,\Sigma]=&\frac{1}{2}\log\det \left(-i\left[G_0^{-1}\right]_{ab}(t_1,t_2)-\mu_{ab}(t_1)\delta(t_1-t_2)+i\Sigma_{ab}(t_1,t_2)\right)\\
    &-\frac{1}{2}\int_\mathcal{C}dt_1dt_2\sum_{a,b}\Big(\Sigma_{ab}(t_1,t_2)G_{ab}(t_1,t_2)+\frac{J_L(t)J_R(t)}{4}G_{ab}(t_1,t_2)^4\Big)~,
\end{split}
\end{align}
where $J_a(t)\equiv J f_a(t)$. The saddle point equations are obtained as $\frac{\delta S_\text{eff}}{\delta G_{ab}}=0$ and $\frac{\delta S_\text{eff}}{\delta \Sigma_{ab}}=0$. The first variation immediately gives
\begin{equation}
    \Sigma_{ab}(t_1,t_2)=-J_a(t_1)J_b(t_2)G_{ab}(t_1,t_2)^3~.
\end{equation}
For the other variation, we use that
\begin{equation}
    \delta(\log\det A)=\Tr\left(A^{-1}\delta A\right)~,
\end{equation}
where the trace is taken in both "$ab$" space and $t_1,t_2$ space. The equation $\frac{\delta S_\text{eff}}{\delta \Sigma_{ab}}=0$ implies that
\begin{equation}
    \left(G_0^{-1}(t_1,t_2)-i\mu(t_1)\delta(t_1-t_2)-\Sigma(t_1,t_2)\right)^{-1}_{ab}=G_{ab}(t_1,t_2)~.
\end{equation}
After inverting,
\begin{equation}
    \left[G_0^{-1}(t_1,t_2)\right]_{ab}-i\mu_{ab}(t_1)\delta(t_1-t_2)-\Sigma_{ab}(t_1,t_2)=\left[G^{-1}(t_1,t_2)\right]_{ab}~.
\label{eq:Dysoninv}
\end{equation}

In order to get the Kadanoff-Baym equations, we convolute from the right and from the left with $G_{ab}$, respectively, to obtain
\begin{align}
\begin{split}
    \int_\mathcal{C}dt\left[G_0^{-1}(t_1,t)\right]_{ac}G_{cb}(t,t_2)-i\mu_{ac}(t_1)G_{cb}(t_1,t_2)-\int_\mathcal{C}dt\Sigma_{ac}(t_1,t)G_{cb}(t,t_2)&=\delta_\mathcal{C}(t_1-t_2)\\
    \int_\mathcal{C}dt~G_{ac}(t_1,t)\left[G_0^{-1}(t,t_2)\right]_{cb}-i G_{ac}(t_1,t_2)\mu_{cb}(t_2)-\int_\mathcal{C}dt~G_{ac}(t_1,t)\Sigma_{cb}(t,t_2)&=\delta_\mathcal{C}(t_1-t_2)~.
\end{split}
\end{align}

In order to rephrase these equations in the Kadanoff-Baym form, we must replace countour integrals by time integrals along both branches $+$ and $-$ of the Schwinger-Keldysh contour. 
Now $t_1$ and $t_2$ can be on any of the two branches. It is convenient to define the greater and lesser Green's functions as
\begin{align}
\begin{split}
    G_{ab}^>(t_1,t_2)&=G_{ab}(t_1^-,t_2^+)\\
    G_{ab}^<(t_1,t_2)&=G_{ab}(t_1^+,t_2^-)~,
\end{split}
\end{align}
where $t_i^-$ lives on the lower contour, and $t_i^+$ lives on the upper contour. More explicitly, since $t_i^-$ is always later than $t_j^+$ in ${\cal C}$
\begin{align}
\begin{split}
    G_{ab}^>(t_1,t_2)&=-\frac{i}{N}\sum_j\langle\chi_a^j(t_1)\chi_b^j(t_2)\rangle\\
    G_{ab}^<(t_1,t_2)&=-\frac{i}{N}\sum_j\langle\chi_b^j(t_2)\chi_a^j(t_1)\rangle~.
\end{split}
\end{align}

From the definitions we have the relation $~G_{ab}^>(t_1,t_2)=-G_{ba}^<(t_2,t_1)$. If we want to get the greater component of $G_{ab}$ we can take $t_1=t_1^-$ and $t_2=t_2^+$. Then, using the expression for $\left[G_0^{-1}(t_1,t)\right]_{ab}$, we get
\begin{align}
\begin{split}
    i\partial_{t_1}G_{ab}^>(t_1,t_2)&=i\mu_{ac}(t_1)G_{cb}^>(t_1,t_2)+\int_\mathcal{C}dt \Sigma_{ac}(t_1,t)G_{cb}(t,t_2)\\
    -i\partial_{t_2}G_{ab}^>(t_1,t_2)&=iG_{ac}^>(t_1,t_2)\mu_{cb}(t_2)+\int_\mathcal{C}dt ~G_{ac}(t_1,t) \Sigma_{cb}(t,t_2)~.
\label{eq:KBcontour}
\end{split}
\end{align}
One can define the retarded, advanced, and Keldysh Green's functions as
\begin{align}
\begin{split}
    G_{ab}^R(t_1,t_2)&=\theta(t_1-t_2)\left[G_{ab}^>(t_1,t_2)-G_{ab}^<(t_1,t_2)\right]\\
    G_{ab}^A(t_1,t_2)&=-\theta(t_2-t_1)\left[G_{ab}^>(t_1,t_2)-G_{ab}^<(t_1,t_2)\right]\\
    G_{ab}^K(t_1,t_2)&=G_{ab}^>(t_1,t_2)+G_{ab}^<(t_1,t_2)~,
    \label{eq:retadvkel}
\end{split}
\end{align}
and similarly for the self-energies,
\begin{align}
\begin{split}
    \Sigma_{ab}^R(t_1,t_2)&=\theta(t_1-t_2)\left[\Sigma_{ab}^>(t_1,t_2)-\Sigma_{ab}^<(t_1,t_2)\right]\\
    \Sigma_{ab}^A(t_1,t_2)&=-\theta(t_2-t_1)\left[\Sigma_{ab}^>(t_1,t_2)-\Sigma_{ab}^<(t_1,t_2)\right]\\
    \Sigma_{ab}^K(t_1,t_2)&=\Sigma_{ab}^>(t_1,t_2)+\Sigma_{ab}^<(t_1,t_2)~.
\end{split}
\end{align}

Then, applying the Langreth rules to rewrite the integral in terms of a single time variable, Eq. \eqref{eq:KBcontour} becomes
\begin{align}
\begin{split}
    i\partial_{t_1}G_{ab}^>(t_1,t_2)&=i\mu_{ac}(t_1)G_{cb}^>(t_1,t_2)+\int_{-\infty}^{\infty}dt~ \Sigma_{ac}^R(t_1,t)G_{cb}^>(t,t_2)+\int_{-\infty}^{\infty}dt~ \Sigma_{ac}^>(t_1,t)G_{cb}^A(t,t_2)\\
    -i\partial_{t_2}G_{ab}^>(t_1,t_2)&=i G_{ac}^>(t_1,t_2)\mu_{cb}(t_2)+\int_{-\infty}^{\infty}dt~ G_{ac}^R(t_1,t)\Sigma_{cb}^>(t,t_2)+\int_{-\infty}^{\infty}dt ~G_{ac}^>(t_1,t)\Sigma_{cb}^A(t,t_2)~.
\label{eq:KBeqsapp}
\end{split}
\end{align}

\setcounter{equation}{0}
\setcounter{appendixsection}{\value{appendixsection}+1} 
\section{Real-time equilibrium equations}\label{app:realtimeeqs}

In this appendix we perform the analytic continuation to real time of the Dyson equations \eqref{eq:SDequilibrium} and \eqref{eq:Sigmaeq}, which we write here again for convenience:
\begin{align}
\begin{split}
    G_{LL}(i\omega_n)&=-\frac{i\omega_n+\Sigma_{LL}}{(i\omega_n+\Sigma_{LL})^2+(i\mu-\Sigma_{LR})^2}\\[10pt]
    G_{LR}(i\omega_n)&=-\frac{i\mu-\Sigma_{LR}}{(i\omega_n+\Sigma_{LL})^2+(i\mu-\Sigma_{LR})^2}~,
\label{eq:SDequilibriumApp}
\end{split}
\end{align}
with
\begin{equation}
    \Sigma_{ab}(\tau)=J^2 G_{ab}(\tau)^3~.    
    \label{eq:SigmaeqApp}
\end{equation}

These equations are derived from the Euclidean action \eqref{eq:Seffeuclidean}:
\begin{align}
\begin{split}
    \frac{S_\text{eff}[G,\Sigma]}{N}=&~-\frac{1}{2}\log\det \left(\delta(\tau-\tau')\left(\delta_{ab}\partial_\tau-\mu \sigma_{ab}^y\right)-\Sigma_{ab}(\tau,\tau')\right)\\
    &+\frac{1}{2}\int d\tau d\tau'\sum_{a,b}\Sigma_{ab}(\tau,\tau')G_{ab}(\tau,\tau')-\frac{J^2}{8}\sum_{a,b}\int d\tau d\tau' \left[G_{ab}(\tau,\tau')\right]^4~.
\label{eq:SeffeuclideanApp}
\end{split}
\end{align}

The analytic continuation $i\omega_n\rightarrow \omega +i\delta$ relates the Matsubara and retarded propagators \cite{Bruus_2004},
\begin{align}
\begin{split}
    G_{LL}^R(\omega)&=-\frac{\omega+i\delta +\Sigma_{LL}}{(\omega+i\delta +\Sigma_{LL})^2+(i\mu-\Sigma_{LR})^2}\\[10pt]
    G_{LR}^R(\omega)&=-\frac{i\mu-\Sigma_{LR}}{(\omega+i\delta +\Sigma_{LL})^2+(i\mu-\Sigma_{LR})^2}~.
\label{eq:SDeqsrealApp}
\end{split}
\end{align}
The equation for $\Sigma$ is written in frequency space before performing the analytic continuation:
\begin{equation}
    \Sigma_{ab}(i\omega_n)=\frac{J^2}{\beta^2}\sum_{n_1,n_2}G_{ab}(i\omega_{n_1})G_{ab}(i\omega_{n_2})G_{ab}(i\omega_n-i\omega_{n_1}-i\omega_{n_2})~.
    \label{eq:SigmaomegaApp}
\end{equation}
If we express the Green's functions through their spectral representation,
\begin{equation}
    G_{LL}(i\omega_k)=-\int d\omega \frac{\rho_{LL}(\omega)}{i\omega_k-\omega},~~~~iG_{LR}(i\omega_k)=\int d\omega \frac{\rho_{LR}(\omega)}{i\omega_k-\omega}~,
\end{equation}
where
\begin{equation}
    \rho_{LL}(\omega)=\frac{1}{\pi}\Im G_{LL}^R(\omega),~~~~\rho_{LR}(\omega)=-\frac{1}{\pi}\Im[iG_{LR}^R(\omega)]
\label{eq:spectdensApp}
\end{equation}
are the spectral functions, Eq. \eqref{eq:SigmaomegaApp} becomes 
\begin{equation}
    \Sigma_{ab}(i\omega_n)=-\tilde{\sigma}_{ab}\frac{J^2}{\beta^2}\int \prod_{i=1}^{3}\left(d\omega_i\rho_{ab}(\omega_i)\right)\sum_{n_1}\frac{1}{i\omega_{n_1}-\omega_1}\sum_{n_2}\frac{1}{i\omega_{n_2}-\omega_2}\frac{1}{i\omega_{n_2}-\Omega}~,
\end{equation}
with $\tilde{\sigma}_{LL}=-1$, $\tilde{\sigma}_{LR}=i$ and $\Omega=i\omega_n-i\omega_{n_1}-\omega_3$. The two Matsubara sums over frequencies can be evaluated\footnote{For instance, the sum over $n_2$ gives $\frac{1}{\beta}\sum\limits_{n_2}\frac{1}{i\omega_{n_2}-\omega_2}\frac{1}{i\omega_{n_2}-\Omega}=\frac{n_F(\omega_2)-n_F(\Omega)}{\omega_2-\Omega}$, where $n_F(\omega)=\frac{1}{e^{\beta \omega}+1}$ is the Fermi distribution function. By using the symmetries of the Fermi distribution function the remaining sum can be simplified and computed as $\sum_{n_1}\frac{1}{i\omega_{n_1}-\omega_1}\frac{1}{i\omega_{n_1}-\tilde{\Omega}}=-\frac{n_F(\omega_1)-n_F(\tilde{\Omega})}{i\omega_n-\omega_1-\omega_2-\omega_3}$, with $\tilde{\Omega}=i\omega_n-\omega_2-\omega_3$.}, and after some manipulations using the Fermi/Bose distribution functions, we obtain
\begin{equation}
    \Sigma_{ab}(i\omega_n)=\tilde{\sigma}_{ab}J^2\int \prod_{i=1}^{3}\left(d\omega_i\rho_{ab}(\omega_i)\right)\frac{\left[n_F(\omega_1)n_F(\omega_2)n_F(\omega_3)+n_F(-\omega_1)n_F(-\omega_2)n_F(-\omega_3)\right]}{i\omega_n-\omega_1-\omega_2-\omega_3}~.
\label{eq:SigmaiomeganApp}
\end{equation}
Now we can analitiycally continue, $i\omega_n\rightarrow \omega+i\delta$, and obtain the retarded self-energy $\Sigma_{ab}^R(\omega)$:
\begin{equation}
    \Sigma_{ab}^R(\omega)=\tilde{\sigma}_{ab}J^2\int \prod_{i=1}^{3}\left(d\omega_i\rho_{ab}(\omega_i)\right)\frac{\left[n_F(\omega_1)n_F(\omega_2)n_F(\omega_3)+n_F(-\omega_1)n_F(-\omega_2)n_F(-\omega_3)\right]}{\omega-\omega_1-\omega_2-\omega_3+i\delta}~.
\label{eq:SigmacontApp}
\end{equation}
Using the identity $\frac{1}{\Omega+i\delta}=-i\int_{0}^{\infty}dt e^{i(\Omega+i\delta)t}$, with $\Omega=\omega-\omega_1-\omega_2-\omega_3$, and defining the "time-dependent occupations"
\begin{equation}
    n_{ab}^s(t)\equiv\int_{-\infty}^{\infty}d\omega \rho_{ab}(\omega)n_F(s\omega)e^{-i\omega t}
\end{equation}
we can write
\begin{equation}
    \Sigma_{ab}^R(\omega)=-i\tilde{\sigma}_{ab}J^2\int_{0}^{\infty}dt e^{i(\omega+i\delta)t}\left[n_{ab}^+(t)^3+n_{ab}^-(t)^3\right]~.
\end{equation}
Finally, noticing that 
\begin{equation}
    n_{ab}^-(t)=\sigma_{ab}\left[n_{ab}^+(t)\right]^*~,
\end{equation}
where $\sigma_{LL}=1$, $\sigma_{LR}=-1$, we can supress the $s=\pm$ label and work with a single function
\begin{equation}
    n_{ab}(t)\equiv\int_{-\infty}^{\infty}d\omega\rho_{ab}(\omega)n_F(\omega)e^{-i\omega t}~.
\label{eq:occupationsApp}
\end{equation}
Then, 
\begin{align}
\begin{split}
    \Sigma_{LL}^R(\omega)&\sim n_{LL}^3(t)+\left[n_{LL}^3(t)\right]^*=2\Re\left[n_{LL}^3(t)\right]\\
    \Sigma_{LR}^R(\omega)&\sim n_{LL}^3(t)-\left[n_{LL}^3(t)\right]^*=2i\Im\left[n_{LR}^3(t)\right]
\end{split}
\end{align}
and we can write explicitly our final expressions for the self-energies:
\begin{align}
\begin{split}
    \Sigma_{LL}^R(\omega)&=2iJ^2\int_{0}^{\infty}dt e^{i(\omega+i\delta)t}\Re\left[n_{LL}^3(t)\right]\\
    \Sigma_{LR}^R(\omega)&=2iJ^2\int_{0}^{\infty}dt e^{i(\omega+i\delta)t}\Im\left[n_{LR}^3(t)\right]~.
\label{eq:SigmaReImApp}
\end{split}
\end{align}

With this, the numerical procedure to find an equilibrium solution consists of the following steps: we start with an initial guess for $G_{LL}(\omega)$, $G_{LR}(\omega)$, which we take to be the free ones (i.e., the ones for $J=0$). From the Schwinger-Dyson equations \eqref{eq:SDeqsrealApp}, these are given by
\begin{align}
\begin{split}
    G_{LL}^{free}&=-\frac{\omega+i\delta}{(\omega+i\delta)^2-\mu^2}\\[10pt]
    G_{LR}^{free}&=-\frac{i\mu}{(\omega+i\delta)^2-\mu^2}~.
\label{eq:freeGFsApp}
\end{split}
\end{align}

From these we compute the spectral densities $\rho_{LL}(\omega)$ and $\rho_{LR}(\omega)$ by virtue of \eqref{eq:spectdensApp}, and we use them to compute the "time-dependent occupations" \eqref{eq:occupationsApp}, which we then plug into \eqref{eq:SigmaReImApp} to get the retarded self-energies $\Sigma_{LL}^R(\omega)$ and $\Sigma_{LR}^R(\omega)$. Finally we plug them in the Schwinger-Dyson equations \eqref{eq:SDeqsrealApp} and obtain new functions $G_{LL}^R(\omega)$, $G_{LR}^R(\omega)$. We repeat this procedure until the solutions converge to the desired accuracy.

Once this is done, we can recover the greater components (in frequency space) as
\begin{equation}
\begin{aligned}
    G_{LL}^>(\omega) & = -in_F(-\omega)\rho_{LL}(\omega)~,\\
    G_{LR}^>(\omega) & = n_F(\omega) \rho_{LR}(\omega)~.
    \label{eq:greaterlesserrho}
\end{aligned}
\end{equation}
Their real time versions are obtained by Fourier transforming. The lesser components are related to the greater ones as $~G_{ab}^>(t_1,t_2)=-G_{ba}^<(t_2,t_1)$.

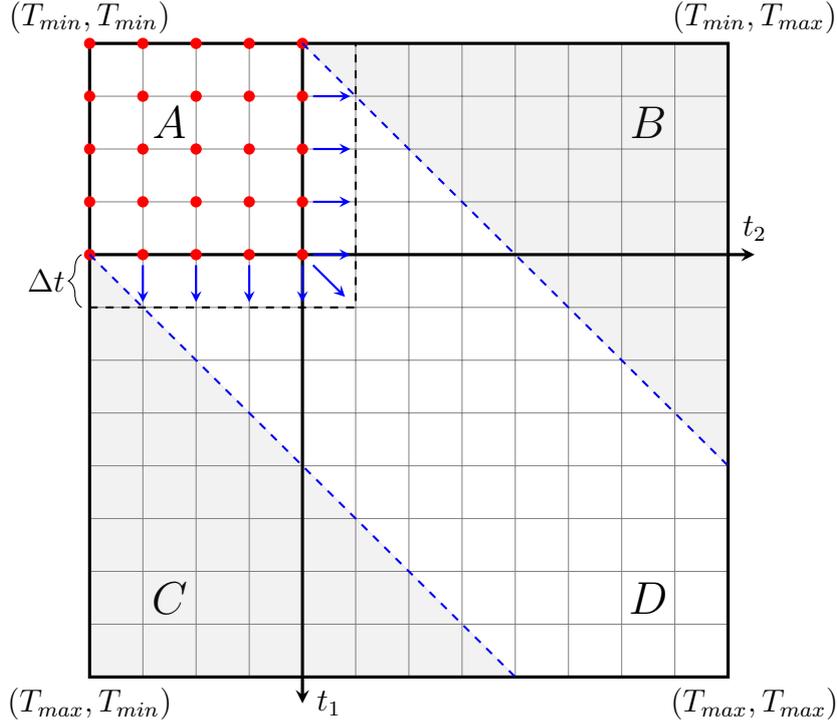
\begin{figure}[H]
    \centering
    \begin{tikzpicture}
    \def\h{0.7cm} 
    \def\Ntot{12}
    \def\Neq{4}
    \draw[step=\h,gray,very thin] (0,0) grid (\Ntot*\h,\Ntot*\h);
    \draw[-stealth,very thick] (\Neq*\h,\Ntot*\h) -- (\Neq*\h,-0.5*\h);
    \draw[-stealth,very thick](0,\Ntot*\h-\Neq*\h) -- (\Ntot*\h+0.5*\h,\Ntot*\h-\Neq*\h);
    \draw[very thick] (0,\Ntot*\h) -- (\Ntot*\h,\Ntot*\h) -- (\Ntot*\h,0) -- (0,0) -- cycle;

    \node at (\Neq*\h+0.5*\h,-0.5*\h) {$t_1$};
    \node at (\Ntot*\h+0.5*\h,\Ntot*\h-\Neq*\h+0.5*\h) {$t_2$};
    \draw[decorate,decoration={brace,amplitude=5pt,mirror,raise=1pt},yshift=0pt] (-0.1*\h,\Ntot*\h-\Neq*\h) -- (-0.1*\h,\Ntot*\h-\Neq*\h-\h) node[midway,xshift=-0.5cm] {$\Delta t$};

    \node at (1.5*\h,\Ntot*\h-1.5*\h) {\Large $A$};
    \node at (\Ntot*\h-1.5*\h,\Ntot*\h-1.5*\h) {\Large $B$};
    \node at (1.5*\h,1.5*\h) {\Large $C$};
    \node at (\Ntot*\h-1.5*\h,1.5*\h) {\Large $D$};

    \node at (0,\Ntot*\h+0.5*\h) {$(T_{min},T_{min})$};
    \node at (0,-0.5*\h) {$(T_{max},T_{min})$};
    \node at (\Ntot*\h+0.5*\h,\Ntot*\h+0.5*\h) {$(T_{min},T_{max})$};
    \node at (\Ntot*\h+0.5*\h,-0.5*\h) {$(T_{max},T_{max})$};

    \draw[dashed,black,thick] (0,\Ntot*\h-\Neq*\h-\h) -- (\Neq*\h+\h,\Ntot*\h-\Neq*\h-\h) -- (\Neq*\h+\h,\Ntot*\h);

    \foreach \i in {0,...,\Neq}{
    \foreach \j in {0,...,\Neq}{
        \node at (\i*\h,\j*\h+\Ntot*\h-\Neq*\h)[red,circle,fill,inner sep=1.5pt]{};
    }
    }

    \foreach \i in {1,...,\Neq}{
    \draw [blue,-stealth,thick](\i*\h,\Ntot*\h-\Neq*\h-0.2*\h) -- (\i*\h,\Ntot*\h-\Neq*\h-0.9*\h);
    \draw [blue,-stealth,thick](\Neq*\h+0.2*\h,\Ntot*\h-\i*\h) -- (\Neq*\h+0.9*\h,\Ntot*\h-\i*\h);
    }
    \draw [blue,-stealth,thick](\Neq*\h+0.2*\h,\Ntot*\h-\Neq*\h-0.2*\h) -- (\Neq*\h+0.8*\h,\Ntot*\h-\Neq*\h-0.8*\h);

    \draw[dashed,blue,thick] (0,\Ntot*\h-\Neq*\h) -- (\Ntot*\h-\Neq*\h,0);
    \draw[dashed,blue,thick] (\Neq*\h,\Ntot*\h) -- (\Ntot*\h,\Neq*\h);

    \draw[fill=gray,opacity=0.1] (0,0) -- (\Ntot*\h-\Neq*\h,0)-- (0,\Ntot*\h-\Neq*\h) -- cycle;
    \draw[fill=gray,opacity=0.1] (\Neq*\h,\Ntot*\h) -- (\Ntot*\h,\Neq*\h)-- (\Ntot*\h,\Ntot*\h) -- cycle;
    
    \end{tikzpicture}
    \caption{Two-times grid used for the integration of the Kadanoff-Baym equations. Time goes from $T_{min}$ to $T_{max}$ with a time step separation $\Delta t$. $T_{min}$ has to be chosen such that $G(T_{min},0),G(0,T_{min})\approx0$, up to some numerical tolerance. The initial values of the Green's functions are the red dots in region A. Then, using the predictor-corrector method (see main text) the equations are solved along the directions of the blue arrows. If the Green's functions decay exponentially (as it is the case in the black hole phase), we can neglect the shaded regions and reduce considerably the computation time.}
    \label{fig:2timesgrid}
\end{figure}
\setcounter{equation}{0}
\setcounter{appendixsection}{\value{appendixsection}+1} 
\section{Numerical methods for the Kadanoff-Baym equations}\label{app:numerics}

In this appendix we provide the details of the numerical method used to solve the Kadanoff-Baym equations \eqref{eq:KBeqs}. For simplicity, we are going to illustrate the method for the case of a single SYK model, \ie, the simpler equations \eqref{eq:KBSYKa}, \eqref{eq:KBSYKb}:
\begin{align}
    \partial_{t_1}G^>(t_1,t_2) & =-i\int_{-\infty}^\infty\h dt_3\left[\Sigma^R(t_1,t_3)G^>(t_3,t_2)+\Sigma^>(t_1,t_3)G^A(t_3,t_2)\right]\equiv F_1^>(t_1,t_2)~,\label{eq:KBSYKaApp}\\
    \partial_{t_2}G^>(t_1,t_2) & = i\int_{-\infty}^\infty\h dt_3\left[G^R(t_1,t_3)\Sigma^>(t_3,t_2)+G^>(t_1,t_3)\Sigma^A(t_3,t_2)\right]\equiv F_2^>(t_1,t_2)~.\label{eq:KBSYKbApp}
\end{align}
There are two equivalent equations for $G^<(t_1,t_2)$, although they will not be needed.

The self-energies are given by
\begin{equation}
    \Sigma^{>,<}(t_1,t_2)=-J(t_1)J(t_2)G^{>,<}(t_1,t_2)^3~,
\end{equation}
and the other functions are defined as
\begin{equation}
\begin{aligned}
    G^R(t_1,t_2)&=\theta(t_1-t_2)\left[G^>(t_1,t_2)-G^<(t_1,t_2)\right]~,\\
    G^A(t_2,t_2)&=-\theta(t_2-t_1)\left[G^>(t_1,t_2)-G^<(t_1,t_2)\right]~,\\
    G^K(t_1,t_2)&=G^>(t_1,t_2)+G^<(t_1,t_2)~,\\
    \Sigma^R(t_1,t_2)&=\theta(t_1-t_2)\left[\Sigma^>(t_1,t_2)-\Sigma^<(t_1,t_2)\right]~,\\
    \Sigma^A(t_1,t_2)&=-\theta(t_2-t_1)\left[\Sigma^>(t_1,t_2)-\Sigma^<(t_1,t_2)\right]~.
\end{aligned}
\end{equation}

The Kadanoff-Baym equations are solved in a two-times grid $(t_1,t_2)$ (Fig. \ref{fig:2timesgrid}) using a predictor-corrector method. Given the equilibrium Green's functions at negative times\footnote{These are obtained independently using the method described in Appendix \ref{app:realtimeeqs}.} (red dots in region A), the method computes the time evolution for positive times. Each time step consists of two parts: a "predictor" part, where the new values are predicted from the previous ones, and a "corrector" part, where they are corrected using the predicted values, and the values at earlier times\footnote{This "double" step method is needed because the unknown functions, namely $G^{>,<}(t_1,t_2)$, appear on both sides of the Kadanoff-Baym equations \eqref{eq:KBSYKaApp}, \eqref{eq:KBSYKbApp}.}. The corrector part can be repeated until the values do not change, up to the desired accuracy. In most of the simulations, one corrector step is enough to get good results.

Notice that, although the integrals in \eqref{eq:KBSYKaApp} run from $-\infty$ to $\infty$, the Heaviside functions of the retarded and advanced functions make the equations to be causal: if the functions are known from $T_{min}$ to $t_1$, $t_2$, the derivative on left hand side of \eqref{eq:KBSYKaApp} at $t_1$, $t_2$ can be obtained. In this way, knowing the equilibrium solutions before the driving is turned on is enough to evolve the system infinitely to the future.

Given the functions from $T_{min}$ up to $(t_m,t_n)$, the predictor and corrector steps, respectively, can be summarized as follows: 
\begin{itemize}
    \item Predictor step: we compute the derivatives at $(t_m,t_n)$ by discretizing the right hand side of \eqref{eq:KBSYKaApp} and computing the integrals numerically. We denote them collectively by $F_{1,2}^{>,<}(t_m,t_n)$. 
    
    Then, the "predicted" value for $G^{>}(t_{n+1},t_m)$ is obtained as
\begin{equation}
    G^{>}(t_{n+1,m})=G^{>}(t_n,t_m)+\Delta t\h F_{1}^{>}(t_n,t_m)~.
\end{equation}

We should repeat the same procedure to obtain $G^{>}(t_n,m+1)$ from \eqref{eq:KBSYKbApp}, and two more times for the two equations for $G^{<}(t_n,t_m)$. However, they can all be obtained from $G^{>}(t_{n+1},t_m)$ as
\begin{equation}
\begin{aligned}
    G^>(t_m,t_{n+1})&=-G^>(t_{n+1},t_m)^*~,\\
    G^<(t_{n+1},t_m)&=-G^>(t_m,t_{n+1})~.
\end{aligned}
\end{equation}
While the first relation is always true, the second one works when we deal with Majorana fermions.

\item Corrector step: In the corrector step, the derivatives $F$ are computed at $(t_{n+1},t_m)$, and they are used to re-calculate, and correct, the predicted values. The values we take are an average of the predicted and corrected ones. In particular,
\begin{equation}
    G^{>}(t_{n+1},t_{m})=G^{>}(t_n,t_m)+\frac{\Delta t}{2}\left[F_1^>(t_{n+1},t_{m})+F_1^>(t_{n},t_{m})\right]~.
\end{equation}

The corrector step can be repeated until a desired convergence is reached.

The diagonal elements are evolved separately. They are obtained as
\begin{equation}
    G^>(t_{n+1},t_{n+1})=G^>(t_n,t_n)+\Delta t\left[F_1^>(t_n,t_n)+F_2^>(t_n,t_n)\right]~.
\end{equation}

In the two coupled model, we perform the same algorithm, but using instead the Kadanoff-Baym equations \eqref{eq:KBeqs} and the different components $LL$, $LR$ of all the functions appearing here.

\end{itemize}

\setcounter{equation}{0}
\setcounter{appendixsection}{\value{appendixsection}+1} 
\section{Time-dependent energy}\label{app:energytdep}

From the time evolution equation of a Majorana field,
\begin{equation}
    \partial_t\chi_a^j=i\left[H,\chi_a^j\right]~,
\end{equation}
we can compute the time derivative of $G_{LL}^>(t_1,t_2)$. As
\begin{equation}
    G_{LL}^>(t_1,t_2)=-\frac{i}{N}\sum_j\langle\chi_L^j(t_1)\chi_L^j(t_2)\rangle~,
\end{equation}
we have
\begin{equation}
    \partial_{t_1}G_{LL}^>(t_1,t_2)=-\frac{i}{N}\sum_j\langle\partial_{t_1}\chi_L^j(t_1)\chi_L^j(t_2)\rangle=\frac{1}{N}\sum_j\langle\left[H,\chi_L^j(t_1)\right]\chi_L^j(t_2)\rangle~.
\end{equation}
Let's take $t_1=t_2=t$ and use
\begin{equation}
    \sum_j\left[H,\chi_L^j(t)\right]\chi_L^j(t)=4H_L(t)+H_{int}(t)~,
\end{equation}
which can be proved by explicitly computing the commutator. Then,
\begin{equation}
    N\partial_tG_{LL}^>(t,t)=4\langle H_L(t)\rangle +\langle H_{int}(t)\rangle~,
\end{equation}
where $\langle H_{int}\rangle= i\mu(t)\sum_j\langle\chi_L^j(t)\chi_R^j(t)\rangle=-\mu(t) NG_{LR}^>(t,t)$. Then, the energy $E_L\equiv \langle H_L(t)\rangle$ is given by
\begin{equation}
    \frac{1}{N}E_L(t)=\frac{1}{4}\partial_t G_{LL}^>(t,t)+\frac{\mu(t)}{4}G_{LR}^>(t,t)~.
\label{eq:expHt}
\end{equation}

We can use the Kadanoff-Baym equations \eqref{eq:KBeqsapp} to remove the time derivative from this expression as
\begin{equation}
    i\partial_{t_1}G_{LL}^>(t_1,t_2)=i\mu(t_1)G_{RL}^>(t_1,t_2)+\int_{-\infty}^{t_1}dt'~ \Sigma_{Lc}^R(t_1,t')G_{cL}^>(t',t_2)+\int_{-\infty}^{t_2}dt'~ \Sigma_{Lc}^>(t_1,t')G_{cL}^A(t',t_2)~,
\end{equation}
where $c$ is summed over $L,R$. We can write each integral as
\begin{align}
    \int_{-\infty}^{t_1}dt'~ \Sigma_{Lc}^R(t_1,t')G_{cL}^>(t',t_2)&=\int_{-\infty}^{t_1}dt'~\left[\Sigma_{Lc}^>(t_1,t')-\Sigma_{Lc}^<(t_1,t')\right]G_{cL}^>(t',t_2)\\
    \int_{-\infty}^{t_2}dt'~ \Sigma_{Lc}^>(t_1,t')G_{cL}^A(t',t_2)&=\int_{-\infty}^{t_2}dt~\Sigma_{Lc}^>(t_1,t')\left[G_{cL}^<(t',t_2)-G_{cL}^>(t',t_2)\right]~.
\end{align}

We can now split the second integral as $\int_{-\infty}^{t_2}\rightarrow \int_{-\infty}^{t_1}+\int_{t_1}^{t_2}$. We can combine $\int_{-\infty}^{t_1}$ with the first integral, and forget about the integral $\int_{t_1}^{t_2}$, since this term will vanish when $t_1=t_2$ (recall \eqref{eq:expHt}).
Combining everything together and expanding the index $c=L,R$, we obtain
\begin{align}
\begin{split}
    i\partial_{t_1}G_{LL}^>(t_1,t_2)=i\mu(t_1)G_{RL}^>(t_1,t_2)+\int_{-\infty}^{t_1} dt'\Bigl[&\Sigma_{LL}^>(t_1,t')G_{LL}^<(t',t_2)-\Sigma_{LL}^<(t_1,t')G_{LL}^>(t',t_2)\\
    +&\Sigma_{LR}^>(t_1,t')G_{RL}^<(t',t_2)-\Sigma_{LR}^<(t_1,t')G_{RL}^>(t',t_2)\Bigr]+...~,
\end{split}
\end{align}
where the dots refer to the terms that will vanish when $t_1=t_2$. We can finally write the self-energies in terms of the Green's functions according to \eqref{eq:Sigmarealt}, use the symmetry $G_{RL}^<(t_1,t_2)=-G_{LR}^>(t_2,t_1)$ and set $t_1=t_2$ to arrive to a final expression for the energy
\begin{equation}
\begin{aligned}
    \frac{1}{N}E_L(t)=-\frac{iJ_L(t)}{4}\int_{-\infty}^{t}dt'\Big[&J_L(t')\left(G_{LL}^>(t,t')^4-G_{LL}^<(t,t')^4\right)\Big.\\
    \Big.+&J_R(t')\left(G_{LR}^>(t,t')^4-G_{LR}^<(t,t')^4\right)\Big]~.
\end{aligned}
\end{equation}

\restoredefaultnumbering
\restoredefaultsectioning

%% file: Part3/Conclusions.tex
\renewcommand{\chapterquote}{\textit{“Well, here at last, dear friends, on the shores of the Sea\\comes the end of our fellowship in Middle-earth.\\Go in peace! I will not say: do not weep; for not all tears are an evil.”} \\[0.8em] \normalfont— J.R.R. Tolkien, \textit{The Return of the King}.}
\chapter{Conclusions}
\markright{\thechapter. Conclusions}

We now arrive at the end of this thesis, which we conclude with a synthesis of its main contributions and a discussion of possible future directions. Although each Chapter includes its own specific conclusions, here we offer a broader synthesis of the results and outlook.

This thesis has explored the non-equilibrium dynamics of various holographic models, with a central focus on systems subject to time-dependent perturbations, focusing mainly on those with periodic drivings, commonly known as Floquet systems.

Two different classes of models have been at the heart of this investigation. The first involves top-down string-theoretic constructions, specifically D3/D5 and D3/D7 brane intersections, coupled to various external gauge fields. The second is based on the SYK model, with particular emphasis on the eternal traversable wormhole proposed by Maldacena and Qi.

To make the thesis self-contained, we began  with a review of the essential background. Starting from the bosonic string, we followed the canonical path to the AdS/CFT correspondence via the dual description of D-branes. After introducing flavor degrees of freedom in this framework, we analyzed the D3/D5 and D3/D7 systems in detail, focusing on their response to different external fields. We then turned to the SYK model and its wormhole construction, presenting both Euclidean and Lorentzian formulations in preparation for studying its out-of-equilibrium behavior.

The second part of the thesis presented the original research, based on the publications \cite{Berenguer_2022,Berenguer_2024,Berenguer_2025_SYK,Berenguer_2025}.

In Chapter \ref{chap:FloquetII}, we studied the D3/D5 system at finite temperature under a time-dependent, rotating electric field. Using a rotating ansatz that eliminates all the time dependence from the Lagrangian, we obtained the non-equilibrium phase diagram of the system, which contains a conductive and an insulating phase. We showed that the presence of vector meson Floquet condensates persists at finite temperature, and we identify, and study in detail, what we call \textit{Floquet suppression points}, where the polarizability is dynamically suppressed at certain discrete frequencies of the external driving. 

We carried out a detailed analysis of the response currents generated by the periodic driving, refining and extending existing results in the literature. Notably, in addition to the numerical results, we obtained several analytic results in certain limits. Among them, we found analytic solutions for the massless embeddings, as well as the small-mass case. The study of both nonlinear and optical conductivities was extended to the finite-temperature regime, again obtaining analytic results in the massless limit.

Despite these successes, the inherently non-equilibrium nature of the system complicates a thermodynamic interpretation of the conductor-insulator transition. This issue, discussed in Section~\ref{sec:problemsnoneq}, highlights the need for more robust real-time frameworks in holography, such as the Skenderis-van Rees or the Glorioso-Crossley-Liu proposal.

In Chapter \ref{chap:HelicalB}, we turned to the D3/D7 intersection in the presence of a helical magnetic field. Although this system is treated in equilibrium, the previous use of a rotating ansatz provided conceptual inspiration. We showed that the helical component can in fact counteract the symmetry-breaking effect of a uniform magnetic field, leading to chiral symmetry restoration.

Chapter \ref{chap:FloquetSYK} focused on the real-time dynamics of the Maldacena–Qi traversable wormhole. We analyzed its response to time-dependent deformations, including periodic drivings of the parameters, and a sudden coupling to a cold bath. In the driven case, we found a set of resonant frequencies at which the system absorbs energy exponentially, eventually closing the wormhole. These resonances coincide with half the conformal spectrum of the undriven model, and we provided additional evidence for a resonance associated with the natural frequency of the boundary graviton, which had not been observed in the large-$N$ equilibrium simulations of the model. Using fine-tuned drivings, we dynamically accessed the unstable "hot wormhole" phase and conducted an in-depth analysis of its fine structure, including the computation of its Lyapunov exponent.

Several directions for future exploration emerge naturally from this thesis. While many have already been noted in their respective chapters, we highlight here the most promising directions. 

The absence of a steady-state thermodynamic framework for non-equilibrium holographic systems remains a challenge. Addressing this could involve implementing and comparing leading proposals such as the Skenderis–van Rees prescription \cite{Skenderis_2008_short,Skenderis_2008_long,vanRees_2009} and the Glorioso–Crossley–Liu approach \cite{Glorioso_2018}. The open-system proposal introduced in \cite{Jana_2020} also offers a compelling direction to explore. Each of these offers complementary perspectives, and their systematic comparison could yield valuable insights.

The analytical and numerical techniques developed to study these constructions can be easily generalized to many other brane setups with different gauge fields in order to obtain different phenomenological models of driven systems, chiral symmetry breaking, or Weyl semimetals in the context of holography.

The traversable wormhole in the SYK model, despite its simplicity compared to the previous models,  opens the door to many unexplored questions. For example, the decay of transmission amplitudes at late times, observed both in the wormhole and hot wormhole phases remains poorly understood from the gravitational side. Our finding that this decay also occurs in the hot wormhole phase hints at a deeper, possibly universal mechanism that deserves further exploration.

On a more technical level, the numerical methods developed to study the coupled model out of equilibrium are highly versatile and can be readily adapted to a broad class of quantum mechanical systems, many of which are expected to demand significantly fewer computational resources than the wormhole phase analyzed here. Only a handful of studies in the literature have undertaken this type of real-time simulation, and those that exist typically fall short of accessing the most interesting regions of the phase diagram due to the exponential growth in computational cost (see Fig. \ref{fig:phasediagramdrivings} and references therein). The combined use of several advanced tools, including the predictor-corrector method detailed in Appendix \ref{app:numerics}, the NESSi library, and the high-performance computing resources provided by CESGA (Centro de Supercomputación de Galicia), has been crucial in achieving these results, which would be difficult to obtain without this infrastructure.

While some of these questions point to long-term directions, the work presented in this thesis shows how holography can serve as a powerful framework for exploring fundamental aspects of gravity and quantum many-body physics, both in- and out-of-equilibrium. The methods developed here will remain valuable tools for advancing our understanding of the interplay between strongly coupled quantum systems and gravitational physics.

\clearpage
\cleardoublepage

\thispagestyle{empty}
\null
\vfill

\begin{flushright}
\begin{minipage}{0.5\textwidth}
\color[gray]{0.6}
\textit{And afterward, everyone dress}\\
\textit{as they best please, and on we go!,}\\
\textit{for all is yet to be done and all is possible.}
\end{minipage}
\begin{minipage}{0.47\textwidth}
\raggedleft
\textit{I en acabat, que cadascú es vesteixi}\\
\textit{com bonament li plagui, i via fora!,}\\
\textit{que tot està per fer i tot és possible.}
\end{minipage}

\vspace{2em}

\normalfont— Miquel Martí i Pol, \textit{Ara mateix}.
\end{flushright}

\vfill
\null

%% file: Part3/Publications.tex
\renewcommand{\chapterquote}{}
\chapter*{List of publications}
\addcontentsline{toc}{chapter}{List of publications}
\pagestyle{noheader}

The Research part of the thesis is based on four publications. Here we provide their details.
\begin{center}
\begin{tabularx}{\textwidth}{ |>{\raggedright\arraybackslash}p{0.25\textwidth}|X| }
\hline
\multicolumn{2}{|c|}{\cellcolor{green!10}
  \begin{tabular}{@{}c@{}}
    \textbf{Holographic Floquet states in low dimensions (II)} \\
    Journal of High Energy Physics \textbf{2022}, 20 (2022)
  \end{tabular}
} \\
\hline
\parbox[c]{\linewidth}{\vspace{1em}\centering\textbf{Authors}\vspace{1em}} &
\parbox[c]{\linewidth}{
\vspace{1em}
Martí Berenguer\(^{1,2}\), Ana Garbayo\(^{1,2}\), Javier Mas\(^{1,2}\) and Alfonso V. Ramallo\(^{1,2}\).\\[1em]
\footnotesize
\(^{1}\) Departamento de Física de Partículas and \\
\(^{2}\) Instituto Galego de Física de Altas Enerxías \\
Universidade de Santiago de Compostela. \\
 E-15782. Santiago de Compostela, Spain.
\vspace{1em}
} \\
\hline
\parbox[c]{\linewidth}{\vspace{1em}\centering\textbf{Included in}\vspace{1em}} &
\parbox[c]{\linewidth}{\vspace{1em}Chapter \ref{chap:FloquetII}. Holographic Floquet states in low dimensions.\vspace{1em}} \\
\hline
\parbox[c]{\linewidth}{\vspace{1em}\centering\textbf{PhD Student\\Contribution}\vspace{1em}} &
\parbox[c]{\linewidth}{\vspace{1em}Collaboration in numerical and analytic computations, discussions, and writing of the article.\vspace{1em}} \\
\hline
\parbox[c]{\linewidth}{\vspace{1em}\centering\textbf{Further\\Information}\vspace{1em}} &
\makecell[l]{{\vspace{-0.15cm}\tiny\transparent{0} This text is invisible and adds some blank space.} \\
\textbf{Journal:} Journal of High Energy Physics \\
\textbf{Publisher:} Springer \\
\textbf{ISSN:} 1029-8479 \\
\textbf{Year of publication:} 2022 \\
\textbf{DOI:} \href{https://doi.org/10.1007/JHEP12(2022)020}{https://doi.org/10.1007/JHEP12(2022)020}  \\
\textbf{Impact factor\(^{*}\):} 5.4 \\
\textbf{Rank in Physics, Particles \& Fields\(^{*}\):} 5/29 (Q1) \\
\\
Article distributed under the terms of the Creative Commons\\ Attribution License (CC-BY 4.0). \\
\\
\(^{*}\)Data corresponding to 2022, from JCR. \vspace{0.5em}
}
\\
\hline
\end{tabularx}
\end{center}

\clearpage
\cleardoublepage
\vspace*{\fill}
\begin{center}
\begin{tabularx}{\textwidth}{ |>{\raggedright\arraybackslash}p{0.25\textwidth}|X| }
\hline
\multicolumn{2}{|c|}{\cellcolor{green!10}
  \begin{tabular}{@{}c@{}}
    \textbf{Floquet SYK wormholes} \\
    Journal of High Energy Physics \textbf{2024}, 106 (2024)
  \end{tabular}
} \\
\hline
\parbox[c]{\linewidth}{\vspace{1em}\centering\textbf{Authors}\vspace{1em}} &
\parbox[c]{\linewidth}{
\vspace{1em}
Martí Berenguer\(^{1,2}\), Anshuman Dey\(^{1,2}\), Javier Mas\(^{1,2}\), Juan Santos-Suárez\(^{1,2}\) and Alfonso V. Ramallo\(^{1,2}\).\\[1em]
\footnotesize
\(^{1}\) Departamento de Física de Partículas and \\
\(^{2}\) Instituto Galego de Física de Altas Enerxías \\
Universidade de Santiago de Compostela. \\
E-15782. Santiago de Compostela, Spain.
\vspace{1em}
} \\
\hline
\parbox[c]{\linewidth}{\vspace{1em}\centering\textbf{Included in}\vspace{1em}} &
\parbox[c]{\linewidth}{\vspace{1em}Chapter \ref{chap:FloquetSYK}. Floquet SYK wormholes.\vspace{1em}} \\
\hline
\parbox[c]{\linewidth}{\vspace{1em}\centering\textbf{PhD Student\\Contribution}\vspace{1em}} &
\parbox[c]{\linewidth}{\vspace{1em}Collaboration in numerical and analytic computations, discussions, and writing of the article.\vspace{1em}} \\
\hline
\parbox[c]{\linewidth}{\vspace{1em}\centering\textbf{Further\\Information}\vspace{1em}} &
\makecell[l]{{\vspace{-0.15cm}\tiny\transparent{0} This text is invisible and adds some blank space.} \\
\textbf{Journal:} Journal of High Energy Physics \\
\textbf{Publisher:} Springer \\
\textbf{ISSN:} 1029-8479 \\
\textbf{Year of publication:} 2024 \\
\textbf{DOI:} \href{https://doi.org/10.1007/JHEP06(2024)106}{https://doi.org/10.1007/JHEP06(2024)106} \\
\textbf{Impact factor\(^{*}\):} 5 \\
\textbf{Rank in Physics, Particles \& Fields\(^{*}\):} 6/31 (Q1) \\
\\
Article distributed under the terms of the Creative Commons\\ Attribution License (CC-BY 4.0). \\
\\
\(^{*}\)Data corresponding to 2023, from JCR. \vspace{0.5em}
}
\\
\hline
\end{tabularx}
\end{center}
\vspace*{\fill}

\clearpage
\cleardoublepage

\vspace*{\fill}
\begin{center}
\begin{tabularx}{\textwidth}{ |>{\raggedright\arraybackslash}p{0.25\textwidth}|X| }
\hline
\multicolumn{2}{|c|}{\cellcolor{green!10}
  \begin{tabular}{@{}c@{}}
    \textbf{Hot wormholes and chaos dynamics in a two-coupled SYK model} \\
    Journal of High Energy Physics \textbf{2025}, 110 (2025)
  \end{tabular}
} \\
\hline
\parbox[c]{\linewidth}{\vspace{1em}\centering\textbf{Authors}\vspace{1em}} &
\parbox[c]{\linewidth}{
\vspace{1em}
Martí Berenguer\(^{1,2}\), Javier Mas\(^{1,2}\), Juan Santos-Suárez\(^{1,2}\) and Alfonso V. Ramallo\(^{1,2}\).\\[1em]
\footnotesize
\(^{1}\) Departamento de Física de Partículas and \\
\(^{2}\) Instituto Galego de Física de Altas Enerxías \\
Universidade de Santiago de Compostela. \\
E-15782. Santiago de Compostela, Spain.
\vspace{1em}
} \\
\hline
\parbox[c]{\linewidth}{\vspace{1em}\centering\textbf{Included in}\vspace{1em}} &
\parbox[c]{\linewidth}{\vspace{1em}Chapter \ref{chap:FloquetSYK}. Floquet SYK wormholes.\vspace{1em}} \\
\hline
\parbox[c]{\linewidth}{\vspace{1em}\centering\textbf{PhD Student\\Contribution}\vspace{1em}} &
\parbox[c]{\linewidth}{\vspace{1em}Proponent of the project. Design of strategy, collaboration in numerical and analytic computations, discussions, and writing of the article.\vspace{1em}} \\
\hline
\parbox[c]{\linewidth}{\vspace{1em}\centering\textbf{Further\\Information}\vspace{1em}} &
\makecell[l]{{\vspace{-0.15cm}\tiny\transparent{0} This text is invisible and adds some blank space.} \\
\textbf{Journal:} Journal of High Energy Physics \\
\textbf{Publisher:} Springer \\
\textbf{ISSN:} 1029-8479 \\
\textbf{Year of publication:} 2025 \\
\textbf{DOI:} \href{https://doi.org/10.1007/JHEP03(2025)110}{https://doi.org/10.1007/JHEP03(2025)110} \\
\textbf{Impact factor\(^{*}\):} 5 \\
\textbf{Rank in Physics, Particles \& Fields\(^{*}\):} 6/31 (Q1) \\
\\
Article distributed under the terms of the Creative Commons\\ Attribution License (CC-BY 4.0). \\
\\
\(^{*}\)Data corresponding to 2023, from JCR. \vspace{0.5em}
}
\\
\hline
\end{tabularx}
\end{center}
\vspace*{\fill}

\clearpage
\cleardoublepage

\vspace*{\fill}
\begin{center}
\begin{tabularx}{\textwidth}{ |>{\raggedright\arraybackslash}p{0.25\textwidth}|X| }
\hline
\multicolumn{2}{|c|}{\cellcolor{green!10}
  \begin{tabular}{@{}c@{}}
    \textbf{Chiral symmetry breaking and restoration by helical magnetic fields}\\ \textbf{in AdS/CFT},
    Journal of High Energy Physics \textbf{2025}, 48 (2025)
  \end{tabular}
} \\
\hline
\parbox[c]{\linewidth}{\vspace{1em}\centering\textbf{Authors}\vspace{1em}} &
\parbox[c]{\linewidth}{
\vspace{1em}
Martí Berenguer\(^{1,2}\), Javier Mas\(^{1,2}\), Masataka Matsumoto\(^{3}\), Keiju Murata\(^{4}\) and Alfonso V. Ramallo\(^{1,2}\).\\[1em]
\footnotesize
\(^{1}\) Departamento de Física de Partículas and \\
\(^{2}\) Instituto Galego de Física de Altas Enerxías \\
Universidade de Santiago de Compostela. \\
E-15782. Santiago de Compostela, Spain.\\
\(^{3}\) Wilczek Quantum Center, School of Physics and Astronomy \\
Shanghai Jiao Tong University \\
Shanghai 200240, China. \\
\(^{4}\) Department of Physics, College of Humanities and Sciences \\
Nihon University \\
Sakurajosui, Tokyo 156-8550, Japan
\vspace{1em}
} \\
\hline
\parbox[c]{\linewidth}{\vspace{1em}\centering\textbf{Included in}\vspace{1em}} &
\parbox[c]{\linewidth}{\vspace{1em}Chapter \ref{chap:HelicalB}. Chiral symmetry breaking and restoration by helical magnetic fields.\vspace{1em}} \\
\hline
\parbox[c]{\linewidth}{\vspace{1em}\centering\textbf{PhD Student\\Contribution}\vspace{1em}} &
\parbox[c]{\linewidth}{\vspace{1em}Collaboration in numerical and analytic computations, discussions, and writing of the article.\vspace{1em}} \\
\hline
\parbox[c]{\linewidth}{\vspace{1em}\centering\textbf{Further\\Information}\vspace{1em}} &
\makecell[l]{{\vspace{-0.15cm}\tiny\transparent{0} This text is invisible and adds some blank space.} \\
\textbf{Journal:} Journal of High Energy Physics \\
\textbf{Publisher:} Springer \\
\textbf{ISSN:} 1029-8479 \\
\textbf{Year of publication:} 2025 \\
\textbf{DOI:} \href{https://doi.org/10.1007/JHEP05(2025)048}{https://doi.org/10.1007/JHEP05(2025)048}  \\
\textbf{Impact factor\(^{*}\):} 5 \\
\textbf{Rank in Physics, Particles \& Fields\(^{*}\):} 6/31 (Q1) \\
\\
Article distributed under the terms of the Creative Commons\\ Attribution License (CC-BY 4.0). \\
\\
\(^{*}\)Data corresponding to 2023, from JCR. \vspace{0.5em}
}
\\
\hline
\end{tabularx}
\end{center}
\vspace*{\fill}

%% file: Part3/Resumo.tex
\renewcommand{\chapterquote}{}
\chapter*{Resumo (en galego)}
\addcontentsline{toc}{chapter}{Resumo (en galego)}

A teoría de cordas foi proposta a finais dos anos 60 como unha posible teoría para a forza nuclear forte. Con todo, a aparición da cromodinámica cuántica (QCD) como un marco máis prometedor para describir as interaccións hadrónicas desviou a atención da teoría de cordas. Sen ser descartada, a teoría de cordas evolucionou cara a unha candidata prometedora para unha teoría consistente da gravidade cuántica, unificando as forzas fundamentais da natureza. Un fito clave nesta evolución foi a formulación da correspondencia AdS/CFT por Juan Maldacena en 1997 \cite{Maldacena_1997}.

A correspondencia AdS/CFT establece unha dualidade entre unha teoría gravitacional en un espazo-tempo asintóticamente Anti-de Sitter (AdS) de dimensión $(d+1)$ e unha teoría cuántica de campos conforme (CFT) que vive na súa fronteira (una variedade en $d$ dimensións). Esta correspondencia, coñecida tamén como dualidade gauge/gravidade, ou dualidade holográfica, revolucionou a física teórica ao proporcionar unha ferramenta poderosa para comprender sistemas cuánticos fortemente acoplados xa que, en xeral, resolver teorías cuánticas de campos fóra do réxime perturbativo resulta extremadamente complexo. Os métodos perturbativos, baseados en expansións en torno a acoplamentos pequenos, son insuficientes cando o acoplamento é forte, pois a serie perturbativa non converge e as técnicas tradicionais, como a teoría de perturbacións, deixan de ser válidas. Isto ocorre en moitos sistemas de interese físico real, desde o plasma de quarks e gluóns ata materiais fortemente acoplados en física da materia condensada.

É aquí onde a correspondencia AdS/CFT demostra a súa utilidade fundamental: ao establecer un mapa entre unha teoría cuántica fortemente acoplada e unha teoría gravitacional clásica ou semiclasica, a dualidade permite trasladar problemas intratables en física cuántica a problemas máis accesibles en gravidade clásica. Por exemplo, as dinámicas non perturbativas dunha CFT fortemente acoplada poden describirse mediante solucións gravitacionais en AdS, onde as ecuacións son máis sinxelas de resolver analiticamente ou numéricamente. Deste xeito, a correspondencia holográfica constitúe un dos escasos métodos capaces de ofrecer unha visión fiable dos efectos non perturbativos en teorías cuánticas de campos, abrindo un novo horizonte para a exploración teórica e a conexión con fenómenos experimentais.

O exemplo mellor entendido desta dualidade é a correspondencia entre a teoría $\mathcal{N}=4$ supersimétrica Yang-Mills (SYM) en catro dimensións e a teoría de cordas de tipo IIB no espazo AdS$_5\times S^5$. Esta dualidade xogou un papel central en moitos desenvolvementos teóricos durante as últimas tres décadas e será central na primeira parte desta tese, xunto con varias das súas deformacións coñecidas.

A correspondencia AdS/CFT formaliza de forma concreta o que se coñece como principio holográfico, formulado por 't Hooft e Susskind na década dos anos 1990 \cite{tHooft_1993,Susskind_1994}: a idea que unha teoría que inclúe gravidade nunha rexión do espazo-tempo pode describirse completamente por unha teoría sen gravidade nunha superficie que delimita esa rexión, e que polo tanto tén unha dimensión menos. Este principio implica que os graos de liberdade "tradicionais" dentro do volume poden ser reinterpretados como graos de liberdade na fronteira. Unha das principais motivacións para formular o principio holográfico fou a observación de Jacob Bekenstein en 1993 que a entropía dos buracos negros depende da área do horizonte de sucesos, e nón do volumo do buraco negro, como es o caso no resto de sistemas termodinámicos coñecidos. A holografía, máis alá de supoñer un cambio conceptual profundo, abre portas a aplicacións prácticas en sistemas non perturbativos, tales como plasmas cuánticos, supercondutores, sistemas fortemente acoplados, ou fenómenos críticos.

Mesmo en situacións nas que a descrición gravitacional exacta é descoñecida, a intuición holográfica demostrou ser útil. Un exemplo notable diso é o modelo Sachdev-Ye-Kitaev (SYK) \cite{KitaevTalk1,KitaevTalk2}, un sistema cuántico de $N$ fermións de Majorana que interaccionan con constantes aleatorias en $0+1$ dimensións. Nun límite adecuado de gran-$N$ e enerxía baixa, o modelo SYK é dual á gravidade de Jackiw-Teitelboim (JT) \cite{Jackiw_1984,Teitelboim_1983}, unha teoría de gravidade dilatónica en dúas dimensións. Esta dualidade fai do modelo SYK un dos poucos exemplos coñecidos dun modelo resoluble de gravidade cuántica e un potente laboratorio teórico para investigar o caos cuántico, a dinámica do entrelazamento e a difusión de información en sistemas fortemente acoplados. A segunda parte desta tese estará dedicada ao estudo do modelo SYK e unha extensión particularmente interesante que admite unha interpretación gravitacional en termos dun buraco de miñoca atravesable \cite{Maldacena_2018}.

Un tema central ao longo desta tese é o estudo da dinámica fora de equilibrio en sistemas cuánticos fortemente acoplados, especialmente baixo excitacións periódicas. A física dos sistemas cuánticos excitados periódicamente foi obxecto de estudo intensivo nos últimos anos (ver \cite{Bukov_2014,Eckardt_2016,Weinberg_2016,Holthaus_2015}, e referencias nas mesmas). Diversas razóns apoian este interese, sendo unha delas de orixe tecnolóxica: a posibilidade de manipular sistemas cuánticos dunha maneira controlada mediante campos externos dependentes do tempo. Este enfoque denomínase en xeral enxeñaría de Floquet \cite{Oka_2018,Rudner_2020}.  A base de todo isto apóiase no teorema de Floquet, un análogo temporal do teorema de Bloch, que describe a evolución de sistemas cuánticos periódicos no tempo. O setup artificial implica principalmente a irradiación do sistema cun laser polarizado circularmente, ou movéndoo ao redor. Con excitacións periódicas adecuadas, creáronse novas fases de materiais cuánticos e xurdiron fenómenos de non equilibrio. Exemplos inclúen a superconductividade inducida pola luz \cite{Fausti_2011,Mitrano_2016}, illantes topolóxicas de Floquet \cite{Oka_2010,Lindner_2010,Kitagawa_2011,Cayssol_2012,Rechtsman_2012,Wang_2013,Jotzu_2014,Dehghani_2015}, e semimetais de Weyl artificiais \cite{Hubener_2016,Zhang_2016,Bucciantini_2016}.

De particular interese é o caso das solucións onde a inxección de enerxía e a disipación equilíbranse, chegando a un estado estacionario fora de equilibrio de tipo Floquet (NESS, do inglés \textit{Non Equilibrium Steady State}). A correspondencia AdS/CFT ofrece potentes técnicas computacionais para estudar estes sistemas desde unha perspectiva complementaria á habitual na física da materia condensada. Ao aplicar métodos holográficos a sistemas fortemente acoplados e dependentes do tempo, pódese acceder a rexións non-perturbaivas que de outro xeito serían difíciles de explorar. Ao mesmo tempo, o crecemento do interese en fenómenos complexos e fora do equilibrio na física de moitos corpos motiva un maior desenvolvemento da correspondencia AdS/CFT máis alá das súas aplicacións tradicionais de equilibrio. Estes desafíos axudan a probar os seus límites e suxiren novas direccións nas que a holografía pode ser extendida, facendo que sexa máis adecuada para captar dinámicas en tempo real en sistemas fortemente acoplados.

O obxectivo desta tese é explorar o papel das excitacións periódicas en dous tipos de sistemas holográficos:

O primeiro tipo implica construcións "top-down" baseadas nas interseccións de branas D3/D5 e D3/D7, onde se introducen campos eléctricos e magnéticos, xeralmente dependentes do tempo. Estes modelos teñen a vantaxe de que a dualidade gravitacional da teoría de campos está explícitamente coñecida dentro da teoría de cordas. Isto permite investigar a resposta do sistema a campos externos en detalle (e nalgúns casos analíticamente), incluso máis alá do réxime da teoría de resposta lineal.

O segundo tipo de sistemas baséase nunha versión do modelo SYK que implica dous copias acopladas, coñecida como o modelo de Maldacena-Qi \cite{Maldacena_2018}, que exhibe comportamento de buraco de miñoca atravesable a baixas temperaturas. A motivación para estudar este problema é dobre. Por un lado, interesa estudar como responde esta fase gravitacional ante perturbacións externas que son periódicas no tempo, cunha esperanza de que isto poida arrollar luz sobre a súa microestrutura gravitacional cuántica. Por outro lado, explórase se as excitacións periódicas poderían contribuír a unha mellor comprensión, e potencial mellora, dos protocolos de teleportación inspirados en buracos de miñoca que recentemente captaron o interese da comunidade de computación cuántica \cite{Brown_2019, Nezami_2021}.

En resumo, esta tese ten como obxectivo afondar na comprensión dos sistemas cuánticos fortemente acoplados e excitados periódicamente, facendo uso das técnicas da correspondencia AdS/CFT e da holografía.

A estrutura da tese divídese en tres partes principais.

A Parte I da tese ofrece unha revisión profunda dos cimentos sobre os que se constrúe. O Capítulo \ref{chap:AdSCFT} ofrece unha introdución amena e detallada á correspondencia AdS/CFT, comezando polo seu punto de orixe: a teoría de cordas. Comezamos repasando os aspectos fundamentais da corda bosónica antes de avanzar cara á teoría de supercordas. Neste contexto, as D-branas emerxen como obxectos fundamentais, onde as cordas abertas terminan, e como nelas xorden teorías de gauge, que desempeñarán un papel central ao longo desta tese. A continuación, abordamos o límite de desacoplamento e os argumentos que levaron a Juan Maldacena, no ano 1997, a formular a correspondencia AdS/CFT. O capítulo conclúe con varios detalles que fan máis precisa a dualidade, comezando polas simetrías en ambos lados da dualidade, a relación entre o acoplamiento forte na teoría gravitatoria e o acoplamiento débil na teoría dual, e cunha visión xeral de como se poden calcular observables da teoría de campos a partir da descrición gravitacional dual.

O Capítulo \ref{chap:Branes} continúa nesta liña, pero xa introducimos os modelos holográficos concretos que se empregarán ao longo das seccións de investigación da tese. Comezamos repasando a incorporación dos graos de liberdade de sabor na correspondencia AdS/CFT, centrando a atención en dous ensamblaxes específicos de branas: as interseccións D3/D7 e D3/D5. Neste contexto, reformulamos o límite de desacoplamento da sección anterior e adaptámolo no caso no que temos grados de liberdade de sabor, para formular a versión xeralizada da dualidade.  Tamén revisamos o marco da renormalización holográfica e o dicionario holográfico, que son ferramentas esenciais para extraer as cantidades da teoría de campos a partir dos datos gravitacionais do volume. Na segunda parte do capítulo, discutimos o impacto da introdución de campos gauge  no volumo das branas de sabor. Repasamos varios aspectos relacionados coa presenza de campos eléctricos na Sección \ref{sec:metallicAdSCFT} e campos magnéticos na Sección \ref{sec:CSB}. Estes resultados proporcionan a base para as análises presentadas nos capítulos \ref{chap:FloquetII} e \ref{chap:HelicalB}.

O Capítulo \ref{chap:SYKwormholes} afástase dos temas dos dous capítulos anteriores e cambia o foco cara aos buracos de miñoca atravesables e ao modelo SYK. Comezamos repasando o mecanismo de Gao-Jafferis-Wall, que mostra como un acoplamento directo entre as fronteiras asintóticas dun espazo AdS pode facer que certos buracos de miñoca sexan atravesables \cite{Gao_2016}. A continuación, presentamos o modelo SYK, destacando as súas características principais e a súa conexión coa gravidade en espazos case-AdS$_2$. Fai especial fincapé no sistema SYK acoplado proposto por Maldacena e Qi, que ofrece unha realización concreta dun buraco de miñoca atravesable \cite{Maldacena_2018}. Finalmente, introducimos o formalismo de Schwinger-Keldysh como ferramenta para estudar a dinámica en tempo real e fóra do equilibrio do modelo, que xogará un papel central nos capítulos de investigación desta tese.

A Parte II contén a investigación orixinal. O Capítulo \ref{chap:FloquetII} presenta o estudo da intersección de branas D3/D5 a temperatura finita baixo un campo eléctrico periódico e dependente do tempo. O sistema presenta un diagrama de fases fora do equilibrio, con fases condutoras e illantes. A excitación externa induce unha corrente rotatoria debido á polarización do baleiro (na fase illante) e ao efecto Schwinger (na fase condutora). Para certos valores particulares da frecuencia de excitación, o campo externo entra en resonancia cos mesóns vectoriais do modelo, e pode producirse unha corrente rotatoria mesmo no límite de campo externo nulo; isto define os \textit{condensados de Floquet de mesóns vectoriais}. Para todas as temperaturas, en determinadas frecuencias intercaladas, atopamos novos estados duais que denominamos \textit{puntos de supresión de Floquet}, nos que a polarización do baleiro desaparece mesmo en presenza dun campo eléctrico. A partir dos datos inferimos que estes estados existen tanto na fase condutora como na illante. No límite sen masa atopamos unha lei de condutividade lineal e instantánea, recuperando os resultados comentados na Sección \ref{subsec:nonlinearsigma}. Tamén examinamos as correntes fotovoltaicas AC e DC como resposta a un campo eléctrico oscilante de proba, e observamos que o aumento da temperatura suprime a corrente fotovoltaica de Hall.

Unha simplificación crucial que permite a resolución do sistema xorde ao pasar a un sistema de referencia xiratorio. Neste sistema, a acción vólvese independente do tempo, e as ecuacións de movemento resultantes son ecuacións diferenciais ordinarias. Isto permítenos aplicar as técnicas desenvolvidas en \cite{Karch_2007}, onde as funcións de un punto obtéñense a partir da condición de que a acción on-shell permaneza real. Aínda que este mecanismo de regularidade no infravermello se utilizou amplamente para configuracións estáticas, a súa aplicación a fontes dependentes do tempo segue a ser limitada. En particular, en dúas dimensións espaciais e para sabores sen masa, sábese que a corrente responde de forma lineal e instantánea ao campo eléctrico: $J(t)=\sigma E(t)$ \cite{Karch_2010}. Os nosos resultados son consistentes con esta observación no réxime sen masa e de masa pequena.

Con todo, antes de entrar no modelo SYK, presentamos o Capítulo \ref{chap:HelicalB}, que continúa as ideas usadas no Capítulo anterior considerando un setup similar para estudar un modelo fenomenolóxico de ruptura de simetría quiral, unha característica fundamental da QCD a baixa enerxía\cite{Bergman_2012}. No marco de AdS/QCD, este fenómeno pode modelarse de maneira robusta. Por exemplo, é ben sabido que un campo magnético uniforme induce a ruptura da simetría quiral \cite{Filev_2007,Filev_2007_2}.

Porén, os experimentos de colisións de ións pesados suxiren que os campos magnéticos xerados neses ambientes non son estáticos nin uniformes. En particular, identificáronse os campos magnéticos helicoidais como unha configuración relevante \cite{Skokov_2009,Voronyuk_2011,Bzdak_2011,Deng_2012}. A interacción entre estes campos magnéticos non uniformes e a dinámica da ruptura e restauración da simetría quiral segue sendo pouco comprendida. Por tanto, desenvolver métodos para incorporar estes campos magnéticos con variación espacial dentro do formalismo de AdS/QCD é fundamental para acadar unha comprensión máis profunda da dinámica da QCD no mundo real.

A teoría de interese neste caso é a $\mathcal{N}=4$ SYM en $(3+1)$ dimensións, con grupo de gauge $SU(N_c)$, acoplada a un número $N_f \ll N_c$ de hiper-multipletes de sabor $\mathcal{N}=2$ na representación fundamental de $SU(N_c)$. Holograficamente, este sistema modélase como $N_f$ D7-branas de proba embebidas na xeometría AdS$_5\times S^5$ xerada por un apilado de $N_c$ D3-branas coincidentes \cite{KarchKatz_2002}.

Adicionalmente, incluímos un campo axial $U(1)$ do tipo $A_j^5=b/2\h\delta_{jz}$, que describe un semimetal de Weyl. A realización holográfica deste escenario foi desenvolvida en \cite{Fadafan_2020}. Analizando os encaixes das branas, obtemos tres tipos de solucións sen masa, que corresponden a tres fases con comportamento distinto na teoría de campos dual. A partir do estudo dos condensados de quarks, da enerxía libre e das correntes eléctricas, atopamos que os campos magnéticos helicoidais poden contrarrestar a ruptura de simetría inducida por un campo uniforme, conducindo o sistema cara á restauración da simetría. Tamén atopamos un efecto análogo ao efecto magnético quiral, no que a corrente é paralela ao campo magnético. Estudamos ademais o caso con masa, e atopamos que a configuración helicoidal é menos eficaz á hora de suprimir a transición de fase de primeira orde que está presente no caso dun campo magnético constante.

Este capítulo desvíase lixeiramente do foco principal desta tese, no sentido de que é o único capítulo no que non tratamos un sistema fóra do equilibrio. Porén, foi o ansatz xiratorio para o campo gauge do Capítulo \ref{chap:FloquetII} o que motivou o campo magnético helicoidal que introducimos aquí.

Finalmente, no Capítulo \ref{chap:FloquetSYK} estudamos a dinámica fóra do equilibrio do modelo de Maldacena-Qi, consistente de dous modelos SYK acoplados, que se conxectura que es dual a un buraco de miñoca eterno e atravesable en AdS$_2$.

Concretamente, inducimos unha oscilación temporal sinusoidal no acoplamento relevante, $\mu$. Tamén exploramos excitacións temporais nas constantes de interacción $J_{L,R}$ de cada modelo SYK por separado. Aínda que estas últimas teñen unha motivación física menos clara (xa que implican unha varianza dependente do tempo na distribución dos acoplos), proporcionan unha visión comparativa útil.

No caso da excitación en $\mu$, o comportamento global é amplamente consistente coas expectativas. Porén, dentro da fase de buraco de miñoca, obsérvase unha amplificación significativa do quecemento cando se excita o sistema en certas frecuencias propias do sistema non excitado. Estas resonancias coinciden coa metade do espectro conforme do modelo sen excitación, e proporcionamos evidencia adicional dunha resonancia asociada coa frecuencia natural do gravitón na fronteira, que non se observara nas simulacións en equilibrio de gran $N$ do modelo.

Os nosos resultados numéricos suxiren que o sistema pode ser levado dinámicamente cara ao unha fase inestable do diagrama de fases, chamado "buraco de miñoca quente". Este réxime é pouco coñecido, e un dos nosos obxectivos é obter nova información sobre as súas propiedades físicas. Por exemplo, aínda que os expoñentes de Lyapunov das dúas fases estables foron calculados en \cite{Nosaka_2020}, os da fase de buraco de miñoca quente permanecían descoñecidos. Usando protocolos de non equilibrio, accedemos a este réxime e calculamos os seus expoñentes de caos.

Isto conséguese mediante dous protocolos fóra do equilibrio. O primeiro, introducido en \cite{Maldacena_2019}, consiste en arrefriar unha solución de buraco negro a alta temperatura mediante o acoplamento a un baño frío. Este proceso de arrefriamento permite unha evolución cuasiestática que transita pola fase inestable. O segundo método consiste na excitación en $\mu$ mencionada anteriormente, que inxecta enerxía nunha solución de buraco de miñoca a baixa temperatura, permitindo así acadar a fase de buraco de miñoca quente. Esta segunda opción ten a vantaxe de preservar as propiedades intrínsecas do sistema orixinal, xa que non require a acción dun baño externo. Ao excitar o parámetro de acoplamento, descubrimos unha estrutura máis rica de posibles estados finais dentro da fase inestable. Estes achados confírmanse mediante un terceiro protocolo fóra do equilibrio que combina os dous anteriores. Ademais, ofrecemos unha interpretación cualitativa dos fenómenos observados empregando a teoría efectiva Schwarziana, que arroxa luz sobre a dinámica desta fase exótica.

Todos estes resultados foron obtenidos cunha combinación de métodos analíticos e numéricos. En varios casos, podemos obter resultados exactos mediante métodos analíticos. Non obstante, moitos dos sistemas considerados, especialmente aqueles que involucran dinámica dependente do tempo, requiren un esforzo numérico significativo. Para iso, empregouse amplamente Wolfram Mathematica, tanto para manipulación simbólica como para a resolución numérica de ecuacións diferenciais.

Na parte da tese dedicada ao estudo do modelo SYK, utilizouse un segundo software, en paralelo con Mathematica. Trátase da biblioteca NESSi (Non-Equilibrium Systems Simulation \cite{Schuler_2019}), adaptada para resolver as ecuacións de Schwinger-Dyson en sistemas fóra de equilibrio. Estos métodos numéricos son altamente versátiles e poden adaptarse facilmente a unha ampla clase de sistemas cuánticos, moitos dos cales se espera que requiran significativamente menos recursos computacionais que a fase de buraco de miñoca analizada aquí. Só un puñado de estudos na literatura teñen realizado este tipo de simulación en tempo real, e os que existen normalmente non alcanzan as rexións máis interesantes do diagrama de fases debido ao crecemento exponencial do custo computacional. O uso combinado de varias ferramentas avanzadas, incluíndo o método predictor-corrector detallado no Apéndice \ref{app:numerics}, a biblioteca NESSi, e os recursos de computación de alto rendemento proporcionados por CESGA (Centro de Supercomputación de Galicia), foi crucial para acadar estes resultados, que serían difíciles de reproducir sen esta infraestrutura.

A tese conclúe cunha parte final que resume os principais resultados e describe posibles direccións para futuras investigacións. Tamén inclúe a lista detallada das catro publicacións nas que está baseada a parte de investigación desta tese.